\pdfoutput=1

\def\CompactJudgments{0}

\def\OPTIONAppendix{0}

\def\OPTIONArxiv{0}
\def\OPTIONLoudLabels{0}
\gdef\OPTIONArxiv{1}

\def\OPTIONConf{1}
\ifnum\OPTIONConf=1%
     \documentclass[
       authorversion=true,
       nonacm=true,
       acmsmall,
       review=false,
       acmthm=true%
       ]{acmart}
     \setcopyright{none}
     \def~{\nobreakspace{}}

\usepackage{srcltx}

\usepackage{jdunfield}
\usepackage{llproof}
\usepackage{rulelinks}
\usepackage[OT2,T1]{fontenc}

\usepackage{relsize}

\usepackage{amsmath,amsthm}
\usepackage{thmtools,thm-restate}

\usepackage{enumerate}

\usepackage{pgf,tikz}
\usetikzlibrary{shapes,arrows,matrix,fit,backgrounds,calc,decorations,decorations.pathmorphing,decorations.markings}

\usepackage{breakurl}
\else
\documentclass[10pt,leqno]{article}
\fi

\ifnum\OPTIONConf=1%
  \setcitestyle{nosort}    %
\else
  \usepackage[authoryear]{natbib}
  \bibpunct{(}{)}{;}{a}{}{,}
\fi

\ifnum\OPTIONConf=1%
\else%
\fi

\usepackage{graphics,pstricks,color}

\usepackage{listings}

\usepackage{pgf,tikz}
\usetikzlibrary{shapes,arrows,matrix,fit,backgrounds,calc,decorations,decorations.pathmorphing}

\usepackage{xspace}
\ifnum\OPTIONConf=1%
\else
    \usepackage{goodcharter}
\fi
\usepackage{multirow}
\usepackage{stmaryrd}
\usepackage{comment}
\usepackage{mathpartir}
\usepackage{framed}
\usepackage{rotating}

\let\MathRightArrow\Rightarrow %
\usepackage{marvosym} %
\def\Rightarrow{\MathRightArrow}

\usepackage{srcltx}
\usepackage{fancyhdr}
\usepackage{enumerate}
\usepackage{relsize}

\usepackage{booktabs}

\def\MathparLineskip{\lineskiplimit=1.0em\lineskip=0.7em plus 0.1em}  %

\usepackage{friendlyappendix}
\usepackage[T1]{fontenc}
\usepackage{pifont}

\usepackage{semantic}

\usepackage{xparse}

\usepackage{mathtools}

\usepackage{tikz-cd} 

\usepackage{array}
\newcolumntype{L}{>{$}l<{$}}

\newcommand{\chkcolor}{dBlue}
\newcommand{\syncolor}{dRed}
\newcommand{\appcolor}{dDkGreen}
\newcommand{\chk}{\mathrel{\mathcolor{\chkcolor}{\Leftarrow}}}
\newcommand{\uncoloredsyn}{{\Rightarrow}}
\newcommand{\syn}{\mathrel{\mathcolor{\syncolor}{\uncoloredsyn}}}

\newcommand{\app}{\mathrel{\mathcolor{\appcolor}{{\uncoloredsyn}\hspace{-1.2ex}{\uncoloredsyn}}}}

\newcommand{\Booltype}{\mathbb{B}}

\newcommand{\downshiftsym}{\downarrow}
\newcommand{\downshift}[1]{{\downshiftsym\!{#1}}}
\newcommand{\upshiftsym}{\uparrow}
\newcommand{\upshift}[1]{{\upshiftsym\!{#1}}}
\newcommand{\pupshift}[1]{{\upharpoonleft\!{#1}}}

\newcommand{\roll}[1]{\textkw{into}({#1})}
\let\into\roll
\newcommand{\unit}{\left\langle\right\rangle}
\newcommand{\thunk}[1]{\left\{{#1}\right\}}
\newcommand{\rec}[2]{\textkw{rec}\;{#1}.\;{#2}}

\newcommand{\Fold}[2]{(\mathsf{fold}_{#1}\;{#2})}

\newcommand{\Ptype}{\hat{P}}
\newcommand{\Btype}{B}
\newcommand{\Const}[1]{\underline{#1}}
\newcommand{\Id}{\mathrm{Id}}

\newcommand{\setof}[1]{{\left\{#1\right\}}}
\newcommand{\comprehend}[2]{\setof{{#1} \;\middle|\; {#2}}}

\newcommand{\alltype}[1]{\AllSym {#1}.\:}
\newcommand{\extype}[1]{\ExistsSym {#1}.\:}
\renewcommand{\implies}{\mathrel{\supset}}
\newcommand{\with}{\mathrel{\land}}

\newcommand{\ground}[1]{{#1}\,{\mathsf{ground}}}

\newcommand{\judgeequiv}[4][]{{#2} \entails {#3} \equiv^{#1} {#4}}
\newcommand{\judgeextract}[5][]{{#2} \entails {#3} \rightsquigarrow^{#1} {#4}\;[{#5}]}
\newcommand{\judgesub}[4][]{{#2} \entails {#3} \leq^{#1} {#4}}
\newcommand{\judgeentail}[2]{{#1} \entails {#2} \;\mathsf{true}}
\newcommand{\judgechkval}[4]{{#1}; {#2} \entails {#3} <= {#4}}
\newcommand{\judgesynhead}[4]{{#1}; {#2} \entails {#3} => {#4}}
\newcommand{\judgechkexp}[4]{{#1}; {#2} \entails {#3} <= {#4}}
\newcommand{\judgesynexp}[4]{{#1}; {#2} \entails {#3} => {#4}}
\newcommand{\judgechkmatch}[5]{{#1}; {#2}; [{#3}] \entails {#4} <= {#5}}
\newcommand{\judgespine}[5]{{#1}; {#2}; [{#4}] \entails {#3} \gg {#5}}

\newcommand{\semideclchkval}[5]{{#1}; {#2} \;{\widetilde{\entails}}\; {#3} <= {#4} \;/\; {#5}}
\newcommand{\semideclsynhead}[4]{{#1}; {#2} \;{\widetilde{\entails}}\; {#3} => {#4}}
\newcommand{\semideclchkexp}[4]{{#1}; {#2} \;{\widetilde{\entails}}\; {#3} <= {#4}}
\newcommand{\semideclsynexp}[4]{{#1}; {#2} \;{\widetilde{\entails}}\; {#3} => {#4}}
\newcommand{\semideclchkmatch}[5]{{#1}; {#2}; [{#3}] \;{\widetilde{\entails}}\; {#4} <= {#5}}
\newcommand{\semideclspine}[6]{{#1}; {#2}; [{#4}] \;{\widetilde{\entails}}\; {#3} \gg {#5} \;/\; {#6}}
\newcommand{\semideclneg}[3]{{#1}; {#2} \;{\widetilde{\lhd}}\; {#3}}

\newcommand{\unrefchkval}[3]{{#1} \entails {#2} <= {#3}}
\newcommand{\unrefsynhead}[3]{{#1} \entails {#2} => {#3}}
\newcommand{\unrefchkexp}[3]{{#1} \entails {#2} <= {#3}}
\newcommand{\unrefsynexp}[3]{{#1} \entails {#2} => {#3}}

\newcommand{\unrefchkmatch}[4]{{#1}; [{#2}] \entails {#3} <= {#4}}
\newcommand{\unrefspine}[4]{{#1}; [{#3}] \entails {#2} \gg {#4}}

\newcommand{\slashsym}{/}
\newcommand{\slashop}{\mathrel{\slashsym}}
\newcommand{\algspine}[7]{{#1}; {#2}; [{#4}] |- {#3} \gg {#5} \slashop {#6} -| {#7}}
\newcommand{\algchkmatch}[5]{{#1}; {#2}; [{#3}] \rhd {#4} <= {#5}}

\newcommand{\algchk}[6]{{#1}; {#2} |- {#3} <= {#4} \slashop {#5} -| {#6}}

\newcommand{\algchkneg}[4]{{#1}; {#2} \rhd {#3} <= {#4}}
\newcommand{\algneg}[3]{{#1}; {#2} \lhd {#3}}
\newcommand{\alginst}[3]{{#1} |- {#2}\;\mathsf{inst} -| {#3}}

\newcommand{\algsynhead}[4]{{#1}; {#2} \rhd {#3} => {#4}}
\newcommand{\algsynexp}[4]{{#1}; {#2} \rhd {#3} => {#4}}

\newcommand{\algpropequivinst}[4]{{#1} |- {#2} \equiv {#3} \;\mathsf{inst} -| {#4}}
\newcommand{\algequiv}[6][]{{#2} \entails {#3} \equiv^{#1} {#4} \slashop {#5} -| {#6}}
\newcommand{\algsubtypesym}{\texttt{\upshape\selectfont<:}}
\newcommand{\algsubtype}{\mathrel{\algsubtypesym}}
\newcommand{\algsub}[6][]{{#2} \entails {#3} \algsubtype^{#1} {#4} \slashop {#5} -| {#6}}

\newcommand{\entailwah}[2]{{#1} \models {#2}}

\newcommand{\resjudgesub}[4][]{{#2} \entails {#3} \algsubtype^{#1} {#4}}

\newcommand{\semideclequiv}[5][]{{#2} \entails {#3} \equiv^{#1} {#4} \slashop {#5}}
\newcommand{\semideclsub}[5][]{{#2} \entails {#3} \algsubtype^{#1} {#4} \slashop {#5}}

\newcommand{\semideclentailwah}[2]{{#1} \rhd {#2}}
\newcommand{\wahequiv}[3]{{#1} \rhd {#2} \leftrightarrow {#3}}
\newcommand{\chiequiv}[4]{{#1}; {#2} \lhd {{#3} \leftrightarrow {#4}}}

\NewDocumentCommand{\judgeunroll}{smmmmmmmm}{
  \IfBooleanTF{#1}{
       {#2}; {#3} \entails \arrayenvl{
       \left\{ {#4} \;\mid\; {#5} \left({#6}\right) =_{#9} {#7} \right\} 
        \\
        \circeq {#8}
      }
    }{
      {#2}; {#3} \entails \left\{ {#4} \;\mid\; {#5} \left({#6}\right) =_{#9} {#7} \right\} 
        \circeq {#8}
    }
}
\newcommand{\unroll}[9]{{#1}; {#2} |- \mathsf{unroll}_{{#4}; {#6}}({#3}; {#5}; {#7}; {#8}) \circeq {#9}}

\newcommand{\unrefunroll}[3]{\entails {#1}\!\left[ \mu{#2} \right] \circeq {#3}}

\newcommand{\judgectx}[2]{{#1} \entails {#2}\; \mathsf{ctx}}
\newcommand{\judgewf}[2]{{#1} \entails {#2}\; \mathsf{wf}}
\newcommand{\judgctx}[1]{{#1}\; \mathsf{ctx}}

\newcommand{\judgeterm}[3]{{#1} \entails {#2} : {#3}}
\newcommand{\judgetp}[3]{{#1} \entails {#2}\; \mathsf{type} {[{#3}}]}
\newcommand{\judgefunctor}[3]{{#1} \entails {#2}\; \mathsf{functor} {[{#3}]}}

\newcommand{\judgealgebra}[5]{{#1}; {#2} \entails {#3} : {#4}({#5}) \Rightarrow {#5}}
\newcommand{\unreftp}[1]{\entails {#1}\; \mathsf{type}}
\newcommand{\unreffunctor}[1]{\entails {#1}\; \mathsf{functor}}

\newcommand{\composeinj}[3]{{#2} \circ \judgeinj{#1} \circeq {#3}}

\newcommand{\ctxoutsym}{\ifnum\CompactJudgments=1%
     \dashv%
  \else%
     \dashv%
  \fi}

\newcommand{\Z}{\mathbb{Z}}

\newcommand{\FV}[1]{\mathsf{FV}(#1)}

\newcommand{\id}{\operatorname{\mathit{id}}}
\newcommand{\fold}[2]{\mathit{fold}_{#1}\;{#2}}
\newcommand{\foldn}[3]{\mathit{fold}^{#1}_{#2}\;{#3}}
\newcommand{\fmap}[2]{\mathit{fmap}\;{#1}\;{#2}}
\newcommand{\minj}[1]{\operatorname{\mathit{inj}}_{#1}}
\newcommand{\minto}{\operatorname{\mathit{into}}}

\newcommand{\kindnat}{\mathbb{N}}

\newcommand{\xsucc}{\mathsf{succ}}
\renewcommand{\succ}[1]{\xsucc \; {#1}}

\newcommand{\unitexp}{\text{\normalfont \tt()}}

\newcommand{\unitty}{\tyname{1}}

\newcommand{\pair}[2]{\langle{#1}, {#2}\rangle}

\newcommand{\fst}[1]{\pi_1\,{#1}}
\newcommand{\snd}[1]{\pi_2\,{#1}}
\newcommand{\wild}{\top}

\newcommand{\injop}[1]{\textkw{inj}_{#1}}

\newcommand{\inj}[1]{\injop{#1}\,}
\newcommand{\inl}{\inj{1}}
\newcommand{\inr}{\inj{2}}
\newcommand{\judgeinj}[1]{\mathsf{inj}_{#1}}

\newcommand{\True}{\mathsf{tt}}
\newcommand{\False}{\mathsf{ff}}

\newcommand{\naive}{na\"ive\xspace}

\newcommand{\xdom}{\mathit{dom}}
\newcommand{\dom}[1]{\xdom(#1)}

\ifnum\OPTIONConf=1%
  \newenvironment{bnfarray}%
             {\begin{center} ~\!\!\begin{array}[t]{lr@{~~}c@{~~}ll}}%
             {\end{array}\end{center}}
\else
             {\begin{center} ~\!\!\begin{array}[t]{lr@{~~}c@{~~}ll}}%
             {\end{array}\end{center}}
\fi

\newcommand{\derives}{\mathrel{::}}

\newcommand{\AllSym}{\forall}
\newcommand{\ExistsSym}{\exists}

\newcommand{\fun}[1]{\lam{#1}}

\newcommand{\Let}[2]{\textkw{let}\;{#1}\,\texttt{=}\,{#2}\textkw{;}\;}

\newcommand{\Return}[1]{\textkw{return}\,{#1}}
\newcommand{\clauses}[4]{\{\clause{{#1}_{#3}}{{#2}_{#3}}\}_{{#3} \in {#4}}}
\newcommand{\match}[2]{\textkw{match}\;{#1}\;{#2}}

\newcommand{\pack}[2]{{\textkw{pack}}{\left(#1, #2\right)}}

\newtheoremstyle{better}%
   {\topsep}%
   {\topsep}%
   {\upshape\slshape}%
   {}%
   {\bfseries}%
   {.}%
   {0.7em}%
   {}%
\theoremstyle{better}
\newtheorem*{lemma*}{Lemma}
\newtheorem*{definition*}{Definition}

\newcommand{\colourpos}{{\mathcolor{dPurple}{\text{\bf+}}}}
\newcommand{\colourneg}{{\mathcolor{dBlue}{\text{\sf\bfseries\selectfont--}}}}

\newcommand{\PrpEquivRule}[1]{Prp\ensuremath{{\equiv}{#1}}}

\newrulecommand{PrpEquivT}{\PrpEquivRule{\True}}
\newrulecommand{PrpEquivF}{\PrpEquivRule{\False}}
\newrulecommand{PrpEquivAnd}{\PrpEquivRule{\land}}
\newrulecommand{PrpEquivOr}{\PrpEquivRule{\lor}}
\newrulecommand{PrpEquivNot}{\PrpEquivRule{\neg}}
\newrulecommand{PrpEquivEq}{\PrpEquivRule{=}}
\newrulecommand{PrpEquivLeq}{\PrpEquivRule{\leq}}

\newcommand{\CtxEquivRule}[1]{Ctx\ensuremath{\equiv}{#1}}
\newrulecommand{CtxEquivEmpty}{\CtxEquivRule{Empty}}
\newrulecommand{CtxEquivVar}{\CtxEquivRule{Var}}
\newrulecommand{CtxEquivProp}{\CtxEquivRule{Prop}}

\newcommand{\FunctorEquivRule}[1]{Functor\ensuremath{\equiv}{#1}}

\newrulecommand{FunEquivConst}{\FunctorEquivRule{Const}}
\newrulecommand{FunEquivId}{\FunctorEquivRule{\ensuremath{\Id}}}
\newrulecommand{FunEquivUnit}{\FunctorEquivRule{\ensuremath{I}}}
\newrulecommand{FunEquivProd}{\FunctorEquivRule{\ensuremath{\otimes}}}
\newrulecommand{FunEquivSum}{\FunctorEquivRule{\ensuremath{\oplus}}}

\newcommand{\TpEquivPosRule}[1]{Tp\ensuremath{{\equiv}^\colourpos{#1}}}
\newcommand{\TpEquivNegRule}[1]{Tp\ensuremath{{\equiv}^\colourneg{#1}}}

\newrulecommand{TpEquivPosWith}{\TpEquivPosRule{{\land}}}
\newrulecommand{TpEquivPosEx}{\TpEquivPosRule{{\exists}}}
\newrulecommand{TpEquivPosFix}{\TpEquivPosRule{\mu}}
\newrulecommand{TpEquivPosVoid}{\TpEquivPosRule{0}}
\newrulecommand{TpEquivPosUnit}{\TpEquivPosRule{1}}
\newrulecommand{TpEquivPosProd}{\TpEquivPosRule{\times}}
\newrulecommand{TpEquivPosSum}{\TpEquivPosRule{+}}
\newrulecommand{TpEquivPosDownshift}{\TpEquivPosRule{\downarrow}}
\newrulecommand{TpEquivNegUpshift}{\TpEquivNegRule{\uparrow}}
\newrulecommand{TpEquivNegImp}{\TpEquivNegRule{{\implies}}}
\newrulecommand{TpEquivNegAll}{\TpEquivNegRule{{\forall}}}
\newrulecommand{TpEquivNegArrow}{\TpEquivNegRule{->}}

\newcommand{\ExtractPosRule}[1]{\ensuremath{{\rightsquigarrow}^\colourpos{#1}}}
\newcommand{\ExtractNegRule}[1]{\ensuremath{{\rightsquigarrow}^\colourneg{#1}}}

\newrulecommand{ExtractStopPos}{\ExtractPosRule{\neq}}
\newrulecommand{ExtractWith}{\ExtractPosRule{\land}}
\newrulecommand{ExtractEx}{\ExtractPosRule{\exists}}
\newrulecommand{ExtractProd}{\ExtractPosRule{\times}}
\newrulecommand{ExtractStopNeg}{\ExtractNegRule{\neq}}
\newrulecommand{ExtractImp}{\ExtractNegRule{\implies}}
\newrulecommand{ExtractAll}{\ExtractNegRule{\forall}}
\newrulecommand{ExtractArrow}{\ExtractNegRule{\to}}

\newcommand{\SubPosRule}[1]{\ensuremath{{\leq}^\colourpos{#1}}}
\newcommand{\SubNegRule}[1]{\ensuremath{{\leq}^\colourneg{#1}}}

\newrulecommand{DeclSubPosRefl}{\SubPosRule{\textsc{Refl}}}
\newrulecommand{DeclSubPosWithL}{\SubPosRule{{\land}\textsc{L}}}
\newrulecommand{DeclSubPosWithR}{\SubPosRule{{\land}\textsc{R}}}
\newrulecommand{DeclSubPosL}{\SubPosRule{{\rightsquigarrow}\textsc{L}}}
\newrulecommand{DeclSubPosExL}{\SubPosRule{{\exists}\textsc{L}}}
\newrulecommand{DeclSubPosExR}{\SubPosRule{{\exists}\textsc{R}}}
\newrulecommand{DeclSubPosFix}{\SubPosRule{\mu}}

\newrulecommand{DeclSubNegRefl}{\SubNegRule{\textsc{Refl}}}
\newrulecommand{DeclSubNegImpL}{\SubNegRule{{\implies}\textsc{L}}}
\newrulecommand{DeclSubNegR}{\SubNegRule{{\rightsquigarrow}\textsc{R}}}
\newrulecommand{DeclSubNegImpR}{\SubNegRule{{\implies}\textsc{R}}}
\newrulecommand{DeclSubNegAllL}{\SubNegRule{{\forall}\textsc{L}}}
\newrulecommand{DeclSubNegAllR}{\SubNegRule{{\forall}\textsc{R}}}
\newrulecommand{DeclSubNegArrow}{\SubNegRule{->}}

\newrulecommand{DeclSubPosVoid}{\SubPosRule{0}}
\newrulecommand{DeclSubPosUnit}{\SubPosRule{1}}
\newrulecommand{DeclSubPosProd}{\SubPosRule{\times}}
\newrulecommand{DeclSubPosSum}{\SubPosRule{+}}
\newrulecommand{DeclSubPosDownshift}{\SubPosRule{\downarrow}}
\newrulecommand{DeclSubNegUpshift}{\SubNegRule{\uparrow}}

\newcommand{\ResSubPosRule}[1]{\ensuremath{{\algsubtype}^\colourpos{#1}}}
\newcommand{\ResSubNegRule}[1]{\ensuremath{{\algsubtype}^\colourneg{#1}}}

\newrulecommand{ResDeclSubPosWithR}{\ResSubPosRule{{\land}\textsc{R}}}
\newrulecommand{ResDeclSubPosExR}{\ResSubPosRule{{\exists}\textsc{R}}}
\newrulecommand{ResDeclSubPosFix}{\ResSubPosRule{\mu}}

\newrulecommand{ResDeclSubNegImpL}{\ResSubNegRule{{\implies}\textsc{L}}}
\newrulecommand{ResDeclSubNegAllL}{\ResSubNegRule{{\forall}\textsc{L}}}
\newrulecommand{ResDeclSubNegArrow}{\ResSubNegRule{->}}

\newrulecommand{ResDeclSubPosVoid}{\ResSubPosRule{0}}
\newrulecommand{ResDeclSubPosUnit}{\ResSubPosRule{1}}
\newrulecommand{ResDeclSubPosProd}{\ResSubPosRule{\times}}
\newrulecommand{ResDeclSubPosSum}{\ResSubPosRule{+}}
\newrulecommand{ResDeclSubPosDownshift}{\ResSubPosRule{\downarrow}}
\newrulecommand{ResDeclSubNegUpshift}{\ResSubNegRule{\uparrow}}

\newcommand{\SemiDeclFunctorEquivRule}[1]{Fun\ensuremath{{\equiv}{\slashop}}{#1}}

\newrulecommand{SemiDeclFunEquivConst}{\SemiDeclFunctorEquivRule{Const}}
\newrulecommand{SemiDeclFunEquivId}{\SemiDeclFunctorEquivRule{\ensuremath{\Id}}}
\newrulecommand{SemiDeclFunEquivUnit}{\SemiDeclFunctorEquivRule{\ensuremath{I}}}
\newrulecommand{SemiDeclFunEquivProd}{\SemiDeclFunctorEquivRule{\ensuremath{\otimes}}}
\newrulecommand{SemiDeclFunEquivSum}{\SemiDeclFunctorEquivRule{\ensuremath{\oplus}}}

\newcommand{\SemiDeclTpEquivPosRule}[1]{Tp\ensuremath{{\equiv}^\colourpos{\slashop}{#1}}}
\newcommand{\SemiDeclTpEquivNegRule}[1]{Tp\ensuremath{{\equiv}^\colourneg{\slashop}{#1}}}

\newrulecommand{SemiDeclTpEquivPosWith}{\SemiDeclTpEquivPosRule{{\land}}}
\newrulecommand{SemiDeclTpEquivPosEx}{\SemiDeclTpEquivPosRule{{\exists}}}
\newrulecommand{SemiDeclTpEquivPosFix}{\SemiDeclTpEquivPosRule{\mu}}
\newrulecommand{SemiDeclTpEquivPosFixInst}{\SemiDeclTpEquivPosRule{\mu}\textsc{Inst}}
\newrulecommand{SemiDeclTpEquivPosFixInstFun}{\SemiDeclTpEquivPosRule{\mu}\textsc{Inst'}}
\newrulecommand{SemiDeclTpEquivPosVoid}{\SemiDeclTpEquivPosRule{0}}
\newrulecommand{SemiDeclTpEquivPosUnit}{\SemiDeclTpEquivPosRule{1}}
\newrulecommand{SemiDeclTpEquivPosProd}{\SemiDeclTpEquivPosRule{\times}}
\newrulecommand{SemiDeclTpEquivPosSum}{\SemiDeclTpEquivPosRule{+}}
\newrulecommand{SemiDeclTpEquivPosDownshift}{\SemiDeclTpEquivPosRule{\downarrow}}
\newrulecommand{SemiDeclTpEquivNegUpshift}{\SemiDeclTpEquivNegRule{\uparrow}}
\newrulecommand{SemiDeclTpEquivNegImp}{\SemiDeclTpEquivNegRule{{\implies}}}
\newrulecommand{SemiDeclTpEquivNegAll}{\SemiDeclTpEquivNegRule{{\forall}}}
\newrulecommand{SemiDeclTpEquivNegArrow}{\SemiDeclTpEquivNegRule{->}}

\newcommand{\SemiDeclSubPosRule}[1]{\ensuremath{{\algsubtype}^\colourpos{\slashop}{#1}}}
\newcommand{\SemiDeclSubNegRule}[1]{\ensuremath{{\algsubtype}^\colourneg{\slashop}{#1}}}

\newrulecommand{SemiDeclSubPosRefl}{\SemiDeclSubPosRule{\textsc{Refl}}}
\newrulecommand{SemiDeclSubPosL}{\SemiDeclSubPosRule{{\rightsquigarrow}\textsc{L}}}
\newrulecommand{SemiDeclSubPosWithL}{\SemiDeclSubPosRule{{\land}\textsc{L}}}
\newrulecommand{SemiDeclSubPosWithR}{\SemiDeclSubPosRule{{\land}\textsc{R}}}
\newrulecommand{SemiDeclSubPosExL}{\SemiDeclSubPosRule{{\exists}\textsc{L}}}
\newrulecommand{SemiDeclSubPosExR}{\SemiDeclSubPosRule{{\exists}\textsc{R}}}
\newrulecommand{SemiDeclSubPosFix}{\SemiDeclSubPosRule{\mu}}

\newrulecommand{SemiDeclSubNegRefl}{\SemiDeclSubNegRule{\textsc{Refl}}}
\newrulecommand{SemiDeclSubNegR}{\SemiDeclSubPosRule{{\rightsquigarrow}\textsc{R}}}
\newrulecommand{SemiDeclSubNegImpL}{\SemiDeclSubNegRule{{\implies}\textsc{L}}}
\newrulecommand{SemiDeclSubNegImpR}{\SemiDeclSubNegRule{{\implies}\textsc{R}}}
\newrulecommand{SemiDeclSubNegAllL}{\SemiDeclSubNegRule{{\forall}\textsc{L}}}
\newrulecommand{SemiDeclSubNegAllR}{\SemiDeclSubNegRule{{\forall}\textsc{R}}}
\newrulecommand{SemiDeclSubNegArrow}{\SemiDeclSubNegRule{->}}

\newrulecommand{SemiDeclSubPosVoid}{\SemiDeclSubPosRule{0}}
\newrulecommand{SemiDeclSubPosUnit}{\SemiDeclSubPosRule{1}}
\newrulecommand{SemiDeclSubPosProd}{\SemiDeclSubPosRule{\times}}
\newrulecommand{SemiDeclSubPosSum}{\SemiDeclSubPosRule{+}}
\newrulecommand{SemiDeclSubPosDownshift}{\SemiDeclSubPosRule{\downarrow}}
\newrulecommand{SemiDeclSubNegUpshift}{\SemiDeclSubNegRule{\uparrow}}

\newcommand{\AlgFunctorEquivRule}[1]{Fun\ensuremath{{\equiv}{\slashop}{-|}}{#1}}

\newrulecommand{AlgFunEquivConst}{\AlgFunctorEquivRule{Const}}
\newrulecommand{AlgFunEquivId}{\AlgFunctorEquivRule{\ensuremath{\Id}}}
\newrulecommand{AlgFunEquivUnit}{\AlgFunctorEquivRule{\ensuremath{I}}}
\newrulecommand{AlgFunEquivProd}{\AlgFunctorEquivRule{\ensuremath{\otimes}}}
\newrulecommand{AlgFunEquivSum}{\AlgFunctorEquivRule{\ensuremath{\oplus}}}

\newrulecommand{PropEquivInst}{PropEquivInst}
\newrulecommand{PropEquivNoInst}{PropEquivNoInst}

\newcommand{\AlgTpEquivPosRule}[1]{Tp\ensuremath{{\equiv}^\colourpos{\slashop}{-|}{#1}}}
\newcommand{\AlgTpEquivNegRule}[1]{Tp\ensuremath{{\equiv}^\colourneg{\slashop}{-|}{#1}}}

\newrulecommand{AlgTpEquivPosWith}{\AlgTpEquivPosRule{{\land}}}
\newrulecommand{AlgTpEquivPosEx}{\AlgTpEquivPosRule{{\exists}}}
\newrulecommand{AlgTpEquivPosFix}{\AlgTpEquivPosRule{\mu}}
\newrulecommand{AlgTpEquivPosFixInst}{\AlgTpEquivPosRule{\mu}\textsc{Inst}}
\newrulecommand{AlgTpEquivPosFixInstFun}{\AlgTpEquivPosRule{\mu}\textsc{Inst'}}
\newrulecommand{AlgTpEquivPosVoid}{\AlgTpEquivPosRule{0}}
\newrulecommand{AlgTpEquivPosUnit}{\AlgTpEquivPosRule{1}}
\newrulecommand{AlgTpEquivPosProd}{\AlgTpEquivPosRule{\times}}
\newrulecommand{AlgTpEquivPosSum}{\AlgTpEquivPosRule{+}}
\newrulecommand{AlgTpEquivPosDownshift}{\AlgTpEquivPosRule{\downarrow}}
\newrulecommand{AlgTpEquivNegUpshift}{\AlgTpEquivNegRule{\uparrow}}
\newrulecommand{AlgTpEquivNegImp}{\AlgTpEquivNegRule{{\implies}}}
\newrulecommand{AlgTpEquivNegAll}{\AlgTpEquivNegRule{{\forall}}}
\newrulecommand{AlgTpEquivNegArrow}{\AlgTpEquivNegRule{->}}

\newcommand{\AlgSubPosRule}[1]{\ensuremath{{\algsubtypesym}^\colourpos{\slashop}{-|}{#1}}}
\newcommand{\AlgSubNegRule}[1]{\ensuremath{{\algsubtypesym}^\colourneg{\slashop}{-|}{#1}}}

\newrulecommand{AlgSubPosRefl}{\AlgSubPosRule{\textsc{Refl}}}
\newrulecommand{AlgSubPosL}{\AlgSubPosRule{{\rightsquigarrow}\textsc{L}}}
\newrulecommand{AlgSubPosWithL}{\AlgSubPosRule{{\land}\textsc{L}}}
\newrulecommand{AlgSubPosWithR}{\AlgSubPosRule{{\land}\textsc{R}}}
\newrulecommand{AlgSubPosExL}{\AlgSubPosRule{{\exists}\textsc{L}}}
\newrulecommand{AlgSubPosExR}{\AlgSubPosRule{{\exists}\textsc{R}}}
\newrulecommand{AlgSubPosFix}{\AlgSubPosRule{\mu}}
\newrulecommand{AlgSubPosFixInst}{\AlgSubPosRule{\mu}\textsc{Inst}}
\newrulecommand{AlgSubPosFixInstFun}{\AlgSubPosRule{\mu}\textsc{Inst'}}

\newrulecommand{AlgSubNegRefl}{\AlgSubNegRule{\textsc{Refl}}}
\newrulecommand{AlgSubNegR}{\AlgSubPosRule{{\rightsquigarrow}\textsc{R}}}
\newrulecommand{AlgSubNegImpL}{\AlgSubNegRule{{\implies}\textsc{L}}}
\newrulecommand{AlgSubNegImpR}{\AlgSubNegRule{{\implies}\textsc{R}}}
\newrulecommand{AlgSubNegAllL}{\AlgSubNegRule{{\forall}\textsc{L}}}
\newrulecommand{AlgSubNegAllR}{\AlgSubNegRule{{\forall}\textsc{R}}}
\newrulecommand{AlgSubNegArrow}{\AlgSubNegRule{->}}

\newrulecommand{AlgSubPosVoid}{\AlgSubPosRule{0}}
\newrulecommand{AlgSubPosUnit}{\AlgSubPosRule{1}}
\newrulecommand{AlgSubPosProd}{\AlgSubPosRule{\times}}
\newrulecommand{AlgSubPosSum}{\AlgSubPosRule{+}}
\newrulecommand{AlgSubPosDownshift}{\AlgSubPosRule{\downarrow}}
\newrulecommand{AlgSubNegUpshift}{\AlgSubNegRule{\uparrow}}

\newcommand{\DeclChkRule}[1]{Decl\ensuremath{{<=}{#1}}}
\newcommand{\DeclSynRule}[1]{Decl\ensuremath{{=>}{#1}}}
\newrulecommand{DeclChkValVar}{\DeclChkRule{\text{Var}}}
\newrulecommand{DeclSynValAnnot}{\DeclSynRule{\text{ValAnnot}}}
\newrulecommand{DeclChkValSub}{\DeclChkRule{\text{Sub}}}
\newrulecommand{DeclChkValUnit}{\DeclChkRule{1}}
\newrulecommand{DeclChkValPair}{\DeclChkRule{\times}}
\newrulecommandONE{DeclChkValIn}{\DeclChkRule{+}\ensuremath{{}_{#1}}}
\newrulecommand{DeclChkValExists}{\DeclChkRule{\exists}}
\newrulecommand{DeclChkValWith}{\DeclChkRule{\land}}
\newrulecommand{DeclChkValFix}{\DeclChkRule{\mu}}
\newrulecommand{DeclChkValDownshift}{\DeclChkRule{\downarrow}}
\newrulecommand{DeclChkExpUpshift}{\DeclChkRule{\uparrow}}
\newrulecommand{DeclChkExpLet}{\DeclChkRule{\textkw{let}}}
\newrulecommand{DeclChkExpMatch}{\DeclChkRule{\textkw{match}}}
\newrulecommand{DeclChkExpLam}{\DeclChkRule{\lambda}}
\newrulecommand{DeclChkExpRec}{\DeclChkRule{\textkw{rec}}}
\newrulecommand{DeclChkExpUnreachable}{\DeclChkRule{}Unreachable}
\newrulecommand{DeclChkExpImplies}{\DeclChkRule{\implies}}
\newrulecommand{DeclChkExpForall}{\DeclChkRule{\forall}}
\newrulecommand{DeclChkExpExtract}{\DeclChkRule{\rightsquigarrow}}
\newrulecommand{DeclSynSpineApp}{\DeclSynRule{\text{App}}}
\newrulecommand{DeclSynExpAnnot}{\DeclSynRule{\text{ExpAnnot}}}
\newrulecommand{DeclSynHeadVar}{\DeclSynRule{\text{Var}}}

\newcommand{\SemiDeclChkRule}[1]{Semi\ensuremath{{<=}{#1}}}
\newcommand{\SemiDeclSynRule}[1]{Semi\ensuremath{{=>}{#1}}}
\newrulecommand{SemiDeclChkValVar}{\SemiDeclChkRule{\text{Var}}}
\newrulecommand{SemiDeclSynValAnnot}{\SemiDeclSynRule{\text{ValAnnot}}}
\newrulecommand{SemiDeclChkValSub}{\SemiDeclChkRule{\text{Sub}}}
\newrulecommand{SemiDeclChkValUnit}{\SemiDeclChkRule{1}}
\newrulecommand{SemiDeclChkValPair}{\SemiDeclChkRule{\times}}
\newrulecommandONE{SemiDeclChkValIn}{\SemiDeclChkRule{+}\ensuremath{{}_{#1}}}
\newrulecommand{SemiDeclChkValExists}{\SemiDeclChkRule{\exists}}
\newrulecommand{SemiDeclChkValWith}{\SemiDeclChkRule{\land}}
\newrulecommand{SemiDeclChkValFix}{\SemiDeclChkRule{\mu}}
\newrulecommand{SemiDeclChkValDownshift}{\SemiDeclChkRule{\downarrow}}
\newrulecommand{SemiDeclChkExpUpshift}{\SemiDeclChkRule{\uparrow}}
\newrulecommand{SemiDeclChkExpLet}{\SemiDeclChkRule{\textkw{let}}}
\newrulecommand{SemiDeclChkExpMatch}{\SemiDeclChkRule{\textkw{match}}}
\newrulecommand{SemiDeclChkExpLam}{\SemiDeclChkRule{\lambda}}
\newrulecommand{SemiDeclChkExpUnreachable}{\SemiDeclChkRule{}Unreachable}
\newrulecommand{SemiDeclChkExpRec}{\SemiDeclChkRule{\textkw{rec}}}
\newrulecommand{SemiDeclChkExpImplies}{\SemiDeclChkRule{\implies}}
\newrulecommand{SemiDeclChkExpForall}{\SemiDeclChkRule{\forall}}
\newrulecommand{SemiDeclChkExpExtract}{\SemiDeclChkRule{\rightsquigarrow}}
\newrulecommand{SemiDeclSynSpineApp}{\SemiDeclSynRule{\text{App}}}
\newrulecommand{SemiDeclSynExpAnnot}{\SemiDeclSynRule{\text{ExpAnnot}}}
\newrulecommand{SemiDeclSynHeadVar}{\SemiDeclSynRule{\text{Var}}}

\newcommand{\UnrefChkRule}[1]{Unref\ensuremath{{<=}{#1}}}
\newcommand{\UnrefSynRule}[1]{Unref\ensuremath{{=>}{#1}}}
\newrulecommand{UnrefChkValVar}{\UnrefChkRule{\text{Var}}}
\newrulecommand{UnrefSynValAnnot}{\UnrefSynRule{\text{ValAnnot}}}
\newrulecommand{UnrefChkValSub}{\UnrefChkRule{\text{Sub}}}
\newrulecommand{UnrefChkValUnit}{\UnrefChkRule{1}}
\newrulecommand{UnrefChkValPair}{\UnrefChkRule{\times}}
\newrulecommandONE{UnrefChkValIn}{\UnrefChkRule{+}\ensuremath{{}_{#1}}}
\newrulecommand{UnrefChkValFix}{\UnrefChkRule{\mu}}
\newrulecommand{UnrefChkValDownshift}{\UnrefChkRule{\downarrow}}
\newrulecommand{UnrefChkExpUpshift}{\UnrefChkRule{\uparrow}}
\newrulecommand{UnrefChkExpLet}{\UnrefChkRule{\textkw{let}}}
\newrulecommand{UnrefChkExpMatch}{\UnrefChkRule{\textkw{match}}}
\newrulecommand{UnrefChkExpLam}{\UnrefChkRule{\lambda}}
\newrulecommand{UnrefChkExpDiverge}{\UnrefChkRule{}Diverge}
\newrulecommand{UnrefChkExpRec}{\UnrefChkRule{\mathsf{rec}}}
\newrulecommand{UnrefSynSpineApp}{\UnrefSynRule{\text{App}}}
\newrulecommand{UnrefSynExpAnnot}{\UnrefSynRule{\text{ExpAnnot}}}
\newrulecommand{UnrefSynHeadVar}{\UnrefSynRule{\text{Var}}}

\newcommand{\AlgChkRule}[1]{Alg\ensuremath{{<=}{#1}}}
\newcommand{\AlgSynRule}[1]{Alg\ensuremath{{=>}{#1}}}
\newrulecommand{AlgChkValVar}{\AlgChkRule{\text{Var}}}
\newrulecommand{AlgSynValAnnot}{\AlgSynRule{\text{ValAnnot}}}
\newrulecommand{AlgChkValSub}{\AlgChkRule{\text{Sub}}}
\newrulecommand{AlgChkValUnit}{\AlgChkRule{1}}
\newrulecommand{AlgChkValPair}{\AlgChkRule{\times}}
\newrulecommandONE{AlgChkValIn}{\AlgChkRule{+}\ensuremath{{}_{#1}}}
\newrulecommand{AlgChkValExists}{\AlgChkRule{\exists}}
\newrulecommand{AlgChkValWith}{\AlgChkRule{\land}}
\newrulecommand{AlgChkValFix}{\AlgChkRule{\mu}}
\newrulecommand{AlgChkValDownshift}{\AlgChkRule{\downarrow}}
\newrulecommand{AlgChkExpUpshift}{\AlgChkRule{\uparrow}}
\newrulecommand{AlgChkExpLet}{\AlgChkRule{\textkw{let}}}
\newrulecommand{AlgChkExpMatch}{\AlgChkRule{\textkw{match}}}
\newrulecommand{AlgChkExpLam}{\AlgChkRule{\lambda}}
\newrulecommand{AlgChkExpUnreachable}{\AlgChkRule{}Unreachable}
\newrulecommand{AlgChkExpRec}{\AlgChkRule{\mathsf{rec}}}
\newrulecommand{AlgChkExpImplies}{\AlgChkRule{\implies}}
\newrulecommand{AlgChkExpForall}{\AlgChkRule{\forall}}
\newrulecommand{AlgChkExpExtract}{\AlgChkRule{\rightsquigarrow}}
\newrulecommand{AlgSynSpineApp}{\AlgSynRule{\text{App}}}
\newrulecommand{AlgSynExpAnnot}{\AlgSynRule{\text{ExpAnnot}}}
\newrulecommand{AlgSynHeadVar}{\AlgSynRule{\text{Var}}}

\newrulecommand{RuleInst}{Inst}
\newrulecommand{RuleNoInst}{NoInst}

\newcommand{\DeclMatchRule}[1]{DeclMatch\ensuremath{#1}}
\newrulecommand{DeclChkMatchEx}{\DeclMatchRule{\exists}}
\newrulecommand{DeclChkMatchWith}{\DeclMatchRule{\land}}
\newrulecommand{DeclChkMatchUnit}{\DeclMatchRule{1}}
\newrulecommand{DeclChkMatchPair}{\DeclMatchRule{\times}}
\newrulecommand{DeclChkMatchSum}{\DeclMatchRule{+}}
\newrulecommand{DeclChkMatchVoid}{\DeclMatchRule{0}}
\newrulecommand{DeclChkMatchFix}{\DeclMatchRule{\mu}}

\newcommand{\SemiDeclMatchRule}[1]{SemiMatch\ensuremath{#1}}
\newrulecommand{SemiDeclChkMatchEx}{\SemiDeclMatchRule{\exists}}
\newrulecommand{SemiDeclChkMatchWith}{\SemiDeclMatchRule{\land}}
\newrulecommand{SemiDeclChkMatchUnit}{\SemiDeclMatchRule{1}}
\newrulecommand{SemiDeclChkMatchPair}{\SemiDeclMatchRule{\times}}
\newrulecommand{SemiDeclChkMatchSum}{\SemiDeclMatchRule{+}}
\newrulecommand{SemiDeclChkMatchVoid}{\SemiDeclMatchRule{0}}
\newrulecommand{SemiDeclChkMatchFix}{\SemiDeclMatchRule{\mu}}

\newcommand{\AlgMatchRule}[1]{AlgMatch\ensuremath{#1}}
\newrulecommand{AlgChkMatchEx}{\AlgMatchRule{\exists}}
\newrulecommand{AlgChkMatchWith}{\AlgMatchRule{\land}}
\newrulecommand{AlgChkMatchUnit}{\AlgMatchRule{1}}
\newrulecommand{AlgChkMatchPair}{\AlgMatchRule{\times}}
\newrulecommand{AlgChkMatchSum}{\AlgMatchRule{+}}
\newrulecommand{AlgChkMatchVoid}{\AlgMatchRule{0}}
\newrulecommand{AlgChkMatchFix}{\AlgMatchRule{\mu}}

\newcommand{\DeclSpineRule}[1]{DeclSpine\ensuremath{#1}}
\newrulecommand{DeclSpineAll}{\DeclSpineRule{\forall}}
\newrulecommand{DeclSpineImplies}{\DeclSpineRule{\implies}}
\newrulecommand{DeclSpineApp}{\DeclSpineRule{}App}
\newrulecommand{DeclSpineNil}{\DeclSpineRule{}Nil}

\newcommand{\SemiDeclSpineRule}[1]{SemiSpine\ensuremath{#1}}
\newrulecommand{SemiDeclSpineAll}{\SemiDeclSpineRule{\forall}}
\newrulecommand{SemiDeclSpineImplies}{\SemiDeclSpineRule{\implies}}
\newrulecommand{SemiDeclSpineApp}{\SemiDeclSpineRule{}App}
\newrulecommand{SemiDeclSpineNil}{\SemiDeclSpineRule{}Nil}

\newcommand{\AlgSpineRule}[1]{AlgSpine\ensuremath{#1}}
\newrulecommand{AlgSpineAll}{\AlgSpineRule{\forall}}
\newrulecommand{AlgSpineImplies}{\AlgSpineRule{\implies}}
\newrulecommand{AlgSpineApp}{\AlgSpineRule{}App}
\newrulecommand{AlgSpineNil}{\AlgSpineRule{}Nil}

\newcommand{\UnrefMatchRule}[1]{UnrefMatch\ensuremath{#1}}
\newrulecommand{UnrefChkMatchUnit}{\UnrefMatchRule{1}}
\newrulecommand{UnrefChkMatchPair}{\UnrefMatchRule{\times}}
\newrulecommand{UnrefChkMatchSum}{\UnrefMatchRule{+}}
\newrulecommand{UnrefChkMatchVoid}{\UnrefMatchRule{0}}
\newrulecommand{UnrefChkMatchFix}{\UnrefMatchRule{\mu}}

\newcommand{\UnrefSpineRule}[1]{UnrefSpine\ensuremath{#1}}
\newrulecommand{UnrefSpineApp}{\UnrefSpineRule{}App}
\newrulecommand{UnrefSpineNil}{\UnrefSpineRule{}Nil}

\newcommand{\SynSubsRule}[1]{{#1}\ensuremath{\sigma}}
\newrulecommand{EmptySyn}{\SynSubsRule{Empty}}
\newrulecommand{IxSyn}{\SynSubsRule{Ix}}
\newrulecommand{PropSyn}{\SynSubsRule{Prop}}
\newrulecommand{ValSyn}{\SynSubsRule{Val}}

\newcommand{\SemSubsRule}[1]{{#1}\ensuremath{\delta}}
\newrulecommand{EmptySem}{\SemSubsRule{Empty}}
\newrulecommand{IxSem}{\SemSubsRule{Ix}}
\newrulecommand{PropSem}{\SemSubsRule{Prop}}
\newrulecommand{ValSem}{\SemSubsRule{Val}}

\newcommand{\XiSubsRule}[1]{{#1}\ensuremath{\xi}}
\newrulecommand{EmptyXi}{\XiSubsRule{Empty}}
\newrulecommand{IxXi}{\XiSubsRule{Ix}}

\newcommand{\UnrefSynSubsRule}[1]{Unref{#1}\ensuremath{\sigma}}
\newrulecommand{UnrefEmptySyn}{\UnrefSynSubsRule{Empty}}
\newrulecommand{UnrefValSyn}{\UnrefSynSubsRule{Val}}

\newcommand{\UnrefSemSubsRule}[1]{Unref{#1}\ensuremath{\delta}}
\newrulecommand{UnrefEmptySem}{\UnrefSemSubsRule{Empty}}
\newrulecommand{UnrefValSem}{\UnrefSemSubsRule{Val}}

\newcommand{\IxRule}[1]{Ix{#1}}
\newrulecommand{IxVar}{\IxRule{Var}}
\newrulecommand{IxConst}{\IxRule{Const}}
\newrulecommand{IxSubtract}{\IxRule{Subtract}}
\newrulecommand{IxAdd}{\IxRule{Add}}
\newrulecommand{IxProd}{\IxRule{Prod}}
\newrulecommandONE{IxProj}{\IxRule{Proj}\ensuremath{{}_{#1}}}
\newrulecommand{IxAnd}{\IxRule{\ensuremath{\land}}}
\newrulecommand{IxOr}{\IxRule{\ensuremath{\lor}}}
\newrulecommand{IxNot}{\IxRule{\ensuremath{\lnot}}}
\newrulecommand{IxEq}{\IxRule{\ensuremath{=}}}
\newrulecommand{IxLeq}{\IxRule{\ensuremath{\leq}}}

\newcommand{\AlgIxRule}[1]{AlgIx{#1}}
\newrulecommand{AlgIxVar}{\AlgIxRule{Var}}
\newrulecommand{AlgIxEVar}{\AlgIxRule{EVar}}
\newrulecommand{AlgIxSolvedEVar}{\AlgIxRule{SolvedEVar}}
\newrulecommand{AlgIxConst}{\AlgIxRule{Const}}
\newrulecommand{AlgIxSubtract}{\AlgIxRule{Subtract}}
\newrulecommand{AlgIxAdd}{\AlgIxRule{Add}}
\newrulecommand{AlgIxProd}{\AlgIxRule{Prod}}
\newrulecommandONE{AlgIxProj}{\AlgIxRule{Proj}\ensuremath{{}_{#1}}}
\newrulecommand{AlgIxAnd}{\AlgIxRule{\ensuremath{\land}}}
\newrulecommand{AlgIxOr}{\AlgIxRule{\ensuremath{\lor}}}
\newrulecommand{AlgIxNot}{\AlgIxRule{\ensuremath{\lnot}}}
\newrulecommand{AlgIxEq}{\AlgIxRule{\ensuremath{=}}}
\newrulecommand{AlgIxLeq}{\AlgIxRule{\ensuremath{\leq}}}

\newrulecommand{PropTrue}{PropTrue}

\newcommand{\DeclPrpRule}[1]{DeclPrp{#1}}
\newrulecommand{DeclPrpT}{\DeclPrpRule{\ensuremath{\True}}}
\newrulecommand{DeclPrpF}{\DeclPrpRule{\ensuremath{\False}}}
\newrulecommand{DeclPrpAnd}{\DeclPrpRule{\ensuremath{\land}}}
\newrulecommand{DeclPrpOr}{\DeclPrpRule{\ensuremath{\lor}}}
\newrulecommand{DeclPrpNot}{\DeclPrpRule{\ensuremath{\lnot}}}
\newrulecommand{DeclPrpEq}{\DeclPrpRule{\ensuremath{=}}}
\newrulecommand{DeclPrpEqVarL}{\DeclPrpRule{\ensuremath{=}VarL}}
\newrulecommand{DeclPrpEqVarR}{\DeclPrpRule{\ensuremath{=}VarR}}
\newrulecommand{DeclPrpLeq}{\DeclPrpRule{\ensuremath{\leq}}}
\newrulecommand{DeclPrpLeqVarL}{\DeclPrpRule{\ensuremath{\leq}VarL}}
\newrulecommand{DeclPrpLeqVarR}{\DeclPrpRule{\ensuremath{\leq}VarR}}
\newrulecommand{DeclPrpSub}{\DeclPrpRule{Sub}}

\newcommand{\AlgPrpRule}[1]{AlgPrp{#1}}
\newrulecommand{AlgPrpT}{\AlgPrpRule{\ensuremath{\True}}}
\newrulecommand{AlgPrpF}{\AlgPrpRule{\ensuremath{\False}}}
\newrulecommand{AlgPrpAnd}{\AlgPrpRule{\ensuremath{\land}}}
\newrulecommand{AlgPrpOr}{\AlgPrpRule{\ensuremath{\lor}}}
\newrulecommand{AlgPrpNot}{\AlgPrpRule{\ensuremath{\lnot}}}
\newrulecommand{AlgPrpEq}{\AlgPrpRule{\ensuremath{=}}}
\newrulecommand{AlgPrpEqVarL}{\AlgPrpRule{\ensuremath{=}VarL}}
\newrulecommand{AlgPrpEqVarR}{\AlgPrpRule{\ensuremath{=}VarR}}
\newrulecommand{AlgPrpLeq}{\AlgPrpRule{\ensuremath{\leq}}}
\newrulecommand{AlgPrpLeqVarL}{\AlgPrpRule{\ensuremath{\leq}VarL}}
\newrulecommand{AlgPrpLeqVarR}{\AlgPrpRule{\ensuremath{\leq}VarR}}
\newrulecommand{AlgPrpSub}{\AlgPrpRule{Sub}}

\newcommand{\DeclTpRule}[1]{DeclTp{#1}}
\newrulecommand{DeclTpVoid}{\DeclTpRule{\ensuremath{0}}}
\newrulecommand{DeclTpUnit}{\DeclTpRule{\ensuremath{1}}}
\newrulecommand{DeclTpSum}{\DeclTpRule{\ensuremath{+}}}
\newrulecommand{DeclTpProd}{\DeclTpRule{\ensuremath{\times}}}
\newrulecommand{DeclTpEx}{\DeclTpRule{\ensuremath{\exists}}}
\newrulecommand{DeclTpDown}{\DeclTpRule{\ensuremath{\downarrow}}}
\newrulecommand{DeclTpWith}{\DeclTpRule{\ensuremath{\land}}}
\newrulecommand{DeclTpFixVar}{\DeclTpRule{\ensuremath{\mu}Var}}
\newrulecommand{DeclTpFix}{\DeclTpRule{\ensuremath{\mu}}}
\newrulecommand{DeclTpAll}{\DeclTpRule{\ensuremath{\forall}}}
\newrulecommand{DeclTpImplies}{\DeclTpRule{\ensuremath{\implies}}}
\newrulecommand{DeclTpArrow}{\DeclTpRule{\ensuremath{\to}}}
\newrulecommand{DeclTpUp}{\DeclTpRule{\ensuremath{\uparrow}}}
\newrulecommand{DeclTpSub}{\DeclTpRule{Sub}}

\newcommand{\AlgTpRule}[1]{AlgTp{#1}}
\newrulecommand{AlgTpVoid}{\AlgTpRule{\ensuremath{0}}}
\newrulecommand{AlgTpUnit}{\AlgTpRule{\ensuremath{1}}}
\newrulecommand{AlgTpSum}{\AlgTpRule{\ensuremath{+}}}
\newrulecommand{AlgTpProd}{\AlgTpRule{\ensuremath{\times}}}
\newrulecommand{AlgTpEx}{\AlgTpRule{\ensuremath{\exists}}}
\newrulecommand{AlgTpDown}{\AlgTpRule{\ensuremath{\downarrow}}}
\newrulecommand{AlgTpWith}{\AlgTpRule{\ensuremath{\land}}}
\newrulecommand{AlgTpFixVar}{\AlgTpRule{\ensuremath{\mu}Var}}
\newrulecommand{AlgTpFixEVar}{\AlgTpRule{\ensuremath{\mu}EVar}}
\newrulecommand{AlgTpFixSolvedEVar}{\AlgTpRule{\ensuremath{\mu}SolvedEVar}}
\newrulecommand{AlgTpFix}{\AlgTpRule{\ensuremath{\mu}}}
\newrulecommand{AlgTpAll}{\AlgTpRule{\ensuremath{\forall}}}
\newrulecommand{AlgTpImplies}{\AlgTpRule{\ensuremath{\implies}}}
\newrulecommand{AlgTpArrow}{\AlgTpRule{\ensuremath{\to}}}
\newrulecommand{AlgTpUp}{\AlgTpRule{\ensuremath{\uparrow}}}
\newrulecommand{AlgTpSub}{\AlgTpRule{Sub}}

\newcommand{\DeclFunctorRule}[1]{DeclFunctor{#1}}
\newrulecommand{DeclFunctorConst}{\DeclFunctorRule{Const}}
\newrulecommand{DeclFunctorId}{\DeclFunctorRule{\ensuremath{\Id}}}
\newrulecommand{DeclFunctorUnit}{\DeclFunctorRule{\ensuremath{I}}}
\newrulecommand{DeclFunctorProd}{\DeclFunctorRule{\ensuremath{\otimes}}}
\newrulecommand{DeclFunctorSum}{\DeclFunctorRule{\ensuremath{\oplus}}}
\newrulecommand{DeclFunctorSub}{\DeclFunctorRule{Sub}}

\newcommand{\AlgFunctorRule}[1]{AlgFunctor{#1}}
\newrulecommand{AlgFunctorConst}{\AlgFunctorRule{Const}}
\newrulecommand{AlgFunctorId}{\AlgFunctorRule{\ensuremath{\Id}}}
\newrulecommand{AlgFunctorUnit}{\AlgFunctorRule{\ensuremath{I}}}
\newrulecommand{AlgFunctorProd}{\AlgFunctorRule{\ensuremath{\otimes}}}
\newrulecommand{AlgFunctorSum}{\AlgFunctorRule{\ensuremath{\oplus}}}
\newrulecommand{AlgFunctorSub}{\AlgFunctorRule{Sub}}

\newcommand{\DeclAlgRule}[1]{DeclAlg{#1}}
\newrulecommand{DeclAlgSum}{\DeclAlgRule{\ensuremath{\oplus}}}
\newrulecommand{DeclAlgUnit}{\DeclAlgRule{\ensuremath{I}}}
\newrulecommand{DeclAlgIdProd}{\DeclAlgRule{\ensuremath{\Id\otimes}}}
\newrulecommand{DeclAlgConstProd}{\DeclAlgRule{Const\ensuremath{\otimes}}}
\newrulecommand{DeclAlgExConstProd}{\DeclAlgRule{Const\ensuremath{\exists\otimes}}}

\newcommand{\AlgAlgRule}[1]{AlgAlg{#1}}
\newrulecommand{AlgAlgSum}{\AlgAlgRule{\ensuremath{\oplus}}}
\newrulecommand{AlgAlgUnit}{\AlgAlgRule{\ensuremath{I}}}
\newrulecommand{AlgAlgIdProd}{\AlgAlgRule{\ensuremath{\Id\otimes}}}
\newrulecommand{AlgAlgConstProd}{\AlgAlgRule{Const\ensuremath{\otimes}}}
\newrulecommand{AlgAlgExConstProd}{\AlgAlgRule{Const\ensuremath{\exists\otimes}}}

\newcommand{\DeclPatRule}[1]{DeclPat{#1}}
\newrulecommandONE{DeclPatNil}{\DeclPatRule{Nil\ensuremath{{}_{#1}}}}
\newrulecommandONE{DeclPatThere}{\DeclPatRule{There\ensuremath{{}_{#1}}}}
\newrulecommandONE{DeclPatHere}{\DeclPatRule{Here\ensuremath{{}_{#1}}}}

\newcommand{\DeclUnrollRule}[1]{Unroll{#1}}
\newrulecommand{DeclUnrollSum}{\DeclUnrollRule{\ensuremath{\oplus}}}
\newrulecommand{DeclUnrollId}{\DeclUnrollRule{\ensuremath{\Id}}}
\newrulecommand{DeclUnrollConstEx}{\DeclUnrollRule{\ensuremath{\exists}}}
\newrulecommand{DeclUnrollConst}{\DeclUnrollRule{Const}}
\newrulecommand{DeclUnrollUnit}{\DeclUnrollRule{\ensuremath{I}}}

\newcommand{\UnrefUnrollRule}[1]{UnrefUnroll{#1}}
\newrulecommand{UnrefUnrollSum}{\UnrefUnrollRule{\ensuremath{\oplus}}}
\newrulecommand{UnrefUnrollId}{\UnrefUnrollRule{\ensuremath{\Id}}}
\newrulecommand{UnrefUnrollConst}{\UnrefUnrollRule{Const}}
\newrulecommand{UnrefUnrollUnit}{\UnrefUnrollRule{\ensuremath{I}}}

\newcommand{\AlgUnrollRule}[1]{AlgUnroll{#1}}
\newrulecommand{AlgUnrollSum}{\AlgUnrollRule{\ensuremath{\oplus}}}
\newrulecommand{AlgUnrollId}{\AlgUnrollRule{\ensuremath{\Id}}}
\newrulecommand{AlgUnrollConstEx}{\AlgUnrollRule{\ensuremath{\exists}}}
\newrulecommand{AlgUnrollConst}{\AlgUnrollRule{Const}}
\newrulecommand{AlgUnrollUnit}{\AlgUnrollRule{\ensuremath{I}}}

\newcommand{\DeclLeadsToRule}[1]{Decl{#1}\ensuremath{\leadsto\downarrow}}
\newrulecommand{DeclLeadsToDown}{\DeclLeadsToRule{\ensuremath{\downarrow}}}
\newrulecommand{DeclLeadsToAnd}{\DeclLeadsToRule{\ensuremath{\land}}}

\newcommand{\ExtRule}[1]{\ensuremath{\longrightarrow}{#1}}
\newrulecommand{ExtEmpty}{\ExtRule{Empty}}
\newrulecommand{ExtVar}{\ExtRule{Var}}
\newrulecommand{ExtPropOne}{\ExtRule{Prop1}}
\newrulecommand{ExtPropTwo}{\ExtRule{Prop2}}
\newrulecommand{ExtUnsolved}{\ExtRule{Unsolved}}
\newrulecommand{ExtSolved}{\ExtRule{Solved}}
\newrulecommand{ExtSolve}{\ExtRule{Solve}}

\newcommand{\RExtRule}[1]{\ensuremath{\tilde\longrightarrow}{#1}}
\newrulecommand{RExtEmpty}{\RExtRule{Empty}}
\newrulecommand{RExtVar}{\RExtRule{Var}}
\newrulecommand{RExtPropOne}{\RExtRule{Prop1}}
\newrulecommand{RExtPropTwo}{\RExtRule{Prop2}}
\newrulecommand{RExtUnsolved}{\RExtRule{Unsolved}}
\newrulecommand{RExtSolved}{\RExtRule{Solved}}
\newrulecommand{RExtSolve}{\RExtRule{Solve}}

\newcommand{\WTrueRule}[1]{\ensuremath{{|=}W}{#1}}
\newrulecommand{WTrueProp}{\WTrueRule{Prp}}
\newrulecommand{WTrueAnd}{\WTrueRule{\ensuremath{\land}}}
\newrulecommand{WTrueImpl}{\WTrueRule{\ensuremath{\implies}}}
\newrulecommand{WTrueAll}{\WTrueRule{\ensuremath{\forall}}}
\newrulecommand{WTruePosSub}{\WTrueRule{\ensuremath{\algsubtype^{+}}}}
\newrulecommand{WTrueNegSub}{\WTrueRule{\ensuremath{\algsubtype^{-}}}}
\newrulecommand{WTruePosEquiv}{\WTrueRule{\ensuremath{\equiv^{+}}}}
\newrulecommand{WTrueNegEquiv}{\WTrueRule{\ensuremath{\equiv^{-}}}}
\newrulecommand{WTruePrpEquiv}{\WTrueRule{Prp\ensuremath{\equiv}}}

\newcommand{\SemiDeclWTrueRule}[1]{\ensuremath{{\rhd}W}{#1}}
\newrulecommand{SemiDeclWTrueProp}{\SemiDeclWTrueRule{Prp}}
\newrulecommand{SemiDeclWTrueAnd}{\SemiDeclWTrueRule{\ensuremath{\land}}}
\newrulecommand{SemiDeclWTrueImpl}{\SemiDeclWTrueRule{\ensuremath{\implies}}}
\newrulecommand{SemiDeclWTrueAll}{\SemiDeclWTrueRule{\ensuremath{\forall}}}
\newrulecommand{SemiDeclWTruePosSub}{\SemiDeclWTrueRule{\ensuremath{\algsubtype^{+}}}}
\newrulecommand{SemiDeclWTrueNegSub}{\SemiDeclWTrueRule{\ensuremath{\algsubtype^{-}}}}
\newrulecommand{SemiDeclWTruePosEquiv}{\SemiDeclWTrueRule{\ensuremath{\equiv^{+}}}}
\newrulecommand{SemiDeclWTrueNegEquiv}{\SemiDeclWTrueRule{\ensuremath{\equiv^{-}}}}
\newrulecommand{SemiDeclWTruePrpEquiv}{\SemiDeclWTrueRule{Prp\ensuremath{\equiv}}}

\newcommand{\WahEquivRule}[1]{\ensuremath{{W}{\leftrightarrow}{#1}}}
\newrulecommand{WahEquivPrp}{\WahEquivRule{}Prp}
\newrulecommand{WahEquivPrpEq}{\WahEquivRule{}PrpEq}
\newrulecommand{WahEquivAnd}{\WahEquivRule{\land}}
\newrulecommand{WahEquivImplies}{\WahEquivRule{\implies}}
\newrulecommand{WahEquivAll}{\WahEquivRule{\forall}}
\newrulecommand{WahEquivPosSub}{\WahEquivRule{\algsubtype^{+}}}
\newrulecommand{WahEquivNegSub}{\WahEquivRule{\algsubtype^{-}}}
\newrulecommand{WahEquivPosEquiv}{\WahEquivRule{\equiv^{+}}}
\newrulecommand{WahEquivNegEquiv}{\WahEquivRule{\equiv^{-}}}

\newcommand{\ChiEquivRule}[1]{\ensuremath{{\chi}{\leftrightarrow}{#1}}}
\newrulecommand{ChiEquivEmpty}{\ChiEquivRule{}Empty}
\newrulecommand{ChiEquivWah}{\ChiEquivRule{W}}
\newrulecommand{ChiEquivChk}{\ChiEquivRule{<=}}

\newcommand{\ChkProblemsRule}[1]{\ensuremath{{\lhd}{#1}}}
\newrulecommand{ChkProblemsEmpty}{\ChkProblemsRule{}Empty}
\newrulecommand{ChkProblemsNegChk}{\ChkProblemsRule{}NegChk}
\newrulecommand{ChkProblemsWah}{\ChkProblemsRule{W}}

\newcommand{\SemiChkProblemsRule}[1]{\ensuremath{{\widetilde{\lhd}}{#1}}}
\newrulecommand{SemiChkProblemsEmpty}{\SemiChkProblemsRule{}Empty}
\newrulecommand{SemiChkProblemsNegChk}{\SemiChkProblemsRule{}NegChk}
\newrulecommand{SemiChkProblemsWah}{\SemiChkProblemsRule{W}}

\newcommand{\LogCtxRule}[1]{LogCtx{\ensuremath{#1}}}
\newrulecommand{LogCtxEmpty}{\LogCtxRule{}Empty}
\newrulecommand{LogCtxVar}{\LogCtxRule{}Var}
\newrulecommand{LogCtxProp}{\LogCtxRule{}Prop}

\newcommand{\ProgCtxRule}[1]{ProgCtx{\ensuremath{#1}}}
\newrulecommand{ProgCtxEmpty}{\ProgCtxRule{}Empty}
\newrulecommand{ProgCtxVar}{\ProgCtxRule{}Var}

\newcommand{\AlgCtxRule}[1]{AlgCtx{\ensuremath{#1}}}
\newrulecommand{AlgCtxEmpty}{\AlgCtxRule{}Empty}
\newrulecommand{AlgCtxUnsolved}{\AlgCtxRule{}Unsolved}
\newrulecommand{AlgCtxSolved}{\AlgCtxRule{}Solved}

\newcommand{\AlgLogCtxRule}[1]{AlgLogCtx{\ensuremath{#1}}}
\newrulecommand{AlgLogCtxEmpty}{\AlgLogCtxRule{}Empty}
\newrulecommand{AlgLogCtxVar}{\AlgLogCtxRule{}Var}
\newrulecommand{AlgLogCtxProp}{\AlgLogCtxRule{}Prop}

\iftrue
\mathlig{|-}{\entails}
\mathlig{-|}{\ctxoutsym}
\mathlig{->}{\arr}
\mathlig{<<}{\langle}
\mathlig{>>}{\rangle}
\fi

\mathlig{<=}{\chk}
\mathlig{=>}{\syn}
\mathlig{=>>}{\app}

\newcommand{\sem}[2]{\left\llbracket{#2}\right\rrbracket_{#1}}
\newcommand{\andtysym}{{\land}}
\newcommand{\andty}{\mathbin{\andtysym}}

\newcommand{\annoexp}[2]{\texttt({#1} : {#2}\texttt)}

\newcommand{\be}{g}
\newcommand{\pa}{r}
\newcommand{\bap}{o}
\newcommand{\Wah}{W}

\newcommand{\clausesym}{\Rightarrow}
\newcommand{\clauseop}{\mathbin{\clausesym}}
\newcommand{\clause}[2]{{#1} \clauseop {#2}}

\newcommand{\matchor}{\ensuremath{\normalfont\,\texttt{|}\hspace{-5.35pt}\texttt{|}\hspace{-5.35pt}\texttt{|}\hspace{-5.35pt}\texttt{|}\hspace{-5.35pt}\texttt{|}\hspace{-5.35pt}\texttt{|}\hspace{-5.35pt}\texttt{|}\hspace{-5.35pt}\texttt{|}\,}}

\newcommand{\ahat}{\hat{a}}
\newcommand{\bhat}{\hat{b}}

\newcommand{\Thetahat}{\hat{\Theta}}

\newcommand{\hypeq}[3]{{#1} : {#2}{=}{#3}}

\newcommand{\judgsubs}[3]{{#1} |- {#2} : {#3}}
\newcommand{\subs}[3]{{#1}:{#2}/{#3}}
\newcommand{\idsubs}[2]{\id_{{#1};{#2}}}

\newcommand{\Set}{\mathbf{Set}}
\newcommand{\Cpo}{\mathbf{Cpo}}
\newcommand{\Cppo}{\mathbf{Cppo}}

\newcommand{\rCpo}{\mathbf{rCpo}}
\newcommand{\rCppo}{\mathbf{rCppo}}

\DeclareMathOperator{\Hom}{Hom}
\DeclareMathOperator{\Ob}{Ob}

\newcommand{\ordsym}[1][]{\sqsubseteq_{#1}}
\newcommand{\ord}[3][]{{#2}\sqsubseteq_{#1}{#3}}
\newcommand{\bott}[1][]{\bot_{#1}}

\newcommand{\sysname}{our system\xspace}
\newcommand{\Sysname}{Our system\xspace}

\newcommand{\unreachable}{\textkw{unreachable}\xspace}
\newcommand{\diverge}{\textkw{diverge}\xspace}

\newcommand{\vectype}[2]{\mathsf{List}{{#1}}{{#2}}}
\newcommand{\nat}{\mathsf{Nat}}
\newcommand{\natsing}[1]{\nat({#1})}
\newcommand{\bool}{\mathsf{Bool}}

\newcommand{\defn}{def.\@\xspace}

\newcommand{\Gee}{\mathcal{G}}
\newcommand{\F}{\mathcal{F}}

\newcommand{\hgt}[1]{{\mathit{hgt}}{(#1)}}

\newcommand{\one}{{\{\bullet\}}}

\newcommand{\erase}[1]{ | #1 | }

\newcommand{\unref}[1]{#1}

\renewcommand{\restriction}{\mathord{\upharpoonright}}

\newcommand{\fvimctx}[2]{\langle {#1} \rangle {#2}}

\newcommand{\extend}[3]{{#1} |- {#2} \longrightarrow {#3}}
\newcommand{\rextend}[3]{{#1} |- {#2}\;{\tilde\longrightarrow}\;{#3}}

\newcommand{\subprob}[2]{{\underline{{#1} \algsubtype^{\pm} {#2}}}}
\newcommand{\negsubprob}[2]{{\underline{{#1} \algsubtype^{-} {#2}}}}
\newcommand{\possubprob}[2]{{\underline{{#1} \algsubtype^{+} {#2}}}}
\newcommand{\eqprob}[2]{{#1} \equiv^{\pm} {#2}}
\newcommand{\negeqprob}[2]{{#1} \equiv^{-} {#2}}
\newcommand{\poseqprob}[2]{{#1} \equiv^{+} {#2}}
\newcommand{\propeqprob}[2]{{#1} \equiv {#2}}

\newcommand{\size}[2][]{{\mathit{sz}}{(#2)}}
\newcommand{\length}[1]{{\mathit{len}}{\left(#1\right)}}
\newcommand{\numlog}[1]{{\#_{\land, \exists, \implies, \forall}}{\left({#1}\right)}}

\newcommand{\dontcare}{\_}

\newcommand{\defconcat}{${-}{,}{-}$~}
\newcommand{\defsem}{$\sem{}{-}$~}
\newcommand{\deferase}{$\erase{{-}}$~}
\newcommand{\deffilterprog}{$\filterprog{{-}}$~}
\newcommand{\defsize}[1][]{$\size[#1]{-}$~}
\newcommand{\defsubst}{$[-]-$~}

\newcommand{\defimpast}{${-}\implies^\ast{-}$~}

\newcommand{\filterprog}[1]{\lfloor{#1}\rfloor}

\newcommand{\sortconsts}[1]{\mathcal{K}_{#1}}

\makeatletter %
\newcommand*\Neg[2][0mu]{\Neginternal{#1}{\negslash}{#2}}

\newcommand*\Neginternal[3]{\mathpalette\Neg@{{#1}{#2}{#3}}}
\newcommand*\Neg@[2]{\Neg@@{#1}#2}
\newcommand*\Neg@@[4]{%
  \mathrel{\ooalign{%
    $\m@th#1#4$\cr
    \hidewidth$\m@th#3{#1}\mkern\muexpr#2*2$\hidewidth\cr
  }}%
}
\newcommand*\negslash[1]{\m@th#1\not\mathrel{\phantom{=}}}
\newcommand*\snegslash[1]{\rotatebox[origin=c]{60}{$\m@th#1-$}}
\makeatother  %
\newcommand{\simple}[2]{{#1}|-{#2}\Neg[1mu]{\rightsquigarrow}}

\newcommand{\judgeequivPf}[5][]{\Pf{#2}{\entails}{{#3} \equiv^{#1} {#4}}{#5}}
\newcommand{\judgeextractPf}[6][]{\Pf{#2}{\entails}{{#3} \rightsquigarrow^{#1} {#4}\;[{#5}]}{#6}}
\newcommand{\simplePf}[3]{\Pf{#1}{|-}{{#2}\Neg[1mu]{\rightsquigarrow}}{#3}}
\newcommand{\judgesubPf}[5][]{\Pf{#2}{\entails}{{#3} \leq^{#1} {#4}}{#5}}
\newcommand{\judgeentailPf}[3]{\Pf{#1}{\entails}{{#2} \;\mathsf{true}}{#3}}
\newcommand{\judgechkvalPf}[5]{\Pf{{#1}; {#2}}{\entails}{{#3} <= {#4}}{#5}}
\newcommand{\judgesynheadPf}[5]{\Pf{{#1}; {#2}}{\entails}{{#3} => {#4}}{#5}}
\newcommand{\judgechkexpPf}[5]{\Pf{{#1}; {#2}}{\entails}{{#3} <= {#4}}{#5}}
\newcommand{\judgesynexpPf}[5]{\Pf{{#1}; {#2}}{\entails}{{#3} => {#4}}{#5}}
\newcommand{\judgechkmatchPf}[6]{\Pf{{#1}; {#2}; [{#3}]}{\entails}{{#4} <= {#5}}{#6}}
\newcommand{\judgespinePf}[6]{\Pf{{#1}; {#2}; [{#4}]}{\entails}{{#3} \gg {#5}}{#6}}

\newcommand{\semideclchkvalPf}[6]{\Pf{{#1}; {#2}}{\widetilde{\entails}}{{#3} <= {#4} \; / \; {#5}}{#6}}
\newcommand{\semideclsynheadPf}[5]{\Pf{{#1}; {#2}}{\widetilde{\entails}}{{#3} => {#4}}{#5}}
\newcommand{\semideclchkexpPf}[5]{\Pf{{#1}; {#2}}{\widetilde{\entails}}{{#3} <= {#4}}{#5}}
\newcommand{\semideclsynexpPf}[5]{\Pf{{#1}; {#2}}{\widetilde{\entails}}{{#3} => {#4}}{#5}}

\newcommand{\semideclspinePf}[7]{\Pf{{#1}; {#2}; [{#4}]}{\widetilde{\entails}}{{#3} \gg {#5} \; / \; {#6}}{#7}}
\newcommand{\semideclnegPf}[4]{\Pf{{#1}; {#2}}{\widetilde{\lhd}}{#3}{#4}}

\newcommand{\unrefchkexpPf}[4]{\Pf{#1}{\entails}{{#2} <= {#3}}{#4}}

\newcommand{\algspinePf}[8]{\Pf{{#1}; {#2}; [{#4}]}{|-}{{#3} \gg {#5} \slashop {#6} -| {#7}}{#8}}

\newcommand{\algchkPf}[7]{\Pf{{#1}; {#2}}{|-}{{#3} <= {#4} \slashop {#5} -| {#6}}{#7}}

\newcommand{\algchknegPf}[5]{\Pf{{#1}; {#2}}{\rhd}{{#3} <= {#4}}{#5}}
\newcommand{\algnegPf}[4]{\Pf{{#1}; {#2}}{\lhd}{#3}{#4}}
\newcommand{\alginstPf}[4]{\Pf{#1}{|-}{{#2}\;\mathsf{inst} -| {#3}}{#4}}

\newcommand{\algsynheadPf}[5]{\Pf{{#1}; {#2}}{\rhd}{{#3} => {#4}}{#5}}
\newcommand{\algsynexpPf}[5]{\Pf{{#1}; {#2}}{\rhd}{{#3} => {#4}}{#5}}

\newcommand{\algequivPf}[7][]{\Pf{#2}{\entails}{{#3} \equiv^{#1} {#4} \slashop {#5} -| {#6}}{#7}}
\newcommand{\algsubPf}[7][]{\Pf{#2}{\entails}{{#3} \algsubtype^{#1} {#4} \slashop {#5} -| {#6}}{#7}}

\newcommand{\entailwahPf}[3]{\Pf{#1}{\models}{#2}{#3}}

\newcommand{\resjudgesubPf}[5][]{\Pf{#2}{\entails}{{#3} \algsubtype^{#1} {#4}}{#5}}

\newcommand{\semideclequivPf}[6][]{\Pf{#2}{\entails}{{#3} \equiv^{#1} {#4} \slashop {#5}}{#6}}
\newcommand{\semideclsubPf}[6][]{\Pf{#2}{\entails}{{#3} \algsubtype^{#1} {#4} \slashop {#5}}{#6}}

\newcommand{\semideclentailwahPf}[3]{\Pf{#1}{\rhd}{#2}{#3}}
\newcommand{\wahequivPf}[4]{\Pf{#1}{\rhd}{{#2} \leftrightarrow {#3}}{#4}}
\newcommand{\chiequivPf}[5]{\Pf{{#1};{#2}}{\lhd}{{#3} \leftrightarrow {#4}}{#5}}

\newcommand{\judgeunrollPf}[9]{\Pf{{#1}; {#2}}{\entails}{\left\{ {#3} \;\mid\; {#4} \left({#5}\right) =_{#8} {#6} \right\} \circeq {#7}}{#9}}

\newcommand{\unrollPf}[9]{\Pf{#1}{|-}{\mathsf{unroll}_{{#3}; {#5}}({#2}; {#4}; {#6}; {#7}) \circeq {#8}}{#9}}

\newcommand{\judgectxPf}[3]{\Pf{#1}{\entails}{{#2}\; \mathsf{ctx}}{#3}}
\newcommand{\judgctxPf}[2]{\Pf{}{}{{#1}\; \mathsf{ctx}}{#2}}
\newcommand{\judgewfPf}[3]{\Pf{#1}{\entails}{{#2}\; \mathsf{wf}}{#3}}

\newcommand{\judgetermPf}[4]{\Pf{#1}{\entails}{{#2} : {#3}}{#4}}
\newcommand{\judgetpPf}[4]{\Pf{#1}{\entails}{{#2}\; \mathsf{type} {[{#3}}]}{#4}}
\newcommand{\judgefunctorPf}[4]{\Pf{#1}{\entails}{{#2}\; \mathsf{functor} {[{#3}]}}{#4}}

\newcommand{\judgealgebraPf}[6]{\Pf{{#1}; {#2}}{\entails}{{#3} : {#4}({#5}) \Rightarrow {#5}}{#6}}

\newcommand{\composeinjPf}[4]{\Pf{{#2} \circ \inj{#1}}{\circeq}{#3}{#4}}

\newcommand{\extendPf}[4]{\Pf{#1}{|-}{{#2} \longrightarrow {#3}}{#4}}
\newcommand{\rextendPf}[4]{\Pf{#1}{|-}{{#2}\;{\tilde\longrightarrow}\;{#3}}{#4}}

\newcommand{\judgsubsPf}[4]{\Pf{#1}{|-}{{#2} : {#3}}{#4}}

\newcommand{\algpropequivinstPf}[5]{\Pf{#1}{|-}{{#2} \equiv {#3} \;\mathsf{inst} -| {#4}}{#5}}

\newcommand{\groundPf}[2]{\Pf{}{}{\ground{#1}}{#2}}

\newcommand{\ordPf}[4][]{\Pf{#2}{\sqsubseteq_{#1}}{#3}{#4}}

\begin{document}
\title%
{%
  Focusing on Refinement Typing\\
}

\ifnum\OPTIONConf=1     %
  \ifnum\OPTIONArxiv=0%
  \else
   \fi
    \author{Dimitrios J.\ Economou}
        \orcid{0000-0001-7920-744X}
        \affiliation{%
          \institution{Queen's University}
          \streetaddress{Goodwin Hall 557}
          \city{Kingston, ON}
          \postcode{K7L 3N6}
          \country{Canada}
        }
        \email{d.economou@queensu.ca}

    \author{Neel Krishnaswami}
        \orcid{0000-0003-2838-5865}
        \affiliation{%
          \institution{University of Cambridge}
          \streetaddress{Computer Laboratory, William Gates Building}
          \city{Cambridge}
          \postcode{CB3 0FD}
          \country{United Kingdom}}
        \email{nk480@cl.cam.ac.uk}

    \author{Jana Dunfield}
        \orcid{0000-0002-3718-3395}
        \affiliation{%
          \institution{Queen's University}
          \streetaddress{Goodwin Hall 557}
          \city{Kingston, ON}
          \postcode{K7L 3N6}
          \country{Canada}
        }
        \email{jd169@queensu.ca}
\fi

\setlength{\pdfpageheight}{\paperheight}
\setlength{\pdfpagewidth}{\paperwidth}

\newcommand{\XXXACM}[1]{\vspace{#1}}  %

\begin{abstract}
  We present a logically principled foundation
  for systematizing,
  in a way that works with any computational effect and evaluation order,
  SMT constraint generation
  seen in refinement type systems
  for functional programming languages.
  By carefully combining a focalized variant of call-by-push-value,
  bidirectional typing,
  and our novel technique of value-determined indexes,
  our system generates solvable SMT constraints without existential (unification) variables.
  We design a polarized subtyping relation allowing us to prove
  our logically focused typing algorithm is sound, complete, and decidable.
  We prove type soundness of our declarative system with respect to
  an elementary domain-theoretic denotational semantics.
  Type soundness %
  implies, relatively simply,
  the total correctness and logical consistency of \sysname.
  The relative ease with which we obtain both algorithmic and semantic results
  ultimately stems from the proof-theoretic technique of focalization.
\end{abstract}



\maketitle

\setcounter{footnote}{0}

\section{Introduction}
\label{sec:introduction}

True, ``well-typed programs cannot `go wrong'\thinspace'' \cite{Milner78},
but only relative to a given semantics
(if the type system is proven sound with respect to it).
Unfortunately,
well-typed programs go wrong,
in many ways that matter,
but about which a conventional type system cannot speak:
divisions by zero, out-of-bounds array accesses, information leaks.
To prove a type system rules out (at compile time) more run-time errors,
its semantics must be refined.
However, there is often not enough type structure with which to express such semantics statically.
So, we must refine our types with more information that tends to be related to programs.
Great care is needed, though,
because incorporating too much information
(such as nonterminating programs themselves,
as may happen in a \emph{dependent} type system,
where program terms may appear in types)
can spoil good properties of the type system,
like type soundness or the decidability of type checking and inference.

Consider the inductive type $\mathsf{List} \; A$ of lists with entries of type $A$.
Such a list is either nil ($[]$) or a term $x$
of type $A$ together with a tail list $\mathit{xs}$ (that is, $x :: \mathit{xs}$).
In a typed functional language like Haskell or OCaml,
the programmer can define such a type by specifying its constructors:
\[
  \arrayenvl{
    \textkw{data} \; \textsf{List} \; A \; \textkw{where} \\
    ~~~~[] : \textsf{List} \; A \\
    ~~~~(::) : A \to \textsf{List} \; A \to \textsf{List} \; A
  }
\]
Suppose we define, by pattern matching, the function \textsf{get},
that takes a list $\mathit{xs}$ and a natural number $y$,
and returns the $y$th element of $\mathit{xs}$
(where the first element is numbered zero):
\[
  \arrayenvl{
    \textsf{get} \; \texttt{[]} \; y = \textkw{error}\text{ ``Out of bounds''} \\
    \textsf{get} \; (\mathit{x} :: \mathit{xs}) \; \textsf{zero} = \mathit{x} \\
    \textsf{get} \; (x :: \mathit{xs}) \; (\xsucc \; y) = \textsf{get} \; \mathit{xs} \; y
  }
\]
A conventional type system has no issue checking \textsf{get}
against, say, the type $\textsf{List} \; A \to \textsf{Nat} \to A$ (for any type $A$),
but \textsf{get} is unsafe because it throws an out-of-bounds error
when the input number is greater than or equal to the length of the input list.
If it should be impossible for \textsf{get} to throw such an error,
then \textsf{get} must have a type
where the input number is restricted to natural numbers
strictly less than the length of the input list.
Ideally, the programmer would simply refine the type of \textsf{get},
while leaving the program alone (except, perhaps, for omitting the first clause).

This,
in contrast to dependent types \cite{Martin-Lof84},
is the chief aim of \emph{refinement types} \cite{Freeman91}:
to increase the expressive power of a pre-existing (unrefined) type system,
while keeping the latter's good properties,
like type soundness and, especially, decidability of typing,
so that programmers are not too burdened
with refactoring their code
or manually providing tedious proofs.
In other words,
the point of refinement types is to increase the expressive power of a given type system
while maintaining high automation (of typing for normal programs),
whereas the point of dependent types is to be maximally expressive
(even with the ambitious aim of expressing all mathematics),
at the cost of automation
(which dependent type system designers may try to increase after the fact of high expressivity).

To refine \textsf{get}'s type so as to rule out, statically,
run-time out-of-bounds errors,
we need to compare numbers against list lengths.
Thus, we refine the type of lists by their length:
$\comprehend{\nu : \textsf{List} \; A}{\textsf{len} \; \nu = n}$,
the type of lists $\nu$ of length $n$.
This type looks a bit worrying, though,
because the measurement, $\textsf{len} \; \nu = n$,
seems to use a recursive program, \textsf{len}.
The structurally recursive 
\[
  \arrayenvl{
    \textsf{len} \; [] = 0 \\
    \textsf{len} \; (x :: \mathit{xs}) = 1 + \textsf{len} \; \mathit{xs}
  }
\]
happens to terminate when applied to lists,
but there is no general algorithm
for deciding whether an arbitrary computation terminates
\cite{Turing36}.
As such,
we would prefer not to use ordinary recursive programs at all
in our type refinements.
Indeed, doing so would seem to violate a phase distinction%
\footnote{A language has a \emph{phase distinction}
if it can distinguish aspects that are relevant at run time
from those that are relevant only at compile time.}
\cite{Moggi89, Harper90-phase-distinction}
between \emph{static} (compile time) specification and \emph{dynamic} (run time) program,
which %
is almost
indispensable for \emph{decidable} typing.

The refinement type system Dependent ML (DML) \cite{XiThesis}
provides a phase distinction in refining ML by an index domain which has no run-time content.%
\footnote{In today's context, ``Refinement ML'' might seem a more appropriate name than Dependent ML.  But when DML was invented, ``refinement types'' referred to \emph{datasort} refinement systems; the abstract of \citet{XiThesis} describes DML as ``another attempt towards refining\dots type systems \dots, following the step of refinement types (Freeman and Pfenning \citeyear{Freeman91}).''}
Type checking and inference in DML is only decidable when
it generates constraints whose satisfiability is decidable.
In practice, DML \emph{did} generate decidable constraints,
but that was not guaranteed by its design.
DML's distinction between indexes and programs allows it to support
refinement types in the presence of computational effects
(such as nontermination, exceptions, and mutable references)
in a relatively %
straightforward manner.
Further, the index-program distinction clarifies how to give a denotational semantics:
a refinement type denotes a subset of what the type's erasure (of indexes) denotes
and a program denotes precisely what its erasure denotes \cite{Mellies15-functors}.
Dependent type systems, by contrast, do not have such an erasure semantics.

It seems liquid type systems \cite{Rondon08:LT, Kawaguchi09, Vazou13, Vazou14}
achieve highly expressive, yet sound and decidable
\emph{recursive refinements} \cite{Kawaguchi09}
of inductive types
by a kind of phase distinction: namely,
by restricting the recursive predicates of specifications
to terminating \emph{measures} (like \textsf{len})
that soundly characterize,
in a theory decidable by off-the-shelf tools like SMT solvers,
the static structure of inductive types.
Unlike DML, liquid typing can, for example,
use the measure of whether a list of natural numbers is in increasing order,
while remaining decidable.
However, liquid typing's lack of index-program distinction
makes it unclear how to give it a denotational semantics,
and has also led to subtleties involving the interaction between effects and evaluation strategy
(we elaborate later in this section and \Secref{sec:overview}).
\citet{Vazou14} appear to provide a denotational semantics in Section 3.3,
but this is not really a denotational semantics in the sense we intend,
because it is defined in terms of an operational semantics
and not a separate and well-established mathematical model (such as domain theory).

Let's return to the $\mathsf{get}$ example.
Following the tradition of index refinement \cite{XiThesis},
we maintain a phase distinction by syntactically distinguishing
index terms, which can safely appear in types,
from program terms, which cannot.
In this approach, we want to check \textsf{get} against a more informative type
\[
  \alltype{l : \kindnat} \underbrace{\comprehend{\nu : \textsf{List} \; A}{\textsf{len} \; \nu = l}}_{\vectype{(A)}{(l)}} \to \comprehend{\nu : \textsf{Nat}}{\nu < l} \to A
\]
quantifying over \emph{indexes} $l$ of \emph{sort} $\kindnat$ (natural numbers)
and requiring the accessing number to be less than $l$.
However, this type isn't quite right,
because \textsf{Nat} is a type and $\kindnat$ is a sort,
so writing ``$\nu < l$'' confounds our phase distinction between programs and indexes.
Instead, the type should look more like
\[
  \alltype{l : \kindnat} \comprehend{\nu : \textsf{List} \; A}{\textsf{len} \; \nu = l} \to \comprehend{\nu : \textsf{Nat}}{\textsf{index} \; \nu < l} \to A
\]
where
\[
  \arrayenvl{
    \textsf{index} \; \textsf{zero} = 0 \\
    \textsf{index} \; (\textsf{succ} \; y) = 1 + \textsf{index} \; y
  }
\]
computes the index term of sort $\kindnat$
that corresponds to a program term of type \textsf{Nat},
by a structural recursion homologous to that of \textsf{len}.
The third clause of \textsf{get} has a nonempty list as input,
so its index (length) must be $1 + m$ for some $m$;
the type checker assumes $\textsf{index} \, (\textsf{succ} \, y) < 1 + m$;
by the aforementioned homology, these constraints are again
satisfied at the recursive call ($\mathsf{index} \, y < m$),
until the second clause returns ($0 < 1 + m'$).
The first clause of \textsf{get} is impossible,
because no natural number is less than zero.
We can therefore
safely remove this clause, or (equivalently) replace \textkw{error} with \unreachable,
which checks against any type under a logically inconsistent context,
such as $l:\kindnat, n:\kindnat, l = 0, n < l$ in this case.
\[
  \arrayenvl{
    \textsf{get} : \alltype{l, n : \kindnat} \comprehend{\nu : \textsf{List} \; A}{\textsf{len} \; \nu = l} \to \comprehend{\nu : \textsf{Nat}}{\textsf{index} \; \nu = n} \land (n < l) \to A \\
    \textsf{get} \; \texttt{[]} \; y = \unreachable
         ~~~~~~~~~~~~~~~~~~~~~~~~~~~~~~~~~~~
         \text{{-}{-} }l=0\text{ and }n:\kindnat\text{ so }n \nless l  \\
    \textsf{get} \; (\mathit{x} :: \mathit{xs}) \; \textsf{zero} = \mathit{x} \\
    \textsf{get} \; (x :: \mathit{xs}) \; (\xsucc \; y) = \textsf{get} \; \mathit{xs} \; y
  }
\]
Applying \textsf{get} to a list and a number determines the indexes $l$ and $n$.
We say that $l$ and $n$ are \emph{value-determined}
(here by applying the function to values).
If (perhaps in a recursive call) \textsf{get} is called with an empty list $[]$
and a natural number,
then $l$ is determined to be $0$,
and since no index that is both negative and a natural number exists,
no out-of-bounds error can arise by calling \textsf{get}.
(Further, because $l : \kindnat$ strictly decreases at recursive calls,
calling $\mathsf{get}$ terminates.)

While this kind of reasoning about \textsf{get}'s type refinement
may seem straightforward,
how do we generalize it to recursion over any algebraic datatype (ADT)?
What are its logical and semantic ingredients?
How do we use these ingredients to concoct a type system
with decidable typing, good (localized) error messages and so on,
while also keeping its metatheory relatively stable or maintainable
under various extensions or different evaluation strategies?

Type systems that can do this kind of reasoning automatically,
especially in a way that can handle any evaluation strategy,
are hard to design correctly.
Indeed,
the techniques used in the original (call-by-value) liquid type system(s) \cite{Rondon08:LT, Kawaguchi09}
had to be modified for Haskell, essentially because of Haskell's
\emph{call-by-name} %
evaluation order \cite{Vazou14}.
The basic issue was that binders can bind in (static) refinement predicates,
which is fine when binders only bind values (as in call-by-value),
but not when they bind computations which may not terminate
(as in call-by-name).
Liquid Haskell regained (operational) type soundness by introducing ad hoc restrictions
that involve approximating whether binders terminate,
and using the refinement logic to verify termination.

We design a foundation on which to build liquid typing features
that allows us to establish clear (semantic) soundness results,
as well as the completeness of a decidable bidirectional typing algorithm.
The main technique facilitating this is focusing,
which we combine with bidirectional typing and value-determined indexes
(the latter being a key ingredient to make measures work).
In other words, this paper is a first step toward reconciling DML and Liquid Haskell,
using the proof-theoretic technique of focusing.

\citet{Andreoli92:focusing} introduced \emph{focusing}
to reduce the search space for proofs (programs) of logical formulas (types),
by exploiting the property that some inference rules are invertible
(the rule's conclusion implies its premises).
In relation to functional programming,
focusing has been used, for example,
to explain features such as pattern matching \cite{Krishnaswami09:pattern-matching},
to explain the interaction between evaluation order and effects \cite{Zeilberger09},
and to reason about contextual program equivalence \cite{Rioux20}.
Focusing has been applied to design a union and intersection refinement typing algorithm \cite{Zeilberger09}.
As far as we know, until now focusing has not been used to design an \emph{index} refinement typing algorithm.
By focusing on the typing of function argument lists and results,
our focused system guarantees that value-determined existential indexes (unification variables)
are solved before passing constraints to an SMT solver.
For example, %
when our system infers a type for
$\mathsf{get} ([3,1,2], 2)$,
we first use the top-level annotation of $\mathsf{get}$ to synthesize the type
\[
  \downshift{\left(\alltype{l, n : \kindnat} \comprehend{\nu : \textsf{List} \; A}{\textsf{len} \; \nu = l} \to \comprehend{\nu : \textsf{Nat}}{\textsf{index} \; \nu = n} \land (n < l) \to \upshift{A}\right)}
\]
(in which we have added polarity shifts $\downshift{-}$ and $\upshift{-}$ arising from focusing).
The downshift $\downshift{-}$ takes a negative type to a positive type of suspended computations.
Second,
we check the argument list ($[3,1,2], 2$) against the negative (universally quantified) type.
The upshift $\upshift{-}$
takes a positive type $A$ to negative type $\upshift{A}$ (computations returning a value of type $A$).
In typechecking the argument list, the upshift signifies the end of a (left) focusing stage,
at which point the first argument value $[3,1,2]$ will have determined $l$ to be $3$
and the second argument value $2$ will have determined $n$ to be $2$,
outputting an SMT constraint without existentials: $2 < 3$.

\citet{Levy04} introduced
the paradigm and calculus \emph{call-by-push-value} (CBPV)
which puts both call-by-name and call-by-value on equal footing
in the storm of computational effects (such as nontermination).
CBPV subsumes both call-by-name (CBN)
and call-by-value (CBV) functional languages,
because it allows us to encode both %
via
type discipline.
In particular, CBPV polarizes types into (positive) value types $P$
and (negative) computation types $N$,
and provides polarity shifts $\upshift{P}$ (negative) and $\downshift{N}$ (positive);
the monads functional programmers use to manage effects arise as the composite $\downshift{\upshift{-}}$.
These polarity shifts are the same as those arising from focusing.
CBPV can be derived logically by way of focalization \cite{EspiritoSanto17},
which we did in \sysname.
Focalized CBPV is a good foundation for a refinement typing algorithm:
designing refinement typing algorithms is challenging and sensitive to effects and evaluation strategy,
so it helps to refine a language that makes evaluation order explicit.
We leverage focusing and our technique of value-determined indexes (a novelty in the DML tradition)
to guarantee (like Liquid Haskell) the generation of SMT-solvable constraints.

Bidirectional typing \cite{Pierce00}
systematizes the difference between input (for example, type checking)
and output (for example, type inference),
and seems to fit nicely with focused systems \cite{Dunfield21-bidir-survey}.
Bidirectional typing has its own practical virtues:
it is easy to implement (if inputs and outputs fit together properly, that is, if the system is well-moded);
it scales well
(to refinement types, higher-rank polymorphism \cite{Dunfield19}, subtyping, effects---and so does CBPV);
it leads to localized error messages;
and it clarifies where type annotations are needed,
typically in reasonable positions (such as at the top level)
that are helpful as machine-checked documentation.
In \sysname, annotations are needed only for recursive functions (to express termination refinements)
and top-level definitions.

A focused and bidirectional approach therefore
appears suitable, both theoretically and practically,
for systematically designing and implementing
an expressive language of type refinement
that can handle any evaluation strategy and effect.
We show that bidirectional typing and logical focusing
work very well together at managing
the complex flow of information pertaining to indexes of recursive data.
In particular,
value-determined existential indexes of input types are solved within focusing stages,
ultimately leading to the output of constraints (and types)
in the quantifier-free fragment solvable by SMT solvers.

\paragraph{Contributions}

Our two key contributions are \emph{both} a declarative/logical/semantic \emph{and} an algorithmic account
of recursive, index-based refinement of algebraic data types.
For the logical account,
we design a declarative type system in a
\emph{bidirectional} and \emph{focused} style,
resulting in a system
with clear denotational semantics and soundness proofs,
and which is convenient for type theorists of programming languages.
The declarative system conjures index solutions to existentials.
For the algorithmic account,
we design a type system similar to the declarative one but solving all existentials,
and prove it is decidable,
as well as sound and complete.
We contribute:

\begin{itemize}
\item A polarized declarative type system, including (polarized) subtyping,
universal types, existential types,
and index refinements with ordinary SMT constraints,
as well as (nullary) recursive predicates on inductive data
(which in ongoing work we are extending to multi-argument measures,
which can express, for example, lists of increasing integer elements).
\item A proof that
  declarative typing is stable under substitution, which requires proving, among other things,
  that subtyping is transitive and typing subsumption is admissible.
\item A clear denotational semantics of the declarative system, based on elementary domain theory.
\item A type soundness proof with respect to our denotational semantics, which implies, relatively easily,
  both the refinement system's logical consistency
  and total correctness---even if the programs are \emph{non-structurally} recursive.
  To prove type soundness, we prove that value-determined indexes are sound:
  that is, semantic values uniquely determine value-determined indexes, semantically speaking
  (in particular, see \Lemref{lem:liq-value-determined-soundness}).
\item A polarized subtyping algorithm, together with proofs that it is sound, complete and decidable.
\item A polarized typing algorithm, together with proofs that it is sound, complete and decidable.
  Completeness relies on the combination of our novel technique of \emph{value-determined} indexes,
  focusing, and bidirectional typing. In particular, \Lemref{lem:liq-typing-solves-val-det}
  implies that all existential variables are solved by the algorithm.
\end{itemize}
We relatively easily obtain
both semantic and algorithmic results for a realistic language
essentially by applying just one technique (based on fundamental logical principles):
focusing.

All proofs are in the appendix.

\section{Overview}
\label{sec:overview}

This paper is a first step toward reconciling Dependent ML and Liquid Haskell.
The main thing we get from DML is the index-program distinction.
Liquid Haskell provides or inspires three things.
First, the observation of difficulties with effects and evaluation order inspired our use of CBPV.
Second, we study (nullary) measures (supporting multi-argument measures is ongoing work).
Third, our technique of value-determined indexes was inspired by the observation
that variables appearing in liquid refinements correspond to inputs or outputs of functions.

Before diving into the details of our type system, %
we give an overview of the central
logical, semantic, and algorithmic issues informing its design.
The main technique we use %
to easily support both semantic and algorithmic results %
is focalization.

\paragraph{Refinement typing, evaluation strategy, and computational effects}

The interactions between
refinement typing (and other fancy typing),
evaluation strategy, and computational effects
are a source of peril.
The combination of parametric polymorphism with
effects is often unsound \citep{Harper91MLcallccUnsound};
the value restriction in Standard ML recovers soundness in the presence of mutable references
by restricting polymorphic generalization to syntactic values
\citep{Wright95:value-restriction}.
The issue was also not peculiar to polymorphism:
\citet{Davies00icfpIntersectionEffects}
discovered that a similar value restriction
recovers type soundness for
intersection refinement types and effects in call-by-value languages.
For union types, \citet{Dunfield03} obtained soundness
by an evaluation context restriction on union elimination.  %

For similar reasons,
Liquid Haskell was also found unsound in practice, and had to be patched;
we adapt an example \cite{Vazou14} demonstrating the discovered unsoundness:
\begin{verbatim}
  diverge :: Nat -> {v:Int | false}
  diverge x = diverge x

  safediv :: n:Nat -> {d:Nat | 0 < d} -> {v:Nat | v <= n}
  safediv n d = if 0 < d then n / d else error "unreachable"

  unsafe :: Nat -> Int
  unsafe x = let notused = diverge 1 in let y = 0 in safediv x y
\end{verbatim}
In typechecking \textkw{unsafe},
we need to check that the type of \textkw{y} (a singleton type of one value: $0$)
is a subtype of \textkw{safediv}'s second argument type
(under the context of the let-binding).
Due to the refinement of the let-bound \textkw{notused},
this subtyping generates a \emph{constraint} or \emph{verification condition}
of the form ``if false is true, then\dots''.
This constraint holds vacuously, implying that \textkw{unsafe} is safe.
But \textkw{unsafe} really is unsafe because Haskell evaluates \emph{lazily}:
since \textkw{notused} is not used, \textkw{diverge} is never called,
and hence \textkw{safediv} divides by zero (and crashes if uncaught).
\citet{Vazou14} recover type soundness and decidable typing
by restricting let-binding and subtyping,
using an operational semantics
to approximate whether or not expressions diverge,
and whether or not terminating terms terminate to a \emph{finite} value.

The value and evaluation context restrictions
seem like ad hoc ways to cope with the failure of simple typing rules
to deal with the interactions between effects and evaluation strategy.
However, \citet{Zeilberger09} explains the value and evaluation context restrictions
in terms of a logical view of refinement typing.
Not only does this perspective explain these restrictions,
it provides theoretical tools
for designing type systems
for functional languages with effects.
At the heart of Zeilberger's approach is the proof-theoretic
technique of \emph{focusing}, which we discuss near the end of this overview.
An important question %
we address
is whether
polarization and focusing can also help us understand Liquid Haskell's
restrictions on let-binding and subtyping:
basically, our let-binding rule requires the bound computation (negative type)
to terminate to a value (positive type).
In other words, focalized systems satisfy any necessary value (and covalue) restrictions by default.
We discuss this further in \Secref{sec:related}.

Focalization can also yield systems with good semantic properties under computational effects,
in particular, variants of call-by-push-value.

\paragraph{Refining call-by-push-value}

Call-by-push-value \cite{Levy04}
subsumes both call-by-value and call-by-name by polarizing
the usual terms and types of the $\lambda$-calculus
into a finer structure that can be used to encode both evaluation strategies
in a way that can accommodate computational effects:
\emph{value} (or \emph{positive}) types
(classifying terms which ``are'', that is, values $v$),
\emph{computation} (or \emph{negative}) types
(classifying terms which ``do'', that is, expressions $e$),
and polarity shifts $\upshift{-}$ and $\downshift{-}$
between them.
An upshift $\upshift{P}$ \emph{lifts} a (positive) value type $P$
up to a (negative) computation type of expressions that compute values (of type $P$).
A downshift $\downshift{N}$ pushes a (negative) computation type $N$
down into a (positive) value type of \emph{thunked}
or \emph{suspended} computations (of type $N$).
We can embed the usual $\lambda$-calculus function type $A \to_\lambda B$
(written with a subscript to distinguish it from the CBPV function type), for example,
into CBPV
(whose function types have the form $P \to N$ for positive $P$ and negative $N$)
so that it behaves like CBV, via the translation $\iota_{\text{CBV}}$
with $\iota_{\text{CBV}}(A \to_\lambda B)
= \downshift{(\iota_{\text{CBV}}(A) \to \upshift{\iota_{\text{CBV}}(B)})}$;
or so that it behaves like CBN, via the translation $\iota_{\text{CBN}}$
with $\iota_{\text{CBN}}(A \to_\lambda B)
= (\downshift{\iota_{\text{CBN}}(A)) \to \iota_{\text{CBN}}(B)}$.

Evaluation order is made explicit by CBPV type discipline.
Therefore, adding a refinement layer on top of CBPV
requires directly and systematically
dealing with the interaction between type refinement and evaluation order.
If we add this layer to CBPV correctly from the very beginning,
then we can be confident that our type refinement system
will be semantically well-behaved when extended with other computational effects.
The semantics of CBPV are well-studied and this helps us establish semantic metatheory.
In later parts of this overview,
we show the practical effects of refining our focalized variant of CBPV,
especially when it comes to algorithmic matters.

\paragraph{Type soundness, totality, and logical consistency}

The unrefined system underlying \sysname
has the computational effect of nontermination
and hence is not total.
To model nontermination,
we give the unrefined system an elementary domain-theoretic denotational semantics.
Semantic type soundness says that a syntactic typing derivation
can be faithfully interpreted as a semantic typing derivation, that is, a morphism in a mathematical category,
in this case a logical refinement of domains.
Semantic type soundness basically corresponds to syntactic type soundness
with respect to a big-step operational semantics.
While we don't provide an operational semantics in this paper,
we do prove a syntactic substitution lemma which would be a key ingredient
to prove that an operational semantics preserves typing
(beta reduction performs syntactic substitution).
The substitution lemma is also helpful to programmers
because it means they can safely perform program transformations and preserve typing.
Because the unrefined system is (a focalized variant of) CBPV,
proving type soundness is relatively straightforward.

In contrast to dependent types,
the denotational semantics of our refined system
is defined in terms of that of its \emph{erasure} (of indexes),
that is, its underlying, unrefined system.
A refined type denotes a logical subset of what its erasure denotes.
An \emph{unrefined} return type $\upshift{P}$ denotes either what $P$ denotes,
or divergence/nontermination.
A \emph{refined} return type $\upshift{P}$ denotes \emph{only} what $P$ denotes.
Therefore, our \emph{refined} type soundness result
implies that our refined system (without a \emph{partial} upshift type) enforces termination.
In we discuss how to extend the refined system (by a partial upshift type)
to permit divergence while keeping type soundness
(which implies partial correctness for partial upshifts).
Type soundness also implies logical consistency, because a logically inconsistent
refinement type denotes the empty set.
We also prove that syntactic substitution is semantically sound,
which would be a main lemma in proving that an operational semantics is equivalent
to our denotational semantics.

In \Secref{sec:semantics},
we discuss these semantic issues in more detail.

\paragraph{Algebraic data types and measures}

A novelty of liquid typing is the use of \emph{measures}:
functions, defined on algebraic data types,
which may be structurally recursive,
but are guaranteed to terminate
and can therefore safely be used to refine the inductive types over which they are defined.
(In this paper, we only consider nullary measures.)

For example, consider the type $\mathsf{BinTree} \; A$ of binary trees
with terms of type $A$ at leaves:
\[
  \arrayenvl{
    \textkw{data} \; \textsf{BinTree} \; A \; \textkw{where} \\
    ~~~~\mathsf{leaf} : A \to \textsf{BinTree} \; A \\
    ~~~~\mathsf{node} : \textsf{BinTree} \; A \to \textsf{BinTree} \; A \to \textsf{BinTree} \; A
  }
\]
Suppose we want to refine $\mathsf{BinTree} \; A$ by the height of trees.
Perhaps the most direct way to specify this
is to measure it using a function \textsf{hgt} defined by structural recursion:
\[
  \arrayenvl{
    \mathsf{hgt} : \mathsf{BinTree} \; A \to \kindnat \\
    \mathsf{hgt} \; \mathsf{leaf} = 0 \\
    \mathsf{hgt} \; (\mathsf{node} \; t \; u) = 1 + \mathsf{max}(\mathsf{hgt}(t), \mathsf{hgt}(u))
  }
\]

As another example, consider an inductive type $\mathsf{Expr}$ of
expressions in a CBV lambda calculus:
\[
  \arrayenvl{
    \textkw{data} \; \mathsf{Expr} \; \textkw{where} \\
    ~~~~\mathsf{var} : \mathsf{Nat} \to \mathsf{Expr} \\
    ~~~~\mathsf{lam} : \mathsf{Nat} \to \mathsf{Expr} \to \mathsf{Expr} \\
    ~~~~\mathsf{app} : \mathsf{Expr} \to \mathsf{Expr} \to \mathsf{Expr}
  }
\]
Measures need not involve recursion.
For example, if we want to refine the type $\mathsf{Expr}$
to expressions $\mathsf{Expr}$ that are values (in the sense of CBV, not CBPV),
then we can use $\mathsf{isval}$:
\[
  \arrayenvl{
    \mathsf{isval} : \mathsf{Expr} \to \mathbb{B} \\
    \mathsf{isval} \; (\mathsf{var} \; z) = \True \\
    \mathsf{isval} \; (\mathsf{lam} \; z \; \mathit{expr}) = \True \\
    \mathsf{isval} \; (\mathsf{app} \; \mathit{expr} \; \mathit{expr'}) = \False
  }
\]

Because $\mathsf{isval}$ isn't recursive and returns indexes,
it's safe to use it to refine $\mathsf{Expr}$ to
expressions that are CBV values:
$\comprehend{\nu : \mathsf{Expr}}{\mathsf{isval} \; \nu = \True}$.
But, as with \textsf{len} (\Secref{sec:introduction}),
we may again be worried about using the seemingly dynamic, recursively defined
$\mathsf{hgt}$ in a static type refinement.
Again,
we need not worry because \textsf{hgt},
like \textsf{len}, is a terminating function
into a decidable logic \cite{BarrettSMT}.
We can use it to specify that, say,
a height function defined by pattern matching on trees of type
$\comprehend{\nu : \mathsf{BinTree} \; A}{\mathsf{hgt} \; \nu = n}$
actually returns (the program value representing)
$n$ for any tree of height $n$.
Given the phase distinction between indexes (like $n$)
and programs,
how do we specify such a function type?
We use refinement type unrolling and singletons.

\paragraph{Unrolling and singletons}
Let's consider a slightly simpler function, \textsf{length},
that takes a list and returns its length:
\[
  \arrayenvl{
    \textsf{length} \; \texttt{[]} = \textsf{zero} \\
    \textsf{length} \; (x :: \mathit{xs}) = \xsucc \; (\textsf{length} \; \mathit{xs})
  }
\]
What should be the type specifying that \textsf{length}
actually returns a list's length?
The proposal
$\alltype{n:\kindnat} \textsf{List}(A)(n) \to \upshift{\textsf{Nat}}$
does not work because \textsf{Nat} has no information about the length $n$.
Something like
$\alltype{n:\kindnat} \textsf{List}(A)(n) \to \upshift{(n : \textsf{Nat})}$,
read literally as returning the index $n$,
would violate our phase distinction between programs and indexes.
Instead, we use a \emph{singleton} type
in the sense of \citet{XiThesis}:
a singleton type contains just those program terms (of the type's erasure),
that correspond to exactly one semantic index.
For example, given $n:\kindnat$,
we define the singleton type $\textsf{Nat}(n)$ (which may also be written $\textsf{Nat} \; n$) by
$\comprehend{\nu : \mathsf{Nat}}{\mathsf{index}\,{\nu} = n}$
where
\[
  \arrayenvl{
    \mathsf{index} : \mathsf{Nat} \to \kindnat \\
    \mathsf{index} \; \mathsf{zero} = 0 \\
    \mathsf{index} \; (\succ{x}) = 1 + \mathsf{index}(x)
  }
\]
specifies the indexes (of sort $\kindnat$) 
corresponding to program values of type $\nat$.

How do we check \textsf{length}
against $\alltype{n:\kindnat} \textsf{List}(A)(n) \to \upshift{(\natsing{n})}$?
In the first clause,
the input \texttt{[]} has type $\textsf{List}(A)(n)$ for some $n$,
but we need to know $n=0$ (and that the index of \textsf{zero} is $0$).
Similarly, we need to know $x :: \mathit{xs}$ has length $n = 1 + n'$
where $n' : \kindnat$ is the length of $\mathit{xs}$.
To generate these constraints,
we use an \emph{unrolling} judgment (\Secref{sec:unrolling})
that unrolls a refined inductive type.
Unrolling puts the type's refinement constraints,
expressed by \emph{asserting} and \emph{existential} types,
in the structurally appropriate positions.
An asserting type is written $P \land \phi$ (read ``$P$ with $\phi$''),
where $P$ is a (positive) type and $\phi$ is an index proposition.
If a term has type $P \land \phi$, then the term has type $P$ and also $\phi$ holds.
(Dual to asserting types, we have the \emph{guarded} type $\phi \implies N$,
which is equivalent to $N$ if $\phi$ holds, but is otherwise useless.)
We use asserting types to express that index equalities like $n=0$
hold for terms of inductive type.
We use existentials to quantify over indexes that characterize the refinements
of recursive subparts of inductive types, like $n'$.
For example, modulo a small difference (see \Secref{sec:unrolling}),
$\textsf{List}(A)(n)$ unrolls to
\[
  \big(
    1 \land (n=0)
  \big)
  +
  \big(A
    \times
    \extype{n':\kindnat} 
        \comprehend{\nu : \mathsf{List} \; A}{\mathsf{len}\,{\nu} =_{} n'}
        \land
        (n = 1 + n')
  \big)
\]
That is, to construct an $A$-list of length $n$,
the programmer (or library designer) can either left-inject a unit value,
\emph{provided the constraint $n=0$ holds},
or right-inject a pair of one $A$ value and a tail list,
\emph{provided that $n'$, the length of the tail list, is $n - 1$}
(the equations
$n = 1 + n'$
and
$n - 1 = n'$
are equivalent).
These index constraints are not a syntactic part of the list itself.
That is,
a term of the above refined type is also a term of the type's \emph{erasure} (of indexes):
\[
  1 + (\erase{A} \times (\mathsf{List} \; \erase{A}))
\]
where $\erase{-}$ erases indexes.
Dual
to \emph{verifying} refined inductive values,
pattern matching on refined inductive values,
such as in the definition of \textsf{length},
allows us to \emph{use} the index refinements
locally in the bodies of the various clauses
for different patterns.
Liquid Haskell similarly extracts refinements of data constructors for use in pattern matching.

The shape of the refinement types generated by our unrolling judgment
(such as the one above)
is a judgmental and refined-ADT version of the fact that
generalized ADTs (GADTs) can be desugared
into types with equalities and existentials that express constraints
of the return types for constructors
\cite{CheneyHinze03:phantom-types, Xi03:guarded}.
It would be tedious and inefficient for the programmer to work directly with terms
of types produced by our unrolling judgment,
but we can implement (in \sysname)
singleton refinements of base types and common functions on them, such as
addition, subtraction, multiplication, division, and the modulo operation
on integers,
and build these into the surface language used by the programmer,
similarly to the implementation of Dependent ML \cite{XiThesis}. 

\paragraph{Inference and subtyping}
For a typed functional language to be practical,
it must support some degree of inference,
especially for function application (to eliminate universal types)
and constructing values (to introduce existential types).
For example, to pass a value to a function,
its type must be compatible with the function's argument type,
but it would be burdensome
to make programmers always have to prove this compatibility.
In our setting, for example, if $x : \textsf{Nat}(3)$
and $f : \downshift{(\alltype{a:\kindnat} \textsf{Nat}(a) \to \upshift{P})}$,
then we would prefer to write $f\;x$ rather than $f\;[3]\;x$,
which would quickly make our programs incredibly---and unnecessarily---verbose.

Omitting index and type annotations, however,
has significant implications. %
In particular, we need a mechanism to instantiate indexes
somewhere in our typing rules:
for example, if
$g : \downshift{(\downshift{(\textsf{Nat}(4+b) \to \upshift{P})} \to N)}$
and
$h : \downshift{((\extype{a:\kindnat} \textsf{Nat}(a)) \to \upshift{P})}$,
then to apply $g$ to $h$,
we need to know $\textsf{Nat}(4+b)$
is compatible with $\extype{a:\kindnat} \textsf{Nat}(a)$,
which requires instantiating the bound $a$ to a term logically equal to $4+b$.
\Sysname does this kind of instantiation via subtyping,
which refinement types naturally give rise to: a type refinement is essentially a subtype of its erasure.
Index instantiations are propagated locally across adjacent nodes in the syntax tree,
similarly to \citet{Pierce00}.
(Liquid typing allows for more inference, including inference of refinements based on templates,
which provides additional convenience for programmers,
but we do not consider this kind of inference in this paper.)

We polarize subtyping into two, mutually recursive,
positive and negative relations
$\judgesub[+]{\Theta}{P}{Q}$ and $\judgesub[-]{\Theta}{N}{M}$
(where $\Theta$ is a \emph{logical context} including index propositions).
The algorithmic versions of these only introduce existential variables
in positive supertypes and negative subtypes,
guaranteeing they can always be solved by indexes
without any existential variables.
We delay checking constraints until the end of certain,
logically designed stages (the \emph{focusing} ones, as we will see),
when all of their existential variables are guaranteed to have been solved.

\paragraph{Value-determined indexes and type well-formedness}

Like DML, we have an index-program distinction,
but unlike DML and like Liquid Haskell, we want to guarantee SMT solvable constraints.
We accomplish this with our technique of value-determined indexes.
To guarantee that our algorithm can always instantiate quantifiers,
we restrict quantification to indexes appearing in certain positions within types:
namely,
those that are uniquely determined
(semantically speaking)
by values of the type,
both according to a measure and before crossing a polarity shift
(which in this case marks the end of a focusing stage).
For example, in
$\comprehend{\nu : \textsf{List} \; A}{\textsf{len}\,{\nu} = b}$,
the index $b$ is uniquely determined by values of that type:
the list $[x, y]$ uniquely determines $b$ to be $2$ (by the length measure).
This value-determinedness restriction on quantification has served to explain
why a similar restriction in the typing algorithm of Flux (Liquid Rust)
seemed to work well in practice \cite{Lehmann23}.

We make this restriction in the type well-formedness judgment,
which outputs a context $\Xi$ tracking value-determined indexes;
well-formed types can only quantify over indexes in $\Xi$.
For example,
$\extype{b:\kindnat} \comprehend{\nu : \textsf{List} \; A}{\textsf{len}\,{\nu} = b}$
is well-formed.
The variables in $\Xi$ also pay attention to type structure:
for example, a value of a product type is a pair of values,
where the first value determines all $\Xi_1$ (for the first type component)
and the second value determines all $\Xi_2$ (second type component),
so the $\Xi$ of the product type is their union $\Xi_1 \cup \Xi_2$.
We also take the union for function types $P \to N$,
because index information flows through argument types toward the return type,
marked by a polarity shift.

By emptying $\Xi$ at shift types,
we prevent lone existential variables from being introduced at a distance,
across polarity shifts.
In practice, this restriction on quantification is not onerous,
because most functional types that programmers use are, in essence, of the form
\[
  \alltype{\cdots} P_1 \to \cdots \to P_n \to \upshift{\extype{\cdots} Q}
\]
where the ``$\forall$'' quantifies over indexes of argument types $P_k$
that are uniquely determined by argument values,
and the ``$\exists$'' quantifies over indexes of the return type
that are determined by (or at least depend on) fully applying the function
and thereby constructing a value to return.
The idea of this restriction was inspired by liquid types
because they implicitly follow it:
variables appearing in liquid type refinements
must ultimately come from arguments $x$ to dependent functions $x\!:\!A \to B$ and their return values
(however, these are not explicitly index variables).

Types that quantify only across polarity shifts
tend to be empty, useless, or redundant.
The ill-formed type $\alltype{n:\kindnat} 1 \to \upshift{\natsing{n}}$
is empty because no function returns all naturals when applied to unit.
A term of ill-formed type $\extype{m:\kindnat} \downshift{(\natsing{m} \to \upshift{\mathsf{Bool}}})$
can only be applied to an unknown number,
which is useless because the number is unknown.
The ill-formed type $\extype{n:\kindnat} \upshift{\downshift{\natsing{n}}}$
is redundant because
it is semantically equivalent to $\downshift{\upshift{\extype{n:\kindnat} \natsing{n}}}$
(which does not quantify across a polarity shift),
and similarly $\alltype{n:\kindnat} \upshift{\downshift{(\natsing{n} \to \upshift{\natsing{n}})}}$
is semantically equivalent to
$\upshift{\downshift{(\alltype{n:\kindnat} \natsing{n} \to \upshift{\natsing{n}})}}$.
Some refinements are \emph{not} value-determined
but useful nonetheless,
such as dimension types \citep{Kennedy94esop,DunfieldThesis}
which statically check that dimensions are used consistently (minutes can be added to minutes,
but not to kilograms) but do not store the dimensions at run time.
In this paper, we do not consider these non-value-determined refinements,
and Liquid Haskell does not support them either.
Our value-determinedness restriction on type well-formedness,
together with focusing,
is very helpful metatheoretically,
because it means that our typing algorithm only introduces---and
is guaranteed to solve---existential variables for indexes
within certain logical stages.
For example,
consider checking a list against
$\extype{b:\kindnat} \comprehend{\nu : \textsf{List} \; A}{\textsf{len}\,{\nu} = b}$.
An existential variable $\bhat$ for $b$ is generated,
and we check the unrolled list against the unrolling of
$\comprehend{\nu : \textsf{List} \; A}{\textsf{len}\,{\nu} = \bhat}$.
A solution to $\bhat$ will be found within value typechecking (the right-focusing stage),
using the invariant that no measure (such as $\textsf{len}$) contains any existential variables.
Similarly, applying a function with universal quantifiers will solve all existential variables
arising from these quantifiers by the end of a left-focusing stage,
which typechecks an argument list.

\paragraph{Focusing, CBPV, and bidirectional typing}

In proof theory,
the technique of \emph{focusing} \cite{Andreoli92:focusing}
exploits invertibility properties of logical formulas (types),
as they appear in inference rules (typing rules),
in order to rule out many redundant proofs.
Having fewer possible proofs makes proof search more tractable.
\citet{Brock-Nannestad15} and \citet{EspiritoSanto17}
study the relation between CBPV and focusing:
each work provides a focused calculus that is essentially the same as CBPV,
``up to the question of $\eta$-expansion'' \cite{Brock-Nannestad15}.
\Sysname is also a focused variant of CBPV;
in fact, it arises from a certain focalization (and bidirectionalization) of ordinary intuitionistic logic.

An inference rule is \emph{invertible} if its conclusion implies its premises.
For example, in intuitionistic logic, the right rule for implication is invertible
because its premise $\Gamma, A |- B$ can be derived from its conclusion $\Gamma |- A \to B$:
\[
  \Infer{(Cut)}
  {\Infer{(Weaken)}{\Infer{\text{(Assume $\to$R conclusion)}}{}{\Gamma |- A \to B}}{\Gamma, A |- A \to B} \\ \Infer{\text{$\to$L}}{\Infer{}{}{\Gamma, A |- A} \\ \Infer{}{}{\Gamma, A, B |- B}}{\Gamma, A, A \to B |- B}}
  {\Gamma, A |- B}
\]
However, both right rules for disjunction, for example, are not invertible,
which we can prove with a counterexample:
$A_1 + A_2 |- A_1 + A_2$
but $A_1 + A_2 \nvdash A_1$ and $A_1 + A_2 \nvdash A_2$.
In a sequent calculus,
\emph{positive} formulas have invertible left rules
and \emph{negative} formulas have invertible right rules.
A \emph{weakly} focused sequent calculus
eagerly applies non-invertible rules as far as possible
(in either \emph{left-} or \emph{right-focusing} stages);
a \emph{strongly} focused sequent calculus does too,
but also eagerly applies invertible rules as far as possible
(in either \emph{left-} or \emph{right-inversion} stages).
There are also \emph{stable} stages (or moments)
in which a decision has to be made between focusing on the left, or on the right
\cite{EspiritoSanto17}.
The decision can be made explicitly via proof terms (specifically, cuts):
in \sysname, a principal task of let-binding, a kind of cut, is to focus on the left
(to process the list of arguments in a bound function application);
and a principal task of pattern matching, another kind of cut, is to focus on the right
(to decompose the value being matched against a pattern).

From a Curry--Howard view, let-binding and pattern matching
are different kinds of cuts.
The cut formula $A$---basically, the type being matched or let-bound---must be
\emph{synthesized} (inferred) as an output
(judgmentally, $\cdots => A$)
from \emph{heads} $h$ (variables and annotated values)
or \emph{bound expressions} $g$
(function application and annotated returner expressions);
and ultimately, the outcomes of these cuts in \sysname are synthesized.
But all other program terms are \emph{checked}
against input types $A$:
judgmentally, $\cdots <= A \cdots$ or $\cdots [A] |- \cdots$.
In this sense,
both our declarative and algorithmic type systems are \emph{bidirectional}
\cite{Dunfield21-bidir-survey}.

In inversion stages, that is, expression typechecking
(where a negative type is on the right of a turnstile)
and pattern matching (where a positive type is on the left of a turnstile),
refinements often need to be \emph{extracted} from types in order to be used.
For example, suppose we want to check the expression $\fun{x} \Return{x}$
against the type $(1 \land \False) \to \upshift{(1 \land \False)}$,
which is semantically equivalent to $\False \implies 1 \to \upshift{(1 \land \False)}$.
We need to extract $\False$ (of $(1 \land \False) \to \cdots$)
and put it in the logical context
so that we can later use it (in typechecking the value $x$)
to verify $\False$ (of $\cdots \upshift{(1 \land \False)}$).
If we instead put $x : 1 \land \False$ in the program context,
then $\False$ would not be usable
(unless we can extract it from the program context,
but its simpler to extract from types as soon as possible
rather than to extend the extraction judgment or to add another one)
and typechecking would fail (it should succeed).
Similarly, since subtyping may be viewed as implication,
index information from positive subtypes or negative supertypes
needs to be extracted for use.
Declaratively, it is okay not to extract eagerly at polarity shifts in subtyping
(the subtyping rules that extract are invertible),
but it seems necessary in the algorithmic system.

Our declarative system includes two focusing stages,
one (value typechecking) for positive types on the right of the turnstile ($|-$),
and the other (spine typing) for negative types on the left.
Our algorithmic system closely mirrors the declarative one,
but does not conjure index instantiations or witnesses
(like $t$ in DeclSpine$\forall$ below),
and instead systematically introduces and solves existential variables
(like solving the existential variable $\ahat$ as $t$ in AlgSpine$\forall$ below),
which we keep in algorithmic contexts $\Delta$.
\begin{mathpar}
  \Infer{DeclSpine$\forall$}
  {
    \judgeterm{\Theta}{t}{\tau}
    \\
    \judgespine{\Theta}{\Gamma}{s}{[t/a]N}{\upshift{P}}
  }
  {
    \judgespine{\Theta}{\Gamma}{s}{\alltype{a:\tau}{N}}{\upshift{P}}
  }
  \and
  \Infer{AlgSpine$\forall$}
  {
    \algspine{\Theta; \Delta, \ahat : \tau}{\Gamma}{s}{[\ahat/a]{N}}{\upshift{P}}{\chi}{\Delta', \hypeq{\ahat}{\tau}{t}}
  }
  {
    \algspine{\Theta; \Delta}{\Gamma}{s}{\alltype{a:\tau}N}{\upshift{P}}{\chi}{\Delta'}
  }
\end{mathpar}
For example, applying a function of type $\alltype{b:\kindnat} \vectype{(\nat)}{(b)} \to \cdots$
to the list $[4,1,2]$ should solve $b$ to an index semantically equal to $3$;
the declarative system guesses an index term (like $3+0$),
but the algorithmic system mechanically solves for it.

Our algorithmic right-focusing judgment has the form
$\algchk{\Theta; \Delta}{\Gamma}{v}{P}{\chi}{\Delta'}$,
where $\chi$ is an output list of typing constraints
and $\Delta'$ is an output context that includes index solutions to existentials.
Similarly, the left-focusing stage is
$\algspine{\Theta; \Delta}{\Gamma}{s}{N}{\upshift{P}}{\chi}{\Delta'}$;
it focuses on decomposing $N$
(the type of the function being applied),
introducing its existential variables for the arguments in the list $s$
(sometimes called a \emph{spine} %
\cite{Cervesato03}),
and outputting its guards to verify at the end of decomposition (an upshift).
These existential variables flow to the right-focusing stage
(value typechecking) and are solved there, possibly via subtyping.
Constraints $\chi$ are only checked right at the end of focusing stages,
when all their existential variables are solved.
For example,
consider our rule for (fully) applying a function $h$
to a list of arguments $s$:
\[
  \Infer{\AlgSpineApp}
  { 
    \algsynhead{\Theta}{\Gamma}{h}{\downshift{N}}
    \\
    \algspine{\Theta; \cdot}{\Gamma}{s}{N}{\upshift{P}}{\chi}{\cdot}
    \\
    \algneg{\Theta}{\Gamma}{\chi}
  }
  {
    \algsynexp{\Theta}{\Gamma}{h(s)}{\upshift{P}}
  }
\]

After synthesizing a thunk type $\downshift{N}$
for the function $h$ we are applying,
we process the entire list of arguments $s$,
until $N$'s return type $\upshift{P}$.
All (existential) value-determined indexes $\Xi_N$ of $N$ are guaranteed to be solved
by the time an upshift is reached,
and these solutions are eagerly applied to constraints $\chi$,
so that $\chi$ does not have existential variables
and is hence SMT-solvable ($\algneg{\Theta}{\Gamma}{\chi}$).
The polarization of CBPV helps guarantee all solutions have no existential variables.
Focusing stages introduce existential variables to input types,
which may appear initially as a positive supertype in the subtyping premise for typechecking (value) variables.
These existential variables are solved using the positive subtype, which never has existential variables.
Dually, negative subtypes may have existential variables, but negative supertypes never do.

\Sysname requires intermediate computations like $h(s)$
to be explicitly named and sequenced by
let-binding
(a kind of $A$-\emph{normal}
\cite{Flanagan93:CompilingWithContinuations}
or \emph{let-normal form}).
Combined with focusing,
this allows us to use
(within the value typechecking stage)
subtyping only in the typing rule for (value) variables.
This makes our subsumption rule syntax-directed,
simplifying and increasing the efficiency of our algorithm.
We nonetheless prove a general subsumption lemma,
which is needed to prove that substitution respects typing,
a %
key syntactic or operational correctness property.

Focusing also gives us pattern matching for free
\cite{Krishnaswami09:pattern-matching}:
from a Curry--Howard view, the left-inversion stage is pattern matching.
The (algorithmic\footnote{We give the algorithmic judgment to note existential variables $\Delta$ do not flow through it, or any of the non-focusing stages.}) left-inversion stage in \sysname is written
$\algchkmatch{\Theta}{\Gamma}{P}{\clauses{\pa}{e}{i}{I}}{N}$:
it decomposes the positive $P$ (representing the pattern being matched)
to the left of the turnstile
(written $\rhd$ to distinguish the algorithmic judgment from
the corresponding declarative judgment, which instead uses $|-$).
\Sysname is ``half'' strongly focused:
we eagerly apply right-invertible but not left-invertible rules.
This makes pattern matching in \sysname
resemble the original presentation of pattern matching in CBPV.
From a Curry--Howard view, increasing the strength of focusing
would permit nested patterns.

A pattern type can have index equality constraints,
such as for refined ADT constructors
(for example, that the length of an empty list is zero)
as output by unrolling.
By using these equality constraints,
we get a coverage-checking algorithm.
For example,
consider checking \textsf{get} (introduced in \Secref{sec:introduction})
against the type
\[
  \alltype{l, k : \kindnat} \comprehend{\nu : \textsf{List} \; A}{\textsf{len} \; \nu = l} \to (\comprehend{\nu : \textsf{Nat}}{\textsf{index} \; \nu = k} \land (k < l)) \to \upshift{A}
\]
At the clause
\[
  \textsf{get} \; \texttt{[]} \; y = \unreachable
\]
we extract a logically inconsistent context ($l:\kindnat, k:\kindnat, l = 0, k < l$),
which entails that \unreachable checks against any type.
Proof-theoretically,
this use of equality
resembles the elimination rule for Girard--Schroeder-Heister equality
\cite{girard-equality, schroeder-heister-equality}.

Bidirectional typing controls the flow of type information.
Focusing in \sysname directs the flow of index information.
The management of the flow of type refinement information,
via the stratification of both focusing and bidirectionality,
makes our algorithmic metatheory highly manageable.

\section{Example: Verifying Mergesort}
\label{sec:example}

We show how to verify a non-structurally recursive mergesort function in \sysname:
namely, that it terminates and returns a list with the same length as the input list
(to verify it returns an ordered list,
we need to extend \sysname with multi-argument measures, which is outside the scope of this paper).
We only consider sorting lists of natural numbers
$\nat$, defined as $\extype{n : \kindnat} \natsing{n}$.
For clarity, and continuity
with Sections \ref{sec:introduction} and \ref{sec:overview},
we sometimes use syntactic sugar %
such as clausal pattern-matching,
combining let-binding with pattern-matching on the let-bound variable,
using ``if-then-else'' rather than pattern-matching on boolean type,
and combining two or more pattern-matching expressions into one with a nested pattern
such as $x :: y :: \mathit{xs}$.

Given type $A$ and $n : \kindnat$,
we define $\vectype{(A)}{(n)}$ by $\comprehend{\nu : \textsf{List} \; A}{\textsf{len} \; \nu = n}$.
Modulo a small difference (see \Secref{sec:unrolling}),
our unrolling judgment unrolls $\textsf{List}(A)(n)$ to
\[
  \big(
    1 \land (n=0)
  \big)
  +
  \big(A
    \times
    \extype{n':\kindnat} 
        \comprehend{\nu : \mathsf{List} \; A}{\mathsf{len}\,{\nu} =_{} n'}
        \land
        (n = 1 + n')
  \big)
\]
which is a refinement of $1 + (\erase{A} \times \mathsf{List} \; \erase{A})$.
This is an unrolling of the inductive type,
not the inductive type itself, so we must roll values of it into the inductive type.
We use syntactic sugar: namely, $[]$ stands for $\roll{\inj{1} \unitexp}$
and $x :: \mathit{xs}$ stands for $\roll{\inj{2} \pair{x}{\mathit{xs}}}$.

Just as we need a natural number type associating natural number program values with natural number indexes,
we need a boolean type of values corresponding to boolean indexes.
To this end, define the measure
\[
  \arrayenvl{
    \mathsf{ixbool} : (1+1) \to \Booltype \\
    \mathsf{ixbool} \; \textkw{true} = \True \\
    \mathsf{ixbool} \; \textkw{false} = \False
  }
\]
Given $b:\Booltype$,
the singleton type of boolean $b$ is
$\mathsf{Bool}(b) = \comprehend{\nu : \bool}{\mathsf{ixbool} \; \nu = b}$.
Our unrolling judgment
(\Secref{sec:unrolling})
outputs the following type,
a refinement of the boolean type encoded as $1 + 1$:
\[
  \big(
    1 \land (b = \True)
  \big)
  +
  \big(
    1 \land (b = \False)
  \big)
\]
We encode $\textkw{true}$
as $\into{\inl{\unitexp}}$ which has type $\bool(\True)$,
and $\textkw{false}$
as $\into{\inr{\unitexp}}$ which has type $\bool(\False)$.
The boolean type $\mathsf{Bool}$ is defined as $\extype{b:\Booltype} \mathsf{Bool}(b)$.

Assume we have the following:
\[
  \renewcommand{\arraystretch}{1.1} \begin{array}[t]{r@{~~}c@{~~}l}
    \mathsf{add} &:& \downshift{(\alltype{m, n : \kindnat} \mathsf{Nat}(m) \to \mathsf{Nat}(n) \to \upshift{\mathsf{Nat}(m+n)})}
    \\
    \mathsf{sub} &:& \downshift({\alltype{m, n : \kindnat} (n \leq m) \implies \mathsf{Nat}(m) \to \mathsf{Nat}(n) \to \upshift{\mathsf{Nat}(m-n)})}
    \\
    \mathsf{div} &:& \downshift{(\alltype{m, n : \kindnat} (n \neq 0) \implies \natsing{m} \to \natsing{n} \to \upshift{\natsing{m \div n}})}
    \\
    \mathsf{lt} &:& \downshift{(\alltype{m, n : \kindnat} \mathsf{Nat}(m) \to \mathsf{Nat}(n) \to \upshift{\extype{b:\Booltype} \mathsf{Bool}(b) \land (b = (m < n))})}
    \\
    \mathsf{len} &:& \downshift{(\alltype{n : \kindnat} \vectype{(\nat)}{(n)} \to \upshift{\mathsf{Nat}(n)})}
    \\
    \mathsf{[]} &:& \vectype{(\nat)}{(0)}
    \\
    (::) &:& \downshift{(\alltype{n : \kindnat} \nat \to \vectype{(\nat)}{(n)} \to \upshift{\vectype{(\nat)}{(1 + n)}})}
 \end{array}
\]
The SMT solver Z3 \cite{Z3}, for example,
supports integer division (and modulo and remainder operators);
internally, these operations are translated to multiplication.
Here, we are considering natural number indexes,
but we can add the constraint $n \geq 0$ (for naturals $n$)
when translating them to integers in an SMT solver such as Z3.
Allowing integer division in general is not SMT decidable,
but for this example, $n$ is always instantiated to a constant ($2$), which is decidable.
Note that Z3 supports division by zero,
but our $\mathsf{div}$ has a guard requiring the divisor to be nonzero ($n \neq 0$),
so we need not consider this.
Division on naturals takes the floor of what would otherwise be a rational number
(for example, $3 \div 2 = 1$).
 
First, we define the function $\mathsf{merge}$ for merging two lists
while preserving order.
It takes two lists as inputs,
but also a natural number representing the sum of their lengths.
Since at least one list decreases in length at recursive calls,
so does the sum of their lengths,
implying the function terminates when applied.
However, in the \emph{refined} system presented in (the figures of) this paper,
to keep things simple,
we provide only one rule for recursive expressions, whose termination metric is $<$ on natural numbers.
Because the system as presented
lacks a rule that supports a termination metric on sums $n_1 + n_2$ of natural numbers,
we need to reify the sum $n = n_1 + n_2$ of the length indexes $n_1$ and $n_2$
into a ghost parameter $n$ of (asserting) singleton type.
However, we emphasize the ghost parameter in this example is not a fundamental limitation of our system,
because our system can be extended to include other termination metrics
such as $<$ on the sum of natural numbers
(we discuss this further in \Secref{sec:typing}).
\begin{align*}
    &\mathsf{merge} : \alltype{n, n_1, n_2 : \kindnat} \natsing{n} \land (n = n_1 + n_2) \to \vectype{(\nat)}{(n_1)} \to \vectype{(\nat)}{(n_2)} &\\
    &~~~~~~~~~~~~~~~~~~~~~~~~~~~~~\to \upshift{\vectype{(\nat)}{(n_1+n_2)}} &\\
    &\mathsf{merge} \; y \; [] \; \mathit{xs}_2 = \Return{\mathit{xs}_2} &\\
    &\mathsf{merge} \; y \; \mathit{xs}_1 \; [] = \Return{\mathit{xs}_1} &\\
    &\mathsf{merge} \; (\succ{y}) \; (x_1 :: \mathit{xs}_1) \; (x_2 :: \mathit{xs}_2) = &\\
    &~~~~~~~~\textkw{if} \; \mathsf{lt}(x_1, x_2) \; \textkw{then} &\\
    &~~~~~~~~~~~~~~~~\textkw{let} \; \mathit{recresult} = \mathsf{merge}(y, \mathit{xs}_1, (x_2 :: \mathit{xs}_2)); &\\
    &~~~~~~~~~~~~~~~~\textkw{let} \; \mathit{result} = x_1 :: \mathit{recresult}; &\\
    &~~~~~~~~~~~~~~~~\Return{\mathit{result}} &\\
    &~~~~~~~~\textkw{else} &\\
    &~~~~~~~~~~~~~~~~\textkw{let} \; \mathit{recresult} = \mathsf{merge}(y, (x_1 :: \mathit{xs}_1), \mathit{xs}_2); &\\
    &~~~~~~~~~~~~~~~~\textkw{let} \; \mathit{result} = x_2 :: \mathit{recresult}; &\\
    &~~~~~~~~~~~~~~~~\Return{\mathit{result}} &
\end{align*}

In a well-typed let-binding $\Let{x}{\be} e$
the bound expression $\be$ is a value-returning computation (that is, has upshift type),
and $e$ is a computation that binds to $x$ the value (of positive type) resulting from computing $\be$.
(Liquid Haskell, lacking CBPV's type distinction between computations and values,
instead approximates whether binders terminate to a value.)
Since $x$ has positive type, we can match it against patterns
(see, for example, the final clause of $\mathsf{split}$, discussed next).

We now define the function $\mathsf{split}$
that takes a list and splits it into two lists.
It is a standard ``every-other'' implementation,
and we have to be a bit careful about the type refinement so as not to be ``off by one''
in the lengths of the resulting lists.
\begin{align*}
  &\mathsf{split} : \alltype{n:\kindnat} \vectype{(\nat)}{(n)} \to \upshift{(\vectype{(\nat)}{((n+1) \div 2)}) \times (\vectype{(\nat)}{(n - ((n+1) \div 2))})} &\\
  &\mathsf{split} \; [] = \Return{\pair{[]}{[]}} &\\
  &\mathsf{split} \; [x] = \Return{\pair{[x]}{[]}} &\\
  &\mathsf{split} \; x_1 :: x_2 :: \mathit{xs} = &\\
  &~~~~\textkw{let} \; \mathit{recresult} = \mathsf{split}(\mathit{xs}); &\\
  &~~~~\textkw{match} \; \mathit{recresult} \; \{ &\\
  &~~~~~~~~\pair{\mathit{xs}_1}{\mathit{xs}_2} \Rightarrow \Return{\pair{x_1 :: \mathit{xs}_1}{x_2 :: \mathit{xs}_2}} &\\
  &~~~~\} &
\end{align*}

We are ready to implement $\mathsf{mergesort}$,
but since we use a ghost parameter,
we need to define an auxiliary function $\mathsf{auxmergesort}$
additionally taking the length of the list being ordered.
We introduce syntactic sugar for a let-binding followed by pattern-matching on its result.
\begin{align*}
    &\mathsf{auxmergesort} : \alltype{n:\kindnat} \natsing{n} \to \vectype{(\nat)}{(n)} \to \upshift{\vectype{(\nat)}{(n)}} &\\
    &\mathsf{auxmergesort} \; y \; [] = \Return{[]} &\\
    &\mathsf{auxmergesort} \; y \; [x] = \Return{[x]} &\\
    &\mathsf{auxmergesort} \; y \; \mathit{xs} = &\\
    &~~~~\textkw{let} \; \pair{\mathit{leftxs}}{\mathit{rightxs}} = \mathsf{split}(\mathit{xs}); &\\
    &~~~~\textkw{let} \; \mathit{lenleftxs} = \mathsf{div}(\succ{y}, 2); &\\
    &~~~~\textkw{let} \; \mathit{lenrightxs} = y - \mathsf{div}(\succ{y},  2); &\\
    &~~~~\textkw{let} \; \mathit{sortleftxs} = \mathsf{auxmergesort}(\mathit{lenleftxs}, \mathit{leftxs}); &\\
    &~~~~\textkw{let} \; \mathit{sortrightxs} = \mathsf{auxmergesort}(\mathit{lenrightxs}, \mathit{rightxs}); &\\
    &~~~~\Return{\mathsf{merge}(y, \mathit{sortleftxs}, \mathit{sortrightxs})} &
\end{align*}

Finally, we define a $\mathsf{mergesort}$
that is verified to terminate
and to return a list of the same length as the input list.
\begin{align*}
  &\mathsf{mergesort} : \alltype{n:\kindnat} \vectype{(\nat)}{(n)} \to \upshift{\vectype{(\nat)}{(n)}} &\\
  &\mathsf{mergesort} \; \mathit{xs} = &\\
  &~~~~\textkw{let} \; \mathit{len} = \mathsf{len}(\mathit{xs}); &\\
  &~~~~\Return{\mathsf{mergesortaux}(\mathit{len}, \mathit{xs})} &
\end{align*}

In the system of this paper,
we cannot verify that $\mathsf{mergesort}$
returns a \emph{sorted} list.
This is because our system lacks multi-argument measures
which can specify relations between indexes of different parts of a data type.
(To handle this,
we are extending \sysname with multi-argument measures,
which is nontrivial and requires a significant degree of additional machinery
outside the scope of this paper.)
But this example is interesting nonetheless,
because $\mathsf{auxmergesort}$ is \emph{not} structurally recursive:
its recursive calls are on lists obtained by splitting the input list roughly in half,
not on the structure of the list ($- :: -$).
Further, it showcases %
the main features of our foundational system,
the declarative specification of which we turn to next, in \Secref{sec:declarative-system}.

\section{Declarative System}
\label{sec:declarative-system}

We present our core declarative calculus and type system.

First, we discuss the syntax of program terms, types, index terms, sorts, functors, algebras, and contexts.
Then, in \Secref{sec:index-sorting-and-verification},
we discuss the index sorting judgment $\Theta |- t : \tau$
and the propositional validity judgment $\judgeentail{\Theta}{\phi}$,
index-level substitution,
and basic logical properties required of the index domain.
In \Secref{sec:well-formedness},
we discuss the well-formedness of
(logical and program) contexts ($\judgctx{\Theta}$ and $\judgectx{\Theta}{\Gamma}$),
types ($\judgetp{\Theta}{A}{\Xi}$),
functors ($\judgefunctor{\Theta}{\mathcal{F}}{\Xi}$),
and algebras ($\judgealgebra{\Xi}{\Theta}{\alpha}{F}{\tau}$).
In \Secref{sec:equivalence},
we discuss judgmental equivalence.
In \Secref{sec:extraction},
we discuss extraction ($\judgeextract{\Theta}{A}{A'}{\Theta'}$).
In \Secref{sec:subtyping},
we discuss subtyping ($\judgesub{\Theta}{A}{B}$).
In \Secref{sec:unrolling},
we discuss the unrolling judgment for refined inductive types.
In \Secref{sec:typing},
we discuss the typing system.
In \Secref{sec:subst},
we extend substitution to that of program values for program variables,
and prove a substitution lemma stating that typing is stable under substitution
(a key operational correctness result).

In \Figref{fig:decl-judgments-presuppositions},
we summarize the key judgments constituting the declarative system.
In the figure, ``pre.'' abbreviates ``presupposes'',
which basically lists the judgments (rightmost column)
we tend to leave implicit in rules defining the given judgment (leftmost column).
Presuppositions also indicate the (input or output) moding of judgments.
For example, on the one hand $\judgechkval{\Theta}{\Gamma}{v}{P}$
presupposes $\judgectx{\Theta}{\Gamma}$ and $\judgetp{\Theta}{P}{\Xi_P}$ for some $\Xi_P$,
where the former presupposes $\judgctx{\Theta}$,
and $\Theta$, $\Gamma$, and $P$
are input-moded;
on the other hand, $\judgesynhead{\Theta}{\Gamma}{h}{P}$
does not presuppose $\judgetp{\Theta}{P}{\Xi_P}$ for some $\Xi_P$,
but rather we must prove that the output-moded $P$ is well-formed
(which is straightforward).
Groups of mutually defined judgments are separated by blank lines. %
\begin{figure}[hptb]
  \begin{tabular}{LlcL}
    \judgctx{\Theta} & (\Secref{sec:well-formedness}) & pre. & \text{no judgment} \\
    \Theta |- t : \tau & (\Secref{sec:index-sorting-and-verification}) & pre. & \judgctx{\Theta} \\
    ~\\
    \judgeentail{\Theta}{\phi} & (\Secref{sec:index-sorting-and-verification}) & pre. & \judgeterm{\Theta}{\phi}{\Booltype} \\
    ~\\
    \judgetp{\Theta}{A}{\Xi} & (\Secref{sec:well-formedness}) & pre. & \judgctx{\Theta} \\
    \judgefunctor{\Theta}{\mathcal{F}}{\Xi} & (\Secref{sec:well-formedness}) & pre. & \judgctx{\Theta} \\
    \judgealgebra{\Xi}{\Theta}{\alpha}{F}{\tau} & (\Secref{sec:well-formedness}) & pre. & \judgctx{\Xi} \text{ and } \judgefunctor{\Theta}{F}{\Xi_F} \\
    ~\\
    \judgeequiv[]{\Theta}{\mathcal{F}}{\mathcal{G}} & (\Secref{sec:equivalence}) & pre. & \judgefunctor{\Theta}{\mathcal{F}}{\Xi_{\mathcal{F}}} \text{ and } \judgefunctor{\Theta}{\mathcal{G}}{\Xi_{\mathcal{G}}} \\
    \judgeequiv[\pm]{\Theta}{A}{B} & (\Secref{sec:equivalence}) & pre. & \judgetp{\Theta}{A}{\Xi_A} \text{ and } \judgetp{\Theta}{B}{\Xi_B} \\
    ~\\
    \judgeextract[\pm]{\Theta}{A}{A'}{\Theta_A} & (\Secref{sec:extraction}) & pre. & \judgetp{\Theta}{A}{\Xi_{A}} \\
    ~\\
    \judgesub[\pm]{\Theta}{A}{B} & (\Secref{sec:subtyping}) & pre. & \judgetp{\Theta}{A}{\Xi_A} \text{ and } \judgetp{\Theta}{B}{\Xi_B} \\
    ~\\
    \unroll{\Xi}{\Theta}{G}{F}{\beta}{\alpha}{\tau}{t}{P} & (\Secref{sec:unrolling}) & pre. & \judgealgebra{\Xi}{\Theta}{\alpha}{F}{\tau} \text{ and } \judgealgebra{\Xi}{\Theta}{\beta}{G}{\tau} \\
    ~ & ~ & ~ & \text{and } \judgeterm{\Theta}{t}{\tau} \\
    ~\\
    \judgectx{\Theta}{\Gamma} & (\Secref{sec:well-formedness}) & pre. & \judgctx{\Theta} \\
    ~\\
    \judgesynhead{\Theta}{\Gamma}{h}{P} & (\Secref{sec:typing}) & pre. & \judgectx{\Theta}{\Gamma} \\
    \judgesynexp{\Theta}{\Gamma}{\be}{\upshift{P}} & (\Secref{sec:typing}) & pre. & \judgectx{\Theta}{\Gamma} \\
    \judgechkval{\Theta}{\Gamma}{v}{P} & (\Secref{sec:typing}) & pre. & \judgectx{\Theta}{\Gamma} \text{ and } \judgetp{\Theta}{P}{\Xi_P} \\
    \judgechkexp{\Theta}{\Gamma}{e}{N} & (\Secref{sec:typing}) & pre. & \judgectx{\Theta}{\Gamma} \text{ and } \judgetp{\Theta}{N}{\Xi_N} \\
    \judgechkmatch{\Theta}{\Gamma}{P}{\clauses{\pa}{e}{i}{I}}{N} & (\Secref{sec:typing}) & pre. & \judgectx{\Theta}{\Gamma} \text{ and } \judgetp{\Theta}{P}{\Xi_P} \text{ and } \judgetp{\Theta}{N}{\Xi_N} \\
    \judgespine{\Theta}{\Gamma}{s}{N}{\upshift{P}} & (\Secref{sec:typing}) & pre. & \judgectx{\Theta}{\Gamma} \text{ and } \judgetp{\Theta}{N}{\Xi_N} \\
  \end{tabular}

  \caption{Declarative judgments and their presuppositions}
  \label{fig:decl-judgments-presuppositions}
\end{figure}

\paragraph{Program terms}
Program terms are defined in \Figref{fig:grammar-program-terms}.
We polarize terms into two main syntactic categories:
expressions (which have negative type) and values (which have positive type).
Program terms  %
are further distinguished
according to whether their principal types are
synthesized (heads and bound expressions)
or checked (spines and patterns).

\begin{figure}[hptb]
  \begin{grammar}
    Program variables  & $x, y, z$
    \\[0.3ex]
    Expressions  & 
    $e$ & \bnfas &
    $\Return{v}
    \bnfalt  \Let{x}{\be}{e}
    \bnfalt  \match{h}{\clauses{\pa}{e}{i}{I}}
    \bnfaltBRK  \fun{x}{e}
    \bnfalt  \rec{x:N}{e}
    \bnfalt  \unreachable
    $
    \\[0.6ex]
    Values  & 
    $v$ & \bnfas &
    $x
    \bnfalt  \unit
    \bnfalt  \pair{v}{v}
    \bnfalt  \inl{v}
    \bnfalt  \inr{v}
    \bnfalt  \roll{v}
    \bnfalt  \thunk{e}
    $
    \\[0.6ex]
    Heads  & 
    $h$ & \bnfas &
    $x
    \bnfalt \annoexp{v}{P}
    $
    \\
    Bound expressions  & 
    $\be$ & \bnfas &
    $h(s)
    \bnfalt \annoexp{e}{\upshift{P}}
    $
    \\[0.6ex]
    Spines  & 
    $s$ & \bnfas &
    $
    \cdot 
    \bnfalt v, s
    $
    \\
    Patterns  & 
    $\pa$ & \bnfas &
    $\roll{x}
    \bnfalt  \unit
    \bnfalt  \pair{x}{y}
    \bnfalt  \inl{x}
    \bnfalt  \inr{x} 
    $
  \end{grammar}

  \caption{Program terms}
  \label{fig:grammar-program-terms}
\end{figure}

Expressions $e$ consist of functions $\fun{x}{e}$,
recursive expressions $\rec{x:N}{e}$,
let-bindings $\Let{x}{g}{e}$,
match expressions $\match{h}{\clauses{\pa}{e}{i}{I}}$,
value returners (or producers) $\Return{v}$,
and an unreachable expression \unreachable (such as an impossible match).
Bound expressions $g$,
which can be let-bound,
consist of
expressions annotated with a returner type $\annoexp{e}{\upshift{P}}$
and applications $h(s)$ of a head $h$ to a spine $s$.
Heads $h$,
which can be applied to a spine or pattern-matched,
consist of variables $x$
and positive-type-annotated values $\annoexp{v}{P}$.
Spines $s$ are lists of values;
we often omit the empty spine $\cdot$,
writing (for example) $v_1, v_2$ instead of $v_1, v_2, \cdot$.
In match expressions, heads are matched against patterns $r$.

Values consist of variables $x$,
the unit value $\unit$,
pairs $\pair{v_1}{v_2}$,
injections into sum type $\inj{k}{v}$ where $k$ is $1$ or $2$,
rollings into inductive type $\roll{v}$,
and thunks (suspended computations) $\thunk{e}$.

\paragraph{Types} 
Types are defined in \Figref{fig:types}.
Types are polarized into positive (value) types $P$
and negative (computation) types $N$.
We write $A$, $B$ and $C$ for types of either polarity.

\begin{figure}[hptb]
  \begin{grammar}
    Positive types  &
    $P,Q$ & \bnfas & $1 \bnfalt P \times Q \bnfalt 0 \bnfalt P + Q \bnfalt{\downshift{N}}
    \bnfaltBRK
    \comprehend{\nu : \mu F}{\Fold{F}{\alpha}\,{\nu} =_\tau t}
    \bnfaltBRK
    \extype{a:\tau}P
    \bnfalt
    P \land \phi
    $
    \\
    Negative types  & 
    $N,M$ & \bnfas
    & $P \to N
    \bnfalt \upshift{P}
    \bnfalt \alltype{a:\tau}{N}
    \bnfalt \phi \implies N
    $
    \\
    Types  & 
    $A,B,C$ & \bnfas
    & $P \bnfalt N$
  \end{grammar}

  \caption{Types}
  \label{fig:types}
\end{figure}

Positive types consist of
the unit type $1$,
products $P_1 \times P_2$,
the void type $0$,
sums $P_1 + P_2$,
downshifts (of negative types; \emph{thunk} types) $\downshift{N}$,
\emph{asserting} types $P \land \phi$
(read ``$P$ with $\phi$''),
index-level existential quantifications $\extype{a:\tau} P$,
and refined inductive types
$\comprehend{\nu : \mu F}{\Fold{F}{\alpha}\,{\nu} =_\tau t}$.
We read $\comprehend{\nu : \mu F}{\Fold{F}{\alpha}\,{\nu} =_\tau t}$
as the type having values $\nu$
of inductive type $\mu F$ (with signature $F$)
such that the (index-level) \emph{measurement} $\Fold{F}{\alpha}\,{\nu} =_\tau t$ holds;
in \Secref{sec:functor-algebra-grammar} and \Secref{sec:semantics},
we explain the metavariables $F$, $\alpha$, $\tau$, and $t$,
as well as what these and the syntactic parts $\mu$ and $\mathsf{fold}$ denote.
Briefly, $\mu$ roughly denotes ``least fixed point of''
and a $\mathsf{fold}$ over $F$ with $\alpha$ (having carrier sort $\tau$)
indicates a measure on the inductive type $\mu F$ into $\tau$.

Negative types consist of
function types $P \to N$,
upshifts (of positive types; \emph{lift} or \emph{returning} types)
$\upshift{P}$ (dual to $\downshift{N}$),
propositionally \emph{guarded} types $\phi \implies N$
(read ``$\phi$ implies $N$''; dual to $P \land \phi$),
and index-level universal quantifications $\alltype{a:\tau} N$
(dual to $\extype{a:\tau} P$).

In $P \land \phi$ and $\phi \implies N$,
the index proposition $\phi$ has no run-time content.
Neither does the $a$ in $\extype{a:\tau} P$ and $\alltype{a:\tau} N$,
nor the recursive refinement predicate $\Fold{F}{\alpha}\,{\nu} =_\tau t$
in $\comprehend{\nu : \mu F}{\Fold{F}{\alpha}\,{\nu} =_\tau t}$.

\paragraph{Index language: sorts, terms, and propositions}
Our type system is parametric in the index domain,
provided the latter has certain (basic) properties.
For \sysname to be decidable, the index domain must be decidable.
It is instructive to work with a specific index domain:
\Figureref{fig:index-domain} defines a 
quantifier-free logic of linear equality, inequality, and arithmetic,
which is decidable \cite{BarrettSMT}.

\begin{figure}[hptb]
  \begin{grammar}
    Sorts  & 
    $\tau$ & \bnfas
    & $\Booltype
    \bnfalt \kindnat
    \bnfalt \Z
    \bnfalt \tau \times \tau
    $
    \\[1em]
    Index variables & $a, b, c$
    \\
    Index terms  & 
    $t$ & \bnfas &
    $a
    \bnfalt n
    \bnfalt t + t
    \bnfalt t - t
    \bnfalt (t,t)
    \bnfalt \fst{t}
    \bnfalt \snd{t}
    \bnfalt \phi
    $
    \\
    Propositions  & 
    $\phi, \psi$ & \bnfas &
    $t = t
    \bnfalt  t \leq t
    \bnfalt  \phi \land \phi
    \bnfalt  \phi \lor \phi
    \bnfalt  \lnot \phi
    \bnfalt  \True
    \bnfalt  \False
    $
  \end{grammar}

  \caption{Index domain}
  \label{fig:index-domain}
\end{figure}

Sorts $\tau$ consist of
booleans $\Booltype$,
natural numbers $\kindnat$,
integers $\Z$,
and products $\tau_1 \times \tau_2$.
Index terms $t$ consist of variables $a$,
numeric constants $n$,
addition $t_1 + t_2$,
subtraction $t_1 - t_2$,
pairs $(t_1, t_2)$,
projections $\fst{t}$ and $\snd{t}$,
and propositions $\phi$.
Propositions $\phi$ (the logic of the index domain)
are built over index terms,
and consist of
equality $t_1 = t_2$,
inequality $t_1 \leq t_2$,
conjunction $\phi_1 \land \phi_2$,
disjunction $\phi_1 \lor \phi_2$,
negation $\lnot \phi$,
trivial truth $\True$, and trivial falsity $\False$.

\subsubsection{Inductive types, functors, and algebras}
\label{sec:functor-algebra-grammar}

We encode algebraic data types (and measures on them)
using their standard semantics.
In the introduction (\Secref{sec:introduction}),
to refine the type of $A$-lists by their length,
we defined a recursive function \textsf{len} over the inductive structure of lists.
Semantically, we characterize this structural recursion
by algebraic folds over polynomial endofunctors;
we design \sysname in line with this semantics.
While this presentation may appear overly abstract for the user,
it should be possible to allow the user to use the same or similar syntax as programs to express measures
if they annotate them as measures in the style of Liquid Haskell.

We express inductive type structure
without reference to constructor names
by syntactic functors resembling
the polynomial functors.
For example (modulo the difference for simplifying unrolling),
we can specify the signature of the inductive type of lists of terms of type $A$
syntactically as a functor
$\Const{1} \oplus (\Const{A} \otimes \Id)$,
where $\Const{C}$ denotes the constant (set) functor
(sending any set to the set denoted by type $C$),
$\Id$ denotes the identity functor (sending any set to itself),
the denotation of $F_1 \otimes F_2$
sends a set $X$ to the product $(\sem{}{F_1} X) \times (\sem{}{F_2} X)$, and
the denotation of $F_1 \oplus F_2$ sends a set $X$
to the disjoint union $(\sem{}{F_1}X) \uplus (\sem{}{F_2}X)$.
The idea is that each component of the sum functor $\oplus$
represents a data constructor, so that (for example)
$\Const{1}$ represents the nullary constructor $[]$,
and $\Const{A}$ represents the head element of a cons cell
which is attached (via $\otimes$) to the recursive tail list represented by $\Id$.

A functor $F$ (\Figref{fig:functors}) is a sum ($\oplus$) of products ($\Ptype$),
which multiply ($\otimes$) base functors ($B$)
consisting of identity functors that represent recursive positions ($\Id$)
and constant functors ($\Const{P}$) at positive type $P$.
The rightmost factor in a product $\Ptype$ is the (product) unit functor $I$.
By convention, $\otimes$ has higher precedence than $\oplus$.
For convenience in specifying functor well-formedness
(appendix \Figref{fig:declwf-fun-alg})
and denotation
(appendix \Figref{fig:denotation-types2}),
$\mathcal{F}$ is a functor $F$ or a base functor $B$.

\begin{figure}[hptb]
  \begin{grammar}
    Functors  &
    $F, G, H$ & \bnfas &
    $\Ptype
    \bnfalt F \oplus F
    $
    \\
    & $\Ptype$ & \bnfas &
    $I
    \bnfalt B \otimes \Ptype
    $
    \\
    & $\Btype$ & \bnfas
    & $\Const{P}
    \bnfalt \Id
    $
    \\
    & $\mathcal{F}$ & \bnfas & $F \bnfalt B$ 
  \end{grammar}

  \caption{Functors}
  \label{fig:functors}
\end{figure}

A direct grammar $F$ for sums of products (of constant and identity functors)
consists of $F \bnfas \Ptype \bnfalt F \oplus F$
and $\Ptype \bnfas B \bnfalt B \otimes \Ptype$
and $\Btype \bnfas \Const{P} \bnfalt \Id$.
The grammar $F \bnfas F \oplus F \bnfalt F \otimes F \bnfalt B$
is semantically equivalent to sums of products,
but syntactically inconvenient,
because it allows writing products of sums.
We do not use either of these grammars,
but rather \Figref{fig:functors},
because it simplifies inductive type unrolling (\Secref{sec:unrolling}).
(In any case, a surface language where data types have named constructors
would have to be elaborated to use one of these grammars.)
These grammars have naturally isomorphic interpretations.
For example, in our functor grammar (\Figref{fig:functors}),
we instead write $\textsf{NatF} = I \oplus (\Id \otimes I)$
(note $I$ is semantically equivalent to $\Const{1}$):
notice that for any set $X$, we have
$\sem{}{I \oplus (\Id \otimes I)} X
  = 1 \uplus (X \times 1)
  \cong 1 \uplus X
  = \sem{}{\Const{1} \oplus \Id} X$.

As we will discuss in \Secref{sec:semantics},
every polynomial endofunctor $F$ has a fixed point $\mu F$
satisfying a recursion principle for defining measures (on $\mu F$)
by folds with algebras.
We define algebras in \Figref{fig:algebras}.
An algebra $\alpha$ is a list of clauses $\clause{p}{t}$
which pattern match on algebraic structure
($p$, $q$, and $o$ are patterns)
and bind variables in index bodies $t$.
Sum algebra patterns $p$ consist of $\inl{p}$ and $\inr{p}$
(which match on sum functors $\oplus$).
Product algebra patterns $q$ consist of
tuples $(o, q)$ (which match on $\otimes$) ending in the unit pattern $\unitexp$
(which match on $I$).
Base algebra patterns $o$
consist of wildcard patterns $\wild$ (which match on constant functors $\Const{P}$),
variable patterns $a$ (which match on the identity functor $\Id$),
and pack patterns $\pack{a}{\bap}$
(which match on \emph{existential} constant functors $\Const{\extype{a:\tau} P}$,
where $a$ is also bound in the bodies $t$ of algebra clauses).

\begin{figure}[hptb]
  \begin{grammar}
    Algebras  & 
    $\alpha, \beta$ & \bnfas
    & $
    \cdot
    \bnfalt
    (\clause{p}{t} \matchor \alpha)
    $
    \\
    Sum algebra patterns & 
    $p$ & \bnfas &
    $\inl{p}
    \bnfalt  \inr{p}
    \bnfalt  q
    $
    \\
    Product algebra patterns  & 
    $q$ & \bnfas & $\unitexp
    \bnfalt  (\bap, q)
    $
    \\
    Base algebra patterns  & 
    $\bap$ & \bnfas & $\wild
    \bnfalt  a
    \bnfalt  \pack{a}{\bap}$
  \end{grammar}

  \caption{Algebras}
  \label{fig:algebras}
\end{figure}

For example, given a type $P$,
consider the functor $I \oplus (\Const{P} \otimes \Id \otimes I)$.
To specify the function $\textsf{length} : \textsf{List}\;P \to \kindnat$
computing the length of a list of values of type $P$,
we write the algebra
$\inl \clause{\unitexp}{0} \matchor \inr \clause{(\wild, (a, \unitexp))}{1+a}$
with which to fold $\textsf{List}\;P$.

With the pack algebra pattern, we can use indexes of an inductive type
in our measures.
For example,
given $a:\kindnat$,
and defining the singleton type $\textsf{Nat}(a)$ as
$\comprehend{\nu:\mu \textsf{NatF}}{\Fold{\textsf{NatF}}{\textsf{ixnat}}\,{\nu} = a}$
where
$\textsf{NatF} = I \oplus \Id \otimes I$
and
$\textsf{ixnat} = \inl \clause{\unitexp}{0} \matchor \inr \clause{(a, \unitexp)}{1+a}$,
consider lists of natural numbers, specified by
$I \oplus \Const{\extype{b:\kindnat} \textsf{Nat}(b)} \otimes \Id \otimes I$.
Folding such a list with the algebra
$\inl \clause{\unitexp}{0} \matchor \inr \clause{(\pack{b}{\wild}, a, \unitexp)}{a + b}$
sums all the numbers in the list.
(For clarity, we updated the definitions in \Secref{sec:overview}
to agree with our grammars as presented in \Figref{fig:functors} and \Figref{fig:algebras}.)

For measures relating indexes
in structurally distinct positions within an inductive type,
in ongoing work
we are extending \sysname with multi-argument measures
by way of higher-order sorts $\tau_1 \Rightarrow \tau_2$.
Doing so would allow us to refine, for example, integer lists
to lists of integers \emph{in increasing order},
because we could then compare the indexed elements of a list.

\paragraph{Contexts}
A \emph{logical context} $\Theta \bnfas \cdot \bnfalt \Theta, a : \tau \bnfalt \Theta, \phi$
is an ordered list of index propositions $\phi$ and (index) variable sortings $a:\tau$
(which may be used in subsequent propositions).
A \emph{program variable context} (or \emph{program context}) $\Gamma \bnfas \cdot \bnfalt \Gamma, x : P$
is a set of (program) variable typings $x:P$.
A value-determined context $\Xi \bnfas \cdot \bnfalt \Theta, a : \tau$ is a set of index sortings $a:\tau$.
In any kind of context, a variable can be declared at most once.

\subsection{Index sorting and propositional validity}
\label{sec:index-sorting-and-verification}

We have a standard index sorting judgment $\judgeterm{\Theta}{t}{\tau}$
(appendix \Figref{fig:ix-sort})
checking that, under context $\Theta$, index term $t$ has sort $\tau$.
For example, $\judgeterm{a:\kindnat}{\lnot (a \leq a+1)}{\Booltype}$.
This judgment does not depend on propositions in $\Theta$,
which only matter when checking propositional validity
(only done in subtyping and program typing).
The operation $\overline{\Theta}$ merely removes all propositions $\phi$ from $\Theta$.

Well-sorted index terms $\Theta |- t : \tau$
denote functions $\sem{}{t} : \sem{}{\overline{\Theta}} \to \sem{}{\tau}$.
For each $\Theta$, define $\sem{}{\Theta}$
as the set of index-level semantic substitutions (defined in this paragraph)
$\comprehend{\delta}{|- \delta : \Theta}$.
For example,
$\sem{3/a}{4+a} = 7$ and $\sem{1/b}{b=1+0} = \one$ (that is, true)
and $\sem{2/a}{a=1} = \emptyset$ (that is, false).
An \emph{index-level semantic substitution} $|- \delta : \Theta$
assigns exactly one semantic index value $d$
to each index variable in $\dom{\Theta}$
such that every proposition $\phi$ in $\Theta$ is true (written $\one$; false is $\emptyset$):
\begin{mathpar}
  \Infer{}
  {
  }
  {
    |- \cdot : \cdot
  }
  \and
  \Infer{}
  {
    |- \delta : \Theta
    \and
    d \in \sem{}{\tau}
    \and
    a \notin \dom{\Theta}
  }
  {
    |- (\delta, d/a) : (\Theta, a:\tau)
  }
  \and
  \Infer{}
  {
    |- \delta : \Theta
    \and
    \sem{\delta}{\phi} = \{\bullet\}
  }
  {
    |- \delta : (\Theta, \phi)
  }
\end{mathpar}

A \emph{propositional validity} or \emph{truth} judgment $\judgeentail{\Theta}{\phi}$,
which is a \emph{semantic} entailment relation,
holds if $\phi$ is valid under $\Theta$, that is,
if $\phi$ is true under every interpretation
of variables in $\Theta$
such that all propositions in $\Theta$ are true.
We say $t$ and $t'$ are \emph{logically equal} under $\Theta$
if $\judgeentail{\Theta}{t = t'}$.

An \emph{index-level syntactic substitution} $\sigma$
is a list of index terms to be substituted for index variables:
$\sigma \bnfas \cdot \bnfalt \sigma, t/a$.
The metaoperation $[\sigma]\mathcal{O}$,
where $\mathcal{O}$ is an index term, program term, or type,
is sequential substitution:
$[\cdot]\mathcal{O} = \mathcal{O}$ and
$[\sigma, t/a]\mathcal{O} = [\sigma]([t/a]\mathcal{O})$,
where $[t/a]\mathcal{O}$ is standard
capture-avoiding (by $\alpha$-renaming) substitution.
Syntactic substitutions (index-level) are typed (``sorted'') in a standard way.
Because
syntactic substitutions substitute terms that may have free variables,
their judgment form includes a context to the left of the turnstile,
in contrast to semantic substitution:
\begin{mathpar}
  ~
  \Infer{}
  {
  }
  {
    \Theta_0 |- \cdot : \cdot
  }
  \and
  \Infer{}
  {
    \Theta_0 |- \sigma : \Theta
    \and
    \Theta_0 |- [\sigma]t : \tau
    \and
    a \notin \dom{\Theta}
  }
  {
    \Theta_0 |- (\sigma, t/a) : (\Theta, a:\tau)
  }
  \and
  \Infer{}
  {
    \Theta_0 |- \sigma : \Theta
    \and
    \judgeentail{\Theta_0}{[\sigma]\phi}
  }
  {
    \Theta_0 |- \sigma : (\Theta, \phi)
  }
\end{mathpar}
Because our substitution operation $[\sigma]-$ applies sequentially,
we type (``sort'') the application of the rest of the substitution to the head
being substituted.
For example, the rule concluding
$\Theta_0 |- (\sigma, t/a) : (\Theta, a:\tau)$
checks that the application $[\sigma]t$ of $\sigma$ to $t$ has sort $\tau$.

The decidability of \sysname depends on the decidability of propositional validity.
Our example index domain is decidable \cite{BarrettSMT}.
\Sysname is parametric in the index domain,
provided the latter has certain properties.
In particular, propositional validity must satisfy the following basic properties
required of a logical theory
($\judgctx{\Theta}$ is logical context well-formedness).
\begin{itemize}
\item \emph{Weaken}: If $\judgctx{\Theta_1, \Theta, \Theta_2}$
  and $\judgeentail{\Theta_1, \Theta_2}{\phi}$,
  then $\judgeentail{\Theta_1, \Theta, \Theta_2}{\phi}$.
\item \emph{Permute}: If $\judgctx{\Theta, \Theta_1}$
  and $\judgctx{\Theta, \Theta_2}$
  and $\judgeentail{\Theta, \Theta_1, \Theta_2, \Theta'}{\phi}$,
  then $\judgeentail{\Theta, \Theta_2, \Theta_1, \Theta'}{\phi}$.
\item \emph{Substitution}: If $\judgeentail{\Theta}{\phi}$ and $\Theta_0 |- \sigma : \Theta$,
  then $\judgeentail{\Theta_0}{[\sigma]\phi}$.
\item \emph{Equivalence}: The relation $\judgeentail{\Theta}{t_1 = t_2}$
  is an equivalence relation.
\item \emph{Assumption}: If $\judgctx{\Theta_1, \phi, \Theta_2}$,
  then $\judgeentail{\Theta_1, \phi, \Theta_2}{\phi}$.
\item \emph{Consequence}:
  If $\judgeentail{\Theta_1}{\psi}$
  and $\judgeentail{\Theta_1, \psi, \Theta_2}{\phi}$,
  then $\judgeentail{\Theta_1, \Theta_2}{\phi}$.
\item \emph{Consistency}: It is not the case that $\judgeentail{\cdot}{\False}$.
\end{itemize}
We also assume
that $\judgeterm{\Theta}{t}{\tau}$ is decidable
and satisfies weakening and substitution.
Our example index domain satisfies all these properties.

\subsection{Well-formedness}
\label{sec:well-formedness}
Context well-formedness $\judgctx{\Theta}$
and $\judgectx{\Theta}{\Gamma}$ (appendix \Figref{fig:declwf-program-ctx}) is straightforward.
For both logical and program context well-formedness, there can be at most one of each variable.
Index terms in well-formed logical contexts must have boolean sort:
\begin{mathpar}
  \Infer{\LogCtxEmpty}
  {}
  {
    \judgctx{\cdot}
  }
  \and
  \Infer{\LogCtxVar}
  {
    \judgctx{\Theta}
    \\
    a \notin \dom{\Theta}
  }
  {
    \judgctx{(\Theta, a:\tau)}
  }
  \and
  \Infer{\LogCtxProp}
  {
    \judgctx{\Theta}
    \\
    \judgeterm{\Theta}{\phi}{\Booltype}
  }
  {
    \judgctx{(\Theta, \phi)}
  }
\end{mathpar}
In well-formed program variable contexts $\judgectx{\Theta}{\Gamma}$,
the types (of program variables) must be well-formed under $\Theta$;
further, we must not be able to extract index information from these types
(in the sense of \Secref{sec:extraction}).
For example, $x : 1 \land \False$ is an ill-formed program context because $\False$ can be extracted,
but $x : \downshift{\upshift{1 \land \False}}$ is well-formed
because nothing under a shift type can be extracted.

Type well-formedness $\judgetp{\Theta}{A}{\Xi}$
(read
``under $\Theta$, type $A$ is well-formed with value-determined indexes $\Xi$'')
has $\Xi$ in output mode,
which tracks index variables appearing in the type $A$
that are uniquely\footnote{Semantically speaking.} determined
by values
of refined inductive types in $A$, particularly by their folds.
(See \Lemref{lem:liq-value-determined-soundness} in \Secref{sec:semantics}.)
Consider the following type well-formedness rule:
\[
  \Infer{\DeclTpFixVar}
  {
    \judgefunctor{\Theta}{F}{\Xi}
    \\
    \judgealgebra{\cdot}{\Theta}{\alpha}{F}{\tau}
    \\
    (b : \tau) \in \Theta
  }
  {
    \judgetp{\Theta}{\comprehend{\nu:\mu F}{\Fold{F}{\alpha}\,{\nu} =_\tau b}}{\Xi \union b:\tau}
  }
\]
The index $b$ is uniquely determined by a value of the conclusion type,
so we add it to $\Xi$.
For example, the value
$\mathsf{one} = \roll{\inj{2}{\pair{\roll{\inj{1}{\unit}}}{\unit}}}$
determines the variable $b$ appearing in the value's type
$\textsf{NatF}(b) = \comprehend{\nu:\mu \textsf{NatF}}{\Fold{\textsf{NatF}}{\textsf{ixnat}}\,{\nu} = b}$
to be one.
(We have a similar rule
where the $b$-position metavariable is not an index variable,
adding nothing to $\Xi$.)
We use set union ($\Xi \union b:\tau$)
as $b$ may already be value-determined in $F$
(that is, $(b:\tau)$ may be in $\Xi$).
The algebra well-formedness premise
$\judgealgebra{\cdot}{\Theta}{\alpha}{F}{\tau}$
requires the algebra $\alpha$ to be closed
(that is, the first context is empty, $\cdot$).
This premise ensures
that existential variables never appear in algebras,
which is desirable because folds with algebras
solve existential variables when typechecking a value
(see \Secref{sec:algorithmic-system}).

Because ultimately $\Xi$ tracks only measure-determined indexes,
\DeclTpFixVar is the only rule that adds to $\Xi$.
The index propositions of asserting and guarded types
do not track anything beyond what is tracked by the types
to which they are connected.
\begin{mathpar}
\Infer{\DeclTpWith}
{
  \judgetp{\Theta}{P}{\Xi}
  \\
  \judgeterm{\Theta}{\phi}{\Booltype}
}
{ \judgetp{\Theta}{P \land \phi}{\Xi} }
\and
\Infer{\DeclTpImplies}
{
  \judgetp{\Theta}{N}{\Xi}
  \\
  \judgeterm{\Theta}{\phi}{\Booltype}
}
{ \judgetp{\Theta}{\phi \implies N}{\Xi} }
\end{mathpar}

We restrict quantification to value-determined index variables
in order to guarantee we can always solve them algorithmically.
For example, in checking $\mathsf{one}$
against the type $\extype{a:\kindnat} \mathsf{Nat}(a)$,
we solve $a$ to an index semantically equal to $1 \in \kindnat$.
If $\judgetp{\Theta, a:\tau}{P}{\Xi}$,
then $\extype{a:\tau} P$ is well-formed if and only if $(a:\tau) \in \Xi$,
and similarly for universal quantification
(which we'll restrict to the argument types of function types;
argument types are positive):
\begin{mathpar}
  \Infer{\DeclTpEx}
  { \judgetp{\Theta, a:\tau}{P}{\Xi, a:\tau} }
  { \judgetp{\Theta}{\extype{a:\tau}{P}}{\Xi} }
  \and
  \Infer{\DeclTpAll}
  { \judgetp{\Theta, a:\tau}{N}{\Xi, a:\tau} }
  { \judgetp{\Theta}{\alltype{a:\tau}{N}}{\Xi} }
\end{mathpar}
We read commas in value-determined contexts such as
$\Xi_1, \Xi_2$ as set union $\Xi_1 \cup \Xi_2$
together with the fact that $\dom{\Xi_1} \sect \dom{\Xi_2} = \emptyset$,
so these rules can be read top-down as removing $a$.

A value of product type is a pair of values,
so we take the union of what each component value determines:
\[
  \Infer{\DeclTpProd}
  { \judgetp{\Theta}{P_1}{\Xi_1} \\ \judgetp{\Theta}{P_2}{\Xi_2} }
  { \judgetp{\Theta}{P_1 \times P_2}{\Xi_1 \union \Xi_2}  }
\]
We also take the union for function types $P \to N$,
because to use a function, due to focusing,
we must provide values for all its arguments.
The $\Xi$ of $\mathsf{Nat}(a) \to \upshift{\mathsf{Nat}(a)}$ is $a:\kindnat$,
so $\alltype{a:\kindnat} \mathsf{Nat}(a) \to \upshift{\mathsf{Nat}(a)}$
is well-formed.
In applying a head of this type to a value,
we must instantiate $a$ to an index semantically
equal to what that value determines;
for example,
if the value is $\mathsf{one}$,
then $a$ gets instantiated to an index semantically equal to $1 \in \kindnat$.

However, a value of sum type is either a left- or right-injected value,
but we don't know which,
so we take the intersection of what each injection determines:
\[
  \Infer{\DeclTpSum}
  { \judgetp{\Theta}{P_1}{\Xi_1} \\ \judgetp{\Theta}{P_2}{\Xi_2} }
  { \judgetp{\Theta}{P_1 + P_2}{\Xi_1 \sect \Xi_2}  }
\]

The unit type $1$ and void (empty) type $0$ both have empty $\Xi$.
We also empty out value-determined indexes at shifts,
preventing certain quantifications over shifts.
For example,
$\alltype{a:\kindnat} \upshift{\textsf{Nat}(a)}$ (which is void anyway)
is not well-formed.
Crucially, we will see that this restriction, together with focusing,
guarantees indexes will be algorithmically solved by the end of certain stages.

We define functor and algebra well-formedness in \Figref{fig:declwf-fun-alg}
of the appendix.

Functor well-formedness $\judgefunctor{\Theta}{\mathcal{F}}{\Xi}$
is similar to type well-formedness:
constant functors output the $\Xi$ of the underlying positive type,
the identity and unit functors $\Id$ and $I$ have empty $\Xi$,
the product functor $B \otimes \hat{P}$ takes the union of the component $\Xi$s,
and the sum functor $F_1 \oplus F_2$ takes the intersection.
The latter two reflect the fact that unrolling inductive types (\Secref{sec:unrolling})
generates $+$ types from $\oplus$ functors and $\times$ types from $\otimes$ functors.
That $I$ has empty $\Xi$ reflects that $1$ (unrolled from $I$) does too,
together with the fact that asserting and guarded types do not affect $\Xi$.

Algebra well-formedness 
$\judgealgebra{\Xi}{\Theta}{\alpha}{F}{\tau}$
(read
``under $\Xi$ and $\Theta$, algebra $\alpha$ is well-formed
and has type $F(\tau) \Rightarrow \tau$'')
has two contexts:
$\Xi$ is for $\alpha$ (in particular, the index bodies of its clauses)
and $\Theta$ is for $F$ (in particular, the positive types of constant functors);
we maintain the invariant that $\Xi \subseteq \Theta$.
We have these separate contexts to prevent existential variables
from appearing in $\alpha$ (as explained with respect to \DeclTpFixVar)
while still allowing them to appear in $F$.
For example, consider
$\extype{b:\kindnat} \comprehend{\nu : \mu F(b)}{\Fold{F(b)}{\alpha}\,{\nu} = n}$
where
$F(b) = (\Const{\textsf{Nat}(b)} \otimes I) \oplus (\Const{\textsf{Nat}(b)} \otimes \Id \otimes I)$
and $\alpha = \inl \clause{(\wild, \unitexp)}{0} \matchor \inr \clause{(\wild, a, \unitexp)}{1+a}$.

Refined inductive type well-formedness initializes the input $\Xi$ to $\cdot$,
but index variables can be bound in the body of an algebra:
\begin{mathpar}
  \Infer{}
        { \judgealgebra{\Xi, a:\tau}{\Theta, a:\tau}{\clause{q}{t}}{\hat{P}}{\tau} }
        { \judgealgebra{\Xi}{\Theta}{\clause{(a, q)}{t}}{(\Id \otimes \hat{P})}{\tau} }
  \and
  \Infer{}
        { \judgealgebra{\Xi, a:\tau'}{\Theta, a:\tau'}{\clause{(\bap, q)}{t}}{(\Const{Q} \otimes \hat{P})}{\tau} }
        { \judgealgebra{\Xi}{\Theta}
          {\clause{(\pack{a}{\bap}, q)}{t}}
          {(\Const{\extype{a:\tau'}{Q}} \otimes \hat{P})}{\tau} }
\end{mathpar}
where the right rule simultaneously binds $a$ in both $t$ and $Q$,
and the left rule only binds $a$ in $t$
(but we add $a$ to both contexts
to maintain the invariant $\Xi \subseteq \Theta$ for inputs $\Xi$ and $\Theta$).
We sort algebra bodies only when a product ends at a unit
(possible by design of the functor grammar),
and merely under $\Xi$; constant functors depend on $\Theta$:
\begin{mathpar}
  \Infer{}
        { \judgeterm{\Xi}{t}{\tau} }
        { \judgealgebra{\Xi}{\Theta}{\clause{\unitexp}{t}}{I}{\tau} }
  \and
  \Infer{}
        { \judgealgebra{\Xi}{\Theta}{\clause{q}{t}}{\hat{P}}{\tau} \\ \judgetp{\Theta}{Q}{\dontcare}}
        { \judgealgebra{\Xi}{\Theta}{\clause{(\wild, q)}{t}}{(\Const{Q} \otimes \hat{P})}{\tau} }
\end{mathpar}

For algebras $\alpha$ of ``type'' $(F_1 \oplus F_2) \; \tau \Rightarrow \tau$,
we use a straightforward judgment $\composeinj{k}{\alpha}{\alpha_k}$
(appendix \Figref{fig:auxpat})
that outputs the $k$th clause $\alpha_k$ of input algebra $\alpha$:
\[
  \Infer{}
  {
    \composeinj{1}{\alpha}{\alpha_1}
    \\
    \composeinj{2}{\alpha}{\alpha_2}
    \\
    \judgealgebra{\Xi}{\Theta}{\alpha_1}{F_1}{\tau}
    \\
    \judgealgebra{\Xi}{\Theta}{\alpha_2}{F_2}{\tau}
  }
  {
    \judgealgebra{\Xi}{\Theta}{\alpha}{(F_1 \oplus F_2)}{\tau}
  }
\]

By restricting the bodies of algebras to \emph{index terms} $t$
and the carriers of our $F$-algebras to
\emph{index sorts} $\tau$,
we uphold the phase distinction:
we can therefore safely refine inductive types by folding them with algebras,
and also manage decidable typing.

\subsection{Equivalence}
\label{sec:equivalence}
We have %
equivalence judgments
for propositions $\judgeequiv[]{\Theta}{\phi}{\psi}$
(appendix \Figref{fig:declpropequiv}),
logical contexts $\judgeequiv[]{\Theta}{\Theta_1}{\Theta_2}$
(appendix \Figref{fig:decllogctxequiv}),
functors $\judgeequiv[]{\Theta}{\mathcal{F}}{\mathcal{G}}$
(appendix \Figref{fig:declfunequiv}),
and types $\judgeequiv[\pm]{\Theta}{A}{B}$
(appendix \Figref{fig:decltpequiv}),
which use $\judgeentail{\Theta}{\phi}$ to verify logical equality of index terms.
Basically, two entities are equivalent
if their respective, structural subparts are equivalent
(under the logical context).
Type/functor equivalence is used in sum and refined ADT subtyping
(type equivalence implies mutual subtyping),
as well as to prove algorithmic completeness (appendix \Lemref{lem:prog-typing-respects-equiv}),
but context equivalence is only used to prove algorithmic completeness
(appendix \Lemref{lem:ctx-equiv-compat}).
However, it should be possible to remove equivalence from the system itself,
by using ``subfunctoring'' and covariant sum subtyping.
For space reasons, we do not show all their rules here (see appendix),
only the ones we think are most likely to surprise.

Refined inductive types are equivalent
only if they use syntactically the same algebra
(but the algebra must be well-formed at both functors $F$ and $G$;
this holds by inversion on the conclusion's
presupposed type well-formedness judgments):
\[
  \Infer{}
  {
    \judgeequiv[]{\Theta}{F}{G}
    \\
    \judgeentail{\Theta}{t = t'}
  }
  {
    \judgeequiv[+]{\Theta}{\comprehend{\nu:\mu F}{\Fold{F}{\alpha}\,\nu =_\tau t}}{\comprehend{\nu:\mu G}{\Fold{G}{\alpha}\,\nu =_\tau t'}}
  }
\]

Two index equality propositions (respectively, two index inequalities) are equivalent if their respective sides are logically equal:
\begin{mathpar}
  \Infer{}
  {
    \judgeentail{\Theta}{t_1 = t_1'}
    \\
    \judgeentail{\Theta}{t_2 = t_2'}
  }
  {
    \judgeequiv[]{\Theta}{(t_1 = t_2)}{(t_1' = t_2')}
  }
  \and
  \Infer{}
  {
    \judgeentail{\Theta}{t_1 = t_1'}
    \\
    \judgeentail{\Theta}{t_2 = t_2'}
  }
  {
    \judgeequiv[]{\Theta}{(t_1 \le t_2)}{(t_1' \le t_2')}
  }
\end{mathpar}

We use logical context equivalence in proving
subsumption admissibility
(see \Secref{sec:subst})
and the completeness of algorithmic typing
(see \Secref{sec:completeness}).
Two logical contexts are judgmentally equivalent under $\Theta$
if they have exactly the same variable sortings (in the same list positions)
and logically equivalent (under $\Theta$)
propositions, in the same order.
The most interesting rule is the one for propositions,
where, in the second premise, we filter out propositions from $\Theta_1$
because we want each respective proposition to be logically equivalent
under the propositions (and indexes) of $\Theta$,
but variables in $\Theta_1$
(or $\Theta_2$) may appear in $\phi_1$ (or $\phi_2$):
\[
  \Infer{}
  {
    \judgeequiv{\Theta}{\Theta_1}{\Theta_2}
    \\
    \judgeequiv{\Theta, \overline{\Theta_1}}{\phi_1}{\phi_2}
  }
  {
    \judgeequiv{\Theta}{(\Theta_1, \phi_1)}{(\Theta_2, \phi_2)}
  }
\]
(Note that it is equivalent to use
$\overline{\Theta_2}$ rather than $\overline{\Theta_1}$
in the second premise above.)

All equivalence judgments satisfy reflexivity
(appendix Lemmas \ref{lem:refl-equiv-prop} and \ref{lem:refl-equiv-tp-fun}),
symmetry
(appendix Lemmas \ref{lem:symmetric-equiv-prop} and \ref{lem:symmetric-equiv-tp-fun}),
and transitivity
(appendix \Lemref{lem:trans-equiv}).

\subsection{Extraction}
\label{sec:extraction}
The judgment $\judgeextract[\pm]{\Theta}{A}{A'}{\Theta_A}$
(\Figref{fig:liq-declextract})
\emph{extracts} quantified variables
and $\land$ and $\implies$ propositions from the type $A$,
outputting the type $A'$ without these, and the context $\Theta_A$ with them.
We call $A'$ and $\Theta_A$ the type and context \emph{extracted} from $A$.
For negative $A$, everything is extracted up to an upshift.
For positive $A$, everything is extracted up to
any connective that is not $\exists$, $\land$, or $\times$.
For convenience in program typing (\Secref{sec:typing}),
$\simple{\Theta}{A}$ abbreviates $\judgeextract{\Theta}{A}{A}{\cdot}$
(we sometimes omit the polarity label from extraction judgments).
If $\simple{\Theta}{A}$, then we say $A$ is \emph{simple}.

\begin{figure}[hptb]
  \judgbox{\judgeextract[\pm]{\Theta}{A}{A'}{\Theta_A}}
          {Under $\Theta$, type $A$ extracts to $A'$ and $\Theta_A$}
\begin{mathpar}
  \Infer{\ExtractStopPos}
  {
    P \neq \exists, \land, \text{ or } \times
  }
  {
    \judgeextract[+]{\Theta}{P}{P}{\cdot}
  }
  \and
  \Infer{\ExtractWith}
  {
    \judgeextract[+]{\Theta}{P}{P'}{\Theta_P}
  }
  {
    \judgeextract[+]{\Theta}{P \land \phi}{P'}{\phi, \Theta_P}
  }
  \and
  \Infer{\ExtractEx}
  {
    \judgeextract[+]{\Theta, a:\tau}{P}{P'}{\Theta_P}
  }
  {
    \judgeextract[+]{\Theta}{\extype{a:\tau}{P}}{P'}{a:\tau, \Theta_P}
  }
  \and
  \Infer{\ExtractProd}
  {
    \judgeextract[+]{\Theta}{P_1}{P_1'}{\Theta_{P_1}}
    \\
    \judgeextract[+]{\Theta}{P_2}{P_2'}{\Theta_{P_2}}
  }
  {
    \judgeextract[+]{\Theta}{P_1 \times P_2}{P_1' \times P_2'}{\Theta_{P_1}, \Theta_{P_2}}
  }
  \\
  \Infer{\ExtractStopNeg}
  {
  }
  {
    \judgeextract[-]{\Theta}{\upshift{P}}{\upshift{P}}{\cdot}
  }
  \and
  \Infer{\ExtractImp}
  {
    \judgeextract[-]{\Theta}{N}{N'}{\Theta_N}
  }
  {
    \judgeextract[-]{\Theta}{\phi \implies N}{N'}{\phi, \Theta_N}
  }
  \and
  \Infer{\ExtractAll}
  {
    \judgeextract[-]{\Theta, a:\tau}{N}{N'}{\Theta_N}
  }
  {
    \judgeextract[-]{\Theta}{\alltype{a:\tau}{N}}{N'}{a:\tau, \Theta_N}
  }
  \and
  \Infer{\ExtractArrow}
  {
    \judgeextract[+]{\Theta}{P}{P'}{\Theta_P}
    \\
    \judgeextract[-]{\Theta}{N}{N'}{\Theta_N}
  }
  {
    \judgeextract[-]{\Theta}{P \to N}{P' \to N'}{\Theta_P, \Theta_N}
  }
\end{mathpar}

\caption{Declarative extraction}
\ifnum\OPTIONAppendix=0
    \label{fig:liq-declextract}
\else
    \label{fig:declextract}
\fi
\end{figure}

\subsection{Subtyping}
\label{sec:subtyping}
Declarative subtyping $\judgesub[\pm]{\Theta}{A}{B}$
is defined in \Figref{fig:liq-declsub}.

\begin{figure}[thbp]
\raggedright
\judgbox{\judgesub[\pm]{\Theta}{A}{B}}{Under $\Theta$, type $A$ is a subtype of $B$}
\begin{mathpar}
  \Infer{\DeclSubPosUnit}
  { }
  {
    \judgesub[+]{\Theta}{1}{1}
  }
  \and
  \Infer{\DeclSubPosVoid}
  { }
  {
    \judgesub[+]{\Theta}{0}{0}
  }
  \\
  \Infer{\DeclSubPosProd}
  {
    \judgesub[+]{\Theta}{P_1}{Q_1}
    \\
    \judgesub[+]{\Theta}{P_2}{Q_2}
  }
  {
    \judgesub[+]{\Theta}{P_1 \times P_2}{Q_1 \times Q_2}
  }
  \and
  \Infer{\DeclSubPosSum}
  {
    \judgeequiv[+]{\Theta}{P_1}{Q_1}
    \\
    \judgeequiv[+]{\Theta}{P_2}{Q_2}
  }
  {
    \judgesub[+]{\Theta}{P_1 + P_2}{Q_1 + Q_2}
  }
  \and
  \Infer{\DeclSubPosL}
  {
    \judgeextract[+]{\Theta}{P}{P'}{\Theta_P}
    \\
    \Theta_P \neq \cdot
    \\
    \judgesub[+]{\Theta, \Theta_P}{P'}{Q}
  }
  {
    \judgesub[+]{\Theta}{P}{Q}
  }
  \\
  \Infer{\DeclSubPosWithR}
  {
    \judgesub[+]{\Theta}{P}{Q}
    \\
    \judgeentail{\Theta}{\phi}
  }
  {
    \judgesub[+]{\Theta}{P}{Q \land \phi}
  }
  \and
  \Infer{\DeclSubPosExR}
  { 
    \judgesub[+]{\Theta}{P}{[t/a]Q} 
    \\
    \judgeterm{\Theta}{t}{\tau} 
  }
  {
    \judgesub[+]{\Theta}{P}{\extype{a:\tau}{Q}}
  }
  \and
  \Infer{\DeclSubPosFix}
  {
    \judgeequiv[]{\Theta}{F}{G}
    \\
    \judgeentail{\Theta}{t = t'}
  }
  {
    \judgesub[+]{\Theta}{\comprehend{\nu:\mu F}{\Fold{F}{\alpha}\,\nu =_\tau t}}{\comprehend{\nu:\mu G}{\Fold{G}{\alpha}\,\nu =_\tau t'}}
  }
  \\
  \Infer{\DeclSubPosDownshift}
  {
    \judgesub[-]{\Theta}{N}{M}
  }
  {
    \judgesub[+]{\Theta}{\downshift{N}}{\downshift{M}}
  }
  \and
  \Infer{\DeclSubNegUpshift}
  {
    \judgesub[+]{\Theta}{P}{Q}
  }
  {
    \judgesub[-]{\Theta}{\upshift{P}}{\upshift{Q}}
  }
  \\
  \Infer{\DeclSubNegImpL}
  {
    \judgesub[-]{\Theta}{N}{M}
    \\
    \judgeentail{\Theta}{\phi}
  }
  {
    \judgesub[-]{\Theta}{\phi \implies N}{M}
  }
  \and
  \Infer{\DeclSubNegAllL}
  {
    \judgesub[-]{\Theta}{[t/a]N}{M} 
    \\ 
    \judgeterm{\Theta}{t}{\tau} 
  }
  {
    \judgesub[-]{\Theta}{\alltype{a:\tau}{N}}{M}
  }
  \and
  \Infer{\DeclSubNegR}
  {
    \judgeextract[-]{\Theta}{M}{M'}{\Theta_M}
    \\
    \Theta_M \neq \cdot
    \\
    \judgesub[-]{\Theta, \Theta_M}{N}{M'}
  }
  {
    \judgesub[-]{\Theta}{N}{M}
  }
  \\
  \Infer{\DeclSubNegArrow}
  {
    \judgesub[+]{\Theta}{Q}{P}
    \\
    \judgesub[-]{\Theta}{N}{M}
  }
  {
    \judgesub[-]{\Theta}{P \to N}{Q \to M}
  }  
\end{mathpar}

\caption{Declarative subtyping}
\ifnum\OPTIONAppendix=0
  \label{fig:liq-declsub}
\else
  \label{fig:declsub}
\fi
\end{figure}

Subtyping is polarized into mutually recursive positive $\judgesub[+]{\Theta}{P}{Q}$
and negative $\judgesub[-]{\Theta}{N}{M}$ relations.
The design of inference rules for subtyping is guided by sequent calculi,
perhaps most clearly seen in the left and right rules pertaining
to quantifiers ($\exists$, $\forall$), asserting types ($\land$),
and guarded types ($\implies$).
This is helpful to establish key properties such as \emph{reflexivity}
and \emph{transitivity} (viewing subtyping as a sequent system,
we might instead say that the structural \emph{identity} and \emph{cut} rules, respectively, are admissible\footnote{%
A proposed inference rule is \emph{admissible} with respect to a system if,
whenever the premises of the proposed rule are derivable,
we can derive the proposed rule's conclusion using the system's inference rules.}%
).
We interpret types as sets with some additional structure (\Secref{sec:semantics}),
but considering only the sets,
we prove that a subtype denotes a subset of the set denoted by any of its supertypes.
That is, membership of a (semantic) value in the subtype \emph{implies}
its membership in any supertype of the subtype.
We may also view subtyping as implication.

Instead of \DeclSubPosL, one might reasonably expect these two rules
(the brackets around the rule names indicate that these rules are \emph{not} in \sysname):
\begin{mathpar}
  \Infer{[{\DeclSubPosWithL}]}
  {\judgesub[+]{\Theta, \phi}{P}{Q} }
  {\judgesub[+]{\Theta}{P \land \phi}{Q} }
  \and
  \Infer{[{\DeclSubPosExL}]}
  {\judgesub[+]{\Theta, a:\tau}{P}{Q} }
  {\judgesub[+]{\Theta}{\extype{a:\tau}{P}}{Q} }
\end{mathpar}
Similarly, one might expect to have [{\DeclSubNegImpR}] and [{\DeclSubNegAllR}], dual to the above rules,
instead of the dual rule \DeclSubNegR.
Reading, for example, the above rule [{\DeclSubPosWithL}] logically and top-down,
if $\Theta$ and $\phi$ implies that $P$ implies $Q$,
then we can infer that $\Theta$ implies that $P$ and $\phi$ implies $Q$.
We can also read rules as a bottom-up decision procedure:
given $P \land \phi$, we know $\phi$, so we can assume it;
given $\extype{a:\tau} P$, we know there exists an index of sort $\tau$ such that $P$,
but we don't have a specific index term.
However, these rules are not powerful enough to derive
reasonable judgments such as
$\judgesub[+]{a:\kindnat}{1 \times (1 \land a=3)}{(1 \land a \geq 3) \times 1}$:
subtyping for the first component requires verifying $a \geq 3$,
which is impossible under no logical assumptions.
But from a logical perspective, $1 \times (1 \land a=3)$
implies $a \geq 3$.
Reading \DeclSubPosL bottom-up, in this case, we extract $a=3$ from the subtype,
which we later use to verify that $a \geq 3$.
The idea is that, for a type in an assumptive position,
it does not matter which product component (products are viewed conjunctively)
or function argument (in \sysname, functions must be fully applied to values)
to which index data is attached.
Moreover, as we'll explain at the end of \Secref{sec:algorithmic-system},
the weaker rules by themselves are incompatible with algorithmic completeness.
We emphasize that we do \emph{not} include
[{\DeclSubPosWithL}], [{\DeclSubPosExL}], [{\DeclSubNegImpR}] or [{\DeclSubNegAllR}]
in the system.

For the unit type and the void type,
rules \DeclSubPosUnit and void \DeclSubPosVoid are simply reflexivity.
Product subtyping \DeclSubPosProd
is covariant subtyping of component types:
a product type is a subtype of another
if each component of the former
is a subtype of the respective component of the latter.
We have covariant shift rules \DeclSubPosDownshift and \DeclSubNegUpshift.
Function subtyping \DeclSubNegArrow is standard:
contravariant (from conclusion to premise, the subtyping direction flips)
in the function type's domain
and covariant in the function type's codomain.

Rule \DeclSubPosWithR and its dual rule \DeclSubNegImpL
verify the validity of the attached proposition.
In rule \DeclSubPosExR and its dual rule \DeclSubNegAllL,
we assume that we can conjure a suitable index term $t$;
in practice (that is, algorithmically),
we must introduce an existential variable $\ahat$
and then solve it.

Rule \DeclSubPosSum says a sum is a subtype of another sum if their
respective subparts are (judgmentally) \emph{equivalent}.
Judgmental equivalence does not use judgmental extraction.
The logical reading of subtyping begins to clarify why
we don't extract anything under a sum connective:
$(1 \land \False) + 1$ does not imply $\False$.
However, using equivalence here is a conservative restriction:
for example, $(1 \land \False) + (1 \land \False)$ does imply $\False$.
Regardless, we don't expect this to be very restrictive in practice
because programmers tend not to work with sum types themselves,
but rather algebraic inductive types (like $\mu F$),
and don't need to directly compare, via subtyping,
(the unrolling of) different such types
(such as the type of lists and the type of natural numbers).

In rule \DeclSubPosFix,
just as in the refined inductive type equivalence rule (\Secref{sec:equivalence}),
a refined inductive type is a subtype of another type
if they have judgmentally equivalent functors,
they use syntactically the same algebra
(that agrees with both subtype and supertype functors),
and the index terms on the right-hand side
of their measurements are equal under the logical context.
As we discuss in \Secref{sec:conclusion},
adding polymorphism to the language
(future work)
might necessitate replacing type and functor equivalence in subtyping
with subtyping and ``subfunctoring''.

In the appendix, we prove that subtyping is
reflexive
(\Lemref{lem:refl-sub})
and transitive
(\Lemref{lem:trans-sub}).

\paragraph{Subtyping and type equivalence}
We prove that type equivalence implies subtyping
(appendix \Lemref{lem:equiv-implies-sub}).
To prove that, we use the fact that
if $\Theta_1$ is logically equivalent to $\Theta_2$
under their prefix context $\Theta$
(judgment $\judgeequiv[]{\Theta}{\Theta_1}{\Theta_2}$)
then we can replace $\Theta_1$ with $\Theta_2$ (and vice versa) in derivations
(appendix \Lemref{lem:ctx-equiv-compat}).
We use
appendix \Lemref{lem:equiv-implies-sub}
to prove subsumption admissibility (\Secref{sec:subst})
and
a subtyping constraint verification transport lemma
(mentioned in \Secref{sec:algorithmic-soundness}).
Conversely, mutual subtyping does not imply type equivalence:
$|- 1 \land \True \leq 1$
and $|- 1 \leq 1 \land \True$
but $|- 1 \not\equiv 1 \land \True$
because the unit type is structurally distinct from an asserting type.

\subsection{Unrolling}
\label{sec:unrolling}

Given $a:\kindnat$,
in \sysname,
the type $\textsf{List} \; P \; a$ of $a$-length lists of elements of type $P$
is defined as
$\comprehend{\nu : \mu \textsf{ListF}_P}{\Fold{\textsf{ListF}_P}{\textsf{lenalg}}\,{\nu} = a}$
where
$\textsf{ListF}_P = I \oplus (\Const{P} \otimes \Id \otimes I)$
and $\textsf{lenalg} = \inl \clause{\unitexp}{0} \matchor \inr \clause{(\wild, (b, \unitexp))}{1 + b}$.
Assuming we have
$\textsf{succ} : \alltype{a:\kindnat} \textsf{Nat}(a) \to \upshift{\textsf{Nat}(1 + a)}$
for incrementing a (program-level) natural number by one,
we define \textsf{length} in \sysname as follows:
\[
  \arrayenvl{
    \textkw{rec} \; \textsf{length} : (\alltype{a:\kindnat} \textsf{List}(P)(a) \to \upshift{\natsing{a}}). \; \lambda x. \; \textkw{match} \; x. \; \{ \\
    ~~~~\roll{x'} \clausesym \textkw{match} \; x' \; \{ \\
    ~~~~~~~~\inl{\unit} \clausesym~~~~~~~~~~~~~~~~~~~~~~~~~~~~~\text{-}\text{- }a=0 \\
    ~~~~~~~~~~~~\Return{\roll{\inl{\unit}}} \\
    ~~~\bnfalt~ \inr{\pair{\wild}{\pair{y}{\unit}}} \clausesym ~~~~~~~~~~~~~~~\text{-}\text{- }a=1+a'\text{ such that }a'\text{ is the length of }y\\
    ~~~~~~~~~~~~\textkw{let} \; z' = \textsf{length}(y) \textkw{;} \\
    ~~~~~~~~~~~~\textkw{let} \; z = \textsf{succ}(z') \textkw{;} \\
    ~~~~~~~~~~~~\Return{z} \\
    ~~~~\} \\
    \}
  }
\]

Checking \textsf{length} %
against its type annotation,
the lambda rule assumes $x : \textsf{List}(P)(a)$
for an arbitrary $a:\kindnat$.
Upon matching $x$ against the pattern $\roll{x'}$,
we know $x'$ should have the \emph{unrolled} type of $\textsf{List}(P)(a)$.
Ignoring refinements,
we know that the erasure of this unrolling should be
a sum type where the left component represents the empty list 
and the right component represents a head element together with a tail list.
However, in order to verify the refinement that \textsf{length}
does what we intend,
we need to know more about the length index associated with $x$---that is, $a$---in
the case where $x$ is nil and in the case where $x$ is a cons cell.
Namely, the unrolling of $\textsf{List}(P)(a)$
should know that $a = 0$ when $x$ is the empty list,
and that $a = 1 + a'$ where $a'$ is the length of the tail of $x$
when $x$ is a nonempty list.
This is the role of the unrolling judgment,
to output just what we need here:
\[
  \judgeunroll*{\cdot}{\cdot}
  {\nu:\textsf{ListF}_P[\mu \textsf{ListF}_P]}
  {\textsf{lenalg}}
  {\textsf{ListF}_P\;\Fold{\textsf{ListF}_P}{\textsf{lenalg}}\,{\nu}}
  {a}
  {(1 \land a = 0)
    +
    \big( P \times (\extype{a':\kindnat}
    {\comprehend{\nu:\mu \textsf{ListF}_P}{\Fold{\textsf{ListF}_P}{\textsf{lenalg}}\,{\nu} =_\kindnat a'}
      \times (1 \land a = 1 + a'))\big)}}
  {\kindnat}
\]
That is, the type of $P$-lists of length $a$ unrolls to either
the unit type $1$ (representing the empty list)
together with the fact that $a$ is $0$,
or the product of $P$ (the type of the head element)
and $P$-lists (representing the tail)
of length $a'$ such that $a'$ is $a$ minus one.

Refined inductive type unrolling
$\judgeunroll{\Xi}{\Theta}{ \nu:G[\mu F] }{\beta}{G\;\Fold{F}{\alpha}\;\nu}{t}{P}{\tau}$,
inspired by work in fibrational dependent type theory \cite{Atkey12},
is defined in \Figref{fig:liq-unroll}.
There are two contexts: $\Xi$ is for $\beta$
and $\Theta$ is for $G$, $F$, and $t$.
Similarly to algebra well-formedness,
we maintain the invariant in unrolling that $\Xi \subseteq \Theta$.
The (non-contextual)
\emph{input} metavariables are $G$, $F$, $\beta$, $\alpha$, $\tau$, and $t$.
The type $P$, called the \emph{unrolled} type, is an output.
As in the list example above,
inductive type unrolling is always initiated
with $\Xi = \cdot$ and $G = F$ and $\beta = \alpha$.

\begin{figure}[htbp]
\raggedright

\judgbox{%
  \judgeunroll{\Xi}{\Theta}{ \nu:G[\mu F] }{\beta}{G\;\Fold{F}{\alpha}\;\nu}{t}{P}{\tau}
}{
  Abbreviated
  \\
  $\unroll{\Xi}{\Theta}{G}{F}{\beta}{\alpha}{\tau}{t}{P}$
}
\vspace{1em}

\begin{mathpar}
  \Infer{\DeclUnrollSum}
  {\arrayenvbl{
      \composeinj{1}{\beta}{\beta_1}
      \\
      \composeinj{2}{\beta}{\beta_2}
    }
    \\
    \arrayenvbl{
      \judgeunroll{\Xi}{\Theta}{\nu:G[\mu F]}{\beta_1}{G\;\Fold{F}{\alpha}\;\nu}{t}{P}{\tau}
      \\
      \judgeunroll{\Xi}{\Theta}{\nu:H[\mu F]}{\beta_2}{H\;\Fold{F}{\alpha}\;\nu}{t}{Q}{\tau}
    }
  }
  { \judgeunroll{\Xi}{\Theta}{\nu:(G \oplus H)[\mu F]}{\beta}{(G \oplus H)\;\Fold{F}{\alpha}\;\nu}{t}{P + Q}{\tau} }
  \and
  \Infer{\DeclUnrollId}
  {
    \judgeunroll{\Xi,a:\tau}{\Theta, a:\tau}{\nu:\hat{P}[\mu F]}
    {(\clause{q}{t'})}{\hat{P}\;\Fold{F}{\alpha}\;\nu}{t}{Q}{\tau} 
  }
  {
    \judgeunroll*{\Xi}{\Theta}{\nu : (\Id\otimes\hat{P})[\mu F]}
    {(\clause{(a,q)}{t'})}
    {(\Id\otimes\hat{P})\;\Fold{F}{\alpha}\;\nu}{t}
    {\extype{a:\tau}
      {\comprehend{\nu:\mu F}{ \Fold{F}{\alpha}\,{\nu} =_\tau a } \times Q}}
    {\tau}
  }
  \and
  \Infer{\DeclUnrollConstEx}
  {
    \judgeunroll{\Xi,a:\tau'}{\Theta,a:\tau'}{\nu:(\Const{Q}\otimes\hat{P})[\mu F]}{(\clause{(\bap,q)}{t'})}{(\Const{Q}\otimes\hat{P})\;\Fold{F}{\alpha}\;\nu}{t}{Q'}{\tau} }
  { \judgeunroll*{\Xi}{\Theta}{\nu :(\Const{\extype{a:\tau'}{Q}}\otimes\hat{P})[\mu F]}{(\clause{(\pack{a}{\bap}, q)}{t'})}{(\Const{\extype{a:\tau'}{Q}}\otimes\hat{P})\;\Fold{F}{\alpha}\;\nu}{t}{\extype{a:\tau'}{Q'}}{\tau} }
  \and
  \Infer{\DeclUnrollConst}
  {
    \judgeunroll{\Xi}{\Theta}{\nu:\hat{P}[\mu F]}{(\clause{q}{t'})}{\hat{P}\;\Fold{F}{\alpha}\;\nu}{t}{Q'}{\tau} }
  { \judgeunroll{\Xi}{\Theta}{\nu :(\Const{Q}\otimes\hat{P})[\mu F]}{(\clause{(\wild,q)}{t'})}{(\Const{Q}\otimes\hat{P})\;\Fold{F}{\alpha}\;\nu}{t}{Q \times Q'}{\tau} }
  \and
  \Infer{\DeclUnrollUnit}
  { }
  { \judgeunroll{\Xi}{\Theta}{\nu:I[\mu F]}{(\clause{\unitexp}{t'})}{I\;\Fold{F}{\alpha}\;\nu}{t}{1 \land (t = t')}{\tau} }
\end{mathpar}

\caption{Unrolling}
\ifnum\OPTIONAppendix=0
  \label{fig:liq-unroll}
\else
  \label{fig:unroll}
\fi
\end{figure}

\DeclUnrollSum unrolls each branch and then sums the resulting types.
\DeclUnrollId
outputs the product of the original inductive type
but with a measurement
given by the recursive result of the fold (over which we existentially quantify),
together with the rest of the unrolling.
\DeclUnrollConstEx pushes the packed index variable $a$ onto the context
and continues unrolling, existentially quantifying over the result;
in the conclusion, $a$ is simultaneously bound in $Q$ and $t'$.
\DeclUnrollConst outputs a product of the type of the constant functor
and the rest of the unrolling.
\DeclUnrollUnit simply outputs the unit type
together with the index term equality given by the (unrolled) measurement.

If our functor and algebra grammars were instead more direct,
like those implicitly used in the introduction (\Secref{sec:introduction})
and overview (\Secref{sec:overview}),
and explicitly discussed in \Secref{sec:functor-algebra-grammar},
then we would have to modify the unrolling judgment,
and it would need two more rules.
We expect everything would still work,
but we prefer having to consider fewer rules when proving metatheory.

\paragraph{Unrolling, equivalence and subtyping}

Substituting judgmentally equivalent types, functors and indexes
for the inputs of unrolling generates an output type
that is both a subtype and supertype of
the original unrolling output:

\begin{lemma}[Unroll to Mutual Subtypes]
  \hfill (\Lemref{lem:unroll-to-mutual-sub} in appendix)\\
  If $\judgeunroll{\Xi}{\Theta}{\nu:G[\mu F]}{\beta}{G\;\Fold{F}{\alpha}\;\nu}{t}{P}{\tau}$\\
  and $\judgeequiv[]{\Theta}{G}{G'}$
  and $\judgeequiv[]{\Theta}{F}{F'}$
  and $\judgeentail{\Theta}{t=t'}$,\\
  then there exists $Q$
  such that $\judgeunroll{\Xi}{\Theta}{\nu:G'[\mu F']}{\beta}{G'\;\Fold{F'}{\alpha}\;\nu}{t'}{Q}{\tau}$\\
  and $\judgesub[+]{\Theta}{P}{Q}$ and $\judgesub[+]{\Theta}{Q}{P}$.
\end{lemma}

We use this to prove subsumption admissibility (see \Secref{sec:subst})
for the cases that involve constructing and pattern matching inductive values.

\subsection{Typing}
\label{sec:typing}
Declarative bidirectional typing rules are given in 
Figures~\ref{fig:liq-decltyping-head-bound-expression},
\ref{fig:liq-decltyping-value-expression},
and \ref{fig:liq-declmatch}.
By careful design,
guided by logical principles,
all typing rules are syntax-directed.
That is,
when deriving a conclusion,
at most one rule is compatible with the syntax of the input program term
and the principal input type.

To manage the interaction between subtyping and program typing,
types in a well-formed (under $\Theta$)
program context $\Gamma$ must be invariant under extraction:
for all $(x:P) \in \Gamma$,
we have $\judgeextract[+]{\Theta}{P}{P}{\cdot}$ (that is, $\simple{\Theta}{P}$).
We maintain this invariant in program typing by extracting
before adding any variable typings to the context.

\begin{figure}[htbp]
\raggedright
\judgbox{\judgesynhead{\Theta}{\Gamma}{h}{P}}
        {Under $\Theta$ and $\Gamma$, head $h$ synthesizes type $P$}
\begin{mathpar}
  \Infer{\DeclSynHeadVar}
      {
        (x : P) \in \Gamma
      }
      {
        \judgesynhead{\Theta}{\Gamma}{x}{P}
      }
  \and
  \Infer{\DeclSynValAnnot}
      {
        \judgetp{\Theta}{P}{\Xi}
        \\
        \judgechkval{\Theta}{\Gamma}{v}{P}
      }
      {
        \judgesynhead{\Theta}{\Gamma}{\annoexp{v}{P}}{P}
      }
\end{mathpar}

\judgbox{\judgesynexp{\Theta}{\Gamma}{\be}{\upshift{P}}}
        {Under $\Theta$ and $\Gamma$, bound expression $\be$ synthesizes type $\upshift{P}$}
\begin{mathpar}
  \Infer{\DeclSynSpineApp}
        { \judgesynhead{\Theta}{\Gamma}{h}{\downshift{N}} \\
          \judgespine{\Theta}{\Gamma}{s}{N}{\upshift{P}} }
        { \judgesynexp{\Theta}{\Gamma}{h(s)}{\upshift{P}} }
  \and
  \Infer{\DeclSynExpAnnot}
        {
          \judgetp{\Theta}{P}{\Xi}
          \\
          \judgechkexp{\Theta}{\Gamma}{e}{\upshift{P}}
        }
        { \judgesynexp{\Theta}{\Gamma}{\annoexp{e}{\upshift{P}}}{\upshift{P}} }
\end{mathpar}

  \caption{Declarative head and bound expression type synthesis}
\ifnum\OPTIONAppendix=0
  \label{fig:liq-decltyping-head-bound-expression}
\else
  \label{fig:decltyping-head-bound-expression}
\fi
\end{figure}

\begin{figure}[htbp]
  \raggedright
\judgbox{\judgechkval{\Theta}{\Gamma}{v}{P}}
        {Under $\Theta$ and $\Gamma$, value $v$ checks against type $P$}
\begin{mathpar}
  \Infer{\DeclChkValVar}
        {
          P \neq \exists, \with
          \\
          (x:Q) \in \Gamma
          \\
          \judgesub[+]{\Theta}{Q}{P}
        }
        {
          \judgechkval{\Theta}{\Gamma}{x}{P}
        }
  \and
  \Infer{\DeclChkValUnit}
        { }
        { \judgechkval{\Theta}{\Gamma}{\unit}{1} }
  \and
  \Infer{\DeclChkValPair}
        { \judgechkval{\Theta}{\Gamma}{v_1}{P_1} \\
          \judgechkval{\Theta}{\Gamma}{v_2}{P_2} } 
        { \judgechkval{\Theta}{\Gamma}{\pair{v_1}{v_2}}{P_1 \times P_2} }
  \and
  \Infer{\DeclChkValIn{k}}
        { \judgechkval{\Theta}{\Gamma}{v}{P_k} }
        { \judgechkval{\Theta}{\Gamma}{\inj{k}{v}}{P_1 + P_2} }
  \and 
  \Infer{\DeclChkValExists}
        {
          \judgechkval{\Theta}{\Gamma}{v}{[t/a]P}
          \\
          \judgeterm{\Theta}{t}{\tau}
        }
        { \judgechkval{\Theta}{\Gamma}{v}{(\extype{a:\tau} P)} }
  \and 
  \Infer{\DeclChkValWith}
        {
          \judgechkval{\Theta}{\Gamma}{v}{P}
          \\
          \judgeentail{\Theta}{\phi}
        }
        { \judgechkval{\Theta}{\Gamma}{v}{P \land \phi} }
  \and
  \Infer{\DeclChkValFix}
        { \judgeunroll{\cdot}{\Theta}{\nu:F[\mu F]}{\alpha}{F\;\Fold{F}{\alpha}\;\nu}{t}{P}{\tau}
          \\ 
          \judgechkval{\Theta}{\Gamma}{v}{P} }
        { \judgechkval{\Theta}{\Gamma}{\roll{v}}
              {\comprehend{\nu:\mu F}{\Fold{F}{\alpha}\,{\nu} =_\tau t}} }
  \and
  \Infer{\DeclChkValDownshift}
        { \judgechkexp{\Theta}{\Gamma}{e}{N} }
        { \judgechkval{\Theta}{\Gamma}{\thunk{e}}{\downshift{N}} }
  \end{mathpar}
  \judgbox{\judgechkexp{\Theta}{\Gamma}{e}{N}}
      {Under $\Theta$ and $\Gamma$, expression $e$ checks against type $N$}
  \begin{mathpar}
  \Infer{\DeclChkExpUpshift}
        { \judgechkval{\Theta}{\Gamma}{v}{P} }
        { \judgechkexp{\Theta}{\Gamma}{\Return{v}}{\upshift{P}} }
  \and 
  \Infer{\DeclChkExpLet}
        { \simple{\Theta}{N} \\
          \judgesynexp{\Theta}{\Gamma}{\be}{\upshift{P}} \\
          \judgeextract[+]{\Theta}{P}{P'}{\Theta_P} \\
          \judgechkexp{\Theta, \Theta_P}{\Gamma, x:P'}{e}{N} }
        { \judgechkexp{\Theta}{\Gamma}{\Let{x}{\be}{e}}{N} }
  \and
  \Infer{\DeclChkExpMatch}
        {
          \simple{\Theta}{N}
          \\
          \judgesynhead{\Theta}{\Gamma}{h}{P}
          \\
          \judgechkmatch{\Theta}{\Gamma}{P}{\clauses{\pa}{e}{i}{I}}{N}
        }
        {
          \judgechkexp{\Theta}{\Gamma}{\match{h}{\clauses{\pa}{e}{i}{I}}}{N}
        }
  \and
  \Infer{\DeclChkExpLam}
        { \simple{\Theta}{P \to N} \\
          \judgechkexp{\Theta}{\Gamma, x:P}{e}{N} }
        { \judgechkexp{\Theta}{\Gamma}{\fun{x}{e}}{P \to N} }
  \and
  \Infer{\DeclChkExpUnreachable}
  {
    \simple{\Theta}{N}
    \\
    \judgeentail{\Theta}{\False}
  }
  {
    \judgechkexp{\Theta}{\Gamma}{\unreachable}{N}
  }
  \and
  \Infer{\DeclChkExpRec}
  {
    \arrayenvbl{
      \simple{\Theta}{N}
      \\
      \judgesub[-]{\Theta}{\alltype{a:\kindnat} M}{N}
    }
    \\
    \judgechkexp{\Theta, a:\kindnat}{\Gamma, x:\downshift{\big(\alltype{a':\kindnat} (a' < a) \implies [a'/a]M}\big)}{e}{M}
  }
  {
    \judgechkexp{\Theta}{\Gamma}{\rec{x : (\alltype{a:\kindnat} M)}{e}}{N}
  }
  \and
  \Infer{\DeclChkExpExtract}
  {
    \judgeextract{\Theta}{N}{N'}{\Theta_N}
    \\
    \Theta_N \neq \cdot
    \\
    \judgechkexp{\Theta, \Theta_N}{\Gamma}{e}{N'}
  }
  {
    \judgechkexp{\Theta}{\Gamma}{e}{N}
  }
\end{mathpar}
  
  \caption{Declarative value and expression type checking}
\ifnum\OPTIONAppendix=0
  \label{fig:liq-decltyping-value-expression}
\else
  \label{fig:decltyping-value-expression}
\fi
\end{figure}

\begin{figure}[htbp]
\raggedright
\judgbox{\judgechkmatch{\Theta}{\Gamma}{P}{\clauses{\pa}{e}{i}{I}}{N}}
        {Under $\Theta$ and $\Gamma$, patterns $r_i$ match against (input) type $P$ \\
          and branch expressions $e_i$ check against type $N$}
\begin{mathpar}
  \Infer{\DeclChkMatchEx}
        {\judgechkmatch{\Theta, a:\tau}{\Gamma}{P}{\clauses{\pa}{e}{i}{I}}{N}}
        {\judgechkmatch{\Theta}{\Gamma}{\extype{a:\tau}{P}}{\clauses{\pa}{e}{i}{I}}{N}}
  \and
  \Infer{\DeclChkMatchWith}
        {\judgechkmatch{\Theta, \phi}{\Gamma}{P}{\clauses{\pa}{e}{i}{I}}{N}}
        {\judgechkmatch{\Theta}{\Gamma}{P \land \phi}{\clauses{\pa}{e}{i}{I}}{N}}
  \and
  \Infer{\DeclChkMatchUnit}
        { \judgechkexp{\Theta}{\Gamma}{e}{N} }
        { \judgechkmatch{\Theta}{\Gamma}{1}{\setof{\clause{\unit}{e}}}{N} }
  \and
  \Infer{\DeclChkMatchPair}
        { \arrayenvb{\judgeextract{\Theta}{P_1}{P_1'}{\Theta_1} \\
          \judgeextract{\Theta}{P_2}{P_2'}{\Theta_2}} \\
          \judgechkexp{\Theta, \Theta_1, \Theta_2}{\Gamma, x_1:P_1', x_2:P_2'}{e}{N} }
        {
          \judgechkmatch{\Theta}{\Gamma}{P_1 \times P_2}
          {\setof{\clause{\pair{x_1}{x_2}}{e}}}{N}
        }
  \and
  \Infer{\DeclChkMatchSum}
        { \arrayenvb{
            \judgeextract{\Theta}{P_1}{P_1'}{\Theta_1} \\
            \judgeextract{\Theta}{P_2}{P_2'}{\Theta_2} \\
          }
          \\
          \arrayenvb{
            \judgechkexp{\Theta, \Theta_1}{\Gamma, x_1:P_1'}{e_1}{N} \\
            \judgechkexp{\Theta, \Theta_2}{\Gamma, x_2:P_2'}{e_2}{N} 
          }
        }
        { \judgechkmatch{\Theta}{\Gamma}{P_1 + P_2}{\setof{ \clause{\inl{x_1}}{e_1} \bnfalt \clause{\inr{x_2}}{e_2}}}{N} }
  \and
  \Infer{\DeclChkMatchVoid}
        { }
        { \judgechkmatch{\Theta}{\Gamma}{0}{\setof{}}{N} }
  \and
  \Infer{\DeclChkMatchFix}
        {
          \judgeunroll{\cdot}{\Theta}{\nu:F[\mu F]}{\alpha}{F\;\Fold{F}{\alpha}\;\nu}{t}{Q}{\tau}
          \\
          \judgeextract{\Theta}{Q}{Q'}{\Theta_Q}
          \\
          \judgechkexp{\Theta, \Theta_Q}{\Gamma, x:Q'}{e}{N}
        }
        {
          \judgechkmatch{\Theta}{\Gamma}{\comprehend{\nu:\mu F}
            {\Fold{F}{\alpha}\,{\nu} =_\tau t}}{\setof{\clause{\roll{x}}{e}}}{N}
        }
\end{mathpar}

\medskip

\judgbox{\judgespine{\Theta}{\Gamma}{s}{N}{\upshift{P}}}
{Under $\Theta$ and $\Gamma$,
  if a head of type $\downshift{N}$ is applied to the spine $s$, \\
  then it will return a result of type $\upshift{P}$}
\begin{mathpar}
  \Infer{\DeclSpineAll}
        {
          \judgeterm{\Theta}{t}{\tau}
          \\
          \judgespine{\Theta}{\Gamma}{s}{[t/a]N}{\upshift{P}}
        }
        {\judgespine{\Theta}{\Gamma}{s}{\alltype{a:\tau}{N}}{\upshift{P}}}
  \and
  \Infer{\DeclSpineImplies}
        {\judgeentail{\Theta}{\phi} \\ \judgespine{\Theta}{\Gamma}{s}{N}{\upshift{P}}  }
        {\judgespine{\Theta}{\Gamma}{s}{\phi \implies N}{\upshift{P}}}
  \and 
  \Infer{\DeclSpineApp}
        {\judgechkval{\Theta}{\Gamma}{v}{Q} \\ \judgespine{\Theta}{\Gamma}{s}{N}{\upshift{P}} }
        {\judgespine{\Theta}{\Gamma}{v,s}{Q \to N}{\upshift{P}}}
  \and 
  \Infer{\DeclSpineNil}
        { }
        {\judgespine{\Theta}{\Gamma}{\cdot}{\upshift{P}}{\upshift{P}}}
\end{mathpar}
  \centering
  
  \caption{Declarative pattern matching and spine typing}
\ifnum\OPTIONAppendix=0
  \label{fig:liq-declmatch}
\else
  \label{fig:declmatch}
\fi
\end{figure}

The judgment $\judgesynhead{\Theta}{\Gamma}{h}{P}$
(\Figref{fig:liq-decltyping-head-bound-expression})
synthesizes the type $P$ from the head $h$.
This judgment is synthesizing,
because it is used in what are,
from a Curry--Howard perspective,
kinds of cut rules:
\DeclSynSpineApp and \DeclChkExpMatch, discussed later.
The synthesized type is the cut type, which does not appear in the conclusion
of \DeclSynSpineApp or \DeclChkExpMatch.
For head variables, we look up the variable's type in the context $\Gamma$ (\DeclSynHeadVar).
For annotated values, we synthesize the annotation (\DeclSynValAnnot).

The judgment $\judgesynexp{\Theta}{\Gamma}{\be}{\upshift{P}}$
(\Figref{fig:liq-decltyping-head-bound-expression})
synthesizes the type $\upshift{P}$ from the bound expression $\be$.
Similarly to the synthesizing judgment for heads,
this judgment is synthesizing because it is used in a cut rule \DeclChkExpLet
(the synthesized type is again the cut type).
Bound expressions only synthesize an upshift
because of their (lone) role in rule \DeclChkExpLet, discussed later.
For an application of a head to a spine (\DeclSynSpineApp, an auxiliary cut rule),
we first synthesize the head's type (which must be a downshift),
and then check the spine against the thunked computation type,
synthesizing the latter's return type.
(Function applications must always be fully applied,
but we can simulate partial application via $\eta$-expansion.
For example, given $x:P_1$ and $h => \downshift{(P_1 \to P_2 \to \upshift{Q})}$,
to partially apply $h$ to $x$ we can write $\fun{y}{\Let{z}{h(x,y)}{\cdots}}$.)
For annotated expressions, we synthesize the annotation (\DeclSynExpAnnot),
which must be an upshift.
If the programmer wants, say, to verify guard constraints in $N$
of an expression $e$ of type $N$ whenever it is run,
then they must annotate it:
$\annoexp{\Return{\thunk{e}}}{\upshift{\downshift{N}}}$.
If an $e$ of type $N$ is intended to be a function to be applied
(as a head to a spine; \DeclSynSpineApp)
only if the guards of $N$ can be verified and the universally quantified indexes of $N$
can be instantiated,
then the programmer must thunk and annotate it:
$\annoexp{\thunk{e}}{\downshift{N}}$.
The two annotation rules have explicit type well-formedness premises
to emphasize that type annotations are provided by the programmer.

The judgment $\judgechkval{\Theta}{\Gamma}{v}{P}$
(\Figref{fig:liq-decltyping-value-expression})
checks the value $v$ against the type $P$.
From a Curry--Howard perspective,
this judgment corresponds to a \emph{right-focusing} stage.
According to rule \DeclChkValExists,
a value checks against an existential type
if there is an index instantiation it checks against
(declaratively, an index is conjured,
but algorithmically we will have to solve for one).
For example, as discussed in \Secref{sec:well-formedness},
checking the program value $\mathsf{one}$ representing $1$
against type $\extype{a:\kindnat} \mathsf{Nat}(a)$ solves $a$
to an index semantically equal to $1$.
According to rule \DeclChkValWith,
a value checks against an asserting type
if it the asserted proposition $\phi$ holds
(and the value checks against the type to which $\phi$ is connected).
Instead of a general value type subsumption rule like
\[
  \Infer{}
  {\judgechkval{\Theta}{\Gamma}{v}{Q} \\ \judgesub[+]{\Theta}{Q}{P}}
  {\judgechkval{\Theta}{\Gamma}{v}{P}}
\]
we restrict subsumption to (value) variables,
and prove that subsumption is admissible
(see Section~\ref{sec:subst}).
This is easier to implement efficiently because the type checker
would otherwise have to guess $Q$ (and possibly need to backtrack),
whereas \DeclChkValVar need only look up the variable.
Further, the $P \neq \exists, \land$ constraint on \DeclChkValVar
means that any top-level $\exists$ or $\land$ constraints
must be verified before subtyping,
eliminating nondeterminism of verifying these in subtyping or typing.
Rule \DeclChkValFix checks the unrolled value against the unrolled inductive type.
Rule \DeclChkValUnit says $\unit$ checks against $1$.
Rule \DeclChkValPair says a pair checks against a product
if each pair component checks against its corresponding factor.
Rule \DeclChkValIn{k} says a value injected into the $k$th position checks against a sum
if it can be checked against the $k$th summand.
Rule \DeclChkValDownshift checks the thunked expression against the computation type $N$
under the given thunk type $\downarrow{N}$.

The judgment $\judgechkexp{\Theta}{\Gamma}{e}{N}$
(\Figref{fig:liq-decltyping-value-expression})
checks the expression $e$ against the type $N$.
From a Curry--Howard perspective,
this judgment is a \emph{right-inversion} stage
with \emph{stable} moments
(\DeclChkExpLet and \DeclChkExpMatch,
which enter left- or right-focusing stages, respectively).
Instead of \DeclChkExpExtract,
one might expect two rules (one for $\forall$ and one for $\implies$)
that simply put the universal index variable or proposition into logical context,
but these alone are less compatible with subsumption admissibility
(see \Secref{sec:subst})
due to the use of extraction in subtyping rule \DeclSubNegR.
However, the idea is still the same:
here we are using indexes, not verifying them as in the dual left-focusing stage.
To reduce \DeclChkExpExtract nondeterminism,
and to enable a formal correspondence
between \sysname and (a variant of) CBPV
(which has a general $\downarrow$ elimination rule),
the other (expression) rules
must check against a \emph{simple} type.
In practice, we eagerly apply (if possible) \DeclChkExpExtract
immediately when type checking an expression;
extracted types are invariant under extraction.

All applications $h(s)$ must be named and sequenced via \DeclChkExpLet,
which we may think of as monadic binding,
and is a key cut rule.
Other computations---annotated returner expressions $\annoexp{e}{\upshift{P}}$---must
also be named and sequenced via \DeclChkExpLet.
It would not make sense to allow arbitrary negative annotations
because that would require verifying constraints and instantiating indexes
that should only be done when the annotated expression is applied,
which does not occur in \DeclChkExpLet itself.

Heads, that is, head variables and annotated values, can be pattern matched
via \DeclChkExpMatch.
From a Curry--Howard perspective,
the rule \DeclChkExpMatch is a cut rule dual to the cut rule \DeclChkExpLet:
the latter binds the result of a computation to a (sequenced) computation,
whereas the former binds the deconstruction of a value to,
and directs control flow of,
a computation.
Rule \DeclChkExpLam is standard (besides the check that $P \to N$ is simple).
Rule \DeclChkExpRec requires an annotation that universally quantifies
over the argument $a$ that must be smaller at each recursive call,
as dictated by its annotation in the last premise: %
$x : \downshift{\big(\alltype{a':\kindnat} (a' < a) \implies [a'/a]M\big)}$
only allows $x$ to be used for $a' < a$,
ensuring that \emph{refined} recursive functions are well-founded (according to $<$ on naturals).
Rule \DeclChkExpUpshift checks that the value being returned
has the positive type under the given returner type ($\uparrow$);
this may be thought of as a monadic return operation.
Rule \DeclChkExpUnreachable says that \unreachable checks against any type,
provided the logical context is inconsistent;
for example, an impossible pattern in pattern matching extracts to an inconsistent context.

Rule \DeclChkExpRec only handles one termination metric, namely $<$ on natural numbers.
This is only to simplify our presentation,
and is not a fundamental limitation of the system.
We can, for example, add a rule that encodes a termination metric $<$ on the sum of two natural numbers:
\[
  \Infer{}
  {
    \simple{\Theta}{N}
    \\
    \judgesub[-]{\Theta}{\alltype{a:\kindnat} \alltype{b:\kindnat} M}{N}
    \\
    \judgechkexp{\Theta, a:\kindnat, b:\kindnat}{\Gamma, x:\downshift{\big(\alltype{a':\kindnat} \alltype{b':\kindnat} (a'+b' < a+b) \implies [a'/a][b'/b]M}\big)}{e}{M}
  }
  {
    \judgechkexp{\Theta}{\Gamma}{\rec{x : (\alltype{a:\kindnat} \alltype{b:\kindnat} M)}{e}}{N}
  }
\]
It is somewhat straightforward to update the metatheory for the system with this rule added.
This rule obviates, for example, the ghost parameter used in the mergesort example of \Secref{sec:example}.
Similarly, one could add rules for other termination metrics, such as lexicographic induction.

The judgment $\judgechkmatch{\Theta}{\Gamma}{P}{\clauses{\pa}{e}{i}{I}}{N}$
(\Figref{fig:liq-declmatch})
decomposes $P$,
according to patterns $\pa_i$
(if $P \neq \land$ or $\exists$, which have no computational content;
if $P = \land$ or $\exists$, the index is put in logical context for use),
and checks that each branch $e_i$ has type $N$.
The rules are straightforward.
Indexes from matching on existential and asserting types
are used, not verified (as in value typechecking);
we deconstruct heads,
and to synthesize a type for a head, its indexes must hold,
so within the pattern matching stage itself, we may assume and use them.
From a Curry--Howard perspective,
this judgment corresponds to a \emph{left-inversion} stage.
However, it is not \emph{strongly} focused, that is,
it does not decompose $P$ eagerly and as far as possible;
therefore, ``stage'' might be slightly misleading.
If our system were more strongly focused, we would have nested patterns,
at least for all positive types except inductive types;
it's unclear how strong focusing on inductive types would work.

The judgment $\judgespine{\Theta}{\Gamma}{s}{N}{\upshift{P}}$
(\Figref{fig:liq-declmatch})
checks the spine $s$ against $N$, synthesizing the return type $\upshift{P}$.
From a Curry--Howard perspective,
this judgment corresponds to a \emph{left-focusing} stage.
The rules are straightforward:
decompose the given $N$,
checking index constraints (\DeclSpineAll and \DeclSpineImplies)
and values (\DeclSpineApp) until an upshift, the return type, is synthesized
(\DeclSpineNil).
Similarly to dual rule \DeclChkValExists,
the declarative rule \DeclSpineAll conjures an index measuring a value,
but in this case an argument value in a spine.
For example, in applying a head of type
$\alltype{a:\kindnat} \mathsf{Nat}(a) \to \upshift{\mathsf{Nat}(a)}$
to the spine with program value $\mathsf{one}$ representing $1$,
we must instantiate $a$ to an index semantically equal to $1$;
we show how this works algorithmically in \Secref{sec:alg-typing}.
All universal quantifiers (in the input type of a spine judgment) are solvable
algorithmically, because in a well-formed return type,
the set of value-determined indexes $\Xi$
is empty.

\subsection{Substitution}
\label{sec:subst}
A key correctness result that we prove is a substitution lemma:
substitution (of index terms for index variables and program values for program variables) preserves typing.
We now extend the index-level syntactic substitutions
(and the sequential substitution operation) introduced
in \Secref{sec:index-sorting-and-verification}.
A \emph{syntactic substitution}
$\sigma \bnfas \cdot \bnfalt \sigma, t/a \bnfalt \sigma, \subs{v}{P}{x}$
is essentially a list of terms to be substituted for variables.
Substitution application $[\sigma]-$ is a sequential substitution
metaoperation on types and terms.
On program terms, it is a kind of \emph{hereditary substitution}\footnote{Typically, hereditary substitution reduces terms after substitution, modifying the syntax tree.} \cite{Watkins04, Pfenning08}
in the sense that,
at head variables
(note the $\mathsf{h}$ superscript in the \Figref{fig:syn-subs-defn} definition;
we elide $\mathsf{h}$ when clear from context),
an annotation is produced if the value and the head variable being replaced by it are not equal---thereby modifying
the syntax tree of the substitutee.
Otherwise, substitution is standard (homomorphic application)
and does not use the value's associated type given in $\sigma$:
see \Figref{fig:syn-subs-defn}.
\begin{figure}[hptb]
  \begin{align*}
    [\subs{v}{P}{x}]^{\mathsf{h}} y &= y \quad(\text{if }y \neq x)\\
    [\subs{v}{P}{x}]^{\mathsf{h}} x &=
                                      \begin{cases}
                                        x &\quad\text{if $v = x$} \\
                                        \annoexp{v}{P} &\quad\text{else}
                                      \end{cases}\\
    [\subs{v}{P}{x}]^{\mathsf{h}} \annoexp{v_0}{P_0} &= \annoexp{[\subs{v}{P}{x}] v_0}{P_0} \\[1em]
    [\subs{v}{P}{x}](h(s)) &= ([\subs{v}{P}{x}]^{\mathsf{h}} h)([\subs{v}{P}{x}]s) \\
    [\subs{v}{P}{x}] \annoexp{e}{\upshift{Q}} &= \annoexp{[\subs{v}{P}{x}] e}{\upshift{Q}} \\
    [\subs{v}{P}{x}]y &= y \quad(\text{if }y \neq x)\\
    [\subs{v}{P}{x}]x &= v \\
    [\subs{v}{P}{x}]\pair{v_1}{v_2} &= \pair{[\subs{v}{P}{x}]v_1}{[\subs{v}{P}{x}]v_2} \\
    \hspace{-0.5em}&\hspace{0.5em}\vdots \\
    [\subs{v}{P}{x}](\match{h}{\clauses{\pa}{e}{i}{I}}) &= \match{\left( [\subs{v}{P}{x}]^{\mathsf{h}}h \right)}{\left( [\subs{v}{P}{x}]\clauses{\pa}{e}{i}{I} \right)} \\
    \hspace{-0.5em}&\hspace{0.5em}\vdots
  \end{align*}
  
  \caption{Definition of syntactic substitution on program terms}
  \label{fig:syn-subs-defn}
\end{figure}

In the definition given in \Figref{fig:syn-subs-defn},
an annotation is not produced if $v = x$
so that $\subs{x}{P}{x}$ is always an \emph{identity} substitution:
that is, $[\subs{x}{P}{x}]^{\mathsf{h}}x = x$.
As usual, we assume variables are $\alpha$-renamed to avoid capture by substitution.

The judgment $\Theta_0; \Gamma_0 |- \sigma : \Theta; \Gamma$
(appendix \Figref{fig:syn-subs})
means that, under $\Theta_0$ and $\Gamma_0$,
we know $\sigma$ is a substitution of index terms and program values
for variables in $\Theta$ and $\Gamma$, respectively.
The key rule of this judgment is for program value entries
(the three elided rules are similar
to the three rules for syntactic substitution typing at index level,
found near the start of \Secref{sec:index-sorting-and-verification},
but adds program contexts $\Gamma$ where appropriate):
\[
  \Infer{}
  {
    \Theta_0; \Gamma_0 |- \sigma : \Theta; \Gamma
    \and
    \judgechkval{\Theta_0}{\Gamma_0}{[\sigma]v}{[\filterprog{\sigma}]P}
    \and
    x \notin \dom{\Gamma}
  }
  {
    \Theta_0; \Gamma_0 |- (\sigma, \subs{v}{P}{x}) : \Theta; \Gamma, x:P
  }
\]
We apply the rest of the syntactic substitution---that is,
the $\sigma$ in the rule---to $v$ and $P$
because the substitution operation is sequential;
$v$ may mention variables in $\Gamma$ and $\Theta$,
and $P$ may mention variables in $\Theta$.
The metaoperation $\filterprog{-}$ filters out program variable entries
(program variables cannot appear in types, functors, algebras or indexes).

That substitution respects typing
is an important correctness property of the type system.
We state only two parts here,
but those of the remaining program typing judgments are similar;
all six parts are mutually recursive.
\begin{lemma}[Syntactic Substitution]
  \label{lem:liq-syn-subs}
  \hfill (\Lemref{lem:syn-subs} in appendix)\\
  Assume $\Theta_0; \Gamma_0 |- \sigma : \Theta; \Gamma$.
  \begin{enumerate}
  \item If $\judgesynhead{\Theta}{\Gamma}{h}{P}$,
      then there exists $Q$
      such that $\judgesub[+]{\Theta_0}{Q}{[\filterprog{\sigma}]P}$ 
      and $\judgesynhead{\Theta_0}{\Gamma_0}{[\sigma]h}{Q}$.
  \item If $\judgechkexp{\Theta}{\Gamma}{e}{N}$,
    then $\judgechkexp{\Theta_0}{\Gamma_0}{[\sigma]e}{[\filterprog{\sigma}]N}$.
  \end{enumerate}
\end{lemma}

In part (1),
if substitution creates a head variable with stronger type,
then the stronger type $Q$ is synthesized.
The proof relies on other structural properties
such as weakening.
It also relies on subsumption admissibility,
which captures what we mean by ``stronger type''.
We show only one part;
the mutually recursive parts for the other five program typing judgments are similar.

\begin{lemma}[Subsumption Admissibility]
  \hfill (\Lemref{lem:subsumption-admissibility} in appendix)\\
  Assume $\judgesub[]{\Theta}{\Gamma'}{\Gamma}$ (pointwise subtyping).
  \begin{enumerate}
  \item If $\judgechkval{\Theta}{\Gamma}{v}{P}$
    and $\judgesub[+]{\Theta}{P}{Q}$,
    then $\judgechkval{\Theta}{\Gamma'}{v}{Q}$.
  \end{enumerate}
\end{lemma}

Subtypes are stronger than supertypes.
That is, if we can check a value
against a type, then we know that it also checks against any of the type's supertypes;
similarly for expressions.
Pattern matching is similar, but it also says we can match on a stronger type.
A head or bound expression
can synthesize a stronger type under a stronger context.
Similarly, with a stronger input type, a spine can synthesize a stronger return type.

\section{Type Soundness}
\label{sec:semantics}

We prove type (and substitution) soundness of the declarative system
with respect to an elementary domain-theoretic denotational semantics.
Refined type soundness implies the refined system's totality and logical consistency.

Refinement type systems refine already-given type systems,
and the soundness of the former depends on that of the latter \cite{Mellies15-functors}.
Thus, the semantics of our refined system
is defined in terms of that of its underlying, unrefined system,
which we discuss in \Sectionref{sec:semantics-unrefined-system}.

\emph{Notation}:
We define the disjoint union $X \uplus Y$ of sets $X$ and $Y$ by
$X \uplus Y = (\{1\} \times X) \cup (\{2\} \times Y)$
and define $\minj{k} : X_k \to X_1 \uplus X_2$ by $\minj{k}(d) = (k, d)$.
Semantic values are usually named $d$, $f$, $g$, or $V$.

\subsection{Unrefined System}
\label{sec:semantics-unrefined-system}

For space reasons,
we do not fully present the unrefined system and its semantics here
(see appendix \Secref{sec:apx-unrefined-system}).
The unrefined system is basically just the refined system
with everything pertaining to indexes erased.
The program terms of the unrefined system have almost the same syntax as those of the refined system,
but an unrefined, recursive expression has no type annotation,
and we replace the expression \unreachable by \diverge,
which stands for an inexhaustive pattern-matching error.
The unrefined system satisfies a substitution lemma
(appendix \Lemref{lem:unref-syn-subs})
similar to that of the refined system,
but its proof is simpler and does not rely on subsumption admissibility,
because the unrefined system has no subtyping.

In CBPV, nontermination is regarded as an effect,
so value and computation types denote different kinds of mathematical things:
predomains and domains, respectively \cite{Levy04},
which are both sets with some structure.
Because we have recursive expressions,
we must model nontermination, an effect.
We use elementary domain theory.
For our (unrefined) system,
we interpret (unrefined) positive types as predomains
and (unrefined) negative types as domains.
The only effect we consider in this paper is nontermination
(though we simulate inexhaustive pattern-matching errors with it);
we take (chain-)complete partial orders (cpo) as predomains,
and pointed (chain-)complete partial orders (cppo) as domains.

\paragraph{Positive types and functors}

The grammar for unrefined positive types
is similar to that for refined positive types,
but lacks asserting and existential types,
and unrefined inductive types $\mu F$ are not refined by predicates.
Unrefined inductive types
use the unrefined functor grammar,
which is the same as the refined functor grammar but uses unrefined types
in constant functors.

\begin{grammar}
    $\unref{P},\unref{Q}$ & \bnfas & $\unref{1} \bnfalt \unref{P} \times \unref{Q} \bnfalt \unref{0} \bnfalt \unref{P} + \unref{Q} \bnfalt \downshift{\unref{N}} \bnfalt \mu \unref{F}$
\end{grammar}

The denotations of unrefined positive types are standard.
We briefly describe their partial orders,
then describe the meaning of functors,
and lastly return to the meaning of inductive types (which involve functors).

We give (the denotation of) $1$ (denoting the distinguished terminal object $\{\bullet\}$) the discrete order
$\{(\bullet, \bullet)\}$.
For
$P \times Q$ (denoting product) we use component-wise order
($\ord[D_1 \times D_2]{(d_1, d_2)}{(d_1', d_2')}$ if $\ord[D_1]{d_1}{d_1'}$ and $\ord[D_2]{d_2}{d_2'}$),
for $0$ (denoting the initial object) we use the empty order,
and for $P + Q$ (denoting coproduct, that is, disjoint union $\uplus$)
we use injection-wise order
($\ord[D_1 \uplus D_2]{\minj{j} d}{\minj{k} d'}$ if $j = k$ and $\ord[D_j]{d}{d'}$).
We give $\downshift{N}$ the order of $N$,
that is,
$\downarrow$ denotes the forgetful functor from the category $\Cppo$ of cppos
and continuous functions to the category $\Cpo$ of cpos and continuous functions.
Finally, $\ord[\sem{}{\mu F}]{V_1}{V_2}$
if $\ord[\sem{}{F}^{k+1} \emptyset]{V_1}{V_2}$ for some $k \in \kindnat$,
inheriting the %
type denotation orders as the functor is applied.

The denotations of unrefined functors are standard
$\Cpo$ endofunctors.
We briefly describe them here,
but full definitions are in appendix \Secref{sec:apx-unrefined-system}.
The sum functor $\oplus$ denotes a functor
that sends a cpo to the disjoint union $\uplus$ of its component applications
(with usual injection-wise order),
and its functorial action is injection-wise.
The product functor $\otimes$ denotes a functor
that sends a cpo to the product $\times$ of its component applications
(with usual component-wise order),
and its functorial action is component-wise.
The unit functor $I$ denotes a functor sending any cpo to $1 = \one$ (discrete order),
and its functorial action sends all morphisms to $\id_\one$.
The constant (type) functor $\Const{P}$ denotes a functor
sending any cpo to the cpo $\sem{}{P}$,
and its functorial action sends all morphisms to the identity $\id_{\sem{}{P}}$
on $\sem{}{P}$.
The identity functor $\Id$
denotes the identity endofunctor on $\Cpo$.
(Forgetting the order structure, functors also denote endofunctors
on the category $\Set$ of sets and functions.)

We now explain the denotational semantics of our inductive types.
Semantically,
we build an inductive type (such as $\sem{}{\mathsf{List} \; A}$),
by repeatedly applying (the denotation of) its functor specification
(such as $\sem{}{\textsf{ListF}_A}$) 
to the initial object $\sem{}{0} = \emptyset$.
For example,
\[
  \sem{}{\textsf{List} \; A} = \bigcup_{k \in \kindnat} \sem{}{\Const{1} \oplus (A \otimes \Id)}^k \emptyset = 1 \uplus \bigg(\sem{}{A} \times \Big(1 \uplus \big(\sem{}{A} \times \cdots\big)\Big)\bigg)
\]
where $1 = \one$
(using the relatively direct functors with more complicated unrolling,
discussed in \Secref{sec:functor-algebra-grammar}).
We denote the nil list $[]$ by $\minj{1} \bullet$,
a list $x :: []$ with one term $x$
by $\minj{2} (\sem{}{x}, \minj{1} \bullet)$,
and so on.
In general, given a (polynomial) $\Set$ (category of sets and functions) endofunctor $F$
(which, for this paper,
will always be the denotation of a well-formed (syntactic) functor,
refined or otherwise),
we define $\mu F = \cup_{k\in\kindnat} F^k \emptyset$.
We then define $\sem{}{\mu F} = \mu \sem{}{F}$.
In \sysname, for every well-formed (unrefined) functor $F$,
the set $\mu \sem{}{F}$
is a \emph{fixed point} of $\sem{}{F}$
(appendix \Lemref{lem:unref-mu-unroll-equal}):
that is, $\sem{}{F} (\mu\sem{}{F}) = \mu \sem{}{F}$
(and similarly for refined functors: appendix \Lemref{lem:mu-unroll-equal}).

\paragraph{Negative types}
The grammar for unrefined negative types has unrefined function types $P \to N$
and unrefined upshifts $\upshift{P}$, with no guarded or universal types.
Unrefined negative types denote cppos.

\begin{grammar}
    $\unref{N}$ & \bnfas & $\unref{P} \to \unref{N} \bnfalt \upshift{\unref{P}}$
\end{grammar}

Function types $P \to N$ denote continuous functions from $\sem{}{P}$ to $\sem{}{N}$
(which we sometimes write as $\sem{}{P} \Rightarrow \sem{}{N}$),
where its order is defined pointwise,
together with the bottom element (the ``point'' of ``\emph{pointed} cpo'')
$\bott[\sem{}{P \to N}]$
that maps every $V \in \sem{}{P}$
to the bottom element $\bott[\sem{}{N}]$ of $\sem{}{N}$
(that is, $\uparrow$ denotes the lift functor from $\Cpo$ to $\Cppo$).
For our purposes, this is equivalent
to lifting $\sem{}{P} \in \Cpo$ to $\Cppo$
and denoting arrow types by \emph{strict} ($\bot$ goes to $\bot$)
continuous functions
so that function types denote $\Cppo$ \emph{exponentials}.

Upshifts $\upshift{P}$ denote $\sem{}{P} \uplus \{\bott[\uparrow]\}$
with the lift order
\[
  \ordsym[\sem{}{\upshift{P}}] = \comprehend{(\minj{1}d, \minj{1}d')}{\ord[\sem{}{P}]{d}{d'}} \cup \comprehend{(\minj{2} \bott[\uparrow], d)}{d \in \sem{}{\upshift{P}}}
\]
and bottom element $\bott[\sem{}{\upshift{P}}] = \minj{2} \bott[\uparrow]$.
We could put, say, $\bullet$ rather than $\bott[\uparrow]$,
but we think the latter is clearer in associating it with the
\emph{bottom} element of upshifts;
or $\bot$ rather than $\bott[\uparrow]$
but we often elide the ``$\sem{}{A}$'' subscript
in $\bott[\sem{}{A}]$ when clear from context.

Appendix \Figref{fig:denotation-unrefined-types-functors}
has the full definition of (unrefined) type and functor denotations.

\paragraph{Well-typed program terms}
We write
$\Gamma |- \mathcal{O} \cdots A$
and
$\Gamma; [B] |- \mathcal{O} \cdots A$
to stand for all six unrefined program typing judgments:
$\unrefsynhead{\Gamma}{h}{P}$ and
$\unrefsynexp{\Gamma}{\be}{\upshift{P}}$ and
$\unrefchkval{\Gamma}{v}{P}$ and
$\unrefchkexp{\Gamma}{e}{N}$ and
$\unrefchkmatch{\Gamma}{P}{\clauses{\pa}{e}{i}{I}}{N}$ and
$\unrefspine{\Gamma}{s}{N}{\upshift{P}}$.

The denotational semantics of well-typed, unrefined program terms of judgmental form
$\Gamma |- \mathcal{O} \cdots A$
or
$\Gamma; [B] |- \mathcal{O} \cdots A$
are continuous functions
$\sem{}{\Gamma} \to \sem{}{A}$
and
$\sem{}{\Gamma} \to \sem{}{B} \to \sem{}{A}$
respectively,
where $\sem{}{\Gamma}$ is the set of all semantic substitutions
$|- \delta : \Gamma$ together with component-wise order.
Similarly to function type denotations,
the bottom element of a $\sem{}{\Gamma} \to \sem{}{N}$
sends every $\delta \in \sem{}{\Gamma}$ to $\bott[\sem{}{N}]$
(equivalently for our purposes,
we can lift source predomains and consider strict continuous functions).
We only interpret typing derivations,
but we often only mention the program term in semantic brackets $\sem{}{-}$.
For example, if $\Gamma |- x => P$,
then $\sem{}{x} = (\delta \in \sem{}{\Gamma}) \mapsto \delta(x)$.
We write the application of the denotation
$\sem{}{E}$ 
of a program term $E$ (typed under $\Gamma$)
to a semantic substitution $\delta \in \sem{}{\Gamma}$ as
$\sem{\delta}{E}$.
We only mention a few of the more interesting cases of the definition
of typing denotations;
for the full definition, see appendix
Figures~\ref{fig:unref-denotation-head-bound},~\ref{fig:unref-denotation-val-exp},
and~\ref{fig:unref-denotation-match-spine}%
.
If $\unrefspine{\Gamma}{v, s}{N}{M}$,
then
\[
  \sem{}{v, s} = (\delta \in \sem{}{\Gamma}) \mapsto (f \mapsto \sem{\delta}{s} (f (\sem{\delta}{v})))
\]
Returner expressions denote monadic returns:
\[
  \sem{\delta}{\Return{v}} = \minj{1} \sem{\delta}{v}
\]
Let-binding denotes monadic binding:
\[
  \sem{\delta}{\Let{x}{\be}{e}}
  = \begin{cases}
    \sem{(\delta, V/x)}{e} &\quad\text{if }\sem{\delta}{\be} = \minj{1} V \\
    \bott[\sem{}{N}] &\quad\text{if }\sem{\delta}{\be} = \minj{2} \bott[\uparrow]
  \end{cases}
\]
A recursive expression denotes a fixed point obtained
by taking the least upper bound ($\sqcup$) of all its successive approximations:
\[
  \sem{\delta}{\unrefchkexp{\Gamma}{\rec{x}{e}}{N}}
  = \bigsqcup_{k \in \kindnat} \left(V \mapsto \sem{\delta, V/x}{\unrefchkexp{\Gamma, x:\downshift{N}}{e}{N}}\right)^k \bott[\sem{}{N}]
\]
In the unrefined system, we include \diverge, to which \unreachable erases
(that is, $\erase{\unreachable} = \diverge$).
We intend \diverge to stand for an undefined body of a pattern-matching clause,
but we interpret this error as divergence to simplify the semantics:
\[
  \sem{\delta}{\unrefchkexp{\Gamma}{\diverge}{N}} = \bott[\sem{}{N}]
\]
The point is that the refined system prevents the error.

We will say more about the semantics of folds in \Secref{sec:semantics-refined-system},
but note that the action of rolling and unrolling syntactic values
is essentially denoted by $d \mapsto d$:
\begin{align*}
  \sem{\delta}{\roll{v}} &= \sem{\delta}{v} \\
  \sem{\delta}{\setof{\clause{\roll{x}}{e}}} &= V\mapsto\sem{\delta,V/x}{e}
\end{align*}
This works due to the fact that unrolling is sound
(roughly, the denotations of each side of ``$\circeq$''
in the unrolling judgment are \emph{equal})
and the fact that $\sem{}{F} (\mu\sem{}{F}) = \mu\sem{}{F}$
(and similarly for the refined system).

\paragraph{Unrefined soundness}
Our proofs of (appendix) \Lemmaref{lem:unrefined-type-soundness}
and (appendix) \Lemmaref{lem:unref-subs-soundness}
use standard techniques in domain theory \cite{GunterTextbook}.

Unrefined type soundness says that a term typed $A$ under $\Gamma$
denotes a continuous function $\sem{}{\Gamma} \to \sem{}{A}$.
We (partly) state (3 out of 6 parts) this in two mutually recursive lemmas
as follows:

\begin{lemma}[Continuous Maps]
  \hfill (\Lemref{lem:continuous-maps} in appendix)\\
  Suppose $|- \delta_1 : \Gamma_1$ and $|- \delta_2 : \Gamma_2$
  and $\judgectx{}{\Gamma_1, y:Q, \Gamma_2}$.
  \begin{enumerate}
  \item If $\unrefsynhead{\Gamma_1, y:Q, \Gamma_2}{h}{P}$,
    then the function
    $\sem{}{Q} \to \sem{}{P}$
    defined by $d \mapsto \sem{\delta_1, d/y, \delta_2}{h}$
    is continuous.
  \item If $\unrefchkexp{\Gamma_1, y:Q, \Gamma_2}{e}{N}$,
    then the function
    $\sem{}{Q} \to \sem{}{N}$
    defined by $d \mapsto \sem{\delta_1, d/y, \delta_2}{e}$
    is continuous.
  \item If $\unrefspine{\Gamma_1, y:Q, \Gamma_2}{s}{N}{\upshift{P}}$,
    then the function
    $\sem{}{Q} \to \sem{}{N} \to \sem{}{\upshift{P}}$
    defined by $d \mapsto \sem{\delta_1, d/y, \delta_2}{s}$
    is continuous.
  \end{enumerate}
\end{lemma}

\begin{lemma}[Unrefined Type Soundness]
  \hfill (\Lemref{lem:unrefined-type-soundness} in appendix)\\
  Assume $|- \delta : \Gamma$.
  \begin{enumerate}
  \item If $\unrefsynhead{\Gamma}{h}{P}$,
    then $\sem{\delta}{\unrefsynhead{\Gamma}{h}{P}} \in \sem{}{P}$.
  \item If $\unrefchkexp{\Gamma}{e}{N}$,
    then $\sem{\delta}{\unrefchkexp{\Gamma}{e}{N}} \in \sem{}{N}$.
  \item If $\unrefspine{\Gamma}{s}{N}{\upshift{P}}$,
    then $\sem{\delta}{\unrefspine{\Gamma}{s}{N}{\upshift{P}}} \in \sem{}{N} \Rightarrow \sem{}{\upshift{P}}$.
  \end{enumerate}
\end{lemma}
The proof of unrefined type soundness is standard,
and uses the well-known fact
that a continuous function in $\Cppo$ has a least fixed point.
Among other things, we also use the fact that
$\mu \sem{}{F}$ is a fixed point of $\sem{}{F}$
(appendix \Lemref{lem:unref-mu-unroll-equal}).
We also use the soundness of unrefined unrolling,
which we didn't mention here
because it's similar to refined unrolling and its soundness,
discussed in the next section.

We interpret an unrefined syntactic substitution (typing derivation)
$\Gamma_0 |- \sigma : \Gamma$
as a continuous function $\sem{}{\Gamma_0} \to \sem{}{\Gamma}$
that takes a $\delta \in \sem{}{\Gamma_0}$
and uses $\delta$ to interpret each of the entries in $\sigma$
(remembering to apply the rest of the syntactic substitution,
because substitution is defined sequentially):
\begin{align*}
  \sem{}{\Gamma_0 |- \cdot : \cdot} &= (\delta \in \sem{}{\Gamma_0}) \mapsto \cdot
  \\
  \sem{}{\Gamma_0 |- \big( \sigma, (\subs{v}{P}{x})\big) : \big(\Gamma, x:P\big)}
     &=
     (\delta \in \sem{}{\Gamma_0}) \mapsto \left((\sem{\delta}{\sigma}), \sem{\delta}{[\sigma]v}/x\right)
\end{align*}
Similarly to typing derivations,
we only consider denotations of typing derivations $\Gamma_0 |- \sigma : \Gamma$
of substitutions,
but often simply write $\sem{}{\sigma}$.

Unrefined substitution soundness says that semantic and syntactic substitution commute:
if $E$ is a program term typed under $\Gamma$
and $\Gamma_0 |- \sigma : \Gamma$ is a substitution,
then $\sem{}{[\sigma]E} = \sem{}{E} \circ \sem{}{\sigma}$.
Here, we partly show how it is stated in the appendix
(1 out of 6 parts):
\begin{lemma}[Unrefined Substitution Soundness]
  \hfill (\Lemref{lem:unref-subs-soundness} in appendix)\\
  Assume $\Gamma_0 |- \sigma : \Gamma$ and $|- \delta : \Gamma_0$.
  \begin{enumerate}
  \item If $\unrefchkexp{\Gamma}{e}{N}$,
    then $\sem{\delta}{\unrefchkexp{\Gamma_0}{[\sigma]e}{N}} = \sem{\sem{\delta}{\sigma}}{\unrefchkexp{\Gamma}{e}{N}}$.
  \end{enumerate}
\end{lemma}

We use unrefined type/substitution soundness
to prove \emph{refined} type/substitution soundness,
discussed next.

\subsection{Refined System}
\label{sec:semantics-refined-system}

\paragraph{Indexes}
For any sort $\tau$, we give its denotation $\sem{}{\tau}$
the discrete order $\sqsubseteq_{\sem{}{\tau}} \; = \comprehend{(d,d)}{d\in\sem{}{\tau}}$,
making it a cpo.

\paragraph{Semantic Substitution}

We introduced semantic substitutions $\delta$ (at the index level)
when discussing propositional validity (\Secref{sec:index-sorting-and-verification}).
Here, they are extended to semantic program values:
\[
  \Infer{}
  {
    |- \delta : \Theta; \Gamma
    \and
    V \in \sem{\filterprog{\delta}}{P}
    \and
    x \notin \dom{\Gamma}
  }
  {
    |- (\delta, V/x) : \Theta; \Gamma, x:P
  }
\]
where \deffilterprog filters out program entries.
\emph{Notation}: we define
$\sem{}{\Theta; \Gamma} = \comprehend{\delta}{|- \delta : \Theta; \Gamma}$.

\paragraph{Erasure}

The \emph{erasure} metaoperation \deferase (appendix \Secref{sec:apx-erasure})
erases all indexes from (refined) types, program terms
(which can have type annotations, but those do not affect program meaning),
and syntactic and semantic substitutions.
For example,
$\erase{\comprehend{\nu : \mu F}{\Fold{F}{\alpha}\,{\nu} =_\tau t}} = \mu\erase{F}$
and $\erase{\alltype{a:\tau} N} = \erase{N}$
and $\erase{P \times Q} = \erase{P} \times \erase{Q}$
and so on.

We use many facts about erasure
to prove refined type/substitution soundness
(appendix lemmas):
\begin{itemize}
\item Refined types denote subsets of what their erasures denote:
  \Lemmaref{lem:type-subset-erasure}.
  Similarly for refined functors and refined inductive types:
  \Lemmaref{lem:functor-subset-erasure} and \Lemmaref{lem:mu-subset-erasure}.
\item The erasure of both types appearing in
  extraction, equivalence, and subtyping judgments results in equal (unrefined) types:
  \Lemmaref{lem:extract-erases-to-equality},
  \Lemmaref{lem:equivalence-erases-to-equality}, and
  \Lemmaref{lem:subtyping-erases-to-equality}.
\item Refined unrolling and typing are sound with respect to their erasure:
  \Lemmaref{lem:unroll-erasure},
  \Lemmaref{lem:erasure-respects-typing}, and
  \Lemmaref{lem:erasure-respects-subs-typing}.
\item Erasure commutes with syntactic and semantic substitution:
  \Lemmaref{lem:erasure-respects-subs} and
  \Lemmaref{lem:erasure-respects-sem-subs}.
\end{itemize}

\paragraph{Types, functors, algebras, and folds}
The denotations of refined types and functors are defined as logical subsets
of the denotations of their erasures
(together with their erasure denotations themselves).
They are defined mutually with the denotations of well-formed algebras.

In appendix \Figref{fig:denotation-types1},
we inductively define the denotations of well-formed types 
$\judgetp{\Theta}{A}{\dontcare}$.
We briefly discuss a few of the cases.
The meaning of an asserting type is the set of refined values
such that the asserted index proposition holds
(read $\one$ as true and $\emptyset$ as false):
\[
  \sem{\delta}{P \andty \phi}
  =
  \comprehend{V \in \sem{}{\erase{P}}}{V \in \sem{\delta}{P} \text{ and } \sem{\delta}{\phi} = \one}
\]
Existential and universal types denote elements of their erasure
such that the relevant index quantification holds:
\begin{align*}
  \sem{\delta}{\extype{a : \tau}{P}}
  &=
    \comprehend{V \in \sem{}{\erase{P}}}{\extype{d \in \sem{}{\tau}}{V \in \sem{\delta, d/a}{P}}}
  \\ 
  \sem{\delta}{\alltype{a:\tau}{N}}
  &=
    \comprehend{f \in \sem{}{\erase{N}}}{\alltype{d \in \sem{}{\tau}}{f \in \sem{\delta,d/a}{N}}}
\end{align*}
Guarded types denote elements of their erasure
such that they are also in the refined type being guarded
if the guard holds ($\one$ means true):
\[
  \sem{\delta}{\phi \implies N} = \comprehend{f \in \sem{}{\erase{N}}}{\text{if } \sem{\delta}{\phi} = \one \text{ then } f \in \sem{\delta}{N}}
\]
The denotation of refined function types $\sem{\delta}{P \to N}$ is \emph{not}
the set $\sem{\delta}{P} \Rightarrow \sem{\delta}{N}$
of (continuous) functions from refined $P$-values to refined $N$-values;
if it were, then type soundness would break:
\begin{align*}
  \sem{\cdot}{\judgechkexp{\cdot}{\cdot}{\fun{x}{\Return{x}}}{(1 \land \False) \to \upshift{1}}}
  &= (\bullet \mapsto \minj{1} \bullet)
\end{align*}
which is not in $(\emptyset \Rightarrow \one \uplus \{\bott[\uparrow]\})$.
Instead, the meaning of a refined function type is a set
(resembling a unary logical relation)
\[
  \comprehend{f \in \sem{}{\erase{P \to N}}}{\alltype{V \in \sem{\delta}{P}}{f(V) \in \sem{\delta}{N}}}
\]
of \emph{unrefined} (continuous) functions that take refined values to refined values.
The meaning of \emph{refined} upshifts enforces termination
(if refined type soundness holds, and we will see it does):
\[
  \sem{\delta}{\upshift{P}} = \comprehend{\minj{1} V}{V \in \sem{\delta}{P}}
\]
Note that divergence $\minj{2} \bott[\uparrow]$ is \emph{not} in the set
$\sem{\delta}{\upshift{P}}$.

In appendix \Figref{fig:denotation-types2},
we inductively define the denotations of well-formed refined functors $F$
and algebras $\alpha$.
The main difference between refined and unrefined functors
is that in refined functors,
constant functors produce subsets of their erasure.
All functors, refined or otherwise,
also (forgetting the partial order structure)
denote endofunctors on the category of sets and functions.
As with our unrefined functors, our refined functors denote
functors with a fixed point (appendix \Lemref{lem:mu-unroll-equal}):
$\sem{\delta}{F} (\mu \sem{\delta}{F}) = \mu \sem{\delta}{F}$.
Moreover, $\mu \sem{\delta}{F}$ satisfies a recursion principle
such that we can inductively define measures on $\mu \sem{\delta}{F}$
via $\sem{\delta}{F}$-algebras
(discussed next).

Categorically, given an endofunctor $F$, an $F$-\emph{algebra}
is an evaluator map $\alpha : F(\tau) \to \tau$
for some carrier set $\tau$.
We may think of this in terms of elementary algebra:
we form algebraic expressions with $F$ and evaluate them with $\alpha$.
A morphism $f$ from algebra $\alpha : F(\tau) \to \tau$
to algebra $\beta : F(\tau') \to \tau'$
is a morphism $f : \tau \to \tau'$ such that
$f \circ \alpha = \beta \circ (F(f))$.
If an endofunctor $F$ has an \emph{initial}\footnote{An object $X$ in a category $\mathbf{C}$ is \emph{initial} if for every
object $Y$ in $\mathbf{C}$, there exists a unique morphism $X \to Y$ in $\mathbf{C}$.}
algebra $\minto : F (\mu F) \to \mu F$,
then it has a recursion principle.
By the recursion principle for $\mu F$,
we can define a recursive function from $\mu F$ to $\tau$
by \emph{folding} $\mu F$ with an $F$-algebra $\alpha : F(\tau) \to \tau$
like so:
\[
  \arrayenvl{
    (\fold{F}{\alpha}) : \mu F \to \tau \\
    (\fold{F}{\alpha}) \; v = \alpha \; \Big(\big(\mathit{fmap} \; F \; (\fold{F}{\alpha})\big) \; \big(\mathit{out}(v)\big)\Big)
  }
\]
where $\mathit{out} : \mu F \to F(\mu F)$,
which by Lambek's lemma exists and is inverse to $\mathit{into}$,
embeds (semantic) inductive values
into the unrolling of the (semantic) inductive type
(we usually elide $\mathit{fmap}$).
Conveniently, in our system and semantics, $\mathit{out}$
is always $d \mapsto d$,
and we almost never explicitly mention it.
Syntactic values $v$ in our system must be rolled
into inductive types---$\roll{v}$---and
this is also how (syntactic) inductive values are pattern-matched
(``applying $\mathit{out}$'' to $\roll{v}$),
but $\roll{-}$ conveniently denotes $d \mapsto d$.

We specify inductive types abstractly
as sums of products
so that they denote polynomial endofunctors more directly.
Polynomial endofunctors always have a ``least'' (initial) fixed point\footnote{This is not the case for all endofunctors. Therefore, not all endofunctors can be said to specify an inductive type. For example, consider the powerset functor.},
and hence specify inductive types, which have a recursion principle.
For example, we specify (modulo the unrolling simplification)
$\mathsf{len} : \mathsf{ListF}_A (\kindnat) \Rightarrow \kindnat$
(\Secref{sec:introduction})
by the (syntactic) algebra
\[
  \alpha
  ~=~
  \inl \clause{\unitexp}{0} \matchor \inr \clause{(\wild, a)}{1 + a}
\]
which denotes the (semantic) algebra
\[
  \sem{}{\alpha} : \underbrace{\sem{}{\textsf{ListF}_A}(\kindnat)}_{1 \uplus (\sem{}{A} \times \kindnat)} \to \kindnat
\]
defined by $\sem{}{\alpha} = [\bullet \mapsto 0, (a, n) \mapsto 1+n]$.
By initiality (the recursion principle), there is a unique function
\[
  \fold{\sem{}{\textsf{ListF}_A}}{\sem{}{\alpha}} : \mu\sem{}{\textsf{ListF}_A} \to \kindnat
\]
such that
$\fold{\sem{}{\textsf{ListF}_A}}{\sem{}{\alpha}} = \sem{}{\alpha} \circ (\sem{}{F} (\fold{\sem{}{\textsf{ListF}_A}}{\sem{}{\alpha}}))$,
which semantically captures $\mathsf{len}$
(\Secref{sec:introduction}).

In \sysname, a refined inductive type is written
$\comprehend{\nu : \mu F}{\Fold{F}{\alpha}\,{\nu} =_\tau t}$,
which looks quite similar to its own %
semantics:
\[
  \sem{\delta}{\comprehend{\nu : \mu F}{\Fold{F}{\alpha}\,{\nu} =_\tau t}}
  ~=~
  \comprehend{V \in \mu \sem{\delta}{F}}{(\fold{\sem{\delta}{F}}{\sem{\delta}{\alpha}}) \; V = \sem{\delta}{t}}
\]
The type $\textsf{List}(A)(n)$
of $A$-lists having length $n : \kindnat$, for example, is defined in our system as:
\[
  \textsf{List}(A)(n) = \comprehend{\nu : \mu \textsf{ListF}_A}{\Fold{\textsf{ListF}_A}{\textsf{lenalg}}\,{\nu} =_{\kindnat} n}
\]
Syntactic types, functors, and algebras
in our system look very similar to their own semantics.

A well-typed algebra $\judgealgebra{\Xi}{\Theta}{\alpha}{F}{\tau}$
denotes a dependent function
$\prod_{\delta \in \sem{}{\Theta}} \sem{\delta}{F}\sem{}{\tau} \to \sem{}{\tau}$.
The definition (appendix \Figref{fig:denotation-types2}) is mostly standard,
but the unit and pack cases could use some explanation.
Because $\Theta$ is for $F$ and $\Xi\ (\subseteq \Theta)$ is for $\alpha$,
we restrict $\delta$ to $\Xi$ at algebra bodies:
\[
  \sem{\delta}{\judgealgebra{\Xi}{\Theta}{\clause{\unitexp}{t}}{I}{\tau}} \bullet
  ~=~
  \sem{\delta\restriction_\Xi}{\judgeterm{\Xi}{t}{\tau}}
\]
The most interesting part of the definition concerns index packing:
\begin{align*}
  \sem{\delta}{\judgealgebra{\Xi}{\Theta}{\clause{(\pack{a}{\bap}, q)}{t}}{(\Const{\extype{a:\tau'}{Q}} \otimes \hat{P})}{\tau}} (V_1,V_2)
  &=  \\
  \hspace{20em}&\hspace{-20em}
                 \arrayenvl{
                 \sem{(\delta,d/a)}{\judgealgebra{\Xi, a:\tau'}{\Theta,a:\tau'}{\clause{(\bap, q)}{t}}{(\Const{Q} \otimes \hat{P})}{\tau}}(V_1, V_2)
  \\
                \text{where }d\in\sem{}{\tau'}\text{ is such that }V_1 \in \sem{\delta,d/a}{Q}
   }
\end{align*}
The pack clause lets us bind the witness $d$ of $\tau'$ in the existential
type $\extype{a:\tau'}{Q}$
to $a$ in the body $t$ of the algebra.
We know $d$ exists since $V_1 \in \sem{\delta}{\extype{a:\tau'}{Q}}$,
but it is not immediate that it is unique.
However, we prove $d$ is uniquely determined by $V_1$;
we call this property the \emph{soundness of value-determined indexes}
(%
all parts are mutually recursive):
\begin{lemma}[Soundness of Value-Determined Indexes]
  \label{lem:liq-value-determined-soundness}
  \hfill (\Lemref{lem:value-determined-soundness} in appendix)\\
  Assume $|- \delta_1 : \Theta$ and $\delta_2 : \Theta$.
  \begin{enumerate}
    \item If $\judgetp{\Theta}{P}{\Xi}$
      and $V \in \sem{\delta_1}{P}$ and $V \in \sem{\delta_2}{P}$,
      then $\delta_1\restriction_\Xi = \delta_2\restriction_\Xi$.
    \item If $\judgefunctor{\Theta}{\mathcal{F}}{\Xi}$
      and $X_1, X_2 \in \Set$
      and $V \in \sem{\delta_1}{\mathcal{F}} X_1$
      and $V \in \sem{\delta_2}{\mathcal{F}} X_2$,
      then $\delta_1\restriction_\Xi = \delta_2\restriction_\Xi$.
    \item If $\judgealgebra{\Xi}{\Theta}{\alpha}{F}{\tau}$
      and $\Xi \subseteq \Theta$
      and $\delta_1\restriction_\Xi = \delta_2\restriction_\Xi$,
      then $\sem{\delta_1}{\alpha} = \sem{\delta_2}{\alpha}$
      on $\sem{\delta_1}{F} \sem{}{\tau} \sect \sem{\delta_2}{F} \sem{}{\tau}$.
  \end{enumerate}
\end{lemma}
Therefore, the $\Xi$ in type and functor well-formedness
really does track index variables that are \emph{uniquely} determined by values,
semantically speaking.

\paragraph{Well-typed program terms}

Appendix \Figref{fig:denotation-program-terms} specifies the denotations
of well-typed refined program terms
in terms of the denotations of their erasure.
The denotation of a refined program term $E$ typed under $(\Theta; \Gamma)$,
at refined semantic substitution $\delta \in \sem{}{\Theta; \Gamma}$,
is the denotation $\sem{\erase{\delta}}{\erase{E}}$
of the (derivation of the) term's erasure $\erase{E}$
at the erased substitution $\erase{\delta}$. For example,
\[
  \sem{}{\judgechkexp{\Theta}{\Gamma}{e}{N}} = (\delta \in \sem{}{\Theta; \Gamma}) \mapsto \sem{\erase{\delta}}{\unrefchkexp{\erase{\Gamma}}{\erase{e}}{\erase{N}}}
\]

\paragraph{Unrolling}

We prove
(appendix \Lemref{lem:unroll-soundness})
that unrolling is sound:

\begin{lemma}[Unrolling Soundness]
  \hfill (\Lemref{lem:unroll-soundness} in appendix)\\
  Assume $|- \delta : \Theta$ and $\Xi \subseteq \Theta$.
  If $\judgeunroll{\Xi}{\Theta}{ \nu:G[\mu F] }{\beta}{G\;\Fold{F}{\alpha}\;\nu}{t}{P}{\tau}$,\\
  then $\comprehend{V \in \sem{\delta}{G} (\mu\sem{\delta}{F})}{\sem{\delta}{\beta} (\sem{\delta}{G}(\fold{\sem{\delta}{F}}{\sem{\delta}{\alpha}})\;V) = \sem{\delta}{t}} = \sem{\delta}{P}$.
\end{lemma}

Due to our definition of algebra denotations (specifically, for the pack pattern),
we use the soundness of value-determined indexes in the pack case of the proof.

\paragraph{Subtyping}

We prove
(appendix \Lemref{lem:type-soundness-sub})
that subtyping is sound:

\begin{lemma}[Soundness of Subtyping]
  \hfill (\Lemref{lem:type-soundness-sub} in appendix)\\
  Assume $|- \delta : \Theta$.
  If $\judgesub[\pm]{\Theta}{A}{B}$,
  then $\sem{\delta}{A} \subseteq \sem{\delta}{B}$.
\end{lemma}

\paragraph{Type soundness}

Denotational-semantic type soundness says that if a program term has type $A$ under $\Theta$ and $\Gamma$,
then the mathematical meaning of that program term
at any interpretation of (that is, semantic environment for) $\Theta$ and $\Gamma$
is an element of the mathematical meaning of $A$ at that interpretation,
that is,
the program term denotes a dependent function
$\prod_{\delta \in \sem{}{\Theta; \Gamma}} \sem{\filterprog{\delta}}{A}$.
This more or less corresponds to proving (operational) type soundness
with respect to a big-step operational semantics.
Refined types pick out subsets of values of unrefined types.
Therefore, by type soundness,
if a program has a refined type, then we have learned something more about that
program than the unrefined system can verify for us.

\begin{theorem}[Type Soundness]
  \label{thm:liq-type-soundness}
  \hfill (\Thmref{thm:type-soundness} in appendix)\\
  Assume $|- \delta : \Theta ; \Gamma$. Then:
  \begin{enumerate}
  \item If $\judgesynhead{\Theta}{\Gamma}{h}{P}$,
    then $\sem{\delta}{h} \in \sem{\filterprog{\delta}}{P}$.
  \item If $\judgesynexp{\Theta}{\Gamma}{\be}{N}$,
    then $\sem{\delta}{\be} \in \sem{\filterprog{\delta}}{N}$.
  \item If $\judgechkval{\Theta}{\Gamma}{v}{P}$,
    then $\sem{\delta}{v} \in \sem{\filterprog{\delta}}{P}$.
  \item If $\judgechkexp{\Theta}{\Gamma}{e}{N}$,
    then $\sem{\delta}{e} \in \sem{\filterprog{\delta}}{N}$.
  \item If $\judgechkmatch{\Theta}{\Gamma}{P}{\clauses{\pa}{e}{i}{I}}{N}$,
    then $\sem{\delta}{\clauses{\pa}{e}{i}{I}} \in \sem{\filterprog{\delta}}{P} \Rightarrow \sem{\filterprog{\delta}}{N}$.
  \item If $\judgespine{\Theta}{\Gamma}{s}{N}{\upshift{P}}$,
    then $\sem{\delta}{s} \in \sem{\filterprog{\delta}}{N} \Rightarrow \sem{\filterprog{\delta}}{\upshift{P}}$.
  \end{enumerate}
\end{theorem}

(All parts are mutually recursive.)
The proof
(appendix \Thmref{thm:type-soundness})
uses the soundness of unrolling and subtyping.
The proof is mostly straightforward.
The hardest case is the one for recursive expressions in part (4), where
we use an \emph{upward closure} lemma---in particular, part (3) below---to show
that the fixed point is in the appropriately refined set:
\begin{lemma}[Upward Closure]
  \hfill (\Lemref{lem:upward-closure} in appendix)\\
  Assume $|- \delta : \Theta$.
  \begin{enumerate}
  \item If $\judgealgebra{\Xi}{\Theta}{\alpha}{F}{\tau}$
    and $\Xi \subseteq \Theta$
    then $\sem{\delta}{\alpha}$ is monotone.
  \item If $\judgefunctor{\Theta}{\mathcal{G}}{\dontcare}$
    and $\judgefunctor{\Theta}{F}{\dontcare}$
    and $k \in \kindnat$\\
    and $V \in \sem{\delta}{\mathcal{G}} (\sem{\delta}{F}^k \emptyset)$
    and $\ord[\sem{}{\erase{\mathcal{G}}} (\sem{}{\erase{F}}^k \emptyset)]{V}{V'}$,\\
    then $V' \in \sem{\delta}{\mathcal{G}} (\sem{\delta}{F}^k \emptyset)$.
  \item If $\judgetp{\Theta}{A}{\dontcare}$
    and $V \in \sem{\delta}{A}$ and $\ord[\sem{}{\erase{A}}]{V}{V'}$,
    then $V' \in \sem{\delta}{A}$.
  \end{enumerate}
\end{lemma}
Out of all proofs in this paper,
the proof of upward closure
(appendix \Lemref{lem:upward-closure})
is a top contender for the most interesting induction metric:
\begin{proof}
  By lexicographic induction on,
  first, $\size{A}$/$\size{F}$ (parts (1), (2) and (3), mutually), and,
  second, $\langle k, \mathcal{G} \text{ structure} \rangle$ (part (2)),
  where $\langle \dots \rangle$ denotes lexicographic order.
\end{proof}
We define the simple size function $\size{-}$,
which is basically a standard structural notion of size,
in appendix \Figref{fig:size}.
This is also the only place, other than unrolling soundness,
where we use the soundness of value-determined indexes
(again for a pack case, in part (1)).

\paragraph{Substitution soundness}

We interpret
a syntactic substitution (typing derivation)
$\Theta_0; \Gamma_0 |- \sigma : \Theta; \Gamma$
as a function $\sem{}{\sigma} : \sem{}{\Theta_0; \Gamma_0} \to \sem{}{\Theta; \Gamma}$
on semantic substitutions (appendix \Defnref{def:den-syn-subs}).
Similarly to the interpretation of unrefined substitution typing derivations,
the interpretation of the head term being substituted
(its typing/sorting subderivation)
pre-applies the rest of the substitution:
\begin{align*}
  \sem{\delta}{\cdot} &= \cdot \\
  \sem{\delta}{\sigma, t/a} &= \sem{\delta}{\sigma}, \sem{\filterprog{\delta}}{[\sigma]t}/a \\
  \sem{\delta}{\sigma, \subs{v}{P}{x}} &= \sem{\delta}{\sigma}, \sem{\delta}{[\sigma]v}/x
\end{align*}
Substitution soundness holds
(appendix \Thmref{thm:subs-soundness}):
if $E$ is a program term typed under $\Theta$ and $\Gamma$,
and $\Theta_0; \Gamma_0 |- \sigma : \Theta; \Gamma$,
then $\sem{}{[\sigma]E} = \sem{}{E} \circ \sem{}{\sigma}$.
(Recall we prove a syntactic substitution lemma: \Lemref{lem:liq-syn-subs}.)
That is, substitution and denotation commute,
or (in other words)
syntactic substitution and semantic substitution are compatible.

\paragraph{Logical consistency, total correctness, and partial correctness}

Our \emph{semantic} type soundness result implies that \sysname is
logically consistent and totally correct.

A logically inconsistent type (for example, $0$ or $\upshift{0}$ or $\upshift{(1 \land \False)}$)
denotes the empty set, which is uninhabited.

\begin{corollary}[Logical Consistency]
  \label{cor:logical-consistency}
  If $\judgechkexp{\cdot}{\cdot}{e}{N}$,
  then $N$ is \emph{logically consistent}, that is, $\sem{\cdot}{N} \neq \emptyset$.
  Similarly,
  if $\judgechkval{\cdot}{\cdot}{v}{P}$,
  then $P$ is logically consistent,
  and so on for the other typing judgments.
\end{corollary}

Proving logical consistency \emph{syntactically},
say,
via progress and preservation lemmas,
would require \emph{also} proving that every reduction sequence eventually terminates
(that is, strong normalization),
which might need a relatively complicated proof using logical relations \cite{Tait67}.

\emph{Total correctness} means that every closed computation
(that is specified as total)
returns a value of the specified type:

\begin{corollary}[Total Correctness]
  If $\judgechkexp{\cdot}{\cdot}{e}{\upshift{P}}$,
  then $\sem{\cdot}{e} \neq \bott[\sem{\cdot}{\upshift{P}}]$,
  that is, $e$ does not diverge.
\end{corollary}
\begin{proof}
  ~\\
  \begin{llproof}
    \Pf{}{|-}{\cdot : \cdot; \cdot}{By \EmptySem}
    \inPf{\sem{\cdot}{e}}{\sem{\filterprog{\cdot}}{\upshift{P}}}{By \Theoremref{thm:liq-type-soundness} (Type Soundness)}
    \eqPf{}{\sem{\cdot}{\upshift{P}}}{By definition of $\filterprog{-}$}
    \eqPf{}{\comprehend{\minj{1} V}{V \in \sem{\cdot}{P}}}{By definition of $\sem{\cdot}{-}$}
  \end{llproof}

  Therefore,
  $\sem{\cdot}{e} \neq \minj{2} \bott[\uparrow] = \bott[\sem{\cdot}{\upshift{P}}]$,
  that is, $e$ terminates (and returns a value).
\end{proof}

\Sysname can be extended to include partiality,
simply by adding a \emph{partial} upshift type connective $\pupshift{P}$ (``partial upshift of $P$''),
with type well-formedness, subtyping and type equivalence rules similar to those of $\upshift{P}$,
and the following two expression typechecking rules.
The first rule introduces the new connective $\pupshift{P}$;
the second rule lacks a termination refinement such as that in \DeclChkExpRec, so it may yield divergence.
\begin{mathpar}
  \Infer{}
  {
    \judgechkval{\Theta}{\Gamma}{v}{P}
  }
  {
    \judgechkexp{\Theta}{\Gamma}{\Return{v}}{\pupshift{P}}
  }
  \and
  \Infer{}
  {
    \judgechkval{\Theta}{\Gamma, x:\downshift{N}}{e}{N}
  }
  {
    \judgechkexp{\Theta}{\Gamma}{\rec{x}{e}}{N}
  }
\end{mathpar}
The meaning of the partial upshift is defined as follows:
\[
  \sem{\delta}{ \judgetp{\Theta}{\pupshift{P}}{\dontcare} }
  = \comprehend{ d \in \sem{}{\upshift{\erase{P}}} }{ \text{if } d \neq \bott[\sem{}{\upshift{\erase{P}}}] \text{ then } d = \minj{1} V \text{ for some } V \in \sem{\delta}{P} }
\]
It is straightforward to update the metatheory to prove
\emph{partial correctness}:
If a closed computation (that is specified as partial) terminates,
then it returns a value of the specified type.
Partial correctness is
a corollary of the updated type soundness result:
if $\judgechkexp{\cdot}{\cdot}{e}{\pupshift{P}}$
and $\sem{\cdot}{e} \neq \bott[\sem{\cdot}{\upshift{P}}]$
then $\sem{\cdot}{e} = \minj{1} V$ and $V \in \sem{\cdot}{P}$.

Adding partiality introduces logical inconsistency,
so we must restate logical consistency for expression typing:
If $\judgechkexp{\cdot}{\cdot}{e}{\upshift{P}}$,
then $\upshift{P}$ is \emph{logically consistent}.

\section{Algorithmic System}
\label{sec:algorithmic-system}

We design our algorithmic system in the spirit of those of \citet{Dunfield13, Dunfield19},
but those systems do not delay constraint verification until all existentials are solved.
The algorithmic rules closely mirror the declarative rules,
except for a few key differences:
\begin{itemize}
\item Whenever a declarative rule conjures an index term,
  the corresponding algorithmic rule adds,
  to a separate (input) algorithmic context $\Delta$,
  an existential variable (written with a hat: $\ahat$) to be solved.
\item As the typing algorithm proceeds, we add index term solutions
  of the existential variables to the output algorithmic context,
  increasing knowledge (see \Secref{sec:context-extension}).
  We eagerly apply index solutions to input types and output constraints,
  and pop them off the output context when out of scope.
\item Whenever a declarative rule checks propositional validity or equivalence
  ($\judgeentail{\Theta}{\phi}$ or $\judgeequiv{\Theta}{\phi}{\psi}$),
  the algorithm delays checking the constraint
  until all existential variables in the propositions
  are solved (at the end of a focusing stage).
  Similarly, subtyping, type equivalence, and expression typechecking
  constraints are delayed until all existential variables are solved.
  When an entity has no existential variables,
  we say that it is \emph{ground}.
\item In subtyping,
  we eagerly extract from assumptive positions immediately under polarity shifts.
\end{itemize}

Syntactically, objects in the algorithmic system are not much different
from corresponding objects of the declarative system.
We extend the grammar for index terms with a production of existential variables,
which we write as an index variable with a hat $\ahat$, $\hat{b}$, or $\hat{c}$:
\[
  t ~\bnfas~ \cdots \bnfalt \ahat
\]
We use this (algorithmic) index grammar everywhere in the algorithmic system,
using the same declarative metavariables.
However, we write algorithmic logical contexts
with a hat: $\hat{\Theta}$.
Algorithmic logical contexts $\hat{\Theta}$ only appear in output mode,
and are like (input) logical contexts $\Theta$,
but propositions listed in them may have existential variables
(its index variable sortings $a:\tau$ are universal).

Constraints are added to the algorithmic system.
\Figureref{fig:constraints} gives grammars for subtyping and typing constraints.
In contrast to DML, the grammar does not include existential constraints.

\begin{figure}[htbp]

  \begin{grammar}
   Subtyping constraints
   &
   $\Wah$ & \bnfas & $
   \phi
   \bnfalt  \propeqprob{\phi}{\psi}
   \bnfalt  \phi \implies \Wah
   \bnfalt  \Wah \land \Wah
   \bnfalt  \forall a:\tau. \Wah
   \bnfaltBRK  \possubprob{P}{Q}
   \bnfalt  \negsubprob{N}{M}
   \bnfalt  \poseqprob{P}{Q}
   \bnfalt  \negeqprob{N}{M}
   $
   \\[0.9ex]
   Typing constraints
   &
   $\chi$ & \bnfas & $\cdot
   \bnfalt  (e <= N), \chi
   \bnfalt  W, \chi
   $
  \end{grammar}

  \caption{Typing and subtyping constraints}
  \label{fig:constraints}
\end{figure}

Checking constraints boils down to checking propositional validity,
$\judgeentail{\Theta}{\phi}$,
which is analogous to checking %
\emph{verification conditions}
in the tradition of imperative program verification initiated by
\citet{Floyd67} and \citet{Hoare69}
(where programs annotated with Floyd--Hoare assertions
are analyzed, generating verification conditions whose validity
implies program correctness).
These propositional validity constraints
are the constraints that can be automatically verified by a theorem prover
such as an SMT solver.
The (algorithmic) $W$
constraint verification judgment is written $\entailwah{\Theta}{W}$
and means that $W$ algorithmically holds under $\Theta$.
Notice that the only context in the judgment is $\Theta$, which has no existential variables:
this reflects the fact
that we delay verifying $W$ until $W$ has no existential variables
(in which case we say $W$ is \emph{ground}).
Similarly, $\algneg{\Theta}{\Gamma}{\chi}$
is the (algorithmic) $\chi$ verification judgment,
meaning all of the constraints in $\chi$ algorithmically
hold under $\Theta$ and $\Gamma$,
and here $\chi$ is also ground (by focusing).

\subsection{Contexts and Substitution}

Algorithmic contexts $\Delta$ are lists of solved or unsolved existential variables,
and are said to be \emph{complete}, and are written as $\Omega$, if they are all solved:
\begin{align*}
  \Delta &~\bnfas~ \cdot \bnfalt \Delta, \ahat : \tau \bnfalt  \Delta, \hypeq{\ahat}{\tau}{t} \\
  \Omega &~\bnfas~ \cdot \bnfalt \Omega, \hypeq{\ahat}{\tau}{t}
\end{align*}
We require solutions $t$ of existential variables $\ahat$ to be well-sorted under
(input) logical contexts $\Theta$, which have no existential variables.
To maintain this invariant that every solution in $\Delta$ is \emph{ground},
that is, has no existential variables,
we exploit type polarity in algorithmic subtyping,
and prevent existential variables from ever appearing in refinement algebras.

We will often treat algorithmic contexts $\Delta$
as substitutions of ground index terms for existential variables $\hat{a}$
in index terms $t$ (including propositions $\phi$), types $A$, functors $\mathcal{F}$,
constraints $W$ and $\chi$,
and output logical contexts $\hat{\Theta}$
(whose propositions may have existential variables).
The definition is straightforward:
homomorphically apply the context to the object $\mathcal{O}$,
and further define $[\Delta]\mathcal{O}$ by induction on $\Delta$.
\begin{align*}
[\cdot]\mathcal{O} &= \mathcal{O} \\
[\Delta, \ahat:\tau]\mathcal{O} &= [\Delta]\mathcal{O} \\
[\Delta, \hypeq{\ahat}{\tau}{t}]\mathcal{O} &= [\Delta]([t/\ahat]\mathcal{O})
\end{align*}
The order of substitutions in the definition of context application above
does not matter because solutions are ground
(we may view $[\Delta]\mathcal{O}$ as simultaneous substitution).
If $\mathcal{O}$ only has existential variables from $\dom{\Omega}$,
then $[\Omega]\mathcal{O}$ is ground.

\subsection{Context Extension}
\label{sec:context-extension}

The algorithmic context extension judgment $\extend{\Theta}{\Delta}{\Delta'}$
says that $\dom{\Delta} = \dom{\Delta'}$
and $\Delta'$ has the same solutions as $\Delta$,
but possibly solves more (that are unsolved in $\Delta$).
All typing and subtyping judgments (under $\Theta$)
that have input and output algorithmic contexts $\Delta$ and $\Delta'$
(respectively)
enjoy the property that they increase index information, that is,
$\extend{\Theta}{\Delta}{\Delta'}$.
If $\extend{\Theta}{\Delta}{\Omega}$, then $\Omega$ \emph{completes} $\Delta$:
it has $\Delta$'s solutions, but also solutions to all of $\Delta$'s unsolved variables.
\begin{mathpar}
  \Infer{}
  {}
  {\extend{\Theta}{\cdot}{\cdot}}
  \and
  \Infer{}
  {\extend{\Theta}{\Delta}{\Delta'}}
  {\extend{\Theta}{\Delta, \ahat:\tau}{\Delta', \ahat:\tau}}
  \and
  \Infer{}
  {\extend{\Theta}{\Delta}{\Delta'}}
  {\extend{\Theta}{\Delta, \hypeq{\ahat}{\tau}{t}}{\Delta', \hypeq{\ahat}{\tau}{t}}}
  \and
  \Infer{}
  {\extend{\Theta}{\Delta}{\Delta'}}
  {\extend{\Theta}{\Delta, \ahat:\tau}{\Delta', \hypeq{\ahat}{\tau}{t}}}
\end{mathpar}

\subsection{Subtyping}
\label{sec:alg-subtyping}
Algorithmic subtyping $\algsub[\pm]{\Theta; \Delta}{A}{B}{\Wah}{\Delta'}$
says that, under logical context $\Theta$ and algorithmic context $\Delta$,
the type $A$ is algorithmically a subtype of $B$ if and only if
output constraint $\Wah$ holds algorithmically
(under suitable solutions including those of $\Delta'$),
outputting index solutions $\Delta'$.
In subtyping and type equivalence, the delayed output constraints $W$
must remember their logical context via $\implies$ and $\forall$.
For example, in checking that $\extype{a:\kindnat} \textsf{Nat}(a) \land (a < 5)$
is a subtype of $\extype{a:\kindnat} \textsf{Nat}(a) \land (a < 10)$,
the output constraint $W$ is $\alltype{a:\kindnat} (a < 5) \implies (a < 10)$.

For space reasons,
we don't present all algorithmic subtyping rules here
(see appendix \Figref{fig:algsubtyping}),
but only enough rules to discuss the key design issues.
Further, we don't present algorithmic equivalence here
(see appendix Figures~\ref{fig:algfunequiv} and~\ref{fig:algtpequiv}),
which is similar to and simpler than algorithmic subtyping.

In algorithmic subtyping, we maintain the invariant
that positive subtypes and negative supertypes are ground.
The rules
\begin{mathpar}
  \runonfontsz{10pt}
  \Infer{}
  { 
    \algsub[+]{\Theta; \Delta, \ahat:\tau}{P}{[\ahat/a] Q}{\Wah}{\Delta', \hypeq{\ahat}{\tau}{t}}
  }
  {
    \algsub[+]{\Theta; \Delta}{P}{\extype{a:\tau} Q}{\Wah}{\Delta'}
  }
  \and
  \Infer{}
  {
    \algsub[-]{\Theta; \Delta,\ahat:\tau}{[\ahat/a]N}{M}{\Wah}{\Delta', \hypeq{\ahat}{\tau}{t}}
  }
  {\algsub[-]{\Theta; \Delta}{\alltype{a:\tau} N}{M}{\Wah}{\Delta'} }
\end{mathpar}
are the only subtyping rules which add existential variables
(to the side not necessarily ground) to be solved
(whereas the declarative system conjures a solution).
We pop off the solution as we have the invariant
that output contexts are eagerly applied to output constraints
and input types.

The rule
\[
  \Infer{\AlgSubPosFixInst}
  {
    \ground{t}
    \\
    \algequiv[]{\Theta; \Delta}{F}{G}{W}{\Delta_1', \ahat:\tau, \Delta_2'}
    \\
    \Delta' = \Delta_1', \hypeq{\ahat}{\tau}{t}, \Delta_2'
  }
  {
    \algsub[+]
    {\Theta; \Delta}
    {\comprehend{\nu:\mu F}{\Fold{F}{\alpha}\,\nu =_\tau t}}
    {\comprehend{\nu:\mu G}{\Fold{G}{\alpha}\,\nu =_\tau \ahat}}
    {
      W \land (t = t)
    }
    {\Delta'}
  }
\]
runs the functor equivalence algorithm
(which outputs constraint $W$ and solutions $\Delta_1', \ahat:\tau, \Delta_2'$),
checks that $\ahat$ does not get solved there,
and then solves $\ahat$ to $t$ (yielding $\Delta'$) after checking that the latter
(which is a subterm of a positive subtype) is ground,
outputting the constraint generated by functor equivalence
together with the equation $t = t$
(the declarative system can conjure a different but logically equal
term for the right-hand side of this equation),
and $\Delta'$.
Alternatively, there is a rule for when $\ahat$ gets solved by functor equivalence,
and a rule where a term that is not an existential variable is in place of $\ahat$.

The rule
\[
  \Infer{}
  {
    \judgeextract{\Theta; \Delta}{M}{M'}{\Thetahat}
  }
  {
    \algsub[+]{\Theta; \Delta}{\downshift{N}}{\downshift{M}}{\big(\Thetahat \implies^\ast \negsubprob{N}{M'}\big)}{\Delta}
  }
\]
extracts $M'$ and $\Thetahat$ from $M$
and delays the resulting negative subtyping constraint $\negsubprob{N}{M'}$,
to be verified under its logical setting $\Thetahat$
(whose propositions,
which were extracted from the side not necessarily ground,
may have existential variables only solved
in value typechecking).
The metaoperation $\implies^\ast$ traverses $\hat{\Theta}$,
creating universal quantifiers from universal variables
and implications from propositions:
\[
\begin{array}[t]{rcl}
  \cdot \implies^\ast W &=& W \\[0.3ex]
  (\hat{\Theta}, \phi) \implies^\ast W &=& \hat{\Theta} \implies^\ast (\phi \implies W) \\[0.3ex]
  (\hat{\Theta}, a:\tau) \implies^\ast W &=& \hat{\Theta} \implies^\ast (\alltype{a:\tau} W)
\end{array}
\]
The dual shift rule is similar.
In the declarative system, \DeclSubPosL and \DeclSubNegR
are invertible,
which means that they can be eagerly applied without getting stuck;
algorithmically, we apply them immediately at polarity shifts,
so the above rule corresponds to an algorithmic combination
of the declarative rules \DeclSubNegR and \DeclSubPosDownshift
(and similarly for its dual rule for $\uparrow$).

For rules with multiple nontrivial premises, such as product subtyping
\[
  \Infer{}
  {
    \algsub[+]{\Theta; \Delta}{P_1}{Q_1}{\Wah_1}{\Delta''}
    \\
    \algsub[+]{\Theta; \Delta''}{P_2}{[\Delta'']Q_2}{\Wah_2}{\Delta'}
  }
  {
    \algsub[+]{\Theta; \Delta}{(P_1 \times P_2)}{(Q_1 \times Q_2)}{[\Delta']\Wah_1 \land \Wah_2}{\Delta'}
  }
\]
we thread solutions through inputs, applying them to the non-ground side
($[\Delta]-$ treats $\Delta$
as a substitution of index solutions for existential variables),
ultimately outputting both delayed constraints.
We maintain the invariant that existential variables in output constraints
are eagerly solved, which is why, for example, $\Delta'$ is applied to $W_1$
in the conclusion of the above rule, but not to $W_2$ (that would be redundant).

\subsection{Typing}
\label{sec:alg-typing}

We now discuss issues specific to algorithmic program typing.

Exploiting polarity, we can restrict the flow of index information
to the right- and left-focusing stages:
in particular,
$\algchk{\Theta; \Delta}{\Gamma}{v}{P}{\chi}{\Delta'}$
and
$\algspine{\Theta; \Delta}{\Gamma}{s}{N}{\upshift{P}}{\chi}{\Delta'}$,
the algorithmic value and spine typechecking judgments.
The input types of these judgments can have existential variables,
and these judgments synthesize constraints and index solutions,
but the algorithmic versions of the other judgments do not;
we judgmentally distinguish the latter by replacing the ``$|-$''
in the declarative judgments with ``$\rhd$''
(for example, $\algsynexp{\Theta}{\Gamma}{\be}{\upshift{P}}$).
Delayed constraints are verified only and immediately after completing a focusing stage,
when all their existential variables are necessarily solved.

\begin{figure}
  \judgbox{\algchk{\Theta; \Delta}{\Gamma}{v}{P}{\chi}{\Delta'}}
  {Under inputs contexts $\Theta$, $\Delta$, and $\Gamma$,
    the value $v$ checks against type $P$,\\
    with output computation constraints $\chi$ and output context $\Delta'$}
  
  \begin{mathpar}
    \Infer{\AlgChkValVar}
    {
      P \neq \exists, \with
      \\
      (x:Q)\in\Gamma
      \\
      \algsub[+]{\Theta; \Delta}{Q}{P}{\Wah}{\Delta'}
    }
    {
      \algchk{\Theta; \Delta}{\Gamma}{x}{P}{\Wah}{\Delta'}
    }
    \and
    \Infer{\AlgChkValUnit}
    {
    }
    {
      \algchk{\Theta; \Delta}{\Gamma}{\unit}{\unitty}{\cdot}{\Delta}
    }
    \and
    \Infer{\AlgChkValPair}
    {
      \algchk{\Theta; \Delta}{\Gamma}{v_1}{P_1}{\chi_1}{\Delta''}
      \\
      \algchk{\Theta; \Delta''}{\Gamma}{v_2}{[\Delta'']P_2}{\chi_2}{\Delta'}
    }
    {
      \algchk{\Theta; \Delta}{\Gamma}{\pair{v_1}{v_2}}
      {(P_1 \times P_2)}{[\Delta']\chi_1, \chi_2}{\Delta'}
    }
    \and
    \Infer{\AlgChkValIn{k}}
    {
      \algchk{\Theta; \Delta}{\Gamma}{v_k}{P_k}{\chi}{\Delta'}
    }
    {
      \algchk{\Theta; \Delta}{\Gamma}{\inj{k}{v_k}}{(P_1 + P_2)}{\chi}{\Delta'}
    }
    \and
    \Infer{\AlgChkValExists}
    {
      \algchk{\Theta; \Delta, \ahat:\tau}{\Gamma}{v}{[\ahat / a]P}{\chi}{\Delta', \hypeq{\ahat}{\tau}{t}}
    }
    {
      \algchk{\Theta; \Delta}{\Gamma}{v}{(\extype{a : \tau} P)}{\chi}{\Delta'}
    }
    \and
    \Infer{\AlgChkValWith}
    {
      \algchk{\Theta; \Delta}{\Gamma}{v}{P}{\chi}{\Delta''}
      \\
      \alginst{\Theta; \Delta''}{[\Delta'']\phi}{\Delta'}
    }
    {
      \algchk{\Theta; \Delta}{\Gamma}{v}{(P \andty \phi)}{([\Delta']\phi, [\Delta']\chi)}{\Delta'}
    }
    \and
    \Infer{\AlgChkValFix}
    {
      \judgeunroll{\cdot}{\Theta; \Delta}{\nu:F[\mu F]}{\alpha}{F\; \Fold{F}{\alpha}\;\nu}{t}{P}{\tau}
      \\
      \algchk{\Theta; \Delta}{\Gamma}{v}{P}{\chi}{\Delta'}
    }
    {
      \algchk{\Theta; \Delta}{\Gamma}{\into{v}}{\comprehend{\nu : \mu F}{\Fold{F}{\alpha}\,{\nu} =_\tau t}}{\chi}{\Delta'}
    }
    \and
    \Infer{\AlgChkValDownshift}
    {
    }
    {
      \algchk{\Theta; \Delta}{\Gamma}{\thunk{e}}{\downshift{N}}{(e <= N)}{\Delta}
    }
  \end{mathpar}

  \judgbox{\algspine{\Theta; \Delta}{\Gamma}{s}{N}{\upshift{P}}{\chi}{\Delta'}}
  {
    Under $\Theta$, $\Delta$, and $\Gamma$,
    passing spine $s$ to a head of type $\downshift{N}$\\
    synthesizes $\upshift{P}$,
    with output constraints $\chi$ and context $\Delta'$
  }
  \begin{mathpar}
    \Infer{\AlgSpineAll}
    {
      \algspine{\Theta; \Delta, \ahat : \tau}{\Gamma}{s}{[\ahat/a]{N}}{\upshift{P}}{\chi}{\Delta', \hypeq{\ahat}{\tau}{t}}
    }
    {
      \algspine{\Theta; \Delta}{\Gamma}{s}{\alltype{a:\tau}N}{\upshift{P}}{\chi}{\Delta'}
    }
    \and
    \Infer{\AlgSpineImplies}
    {
      \algspine{\Theta; \Delta}{\Gamma}{s}{N}{\upshift{P}}{\chi}{\Delta'}
    }
    {
      \algspine{\Theta; \Delta}{\Gamma}{s}{\phi \implies N}{\upshift{P}}{[\Delta']\phi,\chi}{\Delta'}
    }
    \and
    \Infer{\AlgSpineApp}
    {
      \algchk{\Theta; \Delta}{\Gamma}{v}{Q}{\chi}{\Delta''}
      \\
      \algspine{\Theta; \Delta''}{\Gamma}{s}{[\Delta'']N}{\upshift{P}}{\chi'}{\Delta'}
    }
    {
      \algspine{\Theta; \Delta}{\Gamma}{v, s}{Q -> N}{\upshift{P}}{[\Delta']\chi, \chi'}{\Delta'}
    }
    \and
    \Infer{\AlgSpineNil}
    {}
    {
      \algspine{\Theta; \Delta}{\Gamma}{\cdot}{\upshift{P}}{\upshift{P}}{\True}{\Delta}
    }
  \end{mathpar}

  \caption{Algorithmic value and spine typing}
  \label{fig:valuespinealgtyping}
\end{figure}

Consequently, the algorithmic typing judgments for
heads, bound expressions, pattern matching, and expressions
are essentially the same as their declarative versions,
but with a key difference.
Namely, in \AlgSynValAnnot, \AlgSynSpineApp, and \AlgChkExpUpshift
(below, respectively),
focusing stages start with an empty algorithmic context,
outputting ground constraints
(and an empty output context because solutions are eagerly applied),
and a premise is added to verify these constraints:
\begin{mathpar}
  \scalebox{0.85}{
    $\Infer{}
    {
      \algchk{\Theta; \cdot}{\Gamma}{v}{P}{\chi}{\cdot}
      \\
      \algneg{\Theta}{\Gamma}{\chi}
    }
    {
      \algsynhead{\Theta}{\Gamma}{\annoexp{v}{P}}{P}
    }$
  }
  \and
  \scalebox{0.85}{
    $\Infer{}
    { 
      \algsynhead{\Theta}{\Gamma}{h}{\downshift{N}}
      \\
      \algspine{\Theta; \cdot}{\Gamma}{s}{N}{\upshift{P}}{\chi}{\cdot}
      \\
      \algneg{\Theta}{\Gamma}{\chi}
    }
    { \algsynexp{\Theta}{\Gamma}{h(s)}{\upshift{P}} }$
  }
  \and
  \scalebox{0.85}{
    $\Infer{}
    { 
      \algchk{\Theta; \cdot}{\Gamma}{v}{P}{\chi}{\cdot} 
      \\
      \algneg{\Theta}{\Gamma}{\chi}
    }
    {
      \algchkneg{\Theta}{\Gamma}{\Return{v}}{\upshift{P}} 
    }$
  }
\end{mathpar}

Algorithmic typechecking for recursive expressions uses algorithmic subtyping,
which outputs a ground constraint $W$.
Because this $W$ is ground, we can verify it ($\entailwah{\Theta}{W}$) immediately:
\[
  \Infer{}
  {
    \arrayenvl{
      \simple{\Theta}{N}
      \and
      \algsub[-]{\Theta; \cdot}{\alltype{a:\kindnat} M}{N}{W}{\cdot}
      \and
      \entailwah{\Theta}{W}
      \\
      \algchkneg{\Theta, a:\kindnat}{\Gamma, x:\downshift{\big(\alltype{a':\kindnat} (a' < a) \implies [a'/a]M}\big)}{e}{M}
    }
  }
  {
    \algchkneg{\Theta}{\Gamma}{\rec{x : (\alltype{a:\kindnat} M)}{e}}{N}
  }
\]
For the full definition of algorithmic typing,
see appendix
Figures~\ref{fig:algtyping-head-bound-expression},~\ref{fig:algtyping-value-expression}, and~\ref{fig:algtyping-match-spine}.

Besides the instantiation rules
(such as \AlgSubPosFixInst)
for inductive types in algorithmic subtyping and type equivalence,
there are exactly two judgments
($\alginst{\Theta; \Delta}{\phi}{\Delta'}$
and $\algpropequivinst{\Theta; \Delta}{\phi}{\psi}{\Delta'}$)
responsible for inferring index solutions,
both dealing with the output of algorithmic unrolling.
Algorithmic unrolling can output indexes of the form $\ahat = t$ with $t$ ground,
and these equations are solved in either
value typechecking, subtyping, or type equivalence.
In the former two cases,
we can solve $\ahat$ as the algebra body $t$
which is necessarily ground (as discussed in \Secref{sec:well-formedness}).
The judgment
$\alginst{\Theta; \Delta}{\phi}{\Delta'}$
(appendix \Figref{fig:propinst}),
used in \AlgChkValWith,
checks whether $\phi$
has form $\ahat = t$ where $t$ is ground.
If so, then it solves $\ahat$ to $t$ in $\Delta$;
otherwise, it does not touch $\Delta$.
\begin{mathpar}
  \Infer{\RuleInst}
  {
    \Theta |- t : \tau
  }
  {
    \alginst{\Theta; \Delta_1, \ahat:\tau, \Delta_2}{\ahat = t}{\Delta_1, \hypeq{\ahat}{\tau}{t}, \Delta_2}
  }
  \and
  \Infer{\RuleNoInst}
  {
    \phi \text{ not of form } \ahat = t \text{ where } \Theta |- t : \tau
  }
  {
    \alginst{\Theta; \Delta}{\phi}{\Delta}
  }
\end{mathpar}

For example,
suppose a head synthesizes
$\alltype{a:\kindnat} \mathsf{Nat}(a) \to \upshift{\mathsf{Nat}(a)}$
and we wish to apply this head to the spine (containing exactly one argument value)
$\roll{\inj{2}{\pair{\roll{\inj{1}{\unit}}}{\unit}}}$.
We generate a fresh existential variable $\ahat$ for $a$ (rule \AlgSpineAll)
and then check the value against $\textsf{Nat}(\ahat)$ (rule \AlgSpineApp).
(Checking the same value against type $\extype{a:\kindnat} \mathsf{Nat}(a)$
yields the same problem, by dual rule \AlgChkValExists,
and the following solution also works in this case.)
The type $\textsf{Nat}(\ahat)$ has value-determined index $\ahat$
(its $\Xi$ is $\ahat:\kindnat$),
so it is solvable.
We  unroll (rule \AlgChkValFix) $\textsf{Nat}(\ahat)$ to
$\big(
  1 \land (\ahat=0)
\big)
+
\big(
  \extype{a':\kindnat}
      \textsf{Nat}(a')
      \times
      \big(
        1 \land (\ahat = 1 + a')
      \big)
\big)$
and check $\inj{2}{\pair{\roll{\inj{1}{\unit}}}{\unit}}$ against that
($0$ and $1+a'$ are the bodies of the two branches of \textsf{Nat}'s algebra).
In this unrolled type, $\ahat$ is no longer tracked by its $\Xi$,
but we can still solve it.

The value now in question is a right injection, so we must check
$\pair{\roll{\inj{1}{\unit}}}{\unit}$ against
$\extype{a':\kindnat} \textsf{Nat}(a') \times \big( 1 \land (\ahat=1+a')\big)$
(rule \AlgChkValIn{2}).
We generate another fresh existential variable $\ahat'$ in place of $a'$.
We now check the pair using rule \AlgChkValPair.
For the first component, we check $\inj{1}{\unit}$
against the unrolled $\textsf{Nat}(\ahat')$,
which is $1 \land (\ahat' = 0) + \cdots$.
Now we solve $\ahat' = 0$
(rules \AlgChkValIn{1}, \AlgChkValWith, \RuleInst, and \AlgChkValUnit).
This information flows to the type $1 \land (\ahat=1+\ahat')$
against which we need to check the second value component ($\unit$).
By ``this information flows,''
we mean that we apply the context output by type checking the first component,
namely $\ahat:\kindnat, \hypeq{\ahat'}{\kindnat}{0}$
(notice $\ahat$ is not yet solved),
as a substitution to obtain $1 \land (\ahat=1+0)$ for the second premise
of \AlgChkValPair.
The right-hand side of the equation now has no existential variables,
and we solve $\ahat = 1+0 = 1$ (again using \AlgChkValWith), as expected.
It is worth noting that this solving happens entirely within focusing stages.

Values of inductive type may involve program variables,
so existential variables may not be solved by \AlgChkValWith
(and \RuleInst),
but in algorithmic subtyping,
using the same instantiation judgment:
\[
  \Infer{\AlgSubPosWithR}
  {
    \algsub[+]{\Theta; \Delta}{P}{Q}{\Wah}{\Delta''}
    \\
    \alginst{\Theta; \Delta''}{[\Delta'']\phi}{\Delta'}
  }
  {
    \algsub[+]{\Theta; \Delta}{P}{Q \land \phi}{[\Delta']\Wah \land [\Delta']\phi}{\Delta'}
  }
\]
Finally,
if an equation of the form $\ahat = t$ makes its way into type equivalence
(by checking a variable value against a sum type),
then $\ahat$ gets solved, not as $t$,
but rather as the index in the same structural position
(including logical structure)
of the necessarily ground positive type on the left of the equivalence
(see judgment $\algpropequivinst{\Theta; \Delta}{\phi}{\psi}{\Delta'}$
in appendix \Figref{fig:propinst}, used in appendix \Figref{fig:algtpequiv}).
For example,
$\algchk{b:\kindnat; \ahat:\kindnat}{x:(1\land(b=0+0))+\extype{b':\kindnat}\mathsf{Nat}(b')\land(b=b'+0+1)}{x}{P}{\dontcare}{\hypeq{\ahat}{\kindnat}{b}}$
where
\hspace{3.9em}$P = (1\land(\ahat=0))\hspace{1.48em}+\extype{a':\kindnat}\mathsf{Nat}(a')\land(\ahat=a'+1)$.

Next, we cover the algorithmic value and spine typing rules
(\Figref{fig:valuespinealgtyping}) in detail.

\paragraph{Typechecking values}

Because there is no stand-alone algorithmic version of \DeclSubPosL
(recall that, in algorithmic subtyping, we eagerly extract immediately
under polarity shifts),
the rule \AlgChkValVar clarifies why we require types in contexts
to have already been subjected to extraction.
With eager extraction in subtyping under polarity shifts,
but without eager type extraction for program variables,
we would not be able to extract any assumptions from $Q$
in the (algorithmic) subtyping premise.

Rule \AlgChkValExists generates a fresh existential variable
which ultimately gets solved within the same stage.
Its solution is eagerly applied to the input type and output constraints,
so we pop it off of the output context (as it is no longer needed).

Rule \AlgChkValFix unrolls the inductive type,
checks the inductive type's injected value against the unrolled type,
and passes along the constraints and solutions.

Rule \AlgChkValWith delays verifying the validity of the conjoined proposition $\phi$
until it is grounded.
As explained in the example above,
existential variables can be solved via propositions generated by
algorithmic type unrolling.
This is the role of the propositional instantiation judgment used in the second premise:
it simply checks whether the proposition is of the form $\ahat = t$
where $t$ is ground, in which case it solves $\ahat$ as $t$ (rule \RuleInst),
and otherwise it does nothing (rule \RuleNoInst).
If the proposition does solve an existential variable,
then the $[\Delta']\phi$ part of the constraint is a trivial equation,
but $\phi$ could be a non-ground proposition unrelated to unrolling,
in which case $\Delta' = \Delta''$,
whose solutions have not yet been applied to the input $\phi$.

Rule \AlgChkValDownshift does not have a premise
for typechecking the thunked expression (unlike \DeclChkValDownshift).
Instead,
the rule delays this typechecking constraint until its existential variables are solved.
For example, in
\[
  \algchk
    {\cdot; \cdot}
    {\cdot}
    {\pair{\thunk{\Return{\unit}}}{\into{\inj{1}{\unit}}}}
    {\extype{a:\kindnat} (\downshift{\upshift{\big(1 \land (a=0)\big)}}) \times \textsf{Nat}(a)}
    {\chi}
    {\cdot}
\]
the output constraint $\chi$
has $[0/\ahat](\Return{\unit} <= \upshift{\big(1 \land (\ahat=0)\big)})$,
where the index solution $0$ to the $\ahat$ introduced by \AlgChkValExists
is found only in typechecking the second component of the pair.

Rule \AlgChkValUnit says that $\unit$ checks against $1$,
which solves nothing, and there are no further constraints to check.

In rule \AlgChkValPair, we check the first component $v_1$,
threading through solutions found there in checking the second component $v_2$.
Checking the second component can solve further existential variables
in the first component's constraints $\chi_1$,
but solutions are eagerly applied,
so in the conclusion we apply all the solutions only to $\chi_1$.

Rule \AlgChkValIn{k} checks the $k$-injected value $v_k$
against the sum's $k$th component type,
and passes along the constraints and solutions.

\paragraph{Typechecking spines}
Rule \AlgSpineAll, similarly to \AlgChkValExists,
generates a fresh existential variable that ultimately gets solved
in typechecking a value (in this case the spine's corresponding value).
As usual, we pop off the solution because solutions are eagerly applied.

In rule \AlgSpineApp, we typecheck the value $v$, outputting constraints $\chi$
and solutions $\Delta''$.
We thread these solutions through when checking $s$, the rest of the spine,
ultimately outputting constraints $\chi'$ and solutions $\Delta'$.
The context $\Delta'$ may have more solutions than $\Delta''$,
and we eagerly apply solutions, so we need only apply $\Delta'$ to
the first value's constraints $\chi$.

In \AlgSpineImplies, we check the spine $s$
but add the guarding proposition $\phi$
to the list of constraints to verify later
(applying the solutions $\Delta'$ found when checking the spine).

In \AlgSpineNil, nothing gets solved, so we output the input algorithmic context $\Delta$.
Nothing needs to be verified, so we output the trivial constraint $\True$.

\section{Algorithmic Metatheory}
\label{sec:algorithmic-metatheory}

We prove that the algorithmic system
(\Secref{sec:algorithmic-system})
is decidable,
as well as sound and complete with respect to the declarative system
(\Secref{sec:declarative-system}).

\subsection{Decidability}

We have proved that all algorithmic judgments are decidable
(appendix \Secref{sec:apx-decidability}).
Algorithmic constraint verification $\entailwah{\Theta}{W}$ and $\algneg{\Theta}{\Gamma}{\chi}$
boils down to verifying propositional validity $\judgeentail{\Theta}{\phi}$,
which is known to be decidable \cite{BarrettSMT}.
Besides that, our decidability proofs rely on fairly simple metrics for the various algorithmic judgments,
which involve a simple size function
(appendix Figs.~\ref{fig:size} and~\ref{fig:size-prog-chi})
and counting the number of subtyping constraints $W$ in typing constraint lists $\chi$.
We show that, for each algorithmic rule, every premise is smaller 
than the conclusion, according to the metrics.
The most interesting lemmas we use state that the constraints
output by algorithmic equivalence, subtyping and program typing
decrease in size
(appendix Lemmas~\ref{lem:alg-equiv-shrinks-problem},
\ref{lem:alg-sub-shrinks-problem}, and
\ref{lem:program-typing-shrinks-problems}).
For example:

\begin{lemma}[Program Typing Shrinks Constraints]
  \hfill (\Lemref{lem:program-typing-shrinks-problems} in appendix)
  \begin{enumerate}
  \item If $\algchk{\Theta; \Delta}{\Gamma}{v}{P}{\chi}{\Delta'}$,
    then $\size{\chi} \leq \size{v}$.
  \item If $\algspine{\Theta; \Delta}{\Gamma}{s}{N}{\upshift{P}}{\chi}{\Delta'}$,
    then $\size{\chi} \leq \size{s}$.
  \end{enumerate}
\end{lemma}

\subsection{Algorithmic Soundness}
\label{sec:algorithmic-soundness}

We show that the algorithmic system is sound with respect to the declarative system.
Since the algorithmic system is designed
to mimic the judgmental structure of the declarative system,
soundness (and completeness) of the algorithmic system is relatively straightforward
to prove
(the real difficulty lies in designing the overall system to make this the case).

Soundness of algorithmic subtyping says that,
if the subtyping algorithm solves indexes under which its verification conditions hold,
then subtyping holds declaratively under the same solutions:

\begin{theorem}[Soundness of Algorithmic Subtyping]
  \hfill (\Thmref{thm:alg-sub-sound} in appendix)
  \begin{enumerate}
  \item If $\algsub[+]{\Theta; \Delta}{P}{Q}{W}{\Delta'}$
    and $\judgetp{\Theta; \Delta}{Q}{\Xi}$
    and $\ground{P}$
    and $\extend{\Theta}{\Delta'}{\Omega}$
    and $\entailwah{\Theta}{[\Omega]W}$,
    then $\judgesub[+]{\Theta}{P}{[\Omega]Q}$.
  \item If $\algsub[-]{\Theta; \Delta}{N}{M}{W}{\Delta'}$
    and $\judgetp{\Theta; \Delta}{N}{\Xi}$
    and $\ground{M}$
    and $\extend{\Theta}{\Delta'}{\Omega}$
    and $\entailwah{\Theta}{[\Omega]W}$,
    then $\judgesub[-]{\Theta}{[\Omega]N}{M}$.
  \end{enumerate}
\end{theorem}

We prove
(appendix \Thmref{thm:alg-sub-sound})
the soundness of algorithmic subtyping
by way of two intermediate (sound) subtyping systems:
a declarative system that eagerly extracts under shifts,
and a semideclarative system that also eagerly extracts under shifts,
but outputs constraints $W$ in the same way as algorithmic subtyping,
to be checked by the semideclarative judgment $\semideclentailwah{\Theta}{W}$
(that we prove sound with respect to the algorithmic $\entailwah{\Theta}{W}$).
We define a straightforward subtyping constraint equivalence judgment
$\wahequiv{\Theta}{W}{W'}$,
that uses the proposition and type equivalence mentioned in \Secref{sec:equivalence},
to transport semideclarative to algorithmic subtyping constraint verification
(and the other way around for algorithmic subtyping completeness):
if $\semideclentailwah{\Theta}{W}$ and $\wahequiv{\Theta}{W}{W'}$,
then $\semideclentailwah{\Theta}{W'}$
(appendix \Lemref{lem:equiv-respects-entail}).

As a consequence of polarization,
the soundness of head, bound expression, expression, and match typing
can be stated relatively simply.
The typing soundness of values and spines
says that if $\Omega$ completes the algorithm's solutions
such that the algorithm's constraints hold,
then the typing of the value or spine holds declaratively with $\Omega$'s solutions.

\begin{theorem}[Alg.\ Typing Sound]
  \hfill (\Thmref{thm:alg-typing-soundness} in appendix)
  \begin{enumerate}
    \item
      If $\algsynhead{\Theta}{\Gamma}{h}{P}$,
      then $\judgesynhead{\Theta}{\Gamma}{h}{P}$.
    \item
      If $\algsynexp{\Theta}{\Gamma}{\be}{\upshift{P}}$,
      then $\judgesynexp{\Theta}{\Gamma}{\be}{\upshift{P}}$.
    \item
      If $\algchk{\Theta; \Delta}{\Gamma}{v}{P}{\chi}{\Delta'}$
      and $\judgetp{\Theta; \Delta}{P}{\Xi}$
      and $\extend{\Theta}{\Delta'}{\Omega}$
      and $\algneg{\Theta}{\Gamma}{[\Omega]\chi}$,\\
      then $\judgechkval{\Theta}{\Gamma}{[\Omega]v}{[\Omega]P}$.
    \item
      If $\algchkneg{\Theta}{\Gamma}{e}{N}$,
      then $\judgechkexp{\Theta}{\Gamma}{e}{N}$.
    \item
      If $\algchkmatch{\Theta}{\Gamma}{P}{\clauses{\pa}{e}{i}{I}}{N}$,
      then $\judgechkmatch{\Theta}{\Gamma}{P}{\clauses{\pa}{e}{i}{I}}{N}$.
    \item
      If $\algspine{\Theta; \Delta}{\Gamma}{s}{N}{\upshift{P}}{\chi}{\Delta'}$
      and $\judgetp{\Theta; \Delta}{N}{\Xi}$
      and $\extend{\Theta}{\Delta'}{\Omega}$
      and $\algneg{\Theta}{\Gamma}{[\Omega]\chi}$,\\
      then $\judgespine{\Theta}{\Gamma}{[\Omega]s}{[\Omega]N}{[\Omega]\upshift{P}}$.
  \end{enumerate}
\end{theorem}

We prove
(appendix \Thmref{thm:alg-typing-soundness})
algorithmic typing is sound by a straightforward
lexicographic induction on, first, program term structure,
and, second, input type size.
We happen not to use an intermediate system for this proof,
but an intermediate system is very helpful
if not indispensable for proving algorithmic typing completeness,
discussed next.

\subsection{Algorithmic Completeness}
\label{sec:completeness}

We show that the algorithmic system is complete with respect to the declarative system.
The declarative system can conjure index solutions that are different
from the algorithm's solutions,
but they must be equal according to the logical context.
We capture this with \emph{relaxed}
context extension $\rextend{\Theta}{\Delta}{\Delta'}$,
similar to (non-relaxed) context extension ($\extend{\Theta}{\Delta}{\Delta'}$)
but solutions in $\Delta$
may change to equal (under $\Theta$) solutions in $\Delta'$:
\[
  \Infer{}
  {\rextend{\Theta}{\Delta}{\Delta'} \\ \judgeentail{\Theta}{t=t'}}
  {\rextend{\Theta}{\Delta, \hypeq{\ahat}{\tau}{t}}{\Delta', \hypeq{\ahat}{\tau}{t'}}}
\]

Algorithmic completeness
says our subtyping algorithm verifies any declarative subtyping.
Since declarative subtyping does not eagerly extract
from types inside shifts in assumptive positions,
but algorithmic subtyping does,
the conclusion involves extraction from the given ground type.
For example,
the equivalence of the algorithmic solutions $\Delta'$ to the indexes in $\Omega$
for which subtyping declaratively holds
may depend on extracted assumptions like $\Theta_P$ in part (1) just below.

\begin{theorem}[Completeness of Algorithmic Subtyping]
  \hfill (\Thmref{thm:alg-sub-complete} in appendix)
  \begin{enumerate}
  \item If $\judgesub[+]{\Theta}{P}{[\Omega]Q}$
    and $\judgetp{\Theta; \Delta}{Q}{\Xi}$
    and $\ground{P}$
    and $[\Delta]Q = Q$
    and $\rextend{\Theta}{\Delta}{\Omega}$,\\
    then there exist $P'$, $\Theta_P$, $W$, and $\Delta'$
    such that $\algsub[+]{\Theta, \Theta_P; \Delta}{P'}{Q}{W}{\Delta'}$\\
    and $\rextend{\Theta, \Theta_P}{\Delta'}{\Omega}$
    and $\entailwah{\Theta, \Theta_P}{[\Omega]W}$
    and $\judgeextract[+]{\Theta}{P}{P'}{\Theta_P}$.
  \item If $\judgesub[-]{\Theta}{[\Omega]N}{M}$
    and $\judgetp{\Theta; \Delta}{N}{\Xi}$
    and $\ground{M}$
    and $[\Delta]N = N$
    and $\rextend{\Theta}{\Delta}{\Omega}$,\\
    then there exist $M'$, $\Theta_M$, $W$, and $\Delta'$
    such that $\algsub[-]{\Theta, \Theta_M; \Delta}{N}{M'}{W}{\Delta'}$\\
    and $\rextend{\Theta, \Theta_M}{\Delta'}{\Omega}$
    and $\entailwah{\Theta, \Theta_M}{[\Omega]W}$
    and $\judgeextract[-]{\Theta}{M}{M'}{\Theta_M}$.
  \end{enumerate}
\end{theorem}

We prove
(appendix \Thmref{thm:alg-sub-complete})
completeness of algorithmic subtyping
by using, in a similar way, the same intermediate systems used to prove soundness.
However, it's more complicated.
Indexes in semideclarative and algorithmic constraints
may be syntactically different but logically and semantically equal.
More crucially,
to prove the completeness of algorithmic typing with respect to semideclarative typing,
we need to prove that algorithmic subtyping solves all value-determined indexes
of input types that are not necessarily ground:

\begin{lemma}[Sub.\ Solves Val-det.]
  \hfill (\Lemref{lem:sub-solves-val-det} in appendix)
  \begin{enumerate}
  \item
    If $\algsub[+]{\Theta; \Delta}{P}{Q}{W}{\Delta'}$
    and $\ground{P}$
    and $\judgetp{\Theta; \Delta}{Q}{\Xi}$,\\
    then for all $(\ahat : \tau) \in \Xi$,
    there exists $t$ such that $\judgeterm{\Theta}{t}{\tau}$
    and $(\hypeq{\ahat}{\tau}{t}) \in \Delta'$.
  \item
    If $\algsub[-]{\Theta; \Delta}{M}{N}{W}{\Delta'}$
    and $\ground{N}$
    and $\judgetp{\Theta; \Delta}{M}{\Xi}$,\\
    then for all $(\ahat : \tau) \in \Xi$,
    there exists $t$ such that $\judgeterm{\Theta}{t}{\tau}$
    and $(\hypeq{\ahat}{\tau}{t}) \in \Delta'$.
  \end{enumerate}
\end{lemma}

We prove
(appendix \Lemref{lem:sub-solves-val-det})
this by straightforward induction on the given subtyping derivation,
using a similar lemma for type equivalence
(appendix \Lemref{lem:equiv-solves-val-det}).

We use extraction to achieve a complete subtyping algorithm.
For example, the following holds declaratively without extraction
but instead using \DeclSubPosWithL
(this rule is not in our system; see \Secref{sec:subtyping}):
\[
  \judgesub[-]{a:\kindnat, b:\kindnat}
  {\underbrace{[\hypeq{\hat{c}}{\kindnat}{b}]\big(\textsf{Nat}(\hat{c}) \to (\hat{c}=b)
      \implies \upshift{1}\big)}_{\textsf{Nat}(b)\;\to\;(b\;=\;b)\;\implies\;\upshift{\,1}}}
  {\big(\textsf{Nat}(a) \land (a=b)\big) \to \upshift{1}}
\]
However, checking function argument subtyping first,
the non-extractive algorithm solves $\hat{c}$ to $a$ (not $b$)
and outputs a verification condition needing $a=b$
to hold under no logical assumptions,
which is invalid.
Our system instead extracts $a=b$ from the supertype;
the algorithmic solution $a$ for $\hat{c}$
and the declarative choice $b$ for $\hat{c}$
are equal under this assumption ($a=b$).

Finally, we prove the completeness of algorithmic typing.
Like algorithmic typing soundness,
again due to focusing, the head, bound expression, expression, and pattern matching
parts are straightforward to state.
But,
because algorithmic function application may instantiate indexes
different but logically equal to those conjured (semi)declaratively,
bound expressions may algorithmically synthesize a type
(judgmentally) equivalent to the type it synthesizes declaratively.

We introduced logical context equivalence in \Secref{sec:equivalence}.
Other than in proving that type equivalence implies subtyping,
logical context equivalence  is used in proving the completeness of algorithmic typing
(in particular, we effectively use appendix
\Lemref{lem:ctx-equiv-compat}
to swap logically equivalent logical contexts in (semi)declarative typing derivations).
The type $P'$ in the output of the algorithm in part (6) below
can have different index solutions (output $\Delta'$)
that are logically equal (under $\Theta$) to the solutions in $\Omega$
which appear in the declaratively synthesized $P$.
However, $P$ and $P'$ necessarily have the same structure,
so $\judgeequiv[+]{\Theta}{P}{[\Omega]P'}$.
Therefore, a bound expression may (semi)declaratively synthesize
a type that is judgmentally equivalent to the type synthesized algorithmically.
We then extract different but logically equivalent logical contexts
from the (equivalent) types synthesized by a bound expression.

As such, algorithmic typing completeness is stated as follows:

\begin{theorem}[Alg.\ Typing Complete]
  \hfill (\Thmref{thm:alg-typing-complete} in appendix)
  \begin{enumerate}
    \item
      If $\judgesynhead{\Theta}{\Gamma}{h}{P}$,
      then $\algsynhead{\Theta}{\Gamma}{h}{P}$.
    \item
      If $\judgesynexp{\Theta}{\Gamma}{\be}{\upshift{P}}$,
      then there exists $P'$
      such that $\algsynexp{\Theta}{\Gamma}{\be}{\upshift{P'}}$
      and $\judgeequiv[+]{\Theta}{P}{P'}$.
    \item
      If $\judgechkval{\Theta}{\Gamma}{v}{[\Omega]P}$
      and $\judgetp{\Theta; \Delta}{P}{\Xi}$
      and $[\Delta]P = P$
      and $\rextend{\Theta}{\Delta}{\Omega}$,\\
      then there exist $\chi$ and $\Delta'$
      such that $\algchk{\Theta; \Delta}{\Gamma}{v}{P}{\chi}{\Delta'}$\\
      and $\rextend{\Theta}{\Delta'}{\Omega}$
      and $\algneg{\Theta}{\Gamma}{[\Omega]\chi}$.
    \item
      If $\judgechkexp{\Theta}{\Gamma}{e}{N}$,
      then $\algchkneg{\Theta}{\Gamma}{e}{N}$.
    \item
      If $\judgechkmatch{\Theta}{\Gamma}{P}{\clauses{\pa}{e}{i}{I}}{N}$,
      then $\algchkmatch{\Theta}{\Gamma}{P}{\clauses{\pa}{e}{i}{I}}{N}$.
    \item
      If $\judgespine{\Theta}{\Gamma}{s}{[\Omega]N}{\upshift{P}}$
      and $\judgetp{\Theta; \Delta}{N}{\Xi}$
      and $[\Delta]N = N$
      and $\rextend{\Theta}{\Delta}{\Omega}$,\\
      then there exist $P'$, $\chi$, and $\Delta'$
      such that $\algspine{\Theta; \Delta}{\Gamma}{s}{N}{\upshift{P'}}{\chi}{\Delta'}$\\
      and $\rextend{\Theta}{\Delta'}{\Omega}$
      and $\algneg{\Theta}{\Gamma}{[\Omega]\chi}$
      and $\judgeequiv[+]{\Theta}{P}{[\Omega]P'}$.
  \end{enumerate}
\end{theorem}

We prove
(appendix \Thmref{thm:alg-typing-complete})
algorithmic typing completeness by way of an intermediate,
semideclarative typing system,
that is essentially the same as declarative typing in that it conjures indexes,
but differs in a way similar to algorithmic typing:
it outputs constraints $\chi$ and only verifies them
(via semideclarative $\semideclneg{\Theta}{\Gamma}{\chi}$
as opposed to algorithmic $\algneg{\Theta}{\Gamma}{\chi}$)
immediately upon completion of focusing stages.
Similarly to the proof of algorithmic subtyping completeness,
we transport
(appendix \Lemref{lem:probs-equiv-respects-entail})
the semideclarative verification of typing constraints over a straightforward
typing constraint equivalence judgment $\chiequiv{\Theta}{\Gamma}{\chi}{\chi'}$
that uses the subtyping constraint equivalence ($\wahequiv{\Theta}{W}{W'}$)
and type equivalence judgments.

To prove that algorithmic typing is complete
with respect to semideclarative typing,
we use the fact that typing solves all value-determined indexes
in input types of focusing stages.
This fact is similar to the fact that subtyping solves the value-determined indexes
of non-ground types (used in the algorithmic subtyping completeness proof),
but the interaction between value-determined indexes
and unrolling introduces some complexity:
unrolling a refined inductive type does not preserve the type's $\Xi$.
Therefore, we had to split the value typechecking part
into the mutually recursive parts (1) and (2);
part (3) depends on parts (1) and (2) but not vice versa.

\begin{lemma}[Typing Solves Val-det.]
  \label{lem:liq-typing-solves-val-det}
  \hfill (\Lemref{lem:typing-solves-val-det} in appendix)
  \begin{enumerate}
  \item
    Suppose $\Delta = \Delta_1, \ahat:\tau, \Delta_2$.
    If $\judgeunroll{\Xi}{\Theta; \Delta}{\nu:G[\mu F]}{\beta}{G\; \Fold{F}{\alpha}\;\nu}{\ahat}{Q}{\tau}$\\
    and $\judgefunctor{\Theta; \Delta}{G}{\Xi_G}$
    and $(\ahat:\tau) \notin \Xi_G$\\
    and $\algchk{\Theta; \Delta}{\Gamma}{v}{Q}{\chi}{\Delta'}$\\
    and $[\Delta]Q = Q$
    and $\rextend{\Theta}{\Delta'}{\Omega}$
    and $\semideclneg{\Theta}{\Gamma}{[\Omega]\chi}$,\\
    then there exists $t$ such that $\judgeterm{\Theta}{t}{\tau}$
    and $(\hypeq{\ahat}{\tau}{t}) \in \Delta'$.
  \item
    If $\algchk{\Theta; \Delta}{\Gamma}{v}{P}{\chi}{\Delta'}$\\
    and $\judgetp{\Theta; \Delta}{P}{\Xi_P}$
    and $[\Delta]P = P$
    and $\rextend{\Theta}{\Delta'}{\Omega}$
    and $\semideclneg{\Theta}{\Gamma}{[\Omega]\chi}$,\\
    then for all $(\ahat : \tau) \in \Xi_P$,
    there exists $t$ such that $\judgeterm{\Theta}{t}{\tau}$
    and $(\hypeq{\ahat}{\tau}{t}) \in \Delta'$.
  \item
    If $\algspine{\Theta; \Delta}{\Gamma}{s}{N}{\upshift{P}}{\chi}{\Delta'}$\\
    and $\judgetp{\Theta; \Delta}{N}{\Xi_N}$
    and $[\Delta]N = N$
    and $\rextend{\Theta}{\Delta'}{\Omega}$
    and $\semideclneg{\Theta}{\Gamma}{[\Omega]\chi}$,\\
    then for all $(\ahat : \tau) \in \Xi_N$,
    there exists $t$ such that $\judgeterm{\Theta}{t}{\tau}$
    and $(\hypeq{\ahat}{\tau}{t}) \in \Delta'$.
  \end{enumerate}
\end{lemma}

The proof
(appendix \Lemref{lem:typing-solves-val-det})
of part (1) boils down to inversion on the propositional instantiation
judgment $\alginst{\Theta; \Delta}{\phi}{\Delta'}$
in the unit case of unrolling where $\phi$ necessarily has the form $\ahat = t$
with $\ground{t}$, due to the invariant that algebras are ground.

\section{Related Work}
\label{sec:related}

\paragraph{Typing refinement}

As far as we know,
\citet{Constable83} was first to introduce the concept of refinement types
(though not by that name)
as logical subsets of types,
writing \texttt{\{x:A|B\}} for the \emph{subset type}
classifying terms \texttt{x} of type \texttt{A}
that satisfy proposition \texttt{B}.
\citet{Freeman91} introduced type refinement to the programming language ML
via \emph{datasort} refinements---inclusion hierarchies of ML-style
(algebraic, inductive) datatypes---and intersection types for Standard ML:
they showed that full type inference is decidable
under a refinement restriction,
and provided an algorithm based on abstract interpretation.
The dangerous interaction of datasort refinements, intersection types, side effects,
and call-by-value evaluation was first dealt with
by \citet{Davies00icfpIntersectionEffects}
by way of a value restriction on intersection introduction;
they also presented a bidirectional typing algorithm.
Dependent ML (DML)
\cite{XiThesis, Xi99popl}
extended the ML type discipline parametrically by \emph{index} domains.
DML is only decidable modulo decidability of index constraint satisfiability.
DML uses
a bidirectional type system with index refinements for a variant of ML,
capable of checking properties ranging from
in-bound array access \cite{Xi98:ArrayBounds}
to program termination \cite{Xi02:termination}.
DML, similarly to \sysname,
collects constraints from the program and passes them to a constraint solver,
but does not guarantee that they are SMT solvable (unlike \sysname).
This is also the approach of systems like Stardust \cite{Dunfield07:Stardust} 
(which combines datasort and index refinement,
and supports index refinements that are not value-determined, that is, \emph{invaluable refinements},
which we do not consider)
and those with liquid types \cite{Rondon08:LT}.
The latter are based on a Hindley--Milner approach;
typically, Hindley--Damas--Milner inference algorithms
\cite{Hindley1969:Principal, Milner78, Damas82}
generate typing constraints to be verified \cite{Heeren02-generalizing-HM}.

Due to issues with existential index instantiation,
the approach of \citet{XiThesis} (incompletely) translated programs
into a let-normal form \cite{Sabry92-first-ANF} before typing them,
and \citet{DunfieldThesis}
provided a complete let-normal translation for similar issues.
The system in this paper is already let-normal.

\paragraph{Liquid types}

\citet{Rondon08:LT} introduced \emph{logically qualified data types},
that is, \emph{liquid types},
in a system that extends Hindley--Milner 
to infer (by abstract interpretation) refinements based on
built-in or programmer-provided refinement templates or qualifiers.
\citet{Kawaguchi09} introduced \emph{recursive refinement}
via sound and terminating \emph{measures} on algebraic data types;
they also introduced \emph{polymorphic refinement}.
\citet{Vazou13} generalize recursive and polymorphic refinement
into a single, more expressive mechanism: \emph{abstract refinement}.
Our system lacks polymorphism, which we plan to study in future work.
Abstract refinements go together with multi-argument measures because
abstract refinements may be thought of as predicates of higher-order sort
and we can encode multi-argument measures using higher-order sorts.
Extending \sysname with higher-order sorts and abstract refinements is ongoing work.
In future work, it would be interesting to study other features of liquid typing in our setting.
Extending \sysname with liquid inference of refinements, for example,
would require adding Hindley--Milner type inference that creates templates,
as well as mechanisms to solve these templates,
possibly in an initial phase using abstract interpretation.

Unlike DML (and \sysname), liquid type systems do not distinguish index terms from programs.
While this provides simplicity and convenience to the user
(from their perspective, there is just one language),
it makes it relatively difficult to provide liquid type systems a denotational semantics
and to prove soundness results denotationally (rather than operationally),
in contrast to \sysname
(we should be able to recover some of this convenience,
by requiring users, for example, to make measure annotations like Liquid Haskell).
It creates other subtleties such as the fact that annotations for termination metrics in Liquid Haskell
must be internally translated to ghost parameters similar to that used in \Secref{sec:example}.
By contrast, if we extend our system with additional termination metrics,
because these metrics are at the index level, we can obviate such ghost parameters.
Liquid types' lack of index distinction
also makes it trickier to support computational effects and evaluation orders.

Initial work on liquid types \cite{Rondon08:LT, Kawaguchi09} used call-by-value languages,
but Haskell uses lazy evaluation so Liquid Haskell was discovered to be unsound \cite{Vazou14}.
\citet{Vazou14} regained type soundness
by imposing operational-semantic restrictions on subtyping and let-binding.
In their algorithmic subtyping, there is exactly one rule,
$\preceq$-\textsc{Base}-\textsc{D},
which pertains to refinements of base types (integers, booleans and so on)
and inductive data types;
however, these types have a \emph{well-formedness restriction}, namely,
that the refinement predicates have the type of boolean expressions
that reduce to finite values.
But this restriction alone
does not suffice for soundness under laziness and divergence.
As such, their algorithmic typing rule
\textsc{T}-\textsc{Case}-\textsc{D},
which combines let-binding and pattern matching,
uses an operational semantics to approximate whether or not
the bound expression terminates.
If the bound expression might diverge, then so might the entire case expression;
otherwise, it checks each branch in a context
that assumes the expression reduces to a (potentially infinite Haskell) value.

We also have a type well-formedness restriction,
but it is purely syntactic, and only on index quantification,
requiring them to be associated with a fold
that is necessarily decidable
by virtue of a systematic phase distinction between the index level and the program level.
Further, via type polarization,
our let-binding rule requires the bound expression to return a value,
we only allow value types in our program contexts,
and we cannot extract index information across polarity shifts
(such as in a suspended computation).
Therefore, in \sysname,
there is no need to stratify our types
according to an approximate criterion;
rather, we exploit the systemic distinction between positive (value) types
and negative (computation) types,
that \citet{Levy04} designed or discovered to be semantically well-behaved.
We suspect that liquid types' divergence-based stratification
is indirectly grappling with logical polarity.
Because divergence-based stratification is peculiar to the specific effect
of nontermination, it is unclear how their approach
may extend to other effects.
By way of a standard embedding of CBN or CBV into (our focalized variant of) CBPV
we can obtain CBN or CBV subtyping and typing relations
automatically respecting any necessary value or covalue restrictions \cite{Zeilberger09}.
Further, \sysname is already in a good position to handle
the addition of effects other than nontermination.

\paragraph{Contract calculi}
Software \emph{contracts} express program properties
in the same language as the programs themselves;
\citet{Findler02:HigherOrderContracts}
introduced contracts for \emph{run-time}
verification of higher-order functional programs.
These \emph{latent} contracts are not types,
but \emph{manifest} contracts are
\cite{Greenberg10-manifest-contracts}.
Manifest contracts are akin to refinement types.
Indeed, \citet{Vazou13} sketch a proof of type soundness
for a liquid type system by translation from liquid types to the
manifest contract calculus F\textsubscript{H} of \citet{Belo11}.
However, there is no explicit translation back,
from F\textsubscript{H} to liquid typing.
They mention that the translated terms in F\textsubscript{H}
do not have \emph{upcasts} because the latter in F\textsubscript{H}
are logically related to identity functions
if they correspond to static subtyping (as they do in the liquid type system):
an upcast lemma.
Presumably,
this facilitates a translation from F\textsubscript{H} back to liquid types.
However, there are technical problems in F\textsubscript{H}
that break type soundness and the logical consistency of the F\textsubscript{H} contract system;
\citet{Sekiyama17} fix these issues,
resulting in the system $\text{F}^{\hspace{1pt}\sigma}_{\text{H}}$,
but do not consider subtyping and subsumption,
and do not prove an upcast lemma.

\paragraph{Bidirectional typing}

Bidirectional typing \cite{Pierce00}
is a popular way to implement a wide variety of systems,
including dependent types \cite{Coquand96:typechecking-dependent-types,Abel08:MPC},
contextual types \cite{Pientka08:POPL},
and object-oriented languages \cite{Odersky01:ColoredLocal}.
The bidirectional system of \citet{PeytonJones07} supports higher-rank polymorphism.
\citet{Dunfield13}
also present a bidirectional type system for higher-rank polymorphism,
but framed more proof theoretically;
\citet{Dunfield19} extend it to a richer language 
with existentials, indexed types, sums, products, equations over type variables,
pattern matching, polarized subtyping, and principality tracking.
The bidirectional system of this paper
uses logical techniques similar to the systems of Dunfield and Krishnaswami,
but it does not consider polymorphism.
A survey paper
\citep{Dunfield21-bidir-survey}
includes some discussion of bidirectional typing's connections to proof theory.
Basically, good bidirectional systems tend to distinguish checking and synthesizing
terms or proofs according to their form, such as normal or neutral.

\paragraph{Proof theory, polarization, focusing and analyticity}

The concept of polarity most prominent in this paper
dates back to Andreoli's work on focusing for tractable proof search
\cite{Andreoli92:focusing}
and Girard's work on unifying classical, intuitionistic, and linear logic
\cite{Girard93}.
Logical polarity and focusing have been used to explain many common phenomena
in programming languages.
We mentioned in the overview that
\citet{Zeilberger09} explains the value and evaluation context restrictions
in terms of focusing;
and \citet{Krishnaswami09:pattern-matching} explains pattern matching
as (proof terms of) the left-inversion stage of focused systems
(also, that system is bidirectional).
More broadly, \citet{Downen17-phd} discusses many logical dualities
common in programming languages.

\citet{Brock-Nannestad15}
study the relation between polarized intuitionistic logic and CBPV.
They obtain a bidirectionally typed system of natural deduction
related to a variant of the focused sequent calculus
\emph{LJF} \cite{Liang09-LJF}
by $\eta$-expansion (for inversion stages).
\citet{EspiritoSanto17} does a similar study,
but starts with a focused sequent calculus
for intuitionistic logic much like the system of \citet{Simmons12}
(but without positive products),
proves it equivalent to a natural deduction system
(we think the lack of positive products helps establish this equivalence),
and defines, also via $\eta$-expansion, a variant of CBPV
in terms of the natural deduction system.
\Sysname is not in the style of natural deduction, but rather sequent calculus.
We think \sysname relates to CBPV in a similar way---via $\eta$-expansion---but
we do not prove it in this paper,
because we focus on proving type soundness
and algorithmic decidability, soundness and completeness.

\citet{Barendregt83} discovered that a program that typechecks
(in a system with intersection types) using subtyping,
can also be checked without using subtyping,
if the program is sufficiently $\eta$-expanded.
An analogous phenomenon involving identity coercion
was studied by \citet{Zeilberger09} in a focused setting.
Similarly, our ability to place subtyping solely in (value) variable typechecking
is achievable due to the focusing (and let-normality)
of \sysname.

Interpreting Kant,
\citet{Martin-Lof94} considers an \emph{analytic} judgment
to be one that is derivable using information found only in its inputs
(in the sense of the bidirectional modes, input and output).
A \emph{synthetic} judgment, in contrast,
requires us to look beyond the inputs of the judgment in order to find a derivation.
The metatheoretic results for our algorithmic system
demonstrate that our judgments are analytic,
\emph{except} the judgment $\judgeentail{\Theta}{\phi}$,
which is verified by an external SMT solver.
As such, \sysname may be said to be analytic modulo an external SMT solver.
Focusing,
in proper combination with bidirectional typing
(which clarifies where to put type annotations),
let-normality
and value-determinedness,
guarantees that all information needed to generate
verification conditions suitable for an SMT solver
may be found in the inputs to judgments.
In our system,
cut formulas can always be inferred from a type annotation
or by looking up a variable in the program context,
making all our cuts analytic in the sense of \citet{Smullyan69}.

\paragraph{Dependent types}

Dependent types, introduced by \citet{Martin-Lof71,Martin-Lof75},
are
a key conceptual and historical precursor to index refinement types.
Dependent types may depend on arbitrary program terms,
not only terms restricted to indexes.
This is highly expressive, but undecidable in languages with divergence.
The main difference between refinement and dependent typing
is that refinement typing attempts to increase the expressivity
of a highly automatic type system,
whereas dependent typing attempts to increase the automation
of a highly expressive type system.
Semantically, refinement type systems differ from dependent type systems
in that they refine a pre-existing type system,
so that erasure of refinements always preserves typing.

Many dependent type systems impose their own restrictions
for the sake of decidability.
In Cayenne \cite{Augustsson98Cayenne},
typing can only proceed a given number of steps.
All well-typed programs in Epigram \cite{McBride04:ViewFromTheLeft}
are required to terminate so that its type equivalence is decidable.
Epigram, and other systems
\cite{Chen05:CombiningProgramsWithProofs, Licata05:ExplicitEqualityDML},
allow programmers to write explicit proofs of type equivalence.

Systems like ATS \cite{Xi03:ATS}
and $\text{F}^\star$ \cite{Swamy16}
can be thought of as combining refinement and dependent types.
These systems aim to bring the best of both refinement and dependent types,
but ATS is more geared to
practical, effectful functional programming (hence refinement types),
while $\text{F}^\star$ is more geared to formal verification and dependent types.
Unlike \sysname, they allow the programmer to provide proofs.
The overall design of ATS
is closer to \sysname than that of $\text{F}^\star$,
due to its phase distinction between statics and dynamics;
but it allows the programmer to write (in the language itself) proofs
in order to simplify or eliminate constraints for the (external) constraint solver:
Xi calls this \emph{internalized constraint solving}.
It should be possible to internalize constraint solving
to some extent in \sysname.
Liquid Haskell has a similar mechanism called \emph{refinement reflection} \cite{Vazou17-refinement-reflection}
in which programmers can write manual proofs (in Haskell) in cases where automatic
Proof by Logical Evaluation and SMT solving fail.

Both
ATS and $\text{F}^\star$
have a CBV semantics, which is inherently monadic
\cite{Moggi89:XXXComputationalLambda}.
\Sysname is a variant of CBPV, which subsumes both CBV and CBN.
These systems consider effects other than divergence,
like exceptions, mutable state and input/output,
which we hope to add to \sysname in future work;
this should go relatively smoothly because CBPV is inherently monadic.
The system $\text{F}^\star$ allows for termination metrics
other than strong induction on naturals,
such as lexicographic induction,
but we think it would be straightforward to add such metrics to \sysname,
in the way discussed in \Secref{sec:typing}.

\paragraph{Data abstraction and category theory}

Categorically, inductive types are initial algebras of endo\-functors.
We only consider certain \emph{polynomial} endofunctors,
which specify tree-shaped or algebraic data structures.
Objects (in the sense of object-oriented programming) or coinductive types are dual to inductive types
in that, categorically, they are final coalgebras of endofunctors \cite{Cook09}.
A consideration of categorical duality leads us to a natural (perhaps \naive) question:
if we can build a well-behaved system
that refines algebraic data types by algebras,
could it mean anything to refine objects by coalgebras?
We would expect the most direct model of coinductive types would be via negative types,
but working out the details is potential future work.

Our rolled refinement types refine \emph{type} constructors $\mu F$.
\citet{Sekiyama15},
again in work on manifest contracts,
compare this to refining (types of) \emph{data} constructors, and
provide a translation from type constructor to data constructor refinements.
According to \citet{Sekiyama15}
type constructor refinements
(such as our $\comprehend{\nu : \mu F}{\Fold{F}{\alpha} \; \nu =_\tau t}$)
are \emph{easier for the programmer to specify},
but data constructor refinements
(such as the output types of our unrolling judgment)
are \emph{easier to verify automatically}.
\citet{Sekiyama15} say that their translation from type to data constructor refinements
is closely related to the work of \citet{Atkey12}
on refining inductive data in (a fibrational interpretation of)
dependently typed languages.
\citet{Atkey12} provide ``explicit formulas'' computing
inductive characterizations of type constructor refinements.
These semantic formulas resemble our syntactic unrolling judgment,
which may be viewed as a translation from type refinements to data constructor refinements.

\emph{Ornaments} \cite{McBride11-ornaments}
describe how inductive types with different logical or ornamental properties
can be systematically related
using their algebraic and structural commonalities.
Practical work in ornaments seems mostly geared toward code reuse
\cite{Dagand12-ornaments-code-reuse},
code refactoring \cite{Williams17}
and such.
In contrast, this paper focuses on incorporating similar ideas
in a foundational index refinement typing algorithm.

\citet{Mellies15-functors}
provide a categorical theory of type refinement in general,
where functors are considered to be type refinement systems.
This framework is based on Reynolds's distinction
between \emph{intrinsic} (or Church)
and \emph{extrinsic} (or Curry) views of typing
\cite{Reynolds98-book}.
We think that \sysname fits into this framework,
but haven't confirmed it formally.
This is most readily seen in the fact that the semantics of our refined system
is simply the semantics
(\emph{intrinsic} to unrefined typing derivations)
of its erasure of indexes,
which express \emph{extrinsic} properties of (erased) programs.

\section{Conclusion and Future Work}
\label{sec:conclusion}

We have presented a declarative system for index-based recursive refinement typing
(with nullary measures)
that is logically designed,
semantically sound,
and theoretically implementable.
We have proved that our declarative system is sound
with respect to an elementary domain-theoretic denotational semantics,
which implies that \sysname is logically consistent
and totally correct.
We have also presented an algorithmic system
and proved it is decidable, as well as sound and complete
with respect to the declarative system.
Focusing yields CBPV, which already has a clear denotational semantics,
and refining it by an index domain (and by measures in it)
facilitates a semantics in line with the perspective of \citet{Mellies15-functors}.
But focusing (in combination with value-determinedness) also allows for an easy proof of the completeness
of a decidable typing algorithm.
The relative ease with which we demonstrate
(denotational-semantic) soundness and (completeness of) decidable typing for a realistic language
follows from a single, proof-theoretic technique: focusing.

Researchers of liquid typing have laid out an impressive and extensive research program
providing many useful features which would be very interesting to study in our setting.
We plan to add parametric polymorphism in future work,
which goes along with refinement abstraction \cite{Vazou13}.
Refinement abstraction may be thought of as predicates of higher-order sort,
which can also accommodate multi-argument measures
(such as whether a list of naturals is in increasing order).
We are adding multi-argument measures and refinement abstraction in ongoing work.
It is also of great interest to study other features of liquid typing,
like refinement inference with templates and refinement reflection,
though arguably the latter is more closely related to dependent typing.
We also plan to allow the use of multiple measures on inductive types
(so we can specify, for example,
the type of length-$n$ lists of naturals in increasing order).
It would also be interesting to experiment with our value-determinedness technique.
By allowing quantification over indexes in propositions
whose only variable dependencies are \emph{value-determined},
we think we can simulate termination metrics by other ones
(such as $<$ on \emph{sums} of natural numbers by $<$ on natural numbers as such).

In future work, we hope to apply our type refinement system
(or future extensions of it)
to various domains,
from static time complexity analysis \cite{Wang17-TiML}
to resource analysis \cite{Handley19}.
Eventually, we hope to be able to express, for example,
that a program terminates within a worst-case amount of time.
\Sysname is parametric in the index domain,
provided it satisfies some basic properties.
Different index domains may be suitable for different applications.
We also hope to add more effects,
such as input/output and mutable reference cells.
CBPV is built for effects, but our refinement layer may
result in interesting interactions between effects and indexes.

\Sysname may at first seem complicated,
but its metatheoretic proofs are largely straightforward,
if lengthy (at least as presented).
A primary source of this complexity is the proliferation of judgments.
However, having various judgments helps us organize
different forms of knowledge \cite{MartinLof96-judgment}
or (from a Curry--Howard perspective) stages or parts of an implementation
(such as pattern-matching, processing an argument list, and so on).

Our system focuses on the feature of nullary measures of algebraic data types,
and does not include key typing features expected of a realistic programming language
(such as additional effects and polymorphism, which we hope to add in future work).
Adding such expressive features tends to significantly affect the metatheory
and the techniques used to prove it.
We hope to reflect on the development of our proofs
(including those for systems with polymorphism \cite{Dunfield13})
in search of abstractions which may help
designers of practical, general-purpose functional languages
to establish crucial metatheoretic properties.

\begin{acks}
  We thank the anonymous reviewers for their thorough reading and recommendations,
  which helped to improve our paper.
  We also thank Ondrej \citet{Baranovic23} for implementing the system presented in this paper.
  This work was supported in part by the
  \grantsponsor{NSERC}{Natural Sciences and Engineering Research Council of Canada}{http://www.nserc-crsng.gc.ca/index_eng.asp}
  through Discovery Grant \grantnum{NSERC}{RGPIN--2018--04352},
  and also in part by European Research Council (ERC) Consolidator Grant
  for the project “TypeFoundry”, funded under the European Union’s Horizon 2020 Framework
  Programme (grant agreement no. 101002277).
\end{acks}

{%
\bibliography{local}
}

{\small\tableofcontents}
\clearpage

\appendix
\def\OPTIONAppendix{1}
\section{Definitions and Figures}

\subsection{Syntax}

\begin{figure}[htbp]

  \begin{grammar}
    Program variables  & $x, y, z$
    \\
    Expressions  & 
    $e$ & \bnfas &
    $\Return{v}
    \bnfalt  \Let{x}{\be}{e}
    \bnfalt  \match{h}{\clauses{\pa}{e}{i}{I}}
    \bnfalt  \fun{x}{e}
    \bnfaltBRK \rec{x:N}{e}
    \bnfalt  \unreachable
    $
    \\
    Values  & 
    $v$ & \bnfas &
    $x
    \bnfalt  \unit
    \bnfalt  \pair{v}{v}
    \bnfalt  \inl{v}
    \bnfalt  \inr{v}
    \bnfalt  \roll{v}
    \bnfalt  \thunk{e}
    $
    \\
    Heads  & 
    $h$ & \bnfas &
    $x
    \bnfalt \annoexp{v}{P}
    $
    \\
    Bound expressions  & 
    $\be$ & \bnfas &
    $h(s)
    \bnfalt \annoexp{e}{\upshift{P}}
    $
    \\
    Spines  & 
    $s$ & \bnfas &
    $
    \cdot 
    \bnfalt v, s
    $
    \\
    Patterns  & 
    $\pa$ & \bnfas &
    $\roll{x}
    \bnfalt  \unit
    \bnfalt  \pair{x}{y}
    \bnfalt  \inl{x}
    \bnfalt  \inr{x} 
    $
    \\[1em]
    Positive types  &
    $P,Q$ & \bnfas & $1 \bnfalt P \times Q \bnfalt 0 \bnfalt P + Q \bnfalt{\downshift{N}}
    \bnfaltBRK
    \comprehend{\nu : \mu F}{\Fold{F}{\alpha}\,{\nu} =_\tau t}
    \bnfaltBRK
    \extype{a:\tau}P
    \bnfalt
    P \land \phi
    $
    \\
    Negative types  & 
    $N,M$ & \bnfas
    & $P \to N
    \bnfalt \upshift{P}
    \bnfalt \alltype{a:\tau}{N}
    \bnfalt \phi \implies N
    $
    \\
    Types  & 
    $A,B,C$ & \bnfas
    & $P \bnfalt N$
    \\[1em]
    Functors  &
    $F, G, H$ & \bnfas &
    $\Ptype
    \bnfalt F \oplus F
    $
    \\
    & $\Ptype$ & \bnfas &
    $I
    \bnfalt B \otimes \Ptype
    $
    \\
    & $\Btype$ & \bnfas
    & $\Const{P}
    \bnfalt \Id
    $
    \\
    & $\mathcal{F}$ & \bnfas & $F \bnfalt B$ 
    \\[2em]
    Index variables & $a, b, c$
    \\
    Index terms  & 
    $t$ & \bnfas &
    $a
    \bnfalt n
    \bnfalt t + t
    \bnfalt t - t
    \bnfalt (t,t)
    \bnfalt \fst{t}
    \bnfalt \snd{t}
    \bnfalt \phi
    $
    \\
    Propositions  & 
    $\phi, \psi$ & \bnfas &
    $t = t
    \bnfalt  t \leq t
    \bnfalt  \phi \land \phi
    \bnfalt  \phi \lor \phi
    \bnfalt  \lnot \phi
    \bnfalt  \True
    \bnfalt  \False
    $
    \\[1em]
    Algebras  & 
    $\alpha, \beta$ & \bnfas
    & $
    \cdot
    \bnfalt
    (\clause{p}{t} \matchor \alpha)
    $
    \\
    Sum algebra patterns & 
    $p$ & \bnfas &
    $\inl{p}
    \bnfalt  \inr{p}
    \bnfalt  q
    $
    \\
    Product algebra patterns  & 
    $q$ & \bnfas & $\unitexp
    \bnfalt  (\bap, q)
    $
    \\
    Base algebra patterns  & 
    $\bap$ & \bnfas & $\wild
    \bnfalt  a
    \bnfalt  \pack{a}{\bap}$
    \\[1em]
    Sorts  & 
    $\tau$ & \bnfas
    & $\Booltype
    \bnfalt \kindnat
    \bnfalt \Z
    \bnfalt \tau \times \tau
    $
    \\[2em]
    Variable contexts  & 
    $\Gamma$ & \bnfas & $\cdot
    \bnfalt  \Gamma, x:P$
    \\
    Logical contexts  & 
    $\Theta$ & \bnfas & $\cdot
    \bnfalt  \Theta, a:\tau
    \bnfalt  \Theta, \phi$
    \\
    Value-determined contexts  & 
    $\Xi$ & \bnfas & $\cdot
    \bnfalt  \Xi, a:\tau
    $
    \\[2em]
    Algorithmic contexts  & 
    $\Delta$ & \bnfas & $\cdot
    \bnfalt  \Delta, \ahat : \tau
    \bnfalt  \Delta, \hypeq{\ahat}{\tau}{t}
    $
    \\
    Typing constraints
    &
    $\chi$ & \bnfas & $\cdot
    \bnfalt  (e <= N), \chi
    \bnfalt  W, \chi
    $
    \\
    Subtyping constraints
    &
    $\Wah$ & \bnfas & $
    \phi
    \bnfalt  \propeqprob{\phi}{\psi}
    \bnfalt  \phi \implies \Wah
    \bnfalt  \Wah \land \Wah
    \bnfalt  \alltype{a:\tau} \Wah
    \bnfaltBRK  \possubprob{P}{Q}
    \bnfalt  \negsubprob{N}{M}
    \bnfalt  \poseqprob{P}{Q}
    \bnfalt  \negeqprob{N}{M}
    $
  \end{grammar}

  \caption{Syntax}
  \label{fig:grammar}
\end{figure}

\clearpage

\subsection{Judgments and Their Presuppositions}
In \Figref{fig:judgments-presuppositions},
let ``pre.'' abbreviate ``presupposes''.
\begin{figure}[hptb]
  \resizebox{14cm}{!}{%
  \begin{tabular}{LlcL}
    \judgctx{\Theta} & (\Figref{fig:declwf-logical-ctx}) & pre. & \text{no judgment} \\
    \Theta |- t : \tau & (\Figref{fig:ix-sort}) & pre. & \judgctx{\Theta} \\
    \composeinj{k}{\alpha}{\alpha_k} & (\Figref{fig:auxpat}) & pre. & \text{no judgment} \\
    \judgetp{\Theta}{A}{\Xi} & (\Figref{fig:declwf-type}) & pre. & \judgctx{\Theta} \\
    \judgefunctor{\Theta}{\mathcal{F}}{\Xi} & (\Figref{fig:declwf-fun-alg}) & pre. & \judgctx{\Theta} \\
    \judgealgebra{\Xi}{\Theta}{\alpha}{F}{\tau} & (\Figref{fig:declwf-fun-alg}) & pre. & \judgctx{\Xi} \text{ and } \judgefunctor{\Theta}{F}{\Xi_F} \\
    \judgectx{\Theta}{\Gamma} & (\Figref{fig:declwf-program-ctx}) & pre. & \judgctx{\Theta} \\
    \Theta_0; \Gamma_0 |- \sigma : \Theta; \Gamma & (\Figref{fig:syn-subs}) & pre. & \judgectx{\Theta_0}{\Gamma_0} \text{ and } \judgectx{\Theta}{\Gamma} \\
    \judgeentail{\Theta}{\phi} & (\Figref{fig:prop-truth}) & pre. & \judgeterm{\Theta}{\phi}{\Booltype} \\
    \judgeequiv[]{\Theta}{\phi}{\psi} & (\Figref{fig:declpropequiv}) & pre. & \judgeterm{\Theta}{\phi}{\Booltype} \text{ and } \judgeterm{\Theta}{\psi}{\Booltype} \\
    \judgeequiv[]{\Theta}{\Theta_1}{\Theta_2} & (\Figref{fig:decllogctxequiv}) & pre. & \judgctx{\Theta, \Theta_1} \text{ and } \judgctx{\Theta, \Theta_2} \\
    \judgeequiv[]{\Theta}{\mathcal{F}}{\mathcal{G}} & (\Figref{fig:declfunequiv}) & pre. & \judgefunctor{\Theta}{\mathcal{F}}{\Xi_{\mathcal{F}}} \text{ and } \judgefunctor{\Theta}{\mathcal{G}}{\Xi_{\mathcal{G}}} \\
    \judgeequiv[\pm]{\Theta}{A}{B} & (\Figref{fig:decltpequiv}) & pre. & \judgetp{\Theta}{A}{\Xi_A} \text{ and } \judgetp{\Theta}{B}{\Xi_B} \\
    \judgeextract[\pm]{\Theta}{A}{A'}{\Theta_A} & (\Figref{fig:declextract}) & pre. & \judgetp{\Theta}{A}{\Xi_{A}} \\
    \judgesub[\pm]{\Theta}{A}{B} & (\Figref{fig:declsub}) & pre. & \judgetp{\Theta}{A}{\Xi_A} \text{ and } \judgetp{\Theta}{B}{\Xi_B} \\
    \unroll{\Xi}{\Theta}{G}{F}{\beta}{\alpha}{\tau}{t}{P} & (\Figref{fig:unroll}) & pre. & \judgealgebra{\Xi}{\Theta}{\alpha}{F}{\tau} \text{ \dots} \\
    ~ & ~ & ~ & \text{\dots and } \judgealgebra{\Xi}{\Theta}{\beta}{G}{\tau} \text{ and } \judgeterm{\Theta}{t}{\tau} \\
    \judgesynhead{\Theta}{\Gamma}{h}{P} & (\Figref{fig:decltyping-head-bound-expression}) & pre. & \judgectx{\Theta}{\Gamma} \\
    \judgesynexp{\Theta}{\Gamma}{\be}{\upshift{P}} & (\Figref{fig:decltyping-head-bound-expression}) & pre. & \judgectx{\Theta}{\Gamma} \\
    \judgechkval{\Theta}{\Gamma}{v}{P} & (\Figref{fig:decltyping-value-expression}) & pre. & \judgectx{\Theta}{\Gamma} \text{ and } \judgetp{\Theta}{P}{\Xi_P} \\
    \judgechkexp{\Theta}{\Gamma}{e}{N} & (\Figref{fig:decltyping-value-expression}) & pre. & \judgectx{\Theta}{\Gamma} \text{ and } \judgetp{\Theta}{N}{\Xi_N} \\
    \judgechkmatch{\Theta}{\Gamma}{P}{\clauses{\pa}{e}{i}{I}}{N} & (\Figref{fig:declmatch}) & pre. & \judgectx{\Theta}{\Gamma} \text{ and } \judgetp{\Theta}{P}{\Xi_P} \text{ and } \judgetp{\Theta}{N}{\Xi_N} \\
    \judgespine{\Theta}{\Gamma}{s}{N}{\upshift{P}} & (\Figref{fig:declmatch}) & pre. & \judgectx{\Theta}{\Gamma} \text{ and } \judgetp{\Theta}{N}{\Xi_N} \\
    |- \delta : \Theta; \Gamma & (\Figref{fig:sem-subs}) & pre. & \judgectx{\Theta}{\Gamma} \\
    \judgectx{\Theta}{\Delta} & (\Figref{fig:algwf-alg-ctx}) & pre. & \judgctx{\Theta} \\
    \Theta; \Delta |- t : \tau & (\Figref{fig:alg-ix-sort}) & pre. & \judgectx{\Theta}{\Delta} \\
    \judgectx{\Theta; \Delta}{\Thetahat} & (\Figref{fig:algwf-alg-log-ctx}) & pre. & \judgectx{\Theta}{\Delta} \\
    \judgetp{\Theta; \Delta}{A}{\Xi} & (\Figref{fig:algwf-type}) & pre. & \judgectx{\Theta}{\Delta} \\
    \judgefunctor{\Theta; \Delta}{\mathcal{F}}{\Xi} & (\Figref{fig:algwf-fun-alg}) & pre. & \judgectx{\Theta}{\Delta} \\
    \judgealgebra{\Xi}{\Theta; \Delta}{\alpha}{F}{\tau} & (\Figref{fig:algwf-fun-alg}) & pre. & \judgctx{\Xi} \text{ and } \judgefunctor{\Theta; \Delta}{F}{\Xi_F} \\
    \alginst{\Theta; \Delta}{\phi}{\Delta'} & (\Figref{fig:propinst}) & pre. & \judgeterm{\Theta; \Delta}{\phi}{\Booltype} \\
    \algpropequivinst{\Theta; \Delta}{\phi}{\psi}{\Delta'} & (\Figref{fig:propinst}) & pre. & \judgeterm{\Theta; \Delta}{\phi}{\Booltype} \text{ and } \judgeterm{\Theta; \Delta}{\psi}{\Booltype} \\
    \algequiv[]{\Theta; \Delta}{\mathcal{F}}{\mathcal{G}}{W}{\Delta'} & (\Figref{fig:algfunequiv}) & pre. & \judgefunctor{\Theta; \Delta}{\mathcal{F}}{\Xi_{\mathcal{F}}} \text{ and } \judgefunctor{\Theta; \Delta}{\mathcal{G}}{\Xi_{\mathcal{G}}} \\
    \algequiv[+]{\Theta; \Delta}{P}{Q}{W}{\Delta'} & (\Figref{fig:algtpequiv}) & pre. & \judgetp{\Theta}{P}{\Xi_P} \text{ and } \judgetp{\Theta; \Delta}{Q}{\Xi_Q} \\
    \algequiv[-]{\Theta; \Delta}{N}{M}{W}{\Delta'} & (\Figref{fig:algtpequiv}) & pre. & \judgetp{\Theta; \Delta}{N}{\Xi_N} \text{ and } \judgetp{\Theta}{M}{\Xi_M} \\
    \judgeextract[\pm]{\Theta; \Delta}{A}{A'}{\Theta_A} & (\Figref{fig:algextract}) & pre. & \judgetp{\Theta; \Delta}{A}{\Xi_{A}} \\
    \algsub[+]{\Theta; \Delta}{P}{Q}{W}{\Delta'} & (\Figref{fig:algsubtyping}) & pre. & \judgetp{\Theta}{P}{\Xi_P} \text{ and } \judgetp{\Theta; \Delta}{Q}{\Xi_Q} \\
    \algsub[-]{\Theta; \Delta}{N}{M}{W}{\Delta'} & (\Figref{fig:algsubtyping}) & pre. & \judgetp{\Theta; \Delta}{N}{\Xi_N} \text{ and } \judgetp{\Theta}{M}{\Xi_M} \\
    \unroll{\Xi}{\Theta; \Delta}{G}{F}{\beta}{\alpha}{\tau}{t}{P} & (\Figref{fig:alg-unroll-convenient}) & pre. & \judgealgebra{\Xi}{\Theta; \Delta}{\alpha}{F}{\tau} \text{ \dots} \\
    ~ & ~ & ~ & \text{\dots and } \judgealgebra{\Xi}{\Theta; \Delta}{\beta}{G}{\tau} \text{ and } \judgeterm{\Theta; \Delta}{t}{\tau} \\
    \text{(algorithmic typing judgments similar to declarative typing judgments but with ``$\Theta; \Delta$'' for ``$\Theta$'')\hspace{-80em}} & ~ & ~ & ~ \\
    \entailwah{\Theta}{\Wah} & (\Figref{fig:entail-wah}) & pre. & \judgctx{\Theta} \\
    \algneg{\Theta}{\Gamma}{\chi} & (\Figref{fig:entail-wah}) & pre. & \judgectx{\Theta}{\Gamma} \\
    \extend{\Theta}{\Delta}{\Delta'} & (\Figref{fig:algextension}) & pre. & \judgectx{\Theta}{\Delta} \text{ and } \judgectx{\Theta}{\Delta'} \\
    \rextend{\Theta}{\Delta}{\Delta'} & (\Figref{fig:relaxedalgextension}) & pre. & \judgectx{\Theta}{\Delta} \text{ and } \judgectx{\Theta}{\Delta'} \\
    \text{(intermediate judgments in \Secref{sec:intermediate-figs} similar to those of their declarative versions)\hspace{-80em}} & ~ & ~ & ~ \\
  \end{tabular}
  } %

  \caption{Judgments and their (judgmental) presuppositions}
  \label{fig:judgments-presuppositions}
\end{figure}

\clearpage

\subsection{Declarative System}

\begin{figure}[htbp]
  \judgbox{\judgctx{\Theta}}{Logical context $\Theta$ is well-formed}

  \begin{mathpar}
    \Infer{\LogCtxEmpty}
    {}
    {
      \judgctx{\cdot}
    }
    \and
    \Infer{\LogCtxVar}
    {
      \judgctx{\Theta}
      \\
      a \notin \dom{\Theta}
    }
    {
      \judgctx{(\Theta, a:\tau)}
    }
    \and
    \Infer{\LogCtxProp}
    {
      \judgctx{\Theta}
      \\
      \judgeterm{\Theta}{\phi}{\Booltype}
    }
    {
      \judgctx{(\Theta, \phi)}
    }
  \end{mathpar}

  \caption{Declarative logical context well-formedness}
  \label{fig:declwf-logical-ctx}
\end{figure}

\begin{figure}[htbp]
\raggedright

For each base sort $\tau$,
we define the set $\sortconsts{\tau}$ of constant terms of sort $\tau$:
\begin{align*}
  \sortconsts{~} &: \{\Booltype, \kindnat, \Z\} \to \Set \\
  \sortconsts{\Booltype} &= \{\True, \False\} \\
  \sortconsts{\kindnat} &= \{0, 1, 2, \dots\} \\
  \sortconsts{\Z} &= \{\dots, -2, -1, 0, 1, 2, \dots\}
\end{align*}

\judgbox{
  \Theta |- t : \tau
}
{
  Under $\Theta$, index $t$ has sort $\tau$
}
\begin{mathpar}
  \Infer{\IxVar}
    {(a:\tau)\in\Theta}
    {\Theta |- a : \tau}
  \and
  \Infer{\IxConst}
    {t \in \sortconsts{\tau}}
    {\Theta |- t : \tau}
  \and
  \Infer{\IxAdd}
    {
      \tau \in \{\kindnat, \Z\} \\
      \Theta |- t_1 : \tau \\
      \Theta |- t_2 : \tau
    }
    {\Theta |- t_1 + t_2 : \tau}
  \and
  \Infer{\IxSubtract}
    {
      \tau \in \{\kindnat, \Z\} \\
      \Theta |- t_1 : \tau \\
      \Theta |- t_2 : \tau
    }
    {\Theta |- t_1 - t_2 : \tau}
  \and
  \Infer{\IxProd}
    {
      \Theta |- t_1 : \tau_1 \\
      \Theta |- t_2 : \tau_2
    }
    {\Theta |- (t_1, t_2) : \tau_1 \times \tau_2}
  \and
  \Infer{\IxProj{k}}
    { \Theta |- t : \tau_1 \times \tau_2 }
    {\Theta |- \pi_k\;t : \tau_k}
  \and
  \Infer{\IxAnd}
  {
    \judgeterm{\Theta}{\phi_1}{\Booltype}
    \\
    \judgeterm{\Theta}{\phi_2}{\Booltype}
  }
  {
    \judgeterm{\Theta}{\phi_1 \land \phi_2}{\Booltype}
  }
  \and
  \Infer{\IxOr}
  {
    \judgeterm{\Theta}{\phi_1}{\Booltype}
    \\
    \judgeterm{\Theta}{\phi_2}{\Booltype}
  }
  {
    \judgeterm{\Theta}{\phi_1 \lor \phi_2}{\Booltype}
  }
  \and
  \Infer{\IxNot}
  {
    \judgeterm{\Theta}{\phi}{\Booltype}
  }
  {
    \judgeterm{\Theta}{\lnot\phi}{\Booltype}
  }
  \and
  \Infer{\IxEq}
  {
    \Theta |- t_1 : \tau
    \\
    \Theta |- t_2 : \tau
  }
  {
    \judgeterm{\Theta}{t_1 = t_2}{\Booltype}
  }
  \and
  \Infer{\IxLeq}
  {
    \tau \in \{\kindnat, \Z\}
    \\
    \Theta |- t_1 : \tau
    \\
    \Theta |- t_2 : \tau
  }
  {
    \judgeterm{\Theta}{t_1 \leq t_2}{\Booltype}
  }
\end{mathpar}

  \caption{Declarative index sorting}
  \label{fig:ix-sort}
\end{figure}

\begin{figure}[htbp]
\raggedright

\judgbox{
  \composeinj{1}{\alpha}{\alpha_1} \\
  \composeinj{2}{\alpha}{\alpha_2} 
}{
  \vspace{0.5em}\\
  The left pattern of algebra $\alpha$ is $\alpha_1$ \\
  The right pattern of algebra $\alpha$ is $\alpha_2$
}
\begin{mathpar}
  \Infer{\DeclPatNil{1}}
        { }
        { \composeinj{1}{\cdot}{\cdot} }
  \and
  \Infer{\DeclPatThere{1}}
        { \composeinj{1}{\alpha}{\beta} }
        { \composeinj{1}{(\clause{\inr{p}}{t} \matchor \alpha)}{\beta} }
  \and
  \Infer{\DeclPatHere{1}}
        { \composeinj{1}{\alpha}{\beta} }
        { \composeinj{1}{(\clause{\inl{p}}{t} \matchor \alpha)}
          {(\clause{p}{t} \matchor \beta)} }
  \\
  \Infer{\DeclPatNil{2}}
        { }
        { \composeinj{2}{\cdot}{\cdot} }
  \and
  \Infer{\DeclPatThere{2}}
        { \composeinj{2}{\alpha}{\beta} }
        { \composeinj{2}{(\clause{\inl{p}}{t} \matchor \alpha)}{\beta} }
  \and
  \Infer{\DeclPatHere{2}}
        { \composeinj{2}{\alpha}{\beta} }
        { \composeinj{2}{(\clause{\inr{p}}{t} \matchor \alpha)}
          {(\clause{p}{t} \matchor \beta)} }
\end{mathpar}
  
  \caption{Algebra pattern matching}
  \label{fig:auxpat}
\end{figure}

\begin{figure}[htbp]
\raggedright

\judgbox{
    \judgetp{\Theta}{A}{\Xi_A}
  }{
    Under $\Theta$, type $A$ is well-formed, with value-determined indexes $\Xi_A$
}
\begin{mathpar}
  \Infer{\DeclTpVoid}
        { }
        { \judgetp{\Theta}{0}{\cdot}  }
  \and
  \Infer{\DeclTpUnit}
        { }
        { \judgetp{\Theta}{1}{\cdot}  }
  \\
  \Infer{\DeclTpSum}
        { \judgetp{\Theta}{P_1}{\Xi_1} \\ \judgetp{\Theta}{P_2}{\Xi_2} }
        { \judgetp{\Theta}{P_1 + P_2}{\Xi_1 \sect \Xi_2}  }
  \and
  \Infer{\DeclTpProd}
        { \judgetp{\Theta}{P_1}{\Xi_1} \\ \judgetp{\Theta}{P_2}{\Xi_2} }
        { \judgetp{\Theta}{P_1 \times P_2}{\Xi_1 \union \Xi_2}  }
  \and
  \Infer{\DeclTpEx}
        { \judgetp{\Theta, a:\tau}{P}{\Xi, a:\tau} }
        { \judgetp{\Theta}{\extype{a:\tau}{P}}{\Xi} }
  \and
  \Infer{\DeclTpDown}
        { \judgetp{\Theta}{N}{\Xi} }
        { \judgetp{\Theta}{\downshift{N}}{\cdot} }
  \and
  \Infer{\DeclTpWith}
        {
          \judgetp{\Theta}{P}{\Xi}
          \\
          \judgeterm{\Theta}{\phi}{\Booltype}
        }
        { \judgetp{\Theta}{P \land \phi}{\Xi} }
  \\
  \Infer{\DeclTpFixVar}
        {
          \judgefunctor{\Theta}{F}{\Xi}
          \\
          \judgealgebra{\cdot}{\Theta}{\alpha}{F}{\tau}
          \\
          (b : \tau) \in \Theta
        }
        {
          \judgetp{\Theta}{\comprehend{\nu:\mu F}{\Fold{F}{\alpha}\,{\nu} =_\tau b}}{\Xi \union b:\tau}
        }
  \\
  \Infer{\DeclTpFix}
        {
          \judgefunctor{\Theta}{F}{\Xi}
          \\
          \judgealgebra{\cdot}{\Theta}{\alpha}{F}{\tau}
          \\
          \Theta |- t : \tau
          \\
          t \text{ not a variable}
        }
        {
          \judgetp{\Theta}{\comprehend{\nu:\mu F}{\Fold{F}{\alpha}\,{\nu} =_\tau t}}{\Xi}
        }
  \\
  \Infer{\DeclTpAll}
        { \judgetp{\Theta, a:\tau}{N}{\Xi, a:\tau} }
        { \judgetp{\Theta}{\alltype{a:\tau}{N}}{\Xi} }
  \and
  \Infer{\DeclTpImplies}
        {
          \judgetp{\Theta}{N}{\Xi}
          \\
          \judgeterm{\Theta}{\phi}{\Booltype}
        }
        { \judgetp{\Theta}{\phi \implies N}{\Xi} }
  \and
  \Infer{\DeclTpArrow}
        { \judgetp{\Theta}{P}{\Xi_1} \\ \judgetp{\Theta}{N}{\Xi_2} }
        { \judgetp{\Theta}{P \to N}{\Xi_1 \union \Xi_2} }
  \and
  \Infer{\DeclTpUp}
        { \judgetp{\Theta}{P}{\Xi} }
        { \judgetp{\Theta}{\upshift{P}}{\cdot} }
\end{mathpar}

  \centering
  
  \caption{Declarative type well-formedness}
  \label{fig:declwf-type}
\end{figure}

\begin{figure}[htbp]
\judgbox{\judgefunctor{\Theta}{\mathcal{F}}{\Xi}}
      {
        Under $\Theta$, functor $\mathcal{F}$ is well-formed,
        with value-determined indexes $\Xi$
      }
\begin{mathpar}
  \Infer{\DeclFunctorConst}
        { \judgetp{\Theta}{P}{\Xi} } 
        { \judgefunctor{\Theta}{\Const{P}}{\Xi} } 
  \and
  \Infer{\DeclFunctorId}
        {  } 
        { \judgefunctor{\Theta}{\Id}{\cdot} } 
  \and
  \Infer{\DeclFunctorUnit}
        {  } 
        { \judgefunctor{\Theta}{I}{\cdot} } 
  \and
  \Infer{\DeclFunctorProd}
        { \judgefunctor{\Theta}{B}{\Xi_1} \\ \judgefunctor{\Theta}{\hat{P}}{\Xi_2} } 
        { \judgefunctor{\Theta}{B \otimes \hat{P}}{\Xi_1 \union \Xi_2} } 
  \and
  \Infer{\DeclFunctorSum}
        { \judgefunctor{\Theta}{F_1}{\Xi_1} \\ \judgefunctor{\Theta}{F_2}{\Xi_2} } 
        { \judgefunctor{\Theta}{F_1 \oplus F_2}{\Xi_1 \sect \Xi_2} } 
\end{mathpar}

\judgbox{\judgealgebra{\Xi}{\Theta}{\alpha}{F}{\tau}}
      {
           Under $\Xi \subseteq \Theta$, we know $\alpha$ is a well-formed algebra
           of sort $F(\tau) \Rightarrow \tau$,\\
           where $F$ is a well-formed functor under $\Theta$
      }
\begin{mathpar}
  \Infer{\DeclAlgSum}
  {
    \arrayenvb{
      \composeinj{1}{\alpha}{\alpha_1}
      \\
      \composeinj{2}{\alpha}{\alpha_2}
    }
    \\
    \arrayenvb{
      \judgealgebra{\Xi}{\Theta}{\alpha_1}{F_1}{\tau}
      \\
      \judgealgebra{\Xi}{\Theta}{\alpha_2}{F_2}{\tau}
    }
  }
  {
    \judgealgebra{\Xi}{\Theta}{\alpha}{(F_1 \oplus F_2)}{\tau}
  }
  \\
  \Infer{\DeclAlgUnit}
        { \judgeterm{\Xi}{t}{\tau} }
        { \judgealgebra{\Xi}{\Theta}{\clause{\unitexp}{t}}{I}{\tau} }
  \and
  \Infer{\DeclAlgIdProd}
        { \judgealgebra{\Xi, a:\tau}{\Theta, a:\tau}{\clause{q}{t}}{\hat{P}}{\tau} }
        { \judgealgebra{\Xi}{\Theta}{\clause{(a, q)}{t}}{(\Id \otimes \hat{P})}{\tau} }
  \and
  \Infer{\DeclAlgConstProd}
        { \judgealgebra{\Xi}{\Theta}{\clause{q}{t}}{\hat{P}}{\tau} \\ \judgetp{\Theta}{Q}{\dontcare}}
        { \judgealgebra{\Xi}{\Theta}{\clause{(\wild, q)}{t}}{(\Const{Q} \otimes \hat{P})}{\tau} }
  \and
  \Infer{\DeclAlgExConstProd}
        { \judgealgebra{\Xi, a:\tau'}{\Theta, a:\tau'}{\clause{(\bap, q)}{t}}{(\Const{Q} \otimes \hat{P})}{\tau} }
        { \judgealgebra{\Xi}{\Theta}
          {\clause{(\pack{a}{\bap}, q)}{t}}
          {(\Const{\extype{a:\tau'}{Q}} \otimes \hat{P})}{\tau} }
\end{mathpar}
  
  \caption{Declarative well-formedness of functors and algebras}
  \label{fig:declwf-fun-alg}
\end{figure}

\begin{figure}[htbp]
  \judgbox{\judgectx{\Theta}{\Gamma}}{Under logical context $\Theta$, program context $\Gamma$ is well-formed}

  \begin{mathpar}
    \Infer{\ProgCtxEmpty}
    {}
    {
      \judgectx{\Theta}{\cdot}
    }
    \and
    \Infer{\ProgCtxVar}
    {
      \judgectx{\Theta}{\Gamma}
      \\
      \judgetp{\Theta}{P}{\Xi}
      \\
      \judgeextract[+]{\Theta}{P}{P}{\cdot}
      \\
      x \notin \dom{\Gamma}
    }
    {
      \judgectx{\Theta}{\Gamma, x:P}
    }
  \end{mathpar}

  \caption{Program context well-formedness}
  \label{fig:declwf-program-ctx}
\end{figure}

\begin{figure}[htbp]

  \begin{grammar}
    Syntactic substitutions &
    $\sigma$ & \bnfas & $\cdot \bnfalt \sigma, t/a \bnfalt \sigma, \subs{v}{P}{x}$
  \end{grammar}

  \judgbox{\filterprog{\sigma} = \sigma}{Filter out program variable entries}
  \begin{align*}
    \filterprog{\cdot} &= \cdot \\
    \filterprog{\sigma, t/a} &= \filterprog{\sigma}, t/a \\
    \filterprog{\sigma, \subs{v}{P}{x}} &= \filterprog{\sigma}
  \end{align*}
  
  \judgbox{
    \Theta_0; \Gamma_0 |- \sigma : \Theta; \Gamma
  }
  {
    Under $\Theta_0$ and $\Gamma_0$, we know $\sigma$ is a syntactic substitution
    for variables in $\Theta$ and $\Gamma$
  }
  \begin{mathpar}
    \Infer{\EmptySyn}
    {
    }
    {
      \Theta_0; \Gamma_0 |- \cdot : \cdot; \cdot
    }
    \and
    \Infer{\IxSyn}
    {
      \Theta_0; \Gamma_0 |- \sigma : \Theta; \Gamma
      \and
      \Theta_0 |- [\filterprog{\sigma}]t : \tau
      \and
      a \notin \dom{\Theta}
    }
    {
      \Theta_0; \Gamma_0 |- (\sigma, t/a) : \Theta, a:\tau; \Gamma
    }
    \and
    \Infer{\PropSyn}
    {
      \Theta_0; \Gamma_0 |- \sigma : \Theta; \Gamma
      \and
      \judgeentail{\Theta_0}{[\filterprog{\sigma}]\phi}
    }
    {
      \Theta_0; \Gamma_0 |- \sigma : \Theta, \phi; \Gamma
    }
    \and
    \Infer{\ValSyn}
    {
      \Theta_0; \Gamma_0 |- \sigma : \Theta; \Gamma
      \and
      \judgechkval{\Theta_0}{\Gamma_0}{[\sigma]v}{[\filterprog{\sigma}]P}
      \and
      x \notin \dom{\Gamma}
    }
    {
      \Theta_0; \Gamma_0 |- \big(\sigma, (\subs{v}{P}{x}) \big) : \Theta; \Gamma, x:P
    }
  \end{mathpar}

  \caption{Syntactic substitution}
  \label{fig:syn-subs}
\end{figure}

\begin{figure}[htbp]
  \judgbox{
    \judgeentail{\Theta}{\phi}
  }
  {
    Under $\Theta$, proposition $\phi$ is true
  }
  \begin{mathpar}
    \Infer{\PropTrue}
    {
      \text{for all }\delta\text{, if }|- \delta : \Theta\text{ then }
      \sem{\delta}{\phi} = \{\bullet\}
    }
    {\judgeentail{\Theta}{\phi}}
  \end{mathpar}

  \caption{Propositional validity or truth}
  \label{fig:prop-truth}
\end{figure}

\begin{figure}[thbp]
\judgbox{\judgeequiv[]{\Theta}{\phi}{\psi}}
        {Under $\Theta$, propositions $\phi$ and $\psi$ are equivalent}
\begin{mathpar}
  \Infer{\PrpEquivT}
  {}
  {
    \judgeequiv[]{\Theta}{\True}{\True}
  }
  \and
  \Infer{\PrpEquivF}
  {}
  {
    \judgeequiv[]{\Theta}{\False}{\False}
  }
  \and
  \Infer{\PrpEquivAnd}
  {
    \judgeequiv[]{\Theta}{\phi_1}{\psi_1}
    \\
    \judgeequiv[]{\Theta}{\phi_2}{\psi_2}
  }
  {
    \judgeequiv[]{\Theta}{\phi_1 \land \phi_2}{\psi_1 \land \psi_2}
  }
  \and
  \Infer{\PrpEquivOr}
  {
    \judgeequiv[]{\Theta}{\phi_1}{\psi_1}
    \\
    \judgeequiv[]{\Theta}{\phi_2}{\psi_2}
  }
  {
    \judgeequiv[]{\Theta}{\phi_1 \lor \phi_2}{\psi_1 \lor \psi_2}
  }
  \and
  \Infer{\PrpEquivNot}
  {
    \judgeequiv[]{\Theta}{\phi}{\psi}
  }
  {
    \judgeequiv[]{\Theta}{\lnot\phi}{\lnot\psi}
  }
  \and
  \Infer{\PrpEquivEq}
  {
    \judgeentail{\Theta}{t_1 = t_1'}
    \\
    \judgeentail{\Theta}{t_2 = t_2'}
  }
  {
    \judgeequiv[]{\Theta}{(t_1 = t_2)}{(t_1' = t_2')}
  }
  \and
  \Infer{\PrpEquivLeq}
  {
    \judgeentail{\Theta}{t_1 = t_1'}
    \\
    \judgeentail{\Theta}{t_2 = t_2'}
  }
  {
    \judgeequiv[]{\Theta}{(t_1 \le t_2)}{(t_1' \le t_2')}
  }
\end{mathpar}
\caption{Declarative proposition equivalence}
\label{fig:declpropequiv}
\end{figure}

\begin{figure}[htbp]
  \judgbox{
    \judgeequiv{\Theta}{\Theta_1}{\Theta_2}
  }
  {
    Under $\Theta$, logical contexts $\Theta_1$ and $\Theta_2$ are equivalent
  }
  \begin{mathpar}
    \Infer{\CtxEquivEmpty}
    {}
    {
      \judgeequiv{\Theta}{\cdot}{\cdot}
    }
    \and
    \Infer{\CtxEquivVar}
    {
      \judgeequiv{\Theta}{\Theta_1}{\Theta_2}
    }
    {
      \judgeequiv{\Theta}{(\Theta_1, a:\tau)}{(\Theta_2, a:\tau)}
    }
    \and
    \Infer{\CtxEquivProp}
    {
      \judgeequiv{\Theta}{\Theta_1}{\Theta_2}
      \\
      \judgeequiv{\Theta, \overline{\Theta_1}}{\phi_1}{\phi_2}
    }
    {
      \judgeequiv{\Theta}{(\Theta_1, \phi_1)}{(\Theta_2, \phi_2)}
    }
  \end{mathpar}

  \caption{Declarative logical context equivalence}
  \label{fig:decllogctxequiv}
\end{figure}

\begin{figure}[htbp]
\judgbox{\judgeequiv[]{\Theta}{\mathcal{F}}{\mathcal{G}}}
        {Under $\Theta$, functors $\mathcal{F}$ and $\mathcal{G}$ are equivalent}
\begin{mathpar}
  \Infer{\FunEquivConst}
  {
    \judgeequiv[+]{\Theta}{P}{Q}
  }
  {
    \judgeequiv[]{\Theta}{\Const{P}}{\Const{Q}}
  }
  \and 
  \Infer{\FunEquivId}
  {}
  {
    \judgeequiv[]{\Theta}{\Id}{\Id}
  }
  \and 
  \Infer{\FunEquivUnit}
  {}
  {
    \judgeequiv[]{\Theta}{I}{I}
  }
  \and 
  \Infer{\FunEquivProd}
  {
    \judgeequiv[]{\Theta}{B}{B'}
    \\
    \judgeequiv[]{\Theta}{\hat{P}}{\hat{P}'}
  }
  {
    \judgeequiv[]{\Theta}{(B \otimes \hat{P})}{(B' \otimes \hat{P}')}
  }
  \and 
  \Infer{\FunEquivSum}
  {
    \judgeequiv[]{\Theta}{F_1}{G_1}
    \\
    \judgeequiv[]{\Theta}{F_2}{G_2}
  }
  {
    \judgeequiv[]{\Theta}{(F_1 \oplus F_2)}{(G_1 \oplus G_2)}
  }
\end{mathpar}
\caption{Declarative functor equivalence}
\label{fig:declfunequiv}
\end{figure}

\begin{figure}[htbp]
\judgbox{\judgeequiv[\pm]{\Theta}{A}{B}}{Under $\Theta$, types $A$ and $B$ are equivalent}
\begin{mathpar}
  \Infer{\TpEquivPosUnit}
  { }
  {
    \judgeequiv[+]{\Theta}{1}{1}
  }
  \and
  \Infer{\TpEquivPosVoid}
  { }
  {
    \judgeequiv[+]{\Theta}{0}{0}
  }
  \\
  \Infer{\TpEquivPosProd}
  {
    \judgeequiv[+]{\Theta}{P_1}{Q_1}
    \\
    \judgeequiv[+]{\Theta}{P_2}{Q_2}
  }
  {
    \judgeequiv[+]{\Theta}{P_1 \times P_2}{Q_1 \times Q_2}
  }
  \and
  \Infer{\TpEquivPosSum}
  {
    \judgeequiv[+]{\Theta}{P_1}{Q_1}
    \\
    \judgeequiv[+]{\Theta}{P_2}{Q_2}
  }
  {
    \judgeequiv[+]{\Theta}{P_1 + P_2}{Q_1 + Q_2}
  }
  \\
  \Infer{\TpEquivPosWith}
  {
    \judgeequiv[+]{\Theta}{P}{Q}
    \\
    \judgeequiv[]{\Theta}{\phi}{\psi}
  }
  {
    \judgeequiv[+]{\Theta}{P \land \phi}{Q \land \psi}
  }
  \and
  \Infer{\TpEquivPosEx}
  {
    \judgeequiv[+]{\Theta, a:\tau}{P}{Q}
  }
  {
    \judgeequiv[+]{\Theta}{\extype{a:\tau}{P}}{\extype{a:\tau}{Q}}
  }
  \and
  \Infer{\TpEquivPosFix}
  {
    \judgeequiv[]{\Theta}{F}{G}
    \\
    \judgeentail{\Theta}{t = t'}
  }
  {
    \judgeequiv[+]{\Theta}{\comprehend{\nu:\mu F}{\Fold{F}{\alpha}\,\nu =_\tau t}}{\comprehend{\nu:\mu G}{\Fold{G}{\alpha}\,\nu =_\tau t'}}
  }
  \\
  \Infer{\TpEquivPosDownshift}
  {
    \judgeequiv[-]{\Theta}{N}{M}
  }
  {
    \judgeequiv[+]{\Theta}{\downshift{N}}{\downshift{M}}
  }
  \and
  \Infer{\TpEquivNegUpshift}
  {
    \judgeequiv[+]{\Theta}{P}{Q}
  }
  {
    \judgeequiv[-]{\Theta}{\upshift{P}}{\upshift{Q}}
  }
  \\
  \Infer{\TpEquivNegImp}
  {
    \judgeequiv[-]{\Theta}{N}{M}
    \\
    \judgeequiv[]{\Theta}{\psi}{\phi}
  }
  {
    \judgeequiv[-]{\Theta}{\psi \implies N}{\phi \implies M}
  }
  \and
  \Infer{\TpEquivNegAll}
  {
    \judgeequiv[-]{\Theta, a:\tau}{N}{M}
  }
  {
    \judgeequiv[-]{\Theta}{\alltype{a:\tau}{N}}{\alltype{a:\tau}{M}}
  }
  \and
  \Infer{\TpEquivNegArrow}
  {
    \judgeequiv[+]{\Theta}{P}{Q}
    \\
    \judgeequiv[-]{\Theta}{N}{M}
  }
  {
    \judgeequiv[-]{\Theta}{P \to N}{Q \to M}
  }  
\end{mathpar}
\caption{Declarative type equivalence}
\label{fig:decltpequiv}
\end{figure}

\clearpage  %

\subsection{Unrefined System and Its Denotational Semantics}
\label{sec:apx-unrefined-system}

\begin{figure}[htbp]
  \begin{grammar}
    Program variables  & $x, y, z$
    \\
    Expressions  & 
    $e$ & \bnfas &
    $\Return{v}
    \bnfalt  \Let{x}{\be}{e}
    \bnfalt  \match{h}{\clauses{\pa}{e}{i}{I}}
    \bnfalt  \fun{x}{e}
    \bnfaltBRK \rec{x}{e}
    \bnfalt  \diverge
    $
    \\
    Values  & 
    $v$ & \bnfas &
    $x
    \bnfalt  \unit
    \bnfalt  \pair{v}{v}
    \bnfalt  \inl{v}
    \bnfalt  \inr{v}
    \bnfalt  \roll{v}
    \bnfalt  \thunk{e}
    $
    \\
    Heads  & 
    $h$ & \bnfas &
    $x
    \bnfalt \annoexp{v}{P}
    $
    \\
    Bound expressions  & 
    $\be$ & \bnfas &
    $h(s)
    \bnfalt \annoexp{e}{\upshift{P}}
    $
    \\
    Spines  & 
    $s$ & \bnfas &
    $
    \cdot 
    \bnfalt v, s
    $
    \\
    Patterns  & 
    $\pa$ & \bnfas &
    $\roll{x}
    \bnfalt  \unit
    \bnfalt  \pair{x}{y}
    \bnfalt  \inl{x}
    \bnfalt  \inr{x} 
    $
    \\[1em]
    Unrefined positive types  &
    $\unref{P},\unref{Q}$ & \bnfas & $\unref{1} \bnfalt \unref{P} \times \unref{Q} \bnfalt \unref{0} \bnfalt \unref{P} + \unref{Q} \bnfalt \downshift{\unref{N}} \bnfalt \mu \unref{F}
    $
    \\
    Unrefined negative types  & 
    $\unref{N},\unref{M}$ & \bnfas
    & $\unref{P} \to \unref{N}
    \bnfalt \upshift{\unref{P}}
    $
    \\
    Types  & 
    $\unref{A},\unref{B},\unref{C}$ & \bnfas
    & $\unref{P} \bnfalt \unref{N}$
    \\[1em]
    Unrefined functors  &
    $\unref{F}, \unref{G}, \unref{H}$ & \bnfas &
    $\unref{\Ptype}
    \bnfalt \unref{F} \oplus \unref{F}
    $
    \\
    & $\unref{\Ptype}$ & \bnfas &
    $\unref{I}
    \bnfalt \unref{B} \otimes \unref{\Ptype}
    $
    \\
    & $\unref{\Btype}$ & \bnfas
    & $\Const{\unref{P}}
    \bnfalt \unref{\Id}
    $
    \\
    & $\unref{\mathcal{F}}$ & \bnfas & $\unref{F} \bnfalt \unref{B}$ 
  \end{grammar}

  \caption{Syntax of Unrefined System}
  \label{fig:unrefined-types-functors}
\end{figure}

\begin{figure}[htbp]
  \raggedright
  \judgbox{\unrefunroll{G}{F}{P}}{}
  \begin{mathpar}
    \Infer{\UnrefUnrollSum}
    {\unrefunroll{G}{F}{P} \\ \unrefunroll{H}{F}{Q}}
    {\unrefunroll{(G \oplus H)}{F}{P + Q}}
    \and
    \Infer{\UnrefUnrollUnit}
    {}
    {\unrefunroll{I}{F}{1}}
    \and
    \Infer{\UnrefUnrollId}
    {\unrefunroll{\hat{P}}{F}{P}}
    {\unrefunroll{(\Id \otimes \hat{P})}{F}{\mu F \times P}}
    \and
    \Infer{\UnrefUnrollConst}
    {\unrefunroll{\hat{P}}{F}{P}}
    {\unrefunroll{(\Const{Q} \otimes \hat{P})}{F}{Q \times P}}
  \end{mathpar}

  \caption{Unrefined unrolling}
  \label{fig:unref-unroll}
\end{figure}

\begin{figure}[htbp]
\raggedright
\judgbox{\unrefsynhead{\Gamma}{h}{P}}
        {Under $\Gamma$, head $h$ synthesizes type $P$}
\begin{mathpar}
  \Infer{\UnrefSynHeadVar}
      {
        (x : P) \in \Gamma
      }
      {
        \unrefsynhead{\Gamma}{x}{P}
      }
   \and
  \Infer{\UnrefSynValAnnot}
      {
        \unrefchkval{\Gamma}{v}{P}
      }
      {
        \unrefsynhead{\Gamma}{\annoexp{v}{P}}{P}
      }
\end{mathpar}

\judgbox{\unrefsynexp{\Gamma}{\be}{\upshift{P}}}
        {Under $\Gamma$, bound expression $\be$ synthesizes type $\upshift{P}$}
\begin{mathpar}
  \Infer{\UnrefSynSpineApp}
        { \unrefsynhead{\Gamma}{h}{\downshift{N}} \\
          \unrefspine{\Gamma}{s}{N}{\upshift{P}} }
        { \unrefsynexp{\Gamma}{h(s)}{\upshift{P}} }
  \and
  \Infer{\UnrefSynExpAnnot}
        {
          \unrefchkexp{\Gamma}{e}{\upshift{P}}
        }
        { \unrefsynexp{\Gamma}{\annoexp{e}{\upshift{P}}}{\upshift{P}} }
\end{mathpar}

\judgbox{\unrefchkval{\Gamma}{v}{P}}
        {Under $\Gamma$, value $v$ checks against type $P$}
\begin{mathpar}
  \Infer{\UnrefChkValVar}
        { (x:P) \in \Gamma}
        { \unrefchkval{\Gamma}{x}{P} }
  \and
  \Infer{\UnrefChkValUnit}
        { }
        { \unrefchkval{\Gamma}{\unit}{1} }
  \and
  \Infer{\UnrefChkValPair}
        { \unrefchkval{\Gamma}{v_1}{P_1} \\
          \unrefchkval{\Gamma}{v_2}{P_2} } 
        { \unrefchkval{\Gamma}{\pair{v_1}{v_2}}{P_1 \times P_2} }
  \and
  \Infer{\UnrefChkValIn{k}}
        { \unrefchkval{\Gamma}{v}{P_k} }
        { \unrefchkval{\Gamma}{\inj{k}{v}}{P_1 + P_2} }
  \and 
  \Infer{\UnrefChkValFix}
        { \unrefunroll{F}{F}{P}
          \\ 
          \unrefchkval{\Gamma}{v}{P} }
        { \unrefchkval{\Gamma}{\roll{v}}{\mu F} }
  \and
  \Infer{\UnrefChkValDownshift}
        { \unrefchkexp{\Gamma}{e}{N} }
        { \unrefchkval{\Gamma}{\thunk{e}}{\downshift{N}} }
  \end{mathpar}
  \judgbox{\unrefchkexp{\Gamma}{e}{N}}
      {Under $\Gamma$, expression $e$ checks against type $N$}
  \begin{mathpar}
  \Infer{\UnrefChkExpUpshift}
        { \unrefchkval{\Gamma}{v}{P} }
        { \unrefchkexp{\Gamma}{\Return{v}}{\upshift{P}} }
  \and 
  \Infer{\UnrefChkExpLet}
        { \unrefsynexp{\Gamma}{\be}{\upshift{P}} \\
          \unrefchkexp{\Gamma, x:P}{e}{N} }
        { \unrefchkexp{\Gamma}{\Let{x}{\be}{e}}{N} }
  \and
  \Infer{\UnrefChkExpMatch}
        {
          \unrefsynhead{\Gamma}{h}{P}
          \\
          \unrefchkmatch{\Gamma}{P}{\clauses{\pa}{e}{i}{I}}{N}
        }
        {
          \unrefchkexp{\Gamma}{\match{h}{\clauses{\pa}{e}{i}{I}}}{N}
        }
  \and
  \Infer{\UnrefChkExpLam}
        { \unrefchkexp{\Gamma, x:P}{e}{N} }
        { \unrefchkexp{\Gamma}{\fun{x}{e}}{P \to N} }
  \and
  \Infer{\UnrefChkExpDiverge}
  {
  }
  {
    \unrefchkexp{\Gamma}{\diverge}{N}
  }
  \and
  \Infer{\UnrefChkExpRec}
        {
          \unrefchkexp{\Gamma, x:\downshift{N}}{e}{N}
        }
        {
          \unrefchkexp{\Gamma}{\rec{x}{e}}{N}
        }
\end{mathpar}
  
  \caption{Unrefined declarative typing}
  \label{fig:unreftyping}
\end{figure}

\begin{figure}[htbp]
\raggedright
\judgbox{\unrefchkmatch{\Gamma}{P}{\clauses{\pa}{e}{i}{I}}{N}}
        {Under $\Gamma$, patterns $r_i$ match against (input) type $P$ \\
          and branch expressions $e_i$ check against type $N$}
\begin{mathpar}
  \Infer{\UnrefChkMatchUnit}
        { \unrefchkexp{\Gamma}{e}{N} }
        { \unrefchkmatch{\Gamma}{1}{\setof{\clause{\unit}{e}}}{N} }
  \and
  \Infer{\UnrefChkMatchPair}
        { \unrefchkexp{\Gamma, x_1:P_1, x_2:P_2}{e}{N} }
        {
          \unrefchkmatch{\Gamma}{P_1 \times P_2}
          {\setof{\clause{\pair{x_1}{x_2}}{e}}}{N}
        }
  \and
  \Infer{\UnrefChkMatchSum}
        { \unrefchkexp{\Gamma, x_1:P_1}{e_1}{N} \\
          \unrefchkexp{\Gamma, x_2:P_2}{e_2}{N} }
        { \unrefchkmatch{\Gamma}{P_1 + P_2}{\setof{ \clause{\inl{x_1}}{e_1} \bnfalt \clause{\inr{x_2}}{e_2}}}{N} }
  \and
  \Infer{\UnrefChkMatchVoid}
        { }
        { \unrefchkmatch{\Gamma}{0}{\setof{}}{N} }
  \and
  \Infer{\UnrefChkMatchFix}
        {
          \unrefunroll{F}{F}{P}
          \\
          \unrefchkexp{\Gamma, x:P}{e}{N}
        }
        {
          \unrefchkmatch{\Gamma}{\mu F}{\setof{\clause{\roll{x}}{e}}}{N}
        }
\end{mathpar}

\judgbox{\unrefspine{\Gamma}{s}{N}{\upshift{P}}}
{Under $\Gamma$,
  if spine $s$ is applied to a head of type $\downshift{N}$, \\
  then it will return a result of type $\upshift{P}$}
\begin{mathpar}
  \Infer{\UnrefSpineApp}
        {\unrefchkval{\Gamma}{v}{Q} \\ \unrefspine{\Gamma}{s}{N}{\upshift{P}} }
        {\unrefspine{\Gamma}{v,s}{Q \to N}{\upshift{P}}}
  \and 
  \Infer{\UnrefSpineNil}
        { }
        {\unrefspine{\Gamma}{\cdot}{\upshift{P}}{\upshift{P}}}
\end{mathpar}
  \centering
  
  \caption{Unrefined matching and spines}
  \label{fig:unrefmatch}
\end{figure}

\begin{figure}[htbp]

  \begin{grammar}
    Unrefined syntactic substitutions &
    $\sigma$ & \bnfas & $\cdot \bnfalt \sigma, \subs{v}{P}{x}$
  \end{grammar}

  \judgbox{
    \Gamma_0 |- \sigma : \Gamma
  }
  {
    Under $\Gamma_0$, we know $\sigma$ is a syntactic substitution
    for variables in $\Gamma$
  }
  \begin{mathpar}
    \Infer{\UnrefEmptySyn}
    {
    }
    {
      \Gamma_0 |- \cdot : \cdot
    }
    \and
    \Infer{\UnrefValSyn}
    {
      \Gamma_0 |- \sigma : \Gamma
      \and
      \unrefchkval{\Gamma_0}{[\sigma]v}{P}
      \and
      x \notin \dom{\Gamma}
    }
    {
      \Gamma_0 |- \big(\sigma, (\subs{v}{P}{x})\big) : (\Gamma, x:P)
    }
  \end{mathpar}

  \caption{Unrefined syntactic substitution}
  \label{fig:unref-syn-subs}
\end{figure}

\begin{figure}[htbp]

  \begin{grammar}
    Semantic substitutions &
    $\delta$ & \bnfas & $\cdot \bnfalt \delta, V/x$
  \end{grammar}

  \judgbox{
    |- \delta : \Gamma
  }
  {
    We know that
    $\delta$ is a semantic substitution for variables in $\Gamma$
  }
  \begin{mathpar}
    \Infer{\UnrefEmptySem}
    {
    }
    {
      |- \cdot : \cdot
    }
    \and
    \Infer{\UnrefValSem}
    {
      |- \delta : \Gamma
      \and
      V \in \sem{}{P}
      \and
      x \notin \dom{\Gamma}
    }
    {
      |- (\delta, V/x) : (\Gamma, x:P)
    }
  \end{mathpar}

  \begin{align*}
    \sem{}{\Gamma} &= \comprehend{\delta}{|- \delta : \Gamma}\\
  \end{align*}

  \caption{Unrefined semantic substitution}
  \label{fig:unref-sem-subs}
\end{figure}

\begin{figure}[htbp]
  \begin{align*}
    \sem{}{\unref{P}} &: \Cpo \\
    \sem{}{\unref{1}} &= (\one, \ordsym[\one]) \\
    \sem{}{\unref{P}\times\unref{Q}} &= (\sem{}{\unref{P}} \times \sem{}{\unref{Q}}, \ordsym[\sem{}{\unref{P}} \times \sem{}{\unref{Q}}]) \\
       &\text{where }\ord[D_1 \times D_2]{(V_{11}, V_{12})}{(V_{21}, V_{22})}\text{ iff }\ord[D_1]{V_{11}}{V_{12}}\text{ and }\ord[D_2]{V_{21}}{V_{22}}\\
    \sem{}{\unref{0}} &= (\emptyset, \ordsym[\emptyset]) \\
    \sem{}{\unref{P} + \unref{Q}} &= (\sem{}{\unref{P}} \uplus \sem{}{\unref{Q}}, \ordsym[\sem{}{\unref{P}} \uplus \sem{}{\unref{Q}}]) \\
       &\text{where }\ord[D_1 \uplus D_2]{(j, V_{1j})}{(j, V_{2j})}\text{ iff }\ord[D_j]{V_{1j}}{V_{2j}}\\
    \sem{}{\downshift{\unref{N}}} &= (\sem{}{\unref{N}}, \ordsym[\sem{}{\unref{N}}]) \\
    \sem{}{\mu\unref{F}} &= (\cup_{k \in \kindnat} \sem{}{\unref{F}}^k \emptyset, \ordsym[\mu \sem{}{\unref{F}}])\\
       &\text{where }\ord[\mu \sem{}{\unref{F}}]{V_1}{V_2}\text{ iff there exists }k\in\kindnat\text{ such that }\ord[\sem{}{\unref{F}}^{k+1}\emptyset]{V_1}{V_2}\\
       &\text{and }\sqsubseteq_{\sem{}{F_1 \oplus F_2} X} \; = \; \sqsubseteq_{\sem{}{F_1} X \uplus \sem{}{F_2} X}\\
       &\text{and }\sqsubseteq_{\sem{}{B \otimes \hat{P}} X} \; = \; \sqsubseteq_{\sem{}{B} X \times \sem{}{\hat{P}} X}\\
       &\text{and }\sqsubseteq_{\sem{}{\Id} X} \; = \; \sqsubseteq_X\\
       &\text{and }\sqsubseteq_{\sem{}{I} X} \; = \; \sqsubseteq_\one\\
       &\text{and }\sqsubseteq_{\sem{}{\Const{Q}} X} \; = \; \sqsubseteq_{\sem{}{Q}}\\[1em]
    \sem{}{\unref{N}} &: \Cppo \\
    \sem{}{\unref{P} \to \unref{N}} &= (\comprehend{f : \sem{}{\unref{P}} \to \sem{}{\unref{N}}}{f \text{ is continuous}}, \ordsym[\sem{}{\unref{P}} \Rightarrow \sem{}{\unref{N}}], d \mapsto \bott[\sem{}{\unref{N}}])\\
    \sem{}{\upshift{\unref{P}}} &= (\sem{}{\unref{P}} \uplus \{\bott[\uparrow]\}, \comprehend{((1,d),(1,d'))}{\ord[\sem{}{\unref{P}}]{d}{d'}} \cup \comprehend{((2, \bott[\uparrow]), d)}{d \in \sem{}{\upshift{\unref{P}}}}, (2, \bott[\uparrow]))\\[1em]
    \sem{}{\unref{\mathcal{F}}} &: \Cpo \to \Cpo \\
    \sem{}{\unref{F} \oplus \unref{G}} &= X \mapsto \sem{}{\unref{F}} X \uplus \sem{}{\unref{G}} X \\
    \sem{}{\unref{I}} &= X \mapsto \one \\
    \sem{}{\unref{B}\otimes\unref{\Ptype}} &= X \mapsto \sem{}{\unref{B}} X \times \sem{}{\unref{\Ptype}} X \\
    \sem{}{\Const{\unref{P}}} &= X \mapsto \sem{}{\unref{P}} \\
    \sem{}{\unref{\Id}} &= X \mapsto X
    \\[1em]
    \fmap{\sem{}{F_1 \oplus F_2}}{f} &= d \mapsto \begin{cases}
      (1, (\fmap{\sem{}{F_1}}{f})\;d') & \text{if }d=(1,d') \\
      (2, (\fmap{\sem{}{F_2}}{f})\;d') & \text{if }d=(2,d')
    \end{cases} \\
    \fmap{\sem{}{I}}{f} &= \id_\one \\
    \fmap{\sem{}{B\otimes\hat{P}}}{f} &=
                                              (d_1,d_2) \mapsto \left((\fmap{\sem{}{B}}{f})\;d_1
                                              ,(\fmap{\sem{}{\hat{P}}}{f})\;d_2\right) \\
    \fmap{\sem{}{\underline{P}}}{f} &= \id_{\sem{}{P}} \\
    \fmap{\sem{}{\Id}}{f} &= f
  \end{align*}

  \caption{Denotational Semantics of Unrefined Types and Functors}
  \label{fig:denotation-unrefined-types-functors}
\end{figure}

\begin{figure}[htbp]
  \begin{flalign*}
    &\sem{}{\unrefsynhead{\Gamma}{h}{P}} : \sem{}{\Gamma} \to \sem{}{P} &\\
    &\sem{\delta}{x}
      = \delta(x) &\\
    &\sem{\delta}{\annoexp{v}{P}}
      = \sem{\delta}{v}
    &\\[2em]
    &\sem{}{\unrefsynexp{\Gamma}{\be}{\upshift{P}}} : \sem{}{\Gamma} \to \sem{}{\upshift{P}} &\\
    &\sem{\delta}{h(s)}
      = \sem{\delta}{s}\;\sem{\delta}{h} &\\
    &\sem{\delta}{\annoexp{e}{\upshift{P}}}
      = \sem{\delta}{e}
  \end{flalign*}

  \caption{Denotational semantics of unrefined heads $h$ and bound expressions $\be$}
  \label{fig:unref-denotation-head-bound}
\end{figure}

\begin{figure}[htbp]
  \begin{flalign*}
    &\sem{}{\unrefchkval{\Gamma}{v}{P}} : \sem{}{\Gamma} \to \sem{}{P} &\\
    &\sem{\delta}{x} = \delta(x) &\\
    &\sem{\delta}{\unit}
      = \bullet &\\
    &\sem{\delta}{\pair{v_1}{v_2}}
      = ( \sem{\delta}{v_1} , \sem{\delta}{v_2} ) &\\
    &\sem{\delta}{\inj{k}{v}}
      = (k, \sem{\delta}{v}) &\\
    &\sem{\delta}{\roll{v}}
      = \sem{\delta}{v} &\\
    &\sem{\delta}{\thunk{e}}
      = \sem{\delta}{e}
    &\\[2em]
    &\sem{}{\unrefchkexp{\Gamma}{e}{N}} : \sem{}{\Gamma} \to \sem{}{N} &\\
    &\sem{\delta}{\Return{v}}
      = (1, \sem{\delta}{v}) &\\
    &\sem{\delta}{\unrefchkexp{\Gamma}{\Let{x}{\be}{e}}{N}}
      = \begin{cases}
        \sem{(\delta, V/x)}{e} &\quad\text{if }\sem{\delta}{\be} = (1, V) \\
        \bott[\sem{}{N}] &\quad\text{if }\sem{\delta}{\be} = (2, \bott[\uparrow])
        \end{cases} &\\
    &\sem{\delta}{\fun{x}{e}}
      = V \mapsto \sem{(\delta, V/x)}{e} &\\
    &\sem{\delta}{\unrefchkexp{\Gamma}{\diverge}{N}}
      = \bott[\sem{}{N}] &\\
    &\sem{\delta}{\unrefchkexp{\Gamma}{\rec{x}{e}}{N}}
      = \bigsqcup_{k \in \kindnat} \left(V \mapsto \sem{\delta, V/x}{\unrefchkexp{\Gamma, x:\downshift{N}}{e}{N}}\right)^k \bott[\sem{}{N}] &\\
    &\sem{\delta}{\match{h}{\clauses{\pa}{e}{i}{I}}}
      = \sem{\delta}{\clauses{\pa}{e}{i}{I}}
          \sem{\delta}{h} &\\
  \end{flalign*}

  \caption{Denotational semantics of unrefined values $v$ and expressions $e$}
  \label{fig:unref-denotation-val-exp}
\end{figure}

\begin{figure}[htbp]
  \begin{flalign*}
    &\sem{}{\unrefchkmatch{\Gamma}{P}{\clauses{\pa}{e}{i}{I}}{N}} : \sem{}{\Gamma} \to \sem{}{P} \to \sem{}{N} &\\
    &\sem{\delta}{\setof{\clause{\unit}{e}}}
      = \bullet \mapsto \sem{\delta}{e} &\\
    &\sem{\delta}{\setof{\clause{\pair{x_1}{x_2}}{e}}}
      = (V_1, V_2) \mapsto \sem{(\delta,V_1/x_1,V_2/x_2)}{e} &\\
    &\sem{\delta}{\setof{ \clause{\inl{x_1}}{e_1} \bnfalt \clause{\inr{x_2}}{e_2}}}
      = V \mapsto \begin{cases}
          \sem{\delta,V_1/x_1}{e_1} &\quad\text{if }V=(1,V_1) \\
          \sem{\delta,V_2/x_2}{e_2} &\quad\text{if }V=(2,V_2)
        \end{cases} &\\
    &\sem{\delta}{\setof{}}
       = \text{empty function} &\\
    &\sem{\delta}{\setof{\clause{\roll{x}}{e}}}
       = V\mapsto\sem{\delta,V/x}{e}
    \\[2em]
    &\sem{}{\unrefspine{\Gamma}{s}{N}{M}} : \sem{}{\Gamma} \to \sem{}{N} \to \sem{}{M} &\\
    &\sem{\delta}{v, s}
      = f \mapsto \sem{\delta}{s}(f(\sem{\delta}{v})) &\\
    &\sem{\delta}{\cdot}
       = V \mapsto V
  \end{flalign*}

  \caption{Denotational semantics of unrefined match expressions and spines}
  \label{fig:unref-denotation-match-spine}
\end{figure}

\clearpage

\subsection{Erasure}
\label{sec:apx-erasure}
\begin{figure}[htbp]
  \begin{equation*}
    \begin{aligned}[c]
      \Aboxed{\erase{A} &= \text{erasure of }\unref{A}} \\
      \erase{1} &= \unref{1} \\
      \erase{P \times Q} &= \erase{P} \times \erase{Q} \\
      \erase{0} &= \unref{0} \\
      \erase{P + Q} &= \erase{P} + \erase{Q} \\
      \erase{\downshift{N}} &= \downshift{\erase{N}} \\
      \erase{\comprehend{\nu : \mu F}{\Fold{F}{\alpha}\,{\nu} =_\tau t}} &= \mu\erase{F}\\
      \erase{\extype{a:\tau}{P}} &= \erase{P} \\
      \erase{P\land\phi} &= \erase{P} \\
      \erase{P \to N} &= \erase{P} \to \erase{N} \\
      \erase{\upshift{P}} &= \upshift{\erase{P}} \\
      \erase{\alltype{a:\tau}{N}} &= \erase{N} \\
      \erase{\phi \implies N} &= \erase{N}
    \end{aligned}
    \qquad
    \begin{aligned}[c]
      \Aboxed{\erase{\mathcal{F}} &= \text{erasure of }\unref{\mathcal{F}}} \\
      \erase{F \oplus G} &= \erase{F} \oplus \erase{G} \\
      \erase{I} &= \unref{I} \\
      \erase{B \otimes \Ptype} &= \erase{B} \otimes \erase{\Ptype} \\
      \erase{\Const{P}} &= \Const{\erase{P}} \\
      \erase{\Id} &= \unref{\Id}
    \end{aligned}
  \end{equation*}

  \caption{Erasure of Refined Types and Functors to Unrefined Types and Functors}
  \label{fig:erasure}
\end{figure}

\begin{figure}[htbp]
  \begin{align*}
    \erase{x} &= x \\
    \erase{\annoexp{v}{P}} &= \annoexp{\erase{v}}{\erase{P}} \\
    \erase{h(s)} &= \erase{h}(\erase{s}) \\
    \erase{\annoexp{e}{N}} &= \annoexp{\erase{e}}{\erase{N}} \\
    \erase{\cdot} &= \cdot \\
    \erase{v, s} &= \erase{v}, \erase{s} \\
    \erase{\unit} &= \unit \\
    \erase{\pair{v_1}{v_2}} &= \pair{\erase{v_1}}{\erase{v_2}} \\
    \erase{\inl{v}} &= \inl{\erase{v}} \\
    \erase{\inr{v}} &= \inr{\erase{v}} \\
    \erase{\roll{v}} &= \roll{\erase{v}} \\
    \erase{\thunk{e}} &= \thunk{\erase{e}} \\
    \erase{\Return{v}} &= \Return{\erase{v}} \\
    \erase{\Let{x}{\be}{e}} &= \Let{x}{\erase{\be}}{\erase{e}} \\
    \erase{\match{h}{\clauses{\pa}{e}{i}{I}}} &= \match{\erase{h}}{\erase{\clauses{\pa}{e}{i}{I}}} \\
    \erase{\fun{x}{e}} &= \fun{x}{\erase{e}} \\
    \erase{\rec{x:(\alltype{a:\kindnat} M)}{e}} &= \rec{x}{\erase{e}} \\
    \erase{\unreachable} &= \diverge \\
    \erase{\clauses{\pa}{e}{i}{I}} &= \{r_i \clauseop \erase{e_i}\}_{i \in I}
  \end{align*}

  \caption{Erasure of refined program terms}
  \label{fig:erasure-program-terms}
\end{figure}

\begin{figure}[htbp]
  \begin{equation*}
    \begin{aligned}[c]
      \Aboxed{\erase{\sigma} &= \text{erasure of }\unref{\sigma}} \\
      \erase{\cdot} &= \cdot \\
      \erase{\sigma, t/a} &= \erase{\sigma} \\
      \erase{\sigma, \subs{v}{P}{x}} &= \erase{\sigma}, \subs{\erase{v}}{\erase{P}}{x}
    \end{aligned}
    \qquad
    \begin{aligned}[c]
      \Aboxed{\erase{\delta} &= \text{erasure of }\unref{\delta}} \\
      \erase{\cdot} &= \cdot \\
      \erase{\delta, d/a} &= \erase{\delta} \\
      \erase{\delta, V/x} &= \erase{\delta}, V/x \\
    \end{aligned}
  \end{equation*}

  \caption{Substitution erasure}
  \label{fig:subs-erasure}
\end{figure}

\clearpage

\subsection{Denotational Semantics (Refined System)}
\begin{figure}[htbp]

  \begin{grammar}
    Semantic substitutions &
    $\delta$ & \bnfas & $\cdot \bnfalt \delta, d/a \bnfalt \delta, V/x$
  \end{grammar}

  \judgbox{\filterprog{\delta} = \delta}{Filter out program variable entries}
  \begin{align*}
    \filterprog{\cdot} &= \cdot \\
    \filterprog{\delta, d/a} &= \filterprog{\delta}, d/a \\
    \filterprog{\delta, V/x} &= \filterprog{\delta}
  \end{align*}

  \judgbox{
    |- \delta : \Theta; \Gamma
  }
  {
    We know that
    $\delta$ is a semantic substitution for variables in $\Theta$ and $\Gamma$
  }
  \begin{mathpar}
    \Infer{\EmptySem}
    {
    }
    {
      |- \cdot : \cdot; \cdot
    }
    \and
    \Infer{\IxSem}
    {
      |- \delta : \Theta; \Gamma
      \and
      d \in \sem{}{\tau}
      \and
      a \notin \dom{\Theta}
    }
    {
      |- \delta, d/a : \Theta, a:\tau; \Gamma
    }
    \and
    \Infer{\PropSem}
    {
      |- \delta : \Theta; \Gamma
      \and
      \sem{\filterprog{\delta}}{\phi} = \{\bullet\}
    }
    {
      |- \delta : \Theta, \phi; \Gamma
    }
    \and
    \Infer{\ValSem}
    {
      |- \delta : \Theta; \Gamma
      \and
      V \in \sem{\filterprog{\delta}}{P}
      \and
      x \notin \dom{\Gamma}
    }
    {
      |- \delta, V/x : \Theta; \Gamma, x:P
    }
  \end{mathpar}

  \begin{align*}
    \sem{}{\Theta; \Gamma} &= \comprehend{\delta}{|- \delta : \Theta; \Gamma}\\
    \sem{}{\Theta} &= \sem{}{\Theta; \cdot} \\
    \sem{}{\Gamma} &= \sem{}{\cdot; \Gamma}
  \end{align*}

  \caption{Semantic substitutions}
  \label{fig:sem-subs}
\end{figure}

\begin{figure}[htbp]
  \centering

  \begin{align*}
    \sem{}{\judgeterm{\Theta}{t}{\tau}} &: \sem{}{\overline{\Theta}} \to \sem{}{\tau} \\ 
    \sem{\delta}{a} &= \delta(a)\\
    \sem{\delta}{n} &= n \\
    \sem{\delta}{t_1+t_2} &= \sem{\delta}{t_1}+\sem{\delta}{t_2} \\
    \sem{\delta}{t_1-t_2} &= \sem{\delta}{t_1}-\sem{\delta}{t_2} \\
    \sem{\delta}{(t_1,t_2)} &= (\sem{\delta}{t_1}, \sem{\delta}{t_2}) \\
    \sem{\delta}{\pi_k\,{t}} &= \pi_k\sem{\delta}{t} \\
    \sem{\delta}{t = t'}
      &=
      \begin{cases}
        \{\bullet\} &\quad\text{if } \sem{\delta}{t} = \sem{\delta}{t'} \\
        \emptyset &\quad\text{else}
      \end{cases}
    \\
    \sem{\delta}{t \leq t'}
      &=
      \begin{cases}
        \{\bullet\} &\quad\text{if } \sem{\delta}{t} \leq \sem{\delta}{t'} \\
        \emptyset &\quad\text{else}
      \end{cases}
    \\
    \sem{\delta}{\phi_1 \land \phi_2}
      &= \sem{\delta}{\phi_1} \cap \sem{\delta}{\phi_2}
    \\
    \sem{\delta}{\phi_1 \lor \phi_2}
      &= \sem{\delta}{\phi_1} \cup \sem{\delta}{\phi_2}
    \\
    \sem{\delta}{\neg\phi}
      &= \{\bullet\} \setminus \sem{\delta}{\phi}
    \\
    \sem{\delta}{\True}
      &= \{\bullet\}
    \\
    \sem{\delta}{\False}
      &= \emptyset
  \end{align*}
  
  \caption{Denotational semantics of well-formed indexes and propositions}
  \label{fig:denotation-ix-prop}
\end{figure}

\begin{figure}[htbp]
  \centering

  (This figure is part of a mutually recursive definition that includes
  \Figref{fig:denotation-types2}.)

  Given $\judgefunctor{\Theta}{F}{\dontcare}$ and $|- \delta : \Theta$,
  define 
  \[
    \mu\sem{\delta}{F} = \bigcup_{k=0}^{\infty} \sem{\delta}{F}^k \emptyset
  \]

  Given $\judgefunctor{\Theta}{F}{\dontcare}$
  and $\judgealgebra{\cdot}{\Theta}{\alpha}{F}{\tau}$
  and $|- \delta : \Theta$,
  define
  \[
    \fold{\sem{\delta}{F}}{\sem{\delta}{\alpha}} = \sem{\delta}{\alpha} \circ \left( {\sem{\delta}{F}} \; (\fold{\sem{\delta}{F}}{\sem{\delta}{\alpha}}) \right)
  \]

  \begin{align*}
    \Ob(\rCpo) &= \comprehend{(D, R)}{D \in \Cpo \text{ and } R \subseteq D} \\
    \Hom_{\rCpo}((D_1, R_1), (D_2, R_2)) &= \comprehend{f \in \Hom_{\Cpo}(D_1, D_2)}{f(R_1) \subseteq R_2} \\
    &(\rCppo \text{ defined similarly})
    &\\[1em]
    \sem{}{\judgetp{\Theta}{P}{\dontcare}} &: \sem{}{\Theta} \to \rCpo
    \\
    \sem{\delta}{1}
        &=
        \one
    &\\
    \sem{\delta}{P_1 \times P_2}
        &=
        \sem{\delta}{P_1} \times \sem{\delta}{P_2}
    &\\
    \sem{\delta}{0}
        &=
        \emptyset
    &\\
    \sem{\delta}{P_1 + P_2}
        &=
        \sem{\delta}{P_1} \uplus \sem{\delta}{P_2}
    &\\
    \sem{\delta}{ \comprehend{\nu : \mu F}{\Fold{F}{\alpha}\;\nu =_\tau t} }
        &=
        \comprehend{ V \in \mu\sem{}{\erase{F}} }
          {V \in \mu\sem{\delta}{F} \text{ and } (\fold{\sem{\delta}{F}}{\sem{\delta}{\alpha}})\;V = \sem{\delta}{t} }
    &\\
    \sem{\delta}{
        \extype{a : \tau}{P}
        }
        &=
        \comprehend{V \in \sem{}{\erase{P}}}{\extype{d \in \sem{}{\tau}}{V \in \sem{\delta, d/a}{P}}}
    &\\
    \sem{\delta}{
        P \andty \phi
        }
        &=
        \comprehend{V \in \sem{}{\erase{P}}}{V \in \sem{\delta}{P} \text{ and } \sem{\delta}{\phi} = \one}
    &\\
    \sem{\delta}{
        \downshift{N}
        }
        &=
        \sem{\delta}{N}
    &\\[1em]
    \sem{}{\judgetp{\Theta}{N}{\dontcare}} &: \sem{}{\Theta} \to \rCppo
    &\\
    \sem{\delta}{
        \alltype{a:\tau}{N}
        }
        &=
        \comprehend{f \in \sem{}{\erase{N}}}{\alltype{d \in \sem{}{\tau}}{f \in \sem{\delta,d/a}{N}}}
    &\\
    \sem{\delta}{
        P \to N
        }
        &=
        \comprehend{f \in \sem{}{\erase{P \to N}}}{\alltype{V \in \sem{\delta}{P}}{f(V) \in \sem{\delta}{N}}}
    &\\
    \sem{\delta}{
        \phi \implies N
        }
        &=
        \comprehend{f \in \sem{}{\erase{N}}}{\text{if } \sem{\delta}{\phi} = \one \text{ then } f \in \sem{\delta}{N}}
    &\\
    \sem{\delta}{
        \upshift{P}
        }
        &=
        \comprehend{(1, V)}{V \in \sem{\delta}{P}}
  \end{align*} 

  \caption{Denotational semantics of well-formed types (specifying the second component, \ie the refined set; the first component is denotation of erasure)}
  \label{fig:denotation-types1}
\end{figure}

\begin{figure}[htbp]
  \centering

  (This figure is part of a mutually recursive definition that includes
  \Figref{fig:denotation-types1}.)
  
  \begin{align*}
   \sem{}{\judgefunctor{\Theta}{\mathcal{F}}{\Xi}}
       &:
       \sem{}{\Theta} \to \rCpo \to \rCpo
   \\
   \sem{\delta}{F_1 \oplus F_2}
       &=
       X \mapsto \sem{\delta}{F_1} X \uplus \sem{\delta}{F_2} X
   \\
   \sem{\delta}{I}
       &=
       X \mapsto \{ \bullet \}
   \\
   \sem{\delta}{B \otimes \Ptype} 
       &=
       X \mapsto \sem{\delta}{B} X \times \sem{\delta}{\Ptype} X
   \\
   \sem{\delta}{\Const{P}} 
       &=
       X \mapsto \sem{\delta}{P}
   \\
   \sem{\delta}{\Id}
       &=
       X \mapsto X
  \end{align*}
  \begin{flalign*}
   &\sem{}{\judgealgebra{\Xi}{\Theta}{\alpha}{F}{\tau}}
       : \prod_{\delta \in \sem{}{\Theta}} \sem{\delta}{F}\sem{}{\tau} \to \sem{}{\tau}\\
   &\sem{\delta}
     {\judgealgebra{\Xi}{\Theta}{\alpha}{(F_1 \oplus F_2)}{\tau}}
       V
       =
           \begin{cases}
             \sem{\delta}
               {\judgealgebra{\Xi}{\Theta}{\alpha_1}{F_1}{\tau}} V'
                 & \text{if }V = (1, V')\\
             \sem{\delta}
               {\judgealgebra{\Xi}{\Theta}{\alpha_2}{F_2}{\tau}} V'
                 & \text{if }V = (2, V')
             \end{cases}\\
       \hspace{-15em}&\hspace{15em}\text{where }\composeinj{1}{\alpha}{\alpha_1}
       \text{ and }  \composeinj{2}{\alpha}{\alpha_2}
   \\[0.5em]
   &\sem{\delta}{\judgealgebra{\Xi}{\Theta}{\clause{\unitexp}{t}}{I}{\tau}} \bullet
       =
       \sem{\delta\restriction_\Xi}{\judgeterm{\Xi}{t}{\tau}}
   \\[0.5em]
   &\sem{\delta}{\judgealgebra{\Xi}{\Theta}{\clause{(a, q)}{t}}{(\Id \otimes \hat{P})}{\tau}}
       (d, V)
       =
       \sem{(\delta, d/a)}{\judgealgebra{\Xi,a:\tau}{\Theta,a:\tau}{\clause{q}{t}}{\hat{P}}{\tau}}
           V
   \\[0.5em]
   &\sem{\delta}{\judgealgebra{\Xi}{\Theta}{\clause{(\wild, q)}{t}}{(\Const{Q} \otimes \hat{P})}{\tau}} (\wild, V)
       =
       \sem{\delta}{\judgealgebra{\Xi}{\Theta}{\clause{q}{t}}{\hat{P}}{\tau}} V
   \\[0.5em]
   &\sem{\delta}{\judgealgebra{\Xi}{\Theta}{\clause{(\pack{a}{\bap}, q)}{t}}{(\Const{\extype{a:\tau'}{Q}} \otimes \hat{P})}{\tau}} (V_1,V_2)
       =  \\
       \hspace{-10em}&\hspace{10em}\sem{(\delta,d/a)}
           {\judgealgebra{\Xi, a:\tau'}{\Theta,a:\tau'}{\clause{(\bap, q)}{t}}
              {(\Const{Q} \otimes \hat{P})}{\tau}}
         (V_1, V_2) \\
         \hspace{-20em}&\hspace{20em}\text{where }d\in\sem{}{\tau'}\text{ is such that }V_1 \in \sem{\delta,d/a}{Q}
  \end{flalign*} 
  
  \caption{Denotational semantics of well-formed functors and algebras (latter maintains $\Xi \subseteq \Theta$)}
  \label{fig:denotation-types2}
\end{figure}

\begin{figure}[htbp]
  \begin{flalign*}
    &\sem{}{\judgesynhead{\Theta}{\Gamma}{h}{P}} = (\delta \in \sem{}{\Theta; \Gamma}) \mapsto \sem{\erase{\delta}}{\unrefsynhead{\erase{\Gamma}}{\erase{h}}{\erase{P}}} &\\
    &\sem{}{\judgesynexp{\Theta}{\Gamma}{\be}{\upshift{P}}} = (\delta \in \sem{}{\Theta; \Gamma}) \mapsto \sem{\erase{\delta}}{\unrefsynexp{\erase{\Gamma}}{\erase{\be}}{\erase{\upshift{P}}}} &\\
    &\sem{}{\judgechkval{\Theta}{\Gamma}{v}{P}} = (\delta \in \sem{}{\Theta; \Gamma}) \mapsto \sem{\erase{\delta}}{\unrefchkval{\erase{\Gamma}}{\erase{v}}{\erase{P}}} &\\
    &\sem{}{\judgechkexp{\Theta}{\Gamma}{e}{N}} = (\delta \in \sem{}{\Theta; \Gamma}) \mapsto \sem{\erase{\delta}}{\unrefchkexp{\erase{\Gamma}}{\erase{e}}{\erase{N}}} &\\
    &\sem{}{\judgechkmatch{\Theta}{\Gamma}{P}{\clauses{\pa}{e}{i}{I}}{N}} = (\delta \in \sem{}{\Theta; \Gamma}) \mapsto \sem{\erase{\delta}}{\unrefchkmatch{\erase{\Gamma}}{\erase{P}}{\erase{\clauses{\pa}{e}{i}{I}}}{\erase{N}}} \\
    &\sem{}{\judgespine{\Theta}{\Gamma}{s}{N}{M}} = (\delta \in \sem{}{\Theta; \Gamma}) \mapsto \sem{\erase{\delta}}{\unrefspine{\erase{\Gamma}}{\erase{s}}{\erase{N}}{\erase{M}}}
  \end{flalign*}

  \caption{Denotational semantics of refined program terms}
  \label{fig:denotation-program-terms}
\end{figure}

\clearpage

\subsection{Algorithmic System}

\begin{figure}[htbp]
  \judgbox{\judgectx{\Theta}{\Delta}}{Under $\Theta$, algorithmic context $\Delta$ is well-formed}
  \begin{mathpar}
    \Infer{\AlgCtxEmpty}
    {}
    {
      \judgectx{\Theta}{\cdot}
    }
    \and
    \Infer{\AlgCtxUnsolved}
    {
      \judgectx{\Theta}{\Delta}
      \\
      \ahat \notin \dom{\Delta}
    }
    {
      \judgectx{\Theta}{\Delta, \ahat:\tau}
    }
    \and
    \Infer{\AlgCtxSolved}
    {
      \judgectx{\Theta}{\Delta}
      \\
      \judgeterm{\Theta}{t}{\tau}
      \\
      \ahat \notin \dom{\Delta}
    }
    {
      \judgectx{\Theta}{\Delta, \hypeq{\ahat}{\tau}{t}}
    }
  \end{mathpar}

  \caption{Algorithmic context well-formedness}
  \label{fig:algwf-alg-ctx}
\end{figure}

\begin{figure}[htbp]

  \judgbox{
    \Theta; \Delta |- t : \tau
  }
  {
    Under $\Theta$ and $\Delta$, index $t$ has sort $\tau$
  }
  \begin{mathpar}
    \Infer{\AlgIxVar}
    {(a:\tau)\in\Theta}
    {\Theta; \Delta |- a : \tau}
    \and
    \Infer{\AlgIxEVar}
    {(\ahat:\tau)\in\Delta}
    {\Theta; \Delta |- \ahat : \tau}
    \and
    \Infer{\AlgIxSolvedEVar}
    {(\hypeq{\ahat}{\tau}{t}:\tau)\in\Delta}
    {\Theta; \Delta |- \ahat : \tau}
    \and
    \Infer{\AlgIxConst}
    {t \in \sortconsts{\tau}}
    {\Theta; \Delta |- t : \tau}
    \and
    \Infer{\AlgIxAdd}
    {
      \tau \in \{\kindnat, \Z\} \\
      \Theta; \Delta |- t_1 : \tau \\
      \Theta; \Delta |- t_2 : \tau
    }
    {\Theta; \Delta |- t_1 + t_2 : \tau}
    \and
    \Infer{\AlgIxSubtract}
    {
      \tau \in \{\kindnat, \Z\} \\
      \Theta; \Delta |- t_1 : \tau \\
      \Theta; \Delta |- t_2 : \tau
    }
    {\Theta; \Delta |- t_1 - t_2 : \tau}
    \and
    \Infer{\AlgIxProd}
    {
      \Theta; \Delta |- t_1 : \tau_1 \\
      \Theta; \Delta |- t_2 : \tau_2
    }
    {\Theta; \Delta |- (t_1, t_2) : \tau_1 \times \tau_2}
    \and
    \Infer{\AlgIxProj{k}}
    { \Theta; \Delta |- t : \tau_1 \times \tau_2 }
    {\Theta; \Delta |- \pi_k\;t : \tau_k}
   \and
    \Infer{\AlgIxAnd}
    {
      \judgeterm{\Theta; \Delta}{\phi_1}{\Booltype}
      \\
      \judgeterm{\Theta; \Delta}{\phi_2}{\Booltype}
    }
    {
      \judgeterm{\Theta; \Delta}{\phi_1 \land \phi_2}{\Booltype}
    }
    \and
    \Infer{\AlgIxOr}
    {
      \judgeterm{\Theta; \Delta}{\phi_1}{\Booltype}
      \\
      \judgeterm{\Theta; \Delta}{\phi_2}{\Booltype}
    }
    {
      \judgeterm{\Theta; \Delta}{\phi_1 \lor \phi_2}{\Booltype}
    }
    \and
    \Infer{\AlgIxNot}
    {
      \judgeterm{\Theta; \Delta}{\phi}{\Booltype}
    }
    {
      \judgeterm{\Theta; \Delta}{\lnot\phi}{\Booltype}
    }
    \and
    \Infer{\AlgIxEq}
    {
      \Theta; \Delta |- t_1 : \tau
      \\
      \Theta; \Delta |- t_2 : \tau
    }
    {
      \judgeterm{\Theta; \Delta}{t_1 = t_2}{\Booltype}
    }
   \and
    \Infer{\AlgIxLeq}
    {
      \tau \in \{\kindnat, \Z\}
      \\
      \Theta; \Delta |- t_1 : \tau
      \\
      \Theta; \Delta |- t_2 : \tau
    }
    {
      \judgeterm{\Theta; \Delta}{t_1 \leq t_2}{\Booltype}
    }
  \end{mathpar}

  \caption{Algorithmic index sorting}
  \label{fig:alg-ix-sort}
\end{figure}

\begin{figure}[htbp]
  \judgbox{\judgectx{\Theta; \Delta}{\Thetahat}}{Under $\Theta$ and $\Delta$, (output mode) algorithmic logical context $\Thetahat$ is well-formed}
  \begin{mathpar}
    \Infer{\AlgLogCtxEmpty}
    {}
    {
      \judgectx{\Theta; \Delta}{\cdot}
    }
    \and
    \Infer{\AlgLogCtxVar}
    {
      \judgectx{\Theta; \Delta}{\Thetahat}
      \\
      a \notin \dom{\Theta} \cup \dom{\Thetahat}
    }
    {
      \judgectx{\Theta; \Delta}{\Thetahat, a:\tau}
    }
    \and
    \Infer{\AlgLogCtxProp}
    {
      \judgectx{\Theta; \Delta}{\Thetahat}
      \\
      \judgeterm{\Theta, \overline{\Thetahat}; \Delta}{\phi}{\Booltype}
    }
    {
      \judgectx{\Theta; \Delta}{\Thetahat, \phi}
    }
  \end{mathpar}

  \caption{(Output mode) algorithmic logical context well-formedness}
  \label{fig:algwf-alg-log-ctx}
\end{figure}

\begin{figure}[htbp]
\raggedright
\judgbox{
    \judgetp{\Theta; \Delta}{A}{\Xi_A}
  }{
    Under $\Theta$ and $\Delta$, type $A$ is well-formed, with value-determined indexes $\Xi_A$
}
\begin{mathpar}
  \Infer{\AlgTpVoid}
        { }
        { \judgetp{\Theta; \Delta}{0}{\cdot}  }
  \and
  \Infer{\AlgTpUnit}
        { }
        { \judgetp{\Theta; \Delta}{1}{\cdot}  }
  \\
  \Infer{\AlgTpSum}
        { \judgetp{\Theta; \Delta}{P_1}{\Xi_1} \\ \judgetp{\Theta; \Delta}{P_2}{\Xi_2} }
        { \judgetp{\Theta; \Delta}{P_1 + P_2}{\Xi_1 \sect \Xi_2}  }
  \and
  \Infer{\AlgTpProd}
        { \judgetp{\Theta; \Delta}{P_1}{\Xi_1} \\ \judgetp{\Theta; \Delta}{P_2}{\Xi_2} }
        { \judgetp{\Theta; \Delta}{P_1 \times P_2}{\Xi_1 \union \Xi_2}  }
  \and
  \Infer{\AlgTpEx}
        { \judgetp{\Theta, a:\tau; \Delta}{P}{\Xi, a:\tau} }
        { \judgetp{\Theta; \Delta}{\extype{a:\tau}{P}}{\Xi} }
  \and
  \Infer{\AlgTpDown}
        { \judgetp{\Theta; \Delta}{N}{\Xi} }
        { \judgetp{\Theta; \Delta}{\downshift{N}}{\cdot} }
  \and
  \Infer{\AlgTpWith}
        {
          \judgetp{\Theta; \Delta}{P}{\Xi}
          \\
          \judgeterm{\Theta; \Delta}{\phi}{\Booltype}
        }
        { \judgetp{\Theta; \Delta}{P \land \phi}{\Xi} }
  \\
  \Infer{\AlgTpFixVar}
        {
          \judgefunctor{\Theta; \Delta}{F}{\Xi}
          \\
          \judgealgebra{\cdot}{\Theta; \Delta}{\alpha}{F}{\tau}
          \\
          (b : \tau) \in \Theta
        }
        {
          \judgetp{\Theta; \Delta}{\comprehend{\nu:\mu F}{\Fold{F}{\alpha}\,{\nu} =_\tau b}}{\Xi \union b:\tau}
        }
  \\
  \Infer{\AlgTpFixEVar}
  {
    \judgefunctor{\Theta; \Delta}{F}{\Xi}
    \\
    \judgealgebra{\cdot}{\Theta; \Delta}{\alpha}{F}{\tau}
    \\
    (\ahat : \tau) \in \Delta
  }
  {
    \judgetp{\Theta; \Delta}{\comprehend{\nu:\mu F}{\Fold{F}{\alpha}\,{\nu} =_\tau \ahat}}{\Xi \union \ahat:\tau}
  }
  \\
  \Infer{\AlgTpFixSolvedEVar}
  {
    \judgefunctor{\Theta; \Delta}{F}{\Xi}
    \\
    \judgealgebra{\cdot}{\Theta; \Delta}{\alpha}{F}{\tau}
    \\
    (\hypeq{\ahat}{\tau}{t}) \in \Delta
  }
  {
    \judgetp{\Theta; \Delta}{\comprehend{\nu:\mu F}{\Fold{F}{\alpha}\,{\nu} =_\tau \ahat}}{\Xi \union \ahat:\tau}
  }
  \\
  \Infer{\AlgTpFix}
        {
          \judgefunctor{\Theta; \Delta}{F}{\Xi}
          \\
          \judgealgebra{\cdot}{\Theta; \Delta}{\alpha}{F}{\tau}
          \\
          \Theta; \Delta |- t : \tau
          \\
          t \text{ not a variable}
        }
        {
          \judgetp{\Theta; \Delta}{\comprehend{\nu:\mu F}{\Fold{F}{\alpha}\,{\nu} =_\tau t}}{\Xi}
        }
  \\
  \Infer{\AlgTpAll}
        { \judgetp{\Theta, a:\tau; \Delta}{N}{\Xi, a:\tau} }
        { \judgetp{\Theta; \Delta}{\alltype{a:\tau}{N}}{\Xi} }
  \and
  \Infer{\AlgTpImplies}
        {
          \judgeterm{\Theta; \Delta}{\phi}{\Booltype}
          \\
          \judgetp{\Theta; \Delta}{N}{\Xi}
        }
        { \judgetp{\Theta; \Delta}{\phi \implies N}{\Xi} }
  \and
  \Infer{\AlgTpArrow}
        { \judgetp{\Theta; \Delta}{P}{\Xi_1} \\ \judgetp{\Theta; \Delta}{N}{\Xi_2} }
        { \judgetp{\Theta; \Delta}{P \to N}{\Xi_1 \union \Xi_2} }
  \and
  \Infer{\AlgTpUp}
        { \judgetp{\Theta; \Delta}{P}{\Xi} }
        { \judgetp{\Theta; \Delta}{\upshift{P}}{\cdot} }
\end{mathpar}

  \centering
  
  \caption{Algorithmic type well-formedness}
  \label{fig:algwf-type}
\end{figure}

\begin{figure}[htbp]
\judgbox{\judgefunctor{\Theta; \Delta}{\mathcal{F}}{\Xi}}
      {
        Under $\Theta$ and $\Delta$, functor $\mathcal{F}$ is well-formed,
        with value-determined indexes $\Xi$
      }
\begin{mathpar}
  \Infer{\AlgFunctorConst}
        { \judgetp{\Theta; \Delta}{P}{\Xi} } 
        { \judgefunctor{\Theta; \Delta}{\Const{P}}{\Xi} } 
  \and
  \Infer{\AlgFunctorId}
        {  } 
        { \judgefunctor{\Theta; \Delta}{\Id}{\cdot} } 
  \and
  \Infer{\AlgFunctorUnit}
        {  } 
        { \judgefunctor{\Theta; \Delta}{I}{\cdot} } 
  \and
  \Infer{\AlgFunctorProd}
        { \judgefunctor{\Theta; \Delta}{B}{\Xi_1} \\ \judgefunctor{\Theta; \Delta}{\hat{P}}{\Xi_2} } 
        { \judgefunctor{\Theta; \Delta}{B \otimes \hat{P}}{\Xi_1 \union \Xi_2} } 
  \and
  \Infer{\AlgFunctorSum}
        { \judgefunctor{\Theta; \Delta}{F_1}{\Xi_1} \\ \judgefunctor{\Theta; \Delta}{F_2}{\Xi_2} } 
        { \judgefunctor{\Theta; \Delta}{F_1 \oplus F_2}{\Xi_1 \sect \Xi_2} } 
\end{mathpar}

\judgbox{\judgealgebra{\Xi}{\Theta; \Delta}{\alpha}{F}{\tau}}
      {
           Under $\Xi \subseteq \Theta$, we know $\alpha$ is a well-formed algebra
           of sort $F(\tau) \Rightarrow \tau$,\\
           where $F$ is a well-formed functor under $\Theta$ and $\Delta$
      }
\begin{mathpar}
  \Infer{\AlgAlgSum}
  {
    \arrayenvb{
      \composeinj{1}{\alpha}{\alpha_1}
      \\
      \composeinj{2}{\alpha}{\alpha_2}
    }
    \\
    \arrayenvb{
      \judgealgebra{\Xi}{\Theta; \Delta}{\alpha_1}{F_1}{\tau}
      \\
      \judgealgebra{\Xi}{\Theta; \Delta}{\alpha_2}{F_2}{\tau}
    }
  }
  {
    \judgealgebra{\Xi}{\Theta; \Delta}{\alpha}{(F_1 \oplus F_2)}{\tau}
  }
  \\
  \Infer{\AlgAlgUnit}
        { \judgeterm{\Xi}{t}{\tau} }
        { \judgealgebra{\Xi}{\Theta; \Delta}{\clause{\unitexp}{t}}{I}{\tau} }
  \and
  \Infer{\AlgAlgIdProd}
        { \judgealgebra{\Xi, a:\tau}{\Theta, a:\tau; \Delta}{\clause{q}{t}}{\hat{P}}{\tau} }
        { \judgealgebra{\Xi}{\Theta; \Delta}{\clause{(a, q)}{t}}{(\Id \otimes \hat{P})}{\tau} }
  \and
  \Infer{\AlgAlgConstProd}
        { \judgealgebra{\Xi}{\Theta; \Delta}{\clause{q}{t}}{\hat{P}}{\tau} \\ \judgetp{\Theta; \Delta}{Q}{\dontcare}}
        { \judgealgebra{\Xi}{\Theta; \Delta}{\clause{(\wild, q)}{t}}{(\Const{Q} \otimes \hat{P})}{\tau} }
  \and
  \Infer{\AlgAlgExConstProd}
        { \judgealgebra{\Xi, a:\tau'}{\Theta, a:\tau'; \Delta}{\clause{(\bap, q)}{t}}{(\Const{Q} \otimes \hat{P})}{\tau} }
        { \judgealgebra{\Xi}{\Theta; \Delta}
          {\clause{(\pack{a}{\bap}, q)}{t}}
          {(\Const{\extype{a:\tau'}{Q}} \otimes \hat{P})}{\tau} }
\end{mathpar}
  
  \caption{Algorithmic well-formedness of functors and algebras}
  \label{fig:algwf-fun-alg}
\end{figure}

\begin{figure}[thbp]

\judgbox{\algequiv[]{\Theta; \Delta}{\mathcal{F}}{\mathcal{G}}{W}{\Delta'}}
        {Under $\Theta$ and $\Delta$, functors $\mathcal{F}$ and $\mathcal{G}$ are algorithmically equivalent,\\ with output constraint $W$ and output context $\Delta'$}
\begin{mathpar}
  \Infer{\AlgFunEquivConst}
  {
    \algequiv[+]{\Theta; \Delta}{P}{Q}{W}{\Delta'}
  }
  {
    \algequiv[]{\Theta; \Delta}{\Const{P}}{\Const{Q}}{W}{\Delta'}
  }
  \and 
  \Infer{\AlgFunEquivId}
  {}
  {
    \algequiv[]{\Theta; \Delta}{\Id}{\Id}{\True}{\Delta}
  }
  \and 
  \Infer{\AlgFunEquivUnit}
  {}
  {
    \algequiv[]{\Theta; \Delta}{I}{I}{\True}{\Delta}
  }
  \and 
  \Infer{\AlgFunEquivProd}
  {
    \algequiv[]{\Theta; \Delta}{B}{B'}{W_1}{\Delta''}
    \\
    \algequiv[]{\Theta; \Delta''}{\hat{P}}{[\Delta'']\hat{P}'}{W_2}{\Delta'}
  }
  {
    \algequiv[]{\Theta; \Delta}{B \otimes \hat{P}}{B' \otimes \hat{P}'}{[\Delta']W_1 \land W_2}{\Delta'}
  }
  \and 
  \Infer{\AlgFunEquivSum}
  {
    \algequiv[]{\Theta; \Delta}{F_1}{G_1}{W_1}{\Delta''}
    \\
    \algequiv[]{\Theta; \Delta''}{F_2}{[\Delta'']G_2}{W_2}{\Delta'}
  }
  {
    \algequiv[]{\Theta; \Delta}{F_1 \oplus F_2}{G_1 \oplus G_2}{[\Delta']W_1 \land W_2}{\Delta'}
  }
\end{mathpar}

\caption{Algorithmic functor equivalence}
\label{fig:algfunequiv}
\end{figure}

\begin{figure}[htbp]
  \judgbox{\algpropequivinst{\Theta; \Delta}{\phi}{\psi}{\Delta'}}{Under $\Theta$ and $\Delta$, the propositional equivalence $\phi \equiv \psi$ instantiates $\Delta$ to $\Delta'$}
  \begin{mathpar}
    \Infer{\PropEquivInst}
    {
      \judgeterm{\Theta}{t}{\tau}
    }
    {
      \algpropequivinst{\Theta; \Delta_1, \ahat:\tau, \Delta_2}{(t=t')}{(\ahat=t'')}{\Delta_1, \hypeq{\ahat}{\tau}{t}, \Delta_2}
    }
    \\
    \Infer{\PropEquivNoInst}
    {
      \arrayenvbl{
        \psi \text{ not of form } (\ahat = t'') \dots
        \\
        \dots \text{or } \phi \text{ not of form } (t = t') \text{ where } \judgeterm{\Theta}{t}{\tau}
      }
    }
    {
      \algpropequivinst{\Theta; \Delta}{\phi}{\psi}{\Delta}
    }
  \end{mathpar}

  \judgbox{\alginst{\Theta; \Delta}{\phi}{\Delta'}}{Under $\Theta$ and $\Delta$, proposition $\phi$ instantiates $\Delta$ to $\Delta'$}
  \begin{mathpar}
    \Infer{\RuleInst}
    {
      \judgeterm{\Theta}{t}{\tau}
    }
    {
      \alginst{\Theta; \Delta_1, \ahat:\tau, \Delta_2}{\ahat = t}{\Delta_1, \hypeq{\ahat}{\tau}{t}, \Delta_2}
    }
    \and
    \Infer{\RuleNoInst}
    {
      \phi \text{ not of form } \ahat = t \text{ where } \judgeterm{\Theta}{t}{\tau}
    }
    {
      \alginst{\Theta; \Delta}{\phi}{\Delta}
    }
  \end{mathpar}

  \caption{Propositional instantiation}
  \label{fig:propinst}
\end{figure}

\begin{figure}[thbp]
\judgbox{\algequiv[\pm]{\Theta; \Delta}{A}{B}{W}{\Delta'}}{Under $\Theta$ and $\Delta$, types $A$ and $B$ are algorithmically equivalent,\\ with output constraint $W$ and output context $\Delta'$}
\begin{mathpar}
  \Infer{\AlgTpEquivPosUnit}
  { }
  {
    \algequiv[+]{\Theta; \Delta}{1}{1}{\True}{\Delta}
  }
  \and
  \Infer{\AlgTpEquivPosVoid}
  { }
  {
    \algequiv[+]{\Theta; \Delta}{0}{0}{\True}{\Delta}
  }
  \and
  \Infer{\AlgTpEquivPosProd}
  {
    \algequiv[+]{\Theta; \Delta}{P_1}{Q_1}{W_1}{\Delta''}
    \\
    \algequiv[+]{\Theta; \Delta''}{P_2}{[\Delta'']Q_2}{W_2}{\Delta'}
  }
  {
    \algequiv[+]{\Theta; \Delta}{P_1 \times P_2}{Q_1 \times Q_2}{[\Delta']W_1 \land W_2}{\Delta'}
  }
  \and
  \Infer{\AlgTpEquivPosSum}
  {
    \algequiv[+]{\Theta; \Delta}{P_1}{Q_1}{W_1}{\Delta''}
    \\
    \algequiv[+]{\Theta; \Delta''}{P_2}{[\Delta'']Q_2}{W_2}{\Delta'}
  }
  {
    \algequiv[+]{\Theta; \Delta}{P_1 + P_2}{Q_1 + Q_2}{[\Delta']W_1 \land W_2}{\Delta'}
  }
  \and
  \Infer{\AlgTpEquivPosWith}
  {
    \algequiv[+]{\Theta; \Delta}{P}{Q}{W}{\Delta''}
    \\
    \algpropequivinst{\Theta; \Delta''}{\phi}{[\Delta'']\psi}{\Delta'}
  }
  {
    \algequiv[+]{\Theta; \Delta}{P \land \phi}{Q \land \psi}{[\Delta']W \land (\phi \equiv [\Delta']\psi)}{\Delta'}
  }
  \and
  \Infer{\AlgTpEquivPosEx}
  {
    \algequiv[+]{\Theta, a:\tau; \Delta}{P}{Q}{W}{\Delta'}
  }
  {
    \algequiv[+]{\Theta; \Delta}{\extype{a:\tau}{P}}{\extype{a:\tau}{Q}}{\alltype{a:\tau}{W}}{\Delta'}
  }
  \and
  \Infer{\AlgTpEquivPosFix}
  {
    \alltype{\ahat \in \dom{\Delta}}{[\Delta]t' \neq \ahat}
    \\
    \algequiv[]{\Theta; \Delta}{F}{G}{W}{\Delta'}
  }
  {
    \algequiv[+]{\Theta; \Delta}{\comprehend{\nu:\mu F}{\Fold{F}{\alpha}\,\nu =_\tau t}}{\comprehend{\nu:\mu G}{\Fold{G}{\alpha}\,\nu =_\tau t'}}{W \land (t=[\Delta']t')}{\Delta'}
  }
  \and
  \Infer{\AlgTpEquivPosFixInst}
  {
    \ground{t}
    \\
    \algequiv[]{\Theta; \Delta}{F}{G}{W}{\Delta_1', \ahat:\tau, \Delta_2'}
    \\
    \Delta' = \Delta_1', \hypeq{\ahat}{\tau}{t}, \Delta_2'
  }
  {
    \algequiv[+]{\Theta; \Delta}{\comprehend{\nu:\mu F}{\Fold{F}{\alpha}\,\nu =_\tau t}}{\comprehend{\nu:\mu G}{\Fold{G}{\alpha}\,\nu =_\tau \ahat}}{[\Delta']W \land (t=t)}{\Delta'}
  }
  \and
  \Infer{\AlgTpEquivPosFixInstFun}
  {
    \algequiv[]{\Theta; \Delta}{F}{G}{W}{\Delta'}
    \\
    \hypeq{\ahat}{\tau}{t'} \in \Delta'
  }
  {
    \algequiv[+]{\Theta; \Delta}{\comprehend{\nu:\mu F}{\Fold{F}{\alpha}\,\nu =_\tau t}}{\comprehend{\nu:\mu G}{\Fold{G}{\alpha}\,\nu =_\tau \ahat}}{W \land (t=t')}{\Delta'}
  }
  \and
  \Infer{\AlgTpEquivPosDownshift}
  {
  }
  {
    \algequiv[+]{\Theta; \Delta}{\downshift{N}}{\downshift{M}}{\negeqprob{N}{M}}{\Delta}
  }
  \and
  \Infer{\AlgTpEquivNegUpshift}
  {
  }
  {
    \algequiv[-]{\Theta; \Delta}{\upshift{P}}{\upshift{Q}}{\poseqprob{P}{Q}}{\Delta}
  }
  \and
  \Infer{\AlgTpEquivNegImp}
  {
    \algequiv[-]{\Theta; \Delta}{N}{M}{W}{\Delta'}
  }
  {
    \algequiv[-]{\Theta; \Delta}{\psi \implies N}{\phi \implies M}{W \land ([\Delta']\psi \equiv \phi)}{\Delta'}
  }
  \and
  \Infer{\AlgTpEquivNegAll}
  {
    \algequiv[-]{\Theta, a:\tau; \Delta}{N}{M}{W}{\Delta'}
  }
  {
    \algequiv[-]{\Theta; \Delta}{\alltype{a:\tau}{N}}{\alltype{a:\tau}{M}}{\alltype{a:\tau}{W}}{\Delta'}
  }
  \and
  \Infer{\AlgTpEquivNegArrow}
  {
    \algequiv[+]{\Theta; \Delta}{P}{Q}{W_1}{\Delta''}
    \\
    \algequiv[-]{\Theta; \Delta''}{[\Delta'']N}{M}{W_2}{\Delta'}
  }
  {
    \algequiv[-]{\Theta; \Delta}{P \to N}{Q \to M}{[\Delta']W_1 \land W_2}{\Delta'}
  }  
\end{mathpar}

\caption{Algorithmic type equivalence}
\label{fig:algtpequiv}

\end{figure}

\begin{figure}[htbp]
  \judgbox{\judgeextract[\pm]{\Theta; \Delta}{A}{A'}{\Thetahat}}{Under $\Theta$ and $\Delta$, type $A$ extracts to $A'$ and $\Thetahat$}
  \centering 

  ~\\

  Similar to \Figureref{fig:declextract}.

  Propositions in algorithmic logical context $\Thetahat$
  can have existential variables $\ahat$.

  \caption{Algorithmic extraction}
  \label{fig:algextract}
\end{figure}

\begin{figure}[htbp]
  \centering

  \judgbox{\algsub[\pm]{\Theta; \Delta}{A}{B}{\Wah}{\Delta'}}
      {
        Under $\Theta$ and $\Delta$,
        type $A$ is algorithmically a subtype of $B$,\\
        with output constraint $\Wah$ and output context $\Delta'$
      }

\begin{mathpar}
  \Infer{\AlgSubPosVoid}
        { }
        {
          \algsub[+]{\Theta; \Delta}{0}{0}{\True}{\Delta}
        }
  \and
  \Infer{\AlgSubPosUnit}
        { }
        {\algsub[+]{\Theta; \Delta}{1}{1}{\True}{\Delta}}
  \and
  \Infer{\AlgSubPosProd}
        {
          \algsub[+]{\Theta; \Delta}{P_1}{Q_1}{\Wah_1}{\Delta''}
          \\
          \algsub[+]{\Theta; \Delta''}{P_2}{[\Delta'']Q_2}{\Wah_2}{\Delta'}
        }
        {
          \algsub[+]{\Theta; \Delta}{P_1 \times P_2}{Q_1 \times Q_2}{[\Delta']\Wah_1 \land \Wah_2}{\Delta'}
        }
  \and
  \Infer{\AlgSubPosSum}
        {
          \algequiv[+]{\Theta; \Delta}{P_1}{Q_1}{\Wah_1}{\Delta''}
          \\
          \algequiv[+]{\Theta; \Delta''}{P_2}{[\Delta'']Q_2}{\Wah_2}{\Delta'}
        }
        {
          \algsub[+]{\Theta; \Delta}{P_1 + P_2}{Q_1 + Q_2}{[\Delta']\Wah_1 \land \Wah_2}{\Delta'}
        }
  \and
  \Infer{\AlgSubPosWithR}
        {
          \algsub[+]{\Theta; \Delta}{P}{Q}{\Wah}{\Delta''}
          \\
          \alginst{\Theta; \Delta''}{[\Delta'']\phi}{\Delta'}
        }
        {
          \algsub[+]{\Theta; \Delta}{P}{Q \land \phi}{[\Delta']\Wah \land [\Delta']\phi}{\Delta'}
        }
  \and 
  \Infer{\AlgSubPosExR}
        { 
          \algsub[+]{\Theta; \Delta, \ahat:\tau}{P}{[\ahat/a] Q}{\Wah}{\Delta', \hypeq{\ahat}{\tau}{t}}
        }
        {
          \algsub[+]{\Theta; \Delta}{P}{\extype{a:\tau} Q}{\Wah}{\Delta'}
        }
  \and
  \Infer{\AlgSubPosFix}
        {
          \alltype{\ahat \in \dom{\Delta}}{[\Delta]t' \neq \ahat}
          \\
          \algequiv[]{\Theta; \Delta}{F}{G}{W}{\Delta'}
        }
        {
          \algsub[+]
             {\Theta; \Delta}
             {\comprehend{\nu:\mu F}{\Fold{F}{\alpha}\,\nu =_\tau t}}
             {\comprehend{\nu:\mu G}{\Fold{G}{\alpha}\,\nu =_\tau t'}}
             {W \land (t = [\Delta']t')}
             {\Delta'}
        }
  \and
  \Infer{\AlgSubPosFixInst}
        {
          \ground{t}
          \\
          \algequiv[]{\Theta; \Delta}{F}{G}{W}{\Delta_1', \ahat:\tau, \Delta_2'}
          \\
          \Delta' = \Delta_1', \hypeq{\ahat}{\tau}{t}, \Delta_2'
        }
        {
          \algsub[+]
             {\Theta; \Delta}
             {\comprehend{\nu:\mu F}{\Fold{F}{\alpha}\,\nu =_\tau t}}
             {\comprehend{\nu:\mu G}{\Fold{G}{\alpha}\,\nu =_\tau \ahat}}
             {
               [\Delta']W \land (t = t)
             }
             {\Delta'}
        }
  \and
  \Infer{\AlgSubPosFixInstFun}
        {
          \algequiv[]{\Theta; \Delta}{F}{G}{W}{\Delta'}
          \\
          \hypeq{\ahat}{\tau}{t'} \in \Delta'
        }
        {
          \algsub[+]
             {\Theta; \Delta}
             {\comprehend{\nu:\mu F}{\Fold{F}{\alpha}\,\nu =_\tau t}}
             {\comprehend{\nu:\mu G}{\Fold{G}{\alpha}\,\nu =_\tau \ahat}}
             {\Wah \land (t=t')}
             {\Delta'}
        }
  \and
  \Infer{\AlgSubPosDownshift}
        {
          \judgeextract{\Theta; \Delta}{M}{M'}{\Thetahat}
        }
        {
          \algsub[+]{\Theta; \Delta}{\downshift{N}}{\downshift{M}}{\Thetahat \implies^\ast \negsubprob{N}{M'}}{\Delta}
        }
  \and
  \Infer{\AlgSubNegUpshift}
        {
          \judgeextract{\Theta; \Delta}{P}{P'}{\Thetahat}
        }
        {
          \algsub[-]{\Theta; \Delta}{\upshift{P}}{\upshift{Q}}{\Thetahat \implies^\ast \possubprob{P'}{Q}}{\Delta}
        }
  \and
  \Infer{\AlgSubNegImpL}
        {
          \algsub[-]{\Theta; \Delta}{N}{M}{\Wah}{\Delta'}
        }
        {
          \algsub[-]{\Theta; \Delta}{\phi \implies N}{M}
            {\Wah \land [\Delta']\phi}{\Delta'}
        }
  \and
  \Infer{\AlgSubNegAllL}
        {
          \algsub[-]{\Theta; \Delta,\ahat:\tau}{[\ahat/a]N}{M}{\Wah}{\Delta', \hypeq{\ahat}{\tau}{t}}
        }
        {\algsub[-]{\Theta; \Delta}{\alltype{a:\tau} N}{M}{\Wah}{\Delta'} }
  \and
  \Infer{\AlgSubNegArrow}
        {
          \algsub[+]{\Theta; \Delta}{Q}{P}{\Wah_1}{\Delta''}
          \\
          \algsub[-]{\Theta; \Delta''}{[\Delta'']N}{M}{\Wah_2}{\Delta'}
        }
        {
          \algsub[-]{\Theta; \Delta}{P -> N}{Q -> M}
            {[\Delta']\Wah_1 \land \Wah_2}{\Delta'}
        } 
\end{mathpar}

  \caption{Algorithmic subtyping}
  \label{fig:algsubtyping}
\end{figure}

\begin{figure}[htbp]
\raggedright

\judgbox{%
  \judgeunroll{\Xi}{\Theta; \Delta}{ \nu:G[\mu F] }{\beta}{G\;\Fold{F}{\alpha}\;\nu}{t}{P}{\tau}
}{
}
\vspace{1em}

\begin{mathpar}
  \Infer{\AlgUnrollSum}
  {\arrayenvbl{
      \composeinj{1}{\beta}{\beta_1}
      \\
      \composeinj{2}{\beta}{\beta_2}
    }
    \\
    \arrayenvbl{
      \judgeunroll{\Xi}{\Theta; \Delta}{\nu:G[\mu F]}{\beta_1}{G\;\Fold{F}{\alpha}\;\nu}{t}{P}{\tau}
      \\
      \judgeunroll{\Xi}{\Theta; \Delta}{\nu:H[\mu F]}{\beta_2}{H\;\Fold{F}{\alpha}\;\nu}{t}{Q}{\tau}
    }
  }
  { \judgeunroll{\Xi}{\Theta; \Delta}{\nu:(G \oplus H)[\mu F]}{\beta}{(G \oplus H)\;\Fold{F}{\alpha}\;\nu}{t}{P + Q}{\tau} }
  \and
  \Infer{\AlgUnrollId}
  {
    \judgeunroll{\Xi,a:\tau}{\Theta, a:\tau; \Delta}{\nu:\hat{P}[\mu F]}
    {(\clause{q}{t'})}{\hat{P}\;\Fold{F}{\alpha}\;\nu}{t}{Q}{\tau} 
  }
  {
    \judgeunroll*{\Xi}{\Theta; \Delta}{\nu : (\Id\otimes\hat{P})[\mu F]}
    {(\clause{(a,q)}{t'})}
    {(\Id\otimes\hat{P})\;\Fold{F}{\alpha}\;\nu}{t}
    {\extype{a:\tau}
      {\comprehend{\nu:\mu F}{ \Fold{F}{\alpha}\,{\nu} =_\tau a } \times Q}}
    {\tau}
  }
  \and
  \Infer{\AlgUnrollConstEx}
  {
    \judgeunroll{\Xi,a:\tau'}{\Theta,a:\tau'; \Delta}{\nu:(\Const{Q}\otimes\hat{P})[\mu F]}{(\clause{(\bap,q)}{t'})}{(\Const{Q}\otimes\hat{P})\;\Fold{F}{\alpha}\;\nu}{t}{Q'}{\tau} }
  { \judgeunroll*{\Xi}{\Theta; \Delta}{\nu :(\Const{\extype{a:\tau'}{Q}}\otimes\hat{P})[\mu F]}{(\clause{(\pack{a}{\bap}, q)}{t'})}{(\Const{\extype{a:\tau'}{Q}}\otimes\hat{P})\;\Fold{F}{\alpha}\;\nu}{t}{\extype{a:\tau'}{Q'}}{\tau} }
  \and
  \Infer{\AlgUnrollConst}
  {
    \judgeunroll{\Xi}{\Theta; \Delta}{\nu:\hat{P}[\mu F]}{(\clause{q}{t'})}{\hat{P}\;\Fold{F}{\alpha}\;\nu}{t}{Q'}{\tau} }
  { \judgeunroll{\Xi}{\Theta; \Delta}{\nu :(\Const{Q}\otimes\hat{P})[\mu F]}{(\clause{(\wild,q)}{t'})}{(\Const{Q}\otimes\hat{P})\;\Fold{F}{\alpha}\;\nu}{t}{Q \times Q'}{\tau} }
  \and
  \Infer{\AlgUnrollUnit}
  { }
  { \judgeunroll{\Xi}{\Theta; \Delta}{\nu:I[\mu F]}{(\clause{\unitexp}{t'})}{I\;\Fold{F}{\alpha}\;\nu}{t}{1 \land (t = t')}{\tau} }
\end{mathpar}

\caption{Algorithmic unrolling}
\label{fig:alg-unroll}
\end{figure}

\begin{figure}[htbp]
\raggedright

\judgbox{%
  \unroll{\Xi}{\Theta; \Delta}{G}{F}{\beta}{\alpha}{\tau}{t}{P}
}{
}
\vspace{1em}

\begin{mathpar}
  \Infer{\AlgUnrollSum}
  {
    \arrayenvbl
    {
      \composeinj{1}{\beta}{\beta_1}
      \\
      \composeinj{2}{\beta}{\beta_2}
    }
    \\
    \arrayenvbl
    {
      \unroll{\Xi}{\Theta; \Delta}{G_1}{F}{\beta_1}{\alpha}{\tau}{t}{P_1}
      \\
      \unroll{\Xi}{\Theta; \Delta}{G_2}{F}{\beta_2}{\alpha}{\tau}{t}{P_2}
    }
  }
  {
    \unroll{\Xi}{\Theta; \Delta}{G_1 \oplus G_2}{F}{\beta}{\alpha}{\tau}{t}{P_1 + P_2}
  }
  \and
  \Infer{\AlgUnrollId}
  {
    \unroll{\Xi,a:\tau}{\Theta, a:\tau; \Delta}{\hat{P}}{F}{\clause{q}{t'}}{\alpha}{\tau}{t}{Q} 
  }
  {
    \unroll{\Xi}{\Theta; \Delta}{\Id\otimes\hat{P}}{F}{\clause{(a,q)}{t'}}{\alpha}{\tau}{t}{\extype{a:\tau}{\comprehend{\nu:\mu F}{ \Fold{F}{\alpha}\,{\nu} =_\tau a } \times Q}}
  }
  \and
  \Infer{\AlgUnrollConstEx}
  {
    \unroll{\Xi,a:\tau'}{\Theta,a:\tau'; \Delta}{\Const{Q}\otimes\hat{P}}{F}{\clause{(\bap,q)}{t'}}{\alpha}{\tau}{t}{Q'}
  }
  {
    \unroll{\Xi}{\Theta; \Delta}{\Const{\extype{a:\tau'}{Q}}\otimes\hat{P}}{F}{\clause{(\pack{a}{\bap}, q)}{t'}}{\alpha}{\tau}{t}{\extype{a:\tau'}{Q'}}
  }
  \and
  \Infer{\AlgUnrollConst}
  {
    \unroll{\Xi}{\Theta; \Delta}{\hat{P}}{F}{\clause{q}{t'}}{\alpha}{\tau}{t}{Q'}
  }
  {
    \unroll{\Xi}{\Theta; \Delta}{\Const{Q}\otimes\hat{P}}{F}{\clause{(\wild,q)}{t'}}{\alpha}{\tau}{t}{Q \times Q'}
  }
  \and
  \Infer{\AlgUnrollUnit}
  { }
  {
    \unroll{\Xi}{\Theta; \Delta}{I}{F}{\clause{\unitexp}{t'}}{\alpha}{\tau}{t}{1 \land (t = t')}
  }
\end{mathpar}

\caption{Algorithmic unrolling (convenient shorthand)}
\label{fig:alg-unroll-convenient}
\end{figure}

\begin{figure}[htbp]
  \centering

  \judgbox{\algsynhead{\Theta}{\Gamma}{h}{P}}
      {Under $\Theta$ and $\Gamma$,
        head $h$ synthesizes $P$}

      \begin{mathpar}
        \Infer{\AlgSynHeadVar}
          {
            (x : P) \in \Gamma
          }
          {
            \algsynhead{\Theta}{\Gamma}{x}{P}
          }
        \and
        \Infer{\AlgSynValAnnot}
          {
            \judgetp{\Theta; \cdot}{P}{\Xi}
            \\
            \algchk{\Theta; \cdot}{\Gamma}{v}{P}{\chi}{\cdot}
            \\
            \algneg{\Theta}{\Gamma}{\chi}
          }
          {
            \algsynhead{\Theta}{\Gamma}{\annoexp{v}{P}}{P}
          }
      \end{mathpar}

  \judgbox{\algsynexp{\Theta}{\Gamma}{\be}{\upshift{P}}}
      {Under $\Theta$ and $\Gamma$,
        bound expression $\be$ synthesizes $\upshift{P}$}

      \begin{mathpar}
        \Infer{\AlgSynSpineApp}
          { 
            \algsynhead{\Theta}{\Gamma}{h}{\downshift{N}}
            \\
            \algspine{\Theta; \cdot}{\Gamma}{s}{N}{\upshift{P}}{\chi}{\cdot}
            \\
            \algneg{\Theta}{\Gamma}{\chi}
          }
          { \algsynexp{\Theta}{\Gamma}{h(s)}{\upshift{P}} }
        \and
        \Infer{\AlgSynExpAnnot}
          {
            \judgetp{\Theta; \cdot}{P}{\Xi}
            \\
            \algchkneg{\Theta}{\Gamma}{e}{\upshift{P}}
          }
          { \algsynexp{\Theta}{\Gamma}{\annoexp{e}{\upshift{P}}}{\upshift{P}} }
      \end{mathpar}

  \caption{Algorithmic head and bound expression synthesis}
  \label{fig:algtyping-head-bound-expression}
\end{figure}

\begin{figure}[htbp]
  \centering

  \judgbox{\algchk{\Theta; \Delta}{\Gamma}{v}{P}{\chi}{\Delta'}}
        {Under $\Theta$, $\Delta$, and $\Gamma$,
          the value $v$ checks against type $P$,\\
          with output computation constraints $\chi$ and output context $\Delta'$}
  
      \begin{mathpar}
        \Infer{\AlgChkValVar}
            {
              P \neq \exists, \with
              \\
              (x:Q)\in\Gamma
              \\
              \algsub[+]{\Theta; \Delta}{Q}{P}{\Wah}{\Delta'}
            }
            {
              \algchk{\Theta; \Delta}{\Gamma}{x}{P}{\Wah}{\Delta'}
            }
        \and
        \Infer{\AlgChkValUnit}
            {
            }
            {
              \algchk{\Theta; \Delta}{\Gamma}{\unit}{\unitty}{\cdot}{\Delta}
            }
        \and
        \Infer{\AlgChkValPair}
            {
              \algchk{\Theta; \Delta}{\Gamma}{v_1}{P_1}{\chi_1}{\Delta''}
              \\
              \algchk{\Theta; \Delta''}{\Gamma}{v_2}{[\Delta'']P_2}{\chi_2}{\Delta'}
            }
            {
              \algchk{\Theta; \Delta}{\Gamma}{\pair{v_1}{v_2}}
                {(P_1 \times P_2)}{[\Delta']\chi_1, \chi_2}{\Delta'}
            }
        \and
        \Infer{\AlgChkValIn{k}}
            {
              \algchk{\Theta; \Delta}{\Gamma}{v_k}{P_k}{\chi}{\Delta'}
            }
            {
              \algchk{\Theta; \Delta}{\Gamma}{\inj{k}{v_k}}{(P_1 + P_2)}{\chi}{\Delta'}
            }
        \and
        \Infer{\AlgChkValExists}
            {
              \algchk{\Theta; \Delta, \ahat:\tau}{\Gamma}{v}{[\ahat / a]P}{\chi}{\Delta', \hypeq{\ahat}{\tau}{t}}
            }
            {
              \algchk{\Theta; \Delta}{\Gamma}{v}{(\extype{a : \tau} P)}{\chi}{\Delta'}
            }
        \and
        \Infer{\AlgChkValWith}
            {
              \algchk{\Theta; \Delta}{\Gamma}{v}{P}{\chi}{\Delta''}
              \\
              \alginst{\Theta; \Delta''}{[\Delta'']\phi}{\Delta'}
            }
            {
              \algchk{\Theta; \Delta}{\Gamma}{v}{(P \andty \phi)}{([\Delta']\phi, [\Delta']\chi)}{\Delta'}
            }
        \and
        \Infer{\AlgChkValFix}
            {
              \judgeunroll{\cdot}{\Theta; \Delta}{\nu:F[\mu F]}{\alpha}{F\; \Fold{F}{\alpha}\;\nu}{t}{P}{\tau}
              \\
              \algchk{\Theta; \Delta}{\Gamma}{v}{P}{\chi}{\Delta'}
            }
            {
              \algchk{\Theta; \Delta}{\Gamma}{\into{v}}{\comprehend{\nu : \mu F}{\Fold{F}{\alpha}\,{\nu} =_\tau t}}{\chi}{\Delta'}
            }
        \and
        \Infer{\AlgChkValDownshift}
            {
            }
            {
              \algchk{\Theta; \Delta}{\Gamma}{\thunk{e}}{\downshift{N}}{(e <= N)}{\Delta}
            }
      \end{mathpar}

  \judgbox{\algchkneg{\Theta}{\Gamma}{e}{N}}
          {
            Under $\Theta$ and $\Gamma$,
            expression $e$ checks against $N$
          }
  \begin{mathpar}
    \Infer{\AlgChkExpUpshift}
          { 
            \algchk{\Theta; \cdot}{\Gamma}{v}{P}{\chi}{\cdot} 
            \\
            \algneg{\Theta}{\Gamma}{\chi}
          }
          {
            \algchkneg{\Theta}{\Gamma}{\Return{v}}{\upshift{P}} 
          }
    \and 
    \Infer{\AlgChkExpLet}
          {
            \simple{\Theta}{N}
            \\
            \algsynexp{\Theta}{\Gamma}{\be}{\upshift{P}}
            \\
            \judgeextract{\Theta; \cdot}{P}{P'}{\Theta'}
            \\
            \algchkneg{\Theta, \Theta'}{\Gamma, x:P'}{e}{N}
          }
          {
            \algchkneg{\Theta}{\Gamma}{\Let{x}{\be}{e}}{N}
          }
    \and
    \Infer{\AlgChkExpMatch}
          {
            \simple{\Theta}{N}
            \\
            \algsynhead{\Theta}{\Gamma}{h}{P}
            \\
            \algchkmatch{\Theta}{\Gamma}{P}{\clauses{\pa}{e}{i}{I}}{N}
          }
          {
            \algchkneg{\Theta}{\Gamma}{\match{h}{\clauses{\pa}{e}{i}{I}}}{N}
          }
    \and 
    \Infer{\AlgChkExpLam}
          {
            \simple{\Theta}{P \to N}
            \\
            \algchkneg{\Theta}{\Gamma, x:P}{e}{N}
          }
          {
            \algchkneg{\Theta}{\Gamma}{\fun{x}{e}}{P \to N}
          }
    \and
    \Infer{\AlgChkExpUnreachable}
    {
      \simple{\Theta}{N}
      \\
      \judgeentail{\Theta}{\False}
    }
    {
      \algchkneg{\Theta}{\Gamma}{\unreachable}{N}
    }
    \and
    \Infer{\AlgChkExpRec}
    {
      \arrayenvb{
        \simple{\Theta}{N}
        \and
        \algsub[-]{\Theta; \cdot}{\alltype{a:\kindnat} M}{N}{W}{\cdot}
        \and
        \entailwah{\Theta}{W}
      }
      \\
      \algchkneg{\Theta, a:\kindnat}{\Gamma, x:\downshift{\big(\alltype{a':\kindnat} (a' < a) \implies [a'/a]M}\big)}{e}{M}
    }
    {
      \algchkneg{\Theta}{\Gamma}{\rec{x : (\alltype{a:\kindnat} M)}{e}}{N}
    }
    \and
    \Infer{\AlgChkExpExtract}
    {
      \judgeextract{\Theta; \cdot}{N}{N'}{\Theta_N}
      \\
      \Theta_N \neq \cdot
      \\
      \algchkneg{\Theta, \Theta_N}{\Gamma}{e}{N'}
    }
    {
      \algchkneg{\Theta}{\Gamma}{e}{N}
    }
  \end{mathpar}

  \caption{Algorithmic value and expression checking}
  \label{fig:algtyping-value-expression}
\end{figure}

\begin{figure}[htbp]
  \centering
\judgbox{\algchkmatch{\Theta}{\Gamma}{P}{\clauses{\pa}{e}{i}{I}}{N}}
        {Under $\Theta$ and $\Gamma$, patterns $r_i$ match against (input) type $P$ \\
          and branch expressions $e_i$ check against type $N$}
\begin{mathpar}
  \Infer{\AlgChkMatchEx}
        {
          \algchkmatch{\Theta, a:\tau}{\Gamma}{P}{\clauses{\pa}{e}{i}{I}}{N}
        }
        {
          \algchkmatch{\Theta}{\Gamma}{\extype{a:\tau}{P}}{\clauses{\pa}{e}{i}{I}}{N}
        }
  \and
  \Infer{\AlgChkMatchWith}
        {
          \algchkmatch{\Theta, \phi}{\Gamma}{P}{\clauses{\pa}{e}{i}{I}}{N}
        }
        {
          \algchkmatch{\Theta}{\Gamma}{P \land \phi}{\clauses{\pa}{e}{i}{I}}{N}
        }
  \and
  \Infer{\AlgChkMatchUnit}
        {
          \algchkneg{\Theta}{\Gamma}{e}{N}
        }
        {
          \algchkmatch{\Theta}{\Gamma}{1}{\setof{\clause{\unit}{e}}}{N}
        }
  \and
  \Infer{\AlgChkMatchPair}
        {
          \judgeextract{\Theta; \cdot}{P_1}{P_1'}{\Theta_1}
          \\
          \judgeextract{\Theta; \cdot}{P_2}{P_2'}{\Theta_2}
          \\
          \algchkneg{\Theta, \Theta_1, \Theta_2}{\Gamma, x_1:P_1', x_2:P_2'}{e}{N}
        }
        {
          \algchkmatch{\Theta}{\Gamma}{P_1 \times P_2}{\setof{\clause{\pair{x_1}{x_2}}{e}}}{N}
        }
  \and
  \Infer{\AlgChkMatchSum}
        {
          \arrayenvb{
            \judgeextract{\Theta; \cdot}{P_1}{P_1'}{\Theta_1}
            \\
            \judgeextract{\Theta; \cdot}{P_2}{P_2'}{\Theta_2}
          }
          \\
          \arrayenvb{
            \algchkneg{\Theta, \Theta_1}{\Gamma, x_1:P_1'}{e_1}{N}
            \\
            \algchkneg{\Theta, \Theta_2}{\Gamma, x_2:P_2'}{e_2}{N}
          }
        }
        {
          \algchkmatch{\Theta}{\Gamma}{P_1 + P_2}{\setof{ \clause{\inl{x_1}}{e_1} \bnfalt \clause{\inr{x_2}}{e_2}}}{N}
        }
  \and
  \Infer{\AlgChkMatchVoid}
        { }
        {
          \algchkmatch{\Theta}{\Gamma}{0}{\setof{}}{N}
        }
  \and
  \Infer{\AlgChkMatchFix}
        {
          \judgeunroll{\cdot}{\Theta; \cdot}{\nu:F[\mu F]}{\alpha}{F\;\Fold{F}{\alpha}\;\nu}{t}{Q}{\tau}
          \\
          \judgeextract{\Theta; \cdot}{Q}{Q'}{\Theta_Q}
          \\
          \algchkneg{\Theta, \Theta_Q}{\Gamma, x:Q'}{e}{N}
        }
        {
          \algchkmatch{\Theta}{\Gamma}{\comprehend{\nu:\mu F}{\Fold{F}{\alpha}\,{\nu} =_\tau t}}{\setof{\clause{\roll{x}}{e}}}{N}
        }
\end{mathpar}

  \judgbox{\algspine{\Theta; \Delta}{\Gamma}{s}{N}{\upshift{P}}{\chi}{\Delta'}}
        {
          Under $\Theta$, $\Delta$, and $\Gamma$,
          passing spine $s$ to a head of type $\downshift{N}$\\
          synthesizes $\upshift{P}$,
          with output constraints $\chi$ and context $\Delta'$
        }
  \begin{mathpar}
    \Infer{\AlgSpineAll}
        {
          \algspine{\Theta; \Delta, \ahat : \tau}{\Gamma}{s}{[\ahat/a]{N}}{\upshift{P}}{\chi}{\Delta', \hypeq{\ahat}{\tau}{t}}
        }
        {
          \algspine{\Theta; \Delta}{\Gamma}{s}{\alltype{a:\tau}N}{\upshift{P}}{\chi}{\Delta'}
        }
   \and
   \Infer{\AlgSpineImplies}
        {
          \algspine{\Theta; \Delta}{\Gamma}{s}{N}{\upshift{P}}{\chi}{\Delta'}
        }
        {
          \algspine{\Theta; \Delta}{\Gamma}{s}{\phi \implies N}{\upshift{P}}{[\Delta']\phi,\chi}{\Delta'}
        }
    \and
    \Infer{\AlgSpineApp}
        {
          \algchk{\Theta; \Delta}{\Gamma}{v}{Q}{\chi}{\Delta''}
          \\
          \algspine{\Theta; \Delta''}{\Gamma}{s}{[\Delta'']N}{\upshift{P}}{\chi'}{\Delta'}
        }
        {
          \algspine{\Theta; \Delta}{\Gamma}{v, s}{Q -> N}{\upshift{P}}{[\Delta']\chi, \chi'}{\Delta'}
        }
    \and
    \Infer{\AlgSpineNil}
        {}
        {
          \algspine{\Theta; \Delta}{\Gamma}{\cdot}{\upshift{P}}{\upshift{P}}{\True}{\Delta}
        }
  \end{mathpar}

  \caption{Algorithmic match and spine checking}
  \label{fig:algtyping-match-spine}
\end{figure}

\begin{figure}[htbp]
  \judgbox{\entailwah{\Theta}{\Wah}}{Under $\Theta$, constraint $\Wah$ algorithmically holds}
  \begin{mathpar}
    \Infer{\WTrueProp}
    {
      \judgeentail{\Theta}{\phi}
    }
    {
      \entailwah{\Theta}{\phi}
    }
    \and
    \Infer{\WTruePrpEquiv}
    {
      \judgeequiv[]{\Theta}{\phi}{\psi}
    }
    {
      \entailwah{\Theta}{\phi \equiv \psi}
    }
    \\
    \Infer{\WTrueAnd}
    {
      \entailwah{\Theta}{\Wah_1}
      \\
      \entailwah{\Theta}{\Wah_2}
    }
    {
      \entailwah{\Theta}{\Wah_1 \land \Wah_2}
    }
    \and
    \Infer{\WTrueImpl}
    {
      \entailwah{\Theta, \phi}{\Wah}
    }
    {
      \entailwah{\Theta}{\phi \implies \Wah}
    }
    \and
    \Infer{\WTrueAll}
    {
      \entailwah{\Theta,a:\tau}{\Wah}
    }
    {
      \entailwah{\Theta}{\alltype{a:\tau}\Wah}
    }
    \\
    \Infer{\WTruePosSub}
    {
      \algsub[+]{\Theta; \cdot}{P}{Q}{\Wah}{\cdot}
      \\
      \entailwah{\Theta}{\Wah}
    }
    {
      \entailwah{\Theta}{\possubprob{P}{Q}}
    }
    \and
    \Infer{\WTrueNegSub}
    {
      \algsub[-]{\Theta; \cdot}{N}{M}{\Wah}{\cdot}
      \\
      \entailwah{\Theta}{\Wah}
    }
    {
      \entailwah{\Theta}{\negsubprob{N}{M}}
    }
    \\
    \Infer{\WTruePosEquiv}
    {
      \algequiv[+]{\Theta; \cdot}{P}{Q}{\Wah}{\cdot}
      \\
      \entailwah{\Theta}{\Wah}
    }
    {
      \entailwah{\Theta}{\poseqprob{P}{Q}}
    }
    \and
    \Infer{\WTrueNegEquiv}
    {
      \algequiv[-]{\Theta; \cdot}{N}{M}{\Wah}{\cdot}
      \\
      \entailwah{\Theta}{\Wah}
    }
    {
      \entailwah{\Theta}{\negeqprob{N}{M}}
    }
  \end{mathpar}

  \judgbox{\algneg{\Theta}{\Gamma}{\chi}}
  {
    Under $\Theta$ and $\Gamma$,
    constraints $\chi$ algorithmically hold
  }
  \begin{mathpar}
    \Infer{\ChkProblemsEmpty}
    {
    }
    {
      \algneg{\Theta}{\Gamma}{\cdot}
    }
    \and
    \Infer{\ChkProblemsNegChk}
    {
      \algchkneg{\Theta}{\Gamma}{e}{N}
      \\
      \algneg{\Theta}{\Gamma}{\chi}
    }
    {
      \algneg{\Theta}{\Gamma}{(e <= N), \chi}
    }
    \and
    \Infer{\ChkProblemsWah}
    {
      \entailwah{\Theta}{W}
      \\
      \algneg{\Theta}{\Gamma}{\chi}
    }
    {
      \algneg{\Theta}{\Gamma}{W, \chi}
    }
  \end{mathpar}

  \caption{Constraint verification}
  \label{fig:entail-wah}
\end{figure}

\begin{figure}[htbp]
  \judgbox{\extend{\Theta}{\Delta}{\Delta'}}{Under $\Theta$, algorithmic context $\Delta$ extends to $\Delta'$}
  \begin{mathpar}
    \Infer{\ExtEmpty}
    {}
    {\extend{\Theta}{\cdot}{\cdot}}
    \and
    \Infer{\ExtUnsolved}
    {\extend{\Theta}{\Delta}{\Delta'}}
    {\extend{\Theta}{\Delta, \ahat:\tau}{\Delta', \ahat:\tau}}
    \and
    \Infer{\ExtSolved}
    {\extend{\Theta}{\Delta}{\Delta'}}
    {\extend{\Theta}{\Delta, \hypeq{\ahat}{\tau}{t}}{\Delta', \hypeq{\ahat}{\tau}{t}}}
    \and
    \Infer{\ExtSolve}
    {\extend{\Theta}{\Delta}{\Delta'}}
    {\extend{\Theta}{\Delta, \ahat:\tau}{\Delta', \hypeq{\ahat}{\tau}{t}}}
  \end{mathpar}

  \caption{Algorithmic context extension}
  \label{fig:algextension}
\end{figure}

\begin{figure}[htbp]
  \judgbox{\rextend{\Theta}{\Delta}{\Delta'}}{Under $\Theta$, algorithmic context $\Delta$ relaxedly extends to $\Delta'$}
  \begin{mathpar}
    \Infer{\RExtEmpty}
    {}
    {\rextend{\Theta}{\cdot}{\cdot}}
    \and
    \Infer{\RExtUnsolved}
    {\rextend{\Theta}{\Delta}{\Delta'}}
    {\rextend{\Theta}{\Delta, \ahat:\tau}{\Delta', \ahat:\tau}}
    \and
    \Infer{\RExtSolved}
    {\rextend{\Theta}{\Delta}{\Delta'} \\ \judgeentail{\Theta}{t=t'}}
    {\rextend{\Theta}{\Delta, \hypeq{\ahat}{\tau}{t}}{\Delta', \hypeq{\ahat}{\tau}{t'}}}
    \and
    \Infer{\RExtSolve}
    {\rextend{\Theta}{\Delta}{\Delta'}}
    {\rextend{\Theta}{\Delta, \ahat:\tau}{\Delta', \hypeq{\ahat}{\tau}{t}}}
  \end{mathpar}

  \caption{Relaxed algorithmic context extension}
  \label{fig:relaxedalgextension}
\end{figure}

\begin{figure}[htbp]
  \begin{equation*}
    \begin{aligned}[c]
      \size[A]{1}                   &= 0 \\
      \size[A]{P \times Q}          &= \size{P} + \size{Q} + 1 \\
      \size[A]{0}                   &= 0 \\
      \size[A]{P + Q}               &= \size{P} + \size{Q} + 1 \\
      \size[A]{\downshift{N}}       &= \size{N} + 1 \\
      \size[A]{\comprehend{\nu : \mu F}{\Fold{F}{\alpha}\,{\nu} =_\tau t}} &= \size{F} + 1 \\
      \size[A]{\extype{a:\tau}{P}}  &= \size{P} + 1 \\
      \size[A]{P \land \phi}        &= \size{P} + 1 \\
      \size[A]{P \to N}             &= \size{P} + \size{N} + 1 \\
      \size[A]{\upshift{P}}         &= \size{P} + 1 \\
      \size[A]{\alltype{a:\tau}{N}} &= \size{N} + 1 \\
      \size[A]{\phi \implies N}     &= \size{N} + 1
    \end{aligned}
    \qquad
    \begin{aligned}[c]
      \size[\F]{F_1 \oplus F_2}         &= \size{F_1} + \size{F_2} + 1 \\
      \size[\F]{I}                      &= 0 \\
      \size[\F]{B \otimes \Ptype}        &= \size{B} + \size{\Ptype} + 1 \\
      \size[\F]{\Const{P}}              &= \size{P} + 1 \\
      \size[\F]{\Id}                    &= 0 \\
      ~\\
      \size[W]{\phi}                    &= 0 \\
      \size[W]{\propeqprob{\phi}{\psi}} &= 0 \\
      \size[W]{\phi \implies W}         &= \size{W} + 1 \\
      \size[W]{W_1 \land W_2}           &= \size{W_1} + \size{W_2} + 1 \\
      \size[W]{\alltype{a:\tau}{W}}     &= \size{W} + 1 \\
      \size[W]{\subprob{A}{B}}          &= \size{A} + \size{B} + 1 \\
      \size[W]{\eqprob{A}{B}}           &= \size{A} + \size{B} + 1
    \end{aligned}
  \end{equation*}

  The size $\size{\Theta}$ of a logical context $\Theta$ is its length.

  \caption{Size of types $A$, functors $\mathcal{F}$, constraints $W$, and logical contexts $\Theta$}
  \label{fig:size}
\end{figure}

\begin{figure}
  \begin{equation*}
    \begin{aligned}[c]
      \size[v]{x} &= 0 \\
      \size[v]{\unit} &= 0 \\
      \size[v]{\pair{v_1}{v_2}} &= \size{v_1} + \size{v_2} + 1 \\
      \size[v]{\inj{k}{v}} &= \size{v} + 1 \\
      \size[v]{\roll{v}} &= \size{v} + 1 \\
      \size[v]{\thunk{e}} &= \size{e} + 1 \\
      ~\\
      \size[h]{x} &= 0 \\
      \size[h]{\annoexp{v}{P}} &= \size{v} + 1 \\
      ~\\
      \size[e]{\Return{v}} &= \size{v} + 1\\
      \size[e]{\Let{x}{\be}{e}} &= \size{\be} + \size{e} + 1 \\
      \size[e]{\match{h}{\clauses{\pa}{e}{i}{I}}} &= \size{h} + \size{\clauses{\pa}{e}{i}{I}} + 1 \\
      \size[e]{\fun{x}{e}} &= \size{e} + 1 \\
      \size[e]{\unreachable} &= 0 \\
      \size[e]{\rec{x:M}{e}} &= \size{e} + 1 \\
      ~\\
      \size[\be]{h(s)} &= \size{h} + \size{s} + 1 \\
      \size[\be]{\annoexp{e}{N}} &= \size{e} + 1 \\
      ~\\
      \size[s]{\cdot} &= 0 \\
      \size[s]{v, s} &= \size{v} + \size{s} + 1 \\
      ~\\
      \size{\clauses{\pa}{e}{i}{I}} &= 1 + \sum_{i \in I} \size{e_i}
    \end{aligned}
    \qquad
    \begin{aligned}[c]
      \size[\chi]{\cdot} &= 0 \\
      \size[\chi]{W, \chi} &= \size{\chi} \\
      \size[\chi]{(e <= N), \chi} &= \size{e} + \size{\chi} + 1\\
      ~\\
      \size[\sigma]{\cdot} &= 0 \\
      \size[\sigma]{\sigma, \subs{v}{P}{x}} &= \size{\sigma} + \begin{cases} 0 &\quad\text{if }v=x \\ 1 &\quad\text{else} \end{cases} \\
      \size[\sigma]{\sigma, t/a} &= \size{\sigma} + \begin{cases} 0 &\quad\text{if }t=a \\ 1 &\quad\text{else} \end{cases}
    \end{aligned}
  \end{equation*}
  
  \caption{Size of program terms, constraints $\chi$, and substitutions $\sigma$}
  \label{fig:size-prog-chi}
\end{figure}

\clearpage

\subsection{Intermediate Systems for Algorithmic Metatheory}
\label{sec:intermediate-figs}
\begin{figure}[htbp]
\judgbox{\resjudgesub[\pm]{\Theta}{A}{B}}{Under $\Theta$, type $A$ is declaratively a subtype of $B$ (but using eager extraction at shifts)}
\begin{mathpar}
  \Infer{\ResDeclSubPosUnit}
  { }
  {
    \resjudgesub[+]{\Theta}{1}{1}
  }
  \and
  \Infer{\ResDeclSubPosVoid}
  { }
  {
    \resjudgesub[+]{\Theta}{0}{0}
  }
  \\
  \Infer{\ResDeclSubPosProd}
  {
    \resjudgesub[+]{\Theta}{P_1}{Q_1}
    \\
    \resjudgesub[+]{\Theta}{P_2}{Q_2}
  }
  {
    \resjudgesub[+]{\Theta}{P_1 \times P_2}{Q_1 \times Q_2}
  }
  \and
  \Infer{\ResDeclSubPosSum}
  {
    \judgeequiv[+]{\Theta}{P_1}{Q_1}
    \\
    \judgeequiv[+]{\Theta}{P_2}{Q_2}
  }
  {
    \resjudgesub[+]{\Theta}{P_1 + P_2}{Q_1 + Q_2}
  }
  \\
  \Infer{\ResDeclSubPosWithR}
  {
    \resjudgesub[+]{\Theta}{P}{Q}
    \\
    \judgeentail{\Theta}{\phi}
  }
  {
    \resjudgesub[+]{\Theta}{P}{Q \land \phi}
  }
  \and
  \Infer{\ResDeclSubPosExR}
  { 
    \resjudgesub[+]{\Theta}{P}{[t/a]Q} 
    \\
    \judgeterm{\Theta}{t}{\tau} 
  }
  {
    \resjudgesub[+]{\Theta}{P}{\extype{a:\tau}{Q}}
  }
  \and
  \Infer{\ResDeclSubPosFix}
  {
    \judgeequiv[]{\Theta}{F}{G}
    \\
    \judgeentail{\Theta}{t = t'}
  }
  {
    \resjudgesub[+]{\Theta}{\comprehend{\nu:\mu F}{\Fold{F}{\alpha}\,\nu =_\tau t}}{\comprehend{\nu:\mu G}{\Fold{G}{\alpha}\,\nu =_\tau t'}}
  }
  \\
  \Infer{\ResDeclSubPosDownshift}
  {
    \judgeextract[]{\Theta}{M}{M'}{\Theta'}
    \\
    \resjudgesub[-]{\Theta, \Theta'}{N}{M'}
  }
  {
    \resjudgesub[+]{\Theta}{\downshift{N}}{\downshift{M}}
  }
  \and
  \Infer{\ResDeclSubNegUpshift}
  {
    \judgeextract[]{\Theta}{P}{P'}{\Theta'}
    \\
    \resjudgesub[+]{\Theta, \Theta'}{P'}{Q}
  }
  {
    \resjudgesub[-]{\Theta}{\upshift{P}}{\upshift{Q}}
  }
  \\
  \Infer{\ResDeclSubNegImpL}
  {
    \resjudgesub[-]{\Theta}{N}{M} \\ \judgeentail{\Theta}{\phi}
  }
  {
    \resjudgesub[-]{\Theta}{\phi \implies N}{M}
  }
  \and
  \Infer{\ResDeclSubNegAllL}
  {
    \resjudgesub[-]{\Theta}{[t/a]N}{M} 
    \\ 
    \judgeterm{\Theta}{t}{\tau} 
  }
  {\resjudgesub[-]{\Theta}{\alltype{a:\tau}{N}}{M} }
  \and
  \Infer{\ResDeclSubNegArrow}
  {
    \resjudgesub[+]{\Theta}{Q}{P} \\ \resjudgesub[-]{\Theta}{N}{M}
  }
  {
    \resjudgesub[-]{\Theta}{P \to N}{Q \to M}
  }  
\end{mathpar}

\caption{Restricted declarative subtyping}
\label{fig:resdeclsub}

\end{figure}

\begin{figure}[htbp]
\judgbox{\semideclequiv[]{\Theta}{\mathcal{F}}{\mathcal{G}}{W}}
        {Under $\Theta$, functors $\mathcal{F}$ and $\mathcal{G}$ are semideclaratively equivalent,\\that is, declaratively equivalent if output constraint $W$ holds}
\begin{mathpar}
  \Infer{\SemiDeclFunEquivConst}
  {
    \semideclequiv[+]{\Theta}{P}{Q}{W}
  }
  {
    \semideclequiv[]{\Theta}{\Const{P}}{\Const{Q}}{W}
  }
  \and 
  \Infer{\SemiDeclFunEquivId}
  {}
  {
    \semideclequiv[]{\Theta}{\Id}{\Id}{\True}
  }
  \and 
  \Infer{\SemiDeclFunEquivUnit}
  {}
  {
    \semideclequiv[]{\Theta}{I}{I}{\True}
  }
  \and 
  \Infer{\SemiDeclFunEquivProd}
  {
    \semideclequiv[]{\Theta}{B}{B'}{W_1}
    \\
    \semideclequiv[]{\Theta}{\hat{P}}{\hat{P}'}{W_2}
  }
  {
    \semideclequiv[]{\Theta}{B \otimes \hat{P}}{B' \otimes \hat{P}'}{W_1 \land W_2}
  }
  \and 
  \Infer{\SemiDeclFunEquivSum}
  {
    \semideclequiv[]{\Theta}{F_1}{G_1}{W_1}
    \\
    \semideclequiv[]{\Theta}{F_2}{G_2}{W_2}
  }
  {
    \semideclequiv[]{\Theta}{F_1 \oplus F_2}{G_1 \oplus G_2}{W_1 \land W_2}
  }
\end{mathpar}

\caption{Semideclarative functor equivalence}
\label{fig:semideclfunequiv}
\end{figure}

\begin{figure}[htbp]
\judgbox{\semideclequiv[\pm]{\Theta}{A}{B}{W}}{Under $\Theta$, types $A$ and $B$ are semideclaratively equivalent,\\that is, declaratively equivalent if output constraint $W$ holds}
\begin{mathpar}
  \Infer{\SemiDeclTpEquivPosUnit}
  { }
  {
    \semideclequiv[+]{\Theta}{1}{1}{\True}
  }
  \and
  \Infer{\SemiDeclTpEquivPosVoid}
  { }
  {
    \semideclequiv[+]{\Theta}{0}{0}{\True}
  }
  \and
  \Infer{\SemiDeclTpEquivPosProd}
  {
    \semideclequiv[+]{\Theta}{P_1}{Q_1}{W_1}
    \\
    \semideclequiv[+]{\Theta}{P_2}{Q_2}{W_2}
  }
  {
    \semideclequiv[+]{\Theta}{P_1 \times P_2}{Q_1 \times Q_2}{W_1 \land W_2}
  }
  \and
  \Infer{\SemiDeclTpEquivPosSum}
  {
    \semideclequiv[+]{\Theta}{P_1}{Q_1}{W_1}
    \\
    \semideclequiv[+]{\Theta}{P_2}{Q_2}{W_2}
  }
  {
    \semideclequiv[+]{\Theta}{P_1 + P_2}{Q_1 + Q_2}{W_1 \land W_2}
  }
  \and
  \Infer{\SemiDeclTpEquivPosWith}
  {
    \semideclequiv[+]{\Theta}{P}{Q}{W}
  }
  {
    \semideclequiv[+]{\Theta}{P \land \phi}{Q \land \psi}{W \land (\phi \equiv \psi)}
  }
  \and
  \Infer{\SemiDeclTpEquivPosEx}
  {
    \semideclequiv[+]{\Theta, a:\tau}{P}{Q}{W}
  }
  {
    \semideclequiv[+]{\Theta}{\extype{a:\tau}{P}}{\extype{a:\tau}{Q}}{\alltype{a:\tau}{W}}
  }
  \and
  \Infer{\SemiDeclTpEquivPosFix}
  {
    \semideclequiv[]{\Theta}{F}{G}{W}
  }
  {
    \semideclequiv[+]{\Theta}{\comprehend{\nu:\mu F}{\Fold{F}{\alpha}\,\nu =_\tau t}}{\comprehend{\nu:\mu G}{\Fold{G}{\alpha}\,\nu =_\tau t'}}{W \land (t=t')}
  }
  \and
  \Infer{\SemiDeclTpEquivPosDownshift}
  { }
  {
    \semideclequiv[+]{\Theta}{\downshift{N}}{\downshift{M}}{\negeqprob{N}{M}}
  }
  \and
  \Infer{\SemiDeclTpEquivNegUpshift}
  { }
  {
    \semideclequiv[-]{\Theta}{\upshift{P}}{\upshift{Q}}{\poseqprob{P}{Q}}
  }
  \and
  \Infer{\SemiDeclTpEquivNegImp}
  {
    \semideclequiv[-]{\Theta}{N}{M}{W}
  }
  {
    \semideclequiv[-]{\Theta}{\psi \implies N}{\phi \implies M}{W \land (\psi \equiv \phi)}
  }
  \and
  \Infer{\SemiDeclTpEquivNegAll}
  {
    \semideclequiv[-]{\Theta, a:\tau}{N}{M}{W}
  }
  {
    \semideclequiv[-]{\Theta}{\alltype{a:\tau}{N}}{\alltype{a:\tau}{M}}{\alltype{a:\tau}{W}}
  }
  \and
  \Infer{\SemiDeclTpEquivNegArrow}
  {
    \semideclequiv[+]{\Theta}{P}{Q}{W_1}
    \\
    \semideclequiv[-]{\Theta}{N}{M}{W_2}
  }
  {
    \semideclequiv[-]{\Theta}{P \to N}{Q \to M}{W_1 \land W_2}
  }  
\end{mathpar}

\caption{Semideclarative type equivalence}
\label{fig:semidecltpequiv}
\end{figure}

\begin{figure}[htbp]

  \judgbox{\semideclsub[\pm]{\Theta}{A}{B}{\Wah}}
      {
        Under $\Theta$,
        type $A$ is semideclaratively a subtype of $B$,
        with output constraint $\Wah$
      }

\begin{mathpar}
  \Infer{\SemiDeclSubPosVoid}
  {}
  {
    \semideclsub[+]{\Theta}{0}{0}{\True}
  }
  \and
  \Infer{\SemiDeclSubPosUnit}
  { }
  {
    \semideclsub[+]{\Theta}{1}{1}{\True}
  }
  \and
  \Infer{\SemiDeclSubPosProd}
  {
    \semideclsub[+]{\Theta}{P_1}{Q_1}{\Wah_1}
    \\
    \semideclsub[+]{\Theta}{P_2}{Q_2}{\Wah_2}
  }
  {
    \semideclsub[+]{\Theta}{P_1 \times P_2}{Q_1 \times Q_2}{\Wah_1 \land \Wah_2}
  }
  \and
  \Infer{\SemiDeclSubPosSum}
  {
    \semideclequiv[+]{\Theta}{P_1}{Q_1}{\Wah_1}
    \\
    \semideclequiv[+]{\Theta}{P_2}{Q_2}{\Wah_2}
  }
  {
    \semideclsub[+]{\Theta}{P_1 + P_2}{Q_1 + Q_2}{\Wah_1 \land \Wah_2}
  }
  \\
  \Infer{\SemiDeclSubPosWithR}
  {
    \semideclsub[+]{\Theta}{P}{Q}{\Wah}
  }
  {
    \semideclsub[+]{\Theta}{P}{Q \land \phi}{\Wah \land \phi}
  }
  \and
  \Infer{\SemiDeclSubPosExR}
  { 
    \semideclsub[+]{\Theta}{P}{[t/a]Q}{\Wah}
    \\
    \judgeterm{\Theta}{t}{\tau}
  }
  {
    \semideclsub[+]{\Theta}{P}{\extype{a:\tau} Q}{\Wah}
  }
  \\
  \Infer{\SemiDeclSubPosFix}
  {
    \semideclequiv{\Theta}{F}{G}{W}
  }
  {
    \semideclsub[+]{\Theta}{\comprehend{\nu:\mu F}{\Fold{F}{\alpha}\,\nu =_\tau t}}{\comprehend{\nu:\mu G}{\Fold{G}{\alpha}\,\nu =_\tau t'}}{W \land t = t'}
  }
  \\
  \Infer{\SemiDeclSubPosDownshift}
  {
    \judgeextract[-]{\Theta}{M}{M'}{\Theta'}
  }
  {
    \semideclsub[+]{\Theta}{\downshift{N}}{\downshift{M}}{\Theta' \implies^\ast \negsubprob{N}{M'}}
  }
  \and
  \Infer{\SemiDeclSubNegUpshift}
  {
    \judgeextract[+]{\Theta}{P}{P'}{\Theta'}
  }
  {
    \semideclsub[-]{\Theta}{\upshift{P}}{\upshift{Q}}{\Theta' \implies^\ast \possubprob{P'}{Q}}
  }
  \\
  \Infer{\SemiDeclSubNegImpL}
  {
    \semideclsub[-]{\Theta}{N}{M}{\Wah}
  }
  {
    \semideclsub[-]{\Theta}{\phi \implies N}{M}{\Wah \land \phi}
  }
  \and
  \Infer{\SemiDeclSubNegAllL}
  {
    \semideclsub[-]{\Theta}{[t/a]N}{M}{\Wah}
    \\
    \judgeterm{\Theta}{t}{\tau}
  }
  {
    \semideclsub[-]{\Theta}{\alltype{a:\tau} N}{M}{\Wah}
  }
  \\
  \Infer{\SemiDeclSubNegArrow}
  {
    \semideclsub[+]{\Theta}{Q}{P}{\Wah_1}
    \\
    \semideclsub[-]{\Theta}{N}{M}{\Wah_2}
  }
  {
    \semideclsub[-]{\Theta}{P -> N}{Q -> M}{\Wah_1 \land \Wah_2}
  } 
\end{mathpar}

  \caption{Semideclarative subtyping}
  \label{fig:semideclsubtyping}
\end{figure}

\begin{figure}[htbp]
  \raggedright
  \judgbox{\semideclsynhead{\Theta}{\Gamma}{h}{P}}
  {Under $\Theta$ and $\Gamma$, head $h$ semideclaratively synthesizes type $P$}
  \begin{mathpar}
    \Infer{\SemiDeclSynHeadVar}
    {
      (x : P) \in \Gamma
    }
    {
      \semideclsynhead{\Theta}{\Gamma}{x}{P}
    }
    \and
    \Infer{\SemiDeclSynValAnnot}
    {
      \judgetp{\Theta}{P}{\Xi}
      \\
      \semideclchkval{\Theta}{\Gamma}{v}{P}{\chi}
      \\
      \semideclneg{\Theta}{\Gamma}{\chi}
    }
    {
      \semideclsynhead{\Theta}{\Gamma}{\annoexp{v}{P}}{P}
    }
  \end{mathpar}

  \judgbox{\semideclsynexp{\Theta}{\Gamma}{\be}{\upshift{P}}}
  {Under $\Theta$ and $\Gamma$, bound expression $\be$ semideclaratively synthesizes type $\upshift{P}$}
  \begin{mathpar}
    \Infer{\SemiDeclSynSpineApp}
    {
      \semideclsynhead{\Theta}{\Gamma}{h}{\downshift{N}}
      \\
      \semideclspine{\Theta}{\Gamma}{s}{N}{\upshift{P}}{\chi}
      \\
      \semideclneg{\Theta}{\Gamma}{\chi}
    }
    {
      \semideclsynexp{\Theta}{\Gamma}{h(s)}{\upshift{P}}
    }
    \and
    \Infer{\SemiDeclSynExpAnnot}
    {
      \judgetp{\Theta}{P}{\Xi}
      \\
      \semideclchkexp{\Theta}{\Gamma}{e}{\upshift{P}}
    }
    {
      \semideclsynexp{\Theta}{\Gamma}{\annoexp{e}{\upshift{P}}}{\upshift{P}}
    }
  \end{mathpar}

  \caption{Semideclarative head and bound expression type synthesis}
  \label{fig:semidecltyping-head-bound-expression}
\end{figure}

\begin{figure}[htbp]
  \raggedright
  \judgbox{\semideclchkval{\Theta}{\Gamma}{v}{P}{\chi}}
  {Under $\Theta$ and $\Gamma$, value $v$ semideclaratively checks against type $P$,\\with output constraints $\chi$}
  \begin{mathpar}
    \Infer{\SemiDeclChkValVar}
    {
      P \neq \exists, \land 
      \\
      (x:Q) \in \Gamma
      \\
      \semideclsub[+]{\Theta}{Q}{P}{W}
    }
    {
      \semideclchkval{\Theta}{\Gamma}{x}{P}{W}
    }
    \and
    \Infer{\SemiDeclChkValUnit}
    { }
    {
      \semideclchkval{\Theta}{\Gamma}{\unit}{\unitty}{\cdot}
    }
    \and
    \Infer{\SemiDeclChkValPair}
    {
      \semideclchkval{\Theta}{\Gamma}{v_1}{P_1}{\chi_1}
      \\
      \semideclchkval{\Theta}{\Gamma}{v_2}{P_2}{\chi_2}
    } 
    {
      \semideclchkval{\Theta}{\Gamma}{\pair{v_1}{v_2}}{P_1 \times P_2}{\chi_1, \chi_2}
    }
    \and
    \Infer{\SemiDeclChkValIn{k}}
    {
      \semideclchkval{\Theta}{\Gamma}{v}{P_k}{\chi}
    }
    {
      \semideclchkval{\Theta}{\Gamma}{\inj{k}{v}}{P_1 + P_2}{\chi}
    }
    \and 
    \Infer{\SemiDeclChkValExists}
    {
      \semideclchkval{\Theta}{\Gamma}{v}{[t/a]P}{\chi}
      \\
      \judgeterm{\Theta}{t}{\tau}
    }
    {
      \semideclchkval{\Theta}{\Gamma}{v}{(\extype{a:\tau} P)}{\chi}
    }
    \and 
    \Infer{\SemiDeclChkValWith}
    {
      \semideclchkval{\Theta}{\Gamma}{v}{P}{\chi}
      \\
    }
    {
      \semideclchkval{\Theta}{\Gamma}{v}{P \land \phi}{\phi, \chi}
    }
    \and
    \Infer{\SemiDeclChkValFix}
    {
      \judgeunroll{\cdot}{\Theta}{\nu:F[\mu F]}{\alpha}{F\;\Fold{F}{\alpha}\;\nu}{t}{P}{\tau}
      \\ 
      \semideclchkval{\Theta}{\Gamma}{v}{P}{\chi}
    }
    {
      \semideclchkval{\Theta}{\Gamma}{\roll{v}}{\comprehend{\nu:\mu F}{\Fold{F}{\alpha}\,{\nu} =_\tau t}}{\chi}
    }
    \and
    \Infer{\SemiDeclChkValDownshift}
    {
    }
    {
      \semideclchkval{\Theta}{\Gamma}{\thunk{e}}{\downshift{N}}{(e <= N)}
    }
  \end{mathpar}

  \judgbox{\semideclchkexp{\Theta}{\Gamma}{e}{N}}
  {Under $\Theta$ and $\Gamma$, expression $e$ semideclaratively checks against type $N$}
  \begin{mathpar}
    \Infer{\SemiDeclChkExpUpshift}
    {
      \semideclchkval{\Theta}{\Gamma}{v}{P}{\chi}
      \\
      \semideclneg{\Theta}{\Gamma}{\chi}
    }
    {
      \semideclchkexp{\Theta}{\Gamma}{\Return{v}}{\upshift{P}}
    }
    \and 
    \Infer{\SemiDeclChkExpLet}
    {
      \simple{\Theta}{N}
      \\
      \semideclsynexp{\Theta}{\Gamma}{\be}{\upshift{P}}
      \\
      \judgeextract[+]{\Theta}{P}{P'}{\Theta_P}
      \\
      \semideclchkexp{\Theta, \Theta_P}{\Gamma, x:P'}{e}{N}
    }
    {
      \semideclchkexp{\Theta}{\Gamma}{\Let{x}{\be}{e}}{N}
    }
    \and
    \Infer{\SemiDeclChkExpMatch}
    {
      \simple{\Theta}{N}
      \\
      \semideclsynhead{\Theta}{\Gamma}{h}{P}
      \\
      \semideclchkmatch{\Theta}{\Gamma}{P}{\clauses{\pa}{e}{i}{I}}{N}
    }
    {
      \semideclchkexp{\Theta}{\Gamma}{\match{h}{\clauses{\pa}{e}{i}{I}}}{N}
    }
    \and
    \Infer{\SemiDeclChkExpLam}
    {
      \simple{\Theta}{P \to N}
      \\
      \semideclchkexp{\Theta}{\Gamma, x:P}{e}{N}
    }
    {
      \semideclchkexp{\Theta}{\Gamma}{\fun{x}{e}}{P \to N}
    }
    \and
    \Infer{\SemiDeclChkExpUnreachable}
    {
      \simple{\Theta}{N}
      \\
      \judgeentail{\Theta}{\False}
    }
    {
      \semideclchkexp{\Theta}{\Gamma}{\unreachable}{N}
    }
    \and
    \Infer{\SemiDeclChkExpRec}
    {
      \arrayenvb{
        \simple{\Theta}{N}
        \and
        \semideclsub[-]{\Theta}{\alltype{a:\kindnat} M}{N}{W}
        \and
        \semideclentailwah{\Theta}{W}
      }
      \\
      \semideclchkexp{\Theta, a:\kindnat}{\Gamma, x:\downshift{\big(\alltype{a':\kindnat} (a' < a) \implies [a'/a]M\big)}}{e}{M}
    }
    {
      \semideclchkexp{\Theta}{\Gamma}{\rec{x : (\alltype{a:\kindnat} M)}{e}}{N}
    }
    \and
    \Infer{\SemiDeclChkExpExtract}
    {
      \judgeextract{\Theta}{N}{N'}{\Theta_N}
      \\
      \Theta_N \neq \cdot
      \\
      \semideclchkexp{\Theta, \Theta_N}{\Gamma}{e}{N'}
    }
    {
      \semideclchkexp{\Theta}{\Gamma}{e}{N}
    }
  \end{mathpar}
  
  \caption{Semideclarative value and expression type checking}
  \label{fig:semidecltyping-value-expression}
\end{figure}

\begin{figure}[htbp]
  \raggedright
  \judgbox{\semideclchkmatch{\Theta}{\Gamma}{P}{\clauses{\pa}{e}{i}{I}}{N}}
  {Under $\Theta$ and $\Gamma$, patterns $r_i$ match against (input) type $P$\\and branch expressions $e_i$ check against type $N$ (semideclaratively)}
  \begin{mathpar}
    \Infer{\SemiDeclChkMatchEx}
    {
      \semideclchkmatch{\Theta, a:\tau}{\Gamma}{P}{\clauses{\pa}{e}{i}{I}}{N}
    }
    {
      \semideclchkmatch{\Theta}{\Gamma}{\extype{a:\tau}{P}}{\clauses{\pa}{e}{i}{I}}{N}
    }
    \and
    \Infer{\SemiDeclChkMatchWith}
    {
      \semideclchkmatch{\Theta, \phi}{\Gamma}{P}{\clauses{\pa}{e}{i}{I}}{N}
    }
    {
      \semideclchkmatch{\Theta}{\Gamma}{P \land \phi}{\clauses{\pa}{e}{i}{I}}{N}
    }
    \and
    \Infer{\SemiDeclChkMatchUnit}
    {
      \semideclchkexp{\Theta}{\Gamma}{e}{N}
    }
    {
      \semideclchkmatch{\Theta}{\Gamma}{1}{\setof{\clause{\unit}{e}}}{N} 
    }
    \and
    \Infer{\SemiDeclChkMatchPair}
    {
      \arrayenvb{
        \judgeextract{\Theta}{P_1}{P_1'}{\Theta_1}
        \\
        \judgeextract{\Theta}{P_2}{P_2'}{\Theta_2}
      } 
      \\
      \semideclchkexp{\Theta, \Theta_1, \Theta_2}{\Gamma, x_1:P_1', x_2:P_2'}{e}{N} }
    {
      \semideclchkmatch{\Theta}{\Gamma}{P_1 \times P_2}{\setof{\clause{\pair{x_1}{x_2}}{e}}}{N}
    }
    \and
    \Infer{\SemiDeclChkMatchSum}
    { \arrayenvb{
        \judgeextract{\Theta}{P_1}{P_1'}{\Theta_1} \\
        \judgeextract{\Theta}{P_2}{P_2'}{\Theta_2} \\
      }
      \\
      \arrayenvb{
        \semideclchkexp{\Theta, \Theta_1}{\Gamma, x_1:P_1'}{e_1}{N} \\
        \semideclchkexp{\Theta, \Theta_2}{\Gamma, x_2:P_2'}{e_2}{N} 
      }
    }
    {
      \semideclchkmatch{\Theta}{\Gamma}{P_1 + P_2}{\setof{ \clause{\inl{x_1}}{e_1} \bnfalt \clause{\inr{x_2}}{e_2}}}{N}
    }
    \and
    \Infer{\SemiDeclChkMatchVoid}
    { }
    {
      \semideclchkmatch{\Theta}{\Gamma}{0}{\setof{}}{N}
    }
    \and
    \Infer{\SemiDeclChkMatchFix}
    {
      \judgeunroll{\cdot}{\Theta}{\nu:F[\mu F]}{\alpha}{F\;\Fold{F}{\alpha}\;\nu}{t}{Q}{\tau}
      \\
      \judgeextract{\Theta}{Q}{Q'}{\Theta_Q}
      \\
      \semideclchkexp{\Theta, \Theta_Q}{\Gamma, x:Q'}{e}{N}
    }
    {
      \semideclchkmatch{\Theta}{\Gamma}{\comprehend{\nu:\mu F}{\Fold{F}{\alpha}\,{\nu} =_\tau t}}{\setof{\clause{\roll{x}}{e}}}{N}
    }
  \end{mathpar}

  \judgbox{\semideclspine{\Theta}{\Gamma}{s}{N}{\upshift{P}}{\chi}}
  {Under $\Theta$ and $\Gamma$,
    if spine $s$ is applied to a head of type $\downshift{N}$, \\
    it semideclaratively returns a result of type $\upshift{P}$, with output constraints $\chi$}
  \begin{mathpar}
    \Infer{\SemiDeclSpineAll}
    {
      \judgeterm{\Theta}{t}{\tau}
      \\
      \semideclspine{\Theta}{\Gamma}{s}{[t/a]N}{\upshift{P}}{\chi}
    }
    {
      \semideclspine{\Theta}{\Gamma}{s}{\alltype{a:\tau}{N}}{\upshift{P}}{\chi}
    }
    \and
    \Infer{\SemiDeclSpineImplies}
    {
      \semideclspine{\Theta}{\Gamma}{s}{N}{\upshift{P}}{\chi}
    }
    {
      \semideclspine{\Theta}{\Gamma}{s}{\phi \implies N}{\upshift{P}}{\phi, \chi}
    }
    \and 
    \Infer{\SemiDeclSpineApp}
    {
      \semideclchkval{\Theta}{\Gamma}{v}{Q}{\chi}
      \\
      \semideclspine{\Theta}{\Gamma}{s}{N}{\upshift{P}}{\chi'}
    }
    {
      \semideclspine{\Theta}{\Gamma}{v,s}{Q \to N}{\upshift{P}}{\chi, \chi'}
    }
    \and 
    \Infer{\SemiDeclSpineNil}
    { }
    {
      \semideclspine{\Theta}{\Gamma}{\cdot}{\upshift{P}}{\upshift{P}}{\True}
    }
  \end{mathpar}
  \centering
  
  \caption{Semideclarative pattern matching and spine typing}
  \label{fig:semideclmatch}
\end{figure}

\begin{figure}[htbp]
  \judgbox{\semideclentailwah{\Theta}{\Wah}}{Under $\Theta$, constraint $\Wah$ semideclaratively holds}
  \begin{mathpar}
    \Infer{\SemiDeclWTrueProp}
    {
      \judgeentail{\Theta}{\phi}
    }
    {
      \semideclentailwah{\Theta}{\phi}
    }
    \and
    \Infer{\SemiDeclWTruePrpEquiv}
    {
      \judgeequiv[]{\Theta}{\phi}{\psi}
    }
    {
      \semideclentailwah{\Theta}{\phi \equiv \psi}
    }
    \\
    \Infer{\SemiDeclWTrueAnd}
    {
      \semideclentailwah{\Theta}{\Wah_1}
      \\
      \semideclentailwah{\Theta}{\Wah_2}
    }
    {
      \semideclentailwah{\Theta}{\Wah_1 \land \Wah_2}
    }
    \and
    \Infer{\SemiDeclWTrueImpl}
    {
      \semideclentailwah{\Theta, \phi}{\Wah}
    }
    {
      \semideclentailwah{\Theta}{\phi \implies \Wah}
    }
    \and
    \Infer{\SemiDeclWTrueAll}
    {
      \semideclentailwah{\Theta,a:\tau}{\Wah}
    }
    {
      \semideclentailwah{\Theta}{\alltype{a:\tau}\Wah}
    }
    \\
    \Infer{\SemiDeclWTruePosSub}
    {
      \semideclsub[+]{\Theta}{P}{Q}{\Wah}
      \\
      \semideclentailwah{\Theta}{\Wah}
    }
    {
      \semideclentailwah{\Theta}{\possubprob{P}{Q}}
    }
    \and
    \Infer{\SemiDeclWTrueNegSub}
    {
      \semideclsub[-]{\Theta}{N}{M}{\Wah}
      \\
      \semideclentailwah{\Theta}{\Wah}
    }
    {
      \semideclentailwah{\Theta}{\negsubprob{N}{M}}
    }
    \\
    \Infer{\SemiDeclWTruePosEquiv}
    {
      \semideclequiv[+]{\Theta}{P}{Q}{\Wah}
      \\
      \semideclentailwah{\Theta}{\Wah}
    }
    {
      \semideclentailwah{\Theta}{\poseqprob{P}{Q}}
    }
    \and
    \Infer{\SemiDeclWTrueNegEquiv}
    {
      \semideclequiv[-]{\Theta}{N}{M}{\Wah}
      \\
      \semideclentailwah{\Theta}{\Wah}
    }
    {
      \semideclentailwah{\Theta}{\negeqprob{N}{M}}
    }
  \end{mathpar}

  \judgbox{\semideclneg{\Theta}{\Gamma}{\chi}}
  {
    Under $\Theta$ and $\Gamma$,
    constraints $\chi$ hold semideclaratively
  }
  \begin{mathpar}
    \Infer{\SemiChkProblemsEmpty}
    {
    }
    {
      \semideclneg{\Theta}{\Gamma}{\cdot}
    }
    \and
    \Infer{\SemiChkProblemsNegChk}
    {
      \semideclchkexp{\Theta}{\Gamma}{e}{N}
      \\
      \semideclneg{\Theta}{\Gamma}{\chi}
    }
    {
      \semideclneg{\Theta}{\Gamma}{(e <= N), \chi}
    }
    \and
    \Infer{\SemiChkProblemsWah}
    {
      \semideclentailwah{\Theta}{W}
      \\
      \semideclneg{\Theta}{\Gamma}{\chi}
    }
    {
      \semideclneg{\Theta}{\Gamma}{W, \chi}
    }
  \end{mathpar}

  \caption{Semideclarative constraint verification}
  \label{fig:semidecl-entail-wah}
\end{figure}

\begin{figure}[htbp]
  \judgbox{\wahequiv{\Theta}{W}{W'}}{Under $\Theta$, constraints $W$ and $W'$ are equivalent}
  \begin{mathpar}
    \Infer{\WahEquivPrp}
    {
      \judgeequiv[]{\Theta}{\phi}{\psi}
    }
    {
      \wahequiv{\Theta}{\phi}{\psi}
    }
    \and
    \Infer{\WahEquivPrpEq}
    {
      \judgeequiv[]{\Theta}{\phi}{\psi}
      \\
      \judgeequiv[]{\Theta}{\phi'}{\psi'}
    }
    {
      \wahequiv{\Theta}{\phi \equiv \phi'}{\psi \equiv \psi'}
    }
    \and
    \Infer{\WahEquivAnd}
    {
      \wahequiv{\Theta}{W_1}{W_1'}
      \\
      \wahequiv{\Theta}{W_2}{W_2'}
    }
    {
      \wahequiv{\Theta}{W_1 \land W_2}{W_1' \land W_2'}
    }
    \and
    \Infer{\WahEquivImplies}
    {
      \judgeequiv[]{\Theta}{\phi}{\phi'}
      \\
      \wahequiv{\Theta}{W}{W'}
    }
    {
      \wahequiv{\Theta}{\phi \implies W}{\phi' \implies W'}
    }
    \and
    \Infer{\WahEquivAll}
    {
      \wahequiv{\Theta, a:\tau}{W}{W'}
    }
    {
      \wahequiv{\Theta}{\alltype{a:\tau}{W}}{\alltype{a:\tau}{W'}}
    }
    \\
    \Infer{\WahEquivPosSub}
    {
      \judgeequiv[+]{\Theta}{P}{P'}
      \\
      \judgeequiv[+]{\Theta}{Q}{Q'}
    }
    {
      \wahequiv{\Theta}{\possubprob{P}{Q}}{\possubprob{P'}{Q'}}
    }
    \and
    \Infer{\WahEquivNegSub}
    {
      \judgeequiv[-]{\Theta}{N}{N'}
      \\
      \judgeequiv[-]{\Theta}{M}{M'}
    }
    {
      \wahequiv{\Theta}{\possubprob{N}{M}}{\possubprob{N'}{M'}}
    }
    \\
    \Infer{\WahEquivPosEquiv}
    {
      \judgeequiv[+]{\Theta}{P}{P'}
      \\
      \judgeequiv[+]{\Theta}{Q}{Q'}
    }
    {
      \wahequiv{\Theta}{\poseqprob{P}{Q}}{\poseqprob{P'}{Q'}}
    }
    \and
    \Infer{\WahEquivNegEquiv}
    {
      \judgeequiv[-]{\Theta}{N}{N'}
      \\
      \judgeequiv[-]{\Theta}{M}{M'}
    }
    {
      \wahequiv{\Theta}{\negeqprob{N}{M}}{\negeqprob{N'}{M'}}
    }
  \end{mathpar}

  \caption{Constraint equivalence}
  \label{fig:constraint-equivalence}
\end{figure}

\begin{figure}[htbp]
  \judgbox{\chiequiv{\Theta}{\Gamma}{\chi_1}{\chi_2}}{Under $\Theta$ and $\Gamma$, constraint lists $\chi_1$ and $\chi_2$ are equivalent}
  
  \begin{mathpar}
    \Infer{\ChiEquivEmpty}
    {}
    {
      \chiequiv{\Theta}{\Gamma}{\cdot}{\cdot}
    }
    \and
    \Infer{\ChiEquivWah}
    {
      \wahequiv{\Theta}{W_1}{W_2}
      \\
      \chiequiv{\Theta}{\Gamma}{\chi_1}{\chi_2}
    }
    {
      \chiequiv{\Theta}{\Gamma}{W_1, \chi_1}{W_2, \chi_2}
    }
    \and
    \Infer{\ChiEquivChk}
    {
      \judgeequiv[-]{\Theta}{N_1}{N_2}
      \\
      \chiequiv{\Theta}{\Gamma}{\chi_1}{\chi_2}
    }
    {
      \chiequiv{\Theta}{\Gamma}{(e <= N_1), \chi_1}{(e <= N_2), \chi_2}
    }
  \end{mathpar}

  \caption{Equivalence of constraint lists}
  \label{fig:constraint-list-equivalence}
\end{figure}

\clearpage

\subsection{Miscellaneous Definitions}
\label{sec:misc-defs}
\begin{figure}[hptb]
  \begin{tabular}{ll}
    $\sem{}{\Theta_0; \Gamma_0 |- \sigma : \Theta; \Gamma}$ & \Defref{def:den-syn-subs}
    \\[0.6ex]
    $\overline{\Theta}$ & \Defref{def:filter-out-props}
    \\[0.2ex]
    $\idsubs{\Theta}{\Gamma}$ & \Defref{def:id-subs} \\
    $[-]\Xi$ & \Defref{def:Xi-subs} \\
    $\text{Proposition-independent judgments}$ & \Defnref{def:assertion-independent} \\
    $\fvimctx{\sigma}{\Xi}$ & \Defref{def:fvimctx} \\
    $\numlog{A}$ & \Defref{def:numlog} \\
    $\judgesub[+]{\Theta}{\Gamma}{\Gamma'}$ & \Defnref{def:ctx-sub} \\
    $\text{Predomain (cpo)}$ & \Defnref{def:predomain} \\
    $\text{Domain (cppo)}$ & \Defnref{def:domain} \\
    $\text{Continuous function}$ & \Defnref{def:continuous} \\
    $\Cpo$ & \Defnref{def:Cpo} \\
    $\Cppo$ & \Defnref{def:Cppo} \\
    $\mathit{fold}\text{ and }\mathit{fold}^n$ & \Defref{def:fold} \\
    $[\Delta_\Xi/\Xi]-$ & \Defref{def:evar-rename} \\
    $[\Omega_\Xi/\Xi]-$ & \Defref{def:solve-evar-rename} \\
  \end{tabular}

  \caption{Miscellaneous definitions}
  \label{fig:definitions}
\end{figure}

\clearpage

\clearpage

\section{Properties of the Declarative System}

Note that we overload ``$,$'' as syntax
and \defconcat as a list concatenation metaoperation
on logical contexts and semantic substitutions

\begin{definition}[Denotation of Syntactic Substitution]
  \label{def:den-syn-subs}
  Assume $\Dee :: \Theta_0;\Gamma_0 |- \sigma : \Theta; \Gamma$.
  Define $\sem{\delta}{\Dee}$ for any $|- \delta : \Theta_0; \Gamma_0$
  inductively as follows:\\
  $\sem{\delta}{
    \inferrule
    {
    }
    {
      \Theta_0; \Gamma_0 |- \cdot : \cdot; \cdot
    }
  } = \cdot$ \\
  $\sem{\delta}{
      \inferrule
      {
        \Dee_0 :: \Theta_0; \Gamma_0 |- \sigma : \Theta; \Gamma
        \and
        \Theta_0 |- [\filterprog{\sigma}]t : \tau
      }
      {
        \Theta_0; \Gamma_0 |- \sigma, t/a : \Theta, a:\tau; \Gamma
      }
    } = \sem{\delta}{\Dee_0}, \sem{\filterprog{\delta}}{[\filterprog{\sigma}]t}/a$ \\
  $\sem{\delta}{
    \inferrule
    {
      \Dee_0 :: \Theta_0; \Gamma_0 |- \sigma : \Theta; \Gamma
      \and
      \judgeentail{\Theta_0}{[\filterprog{\sigma}]\phi}
    }
    {
      \Theta_0; \Gamma_0 |- \sigma : \Theta, \phi; \Gamma
    }
  } = \sem{\delta}{\Dee_0}$ \\
  $\sem{\delta}{
    \inferrule
    {
      \Dee_0 :: \Theta_0; \Gamma_0 |- \sigma : \Theta; \Gamma
      \and
      \E :: \judgechkval{\Theta_0}{\Gamma_0}{[\sigma]v}{[\filterprog{\sigma}]P}
    }
    {
      \Theta_0; \Gamma_0 |- \sigma, \subs{v}{P}{x} : \Theta; \Gamma, x:P
    }
  } = \sem{\delta}{\Dee_0}, \sem{\delta}{\E}/x$ \\
  We may also write $\sem{\delta}{\Theta_0;\Gamma_0 |- \sigma : \Theta; \Gamma}$,
  if the derivation $\Dee$ is clear from context, 
  or $\sem{\delta}{\sigma}$
  if $\Dee, \Theta_0, \Gamma_0, \Theta$, and $\Gamma$ are clear from context.
  As we proceed, we will prove that this definition is sound:
  \[
    \sem{}{\Theta_0; \Gamma_0 |- \sigma : \Theta; \Gamma} 
    : \underbrace{\sem{}{\Theta_0; \Gamma_0}}_{\comprehend{\delta}{|- \delta : \Theta_0; \Gamma_0}}
    \to \underbrace{\sem{}{\Theta; \Gamma}}_{\comprehend{\delta}{|- \delta : \Theta; \Gamma}}
  \]
  In particular, see
  Lemmas~\ref{lem:type-soundness-subs-ix},~\ref{lem:type-soundness-subs-ix-prop},
  and~\ref{lem:type-soundness-subs}.
\end{definition}

Substitution can alter the set of value-determined indexes.
To reflect this,
we define substitution on value-determined contexts $\Xi$ as follows.

\begin{definition}[Substitution on $\Xi$]
  \label{def:Xi-subs}
  Let $\sigma$ be a substitution.
  Define $[\sigma]\Xi$ inductively as follows:
  \begin{align*}
    [\sigma](\cdot) &= \cdot \\
    [\sigma](\Xi,a:\tau) &=
      \begin{cases}
        [\sigma]\Xi, [\sigma]a : \tau &\text{if }[\sigma]a\text{ is a variable} \\
        [\sigma]\Xi                   &\text{else}
      \end{cases}
  \end{align*}
\end{definition}

We often implicitly use the following lemma
($\filterprog{-}$ filters out program variable entries).

\begin{lemma}[Filter Out Prog.\ Vars.\ Syn.]
  \label{lem:filter-out-prog-vars-syn}
  If $\Theta_0; \Gamma_0 |- \sigma : \Theta; \Gamma$,
  then $\Theta_0 |- \filterprog{\sigma} : \Theta$.
\end{lemma}
\begin{proof}
  By structural induction on the given syntactic substitution typing derivation.
  (Use fact that $\filterprog{-}$ is idempotent.)
\end{proof}

\begin{definition}[Filter Out Propositions]
  \label{def:filter-out-props}
  For any logical context $\Theta$,
  define $\overline{\Theta}$ inductively as:
  \begin{align*}
    \overline{\cdot} &= \cdot \\
    \overline{\Theta, a:\tau} &= \overline{\Theta}, a:\tau \\
    \overline{\Theta, \phi} &= \overline{\Theta}
  \end{align*}
\end{definition}

\begin{lemma}[Filter Out Propositions]
  \label{lem:filter-out-props}
  ~
  \begin{enumerate}
  \item If $|- \delta : \Theta; \Gamma$, then $|- \delta : \overline{\Theta}; \Gamma$.
  \item If $\Theta_0; \Gamma_0 |- \sigma : \Theta$, then $\Theta_0 |- \sigma : \overline{\Theta}; \Gamma$.
  \end{enumerate}
\end{lemma}
\begin{proof}
  Each part is proved by structural induction on the given derivation.
\end{proof}

\begin{definition}[Id.\ Substitution]
  \label{def:id-subs}
  The \emph{identity substitution} on $\Theta$ and $\Gamma$,
  written $\idsubs{\Theta}{\Gamma}$, is defined inductively as follows:
  \begin{align*}
    \idsubs{\cdot}{\cdot} &= \cdot \\
    \idsubs{\Theta,a:\tau}{\cdot} &= \idsubs{\Theta}{\cdot}, a/a \\
    \idsubs{\Theta,\phi}{\cdot} &= \idsubs{\Theta}{\cdot} \\
    \idsubs{\Theta}{\Gamma,x:P} &= \idsubs{\Theta}{\Gamma}, x:P/x \\
  \end{align*}
\end{definition}

\begin{lemma}[Id.\ Subst.]
  \label{lem:id-subs-id}
  The substitution $\idsubs{\Theta}{\Gamma}$ is an identity substitution, \ie,
  $[\idsubs{\Theta}{\Gamma}]\mathcal{O} = \mathcal{O}$
  for every object $\mathcal{O}$ to which it may be applied.
\end{lemma}
\begin{proof}
  Straightforward.
\end{proof}

\begin{lemma}[Sorting Weakening]
  \label{lem:sorting-weakening}
  Assume $\judgctx{(\Theta_1, \Theta_0, \Theta_2)}$.
  If $\judgeterm{\Theta_1, \Theta_2}{t}{\tau}$,\\
  then $\judgeterm{\Theta_1, \Theta_0, \Theta_2}{t}{\tau}$
  by a derivation of equal height.
\end{lemma}
\begin{proof}
  By structural induction on the given sorting derivation.
\end{proof}

\begin{lemma}[Ix.\ Meaning Weak.\ Invariant]
  \label{lem:ix-meaning-weakening-invariant}
  Assume $|- \delta : \Theta_1, \Theta_2$ and $|- \delta' : \Theta_1, \Theta, \Theta_2$
  and $\delta'\restriction_{\Theta_1, \Theta_2} = \delta$.
  If $\judgeterm{\Theta_1, \Theta_2}{t}{\tau}$, 
  then
  $\sem{\delta}{\judgeterm{\Theta_1, \Theta_2}{t}{\tau}}
    = \sem{\delta'}{\judgeterm{\Theta_1, \Theta, \Theta_2}{t}{\tau}}$
  (for any such weakened derivation).
\end{lemma}
\begin{proof}
  By structural induction on the derivation of
  $\judgeterm{\Theta_1, \Theta_2}{t}{\tau}$.
\end{proof}

\begin{lemma}[Prop.\ Truth Weakening]
  \label{lem:prop-truth-weakening}
  Assume $\judgctx{\Theta_1, \Theta_0, \Theta_2}$.
  If $\judgeentail{\Theta_1, \Theta_2}{\phi}$,\\
  then $\judgeentail{\Theta_1, \Theta_0, \Theta_2}{\phi}$
  by a derivation of equal height.
\end{lemma}
\begin{proof}
  Assume $|- \delta : \Theta_1, \Theta_0, \Theta_2$.

  \begin{llproof}
    \judgeentailPf{\Theta_1, \Theta_2}{\phi}{Given}
    \judgetermPf{\Theta_1, \Theta_2}{\phi}{\Booltype}{Presupposed derivation}
    \judgetermPf{\Theta_1, \Theta_0, \Theta_2}{\phi}{\Booltype}{By \Lemmaref{lem:sorting-weakening}}
    \eqPf{\sem{\delta}{\phi}}{\sem{\delta\restriction_{\Theta_1, \Theta_2}}{\phi}}{By \Lemmaref{lem:ix-meaning-weakening-invariant}}
    \eqPf{}{\one}{By inversion on $\judgeentail{\Theta_1, \Theta_2}{\phi}$}
    \judgeentailPf{\Theta_1, \Theta_0, \Theta_2}{\phi}{By \PropTrue}
  \end{llproof} 
\end{proof}

\begin{lemma}[Sorting Soundness]
  \label{lem:type-soundness-ix}
  If $\Theta |- t : \tau$,
  then $\sem{\delta}{t} \in \sem{}{\tau}$
  for all $|- \delta : \overline{\Theta}$.
\end{lemma}
\begin{proof}
  By structural induction on the given index sorting derivation $\Theta |- t : \tau$.
\end{proof}

\begin{lemma}[Ix.\ Subst.\ Typing Sound]
  \label{lem:type-soundness-subs-ix}
  Assume $\Theta_0;\cdot |- \sigma : \Theta; \cdot$.
  If $\overline{\Theta} = \Theta$,
  then $|- \sem{\delta}{\sigma} : \Theta; \cdot$
  for all $|- \delta : \overline{\Theta_0}; \cdot$.
\end{lemma}
\begin{proof}
  By structural induction on the given substitution typing derivation.
  Assume $|- \delta : \overline{\Theta_0}$.
  \begin{itemize}
    \DerivationProofCase{\EmptySyn}
    {
    }
    {
      \Theta_0; \cdot |- \cdot : \cdot; \cdot
    }
    \begin{llproof}
      \Pf{\sem{\delta}{\cdot} = \cdot}{}{}{By def.}
      \Pf{|- \cdot : \cdot;\cdot}{}{}{By \EmptySem}
    \end{llproof}
    \DerivationProofCase{\IxSyn}
    {
      \Theta_0; \cdot |- \sigma' : \Theta'; \cdot
      \\
      \Theta_0 |- [\sigma']t : \tau
    }
    {
      \Theta_0; \cdot |- \sigma', t/a : \Theta', a:\tau; \cdot
    }
    \begin{llproof}
      \Pf{\Theta_0; \cdot |- \sigma' : \Theta'; \cdot}{}{}{Subderivation}
      \Pf{|- \sem{\delta}{\sigma'} : \Theta'; \cdot}{}{}{By i.h.}
      \Pf{\Theta_0 |- [\sigma']t : \tau}{}{}{Subderivation}
      \Pf{\sem{\delta}{[\sigma']t} \in \sem{}{\tau}}{}{}{By~\Lemref{lem:type-soundness-ix}}
      \Pf{|- \underbrace{\sem{\delta}{\sigma'}, \sem{\delta}{[\sigma']t}/a}_{\sem{\delta}{\sigma}} : \underbrace{\Theta', a:\tau}_\Theta; \cdot}{}{}{By \IxSem}
    \end{llproof}
    \ProofCaseRule{\PropSyn}
    Impossible
    (because $\overline{\Theta} = \Theta$, there are no propositions in $\Theta$).
    \ProofCaseRule{\ValSyn}
    Impossible (because $\Gamma = \cdot$).
    \qedhere
  \end{itemize}
\end{proof}

\begin{lemma}[Index Syntactic Substitution]
  \label{lem:syn-subs-ix}
  If $\Theta_0 |- \sigma : \Theta$
  and $\judgeterm{\Theta}{t}{\tau}$,
  then $\judgeterm{\Theta_0}{[\sigma]t}{\tau}$.
\end{lemma}
\begin{proof}
  By structural induction on the sorting derivation.
\end{proof}

\begin{lemma}[Index Substitution Soundness]
  \label{lem:subs-soundness-ix}
  Assume $\Theta_0;\cdot |- \sigma : \Theta; \cdot$.
  If $\judgeterm{\Theta}{t}{\tau}$,
  then
  $\sem{\delta}{[\sigma]t}
  = \sem{\sem{\delta}{\sigma}}{t}$
  for all $|- \delta : \overline{\Theta_0}; \cdot$.
\end{lemma}
\begin{proof}
  Assume $\judgeterm{\Theta}{t}{\tau}$
  and $\judgeterm{\Theta_0}{[\sigma]t}{\tau}$.

  Suppose $|- \delta : \overline{\Theta_0}$.
  By \Lemmaref{lem:filter-out-props},
  $\Theta_0;\cdot |- \sigma : \overline{\Theta}; \cdot$.
  By \Lemmaref{lem:type-soundness-subs-ix},
  $|- \sem{\delta}{\sigma} : \overline{\Theta}$,
  so we can apply $\sem{}{t}$ to $\sem{\delta}{\sigma}$.

  Proceed by structural induction on the given sorting derivation
  $\judgeterm{\Theta}{t}{\tau}$,
  considering cases for its concluding rule.
  Each case is straightforward.
  The \IxVar case uses Definition~\ref{def:den-syn-subs}.
\end{proof}

\begin{lemma}[Index Subst.\ Typing Sound]
  \label{lem:type-soundness-subs-ix-prop}
  Assume $\Dee :: \Theta_0;\cdot |- \sigma : \Theta; \cdot$.
  Then $|- \sem{\delta}{\Dee} : \Theta; \cdot$
  for all $|- \delta : \Theta_0; \cdot$.
\end{lemma}
\begin{proof}
  By structural induction on the derivation $\Dee$.
  Consider cases for the rule concluding $\Dee$:
  \begin{itemize}
    \ProofCaseRule{\EmptySyn}
    Similar to \EmptySyn case of \Lemref{lem:type-soundness-subs-ix}.
    \ProofCaseRule{\IxSyn}
    Similar to \IxSyn case of \Lemref{lem:type-soundness-subs-ix}.
    \DerivationProofCase{\PropSyn}
    {
      \E :: \Theta_0; \cdot |- \sigma : \Theta'; \cdot
      \\
      \judgeentail{\Theta_0}{[\sigma]\phi}
    }
    {
      \Dee :: \Theta_0; \cdot |- \sigma : \Theta', \phi; \cdot
    }
    \begin{llproof}
      \Pf{\E :: \Theta_0; \cdot |- \sigma : \Theta'; \cdot}{}{}{Subderivation}
      \Pf{|- \sem{\delta}{\E} : \Theta';\cdot}{}{}{By i.h.}
      \Pf{\judgeentail{\Theta_0}{[\sigma]\phi}}{}{}{By inversion}
      \Pf{\text{for all} |- \delta' : \Theta_0; \Gamma \text{, }\sem{\delta'}{[\sigma]\phi} = \{\bullet\}}{}{}{By inversion on \PropTrue}
      \Pf{|- \delta : \Theta_0; \cdot}{}{}{Assumption}
      \Pf{\sem{\delta}{[\sigma]\phi} = \{\bullet\}}{}{}{By above two lines (for $\sigma$ appearing in what $\E$ derives)}
      \Pf{\sem{\delta}{[\sigma]\phi} = \sem{\sem{\delta}{\E}}{\phi}}{}{}{By \Lemmaref{lem:subs-soundness-ix}}
      \Pf{\sem{\sem{\delta}{\E}}{\phi} = \{\bullet\}}{}{}{By transitivity of equality}
      \Pf{|- \underbrace{\sem{\delta}{\E}}_{\sem{\delta}{\Dee}} : \Theta', \phi; \cdot}{}{}{By \PropSem}
    \end{llproof}
    \ProofCaseRule{\ValSyn}
    Impossible.
    \qedhere
  \end{itemize}
\end{proof}

\begin{lemma}[Stratify Sem.\ Subs. 1]
  \label{lem:stratify-sem-subs-1}
  If $|- \delta : \Theta, \Theta_1$,
  then there exist $\delta'$ and $\delta_1$\\
  such that $\delta = \delta', \delta_1$
  and $|- \delta' : \Theta$
  and $|- \delta', \delta_1 : \Theta, \Theta_1$.
\end{lemma}
\begin{proof}
  By structural induction on $\Theta_1$.
\end{proof}

\begin{lemma}[Stratify Sem.\ Subs.]
  \label{lem:stratify-sem-subs}
  If $|- \delta : \Theta, \Theta_1, \dots, \Theta_n$,
  then there exist $\delta', \delta_1, \dots, \delta_n$\\
  such that $\delta = \delta', \delta_1, \dots, \delta_n$
  and $|- \delta' : \Theta$
  and $|- \delta', \delta_1 : \Theta, \Theta_1$
  and \ldots\ and
  $|- \delta', \delta_1, \dots, \delta_n : \Theta, \Theta_1, \dots, \Theta_n$.
\end{lemma}
\begin{proof}
  By induction on $n$.
  The $n=0$ case is immediate.
  
  Suppose $|- \delta : \Theta, \Theta_1, \dots, \Theta_k, \Theta_{k+1}$.
  By \Lemmaref{lem:stratify-sem-subs-1}, there exist $\tilde{\delta}$ and $\delta_{k+1}$
  such that $|- \tilde{\delta} : \Theta, \Theta_1, \dots, \Theta_k$
  and $|- \tilde{\delta}, \delta_{k+1} : \Theta, \Theta_1, \dots, \Theta_k, \Theta_{k+1}$.
  By the \ih, there exist $\delta', \delta_1, \dots, \delta_k$
  such that $\tilde{\delta} = \delta', \delta_1, \dots, \delta_k$
  and $|- \delta' : \Theta$
  and $|- \delta', \delta_1 : \Theta, \Theta_1$
  and \ldots\ and
  $|- \delta', \delta_1, \dots, \delta_k : \Theta, \Theta_1, \dots, \Theta_k$.
  Rewriting a semantic substitution above with the equation just obtained, we have
  $|- \delta', \delta_1, \dots, \delta_k, \delta_{k+1} : \Theta, \Theta_1, \dots, \Theta_k, \Theta_{k+1}$. \qedhere
\end{proof}

\begin{lemma}[Sem.\ Subs.\ Length]
  \label{lem:sem-subs-length}
  If $|- \delta : \Theta$, then $\length{\delta} = \length{\overline{\Theta}}$.
\end{lemma}
\begin{proof}
  By structural induction on the given semantic substitution derivation.
\end{proof}

\begin{lemma}[Sem.\ Subs.\ Entry]
  \label{lem:sem-subs-deep-entry}
  ~
  \begin{enumerate}
  \item If $|- \delta_1 : \Theta_1$ and $|- \delta_1, \delta_2 : \Theta_1, \Theta_2$
    and $d \in \sem{}{\tau}$ and $a$ is fresh,
    then $|- \delta_1, d/a, \delta_2 : \Theta_1, a:\tau, \Theta_2$.
  \item If $|- \delta_1 : \Theta_1$ and $|- \delta_1, \delta_2 : \Theta_1, \Theta_2$
    and $\judgeterm{\Theta_1}{\phi}{\Booltype}$ and $\sem{\delta_1}{\phi} = 1$,
    then $|- \delta_1, \delta_2 : \Theta_1, \phi, \Theta_2$.
  \end{enumerate}
\end{lemma}
\begin{proof}
  Each part is proved by structural induction on $\Theta_2$.
\end{proof}

\begin{lemma}[Permute Sem.\ Subs.]
  \label{lem:permute-sem-subs}
  Assume $\judgctx{(\Theta, \Theta_1)}$
  and $\judgctx{(\Theta, \Theta_2)}$.\\
  If $|- \delta' : \Theta$
  and $|- \delta', \delta_1 : \Theta, \Theta_1$
  and $|- \delta', \delta_1, \delta_2 : \Theta, \Theta_1, \Theta_2$
  and $|- \delta', \delta_1, \delta_2, \delta_f : \Theta, \Theta_1, \Theta_2, \Theta_f$,\\
  then $|- \delta', \delta_2, \delta_1, \delta_f : \Theta, \Theta_2, \Theta_1, \Theta_f$.
\end{lemma}
\begin{proof}
  By lexicographic induction on,
  first, the structure of $\Theta_f$, and,
  second, the structure of $\Theta_1$.
  We case analyze $\Theta_f$, and then $\Theta_1$ in the $\Theta_f = \cdot$ case.
  \qedhere
\end{proof}

\begin{lemma}[Meaning Permute Invariant]
  \label{lem:meaning-permute-invariant}
  Assume $|- \delta, \delta_1, \delta_2, \delta_f : \overline{\Theta, \Theta_1, \Theta_2, \Theta_f}$
  and $|- \delta, \delta_2, \delta_1, \delta_f : \overline{\Theta, \Theta_2, \Theta_1, \Theta_f}$.
  If $\judgeterm{\Theta, \Theta_1, \Theta_2, \Theta_f}{t}{\tau}$
  and $\judgeterm{\Theta, \Theta_2, \Theta_1, \Theta_f}{t}{\tau}$,
  then $\sem{\delta, \delta_1, \delta_2, \delta_f}{t} = \sem{\delta, \delta_2, \delta_1, \delta_f}{t}$.
\end{lemma}
\begin{proof}
  By \Lemmaref{lem:type-soundness-ix},
  $\sem{}{\judgeterm{\Theta, \Theta_1, \Theta_2, \Theta_f}{t}{\tau}} : \sem{}{\overline{\Theta, \Theta_1, \Theta_2, \Theta_f}} \to \sem{}{\tau}$,
  and similarly for the permuted sorting derivation.
  We have
  $\sem{\delta, \delta_1, \delta_2, \delta_f}{t} = \sem{\delta, \delta_2, \delta_1, \delta_f}{t}$
  because looking up the value of an index variable
  in a (proposition-filtered) semantic substitution
  is independent of the latter's order.
\end{proof}

\subsection{Index Domain Properties}

\begin{lemma}[Prop.\ Truth Equiv.\ Relation]
  \label{lem:equivassert}
  ~
  \begin{enumerate}
    \item If $\judgeterm{\Theta}{t}{\tau}$, then $\judgeentail{\Theta}{t = t}$.
    \item If $\judgeentail{\Theta}{t_1=t_2}$, then $\judgeentail{\Theta}{t_2=t_1}$.
    \item If $\judgeentail{\Theta}{t_1=t_2}$ and $\judgeentail{\Theta}{t_2=t_3}$,
      then $\judgeentail{\Theta}{t_1 = t_3}$.
  \end{enumerate}
\end{lemma}
\begin{proof}
  Follows from the fact that
  the underlying equality used in the definition of $\sem{\delta}{\phi}$
  (\Figureref{fig:denotation-ix-prop}) is an equivalence relation.
\end{proof}

\begin{lemma}[Assumption]
  \label{lem:propvar}
  If $\judgctx{\Theta_1, \phi, \Theta_2}$, then $\judgeentail{\Theta_1, \phi, \Theta_2}{\phi}$.
\end{lemma}
\begin{proof}
  Suppose $|- \delta : \Theta_1, \phi, \Theta_2$.
  By \Lemmaref{lem:stratify-sem-subs}, $\delta = \delta_1, \delta_2$
  and $|- \delta_1 : \Theta_1, \phi$.
  By inversion on \PropSem, $\sem{\delta_1}{\phi} = \{\bullet\}$.
  By \Lemmaref{lem:ix-meaning-weakening-invariant},
  $\sem{\delta}{\phi} = \{\bullet\}$.
  By \PropTrue, $\judgeentail{\Theta_1, \phi, \Theta_2}{\phi}$.
\end{proof}

\begin{lemma}[Consequence]
  \label{lem:consequence}
  If $\judgeentail{\Theta_1,\psi,\Theta_2}{\phi}$
  and $\judgeentail{\Theta_1}{\psi}$,\\
  then $\judgeentail{\Theta_1,\Theta_2}{\phi}$.
\end{lemma}
\begin{proof}
  By inversion on $\judgeentail{\Theta_1,\psi,\Theta_2}{\phi}$,
  for all $|- \delta : \Theta_1,\psi,\Theta_2$,
  we have $\sem{\delta}{\phi} = \one$.
  Suppose $|- \delta : \Theta_1,\Theta_2$.
  By \Lemmaref{lem:stratify-sem-subs}, there exist $\delta_1, \delta_2$
  such that $\delta = \delta_1, \delta_2$
  and $|- \delta_1 : \Theta_1$
  and $|- \delta_1, \delta_2 : \Theta_1, \Theta_2$;
  these together with the given $\judgeentail{\Theta_1}{\psi}$
  yields $|- \delta : \Theta_1,\psi,\Theta_2$
  by \Lemmaref{lem:sem-subs-deep-entry}.
  Eliminating the above ``for all'', we have $\sem{\delta}{\phi'} = \one$.
  By \PropTrue, $\judgeentail{\Theta_1,\Theta_2}{\phi}$.
\end{proof}

\begin{lemma}[Subs.\ Respects Prop.\ Truth]
  \label{lem:syn-subs-prop-true}
  If $\Theta_0 |- \sigma : \Theta$
  and $\judgeentail{\Theta}{\phi}$,
  then $\judgeentail{\Theta_0}{[\sigma]\phi}$.
\end{lemma}
\begin{proof}
  Only one rule can conclude $\judgeentail{\Theta}{\phi}$:
  \begin{itemize}
    \DerivationProofCase{\PropTrue}
    {
      \text{for all }\delta\text{, if }|- \delta : \Theta\text{ then }
      \sem{\delta}{\phi} = \{\bullet\}
    }
    {\judgeentail{\Theta}{\phi}}
    \begin{llproof}
      \Pf{\text{for all }\delta\text{, if }|- \delta : \Theta\text{ then }
        \sem{\delta}{\phi} = \{\bullet\}}{}{}{Premise}
      \Pf{|- \delta : \Theta_0}{}{}{Suppose}
      \Pf{\Theta_0 |- \sigma : \Theta}{}{}{Given}
      \Pf{|- \sem{\delta}{\sigma} : \Theta}{}{}{By Lemma~\ref{lem:type-soundness-subs-ix-prop}}
      \Pf{\sem{\sem{\delta}{\sigma}}{\phi} = \{\bullet\}}{}{}{By lines above}
      \Pf{\sem{\delta}{[\sigma]\phi} = \sem{\sem{\delta}{\sigma}}{\phi}}{}{}{By Lemma~\ref{lem:subs-soundness-ix}}
      \Pf{\sem{\delta}{[\sigma]\phi} = \{\bullet\}}{}{}{By equations above}
      \Pf{\judgeentail{\Theta_0}{[\sigma]\phi}}{}{}{By \PropTrue ($\delta$ was arbitrary)}
    \end{llproof}
    \qedhere
  \end{itemize}
\end{proof}

\begin{lemma}[Subst.\ Inconsistent]
  \label{lem:subst-inconsistent}
  If $\judgsubs{\Theta_0}{\sigma}{\Theta}$ and $\judgeentail{\Theta}{\False}$,
  then $\judgeentail{\Theta_0}{\False}$.
\end{lemma}
\begin{proof}
  Follows from \Lemmaref{lem:syn-subs-prop-true}
  and $[\sigma]\False = \False$ (by definition of substitution).
\end{proof}

\subsection{Substitution Properties}

\begin{lemma}[Ix.-Level \ Subs.\ Weakening]
  \label{lem:ix-syn-subs-weakening}
  If $\Theta_1,\Theta_2; \cdot |- \sigma : \Theta; \cdot$
  and $\judgctx{\Theta_1, \Theta_0, \Theta_2}$,\\
  then $\Theta_1,\Theta_0,\Theta_2; \cdot |- \sigma : \Theta; \cdot$.
\end{lemma}
\begin{proof}
  By induction on the derivation.
  Consider cases for the final rule of the derivation:
  \begin{itemize}
    \DerivationProofCase{\EmptySyn}
    {
    }
    {
      \Theta_1,\Theta_2; \cdot |- \cdot : \cdot; \cdot
    }
    \begin{llproof}
      \Pf{\Theta_1,\Theta_0,\Theta_2; \cdot |- \cdot : \cdot; \cdot}{}{}{By \EmptySyn}
    \end{llproof}

    \DerivationProofCase{\IxSyn}
    {
      \Theta_1,\Theta_2; \cdot |- \sigma : \Theta; \cdot
      \and
      \Theta_1,\Theta_2 |- [\filterprog{\sigma}]t : \tau
      \and
      a \notin \dom{\Theta}
    }
    {
      \Theta_1,\Theta_2; \cdot |- \sigma, t/a : \Theta, a:\tau; \cdot
    }
    \begin{llproof}
      \Pf{\Theta_1,\Theta_2; \cdot |- \sigma : \Theta; \cdot}{}{}{Subderivation}
      \Pf{\Theta_1,\Theta_0,\Theta_2; \cdot |- \sigma : \Theta; \cdot}{}{}{By i.h.}
      \Pf{\Theta_1,\Theta_2 |- [\filterprog{\sigma}]t : \tau}{}{}{Subderivation}
      \Pf{\Theta_1,\Theta_0,\Theta_2 |- [\filterprog{\sigma}]t : \tau}{}{}{By \Lemmaref{lem:sorting-weakening}}
      \Pf{\Theta_1,\Theta_0,\Theta_2; \cdot |- \sigma, t/a : \Theta, a:\tau; \cdot}{}{}{By \IxSyn}
    \end{llproof}

    \ProofCaseRule{\PropSyn} Similar to case for \IxSyn.

    \ProofCaseRule{\ValSyn} Similar to case for \IxSyn.
    \qedhere
  \end{itemize}
\end{proof}

\begin{lemma}[Id.\ Subst.\ Typing]
  \label{lem:id-subs-typing}
  For all $\Theta$, we have
  $\Theta;\cdot |- \idsubs{\Theta}{\cdot} : \Theta;\cdot$.
\end{lemma}
\begin{proof}
  By structural induction on $\Theta$.
  Use \Lemmaref{lem:ix-syn-subs-weakening},
  \Lemmaref{lem:id-subs-id}
  and \Lemmaref{lem:consequence}.
\end{proof}

\begin{lemma}
  \label{lem:union-subs}
  For all $\sigma$, $\Xi_1$, and $\Xi_2$,
  we have $[\sigma]\Xi_1 \cup [\sigma]\Xi_2 = [\sigma](\Xi_1 \cup \Xi_2)$.
\end{lemma}
\begin{proof}
  If $(a:\tau) \in [\sigma](\Xi_1 \cup \Xi_2)$,  then there exists $(a':\tau)$
  in either $\Xi_1$ or $\Xi_2$ such that $[\sigma]a' = a$,
  i.e. $(a:\tau) \in [\sigma]\Xi_1 \cup [\sigma]\Xi_2$; hence
  $[\sigma](\Xi_1 \cup \Xi_2) \subseteq [\sigma]\Xi_1 \cup [\sigma]\Xi_2$.
  Similarly, $[\sigma](\Xi_1 \cup \Xi_2) \supseteq [\sigma]\Xi_1 \cup [\sigma]\Xi_2$.
\end{proof}

\begin{lemma}
  \label{lem:intersection-subs}
  For all $\sigma$, $\Xi_1$, and $\Xi_2$,
  we have $[\sigma]\Xi_1 \cap [\sigma]\Xi_2 \supseteq [\sigma](\Xi_1 \cap \Xi_2)$.
\end{lemma}
\begin{proof}
  Suppose $(a:\tau) \in [\sigma](\Xi_1 \cap \Xi_2)$.
  Then there exists $(a':\tau)$ in both $\Xi_1$ and $\Xi_2$
  such that $[\sigma]a' = a$, so $(a:\tau) \in [\sigma]\Xi_2 \cap [\sigma]\Xi_2$.
\end{proof}

\begin{lemma}[Substitution Monotonicity]
  \label{lem:monotone-subs}
  If $\Xi' \supseteq \Xi$, then $[\sigma]\Xi' \supseteq [\sigma]\Xi$.
\end{lemma}
\begin{proof}
  Suppose $\Xi' \supseteq \Xi$. Let $(a':\tau) \in [\sigma]\Xi$.
  Then there exists $(a:\tau) \in \Xi$ such that $[\sigma]a = a'$.
  Because $\Xi' \supseteq \Xi$ and $(a:\tau)\in\Xi$, we have $(a:\tau) \in \Xi'$.
  We need to show $(a':\tau)\in [\sigma]\Xi'$,
  i.e. that there exists $(b:\tau)\in\Xi'$ such that $[\sigma]b = a'$;
  but we have just shown that $(a:\tau)$ is such an element.
\end{proof}

\begin{lemma}[Barendregt's Substitution]
  \label{lem:barendregt}
  Assume $\Theta_0;\cdot |- \sigma : \Theta;\cdot$ and $\Theta |- t : \tau$\\
  and $a \notin \dom{\Theta_0}$.
  \begin{enumerate}
  \item If $\judgetp{\Theta,a:\tau}{A}{\Xi}$,
    then $[\sigma]([t/a]A) = [[\sigma]t/a]([\sigma]A)$.
  \item If $\judgefunctor{\Theta,a:\tau}{\mathcal{F}}{\Xi}$,
    then $[\sigma]([t/a]\mathcal{F}) = [[\sigma]t/a]([\sigma]\mathcal{F})$.
  \end{enumerate}
\end{lemma}
\begin{proof}
  By mutual induction on the structure of the given well-formedness derivation.
  (Use similar but unstated Barendregt's substitution lemma
  for index sorting derivations.)
\end{proof}

\begin{lemma}[Algebra Pattern-Picking Substitution]
  \label{lem:alg-pat-subs}
  If $\composeinj{k}{\alpha}{\alpha_k}$,\\
  then $\composeinj{k}{[\sigma]\alpha}{[\sigma]\alpha_k}$ for all $\sigma$.
\end{lemma}
\begin{proof}
  By induction on the derivation of $\composeinj{k}{\alpha}{\alpha_k}$.
\end{proof}

\begin{lemma}[Id.\ Prepend]
  \label{lem:id-prepend}
  If $\Theta_0; \cdot |- \sigma : \Theta; \cdot$
  then $\Theta_0; \cdot |- \idsubs{\Theta_0}{\cdot}, \sigma : \Theta_0, \Theta; \cdot$.
\end{lemma}
\begin{proof}
  By structural induction on $\Theta_0; \cdot |- \sigma : \Theta; \cdot$.
  Use \Lemmaref{lem:id-subs-id},
  \Lemmaref{lem:syn-subs-prop-true},
  \Lemmaref{lem:syn-subs-ix},
  \Lemmaref{lem:id-subs-typing},
  the definition of substitution,
  and properties of appending lists.
\end{proof}

\begin{lemma}[Subst.\ Append]
  \label{lem:subs-append}
  If $\Theta_0 |- \sigma_1 : \Theta_1$
  and $\Theta_0 |- \sigma_2 : \Theta_2$
  and $\judgctx{(\Theta_1, \Theta_2)}$
  and $\judgctx{(\Theta_0, \Theta_1)}$,\\
  then $\Theta_0 |- \sigma_1, \sigma_2 : \Theta_1, \Theta_2$.
\end{lemma}
\begin{proof}
  By structural induction on $\Theta_0 |- \sigma_2 : \Theta_2$.
  We show one case:
  \begin{enumerate}
    \DerivationProofCase{\IxSyn}
    {
      \Theta_0 |- \sigma_2' : \Theta_2'
      \and
      \Theta_0 |- [\sigma_2']t : \tau
      \and
      a \notin \dom{\Theta'}
    }
    {
      \Theta_0 |- \sigma_2', t/a : \Theta_2', a:\tau
    }
    \begin{llproof}
      \judgsubsPf{\Theta_0}{\sigma_2'}{\Theta_2'}{Subderivation}
      \judgsubsPf{\Theta_0}{\sigma_1, \sigma_2'}{\Theta_1, \Theta_2'}{By \ih}
      \judgetermPf{\Theta_0}{[\sigma_2']t}{\tau}{Premise}
      \judgctxPf{(\Theta_0, \Theta_1)}{Given}
      \eqPf{\dom{\Theta_0} \cap \dom{\Theta_1}}{\emptyset}{By inversion}
      \eqPf{[\sigma_1, \sigma_2']t}{[\sigma_1]([\sigma_2']t)}{Straightforward}
      \eqPf{}{[\sigma_2']t}{$\dom{\Theta_0} \cap \dom{\Theta_1} = \emptyset$}
      \judgetermPf{\Theta_0}{[\sigma_1, \sigma_2']t}{\tau}{By equality}
      \judgsubsPf{\Theta_0}{\sigma_1, \sigma_2}{\Theta_1, \Theta_2}{By \IxSyn}
    \end{llproof} 
    \qedhere
  \end{enumerate}
\end{proof}

\begin{lemma}[Sem.\ Subst.\ Append]
  \label{lem:sem-subs-append}
  If $|- \delta, \delta_1 : \Theta, \Theta_1$
  and $|- \delta, \delta_2 : \Theta, \Theta_2$
  and $|- \delta : \Theta$,
  then $|- \delta, \delta_1, \delta_2 : \Theta, \Theta_1, \Theta_2$.
\end{lemma}
\begin{proof}
  By structural induction on $\Theta_2$.
\end{proof}

\begin{lemma}[Syn.\ Subs.\ Entry]
  \label{lem:syn-subs-deep-entry}
  ~
  \begin{enumerate}
  \item If $\Theta_0 |- \sigma_1 : \Theta_1$
    and $\Theta_0 |- \sigma_1, \sigma_2 : \Theta_1, [t/a]\Theta_2$
    and $\judgeterm{\Theta_0}{t}{\tau}$ and $a$ is fresh,\\
    then $\Theta_0 |- \sigma_1, t/a, \sigma_2 : \Theta_1, a:\tau, \Theta_2$.
  \item If $\Theta_0 |- \sigma_1 : \Theta_1$
    and $\Theta_0 |- \sigma_1, \sigma_2 : \Theta_1, \Theta_2$
    and $\judgeentail{\Theta_0}{[\sigma_1]\phi}$,\\
    then $\Theta_0 |- \sigma_1, \sigma_2 : \Theta_1, \phi, \Theta_2$.
  \end{enumerate}
\end{lemma}
\begin{proof}
  Each part is proved by structural induction on $\Theta_2$.
\end{proof}

\subsection{Structural Properties}

\begin{definition}[Proposition-Independent]
  \label{def:assertion-independent}
  We say that a judgement of the form $\Theta |- \mathcal{J}$
  is \emph{proposition-independent} if
  $\Theta_1,\phi,\Theta_2 |- \mathcal{J}$ implies $\Theta_1,\Theta_2 |- \mathcal{J}$.
  Similarly, $\Xi; \Theta |- \mathcal{J}$ is proposition-independent
  if $\Xi; \Theta_1, \phi, \Theta_2 |- \mathcal{J}$
  implies $\Xi; \Theta_1, \Theta_2 |- \mathcal{J}$.
\end{definition}

\begin{lemma}[Proposition-Independent Judgement]
  \label{lem:assertion-independent}
  The following judgement forms are proposition-independent
  (Definition~\ref{def:assertion-independent}):
  \begin{enumerate}
    \item $\Theta |- t : \tau$
    \item $\judgetp{\Theta}{A}{\Xi}$
    \item $\judgetp{\Theta}{\mathcal{F}}{\Xi}$
    \item $\judgealgebra{\Xi}{\Theta}{\alpha}{F}{\tau}$
    \item $\judgeunroll{\Xi}{\Theta}{ \nu:G[\mu F] }{\beta}{G\;\Fold{F}{\alpha}\;\nu}{t}{P}{\tau}$
    \item $\judgeextract[\pm]{\Theta}{A}{A'}{\Theta'}$
  \end{enumerate}
\end{lemma}
\begin{proof}
  Each part is proved by induction on the derivation;
  parts (2), (3), and (4) mutually. \qedhere
\end{proof}

\begin{lemma}[Ix.-Level Weakening]
  \label{lem:ix-level-weakening}
  Assume $\judgctx{(\Theta_1, \Theta_0, \Theta_2)}$.
  \begin{enumerate}
    \item If
      $\Dee :: \judgetp{\Theta_1,\Theta_2}{A}{\Xi}$,
      then $\judgetp{\Theta_1,\Theta_0,\Theta_2}{A}{\Xi}$
      by a derivation of equal height.
    \item If
      $\Dee :: \judgefunctor{\Theta_1,\Theta_2}{\mathcal{F}}{\Xi}$,
      then $\judgefunctor{\Theta_1,\Theta_0,\Theta_2}{\mathcal{F}}{\Xi}$
      by a derivation of equal height.
    \item If
      $\Dee :: \judgealgebra{\Xi}{\Theta_1,\Theta_2}{\alpha}{F}{\tau}$,
      then $\judgealgebra{\Xi}{\Theta_1,\Theta_0,\Theta_2}{\alpha}{F}{\tau}$
      by a derivation of equal height.

    \item If $\Dee :: \judgeequiv{\Theta_1, \Theta_2}{\phi}{\psi}$,
      then $\judgeequiv{\Theta_1, \Theta_0, \Theta_2}{\phi}{\psi}$
      by a derivation of equal height.
    \item If $\Dee :: \judgeequiv{\Theta_1, \Theta_2}{\Theta}{\Theta'}$,
      then $\judgeequiv{\Theta_1, \Theta_0, \Theta_2}{\Theta}{\Theta'}$
      by a derivation of equal height.
    \item If $\Dee :: \judgeequiv[]{\Theta_1, \Theta_2}{\mathcal{F}}{\mathcal{G}}$,
      then $\judgeequiv[]{\Theta_1, \Theta_0, \Theta_2}{\mathcal{F}}{\mathcal{G}}$
      by a derivation of equal height.
    \item If $\Dee :: \judgeequiv[\pm]{\Theta_1, \Theta_2}{A}{B}$,
      then $\judgeequiv[\pm]{\Theta_1, \Theta_0, \Theta_2}{A}{B}$
      by a derivation of equal height.

    \item If
      $\Dee :: \judgeextract[\pm]{\Theta_1, \Theta_2}{A}{A'}{\Theta'}$,
      then 
      $\judgeextract[\pm]{\Theta_1, \Theta_0, \Theta_2}{A}{A'}{\Theta'}$
      by a derivation of equal height.

    \item If
      $\Dee :: \judgesub[\pm]{\Theta_1, \Theta_2}{A}{B}$,
      then
      $\judgesub[\pm]{\Theta_1, \Theta_0, \Theta_2}{A}{B}$
      by a derivation of equal height.

    \item If
      $\Dee :: \judgeunroll{\Xi}{\Theta_1,\Theta_2}{ \nu:G[\mu F] }{\beta}
      {G\;\Fold{F}{\alpha}\;\nu}{t}{P}{\tau}$,\\
      then
      $\judgeunroll{\Xi}{\Theta_1,\Theta_0,\Theta_2}{ \nu:G[\mu F] }{\beta}
      {G\;\Fold{F}{\alpha}\;\nu}{t}{P}{\tau}$
      by a derivation of equal height.

    \item If
      $\Dee :: \judgesynhead{\Theta_1, \Theta_2}{\Gamma}{h}{P}$,
      then
      $\judgesynhead{\Theta_1, \Theta_0, \Theta_2}{\Gamma}{h}{P}$
      by a derivation of equal height.
    \item If
      $\Dee :: \judgesynexp{\Theta_1, \Theta_2}{\Gamma}{\be}{\upshift{P}}$,
      then
      $\judgesynexp{\Theta_1, \Theta_0, \Theta_2}{\Gamma}{\be}{\upshift{P}}$
      by a derivation of equal height.
    \item If
      $\Dee :: \judgechkval{\Theta_1, \Theta_2}{\Gamma}{v}{P}$,
      then
      $\judgechkval{\Theta_1, \Theta_0, \Theta_2}{\Gamma}{v}{P}$
      by a derivation of equal height.
    \item If
      $\Dee :: \judgechkexp{\Theta_1, \Theta_2}{\Gamma}{e}{N}$,
      then
      $\judgechkexp{\Theta_1, \Theta_0, \Theta_2}{\Gamma}{e}{N}$
      by a derivation of equal height.
    \item If 
      $\Dee :: \judgechkmatch{\Theta_1, \Theta_2}{\Gamma}{P}{\clauses{\pa}{e}{i}{I}}{N}$,
      then
      $\judgechkmatch{\Theta_1, \Theta_0, \Theta_2}{\Gamma}{P}{\clauses{\pa}{e}{i}{I}}{N}$
      by a derivation of equal height.
    \item If
      $\Dee :: \judgespine{\Theta_1, \Theta_2}{\Gamma}{s}{N}{\upshift{P}}$,
      then 
      $\judgespine{\Theta_1, \Theta_0, \Theta_2}{\Gamma}{s}{N}{\upshift{P}}$
        by a derivation of equal height.
  \end{enumerate}
\end{lemma}
\begin{proof}
  Each part is proved by structural induction on the given derivation $\Dee$;
  parts (2), (3), and (4) mutually;
  parts (8) and (9) mutually;
  parts (13), (14), (15), (16), (17), and (18) mutually.
\end{proof}

\begin{lemma}[Ix.-Level \ Subs.\ Weakening]
  \label{lem:syn-subs-weakening}
  If $\Theta_1,\Theta_2; \Gamma_0 |- \sigma : \Theta; \Gamma$
  and $\judgctx{\Theta_1, \Theta_0, \Theta_2}$,\\
  then $\Theta_1,\Theta_0,\Theta_2; \Gamma_0 |- \sigma : \Theta; \Gamma$.
\end{lemma}
\begin{proof}
  Similar to \Lemmaref{lem:ix-syn-subs-weakening},
  but the \ValSyn case uses (the value typechecking part of)
  \Lemmaref{lem:ix-level-weakening}.
\end{proof}

\begin{lemma}[Permute Ctx.\ WF]
  \label{lem:permute-ctx-wf}
  Assume $\judgctx{(\Theta, \Theta_1)}$ and $\judgctx{(\Theta, \Theta_2)}$. Then:
  \begin{enumerate}
  \item We have $\judgctx{(\Theta, \Theta_1, \Theta_2)}$.
  \item We have $\judgctx{(\Theta, \Theta_2, \Theta_1)}$.
  \end{enumerate}
\end{lemma}
\begin{proof}
  ~
  \begin{enumerate}
  \item
    By structural induction on $\Theta_2$, case analyzing its structure.
    The $\Theta_2 = \Theta_2', \phi$ case uses \Lemmaref{lem:sorting-weakening}.
  \item
    Similarly to part (1), by structural induction on $\Theta_1$. \qedhere
  \end{enumerate}
\end{proof}

\begin{lemma}[Index Id.\ Subs.\ Extension]
  \label{lem:id-subs-ext-ix}
  Assume $\Theta_0;\Gamma_0 |- \sigma : \Theta; \Gamma$.
  If $a \notin \dom{\Theta} \cup \dom{\Theta_0}$,
  then $\Theta_0,a:\tau;\Gamma_0 |- \sigma,a/a : \Theta,a:\tau;\Gamma$.
\end{lemma}
\begin{proof}
  ~\\
  \begin{llproof}
    \Pf{\Theta_0;\Gamma_0 |- \sigma : \Theta;\Gamma}{}{}{Given}
    \Pf{\Theta_0,a:\tau;\Gamma_0 |- \sigma : \Theta;\Gamma}{}{}{By \Lemmaref{lem:syn-subs-weakening}}
    \Pf{\Theta_0,a:\tau |- a : \tau}{}{}{By \IxVar}
    \Pf{\Theta_0,a:\tau |- [\filterprog{\sigma}]a : \tau}{}{}{$a \notin \dom{\Theta}$}
    \Pf{\Theta_0,a:\tau;\Gamma_0 |- \sigma,a/a : \Theta,a:\tau;\Gamma}{}{}{By \IxSyn}
  \end{llproof}
\end{proof}

\begin{lemma}[Proposition Identity Substitution Extension]
  \label{lem:id-subs-ext-prop}
  Assume $\Theta_0;\Gamma_0 |- \sigma : \Theta; \Gamma$.
  If $\judgeterm{\Theta}{\phi}{\Booltype}$,
  then $\Theta_0,[\filterprog{\sigma}]\phi;\Gamma_0 |- \sigma : \Theta,\phi;\Gamma$.
\end{lemma}
\begin{proof}
  Assume $\judgeterm{\Theta}{\phi}{\Booltype}$.\\
  \begin{llproof}
    \Pf{\judgeterm{\Theta_0}{[\filterprog{\sigma}]\phi}{\Booltype}}{}{}{By \Lemmaref{lem:syn-subs-ix}}
    \Pf{\Theta_0,[\filterprog{\sigma}]\phi;\Gamma_0 |- \sigma : \Theta;\Gamma}{}{}{By \Lemmaref{lem:syn-subs-weakening}}
    \Pf{\judgeentail{\Theta_0,[\filterprog{\sigma}]\phi}{[\filterprog{\sigma}]\phi}}{}{}{By \Lemmaref{lem:propvar}}
    \Pf{\Theta_0,[\filterprog{\sigma}]\phi;\Gamma_0 |- \sigma : \Theta,\phi;\Gamma}{}{}{By \PropSyn}
  \end{llproof}
\end{proof}

\begin{lemma}[Ix.-Level Id.\ Subs.\ Extension]
  \label{lem:id-subs-ext-ix-level}
  Assume $\Theta_0; \Gamma_0 |- \sigma : \Theta; \Gamma$.\\
  If $\judgctx{(\Theta, \Theta')}$ and $\judgctx{(\Theta_0, \Theta')}$,
  then $\Theta_0, [\filterprog{\sigma}]\Theta'; \Gamma_0 |- \sigma, \id_{\Theta'} : \Theta, \Theta'; \Gamma$.
\end{lemma}
\begin{proof}
  By structural induction on the given substitution typing derivation,
  analyzing cases for the structure of $\Theta'$.
  The $\Theta' = \Theta_1', a:\tau$ case uses \Lemmaref{lem:id-subs-ext-ix}.
  The $\Theta' = \Theta_1', \phi$ case uses \Lemmaref{lem:id-subs-ext-prop}. \qedhere
\end{proof}

\begin{definition}[Subst.\ FV Image]
  \label{def:fvimctx}
  Given $\Theta_0; \Gamma_0 |- \sigma : \Theta; \Gamma$ and $\Xi \subseteq \Theta$,
  define
  \[
    \fvimctx{\sigma}{\Xi} = \bigcup_{(a:\wild)\in\Xi} \bigcup_{b \in \FV{[\sigma]a}} \{b:\Theta_0(b)\}
  \]
\end{definition}

\begin{lemma}[Subst.\ FV Image Subset]
  \label{lem:fvimctx-subset}
  If $\Theta_0; \Gamma_0 |- \sigma : \Theta; \Gamma$ and $\Xi \subseteq \Theta$,
  then $\fvimctx{\sigma}{\Xi} \subseteq \Theta_0$.
\end{lemma}
\begin{proof}
  Follows from \Definitionref{def:fvimctx}.
\end{proof}

\begin{lemma}[Ix.\ Ctx.\ Strengthen]
  \label{lem:ix-ctx-strengthen}
  If $\judgeterm{\Theta}{t}{\tau}$,
  then $\judgeterm{\cup_{a \in \FV{t}} \{a : \Theta(a)\}}{t}{\tau}$.
\end{lemma}
\begin{proof}
  By structural induction on the given sorting derivation.
\end{proof}

\begin{lemma}[Ix.\ Subst.\ FV Image Subset]
  \label{lem:fvimctx-ix-subset}
  If $\Theta_0; \cdot |- \sigma : \Theta; \cdot$
  and $\Xi \subseteq \Theta$
  and $\judgeterm{\Xi}{t}{\tau}$,\\
  then $\cup_{a\in\FV{[\sigma]t}}\{a:\Theta_0(a)\} \subseteq \fvimctx{\sigma}{\Xi}$.
\end{lemma}
\begin{proof}
  By structural induction on the given sorting derivation,
  and using \Definitionref{def:fvimctx}.
\end{proof}

\begin{lemma}[Strong Index Substitution]
  \label{lem:strong-ix-subs}
  If $\Theta_0; \cdot |- \sigma : \Theta; \cdot$
  and $\Xi \subseteq \Theta$
  and $\judgeterm{\Xi}{t}{\tau}$,\\
  then $\judgeterm{\fvimctx{\sigma}{\Xi}}{[\sigma]t}{\tau}$.
\end{lemma}
\begin{proof} ~\\
  \begin{llproof}
    \judgetermPf{\Theta}{t}{\tau}{By repeated \Lemmaref{lem:sorting-weakening}}
    \judgetermPf{\Theta_0}{[\sigma]t}{\tau}{By \Lemmaref{lem:syn-subs-ix}}
    \judgetermPf{\cup_{a \in \FV{[\sigma]t}} \{a:\Theta_0(a)\}}{[\sigma]t}{\tau}{By \Lemmaref{lem:ix-ctx-strengthen}}
    \Pf{\cup_{a\in\FV{[\sigma]t}} \{a:\Theta_0(a)\}}{\subseteq}{\fvimctx{\sigma}{\Xi}}{By \Lemmaref{lem:fvimctx-ix-subset}}
    \judgetermPf{\fvimctx{\sigma}{\Xi}}{[\sigma]t}{\tau}{By repeated \Lemmaref{lem:sorting-weakening}}
  \end{llproof}
\end{proof}

\begin{lemma}[Id.\ FV Image]
  \label{lem:fvimctx-id}
  If $\Xi |- \id_\Xi : \Xi$, then $\fvimctx{\id_\Xi}{\Xi} = \Xi$.
\end{lemma}
\begin{proof}
  Straightforward.
\end{proof}

\begin{lemma}
  \label{lem:fvimctx-id-convenient}
  If $\Theta_0;\cdot|-\sigma:\Theta;\cdot$
  and $\Theta_0,a:\tau;\cdot|-\sigma,a/a:\Theta,a:\tau;\cdot$
  and $\Xi \subseteq \Theta$,\\
  then $\fvimctx{\sigma,a/a}{(\Xi,a:\tau)} = \fvimctx{\sigma}{\Xi}, a:\tau$.
\end{lemma}
\begin{proof}
  \begin{align*}
    \hspace{2em}&\hspace{-2em}\fvimctx{\sigma,a/a}{(\Xi,a:\tau)}\\
    &= \bigcup_{(a':\wild)\in(\Xi,a:\tau)} \bigcup_{b\in\FV{[\sigma,a/a]a'}} \{b:(\Theta_0,a:\tau)(b)\}
      &&\quad\text{By \Defnref{def:fvimctx}} \\
    &= \left(
      \bigcup_{(a':\wild)\in\Xi} \bigcup_{b\in\FV{[\sigma,a/a]a'}} \{b : (\Theta_0,a:\tau)(b)\}
      \right)
      \cup (a:\tau)
      &&\quad\text{By set theory} \\
    &= \left(
      \bigcup_{(a':\wild)\in\Xi} \bigcup_{b\in\FV{[\sigma]a'}} \{b : (\Theta_0,a:\tau)(b)\}
      \right)
      \cup (a:\tau)
      &&\quad\text{Because } a \notin \dom{\Xi}\\
    &= \left(
      \bigcup_{(a':\wild)\in\Xi} \bigcup_{b\in\FV{[\sigma]a'}} \{b : \Theta_0(b)\}
      \right)
      \cup (a:\tau)
      &&\quad\text{Because }a\notin\dom{\Theta_0}\cup\dom{\Xi}\\
    &= (\fvimctx{\sigma}{\Xi}), (\fvimctx{a/a}{(a:\tau)})
      &&\quad\text{By \Defnref{def:fvimctx}} \\
    &= \fvimctx{\sigma}{\Xi}, a:\tau
      &&\quad\text{By \Lemmaref{lem:fvimctx-id}}
  \end{align*}
  where the last line uses the fact that $a:\tau |- a/a : a:\tau$.
\end{proof}

\begin{lemma}[WF Syn.\ Substitution]
  \label{lem:syn-subs-tp-fun-alg}
  Assume $\Theta_0; \cdot |- \sigma : \Theta; \cdot$. Then:
  \begin{enumerate}
  \item If $\judgetp{\Theta}{A}{\Xi}$,
    then there exists $\Xi'$
    such that $\Xi' \supseteq [\sigma]\Xi$
    and $\judgetp{\Theta_0}{[\sigma]A}{\Xi'}$.
  \item If $\judgefunctor{\Theta}{\mathcal{F}}{\Xi}$,
    then there exists $\Xi'$
    such that $\Xi' \supseteq [\sigma]\Xi$
    and $\judgefunctor{\Theta_0}{[\sigma]\mathcal{F}}{\Xi'}$.
  \item If $\judgealgebra{\Xi}{\Theta}{\alpha}{F}{\tau}$ and $\Xi \subseteq \Theta$,
    then $\judgealgebra{\fvimctx{\sigma}{\Xi}}{\Theta_0}{[\sigma]\alpha}{([\sigma]F)}{\tau}$.
  \end{enumerate}
\end{lemma}
\begin{proof}
  By mutual induction on the structure of the given well-formedness derivation.
  \begin{enumerate}
  \item Assume $\Dee::\judgetp{\Theta}{A}{\Xi}$.
    Consider cases for $\Dee$'s concluding rule:
    \begin{itemize}
      \DerivationProofCase{\DeclTpVoid}
      { }
      { \judgetp{\Theta}{0}{\cdot}  }
      \begin{llproof}
        \Pf{\judgetp{\Theta_0}{0}{\cdot}}{}{}{By \DeclTpVoid}
        \Pf{\judgetp{\Theta_0}{[\sigma]0}{\underbrace{[\sigma]\cdot}_\cdot}}{}{}{By def. of subst.}
        \Pf{\cdot \supseteq \cdot}{}{}{Set theory}
      \end{llproof}

      \ProofCaseRule{\DeclTpUnit} Similar to case for \DeclTpVoid.

      \DerivationProofCase{\DeclTpSum}
      { \judgetp{\Theta}{P_1}{\Xi_1} \\ \judgetp{\Theta}{P_2}{\Xi_2} }
      { \judgetp{\Theta}{P_1 + P_2}{\Xi_1 \sect \Xi_2}  }
      \begin{llproof}
        \Pf{\judgetp{\Theta_0}{[\sigma]P_1}{\Xi_1'}}{}{}{By i.h.}
        \Pf{\Xi_1' \supseteq [\sigma]\Xi_1}{}{}{\ditto}
        \Pf{\judgetp{\Theta_0}{[\sigma]P_2}{\Xi_2'}}{}{}{By i.h.}
        \Pf{\Xi_2' \supseteq [\sigma]\Xi_2}{}{}{\ditto}
        \Pf{\judgetp{\Theta_0}{[\sigma]P_1 + [\sigma]P_2}{\Xi_1' \sect \Xi_2'}}{}{}{By \DeclTpSum}
        \Pf{\judgetp{\Theta_0}{[\sigma](P_1 + P_2)}{\Xi_1' \sect \Xi_2'}}{}{}{By def. of subst.}
        \decolumnizePf
        \Pf{\Xi_1' \sect \Xi_2'}{\supseteq}{[\sigma]\Xi_1 \sect [\sigma]\Xi_2}{Set theory}
        \Pf{}{\supseteq}{[\sigma](\Xi_1 \sect \Xi_2)}{By Lemma~\ref{lem:intersection-subs}}
      \end{llproof}
      
      \DerivationProofCase{\DeclTpProd}
      { \judgetp{\Theta}{P_1}{\Xi_1} \\ \judgetp{\Theta}{P_2}{\Xi_2} }
      { \judgetp{\Theta}{P_1 \times P_2}{\Xi_1 \union \Xi_2}  }
      \begin{llproof}
        \Pf{\judgetp{\Theta_0}{[\sigma]P_k}{\Xi_k'}}{}{}{By \ih (twice)}
        \Pf{\Xi_k' \supseteq [\sigma]\Xi_k}{}{}{\ditto}
        \Pf{\judgetp{\Theta_0}{[\sigma]P_1 \times [\sigma]P_2}{\Xi_1' \union \Xi_2'}}{}{}{By \DeclTpProd}
        \Pf{\judgetp{\Theta_0}{[\sigma](P_1 \times P_2)}{\Xi_1' \union \Xi_2'}}{}{}{By def. of subst.}
        \decolumnizePf
        \Pf{\Xi_1' \union \Xi_2'}{\supseteq}{[\sigma]\Xi_1 \union [\sigma]\Xi_2}{Set theory}
        \Pf{}{=}{[\sigma](\Xi_1 \union \Xi_2)}{By Lemma~\ref{lem:union-subs}}
      \end{llproof}

      \DerivationProofCase{\DeclTpEx}
      { \judgetp{\Theta, a:\tau}{P}{\Xi, a:\tau} }
      { \judgetp{\Theta}{\extype{a:\tau}{P}}{\Xi} }
      \begin{llproof}
        \judgsubsPf{\Theta_0,a:\tau;\cdot}{\sigma,a/a}{\Theta,a:\tau;\cdot}{By \Lemmaref{lem:id-subs-ext-ix}}
        \judgetpPf{\Theta, a:\tau}{P}{\Xi, a:\tau}{Subderivation}
        \judgetpPf{\Theta_0, a:\tau}{[\sigma,a/a]P}{\Xi''}{By i.h.}
        \Pf{\Xi''}{\supseteq}{[\sigma,a/a](\Xi, a:\tau)}{\ditto}
        \Pf{}{=}{[\sigma,a/a]\Xi, a:\tau}{By \Defnref{def:Xi-subs}}
        \Pf{}{=}{[\sigma]\Xi, a:\tau}{$a \notin \dom{\Xi}$}
        \Pf{\Xi''}{=}{\Xi', a:\tau}{Follows from above $\supseteq$}
        \Hand\Pf{\Xi'}{\supseteq}{[\sigma]\Xi}{\ditto}
        \judgetpPf{\Theta_0, a:\tau}{[\sigma]P}{\Xi', a:\tau}{By identity subst. and def. of subst.}
        \judgetpPf{\Theta_0}{\extype{a:\tau}{[\sigma]P}}{\Xi'}{By \DeclTpEx}
        \judgetpPf{\Theta_0}{[\sigma]\extype{a:\tau}{P}}{\Xi'}{By def. of subst.}
      \end{llproof}

      \DerivationProofCase{\DeclTpDown}
      { \judgetp{\Theta}{N}{\Xi} }
      { \judgetp{\Theta}{\downshift{N}}{\cdot} }
      \begin{llproof}
        \Pf{\judgetp{\Theta_0}{[\sigma]N}{\dontcare}}{}{}{By i.h.}
        \Pf{\judgetp{\Theta_0}{[\sigma]\downshift{N}}{\cdot}}{}{}{By \DeclTpDown and def. of subst.}
        \Pf{\cdot = [\sigma]\cdot}{}{}{By \defn}
      \end{llproof}

      \DerivationProofCase{\DeclTpWith}
      {
        \judgetp{\Theta}{P}{\Xi}
        \\
        \judgeterm{\Theta}{\phi}{\Booltype}
      }
      { \judgetp{\Theta}{P \land \phi}{\Xi} }
      \begin{llproof}
        \Pf{\judgetp{\Theta_0}{[\sigma]P}{\Xi'}}{}{}{By i.h.}
        \Hand\Pf{\Xi' \supseteq [\sigma]\Xi}{}{}{\ditto}
        \Pf{\judgeterm{\Theta_0}{[\sigma]\phi}{\Booltype}}{}{}{By \Lemmaref{lem:syn-subs-ix}}
        \Pf{\judgetp{\Theta}{[\sigma](P \land \phi)}{\Xi'}}{}{}{By \DeclTpWith and def. of subst.}
      \end{llproof}

      \DerivationProofCase{\DeclTpFixVar}
      {
        \judgefunctor{\Theta}{F}{\Xi_0}
        \\
        \judgealgebra{\cdot}{\Theta}{\alpha}{F}{\tau}
        \\
        (b : \tau) \in \Theta
      }
      {
        \judgetp{\Theta}{\comprehend{\nu:\mu F}{\Fold{F}{\alpha}\,{\nu} =_\tau b}}{\Xi_0 \union b:\tau}
      }
      \begin{llproof}
        \Pf{\judgefunctor{\Theta_0}{[\sigma]F}{\Xi_0'}}{}{}{By i.h.}
        \Pf{\Xi_0' \supseteq [\sigma]\Xi_0}{}{}{\ditto}
        \Pf{\judgealgebra{\cdot}{\Theta_0}{[\sigma]\alpha}{([\sigma]F)}{\tau}}{}{}{By i.h. (and $\cdot = \cup_{(a:\tau)\in\cdot} \FV{[\sigma]a}$)}
        \Pf{\Theta_0 |- [\sigma]b : \tau}{}{}{By inversion}
      \end{llproof}
      \begin{itemize}
      \item \textbf{Case} $[\sigma]b$ is a variable:\\
        \begin{llproof}
          \Pf{([\sigma]b:\tau) \in \Theta_0}{}{}{By inversion on \IxVar}
          \decolumnizePf
          \Pf{\judgetp{\Theta_0}{\comprehend{\nu:\mu [\sigma]F}{\Fold{[\sigma]F}{[\sigma]\alpha}\,{\nu} =_\tau [\sigma]b}}{\Xi_0' \union [\sigma]b:\tau}}{}{}{By \DeclTpFixVar}
          \Pf{\judgetp{\Theta_0}{\comprehend{\nu:\mu [\sigma]F}{\Fold{[\sigma]F}{[\sigma]\alpha}\,{\nu} =_\tau [\sigma]b}}{\Xi_0' \union [\sigma](b:\tau)}}{}{}{By \Defnref{def:Xi-subs}}
          \decolumnizePf
          \Pf{\Xi_0' \union [\sigma](b:\tau)}{\supseteq}{[\sigma]\Xi \union [\sigma](b:\tau)}{By above $\supseteq$ and set theory}
          \Pf{}{=}{[\sigma](\Xi \union b:\tau)}{By \Lemref{lem:union-subs}}
        \end{llproof}
      \item \textbf{Case} $[\sigma]b$ is not a variable:\\
        First notice that
        \begin{align*}
          [\sigma](\Xi \union b:\tau)
          &= [\sigma]\Xi \union [\sigma](b:\tau) &&\quad\text{By \Lemref{lem:union-subs}} \\
          &= [\sigma]\Xi \union \cdot &&\quad\text{By \Defnref{def:Xi-subs}} \\
          &= [\sigma]\Xi &&\quad\text{Set theory}
        \end{align*}
        So, $\Xi_0' \supseteq [\sigma](\Xi \cup b:\tau)$.
        Now,\\
        \begin{llproof}
          \Pf{\judgetp{\Theta_0}{\comprehend{\nu:\mu [\sigma]F}{\Fold{[\sigma]F}{[\sigma]\alpha}\,{\nu} =_\tau [\sigma]b}}{\Xi_0'}}{}{}{By \DeclTpFix}
          \Pf{\judgetp{\Theta_0}{[\sigma]\comprehend{\nu:\mu F}{\Fold{F}{\alpha}\,{\nu} =_\tau b}}{\Xi_0'}}{}{}{By \defn of subst.}
        \end{llproof}
      \end{itemize}

      \DerivationProofCase{\DeclTpFix}
      {
        \judgefunctor{\Theta}{F}{\Xi}
        \\
        \judgealgebra{\cdot}{\Theta}{\alpha}{F}{\tau}
        \\
        \Theta |- t : \tau
        \\
        t \text{ not a variable}
      }
      {
        \judgetp{\Theta}{\comprehend{\nu:\mu F}{\Fold{F}{\alpha}\,{\nu} =_\tau t}}{\Xi}
      }
      \begin{llproof}
        \judgefunctorPf{\Theta_0}{[\sigma]F}{\Xi'}{By \ih}
        \Hand\Pf{\Xi'}{\supseteq}{[\sigma]\Xi}{\ditto}
        \judgealgebraPf{\cdot}{\Theta_0}{[\sigma]\alpha}{([\sigma]F)}{\tau}{By \ih}
        \decolumnizePf
        \judgsubsPf{\Theta_0}{[\sigma]t}{\tau}{By \Lemmaref{lem:syn-subs-ix}}
        \Pf{}{[\sigma]t}{\text{ not a variable}}{By \defn of subst.}
        \trailingjust{(and $t$ not a variable)}
        \judgetpPf{\Theta_0}{[\sigma]\comprehend{\nu:\mu F}{\Fold{F}{\alpha}\,{\nu} =_\tau t}}{\Xi'}{By \DeclTpFix and \defn of subst.}
      \end{llproof}

      \ProofCaseRule{\DeclTpAll} Similar to case for dual rule \DeclTpEx.

      \ProofCaseRule{\DeclTpImplies} Similar to case for dual rule \DeclTpWith.
      
      \ProofCaseRule{\DeclTpArrow} Similar to case for \DeclTpProd.
      
      \ProofCaseRule{\DeclTpUp} Similar to case for dual rule \DeclTpDown.
    \end{itemize}
  \item Similar to the type well-formedness part.
  \item Assume $\Dee :: \judgealgebra{\Xi}{\Theta}{\alpha}{F}{\tau}$
    and $\Xi \subseteq \Theta$.
    Consider cases for $\Dee$'s concluding rule:
    \begin{itemize}
      \DerivationProofCase{\DeclAlgSum}
      { \arrayenvb{\composeinj{1}{\alpha}{\alpha_1} \\ \composeinj{2}{\alpha}{\alpha_2}} \\
        \arrayenvb{\judgealgebra{\Xi}{\Theta}{\alpha_1}{F_1}{\tau} \\
        \judgealgebra{\Xi}{\Theta}{\alpha_2}{F_2}{\tau}} }
      { \judgealgebra{\Xi}{\Theta}{\alpha}{(F_1 \oplus F_2)}{\tau} }
      \begin{llproof}
        \Pf{\composeinj{k}{\alpha}{\alpha_k}}{}{}{Subderivations}
        \Pf{\composeinj{k}{([\sigma]\alpha)}{[\sigma]\alpha_k}}{}{}{By Lemma~\ref{lem:alg-pat-subs}}
        \Pf{\fvimctx{\sigma}{\Xi} \subseteq \Theta_0}{}{}{By \Lemmaref{lem:fvimctx-subset}}
        \Pf{\judgealgebra{\fvimctx{\sigma}{\Xi}}{\Theta_0}{[\sigma]\alpha_k}{[\sigma]F_k}{\tau}}{}{}{By i.h.}
        \Pf{\judgealgebra{\fvimctx{\sigma}{\Xi}}{\Theta_0}{[\sigma]\alpha}{([\sigma]F_1 \oplus [\sigma]F_2)}{\tau}}{}{}{By \DeclAlgSum}
        \Pf{\judgealgebra{\fvimctx{\sigma}{\Xi}}{\Theta_0}{[\sigma]\alpha}{([\sigma](F_1 \oplus F_2))}{\tau}}{}{}{By definition of substitution}
      \end{llproof}

      \DerivationProofCase{\DeclAlgUnit}
      { \judgeterm{\Xi}{t}{\tau} }
      { \judgealgebra{\Xi}{\Theta}{\clause{\unitexp}{t}}{I}{\tau} }
      \begin{llproof}
        \Pf{\fvimctx{\sigma}{\Xi} |- [\sigma]t : \tau}{}{}{By \Lemmaref{lem:strong-ix-subs}}
        \Pf{\judgealgebra{\fvimctx{\sigma}{\Xi}}{\Theta_0}{\clause{\unitexp}{[\sigma]t}}{I}{\tau}}{}{}{By \DeclAlgUnit}
        \Pf{\judgealgebra{\fvimctx{\sigma}{\Xi}}{\Theta_0}{[\sigma](\clause{\unitexp}{t})}{([\sigma]I)}{\tau}}{}{}{By def. of subst.}
      \end{llproof}

      \DerivationProofCase{\DeclAlgIdProd}
      { \judgealgebra{\Xi,a:\tau}{\Theta, a:\tau}{\clause{q}{t}}{\hat{P}}{\tau} }
      { \judgealgebra{\Xi}{\Theta}{\clause{(a, q)}{t}}{(\Id \otimes \hat{P})}{\tau} }
      \begin{llproof}
        \Pf{\Theta_0,a:\tau;\cdot |- \sigma,a/a : \Theta,a:\tau;\cdot}{}{}{By \Lemmaref{lem:id-subs-ext-ix}}
        \decolumnizePf
        \Pf{\Xi \subseteq \Theta}{}{}{Given}
        \Pf{\Xi, a:\tau \subseteq \Theta, a:\tau}{}{}{Obvious}
        \Pf{\judgealgebra{\fvimctx{\sigma,a/a}{(\Xi,a:\tau)}}{\Theta_0, a:\tau}{[\sigma](\clause{q}{t})}{([\sigma]\hat{P})}{\tau}}{}{}{By i.h.}
        \Pf{\judgealgebra{\fvimctx{\sigma}{\Xi},a:\tau}{\Theta_0, a:\tau}{[\sigma](\clause{q}{t})}{([\sigma]\hat{P})}{\tau}}{}{}{By \Lemref{lem:fvimctx-id-convenient}}
        \Pf{\judgealgebra{\fvimctx{\sigma}{\Xi},a:\tau}{\Theta_0, a:\tau}{\clause{q}{[\sigma]t}}{([\sigma]\hat{P})}{\tau}}{}{}{By def. of subst.}
        \Pf{\judgealgebra{\fvimctx{\sigma}{\Xi}}{\Theta_0}{\clause{(a, q)}{[\sigma]t}}{(\Id \otimes [\sigma]\hat{P})}{\tau}}{}{}{By \DeclAlgIdProd}
        \Pf{\judgealgebra{\fvimctx{\sigma}{\Xi}}{\Theta_0}{[\sigma](\clause{(a, q)}{t})}{([\sigma](\Id \otimes \hat{P}))}{\tau}}{}{}{By def. of subst.}
      \end{llproof}

      \DerivationProofCase{\DeclAlgConstProd}
      { \judgealgebra{\Xi}{\Theta}{\clause{q}{t}}{\hat{P}}{\tau} \\ \judgetp{\Theta}{Q}{\dontcare}}
      { \judgealgebra{\Xi}{\Theta}{\clause{(\wild, q)}{t}}{(\Const{Q} \otimes \hat{P})}{\tau} }
      \begin{llproof}
        \Pf{\judgetp{\Theta}{Q}{\dontcare}}{}{}{Subderivation}
        \Pf{\judgetp{\Theta_0}{[\sigma]Q}{\dontcare}}{}{}{By \ih}
        \Pf{\judgealgebra{\Xi}{\Theta}{\clause{q}{t}}{\hat{P}}{\tau}}{}{}{Subderivation}
        \Pf{\judgealgebra{\fvimctx{\sigma}{\Xi}}{\Theta_0}{\clause{q}{[\sigma]t}}{([\sigma]\hat{P})}{\tau}}{}{}{By i.h. and \defn of subst.}
        \Pf{\judgealgebra{\fvimctx{\sigma}{\Xi}}{\Theta_0}{\clause{(\wild, q)}{[\sigma]t}}{(\Const{[\sigma]Q} \otimes [\sigma]\hat{P})}{\tau}}{}{}{By \DeclAlgConstProd}
        \Pf{\judgealgebra{\fvimctx{\sigma}{\Xi}}{\Theta_0}{[\sigma](\clause{(\wild, q)}{t})}{([\sigma](\Const{Q} \otimes \hat{P}))}{\tau}}{}{}{By \defn of $[-]-$}
      \end{llproof}

      \DerivationProofCase{\DeclAlgExConstProd}
      {
        \judgealgebra{\Xi, a:\tau'}{\Theta, a:\tau'}{\clause{(\bap, q)}{t}}{(\Const{Q} \otimes \hat{P})}{\tau}
      }
      {
        \judgealgebra{\Xi}{\Theta}
        {\clause{(\pack{a}{\bap}, q)}{t}}
        {(\Const{\extype{a:\tau'}{Q}} \otimes \hat{P})}{\tau}
      }
      By \Lemmaref{lem:id-subs-ext-ix},
      \[
        \Theta_0,a:\tau';\cdot |- \sigma, a/a : \Theta,a:\tau';\cdot
      \]
      By i.h. and definition of substitution,
      \[
        \judgealgebra{\fvimctx{\sigma,a/a}{(\Xi,a:\tau')}}{\Theta_0, a:\tau'}{\clause{(\bap, q)}{[\sigma,a/a]t}}{(\Const{[\sigma,a/a]Q} \otimes [\sigma,a/a]\hat{P})}{\tau}
      \]
      By \Lemref{lem:fvimctx-id-convenient},
      \[
        \judgealgebra{\fvimctx{\sigma}{\Xi},a:\tau'}{\Theta_0, a:\tau'}{\clause{(\bap, q)}{[\sigma,a/a]t}}{(\Const{[\sigma,a/a]Q} \otimes [\sigma,a/a]\hat{P})}{\tau}
      \]
      By \DeclAlgExConstProd,
      \[
        \judgealgebra{\fvimctx{\sigma}{\Xi}}{\Theta_0}
        {\clause{(\pack{a}{\bap}, q)}{[\sigma,a/a]t}}
        {(\Const{\extype{a:\tau'}{[\sigma,a/a]Q}} \otimes [\sigma,a/a]\hat{P})}{\tau}
      \]
      By definition of substitution,
      \[
        \judgealgebra{\fvimctx{\sigma}{\Xi}}{\Theta_0}
        {[\sigma](\clause{(\pack{a}{\bap}, q)}{t})}
        {([\sigma](\Const{\extype{a:\tau'}{Q}} \otimes \hat{P}))}{\tau}
      \]
      \qedhere
    \end{itemize}
  \end{enumerate}
\end{proof}

We sometimes implicitly use the following lemma.

\begin{lemma}[Decl.\ Unrolling Output WF]
  \label{lem:decl-unroll-output-wf}
  If $\judgeunroll{\Xi}{\Theta}{\nu:G[\mu F]}{\beta}{G\; \Fold{F}{\alpha}\;\nu}{t}{P}{\tau}$,
  then there exists $\Xi'$ such that $\judgetp{\Theta}{P}{\Xi'}$.
\end{lemma}
\begin{proof}
  By structural induction on the given unrolling derivation.
\end{proof}

\begin{lemma}[Unrolling Syntactic Substitution]
  \label{lem:syn-subs-unroll}
  ~\\
  If $\Theta_0; \cdot |- \sigma : \Theta; \cdot$
  and $\Xi \subseteq \Theta$
  and $\judgeunroll{\Xi}{\Theta}{ \nu:G[\mu F] }{\beta}{G\;\Fold{F}{\alpha}\;\nu}{t}{P}{\tau}$,
  \\
  then $\judgeunroll{\fvimctx{\sigma}{\Xi}}{\Theta_0}{ \nu:([\sigma]G)[\mu [\sigma]F] }{([\sigma]\beta)}
  {([\sigma]G)\;\Fold{[\sigma]F}{[\sigma]\alpha}\;\nu}{[\sigma]t}{[\sigma]P}{\tau}$.
\end{lemma}
\begin{proof}
  By induction on the derivation of the unrolling judgement,
  considering cases for its concluding rule.
  Note that all six of the necessary ``substituted'' presupposed derivations,
  for the metavariables $G$, $F$, $\beta$, $\alpha$, $t$, and $P$,
  hold by \Lemmaref{lem:syn-subs-tp-fun-alg} and \Lemmaref{lem:syn-subs-ix}.
  \begin{itemize}
    \DerivationProofCase{\DeclUnrollSum}
    {\arrayenvbl{
        \composeinj{1}{\beta}{\beta_1} \\
        \composeinj{2}{\beta}{\beta_2}
      }
      \\
      \arrayenvbl{
        \judgeunroll{\Xi}{\Theta}{\nu:G[\mu F]}{\beta_1}{G\;\Fold{F}{\alpha}\;\nu}{t}{P}{\tau} \\
        \judgeunroll{\Xi}{\Theta}{\nu:H[\mu F]}{\beta_2}{H\;\Fold{F}{\alpha}\;\nu}{t}{Q}{\tau}
      }
    }
    { \judgeunroll{\Xi}{\Theta}{\nu:(G \oplus H)[\mu F]}{\beta}{(G \oplus H)\;\Fold{F}{\alpha}\;\nu}{t}{P + Q}{\tau} }
    By \Lemmaref{lem:alg-pat-subs},
    $\composeinj{k}{([\sigma]\beta)}{[\sigma]\beta_k}$.\\
    By i.h.,
    \[
      \judgeunroll{\fvimctx{\sigma}{\Xi}}{\Theta_0}{\nu:([\sigma]G)[\mu ([\sigma]F)]}{[\sigma]\beta_1}{([\sigma]G)\;\Fold{[\sigma]F}{[\sigma]\alpha}\;\nu}{[\sigma]t}{[\sigma]P}{\tau}
    \]
    By i.h.,
    \[
      \judgeunroll{\fvimctx{\sigma}{\Xi}}{\Theta_0}{\nu:([\sigma]H)[\mu ([\sigma]F)]}{[\sigma]\beta_2}{([\sigma]H)\;\Fold{[\sigma]F}{[\sigma]\alpha}\;\nu}{[\sigma]t}{[\sigma]Q}{\tau}
    \]
    By \DeclUnrollSum,
    \[
      \judgeunroll*{\fvimctx{\sigma}{\Xi}}{\Theta_0}{\nu:([\sigma]G \oplus [\sigma]H)[\mu [\sigma]F]}
      {[\sigma]\beta}
      {([\sigma]G \oplus [\sigma]H)\;\Fold{[\sigma]F}{[\sigma]\alpha}\;\nu}
      {[\sigma]t}{[\sigma]P + [\sigma]Q}{\tau}
    \]
    By definition of substitution,
    \[
      \judgeunroll*{\fvimctx{\sigma}{\Xi}}{\Theta_0}{\nu:([\sigma](G \oplus H))[\mu [\sigma]F]}
      {[\sigma]\beta}
      {([\sigma](G \oplus H))\;\Fold{[\sigma]F}{[\sigma]\alpha}\;\nu}
      {[\sigma]t}{[\sigma](P + Q)}{\tau}
    \]

    \DerivationProofCase{\DeclUnrollId}
    {
      \judgeunroll{\Xi, a:\tau}{\Theta, a:\tau}{\nu:\hat{P}[\mu F]}
      {(\clause{q}{t'})}{\hat{P}\;\Fold{F}{\alpha}\;\nu}{t}{Q}{\tau} 
    }
    {
      \judgeunroll*{\Xi}{\Theta}{\nu : (\Id\otimes\hat{P})[\mu F]}
      {(\clause{(a,q)}{t'})}
      {(\Id\otimes\hat{P})\;\Fold{F}{\alpha}\;\nu}{t}
      {\extype{a:\tau}
        {\comprehend{\nu:\mu F}{ \Fold{F}{\alpha}\,{\nu} =_\tau a } \times Q}}
      {\tau}
    }
    By \Lemmaref{lem:id-subs-ext-ix},
    \[
      \Theta_0,a:\tau;\cdot |- \sigma,a/a : \Theta,a:\tau;\cdot
    \]
    By i.h. and def. of subst. (applying $\sigma,a/a$ same as applying $\sigma$),
    \[
      \judgeunroll*{\fvimctx{\sigma,a/a}{(\Xi,a:\tau)}}{\Theta_0, a:\tau}
      {\nu:([\sigma]\hat{P})[\mu [\sigma]F]}
      {(\clause{q}{[\sigma]t'})}
      {([\sigma]\hat{P})\;\Fold{[\sigma]F}{[\sigma]\alpha}\;\nu}
      {[\sigma]t}{[\sigma]Q}{\tau} 
    \]
    By \Lemref{lem:fvimctx-id-convenient},
    \[
      \judgeunroll{\fvimctx{\sigma}{\Xi}, a:\tau}{\Theta_0, a:\tau}
      {\nu:([\sigma]\hat{P})[\mu [\sigma]F]}
      {(\clause{q}{[\sigma]t'})}
      {([\sigma]\hat{P})\;\Fold{[\sigma]F}{[\sigma]\alpha}\;\nu}
      {[\sigma]t}{[\sigma]Q}{\tau} 
    \]
    By \DeclUnrollId,
    \[
      \judgeunroll*{\fvimctx{\sigma}{\Xi}}{\Theta_0}{\nu : (\Id\otimes[\sigma]\hat{P})[\mu [\sigma]F]}
      {(\clause{(a,q)}{[\sigma]t'})}
      {(\Id\otimes[\sigma]\hat{P})\;\Fold{[\sigma]F}{[\sigma]\alpha}\;\nu}{[\sigma]t}
      {\extype{a:\tau}
        {\comprehend{\nu:\mu [\sigma]F}{ \Fold{[\sigma]F}{[\sigma]\alpha}\,{\nu} =_\tau a } \times [\sigma]Q}}
      {\tau}
    \]
    By definition of substitution,
    \[
      \judgeunroll*{\fvimctx{\sigma}{\Xi}}{\Theta_0}{\nu : ([\sigma](\Id\otimes\hat{P}))[\mu [\sigma]F]}
      {([\sigma](\clause{(a,q)}{t'}))}
      {([\sigma](\Id\otimes\hat{P}))\;\Fold{[\sigma]F}{[\sigma]\alpha}\;\nu}{[\sigma]t}
      {[\sigma](\extype{a:\tau}
        {\comprehend{\nu:\mu F}{ \Fold{F}{\alpha}\,{\nu} =_\tau a } \times Q})}
      {\tau}
    \]

    \DerivationProofCase{\DeclUnrollConst}
    {
      \judgeunroll{\Xi}{\Theta}{\nu:\hat{P}[\mu F]}{(\clause{q}{t'})}{\hat{P}\;\Fold{F}{\alpha}\;\nu}{t}{Q'}{\tau} }
    { \judgeunroll*{\Xi}{\Theta}{\nu :(\Const{Q}\otimes\hat{P})[\mu F]}{(\clause{(\wild,q)}{t'})}{(\Const{Q}\otimes\hat{P})\;\Fold{F}{\alpha}\;\nu}{t}{Q \times Q'}{\tau} }
    \begin{llproof}
      \Pf{\judgefunctor{\Theta}{\Const{Q} \otimes \hat{P}}{\dontcare}}{}{}{Presupposed derivation}
      \Pf{\judgetp{\Theta}{Q}{\dontcare}}{}{}{By inversion}
      \Pf{\judgetp{\Theta_0}{[\sigma]Q}{\dontcare}}{}{}{By \Lemmaref{lem:syn-subs-tp-fun-alg}}
    \end{llproof}\\
    By i.h. (and def. of subst. for the algebra),
    \[
      \judgeunroll{\fvimctx{\sigma}{\Xi}}{\Theta_0}{\nu:([\sigma]\hat{P})[\mu [\sigma]F]}{(\clause{q}{[\sigma]t'})}{([\sigma]\hat{P})\;\Fold{[\sigma]F}{[\sigma]\alpha}\;\nu}{[\sigma]t}{[\sigma]Q'}{\tau}
    \]
    By \DeclUnrollConst and definition of substitution,
    \[
      \judgeunroll*{\fvimctx{\sigma}{\Xi}}{\Theta_0}{\nu :([\sigma](\Const{Q}\otimes\hat{P}))[\mu [\sigma]F]}{([\sigma](\clause{(\wild,q)}{t'}))}{([\sigma](\Const{Q}\otimes\hat{P}))\;\Fold{[\sigma]F}{[\sigma]\alpha}\;\nu}{[\sigma]t}{[\sigma](Q \times Q')}{\tau}
    \]

    \ProofCaseRule{\DeclUnrollConstEx}
    Similar to case for \DeclUnrollId.
    
    \DerivationProofCase{\DeclUnrollUnit}
    { }
    { \judgeunroll{\Xi}{\Theta}{\nu:I[\mu F]}{(\clause{\unitexp}{t'})}{I\;\Fold{F}{\alpha}\;\nu}{t}{1 \land (t = t')}{\tau} }
    \begin{llproof}
      \Pf{\judgefunctor{\Theta}{F}{\Xi'}}{}{}{Presupposed derivation}
      \Pf{\judgefunctor{\Theta_0}{[\sigma]F}{\dontcare}}{}{}{By \Lemmaref{lem:syn-subs-tp-fun-alg}}
      \Pf{\judgealgebra{\Xi}{\Theta}{\clause{\unitexp}{t'}}{I}{\tau}}{}{}{Presupposed derivation}
      \Pf{\judgealgebra{\fvimctx{\sigma}{\Xi}}{\Theta_0}{\clause{\unitexp}{[\sigma]t'}}{I}{\tau}}{}{}{By Lemma~\ref{lem:syn-subs-tp-fun-alg} and def. of subst.}
      \Pf{\Theta_0 |- [\sigma]t' : \tau}{}{}{By inversion on \DeclAlgUnit}
      \Pf{\judgealgebra{\cdot}{\Theta}{\alpha}{F}{\tau}}{}{}{Presupposed derivation}
      \Pf{\judgealgebra{\fvimctx{\sigma}{\cdot}}{\Theta_0}{[\sigma]\alpha}{([\sigma]F)}{\tau}}{}{}{By Lemma~\ref{lem:syn-subs-tp-fun-alg}}
      \Pf{\judgealgebra{\cdot}{\Theta_0}{[\sigma]\alpha}{([\sigma]F)}{\tau}}{}{}{By \Defnref{def:fvimctx}}
      \Pf{\Theta |- t : \tau}{}{}{Presupposed derivation}
      \Pf{\Theta_0 |- [\sigma]t : \tau}{}{}{By Lemma~\ref{lem:syn-subs-ix}}
    \end{llproof}\\
    By \DeclUnrollUnit,
    \[
      \judgeunroll{\fvimctx{\sigma}{\Xi}}{\Theta_0}{\nu:I[\mu [\sigma]F]}{(\clause{\unitexp}{[\sigma]t'})}{I\;\Fold{[\sigma]F}{[\sigma]\alpha}\;\nu}{[\sigma]t}{1 \land ([\sigma]t = [\sigma]t')}{\tau}
    \]
    By definition of substitution,
    \[
      \judgeunroll{\fvimctx{\sigma}{\Xi}}{\Theta_0}{\nu:([\sigma]I)[\mu [\sigma]F]}{([\sigma](\clause{\unitexp}{t'}))}{[\sigma]I\;\Fold{[\sigma]F}{[\sigma]\alpha}\;\nu}{[\sigma]t}{[\sigma](1 \land (t = t'))}{\tau}
      \qedhere
    \]
  \end{itemize}
\end{proof}

\begin{lemma}[Prop.\ Equiv.\ Syn.\ Subs.]
  \label{lem:syn-subs-equiv-prop}
  Assume $\Theta_0 |- \sigma : \Theta$.
  If $\judgeequiv[]{\Theta}{\phi}{\psi}$,
  then $\judgeequiv[]{\Theta_0}{[\sigma]\phi}{[\sigma]\psi}$.
\end{lemma}
\begin{proof}
  By structural induction on the propositional equivalence derivation,
  and case analysis of its concluding rule.
  The \PrpEquivEq and \PrpEquivLeq cases use \Lemmaref{lem:syn-subs-prop-true}.
\end{proof}

\begin{lemma}[Tp./Func.\ Equiv.\ Syn.\ Subs.]
  \label{lem:syn-subs-equiv-tp-fun}
  Assume $\Theta_0; \cdot |- \sigma : \Theta; \cdot$.
  \begin{enumerate}
    \item If $\Dee \derives \judgeequiv[\pm]{\Theta}{A}{B}$,
      then there exists
      $\Dee' \derives \judgeequiv[\pm]{\Theta_0}{[\sigma]A}{[\sigma]B}$
      such that $\hgt{\Dee'} \leq \hgt{\Dee}$.
    \item If $\Dee \derives \judgeequiv[]{\Theta}{\mathcal{F}}{\mathcal{G}}$,
      then there exists
      $\Dee' \derives \judgeequiv[]{\Theta_0}{[\sigma]\mathcal{F}}{[\sigma]\mathcal{G}}$
      such that $\hgt{\Dee'} \leq \hgt{\Dee}$.
  \end{enumerate}
\end{lemma}
\begin{proof}
  By mutual induction on the equivalence derivation,
  and case analysis of its concluding rule,
  using \Lemmaref{lem:syn-subs-equiv-prop}
  and \Lemmaref{lem:syn-subs-prop-true}
  as needed.
\end{proof}

\begin{lemma}[Extraction Syn.\ Subs.]
  \label{lem:syn-subs-extract}
  If $\Theta_0; \cdot |- \sigma : \Theta, \cdot$
  and $\judgeextract{\Theta}{A}{A'}{\Theta'}$,\\
  then $\judgeextract{\Theta_0}{[\sigma]A}{[\sigma]A'}{[\sigma]\Theta'}$
  by a derivation of lesser or equal height.
\end{lemma}
\begin{proof}
  By structural induction on the given extraction derivation.
\end{proof}

\begin{lemma}[Sub.\ Syn.\ Subs.]
  \label{lem:syn-subs-sub}
  Assume $\Theta_0; \cdot |- \sigma : \Theta; \cdot$.
  If $\Dee \derives \judgesub[\pm]{\Theta}{A}{B}$,
  then there exists $\Dee' \derives \judgesub[\pm]{\Theta_0}{[\sigma]A}{[\sigma]B}$
  such that $\hgt{\Dee'} \leq \hgt{\Dee}$.
\end{lemma}
\begin{proof}
  We only show the construction of the derivation $\Dee'$.
  The fact that $\hgt{\Dee'} \leq \hgt{\Dee}$ is readily apparent from context:
  for each case where we use the \ih, we get that height condition for
  the subderivations to which we reapply only the corresponding subtyping rule
  of that case.
  \begin{itemize}
    \item \textbf{Cases} \DeclSubPosVoid and \DeclSubPosUnit: Straightforward.

    \DerivationProofCase{\DeclSubPosProd}
    {
      \judgesub[+]{\Theta}{P_1}{Q_1}
      \\
      \judgesub[+]{\Theta}{P_2}{Q_2}
    }
    {\judgesub[+]{\Theta}{P_1 \times P_2}{Q_1 \times Q_2}}
    \begin{llproof}
      \Pf{\judgesub[+]{\Theta}{P_k}{Q_k}}{}{}{Subderivations}
      \Pf{\judgesub[+]{\Theta_0}{[\sigma]P_k}{[\sigma]Q_k}}{}{}{By \ih (twice)}
      \Pf{\judgesub[+]{\Theta_0}{[\sigma](P_1 \times P_2)}{[\sigma](Q_1 \times Q_2)}}{}{}{By \DeclSubPosProd and def. of subst.}
    \end{llproof}

    \ProofCaseRule{\DeclSubPosSum}
    Similar to case for \DeclSubPosProd,
    but uses \Lemmaref{lem:syn-subs-equiv-tp-fun}.

    \DerivationProofCase{\DeclSubPosL}
    {
      \judgeextract{\Theta}{P}{P'}{\Theta'}
      \\
      \Theta' \neq \cdot
      \\
      \judgesub[+]{\Theta, \Theta'}{P'}{Q}
    }
    {
      \judgesub[+]{\Theta}{P}{Q}
    }
    \begin{llproof}
      \judgsubsPf{\Theta_0; \cdot}{\sigma}{\Theta; \cdot}{Given}
      \judgeextractPf{\Theta}{P}{P'}{\Theta'}{Subderivation}
      \judgeextractPf{\Theta_0}{[\sigma]P}{[\sigma]P'}{[\sigma]\Theta'}{By \Lemmaref{lem:syn-subs-extract}}
      \judgsubsPf{\Theta_0, [\sigma]\Theta'; \cdot}{\sigma, \id_{\Theta'}}{\Theta, \Theta'; \cdot}{By \Lemmaref{lem:id-subs-ext-ix-level}}
      \judgesubPf[+]{\Theta, \Theta'}{P'}{Q}{Subderivation}
      \judgesubPf[+]{\Theta, [\sigma]\Theta'}{[\sigma, \id_{\Theta'}]P'}{[\sigma, \id_{\Theta'}]Q}{By \ih}
      \judgesubPf[+]{\Theta, [\sigma]\Theta'}{[\sigma]P'}{[\sigma]Q}{By \Lemmaref{lem:id-subs-id}}
      \judgesubPf[+]{\Theta}{[\sigma]P}{[\sigma]Q}{By \DeclSubPosL}
    \end{llproof} 

    \DerivationProofCase{\DeclSubPosWithR}
    {\judgesub[+]{\Theta}{P}{Q} \\ \judgeentail{\Theta}{\phi} }
    {\judgesub[+]{\Theta}{P}{Q \land \phi} }
    \begin{llproof}
      \Pf{\judgesub[+]{\Theta_0}{[\sigma]P}{[\sigma]Q}}{}{}{By i.h.}
      \Pf{\judgeentail{\Theta_0}{[\sigma]\phi}}{}{}{By \Lemmaref{lem:syn-subs-prop-true}}
      \Pf{\judgesub[+]{\Theta_0}{[\sigma]P}{[\sigma](Q \land \phi)}}{}{}{By \DeclSubPosWithR and def. of subst.}
    \end{llproof}

    \DerivationProofCase{\DeclSubPosExR}
    { 
      \judgesub[+]{\Theta}{P}{[t/a]Q} 
      \\
      \judgeterm{\Theta}{t}{\tau} 
    }
    {\judgesub[+]{\Theta}{P}{\extype{a:\tau}{Q}} }
    \begin{llproof}
      \Pf{\judgesub[+]{\Theta}{P}{[t/a]Q}}{}{}{Subderivation}
      \Pf{\judgesub[+]{\Theta_0}{[\sigma]P}{[\sigma]([t/a]Q)}}{}{}{By i.h.}
      \Pf{[\sigma]([t/a]Q) = [[\sigma]t/a]([\sigma]Q)}{}{}{By \Lemmaref{lem:barendregt}}
      \Pf{\judgesub[+]{\Theta_0}{[\sigma]P}{[[\sigma]t/a]([\sigma]Q)}}{}{}{By above two lines}
      \Pf{\judgeterm{\Theta_0}{[\sigma]t}{\tau}}{}{}{By \Lemmaref{lem:syn-subs-ix}}
      \Pf{\judgesub[+]{\Theta_0}{[\sigma]P}{\extype{a:\tau}{[\sigma]Q}}}{}{}{By \DeclSubPosExR}
      \Pf{\judgesub[+]{\Theta_0}{[\sigma]P}{[\sigma](\extype{a:\tau}{Q}})}{}{}{By def. of subst.}
    \end{llproof}

    \DerivationProofCase{\DeclSubPosFix}
    { \judgeequiv[]{\Theta}{F}{G} \\ \judgeentail{\Theta}{t = t'} }
    { \judgesub[+]{\Theta}{\comprehend{\nu:\mu F}{\Fold{F}{\alpha}\,\nu =_\tau t}}{\comprehend{\nu:\mu G}{\Fold{G}{\alpha}\,\nu =_\tau t'}} }
    \begin{llproof}
      \Pf{\judgeequiv[]{\Theta}{F}{G}}{}{}{Subderivation}
      \Pf{\judgeequiv[]{\Theta_0}{[\sigma]F}{[\sigma]G}}{}{}{By \Lemmaref{lem:syn-subs-equiv-tp-fun}}
      \Pf{\judgeentail{\Theta}{t = t'}}{}{}{By inversion on \DeclSubPosFix}
      \Pf{\judgeentail{\Theta_0}{[\sigma](t = t')}}{}{}{By \Lemref{lem:syn-subs-prop-true}}
      \Pf{\judgeentail{\Theta_0}{[\sigma]t = [\sigma]t'}}{}{}{By def. of subst.}
    \end{llproof}\\
    By \DeclSubPosFix,
    \begin{align*}
      \Theta_0 |-&
                 \comprehend{\nu:\mu ([\sigma]F)}
                 {\Fold{[\sigma]F}{\alpha}\,\nu =_\tau [\sigma]t} \\
               \leq^{+}&
                 \comprehend{\nu:\mu ([\sigma]G)}
                 {\Fold{[\sigma]G}{\alpha}\,\nu =_\tau [\sigma]t'}
    \end{align*}
    By definition of substitution,
    \begin{align*}
      \Theta_0 |-&
                   [\sigma]\comprehend{\nu:\mu F}
                   {\Fold{F}{\alpha}\,\nu =_\tau t} \\
      \leq^{+}&
                [\sigma]\comprehend{\nu:\mu G}
                {\Fold{G}{\alpha}\,\nu =_\tau t'}
    \end{align*}

    \DerivationProofCase{\DeclSubPosDownshift}
    {
      \judgesub[-]{\Theta}{N}{M}
    }
    {\judgesub[+]{\Theta}{\downshift{N}}{\downshift{M}}}
    \begin{llproof}
      \Pf{\judgesub[-]{\Theta_0}{[\sigma]N}{[\sigma]M}}{}{}{By i.h.}
      \Pf{\judgesub[+]{\Theta_0}{[\sigma]\downshift{N}}{[\sigma]\downshift{M}}}{}{}{By \DeclSubPosDownshift and def. of subst.}
    \end{llproof}

    \ProofCaseRule{\DeclSubNegUpshift}
    Similar to case for dual rule \DeclSubPosDownshift.

    \ProofCaseRule{\DeclSubNegImpL}
    Similar to case for dual rule \DeclSubPosWithR.

    \ProofCaseRule{\DeclSubNegAllL}
    Similar to case for dual rule \DeclSubPosExR.
    
    \ProofCaseRule{\DeclSubNegR}
    Similar to case for dual rule \DeclSubPosL.
    
    \ProofCaseRule{\DeclSubNegArrow}
    Similar to case for \DeclSubPosProd.
    \qedhere
  \end{itemize}
\end{proof}

\subsection{Extraction Properties}

\begin{definition}[Num.\ of Index Connectives]
  \label{def:numlog}
  Given $\judgetp{\Theta}{A}{\dontcare}$,
  define $\numlog{A}$
  by the total number of $\land$, $\exists$, $\implies$, and $\forall$ connectives
  in $A$,
  not counting any such connectives in functors of inductive types.
\end{definition}

\begin{lemma}[Extraction Determinism]
  \label{lem:extract-determinism}
  If $\judgetp{\Theta}{A}{\Xi}$,\\
  then there exist unique (up to $\alpha$-equivalence) $A'$ and $\Theta'$
  such that $\judgeextract[\pm]{\Theta}{A}{A'}{\Theta'}$.
\end{lemma}
\begin{proof}
  By structural induction on the type well-formedness derivation.
\end{proof}

\begin{lemma}[Extract to Ctx.\ WF]
  \label{lem:extract-to-ctx-wf}
  If $\judgeextract[\pm]{\Theta}{A}{A'}{\Theta'}$,
  then $\judgctx{(\Theta, \Theta')}$.
\end{lemma}
\begin{proof}
  By structural induction on the extraction derivation.
  The \ExtractWith and \ExtractImp cases use \Lemmaref{lem:sorting-weakening}.
  The \ExtractProd and \ExtractArrow cases uses \Lemref{lem:permute-ctx-wf}.
\end{proof}

\begin{lemma}[Extract to Type WF]
  \label{lem:extract-to-type-wf}
  If $\judgeextract[\pm]{\Theta}{A}{A'}{\Theta'}$,
  then there exists $\Xi$ such that $\judgetp{\Theta, \Theta'}{A'}{\Xi}$.
\end{lemma}
\begin{proof}
  By structural induction on the extraction derivation.
  The (presupposed) context well-formedness derivation $\judgctx{(\Theta, \Theta')}$
  is obtained via \Lemmaref{lem:extract-to-ctx-wf}.
  The \ExtractWith, \ExtractProd, \ExtractImp, and \ExtractArrow cases
  use \Lemmaref{lem:ix-level-weakening}.
\end{proof}

\begin{lemma}[Extraction Size Eqn.]
  \label{lem:extract-size-eqn}
  If $\judgeextract[\pm]{\Theta}{A}{A'}{\Theta'}$,
  then $\size{A} = \size{A'} + \size{\Theta'}$.
\end{lemma}
\begin{proof}
  By structural induction on the given extraction derivation.
\end{proof}

\begin{lemma}[Extraction Decreases Size]
  \label{lem:extract-decreases-size}
  If $\judgeextract[\pm]{\Theta}{A}{A'}{\Theta'}$,
  then $\size{A'} \leq \size{A}$.
\end{lemma}
\begin{proof}
  Follows from \Lemmaref{lem:extract-size-eqn}.
\end{proof}

\begin{lemma}[Extraction Terminates]
  \label{lem:extract-terminates}
  If $\judgeextract[\pm]{\Theta}{A}{A'}{\Theta'}$,
  then $\judgeextract[\pm]{\Theta, \Theta'}{A'}{A'}{\cdot}$.
\end{lemma}
\begin{proof}
  By structural induction on the extraction derivation.
  The \ExtractWith, \ExtractProd, \ExtractImp, and \ExtractArrow cases
  use \Lemmaref{lem:ix-level-weakening}.
\end{proof}

\begin{lemma}[Extraction Disjunction]
  \label{lem:extract-disjunction}
  If $\judgeextract[\pm]{\Theta}{A}{A'}{\Theta'}$,
  then either:
  \begin{enumerate}[(a)]
    \item $A' = A$ and $\Theta' = \cdot$; or
    \item $A' \neq A$ and $\Theta' \neq \cdot$.
  \end{enumerate}
\end{lemma}
\begin{proof}
  By structural induction on the extraction derivation.
\end{proof}

\begin{lemma}[Shrinking Extraction]
  \label{lem:shrinking-extract}
  If $\judgeextract[\pm]{\Theta}{A}{A'}{\Theta'}$ and $\Theta' \neq \cdot$,\\
  then $\numlog{A'} < \numlog{A}$.
\end{lemma}
\begin{proof}
  By structural induction on the extraction derivation.
\end{proof}

\subsection{Subtyping Properties}

\begin{definition}
  \label{def:ctx-sub}
  Define the judgement $\judgesub[+]{\Theta}{\Gamma}{\Gamma'}$ by pointwise subtyping
  of the variables' types, assuming $\dom{\Gamma} = \dom{\Gamma'}$
  and $\Gamma$ and $\Gamma'$ are in the same order.
\end{definition}

\begin{lemma}[Permute Context]
  \label{lem:permute-ctx}
  Assume $\judgctx{(\Theta, \Theta_1)}$ and $\judgctx{(\Theta, \Theta_2)}$. Then:
  \begin{enumerate}
  \item
    If
    $\judgeterm{\Theta, \Theta_1, \Theta_2, \Theta_f}{t}{\tau}$,
    then
    $\judgeterm{\Theta, \Theta_2, \Theta_1, \Theta_f}{t}{\tau}$.
  \item
    If
    $\judgeentail{\Theta, \Theta_1, \Theta_2, \Theta_f}{\phi}$,
    then
    $\judgeentail{\Theta, \Theta_2, \Theta_1, \Theta_f}{\phi}$.
  \item
    If
    $\judgeequiv[]{\Theta, \Theta_1, \Theta_2, \Theta_f}{\phi}{\psi}$,
    then
    $\judgeequiv[]{\Theta, \Theta_2, \Theta_1, \Theta_f}{\phi}{\psi}$.
  \item
    If
    $\judgeequiv[]{\Theta, \Theta_1, \Theta_2, \Theta_f}{\mathcal{F}}{\mathcal{G}}$,
    then
    $\judgeequiv[]{\Theta, \Theta_2, \Theta_1, \Theta_f}{\mathcal{F}}{\mathcal{G}}$.
  \item
    If
    $\judgeequiv[\pm]{\Theta, \Theta_1, \Theta_2, \Theta_f}{A}{B}$,
    then
    $\judgeequiv[\pm]{\Theta, \Theta_2, \Theta_1, \Theta_f}{A}{B}$.
  \item
    If
    $\judgeextract[\pm]{\Theta, \Theta_1, \Theta_2, \Theta_f}{A}{A'}{\Theta'}$
    then
    $\judgeextract[\pm]{\Theta, \Theta_2, \Theta_1, \Theta_f}{A}{A'}{\Theta'}$
  \item
    If
    $\judgesub[\pm]{\Theta, \Theta_1, \Theta_2, \Theta_f}{A}{B}$,
    then
    $\judgesub[\pm]{\Theta, \Theta_2, \Theta_1, \Theta_f}{A}{B}$.
  \end{enumerate}
\end{lemma}
\begin{proof}
  ~
  \begin{enumerate}
  \item By structural induction on the sorting derivation.

  \item 
    By inversion on \PropTrue,
    for all $|- \delta : \Theta, \Theta_1, \Theta_2, \Theta_f$,
    we have $\sem{\delta}{\phi} = 1$.\\
    Suppose $|- \delta : \Theta, \Theta_2, \Theta_1, \Theta_f$.\\
    \begin{llproof}
      \Pf{\delta}{=}{\delta', \delta_2, \delta_1, \delta_f}{By \Lemmaref{lem:stratify-sem-subs}}
      \Pf{}{|-}{\delta' : \Theta}{\ditto}
      \Pf{}{|-}{\delta', \delta_2 : \Theta, \Theta_2}{\ditto}
      \Pf{}{|-}{\delta', \delta_2, \delta_1 : \Theta, \Theta_2, \Theta_1}{\ditto}
      \Pf{}{|-}{\delta', \delta_2, \delta_1, \delta_f : \Theta, \Theta_2, \Theta_1, \Theta_f}{\ditto}
      \judgctxPf{(\Theta, \Theta_1)}{Given}
      \judgctxPf{(\Theta, \Theta_2)}{Given}
      \Pf{}{|-}{\delta', \delta_1, \delta_2, \delta_f : \Theta, \Theta_1, \Theta_2, \Theta_f}{By \Lemmaref{lem:permute-sem-subs}}
      \judgeentailPf{\Theta, \Theta_1, \Theta_2, \Theta_f}{\phi}{Given}
      \judgetermPf{\Theta, \Theta_1, \Theta_2, \Theta_f}{\phi}{\Booltype}{Presupposed derivation}
      \judgetermPf{\Theta, \Theta_2, \Theta_1, \Theta_f}{\phi}{\Booltype}{By part (1)}
      \Pf{\sem{\delta', \delta_2, \delta_1, \delta_f}{\phi}}{=}{\sem{\delta', \delta_1, \delta_2, \delta_f}{\phi}}{By \Lemmaref{lem:meaning-permute-invariant}}
      \trailingjust{with \Lemmaref{lem:filter-out-props}}
      \Pf{}{=}{1}{Eliminate above ``for all''}
      \judgeentailPf{\Theta, \Theta_2, \Theta_1, \Theta_f}{\phi}{By \PropTrue ($\delta$ is arbitrary)}
    \end{llproof}

  \item By structural induction on the given proposition equivalence derivation,
    using part (2) as needed.

  \item By structural induction on the given functor equivalence derivation,
    mutually with part (5).

  \item By structural induction on the given type equivalence derivation,
    mutually with part (4),
    and using parts (2) and (3) as necessary.

  \item By structural induction on the given extraction derivation.

  \item By structural induction on the given subtyping derivation.
    We case analyze the rule concluding the derivation.
    Each case is straightforward.
    The \DeclSubPosSum case uses part (5).
    The \DeclSubPosFix case uses parts (2) and (4).
    The \DeclSubPosWithR and \DeclSubNegImpL case uses part (2).
    The \DeclSubPosExR and \DeclSubNegAllL case uses part (1).
    The \DeclSubPosL and \DeclSubNegR cases are the least straightforward.
    We show the former; the latter is similar.
    \begin{itemize}
      \DerivationProofCase{\DeclSubPosL}
      {
        \judgeextract[+]{\Theta, \Theta_1, \Theta_2, \Theta_f}{P}{P'}{\Theta'}
        \\
        \Theta' \neq \cdot
        \\
        \judgesub[+]{\Theta, \Theta_1, \Theta_2, \Theta_f, \Theta'}{P'}{Q}
      }
      {
        \judgesub[+]{\Theta, \Theta_1, \Theta_2, \Theta_f}{P}{Q}
      }
      \begin{llproof}
        \judgeextractPf[+]{\Theta, \Theta_1, \Theta_2, \Theta_f}{P}{P'}{\Theta'}{Subderivation}
        \judgeextractPf[+]{\Theta, \Theta_2, \Theta_1, \Theta_f}{P}{P'}{\Theta'}{By part (6)}
        \judgctxPf{(\Theta, \Theta_1)}{Given}
        \judgctxPf{(\Theta, \Theta_2)}{Given}
        \judgesubPf[+]{\Theta, \Theta_1, \Theta_2, \Theta_f, \Theta'}{P'}{Q}{Subderivation}
        \judgesubPf[+]{\Theta, \Theta_2, \Theta_1, \Theta_f, \Theta'}{P'}{Q}{By \ih}
        \judgesubPf[+]{\Theta, \Theta_2, \Theta_1, \Theta_f}{P}{Q}{By \DeclSubPosL}
      \end{llproof} 
      \qedhere
    \end{itemize}
  \end{enumerate}
\end{proof}

\begin{lemma}[Reflexive Verification]
  \label{lem:reflexive-verification}
  ~
  \begin{enumerate}
  \item
    If $\Dee :: \judgeextract[+]{\Theta}{P}{P'}{\Theta'}$
    and $\judgesub[+]{\Theta, \Theta'}{P'}{P'}$,
    then $\judgesub[+]{\Theta, \Theta'}{P'}{P}$.
  \item
    If $\Dee :: \judgeextract[-]{\Theta}{M}{M'}{\Theta'}$
    and $\judgesub[-]{\Theta, \Theta'}{M'}{M'}$,
    then $\judgesub[-]{\Theta, \Theta'}{M}{M'}$.
  \end{enumerate}
\end{lemma}
\begin{proof}
  By induction on $\hgt{\Dee}$.
  \begin{enumerate}
  \item 
    \begin{itemize}
      \ProofCaseRule{\ExtractStopPos}
      Immediate.

      \DerivationProofCase{\ExtractWith}
      {
        \Dee_0 :: \judgeextract[+]{\Theta}{P_0}{P'}{\Theta_0'}
      }
      {
        \Dee :: \judgeextract[+]{\Theta}{P_0 \land \phi}{P'}{\phi, \Theta_0'}
      }
      \begin{llproof}
        \judgeextractPf[+]{\Theta}{P_0}{P'}{\Theta_0'}{Subderivation}
        \judgeextractPf[+]{\Dee_0' :: \Theta, \phi}{P_0}{P'}{\Theta_0'}{By \Lemmaref{lem:ix-level-weakening}}
        \Pf{\hgt{\Dee_0'}}{=}{\hgt{\Dee_0}}{\ditto}
        \Pf{}{<}{\hgt{\Dee}}{Subderivation}
        \judgesubPf[+]{\Theta, \phi, \Theta_0'}{P}{P'}{Given}
        \judgesubPf[+]{\Theta, \phi, \Theta_0'}{P}{P_0}{By \ih}
        \judgeentailPf{\Theta, \phi, \Theta_0'}{\phi}{By \Lemmaref{lem:propvar}}
        \judgesubPf[+]{\Theta, \phi, \Theta_0'}{P}{P_0 \land \phi}{By \DeclSubPosWithR}
      \end{llproof} 

      \DerivationProofCase{\ExtractEx}
      {
        \judgeextract[+]{\Theta, a:\tau}{P_0}{P'}{\Theta_0'}
      }
      {
        \judgeextract[+]{\Theta}{\extype{a:\tau}{P_0}}{P'}{a:\tau, \Theta_0'}
      }
      \begin{llproof}
        \judgeextractPf[+]{\Theta, a:\tau}{P_0}{P'}{\Theta_0'}{Subderivation}
        \judgesubPf[+]{\Theta, a:\tau, \Theta_0'}{P}{P'}{Given}
        \judgesubPf[+]{\Theta, a:\tau, \Theta_0'}{P}{P_0}{By \ih}
        \judgesubPf[+]{\Theta, a:\tau, \Theta_0'}{P}{[a/a]P_0}{Identity substitution}
        \judgetermPf{\Theta, a:\tau, \Theta_0'}{a}{\tau}{By \IxVar}
        \judgesubPf[+]{\Theta, a:\tau, \Theta_0'}{P}{\extype{a:\tau}{P_0}}{By \DeclSubPosExR}
      \end{llproof}

      \DerivationProofCase{\ExtractProd}
      {
        \Dee_1 :: \judgeextract[+]{\Theta}{P_1}{P_1'}{\Theta_1}
        \\
        \Dee_2 :: \judgeextract[+]{\Theta}{P_2}{P_2'}{\Theta_2}
      }
      {
        \Dee :: \judgeextract[+]{\Theta}{P_1 \times P_2}{P_1' \times P_2'}{\Theta_1, \Theta_2}
      }
      \begin{llproof}
        \judgesubPf[+]{\Theta, \Theta_1, \Theta_2}{P_1' \times P_2'}{P_1' \times P_2'}{Given}
        \judgeextractPf[+]{\Theta, \Theta_1, \Theta_2}{P_1' \times P_2'}{P_1' \times P_2'}{\cdot}{By \Lemmaref{lem:extract-terminates}}
        \judgesubPf[+]{\Theta, \Theta_1, \Theta_2}{P_1'}{P_1'}{By inversion on \DeclSubPosProd (\DeclSubPosL impossible)}
        \judgesubPf[+]{\Theta, \Theta_1, \Theta_2}{P_2'}{P_2'}{\ditto}
        \judgeextractPf[+]{\Dee_1 :: \Theta}{P_1}{P_1'}{\Theta_1}{Subderivation}
        \judgeextractPf[+]{\Dee_2 :: \Theta}{P_2}{P_2'}{\Theta_2}{Subderivation}
        \judgeextractPf[+]{\Dee_2' :: \Theta, \Theta_1}{P_2}{P_2'}{\Theta_2}{By \Lemmaref{lem:ix-level-weakening}}
        \Pf{\hgt{\Dee_2'}}{=}{\hgt{\Dee_2}}{\ditto}
        \Pf{}{<}{\hgt{\Dee}}{Subderivation}
        \judgesubPf[+]{\Theta, \Theta_1, \Theta_2}{P_2'}{P_2}{By \ih}
        \judgctxPf{(\Theta, \Theta_1)}{By \Lemmaref{lem:extract-to-ctx-wf}}
        \judgctxPf{(\Theta, \Theta_2)}{By \Lemmaref{lem:extract-to-ctx-wf}}
        \judgesubPf[+]{\Theta, \Theta_2, \Theta_1}{P_1'}{P_1'}{By \Lemmaref{lem:permute-ctx}}
        \judgeextractPf[+]{\Dee_1' :: \Theta, \Theta_2}{P_1}{P_1'}{\Theta_1}{By \Lemmaref{lem:ix-level-weakening}}
        \Pf{\hgt{\Dee_1'}}{=}{\hgt{\Dee_1}}{\ditto}
        \Pf{}{<}{\hgt{\Dee}}{Subderivation}
        \judgesubPf[+]{\Theta, \Theta_2, \Theta_1}{P_1'}{P_1}{By \ih}
        \judgesubPf[+]{\Theta, \Theta_1, \Theta_2}{P_1'}{P_1}{By \Lemmaref{lem:permute-ctx}}
        \judgesubPf[+]{\Theta, \Theta_1, \Theta_2}{P_1' \times P_2'}{P_1 \times P_2}{By \DeclSubPosProd}
      \end{llproof} 
    \end{itemize}

  \item
    Similar to part (1). \qedhere
  \end{enumerate}
\end{proof}

\begin{lemma}[Prop.\ Equiv.\ Consequence]
  \label{lem:prop-equiv-consequence}
  If $\judgeequiv[]{\Theta_1,\phi,\Theta_2}{\psi_1}{\psi_2}$
  and $\judgeentail{\Theta_1}{\phi}$,\\
  then $\judgeequiv[]{\Theta_1,\Theta_2}{\psi_1}{\psi_2}$.
\end{lemma}
\begin{proof}
  By structural induction on the propositional equivalence derivation.
  The \PrpEquivEq and \PrpEquivLeq cases use \Lemmaref{lem:consequence}.
\end{proof}

\begin{lemma}[Tp./Func.\ Equiv.\ Consequence]
  \label{lem:tp-fun-equiv-consequence}
  ~
  \begin{enumerate}
  \item
    If $\judgeequiv[\pm]{\Theta_1,\phi,\Theta_2}{A}{B}$
    and $\judgeentail{\Theta_1}{\phi}$,
    then $\judgeequiv[\pm]{\Theta_1,\Theta_2}{A}{B}$.
  \item
    If $\judgeequiv[]{\Theta_1,\phi,\Theta_2}{\mathcal{F}}{\mathcal{G}}$
    and $\judgeentail{\Theta_1}{\phi}$,
    then $\judgeequiv[]{\Theta_1,\Theta_2}{\mathcal{F}}{\mathcal{G}}$.
  \end{enumerate}
\end{lemma}
\begin{proof}
  Similar to \Lemmaref{lem:subtyping-consequence}, but simpler.
  The \TpEquivPosWith and \TpEquivNegImp cases
  use \Lemmaref{lem:prop-equiv-consequence}.
\end{proof}

\begin{lemma}[Subtyping Consequence]
  \label{lem:subtyping-consequence}
  If $\judgesub[\pm]{\Theta_1,\phi,\Theta_2}{A}{B}$
  and $\judgeentail{\Theta_1}{\phi}$,\\
  then $\judgesub[\pm]{\Theta_1,\Theta_2}{A}{B}$.
\end{lemma}
\begin{proof}
  By induction on the subtyping derivation,
  analyzing cases for the latter's concluding rule.
  Use \Lemmaref{lem:consequence} when necessary
  (e.g., for the \DeclSubPosExR case).
  Use \Lemmaref{lem:assertion-independent}
  to remove $\phi$ from logical contexts of
  sorting, well-formedness, and extraction derivations.
  The \DeclSubPosSum and \DeclSubPosFix cases use
  \Lemmaref{lem:tp-fun-equiv-consequence}.
\end{proof}

\begin{lemma}[Prop.\ Equiv.\ Reflexivity]
  \label{lem:refl-equiv-prop}
  If $\judgeterm{\Theta}{\phi}{\Booltype}$, then $\judgeequiv[]{\Theta}{\phi}{\phi}$.
\end{lemma}
\begin{proof}
  By structural induction on $\phi$.
  The $\phi = (t = t')$ and $\phi = (t \leq t')$ cases use \Lemmaref{lem:equivassert}.
\end{proof}

\begin{lemma}[Tp./Func.\ Equiv.\ Reflexivity]
  \label{lem:refl-equiv-tp-fun}
  ~
  \begin{enumerate}
    \item If $\judgetp{\Theta}{A}{\Xi}$, then $\judgeequiv[\pm]{\Theta}{A}{A}$.
    \item If $\judgefunctor{\Theta}{\mathcal{F}}{\Xi}$,
      then $\judgeequiv[\pm]{\Theta}{\mathcal{F}}{\mathcal{F}}$.
  \end{enumerate}
\end{lemma}
\begin{proof}
  By mutual induction on the structure of $A$ and $\mathcal{F}$.
  For part (1),
  the $A = P \land \phi$ and $A = \phi \implies N$ cases
  use \Lemmaref{lem:refl-equiv-prop},
  and the $A = \comprehend{\nu:\mu F}{\Fold{F}{\alpha}\,{\nu} =_\tau t}$ case
  uses \Lemmaref{lem:equivassert}.
\end{proof}

\begin{lemma}[Subtyping Reflexivity]
  \label{lem:refl-sub}
  If $\judgetp{\Theta}{A}{\Xi}$, then $\judgesub[\pm]{\Theta}{A}{A}$.
\end{lemma}
\begin{proof}
  By induction on $\size{A}$. We case analyze the structure of $A$:
  \begin{itemize}
    \item \textbf{Cases} $A = 0$ and $A = 1$: Apply relevant rule.
    \item \textbf{Case} $A = P_1 \times P_2$:\\
      \begin{llproof}
        \judgetpPf{\Theta}{P_k}{\dontcare}{By inversion on type WF}
        \judgesubPf[+]{\Theta}{P_k}{P_k}{By i.h.}
        \judgesubPf[+]{\Theta}{P_1 \times P_2}{P_1 \times P_2}{By \DeclSubPosProd}
      \end{llproof}
    \item \textbf{Case} $A = P_1 + P_2$:
      Similar to case for $A = P_1 \times P_2$,
      but uses \Lemmaref{lem:refl-equiv-tp-fun}.
    \item \textbf{Case} $A = \extype{a:\tau}{P}$:\\
      \begin{llproof}
        \judgeextractPf[+]{\Theta}{\extype{a:\tau}{P}}{P'}{\Theta'}{By \Lemmaref{lem:extract-determinism}}
        \judgetpPf{\Theta, \Theta'}{P'}{\dontcare}{By \Lemmaref{lem:extract-to-type-wf}}
        \judgeextractPf[+]{\Theta, a:\tau}{P}{P'}{\dontcare}{By inversion on \ExtractEx}
        \Pf{\size{P'}}{\leq}{\size{P}}{By \Lemmaref{lem:extract-decreases-size}}
        \Pf{}{<}{\size{P} + 1}{Obvious}
        \Pf{}{=}{\size{\extype{a:\tau}{P}}}{By \defn}
        \judgesubPf[+]{\Theta, \Theta'}{P'}{P'}{By i.h.}
        \judgesubPf[+]{\Theta, \Theta'}{P'}{\extype{a:\tau}{P}}{By \Lemmaref{lem:reflexive-verification}}
        \judgesubPf[+]{\Theta}{\extype{a:\tau}{P}}{\extype{a:\tau}{P}}{By \DeclSubPosL}
      \end{llproof}
    \item \textbf{Case}
      $A = \comprehend{\nu:\mu F}{\Fold{F}{\alpha}\,\nu =_\tau t}$:
      Suppose $t$ is not a variable.\\
      \begin{llproof}
        \judgefunctorPf{\Theta}{F}{\dontcare}{By inversion on type WF}
        \judgetermPf{\Theta}{t}{\tau}{\ditto}
        \judgeequivPf{\Theta}{F}{F}{By \Lemref{lem:refl-equiv-tp-fun}}
        \judgeentailPf{\Theta}{t=t}{By \Lemref{lem:equivassert}}
        \judgesubPf[+]{\Theta}{\comprehend{\nu:\mu F}{\Fold{F}{\alpha}\,\nu =_\tau t}}{\comprehend{\nu:\mu F}{\Fold{F}{\alpha}\,\nu =_\tau t}}{By \DeclSubPosFix}
      \end{llproof}
      ~\\
      The subcase where $t$ is a variable is similar.
    \item \textbf{Case} $A = \downshift{N}$:\\
      \begin{llproof}
        \judgetpPf{\Theta}{N}{\Xi'}{By inversion on \DeclTpDown}
        \judgesubPf[-]{\Theta}{N}{N}{By i.h.}
        \judgesubPf[+]{\Theta}{\downshift{N}}{\downshift{N}}{By \DeclSubPosDownshift}
      \end{llproof}
    \item \textbf{Case} $A = P \land \phi$:
      Similar to case for $A = \extype{a:\tau}{P}$.
    \item \textbf{Case} $A = \alltype{a:\tau}{N}$:
      Similar to case for $A = \extype{a:\tau}{P}$.
    \item \textbf{Case} $A = \phi \implies N$:
      Similar to case for $A = P \land \phi$.
    \item \textbf{Case} $A = P \to N$:\\
      \begin{llproof}
        \Pf{\judgetp{\Theta}{P}{\Xi_1}}{}{}{By inversion on \DeclTpArrow}
        \Pf{\judgetp{\Theta}{N}{\Xi_2}}{}{}{By inversion on \DeclTpArrow}
        \Pf{\judgesub[+]{\Theta}{P}{P}}{}{}{By i.h.}
        \Pf{\judgesub[-]{\Theta}{N}{N}}{}{}{By i.h.}
        \Pf{\judgesub[-]{\Theta}{P \to N}{P \to N}}{}{}{By \DeclSubNegArrow}
      \end{llproof}
    \item \textbf{Case} $A = \upshift{P}$: Similar to case for $A = \downshift{N}$.
      \qedhere
  \end{itemize}
\end{proof}

\begin{lemma}[Equiv.\ Transitivity]
  \label{lem:trans-equiv}
  ~
  \begin{enumerate}
  \item If $\judgeequiv[]{\Theta}{\phi}{\psi'}$
    and $\judgeequiv[]{\Theta}{\psi'}{\psi}$,
    then $\judgeequiv[]{\Theta}{\phi}{\psi}$.
  \item If $\judgeequiv[]{\Theta}{\mathcal{F}}{\mathcal{F}''}$
    and $\judgeequiv[]{\Theta}{\mathcal{F}''}{\mathcal{F}'}$,
    then $\judgeequiv[]{\Theta}{\mathcal{F}}{\mathcal{F'}}$.
  \item If $\judgeequiv[\pm]{\Theta}{A}{C}$ and $\judgeequiv[\pm]{\Theta}{C}{B}$,
    then $\judgeequiv[\pm]{\Theta}{A}{B}$.
  \end{enumerate}
\end{lemma}
\begin{proof}
  Each part is proved by induction on the sum of the heights of the given derivations.
  Parts (2) and (3) are proved mutually.
  \begin{enumerate}
  \item Straightforward.
    The \PrpEquivEq and \PrpEquivLeq cases use \Lemmaref{lem:equivassert}.
  \item Straightforward.
  \item Straightforward.
    The \TpEquivPosFix cases uses \Lemmaref{lem:equivassert}. \qedhere
  \end{enumerate}
\end{proof}

\begin{lemma}[Sub.\ Extraction Inversion]
  \label{lem:sub-extract-inversion}
  ~
  \begin{enumerate}
  \item
    If $\judgesub[+]{\Theta}{P}{Q}$,
    then $\judgeextract[+]{\Theta}{P}{P'}{\Theta'}$
    and $\judgesub[+]{\Theta, \Theta'}{P'}{Q}$.
  \item
    If $\judgesub[-]{\Theta}{N}{M}$,
    then $\judgeextract[-]{\Theta}{M}{M'}{\Theta'}$
    and $\judgesub[-]{\Theta, \Theta'}{N}{M'}$.
  \end{enumerate}
\end{lemma}
\begin{proof}
  By mutual induction on the structure of the subtyping derivation.
  \begin{enumerate}
  \item We case analyze rules concluding the subtyping derivation:
    \begin{itemize}
      \ProofCaseRule{\DeclSubPosUnit}
      ~\\
      \begin{llproof}
        \judgeextractPf[+]{\Theta}{1}{1}{\cdot}{By \ExtractStopPos}
        \judgesubPf[+]{\Theta, \underbrace{\cdot}_{\Theta'}}{\underbrace{1}_{P'}}{\underbrace{1}_P}{Given}
      \end{llproof}

      \ProofCaseRule{\DeclSubPosVoid}
      Similar to \DeclSubPosUnit case.

      \DerivationProofCase{\DeclSubPosProd}
      {
        \judgesub[+]{\Theta}{P_1}{Q_1}
        \\
        \judgesub[+]{\Theta}{P_2}{Q_2}
      }
      {
        \judgesub[+]{\Theta}{P_1 \times P_2}{Q_1 \times Q_2}
      }
      \begin{llproof}
        \judgesubPf[+]{\Theta}{P_1}{Q_1}{Given}
        \judgesubPf[+]{\Theta}{P_2}{Q_2}{Given}
        \judgesubPf[+]{\Theta, \Theta_1}{P_1'}{Q_1}{By \ih}
        \judgeextractPf[+]{\Theta}{P_1}{P_1'}{\Theta_1}{\ditto}
        \judgesubPf[+]{\Theta, \Theta_2}{P_2'}{Q_2}{By \ih}
        \judgeextractPf[+]{\Theta}{P_2}{P_2'}{\Theta_2}{\ditto}
        \judgesubPf[+]{\Theta, \Theta_1, \Theta_2}{P_1'}{Q_1}{By \Lemmaref{lem:ix-level-weakening}}
        \judgesubPf[+]{\Theta, \Theta_1, \Theta_2}{P_2'}{Q_2}{By \Lemmaref{lem:ix-level-weakening}}
        \judgesubPf[+]{\Theta, \Theta_1, \Theta_2}{P_1' \times P_2'}{Q_1 \times Q_2}{By \DeclSubPosProd}
        \judgeextractPf[+]{\Theta}{P_1 \times P_2}{P_1' \times P_2'}{\Theta_1, \Theta_2}{By \ExtractProd}
      \end{llproof} 

      \ProofCaseRule{\DeclSubPosSum}
      Similar to \DeclSubPosUnit case.

      \DerivationProofCase{\DeclSubPosL}
      {
        \judgeextract[+]{\Theta}{P}{P'}{\Theta'}
        \\
        \Theta' \neq \cdot
        \\
        \judgesub[+]{\Theta, \Theta'}{P'}{Q}
      }
      {
        \judgesub[+]{\Theta}{P}{Q}
      }
      \begin{llproof}
        \judgeextractPf[+]{\Theta}{P}{P'}{\Theta'}{Subderivation}
        \judgesubPf[+]{\Theta, \Theta'}{P'}{Q}{Subderivation}
      \end{llproof}

      \DerivationProofCase{\DeclSubPosWithR}
      {
        \judgesub[+]{\Theta}{P}{Q_0}
        \\
        \judgeentail{\Theta}{\phi}
      }
      {
        \judgesub[+]{\Theta}{P}{Q_0 \land \phi}
      }
      \begin{llproof}
        \judgesubPf[+]{\Theta}{P}{Q_0}{Subderivation}
        \judgesubPf[+]{\Theta, \Theta'}{P'}{Q_0}{By \ih}
        \Hand\judgeextractPf[+]{\Theta}{P}{P'}{\Theta'}{\ditto}
        \judgeentailPf{\Theta}{\phi}{Subderivation}
        \judgeentailPf{\Theta, \Theta'}{\phi}{By \Lemmaref{lem:prop-truth-weakening}}
        \Hand\judgesubPf[+]{\Theta, \Theta'}{P'}{Q_0 \land \phi}{By \DeclSubPosWithR}
      \end{llproof} 

      \ProofCaseRule{\DeclSubPosExR}
      Similar to \DeclSubPosWithR case.

      \ProofCaseRule{\DeclSubPosFix}
      Similar to \DeclSubPosUnit case.

      \ProofCaseRule{\DeclSubPosDownshift}
      Similar to \DeclSubPosUnit case.
    \end{itemize}

  \item
    Similar to part (1).
    For each case with a dual positive rule (\ie, all but $\DeclSubNegArrow$),
    the proof is similar to its dual rule;
    the \DeclSubNegArrow case is similar to the \DeclSubPosProd case.
    \qedhere
  \end{enumerate}
\end{proof}

\begin{lemma}[Undo Subst.]
  \label{lem:subs-typing-undo-subs}
  If $\judgeterm{\Theta_1}{t}{\tau}$
  and $\Theta_1, \Theta_2 |- \sigma : [t/a]\Theta$
  and $\dom{\Theta_1} \sect \dom{\Theta} = \emptyset$,\\
  then $\Theta_1, \Theta_2 |- t/a, \sigma : a:\tau, \Theta$.
\end{lemma}
\begin{proof}
  By structural induction on the given substitution typing derivation;
  we case analyze its concluding rule.
  The \PropSyn case uses the fact that $\dom{\Theta_1} \sect \dom{\Theta} = \emptyset$.
\end{proof}

\begin{lemma}[Sub.\ Instantiate]
  \label{lem:sub-instantiate}
  ~
  \begin{enumerate}
  \item
    If $\Dee :: \judgesub[+]{\Theta}{P}{Q}$
    and $\judgeextract[+]{\Theta}{P}{P'}{\Theta_P}$
    and $\judgeextract[+]{\Theta}{Q}{Q'}{\Theta_Q}$,\\
    then there exists $\Theta, \Theta_P |- \sigma : \Theta_Q$
    and $\Dee' :: \judgesub[+]{\Theta, \Theta_P}{P'}{[\sigma]Q'}$
    such that $\hgt{\Dee'} \leq \hgt{\Dee}$.
  \item
    If $\Dee :: \judgesub[-]{\Theta}{N}{M}$
    and $\judgeextract[-]{\Theta}{N}{N'}{\Theta_N}$
    and $\judgeextract[-]{\Theta}{M}{M'}{\Theta_M}$,\\
    then there exists $\Theta, \Theta_M |- \sigma : \Theta_N$
    and $\Dee' :: \judgesub[-]{\Theta, \Theta_M}{[\sigma]N'}{M'}$
    such that $\hgt{\Dee'} \leq \hgt{\Dee}$.
  \end{enumerate}
\end{lemma}
\begin{proof}
  By mutual induction on the structure of $\Dee$.
  We only show the proof of the first part; the second part is similar.
  Note that we use the convention that extracted variables are fresh.
  \begin{enumerate}
  \item
    \begin{itemize}
      \ProofCaseRule{\DeclSubPosUnit}
      By inversion on \ExtractStopPos,
      we have $P' = P$, $Q' = Q$, and $\Theta_P = \cdot = \Theta_Q$.
      Choose $\sigma = \cdot$.
    \item \textbf{Cases} \DeclSubPosVoid, \DeclSubPosSum, \DeclSubPosFix, \DeclSubPosDownshift:
      Similarly to the \DeclSubPosUnit case, choose $\sigma = \cdot$.
      \DerivationProofCase{\DeclSubPosProd}
      {
        \judgesub[+]{\Theta}{P_1}{Q_1}
        \\
        \judgesub[+]{\Theta}{P_2}{Q_2}
      }
      {
        \judgesub[+]{\Theta}{P_1 \times P_2}{Q_1 \times Q_2}
      }
      \begin{llproof}
        \judgeextractPf[+]{\Theta}{P_1 \times P_2}{P'}{\Theta_P}{Given}
        \Pf{P'}{=}{P_1' \times P_2'}{By inversion}
        \Pf{\Theta_P}{=}{\Theta_P', \Theta_P''}{\ditto}
        \judgeextractPf[+]{\Theta}{P_1}{P_1'}{\Theta_P'}{\ditto}
        \judgeextractPf[+]{\Theta}{P_2}{P_2'}{\Theta_P''}{\ditto}
        \judgeextractPf[+]{\Theta}{Q_1 \times Q_2}{Q'}{\Theta_Q}{Given}
        \Pf{Q'}{=}{Q_1' \times Q_2'}{By inversion}
        \Pf{\Theta_Q}{=}{\Theta_Q', \Theta_Q''}{\ditto}
        \judgeextractPf[+]{\Theta}{Q_1}{Q_1'}{\Theta_Q'}{\ditto}
        \judgeextractPf[+]{\Theta}{Q_2}{Q_2'}{\Theta_Q''}{\ditto}
        \proofsep
        \judgesubPf[+]{\Theta}{P_1}{Q_1}{Subderivation}
        \Pf{\Theta, \Theta_P'}{|-}{\sigma_1 : \Theta_Q'}{By \ih}
        \judgesubPf[+]{\Theta, \Theta_P'}{P_1'}{[\sigma_1]Q_1'}{\ditto}
        \judgesubPf[+]{\Theta}{P_2}{Q_2}{Subderivation}
        \Pf{\Theta, \Theta_P''}{|-}{\sigma_2 : \Theta_Q''}{By \ih}
        \judgesubPf[+]{\Theta, \Theta_P''}{P_2'}{[\sigma_2]Q_2'}{\ditto}
        \proofsep
        \Pf{\Theta, \Theta_P', \Theta_P''}{|-}{\sigma_1 : \Theta_Q'}{By \Lemref{lem:syn-subs-weakening}}
        \Pf{\Theta, \Theta_P', \Theta_P''}{|-}{\sigma_2 : \Theta_Q''}{By \Lemref{lem:syn-subs-weakening}}
        \Hand\Pf{\Theta, \Theta_P', \Theta_P''}{|-}{\sigma_1, \sigma_2 : \Theta_Q', \Theta_Q''}{By \Lemmaref{lem:subs-append}}
        \proofsep
        \judgetpPf{\Theta, \Theta_Q'}{Q_1'}{\dontcare}{By \Lemmaref{lem:extract-to-type-wf}}
        \judgetpPf{\Theta, \Theta_P'}{[\sigma_1]Q_1'}{\dontcare}{By \Lemmaref{lem:syn-subs-tp-fun-alg}}
        \Pf{[\sigma_1, \sigma_2]Q_1'}{=}{[\sigma_1]([\sigma_2]Q_1')}{By subst.\ property}
        \Pf{}{=}{[[\sigma_1]\sigma_2]([\sigma_1]Q_1')}{By subst.\ property}
        \Pf{}{=}{[\sigma_2]([\sigma_1]Q_1')}{$\because \dom{\Theta_Q'} \cap \dom{\Theta_P''} = \emptyset$}
        \Pf{}{=}{[\sigma_1]Q_1'}{$\because \dom{\Theta_Q'} \cap \dom{\Theta_P'} = \emptyset$}
        \proofsep
        \judgetpPf{\Theta, \Theta_Q''}{Q_2'}{\dontcare}{By \Lemmaref{lem:extract-to-type-wf}}
        \Pf{[\sigma_1, \sigma_2]Q_2'}{=}{[\sigma_1]([\sigma_2]Q_2')}{By subst.\ property}
        \Pf{}{=}{[[\sigma_1]\sigma_2]([\sigma_1]Q_2')}{By subst.\ property}
        \Pf{}{=}{[\sigma_2]([\sigma_1]Q_2')}{$\because \dom{\Theta_Q'} \cap \dom{\Theta_P''} = \emptyset$}
        \Pf{}{=}{[\sigma_2]Q_2'}{$\because \dom{\Theta_Q'} \cap \dom{\Theta_Q''} = \emptyset$}
        \proofsep
        \judgesubPf[+]{\Theta, \Theta_P', \Theta_P''}{P_1'}{[\sigma_1]Q_1'}{By \Lemmaref{lem:ix-level-weakening}}
        \judgesubPf[+]{\Theta, \Theta_P', \Theta_P''}{P_1'}{[\sigma_1, \sigma_2]Q_1'}{By above equations}
        \judgesubPf[+]{\Theta, \Theta_P', \Theta_P''}{P_2'}{[\sigma_2]Q_2'}{By \Lemmaref{lem:ix-level-weakening}}
        \judgesubPf[+]{\Theta, \Theta_P', \Theta_P''}{P_2'}{[\sigma_1, \sigma_2]Q_2'}{By above equations}
        \judgesubPf[+]{\Theta, \Theta_P', \Theta_P''}{P_1' \times P_2'}{[\sigma_1, \sigma_2]Q_1' \times [\sigma_1, \sigma_2]Q_2'}{By \DeclSubPosProd}
        \Hand\judgesubPf[+]{\Theta, \Theta_P', \Theta_P''}{P_1' \times P_2'}{[\sigma_1, \sigma_2](Q_1' \times Q_2')}{By \defn of $[-]-$}
      \end{llproof}
      \DerivationProofCase{\DeclSubPosL}
      {
        \judgeextract[+]{\Theta}{P}{P''}{\Theta''}
        \\
        \Theta'' \neq \cdot
        \\
        \judgesub[+]{\Theta, \Theta''}{P''}{Q}
      }
      {
        \judgesub[+]{\Theta}{P}{Q}
      }
      \begin{llproof}
        \judgeextractPf[+]{\Theta}{P}{P'}{\Theta_P}{Given}
        \judgeextractPf[+]{\Theta}{P}{P''}{\Theta''}{Subderivation}
        \judgsubsPf{\Theta, \Theta_P}{\id_\Theta, \sigma}{\Theta, \Theta''}{By \Lemmaref{lem:extract-determinism}}
        \Pf{[\sigma]P''}{=}{P'}{\ditto}
        \judgesubPf[+]{\Theta, \Theta''}{P''}{Q}{Subderivation}
        \judgesubPf[+]{\Theta, \Theta_P}{[\sigma]P''}{[\sigma]Q}{By \Lemmaref{lem:syn-subs-sub}}
        \trailingjust{(by a derivation of smaller or equal height)}
        \judgesubPf[+]{\Theta, \Theta_P}{P'}{[\sigma]Q}{By equation}
        \judgesubPf[+]{\Theta, \Theta_P}{P'}{Q}{$\dom{\Theta} \cap \dom{\Theta''} = \emptyset$}
        \judgeextractPf[+]{\Theta, \Theta_P}{P'}{P'}{\cdot}{By \Lemmaref{lem:extract-terminates}}
        \judgeextractPf[+]{\Theta}{Q}{Q'}{\Theta_Q}{Given}
        \judgeextractPf[+]{\Theta, \Theta_P}{Q}{Q'}{\Theta_Q}{By \Lemmaref{lem:ix-level-weakening}}
        \Pf{\Theta, \Theta_P, \cdot}{|-}{\sigma : \Theta_Q}{By \ih}
        \judgesubPf[+]{\Theta, \Theta_P, \cdot}{P'}{[\sigma]Q'}{\ditto}
        \Hand\Pf{\Theta, \Theta_P}{|-}{\sigma : \Theta_Q}{By property of list append}
        \Hand\judgesubPf[+]{\Theta, \Theta_P}{P'}{[\sigma]Q'}{By property of list append}
      \end{llproof}
      \DerivationProofCase{\DeclSubPosWithR}
      {
        \judgesub[+]{\Theta}{P}{Q_0}
        \\
        \judgeentail{\Theta}{\phi}
      }
      {
        \judgesub[+]{\Theta}{P}{Q_0 \land \phi}
      }
      \begin{llproof}
        \judgeextractPf[+]{\Theta}{P}{P'}{\Theta_P}{Given}
        \judgeextractPf[+]{\Theta}{Q_0 \land \phi}{Q'}{\Theta_Q}{Given}
        \Pf{\Theta_Q}{=}{\phi, \Theta_{Q_0}}{By inversion}
        \judgeextractPf[+]{\Theta}{Q_0}{Q'}{\Theta_{Q_0}}{\ditto}
        \judgesubPf[+]{\Theta}{P}{Q_0}{Subderivation}
        \Pf{\Theta, \Theta_P}{|-}{\sigma : \Theta_{Q_0}}{By \ih}
        \Hand\judgesubPf[+]{\Theta, \Theta_P}{P'}{[\sigma]Q'}{\ditto}
        \judgeentailPf{\Theta}{\phi}{Subderivation}
        \judgeentailPf{\Theta, \Theta_P}{\underbrace{\phi}_{[\cdot]\phi}}{By \Lemmaref{lem:prop-truth-weakening}}
        \Pf{\Theta, \Theta_P}{|-}{\cdot : \cdot}{By \EmptySyn}
        \Hand\Pf{\Theta, \Theta_P}{|-}{\sigma : \underbrace{\phi, \Theta_{Q_0}}_{\Theta_Q}}{By \Lemmaref{lem:syn-subs-deep-entry}}
      \end{llproof}
      \DerivationProofCase{\DeclSubPosExR}
      { 
        \judgesub[+]{\Theta}{P}{[t/a]Q_0} 
        \\
        \judgeterm{\Theta}{t}{\tau} 
      }
      {
        \judgesub[+]{\Theta}{P}{\extype{a:\tau}{Q_0}}
      }
      Similar to the \DeclSubPosWithR case.\\
      \begin{llproof}
        \judgesubPf[+]{\Theta}{P}{[t/a]Q_0}{Subderivation}
        \judgeextractPf[+]{\Theta}{P}{P'}{\Theta_P}{Given}
        \judgeextractPf[+]{\Theta}{\extype{a:\tau}{Q_0}}{Q'}{\Theta_Q}{Given}
        \Pf{\Theta_Q}{=}{a:\tau, \Theta_{Q_0}}{By inversion}
        \judgeextractPf[+]{\Theta, a:\tau}{Q_0}{Q'}{\Theta_{Q_0}}{\ditto}
        \Pf{\Theta}{|-}{\id_\Theta : \Theta}{By \Lemmaref{lem:id-subs-typing}}
        \judgetermPf{\Theta}{t}{\tau}{Subderivation}
        \judgetermPf{\Theta}{[\id_\Theta]t}{\tau}{Identity substitution}
        \Pf{\Theta}{|-}{\id_\Theta, t/a : \Theta, a:\tau}{By \IxSyn}
        \judgeextractPf[+]{\Theta}{[t/a]Q_0}{[t/a]Q'}{[t/a]\Theta_{Q_0}}{By \Lemmaref{lem:syn-subs-extract}}
        \Pf{\Theta, \Theta_P}{|-}{\sigma : [t/a]\Theta_{Q_0}}{By \ih}
        \judgesubPf[+]{\Theta, \Theta_P}{P'}{[\sigma]([t/a]Q')}{\ditto}
        \judgesubPf[+]{\Theta, \Theta_P}{P'}{[[\sigma]t/a]([\sigma]Q')}{By \Lemmaref{lem:barendregt}}
        \judgctxPf{(\Theta, [t/a]\Theta_{Q_0})}{By \Lemmaref{lem:extract-to-ctx-wf}}
        \Pf{\emptyset}{=}{\dom{\Theta} \sect \dom{[t/a]\Theta_{Q_0}}}{By inversion on ctx. WF}
        \judgesubPf[+]{\Theta, \Theta_P}{P'}{[t/a]([\sigma]Q')}{$[\sigma]t = t$}
        \trailingjust{by above equation and $\FV{t} \subseteq \dom{\Theta}$}
        \Hand\judgesubPf[+]{\Theta, \Theta_P}{P'}{[t/a, \sigma]Q'}{By \defn of substitution}
        \judgetermPf{\Theta, \Theta_P}{t}{\tau}{By \Lemmaref{lem:sorting-weakening}}
        \Hand\Pf{\Theta, \Theta_P}{|-}{t/a, \sigma : a:\tau, \Theta_{Q_0}}{By \Lemmaref{lem:subs-typing-undo-subs}}
      \end{llproof}
    \end{itemize}
  \item Similar to part (1).
    For each case with a dual positive rule (\ie, all but $\DeclSubNegArrow$),
    the proof is similar to its dual rule;
    the \DeclSubNegArrow case is similar to the \DeclSubPosProd case.
    \qedhere
  \end{enumerate}
\end{proof}

\begin{lemma}[Num.\ Log.\ Subs.\ Invariant]
  \label{lem:numlog-subs-invariant}
  If $\judgetp{\Theta}{A}{\dontcare}$ and $\Theta_0 |- \sigma : \Theta$,\\
  then $\numlog{[\sigma]A} = \numlog{A}$.
\end{lemma}
\begin{proof}
  By structural induction on the type well-formedness derivation.
\end{proof}
 
\begin{lemma}[Subtyping Transitivity]
  \label{lem:trans-sub}
  ~
  \begin{enumerate}
  \item If $\Dee_1 :: \judgesub[+]{\Theta}{P}{\widetilde{Q}}$
    and $\Dee_2 :: \judgesub[+]{\Theta}{\widetilde{Q}}{Q}$,
    then $\judgesub[+]{\Theta}{P}{Q}$.
  \item If $\Dee_1 :: \judgesub[-]{\Theta}{N}{\widetilde{N}}$
    and $\Dee_2 :: \judgesub[-]{\Theta}{\widetilde{N}}{M}$,
    then $\judgesub[-]{\Theta}{N}{M}$.
  \end{enumerate}
\end{lemma}
\begin{proof}
  By lexicographic induction on,
  first, the total number of $\land$, $\exists$, $\implies$, and $\forall$ connectives
  in the cut formula, \ie $\numlog{\widetilde{Q}}$ or $\numlog{\widetilde{N}}$; and,
  second, the total height of the subtyping derivations.
  \begin{enumerate}
  \item
    We first consider the case where \DeclSubPosL concludes $\Dee_1$.
    Then we case analyze rules concluding $\Dee_2$,
    without needing to consider any subcases where \DeclSubPosL concludes $\Dee_1$
    (this is what we will mean by ``the only other possible subcase \ldots'').
    \begin{itemize}
      \DerivationProofCase{\DeclSubPosL}
      {
        \judgeextract[+]{\Theta}{P}{P'}{\Theta'}
        \\
        \Theta' \neq \cdot
        \\
        \judgesub[+]{\Theta, \Theta'}{P'}{\widetilde{Q}}
      }
      {
        \Dee_1 :: \judgesub[+]{\Theta}{P}{\widetilde{Q}}
      }
      \begin{llproof}
        \judgesubPf[+]{\Theta, \Theta'}{P'}{\widetilde{Q}}{Subderivation}
        \judgesubPf[+]{\Theta}{\widetilde{Q}}{Q}{Given}
        \judgesubPf[+]{\Theta, \Theta'}{\widetilde{Q}}{Q}{By \Lemmaref{lem:ix-level-weakening}}
        \judgesubPf[+]{\Theta, \Theta'}{P'}{Q}{By \ih}
        \judgeextractPf[+]{\Theta}{P}{P'}{\Theta'}{Subderivation}
        \Pf{\Theta'}{\neq}{\cdot}{Subderivation}
        \judgesubPf[+]{\Theta}{P}{Q}{By \DeclSubPosL}
      \end{llproof} 

      \DerivationProofCase{\DeclSubPosL}
      {
        \judgeextract[+]{\Theta}{\widetilde{Q}}{\widetilde{Q}'}{\Theta'}
        \\
        \Theta' \neq \cdot
        \\
        \judgesub[+]{\Theta, \Theta'}{\widetilde{Q}'}{Q}
      }
      {
        \Dee_2 :: \judgesub[+]{\Theta}{\widetilde{Q}}{Q}
      }
      \begin{llproof}
        \judgesubPf[+]{\Theta}{P}{\widetilde{Q}}{Given}
        \judgeextractPf[+]{\Theta}{P}{P'}{\Theta_P}{By \Lemmaref{lem:sub-extract-inversion}}
        \judgesubPf[+]{\Theta, \Theta_P}{P'}{\widetilde{Q}}{\ditto}
        \judgeextractPf[+]{\Theta}{\widetilde{Q}}{\widetilde{Q}'}{\Theta'}{Subderivation}
        \Pf{\Theta, \Theta_P}{|-}{\sigma : \Theta'}{By \Lemmaref{lem:sub-instantiate}}
        \judgesubPf[+]{\Theta, \Theta_P}{P'}{[\sigma]\widetilde{Q}'}{\ditto}
        \Pf{\Theta}{|-}{\id_\Theta : \Theta}{By \Lemmaref{lem:id-subs-typing}}
        \Pf{\Theta, \Theta_P}{|-}{\id_\Theta : \Theta}{By \Lemmaref{lem:syn-subs-weakening}}
        \Pf{\Theta, \Theta_P}{|-}{\id_\Theta, \sigma : \Theta, \Theta'}{By \Lemmaref{lem:id-prepend}}
        \judgesubPf[+]{\Theta, \Theta'}{\widetilde{Q}'}{Q}{Subderivation}
        \judgesubPf[+]{\Theta, \Theta_P}{[\id_\Theta, \sigma]\widetilde{Q}'}{[\id_\Theta, \sigma]Q}{By \Lemmaref{lem:syn-subs-sub}}
        \judgesubPf[+]{\Theta, \Theta_P}{[\sigma]\widetilde{Q}'}{[\sigma]Q}{By \Lemmaref{lem:id-subs-id}}
        \judgesubPf[+]{\Theta, \Theta_P}{[\sigma]\widetilde{Q}'}{Q}{$\because \dom{\Theta'} \sect \FV{Q} = \emptyset$}
        \Pf{\numlog{[\sigma]\widetilde{Q}'}}{=}{\numlog{\widetilde{Q}'}}{By \Lemref{lem:numlog-subs-invariant}}
        \Pf{}{<}{\numlog{\widetilde{Q}}}{By \Lemmaref{lem:shrinking-extract}}
        \judgesubPf[+]{\Theta, \Theta_P}{P'}{Q}{By \ih}
      \end{llproof} 
      ~\\
      By \Lemmaref{lem:extract-disjunction},
      either (a) $P' = P$ and $\Theta_P = \cdot$,
      or (b) $P' \neq P$ and $\Theta_P \neq \cdot$.
      If (a), then we are done.
      If (b), then apply \DeclSubPosL.
      
      \DerivationProofCase{\DeclSubPosUnit}
      { }
      {
        \Dee_2 :: \judgesub[+]{\Theta}{1}{1}
      }
      The only other possible subcase for rule concluding $\Dee_1$ is \DeclSubPosUnit.
      Simply apply \DeclSubPosUnit.
      
      \DerivationProofCase{\DeclSubPosVoid}
      { }
      {
        \Dee_2 :: \judgesub[+]{\Theta}{0}{0}
      }
      Similar to \DeclSubPosUnit case.

      \DerivationProofCase{\DeclSubPosProd}
      {
        \judgesub[+]{\Theta}{\widetilde{Q_1}}{Q_1}
        \\
        \judgesub[+]{\Theta}{\widetilde{Q_2}}{Q_2}
      }
      {
        \Dee_2 :: \judgesub[+]{\Theta}{\widetilde{Q_1} \times \widetilde{Q_2}}{Q_1 \times Q_2}
      }
      The only other possible subcase for rule concluding $\Dee_1$ is \DeclSubPosProd.
      For each pair of relevant subderivations, apply the \ih
      (for each, $\widetilde{Q}$ loses a $\times$),
      and then apply \DeclSubPosProd.

      \DerivationProofCase{\DeclSubPosSum}
      {
        \judgeequiv[+]{\Theta}{\widetilde{Q_1}}{Q_1}
        \\
        \judgeequiv[+]{\Theta}{\widetilde{Q_2}}{Q_2}
      }
      {
        \Dee_2 :: \judgesub[+]{\Theta}{\widetilde{Q_1} + \widetilde{Q_2}}{Q_1 + Q_2}
      }
      The only other possible subcase for rule concluding $\Dee_1$ is \DeclSubPosSum.
      For each pair of relevant subderivations, use \Lemmaref{lem:trans-equiv},
      and then apply \DeclSubPosSum.

      \DerivationProofCase{\DeclSubPosWithR}
      {
        \judgesub[+]{\Theta}{\widetilde{Q}}{Q_0}
        \\
        \judgeentail{\Theta}{\phi}
      }
      {
        \Dee_2 :: \judgesub[+]{\Theta}{\widetilde{Q}}{Q_0 \land \phi}
      }
      \begin{llproof}
        \judgesubPf[+]{\Theta}{\widetilde{Q}}{Q_0}{Subderivation}
        \judgesubPf[+]{\Theta}{P}{\widetilde{Q}}{Given}
        \judgesubPf[+]{\Theta}{P}{Q_0}{By \ih}
        \judgeentailPf{\Theta}{\phi}{Subderivation}
        \judgesubPf[+]{\Theta}{P}{Q_0 \land \phi}{By \DeclSubPosWithR}
      \end{llproof}

      \DerivationProofCase{\DeclSubPosExR}
      { 
        \judgesub[+]{\Theta}{\widetilde{Q}}{[t/a]Q} 
        \\
        \judgeterm{\Theta}{t}{\tau} 
      }
      {
        \Dee_2 :: \judgesub[+]{\Theta}{\widetilde{Q}}{\extype{a:\tau}{Q}}
      }
      Similar to case where \DeclSubPosWithR concludes $\Dee_2$.

      \DerivationProofCase{\DeclSubPosFix}
      {
        \judgeequiv[]{\Theta}{F}{G}
        \\
        \judgeentail{\Theta}{t = t'}
      }
      {
        \Dee_2 :: \judgesub[+]{\Theta}{\comprehend{\nu:\mu F}{\Fold{F}{\alpha}\,\nu =_\tau t}}{\comprehend{\nu:\mu G}{\Fold{G}{\alpha}\,\nu =_\tau t'}}
      }
      The only other possible subcase for rule concluding $\Dee_1$ is \DeclSubPosFix.
      Use \Lemmaref{lem:trans-equiv} for the two functor equivalence subderivations,
      use \Lemmaref{lem:equivassert} for the two propositional truth derivations,
      and then apply \DeclSubPosFix.

      \DerivationProofCase{\DeclSubPosDownshift}
      {
        \judgesub[-]{\Theta}{\widetilde{N}}{M}
      }
      {
        \Dee_2 :: \judgesub[+]{\Theta}{\downshift{\widetilde{N}}}{\downshift{M}}
      }
      The only other possible subcase for rule concluding $\Dee_1$
      is \DeclSubPosDownshift.
      Apply the \ih for the two negative subtyping subderivations,
      and then apply \DeclSubPosDownshift.
    \end{itemize}

  \item
    Similar to part (1).
    We first consider the case where \DeclSubNegR concludes $\Dee_2$.
    We then case analyze rules concluding $\Dee_1$:
    each case that has a dual rule (\ie, all but \DeclSubNegArrow)
    is similar to its dual rule;
    the \DeclSubNegArrow case is similar to the \DeclSubPosProd case.
    \qedhere
  \end{enumerate}
\end{proof}

\subsection{Subsumption Admissibility and Substitution Lemma}

\begin{lemma}[Prog.-Level Weakening]
  \label{lem:prog-level-weakening}
  Assume $\judgectx{\Theta}{\Gamma_1, \Gamma, \Gamma_2}$.
  \begin{enumerate}
    \item If
      $\Dee :: \judgesynhead{\Theta}{\Gamma_1,\Gamma_2}{h}{P}$,
      then
      $\judgesynhead{\Theta}{\Gamma_1,\Gamma,\Gamma_2}{h}{P}$
      by a derivation of equal height.
    \item If
      $\Dee :: \judgesynexp{\Theta}{\Gamma_1,\Gamma_2}{\be}{\upshift{P}}$,
      then
      $\judgesynexp{\Theta}{\Gamma_1,\Gamma,\Gamma_2}{\be}{\upshift{P}}$
      by a derivation of equal height.
    \item If
      $\Dee :: \judgechkval{\Theta}{\Gamma_1,\Gamma_2}{v}{P}$,
      then
      $\judgechkval{\Theta}{\Gamma_1,\Gamma,\Gamma_2}{v}{P}$
      by a derivation of equal height.
    \item If
      $\Dee :: \judgechkexp{\Theta}{\Gamma_1,\Gamma_2}{e}{N}$,
      then
      $\judgechkexp{\Theta}{\Gamma_1,\Gamma,\Gamma_2}{e}{N}$
      by a derivation of equal height.
    \item If 
      $\Dee :: \judgechkmatch{\Theta}{\Gamma_1,\Gamma_2}{P}{\clauses{\pa}{e}{i}{I}}{N}$,
      then
      $\judgechkmatch{\Theta}{\Gamma_1,\Gamma,\Gamma_2}{P}{\clauses{\pa}{e}{i}{I}}{N}$
      by a derivation of equal height.
    \item If
      $\Dee :: \judgespine{\Theta}{\Gamma_1,\Gamma_2}{s}{N}{\upshift{P}}$,
      then 
      $\judgespine{\Theta}{\Gamma_1,\Gamma,\Gamma_2}{s}{N}{\upshift{P}}$
        by a derivation of equal height.
  \end{enumerate}
\end{lemma}
\begin{proof}
  By mutual induction on the structure of the given derivation $\Dee$.
\end{proof}

\begin{lemma}[Prog.-Level \ Subs.\ Weakening]
  \label{lem:prog-level-syn-subs-weakening}
  Assume $\judgectx{\Theta}{\Gamma_1, \Gamma_0, \Gamma_2}$.
  If $\Theta; \Gamma_1, \Gamma_2 |- \sigma : \Theta; \Gamma$,
  then $\Theta; \Gamma_1, \Gamma_0, \Gamma_2 |- \sigma : \Theta; \Gamma$.
\end{lemma}
\begin{proof}
  By structural induction on the given substitution typing derivation,
  case analyzing its concluding rule.
  Similar to \Lemmaref{lem:syn-subs-weakening}.
  The \ValSyn case uses \Lemmaref{lem:prog-level-weakening}.
\end{proof}

\begin{lemma}[Id.\ Subst.\ Extension]
  \label{lem:id-subs-ext-prog}
  Assume $\Theta_0;\Gamma_0 |- \sigma : \Theta; \Gamma$.
  If $\judgetp{\Theta}{P}{\Xi}$ and $P \neq \exists, \land$
  and $x \notin \dom{\Gamma} \cup \dom{\Gamma_0}$,
  then $\Theta_0;\Gamma_0,x:[\filterprog{\sigma}]P|-\sigma,\subs{x}{P}{x}:\Theta;\Gamma,x:P$.
\end{lemma}
\begin{proof}
  Assume $\judgetp{\Theta}{P}{\Xi}$ and $x \notin \dom{\Gamma} \cup \dom{\Gamma_0}$.\\
  \begin{llproof}
    \Pf{\judgetp{\Theta_0}{[\filterprog{\sigma}]P}{\dontcare}}{}{}{By \Lemmaref{lem:syn-subs-tp-fun-alg}}
    \Pf{\Theta_0;\Gamma_0,x:[\filterprog{\sigma}]P|-\sigma:\Theta;\Gamma}{}{}{By \Lemmaref{lem:prog-level-syn-subs-weakening}}
    \Pf{\judgesub[+]{\Theta_0}{[\filterprog{\sigma}]P}{[\filterprog{\sigma}]P}}{}{}{By \Lemmaref{lem:refl-sub}}
    \Pf{P \neq \exists, \land}{}{}{Given}
    \Pf{[\filterprog{\sigma}]P \neq \exists, \land}{}{}{Straightforward}
    \Pf{\judgechkval{\Theta_0}{\Gamma_0,x:[\filterprog{\sigma}]P}{\underbrace{x}_{[\sigma]x}}{[\filterprog{\sigma}]P}}{}{}{By \DeclChkValVar}
    \Pf{\Theta_0;\Gamma_0,x:[\filterprog{\sigma}]P|-\sigma,\subs{x}{P}{x}:\Theta;\Gamma,x:P}{}{}{By \ValSyn \qedhere}
  \end{llproof}
\end{proof}

\begin{lemma}[Typing Consequence]
  \label{lem:typing-consequence}
  Assume $\judgeentail{\Theta_1}{\phi}$. Then:
  \begin{enumerate}
  \item If
    $\judgesynhead{\Theta_1,\phi,\Theta_2}{\Gamma}{h}{P}$,
    then
    $\judgesynhead{\Theta_1,\Theta_2}{\Gamma}{h}{P}$.
  \item If
    $\judgesynexp{\Theta_1,\phi,\Theta_2}{\Gamma}{\be}{\upshift{P}}$,
    then
    $\judgesynexp{\Theta_1,\Theta_2}{\Gamma}{\be}{\upshift{P}}$.
  \item If
    $\judgechkval{\Theta_1,\phi,\Theta_2}{\Gamma}{v}{P}$,
    then
    $\judgechkval{\Theta_1,\Theta_2}{\Gamma}{v}{P}$.
  \item If
    $\judgechkexp{\Theta_1,\phi,\Theta_2}{\Gamma}{e}{N}$,
    then
    $\judgechkexp{\Theta_1,\Theta_2}{\Gamma}{e}{N}$.
  \item If 
    $\judgechkmatch{\Theta_1,\phi,\Theta_2}{\Gamma}{P}{\clauses{\pa}{e}{i}{I}}{N}$
    then
    $\judgechkmatch{\Theta_1,\Theta_2}{\Gamma}{P}{\clauses{\pa}{e}{i}{I}}{N}$.
  \item If
    $\judgespine{\Theta_1,\phi,\Theta_2}{\Gamma}{s}{N}{\upshift{P}}$,
    then
    $\judgespine{\Theta_1,\Theta_2}{\Gamma}{s}{N}{\upshift{P}}$.
  \end{enumerate}
\end{lemma}
\begin{proof}
  By mutual induction on the program typing derivation,
  and case analysis on the rule concluding the latter,
  using \Lemmaref{lem:subtyping-consequence}
  for the \DeclChkValVar case of the value typing part
  and for the \DeclChkExpRec case of the expression typing part,
  using \Lemmaref{lem:consequence} when necessary
  (e.g., for the \DeclChkValWith case of the value typing part),
  and using
  \Lemmaref{lem:assertion-independent}
  to remove $\phi$ from (logical contexts of)
  index sorting, well-formedness, unrolling, and extraction subderivations,
  as needed. \qedhere
\end{proof}

\begin{lemma}[Index-Level Substitution]
  \label{lem:ix-syn-subs}
  Assume $\Theta_0; \cdot |- \sigma : \Theta; \cdot$. Then:
  \begin{enumerate}
  \item If $\Dee \derives\judgesynhead{\Theta}{\Gamma}{h}{P}$,\\
    then there exists $\Dee' \derives \judgesynhead{\Theta_0}{[\sigma]\Gamma}{[\sigma]h}{[\sigma]P}$
    such that $\hgt{\Dee'} \leq \hgt{\Dee}$.
  \item If $\Dee \derives \judgesynexp{\Theta}{\Gamma}{\be}{\upshift{P}}$,\\
    then there exists $\Dee' \derives \judgesynexp{\Theta_0}{[\sigma]\Gamma}{[\sigma]\be}{[\sigma]\upshift{P}}$
    such that $\hgt{\Dee'} \leq \hgt{\Dee}$.
  \item If $\Dee \derives \judgechkval{\Theta}{\Gamma}{v}{P}$,\\
    then there exists $\Dee' \derives \judgechkval{\Theta_0}{[\sigma]\Gamma}{[\sigma]v}{[\sigma]P}$
    such that $\hgt{\Dee'} \leq \hgt{\Dee}$.
  \item If $\Dee \derives \judgechkexp{\Theta}{\Gamma}{e}{N}$,\\
    then there exists $\Dee' \derives \judgechkexp{\Theta_0}{[\sigma]\Gamma}{[\sigma]e}{[\sigma]N}$
    such that $\hgt{\Dee'} \leq \hgt{\Dee}$.
  \item If $\Dee \derives \judgechkmatch{\Theta}{\Gamma}{P}{\clauses{\pa}{e}{i}{I}}{N}$,\\
    then there exists
    $\Dee' \derives \judgechkmatch{\Theta_0}{[\sigma]\Gamma}{[\sigma]P}
      {[\sigma]\clauses{\pa}{e}{i}{I}}{[\sigma]N}$\\
      such that $\hgt{\Dee'} \leq \hgt{\Dee}$.
  \item If $\Dee \derives \judgespine{\Theta}{\Gamma}{s}{N}{\upshift{P}}$,\\
    then there exists
    $\Dee' \derives \judgespine{\Theta_0}{[\sigma]\Gamma}{[\sigma]s}{[\sigma]N}{[\sigma]\upshift{P}}$
    such that $\hgt{\Dee'} \leq \hgt{\Dee}$.
  \end{enumerate}
\end{lemma}
\begin{proof}
  By mutual induction on the structure of the program typing derivation.
  Note that, given $\Theta;\Gamma |- \mathcal{J}$ or $\Theta;\Gamma;[P]|-\mathcal{J}$,
  for every $(x:A)\in\Gamma$ where $\judgetp{\Theta}{A}{\Xi}$
  we have $\judgetp{\Theta_0}{[\sigma]A}{\dontcare}$
  by \Lemmaref{lem:syn-subs-tp-fun-alg}.
  The height condition holds immediately in the base cases;
  it holds in the inductive cases by the \ih and the fact that
  we only re-apply the same rule as the given case.
  \begin{enumerate}
  \item
    \begin{itemize}
      \DerivationProofCase{\DeclSynHeadVar}
      {
        (x : P) \in \Gamma
      }
      {
        \judgesynhead{\Theta}{\Gamma}{x}{P}
      }
      \begin{llproof}
        \Pf{(x:P)\in\Gamma}{}{}{Subderivation}
        \Pf{(x:[\sigma]P)\in[\sigma]\Gamma}{}{}{By definition}
        \Pf{\judgesynhead{\Theta_0}{[\sigma]\Gamma}{x}{[\sigma]P}}{}{}{By \DeclSynHeadVar}
        \Pf{\judgesynhead{\Theta_0}{[\sigma]\Gamma}{[\sigma]x}{[\sigma]P}}{}{}{$x\notin\dom{\sigma}$}
      \end{llproof}

      \DerivationProofCase{\DeclSynValAnnot}
      {
        \judgetp{\Theta}{P}{\Xi}
        \\
        \judgechkval{\Theta}{\Gamma}{v_0}{P}
      }
      {
        \judgesynhead{\Theta}{\Gamma}{\annoexp{v_0}{P}}{P}
      }
      \begin{llproof}
        \Pf{\judgetp{\Theta_0}{[\sigma]P}{\dontcare}}{}{}{By \Lemmaref{lem:syn-subs-tp-fun-alg}}
        \Pf{\judgechkval{\Theta_0}{[\sigma]\Gamma}{[\sigma]v_0}{[\sigma]P}}{}{}{By i.h.}
        \Pf{\judgesynhead{\Theta_0}{[\sigma]\Gamma}{\annoexp{[\sigma]v_0}{[\sigma]P}}{[\sigma]P}}{}{}{By \DeclSynValAnnot}
        \Pf{\judgesynhead{\Theta_0}{[\sigma]\Gamma}{[\sigma]\annoexp{v_0}{P}}{[\sigma]P}}{}{}{By def. of subst.}
      \end{llproof} 

    \end{itemize}

  \item
    \begin{itemize}
      \DerivationProofCase{\DeclSynSpineApp}
      { \judgesynhead{\Theta}{\Gamma}{h}{\downshift{N}} \\
        \judgespine{\Theta}{\Gamma}{s}{N}{\upshift{P}} }
      { \judgesynexp{\Theta}{\Gamma}{h(s)}{\upshift{P}} }
      \begin{llproof}
        \Pf{\Theta_0;\cdot|-\sigma:\Theta;\cdot}{}{}{Given}
        \Pf{\judgesynhead{\Theta}{\Gamma}{h}{\downshift{N}}}{}{}{Subderivation}
        \Pf{\judgesynhead{\Theta_0}{[\sigma]\Gamma}{[\sigma]h}{[\sigma]\downshift{N}}}{}{}{By \ih}
        \Pf{\judgesynhead{\Theta_0}{[\sigma]\Gamma}{[\sigma]h}{\downshift{[\sigma]N}}}{}{}{By \defn of subst.}
        \Pf{\judgespine{\Theta}{\Gamma}{s}{N}{\upshift{P}}}{}{}{Subderivation}
        \Pf{\judgespine{\Theta_0}{[\sigma]\Gamma}{[\sigma]s}{[\sigma]N}{\upshift{[\sigma]P}}}{}{}{By i.h. and def. of subst.}
        \Pf{\judgesynexp{\Theta_0}{\Gamma_0}{[\sigma](h(s))}{[\sigma]\upshift{P}}}{}{}{By \DeclSynSpineApp and def. of subst.}
      \end{llproof}

      \ProofCaseRule{\DeclSynExpAnnot}
      Apply i.h. to subderivations, reapply rule, use definition of substitution.
    \end{itemize}

  \item
    \begin{itemize}
      \DerivationProofCase{\DeclChkValVar}
      { P \neq \exists, \land \\ (x:Q) \in \Gamma \\ \judgesub[+]{\Theta}{Q}{P} }
      { \judgechkval{\Theta}{\Gamma}{x}{P} }
      \begin{llproof}
        \Pf{(x:Q) \in \Gamma}{}{}{Subderivation}
        \Pf{(x:[\sigma]Q) \in [\sigma]\Gamma}{}{}{By definition}
        \Pf{\judgesub[+]{\Theta}{Q}{P}}{}{}{Subderivation}
        \Pf{\judgesub[+]{\Theta_0}{[\sigma]Q}{[\sigma]P}}{}{}{By \Lemmaref{lem:syn-subs-sub}}
        \Pf{P}{\neq}{\exists, \land}{Premise}
        \Pf{[\sigma]P}{\neq}{\exists, \land}{Straightforward}
        \Pf{\judgechkval{\Theta_0}{[\sigma]\Gamma}{x}{[\sigma]P}}{}{}{By \DeclChkValVar}
        \Pf{\judgechkval{\Theta_0}{[\sigma]\Gamma}{[\sigma]x}{[\sigma]P}}{}{}{$x\notin\dom{\sigma}$}
      \end{llproof} 

      \item \textbf{Cases} \DeclChkValUnit, \DeclChkValPair, \DeclChkValIn{k}:
        Straightforward.

      \DerivationProofCase{\DeclChkValExists}
      {
        \judgechkval{\Theta}{\Gamma}{v}{[t/a]P'}
        \\
        \judgeterm{\Theta}{t}{\tau}
      }
      { \judgechkval{\Theta}{\Gamma}{v}{(\extype{a:\tau} P')} }
      \begin{llproof}
        \Pf{\judgechkval{\Theta_0}{[\sigma]\Gamma}{[\sigma]v}{[\sigma]([t/a]P')}}{}{}{By i.h.}
        \Pf{\judgechkval{\Theta_0}{[\sigma]\Gamma}{[\sigma]v}{[[\sigma]t/a]([\sigma]P')}}{}{}{By \Lemmaref{lem:barendregt}}
        \Pf{\judgeterm{\Theta_0}{[\sigma]t}{\tau}}{}{}{By \Lemmaref{lem:syn-subs-ix}}
        \Pf{\judgechkval{\Theta_0}{[\sigma]\Gamma}{[\sigma]v}{[\sigma](\extype{a:\tau} P')}}{}{}{By \DeclChkValExists and def. of subst.}
      \end{llproof}

      \DerivationProofCase{\DeclChkValWith}
      {
        \judgechkval{\Theta}{\Gamma}{v}{P'}
        \\
        \judgeentail{\Theta}{\phi}
      }
      { \judgechkval{\Theta}{\Gamma}{v}{P' \land \phi} }
      \begin{llproof}
        \Pf{\Theta_0;\cdot|-\sigma:\Theta;\cdot}{}{}{Given}
        \Pf{\judgechkval{\Theta}{\Gamma}{v}{P'}}{}{}{Subderivation}
        \Pf{\judgechkval{\Theta_0}{[\sigma]\Gamma}{[\sigma]v}{[\sigma]P'}}{}{}{By i.h.}
        \Pf{\judgeentail{\Theta}{\phi}}{}{}{Subderivation}
        \Pf{\judgeentail{\Theta_0}{[\sigma]\phi}}{}{}{By \Lemmaref{lem:syn-subs-prop-true}}
        \Pf{\judgechkval{\Theta_0}{[\sigma]\Gamma}{[\sigma]v}{[\sigma](P' \land \phi)}}{}{}{By \DeclChkValWith and def. of subst.}
      \end{llproof}

      \DerivationProofCase{\DeclChkValFix}
      { \judgeunroll{\cdot}{\Theta}{\nu:F[\mu F]}{\alpha}{F\;\Fold{F}{\alpha}\;\nu}{t}{P'}{\tau}
        \\ 
        \judgechkval{\Theta}{\Gamma}{v'}{P'} }
      { \judgechkval{\Theta}{\Gamma}{\roll{v'}}
        {\comprehend{\nu:\mu F}{\Fold{F}{\alpha}\,{\nu} =_\tau t}} }
      By \Lemmaref{lem:syn-subs-unroll},
      \[
        \judgeunroll{\cdot}{\Theta_0}{\nu:([\sigma]F)[\mu [\sigma]F]}{[\sigma]\alpha}{([\sigma]F)\;\Fold{[\sigma]F}{[\sigma]\alpha}\;\nu}{[\sigma]t}{[\sigma]P'}{\tau}
      \]
      By the induction hypothesis,
      \[
        \judgechkval{\Theta_0}{[\sigma]\Gamma}{[\sigma]v'}{[\sigma]P'}
      \]
      By \DeclChkValWith and def. of subst.
      \[
        \judgechkval{\Theta_0}{[\sigma]\Gamma}{[\sigma]\roll{v'}}
        {[\sigma]\comprehend{\nu:\mu F}{\Fold{F}{\alpha}\,{\nu} =_\tau t}}
      \]

      \ProofCaseRule{\DeclChkValDownshift}
      Straightforward.
    \end{itemize}
  \item
    \begin{itemize}
      \ProofCaseRule{\DeclChkExpUpshift}:
      Straightforward.

      \DerivationProofCase{\DeclChkExpLet}
      { \simple{\Theta}{N} \\
        \judgesynexp{\Theta}{\Gamma}{\be}{\upshift{P}} \\
        \judgeextract[+]{\Theta}{P}{P'}{\Theta'} \\
        \judgechkexp{\Theta, \Theta'}{\Gamma, x:P'}{e'}{N} }
      { \judgechkexp{\Theta}{\Gamma}{\Let{x}{\be}{e'}}{N} }
      \begin{llproof}
        \Pf{\Theta_0;\cdot|-\sigma:\Theta;\cdot}{}{}{Given}
        \Pf{\simple{\Theta}{N}}{}{}{Subderivation}
        \Pf{\simple{\Theta_0}{[\sigma]N}}{}{}{By \Lemmaref{lem:syn-subs-extract}}
        \Pf{\judgesynexp{\Theta}{\Gamma}{\be}{\upshift{P}}}{}{}{Subderivation}
        \Pf{\judgesynexp{\Theta_0}{[\sigma]\Gamma}{[\sigma]\be}{\upshift{[\sigma]P}}}{}{}{By i.h. and def. of subst.}
        \Pf{\judgeextract[+]{\Theta}{P}{P'}{\Theta'}}{}{}{Subderivation}
        \Pf{\judgeextract[+]{\Theta_0}{[\sigma]P}{[\sigma]P'}{[\sigma]\Theta'}}{}{}{By \Lemmaref{lem:syn-subs-extract}}
        \Pf{\judgechkexp{\Theta, \Theta'}{\Gamma, x:P'}{e'}{N}}{}{}{Subderivation}
        \Pf{\Theta_0, [\sigma]\Theta'; \cdot|-\sigma, \id_{\Theta'} :\Theta, \Theta'; \cdot}{}{}{By \Lemmaref{lem:id-subs-ext-ix-level}}
        \Pf{\judgechkexp{\Theta_0, [\sigma]\Theta'}{[\sigma]\Gamma, x:[\sigma]P'}{[\sigma]e'}{[\sigma]N}}{}{}{By \ih and \defn of subst.\ and $\id_{\Theta'}$}
        \Pf{\judgechkexp{\Theta_0}{[\sigma]\Gamma}{[\sigma](\Let{x}{\be}{e'})}{[\sigma]N}}{}{}{By \DeclChkExpLet and def. of subst.}
      \end{llproof}

      \DerivationProofCase{\DeclChkExpMatch}
      {
        \simple{\Theta}{N}
        \\
        \judgesynhead{\Theta}{\Gamma}{h}{P}
        \\
        \judgechkmatch{\Theta}{\Gamma}{P}{\clauses{\pa}{e}{i}{I}}{N}
      }
      {
        \judgechkexp{\Theta}{\Gamma}{\match{h}{\clauses{\pa}{e}{i}{I}}}{N}
      }
      \begin{llproof}
        \Pf{\Theta_0;\cdot|-\sigma:\Theta;\cdot}{}{}{Given}
        \Pf{\simple{\Theta}{N}}{}{}{Subderivation}
        \Pf{\simple{\Theta_0}{[\sigma]N}}{}{}{By \Lemmaref{lem:syn-subs-extract}}
        \Pf{\judgesynhead{\Theta}{\Gamma}{h}{P}}{}{}{Subderivation}
        \Pf{\judgesynhead{\Theta_0}{[\sigma]\Gamma}{[\sigma]h}{[\sigma]P}}{}{}{By i.h.}
        \Pf{\judgechkmatch{\Theta}{\Gamma}{P}{\clauses{\pa}{e}{i}{I}}{N}}{}{}{Subderivation}
        \Pf{\judgechkmatch{\Theta_0}{[\sigma]\Gamma}{[\sigma]P}{[\sigma]\clauses{\pa}{e}{i}{I}}{[\sigma]N}}{}{}{By i.h.}
        \Pf{\judgechkexp{\Theta_0}{[\sigma]\Gamma}{[\sigma](\match{h}{\clauses{\pa}{e}{i}{I}})}{[\sigma]N}}{}{}{By \DeclChkExpMatch and def. of subst.}
      \end{llproof}

      \DerivationProofCase{\DeclChkExpLam}
      { \simple{\Theta}{P \to N'} \\ \judgechkexp{\Theta}{\Gamma, x:P}{e'}{N'} }
      { \judgechkexp{\Theta}{\Gamma}{\fun{x}{e'}}{P \to N'} }
      \begin{llproof}
        \Pf{\Theta_0;\cdot |- \sigma : \Theta;\cdot}{}{}{Given}
        \Pf{\simple{\Theta}{P \to N'}}{}{}{Subderivation}
        \Pf{\simple{\Theta_0}{[\sigma](P \to N')}}{}{}{By \Lemmaref{lem:syn-subs-extract}}
        \Pf{\judgechkexp{\Theta}{\Gamma, x:P}{e'}{N'}}{}{}{Subderivation}
        \Pf{\judgechkexp{\Theta_0}{[\sigma]\Gamma, x:[\sigma]P}{[\sigma]e'}{[\sigma]N'}}{}{}{By \ih and \defn of subst.}
        \Pf{\judgechkexp{\Theta_0}{[\sigma]\Gamma}{[\sigma](\fun{x}{e'})}{[\sigma](P \to N')}}{}{}{By \DeclChkExpLam and def. of subst.}
      \end{llproof}

      \DerivationProofCase{\DeclChkExpRec}
      {
        \arrayenvb{
          \simple{\Theta}{N}
          \\
          \judgesub[-]{\Theta}{\alltype{a:\kindnat} M}{N}
        }
        \\
        \judgechkexp{\Theta, a:\kindnat}{\Gamma, x:\downshift{\alltype{a':\kindnat} a' < a \implies [a'/a]M}}{e_0}{M}
      }
      {
        \judgechkexp{\Theta}{\Gamma}{\rec{x : (\alltype{a:\kindnat} M)}{e_0}}{N}
      }
      \begin{llproof}
        \judgsubsPf{\Theta_0;\cdot}{\sigma}{\Theta;\cdot}{Given}
        \simplePf{\Theta}{N}{Premise}
        \simplePf{\Theta_0}{[\sigma]N}{By \Lemmaref{lem:syn-subs-extract}}
        \proofsep
        \judgesubPf[-]{\Theta}{\alltype{a:\kindnat} M}{N}{Premise}
        \judgesubPf[-]{\Theta_0}{[\sigma](\alltype{a:\kindnat} M)}{[\sigma]N}{By \Lemmaref{lem:syn-subs-sub}}
        \judgesubPf[-]{\Theta_0}{\alltype{a:\kindnat} [\sigma]M}{[\sigma]N}{By \defn of \defsubst}
        \decolumnizePf
        \judgechkexpPf{\Theta, a:\kindnat}{\Gamma, x:\downshift{\alltype{a':\kindnat} a' < a \implies [a'/a]M}}{e_0}{M}{Subderivation}
        \judgsubsPf{\Theta_0,a:\tau;\cdot}{\sigma,a/a}{\Theta,a:\tau;\cdot}{By \Lemref{lem:id-subs-ext-ix}}
        \decolumnizePf
        \judgechkexpPf{\Theta_0, a:\kindnat}{[\sigma, a/a](\Gamma, x:\downshift{\alltype{a':\kindnat} a' < a \implies [a'/a]M})}{[\sigma,a/a]e_0}{[\sigma,a/a]M}{By \ih}
        \decolumnizePf
        \judgechkexpPf{\Theta_0, a:\kindnat}{[\sigma]\Gamma, x:\downshift{\alltype{a':\kindnat} a' < a \implies [\sigma]([a'/a]M)})}{[\sigma]e_0}{[\sigma]M}{By \defn of \defsubst}
        \judgechkexpPf{\Theta_0, a:\kindnat}{[\sigma]\Gamma, x:\downshift{\alltype{a':\kindnat} a' < a \implies [[\sigma]a'/a]([\sigma]M)})}{[\sigma]e_0}{[\sigma]M}{By subst.\ property}
        \judgechkexpPf{\Theta_0, a:\kindnat}{[\sigma]\Gamma, x:\downshift{\alltype{a':\kindnat} a' < a \implies [a'/a]([\sigma]M)})}{[\sigma]e_0}{[\sigma]M}{By convention}
        \decolumnizePf
        \judgechkexpPf{\Theta_0}{[\sigma]\Gamma}{\rec{x:(\alltype{a:\kindnat} [\sigma]M)}{[\sigma]e_0}}{[\sigma]N}{By \DeclChkExpRec}
        \judgechkexpPf{\Theta_0}{[\sigma]\Gamma}{[\sigma](\rec{x:(\alltype{a:\kindnat} M)}{e_0})}{[\sigma]N}{By \defn of \defsubst}
      \end{llproof} 

      \DerivationProofCase{\DeclChkExpExtract}
      {
        \judgeextract{\Theta}{N}{N'}{\Theta'}
        \\
        \Theta' \neq \cdot
        \\
        \judgechkexp{\Theta, \Theta'}{\Gamma}{e}{N'}
      }
      {
        \judgechkexp{\Theta}{\Gamma}{e}{N}
      }
      \begin{llproof}
        \Pf{\Theta_0;\cdot |- \sigma : \Theta;\cdot}{}{}{Given}
        \Pf{\judgeextract[+]{\Theta}{N}{N'}{\Theta'}}{}{}{Subderivation}
        \Pf{\judgeextract[+]{\Theta_0}{[\sigma]N}{[\sigma]N'}{[\sigma]\Theta'}}{}{}{By \Lemmaref{lem:syn-subs-extract}}
        \Pf{\Theta_0, [\sigma]\Theta'; \cdot|-\sigma, \id_{\Theta'} :\Theta, \Theta'; \cdot}{}{}{By \Lemmaref{lem:id-subs-ext-ix-level}}
        \Pf{\judgechkexp{\Theta, \Theta'}{\Gamma}{e}{N'}}{}{}{Subderivation}
        \Pf{\judgechkexp{\Theta_0, [\sigma]\Theta'}{[\sigma]\Gamma}{[\sigma]e}{[\sigma]N'}}{}{}{By \ih and \defn of subst.\ and $\id_{\Theta'}$}
        \Pf{\judgechkexp{\Theta_0}{[\sigma]\Gamma}{[\sigma]e}{[\sigma](P \to N')}}{}{}{By \DeclChkExpExtract}
      \end{llproof} 

      \ProofCaseRule{\DeclChkExpUnreachable}
      Straightforward. Use \Lemmaref{lem:subst-inconsistent}.
    \end{itemize}

  \item
    \begin{itemize}
      \DerivationProofCase{\DeclChkMatchEx}
      {\judgechkmatch{\Theta, a:\tau}{\Gamma}{P'}{\clauses{\pa}{e}{i}{I}}{N}}
      {\judgechkmatch{\Theta}{\Gamma}{\extype{a:\tau}{P'}}{\clauses{\pa}{e}{i}{I}}{N}}
      \begin{llproof}
        \Pf{\Theta_0;\cdot}{|-}{\sigma : \Theta;\cdot}{Given}
        \Pf{\Theta_0,a:\tau;\cdot}{|-}{\sigma,a/a : \Theta,a:\tau;\cdot}{By Lemma~\ref{lem:id-subs-ext-ix}}
        \judgechkmatchPf{\Theta, a:\tau}{\Gamma}{P'}{\clauses{\pa}{e}{i}{I}}{N}{Subderivation}
        \judgechkmatchPf{\Theta_0, a:\tau}{[\sigma,a/a]\Gamma}{[\sigma,a/a]P'}{[\sigma,a/a]\clauses{\pa}{e}{i}{I}}{[\sigma,a/a]N}{By \ih}
        \judgechkmatchPf{\Theta_0, a:\tau}{[\sigma]\Gamma}{[\sigma]P'}{[\sigma]\clauses{\pa}{e}{i}{I}}{[\sigma]N}{By Lemma~\ref{lem:id-subs-id}}
        \judgechkmatchPf{\Theta_0}{[\sigma]\Gamma}{\extype{a:\tau}{[\sigma]P'}}{[\sigma]\clauses{\pa}{e}{i}{I}}{[\sigma]N}{By \DeclChkMatchEx}
        \judgechkmatchPf{\Theta_0}{[\sigma]\Gamma}{[\sigma]\extype{a:\tau}{P'}}{[\sigma]\clauses{\pa}{e}{i}{I}}{[\sigma]N}{By \defn of subst.}
      \end{llproof}

      \ProofCaseRule{\DeclChkMatchWith}
      Similar to case for \DeclChkMatchEx.

      \DerivationProofCase{\DeclChkMatchUnit}
      { \judgechkexp{\Theta}{\Gamma}{e}{N} }
      { \judgechkmatch{\Theta}{\Gamma}{1}{\setof{\clause{\unit}{e}}}{N} }
      \begin{llproof}
        \Pf{\Theta_0;\cdot |- \sigma : \Theta;\cdot}{}{}{Given}
        \Pf{\judgechkexp{\Theta}{\Gamma}{e}{N}}{}{}{Subderivation}
        \Pf{\judgechkexp{\Theta_0}{[\sigma]\Gamma}{[\sigma]e}{[\sigma]N}}{}{}{By i.h.}
        \Pf{\judgechkmatch{\Theta_0}{[\sigma]\Gamma}{1}{\setof{\clause{\unit}{[\sigma]e}}}{[\sigma]N}}{}{}{By \DeclChkMatchUnit}
        \Pf{\judgechkmatch{\Theta_0}{[\sigma]\Gamma}{[\sigma]1}{[\sigma]\setof{\clause{\unit}{e}}}{[\sigma]N}}{}{}{By def. of subst.}
      \end{llproof}

      \DerivationProofCase{\DeclChkMatchPair}
      { \arrayenvb{\judgeextract[+]{\Theta}{P_1}{P_1'}{\Theta_1} \\
        \judgeextract[+]{\Theta}{P_2}{P_2'}{\Theta_2}} \\
        \judgechkexp{\Theta, \Theta_1, \Theta_2}{\Gamma, x_1:P_1', x_2:P_2'}{e}{N} }
      {
        \judgechkmatch{\Theta}{\Gamma}{P_1 \times P_2}
        {\setof{\clause{\pair{x_1}{x_2}}{e}}}{N}
      }
      \begin{llproof}
        \Pf{\Theta_0;\cdot}{|-}{\sigma : \Theta;\cdot}{Given}
        \judgeextractPf[+]{\Theta}{P_1}{P_1'}{\Theta_1}{Subderivation}
        \judgeextractPf[+]{\Theta}{P_2}{P_2'}{\Theta_2}{Subderivation}
        \judgeextractPf{\Theta_0}{[\sigma]P_1}{[\sigma]P_1'}{[\sigma]\Theta_1}{By \Lemref{lem:syn-subs-extract}}
        \judgeextractPf{\Theta_0}{[\sigma]P_2}{[\sigma]P_2'}{[\sigma]\Theta_2}{By \Lemref{lem:syn-subs-extract}}
        \Pf{\Theta_0, [\sigma](\Theta_1, \Theta_2);\cdot}{|-}{\sigma, \id_{\Theta_1, \Theta_2} : \Theta, \Theta_1, \Theta_2;\cdot}{By \Lemref{lem:id-subs-ext-ix-level}}
        \Pf{\Theta_0, [\sigma]\Theta_1, [\sigma]\Theta_2;\cdot}{|-}{\sigma, \id_{\Theta_1, \Theta_2} : \Theta, \Theta_1, \Theta_2;\cdot}{By \defn of subst.}
        \judgechkexpPf{\Theta, \Theta_1, \Theta_2}{\Gamma, x_1:P_1', x_2:P_2'}{e}{N}{Subderivation}
        \decolumnizePf
        \judgechkexpPf{\Theta_0, [\sigma]\Theta_1, [\sigma]\Theta_2}{[\sigma]\Gamma, x_1:[\sigma]P_1', x_2:[\sigma]P_2'}{[\sigma]e}{[\sigma]N}{By i.h. and \defn}
        \decolumnizePf
        \judgechkmatchPf{\Theta_0}{[\sigma]\Gamma}{[\sigma]P_1 \times [\sigma]P_2}{\setof{\clause{\pair{x_1}{x_2}}{[\sigma]e}}}{[\sigma]N}{By \DeclChkMatchPair}
        \judgechkmatchPf{\Theta_0}{[\sigma]\Gamma}{[\sigma](P_1 \times P_2)}{[\sigma]\setof{\clause{\pair{x_1}{x_2}}{e}}}{[\sigma]N}{By def. of subst.}
      \end{llproof}

      \ProofCaseRule{\DeclChkMatchSum}
      Similar to case for \DeclChkMatchPair.

      \DerivationProofCase{\DeclChkMatchVoid}
      { }
      { \judgechkmatch{\Theta}{\Gamma}{0}{\setof{}}{N} }
      \begin{llproof}
        \Pf{\judgechkmatch{\Theta_0}{[\sigma]\Gamma}{0}{\setof{}}{[\sigma]N}}{}{}{By \DeclChkMatchVoid}
        \Pf{\judgechkmatch{\Theta_0}{[\sigma]\Gamma}{[\sigma]0}{[\sigma]\setof{}}{[\sigma]N}}{}{}{By def. of subst.}
      \end{llproof}

      \DerivationProofCase{\DeclChkMatchFix}
      {
        \arrayenvb{\judgeunroll{\cdot}{\Theta}{\nu:F[\mu F]}{\alpha}{F\;\Fold{F}{\alpha}\;\nu}{t}{P}{\tau} \\ \judgeextract[+]{\Theta}{P}{P'}{\Theta'}}
        \\
        \judgechkexp{\Theta'}{\Gamma, x:P'}{e}{N}
      }
      {
        \judgechkmatch{\Theta}{\Gamma}{\comprehend{\nu:\mu F}
          {\Fold{F}{\alpha}\,{\nu} =_\tau t}}{\setof{\clause{\roll{x}}{e}}}{N}
      }
        We are given $\Theta_0;\cdot |- \sigma : \Theta;\cdot$.
        Subderivation:
        \[
          \judgeunroll{\cdot}{\Theta}{\nu:F[\mu F]}{\alpha}{F\;\Fold{F}{\alpha}\;\nu}{t}{P}{\tau}
        \]
        By \Lemmaref{lem:syn-subs-unroll},
        \[
          \judgeunroll{\cdot}{\Theta_0}{\nu:[\sigma]F[\mu [\sigma]F]}{[\sigma]\alpha}{[\sigma]F\;\Fold{[\sigma]F}{[\sigma]\alpha}\;\nu}{[\sigma]t}{[\sigma]P}{\tau}
        \]
        By \Lemmaref{lem:syn-subs-extract} on the extraction subderivation,
        \[
          \judgeextract[+]{\Theta_0}{[\sigma]P}{[\sigma]P'}{[\sigma]\Theta'}
        \]
        By \Lemmaref{lem:id-subs-ext-ix-level},
        $\Theta_0, [\sigma]\Theta';\cdot |- \sigma, \id_{\Theta'} : \Theta, \Theta';\cdot$.
        Consider the subderivation $\judgechkexp{\Theta, \Theta'}{\Gamma, x:P'}{e}{N}$;
        by the \ih and def. of subst. and $\id_{\Theta'}$,
        \[
          \judgechkexp{\Theta_0, [\sigma]\Theta'}{[\sigma]\Gamma, x:[\sigma]P'}{[\sigma]e}{[\sigma]N}
        \]
        By \DeclChkMatchFix,
        \[
          \judgechkmatch{\Theta_0}{[\sigma]\Gamma}{\comprehend{\nu:\mu [\sigma]F}{\Fold{[\sigma]F}{[\sigma]\alpha}\,{\nu} =_\tau [\sigma]t}}{\setof{\clause{\roll{x}}{[\sigma]e}}}{[\sigma]N}
        \]
        By definition of substitution,
        \[
          \judgechkmatch{\Theta_0}{[\sigma]\Gamma}{[\sigma]\comprehend{\nu:\mu F}{\Fold{F}{\alpha}\,{\nu} =_\tau t}}{[\sigma]\setof{\clause{\roll{x}}{e}}}{[\sigma]N}
        \]
      \end{itemize}

  \item
    \begin{itemize}
      \DerivationProofCase{\DeclSpineAll}
      {
        \judgeterm{\Theta}{t}{\tau}
        \\
        \judgespine{\Theta}{\Gamma}{s}{[t/a]N'}{\upshift{P}}
      }
      {\judgespine{\Theta}{\Gamma}{s}{(\alltype{a:\tau}{N'})}{\upshift{P}}}
      \begin{llproof}
        \Pf{\Theta_0;\cdot |- \sigma : \Theta;\cdot}{}{}{Given}
        \Pf{\judgespine{\Theta}{\Gamma}{s}{[t/a]N'}{\upshift{P}}}{}{}{Subderivation}
        \Pf{\judgespine{\Theta_0}{[\sigma]\Gamma}{[\sigma]s}{[\sigma]([t/a]N')}{[\sigma]\upshift{P}}}{}{}{By i.h.}
        \Pf{\judgespine{\Theta_0}{[\sigma]\Gamma}{[\sigma]s}{[[\sigma]t/a]([\sigma]N')}{[\sigma]\upshift{P}}}{}{}{By \Lemmaref{lem:barendregt}}
        \Pf{\judgeterm{\Theta}{t}{\tau}}{}{}{By inversion}
        \Pf{\judgeterm{\Theta_0}{[\sigma]t}{\tau}}{}{}{By Lemma~\ref{lem:syn-subs-ix}}
        \Pf{\judgespine{\Theta_0}{[\sigma]\Gamma}{[\sigma]s}{[\sigma](\alltype{a:\tau}{N'})}{[\sigma]\upshift{P}}}{}{}{By \DeclSpineAll and def. of subst.}
      \end{llproof}

      \DerivationProofCase{\DeclSpineImplies}
      {\judgeentail{\Theta}{\phi} \\ \judgespine{\Theta}{\Gamma}{s}{N'}{\upshift{P}}  }
      {\judgespine{\Theta}{\Gamma}{s}{(\phi \implies N')}{\upshift{P}}}
      \begin{llproof}
        \Pf{\judgespine{\Theta}{\Gamma}{s}{N'}{\upshift{P}}}{}{}{Subderivation}
        \Pf{\judgespine{\Theta_0}{[\sigma]\Gamma}{[\sigma]s}{[\sigma]N'}{[\sigma]\upshift{P}}}{}{}{By i.h.}
        \Pf{\judgeentail{\Theta}{\phi}}{}{}{By inversion}
        \Pf{\judgeentail{\Theta_0}{[\sigma]\phi}}{}{}{By \Lemmaref{lem:syn-subs-prop-true}}
        \Pf{\judgespine{\Theta_0}{[\sigma]\Gamma}{[\sigma]s}{[\sigma](\phi \implies N')}{[\sigma]\upshift{P}}}{}{}{By \DeclSpineImplies and def. of subst.}
      \end{llproof}

      \item \textbf{Cases} \DeclSpineApp and \DeclSpineNil:
        Straightforward.
        \qedhere
    \end{itemize}
  \end{enumerate}
\end{proof}

\begin{corollary}[Index-Level Substitution]
  \label{cor:ix-syn-subs}
  Assume $\Theta,\Theta_0; \cdot |- \sigma : \Theta'; \cdot$. Then:
  \begin{enumerate}
  \item If $\Dee \derives\judgesynhead{\Theta, \Theta'}{\Gamma}{h}{P}$,\\
    then there exists $\Dee' \derives \judgesynhead{\Theta, \Theta_0}{[\sigma]\Gamma}{[\sigma]h}{[\sigma]P}$
    such that $\hgt{\Dee'} \leq \hgt{\Dee}$.
  \item If $\Dee \derives \judgesynexp{\Theta, \Theta'}{\Gamma}{\be}{\upshift{P}}$,\\
    then there exists $\Dee' \derives \judgesynexp{\Theta, \Theta_0}{[\sigma]\Gamma}{[\sigma]\be}{[\sigma]\upshift{P}}$
    such that $\hgt{\Dee'} \leq \hgt{\Dee}$.
  \item If $\Dee \derives \judgechkval{\Theta, \Theta'}{\Gamma}{v}{P}$,\\
    then there exists $\Dee' \derives \judgechkval{\Theta, \Theta_0}{[\sigma]\Gamma}{[\sigma]v}{[\sigma]P}$
    such that $\hgt{\Dee'} \leq \hgt{\Dee}$.
  \item If $\Dee \derives \judgechkexp{\Theta, \Theta'}{\Gamma}{e}{N}$,\\
    then there exists $\Dee' \derives \judgechkexp{\Theta, \Theta_0}{[\sigma]\Gamma}{[\sigma]e}{[\sigma]N}$
    such that $\hgt{\Dee'} \leq \hgt{\Dee}$.
  \item If $\Dee \derives \judgechkmatch{\Theta, \Theta'}{\Gamma}{P}{\clauses{\pa}{e}{i}{I}}{N}$,\\
    then there exists
    $\Dee' \derives \judgechkmatch{\Theta, \Theta_0}{[\sigma]\Gamma}{[\sigma]P}
      {[\sigma]\clauses{\pa}{e}{i}{I}}{[\sigma]N}$\\
      such that $\hgt{\Dee'} \leq \hgt{\Dee}$.
  \item If $\Dee \derives \judgespine{\Theta, \Theta'}{\Gamma}{s}{N}{\upshift{P}}$,\\
    then there exists
    $\Dee' \derives \judgespine{\Theta, \Theta_0}{[\sigma]\Gamma}{[\sigma]s}{[\sigma]N}{[\sigma]\upshift{P}}$
    such that $\hgt{\Dee'} \leq \hgt{\Dee}$.
  \end{enumerate}
\end{corollary}
\begin{proof}
  We only show the first part; the other parts are similar.\\
  \begin{llproof}
    \judgsubsPf{\Theta, \Theta_0}{\sigma}{\Theta'}{Given}
    \judgsubsPf{\Theta}{\id_\Theta}{\Theta}{By \Lemmaref{lem:id-subs-typing}}
    \judgsubsPf{\Theta, \Theta_0}{\id_\Theta}{\Theta}{By \Lemmaref{lem:syn-subs-weakening}}
    \judgsubsPf{\Theta, \Theta_0}{\id_\Theta, \sigma}{\Theta, \Theta'}{By \Lemmaref{lem:id-prepend}}
    \judgesynheadPf{\Dee \derives \Theta, \Theta'}{\Gamma}{h}{P}{Given}
    \judgesynheadPf{\Dee' \derives \Theta, \Theta_0}{[\id_\Theta, \sigma]\Gamma}{[\id_\Theta, \sigma]h}{[\id_\Theta, \sigma]P}{By \Lemmaref{lem:ix-syn-subs}}
    \Pf{\hgt{\Dee'}}{\leq}{\hgt{\Dee}}{\ditto}
    \judgesynheadPf{\Dee' \derives \Theta, \Theta_0}{[\sigma]\Gamma}{[\sigma]h}{[\sigma]P}{By \Lemmaref{lem:id-subs-id} \qedhere}
  \end{llproof} 
\end{proof}

\begin{lemma}[Unroll to Equiv.\ Types]
  \label{lem:unroll-to-equiv}
  ~\\
  If $\judgeunroll{\Xi}{\Theta}{\nu:G[\mu F]}{\beta}{G\;\Fold{F}{\alpha}\;\nu}{t}{P}{\tau}$\\
  and $\judgeequiv[]{\Theta}{G}{G'}$
  and $\judgeequiv[]{\Theta}{F}{F'}$
  and $\judgeentail{\Theta}{t=t'}$,\\
  then there exists $Q$
  such that $\judgeunroll{\Xi}{\Theta}{\nu:G'[\mu F']}{\beta}{G'\;\Fold{F'}{\alpha}\;\nu}{t'}{Q}{\tau}$\\
  and $\judgeequiv[+]{\Theta}{P}{Q}$.
\end{lemma}
\begin{proof}
  By structural induction on the unrolling derivation.
  \begin{itemize}
    \DerivationProofCase{\DeclUnrollSum}
    {\arrayenvbl{
        \composeinj{1}{\beta}{\beta_1}
        \\
        \composeinj{2}{\beta}{\beta_2}
      }
      \\
      \arrayenvbl{
        \judgeunroll{\Xi}{\Theta}{\nu:G_1[\mu F]}{\beta_1}{G_1\;\Fold{F}{\alpha}\;\nu}{t}{P_1}{\tau}
        \\
        \judgeunroll{\Xi}{\Theta}{\nu:G_2[\mu F]}{\beta_2}{G_2\;\Fold{F}{\alpha}\;\nu}{t}{P_2}{\tau}
      }
    }
    { \judgeunroll{\Xi}{\Theta}{\nu:(G_1 \oplus G_2)[\mu F]}{\beta}{(G_1 \oplus G_2)\;\Fold{F}{\alpha}\;\nu}{t}{P_1 + P_2}{\tau} }
    \begin{llproof}
      \judgeentailPf{\Theta}{t=t'}{Given}
      \judgeequivPf{\Theta}{G_1 \oplus G_2}{G'}{Given}
      \Pf{G'}{=}{G_1' \oplus G_2'}{By inversion}
      \judgeequivPf{\Theta}{G_1}{G_1'}{\ditto}
      \judgeequivPf{\Theta}{G_2}{G_2'}{\ditto}
      \judgeequivPf[]{\Theta}{F}{F'}{Given}
      \judgeunrollPf{\Xi}{\Theta}{\nu:G_k[\mu F]}{\beta_k}{G_k\;\Fold{F}{\alpha}\;\nu}{t}{P_k}{\tau}{Subderivations}
      \judgeunrollPf{\Xi}{\Theta}{\nu:G_k'[\mu F']}{\beta_k}{G_k'\;\Fold{F'}{\alpha}\;\nu}{t'}{Q_k}{\tau}{By \ih (twice)}
      \judgeequivPf[+]{\Theta}{P_k}{Q_k}{\ditto}
      \Hand\judgeequivPf[+]{\Theta}{P_1+P_2}{Q_1+Q_2}{By \TpEquivPosSum}
      \composeinjPf{k}{\beta}{\beta_k}{Subderivations}
      \decolumnizePf
      \Hand\judgeunrollPf{\Xi}{\Theta}{\nu:(G_1' \oplus G_2')[\mu F']}{\beta}{(G_1' \oplus G_2')\;\Fold{F'}{\alpha}\;\nu}{t'}{Q_1 + Q_2}{\tau}{By \DeclUnrollSum}
    \end{llproof}

    \DerivationProofCase{\DeclUnrollId}
    {
      \judgeunroll{\Xi, a:\tau}{\Theta, a:\tau}{\nu:\hat{P}[\mu F]}{(\clause{q}{\tilde{t}})}{\hat{P}\;\Fold{F}{\alpha}\;\nu}{t}{P'}{\tau} 
    }
    {
      \judgeunroll*{\Xi}{\Theta}{\nu : (\Id\otimes\hat{P})[\mu F]}
      {(\clause{(a,q)}{\tilde{t}})}
      {(\Id\otimes\hat{P})\;\Fold{F}{\alpha}\;\nu}{t}
      {\extype{a:\tau}
        {\comprehend{\nu:\mu F}{ \Fold{F}{\alpha}\,{\nu} =_\tau a } \times P'}}
      {\tau}
    }
    \begin{llproof}
      \judgeentailPf{\Theta}{t=t'}{Given}
      \judgeentailPf{\Theta,a:\tau}{t=t'}{By \Lemref{lem:prop-truth-weakening}}
      \judgeequivPf{\Theta}{\Id \otimes \hat{P}}{G'}{Given}
      \Pf{G'}{=}{\Id \otimes \hat{P}'}{By inversion}
      \judgeequivPf{\Theta}{\hat{P}}{\hat{P}'}{\ditto}
      \judgeequivPf{\Theta, a:\tau}{\hat{P}}{\hat{P}'}{By \Lemref{lem:ix-level-weakening}}
      \judgeunrollPf{\Xi, a:\tau}{\Theta, a:\tau}{\nu:\hat{P}[\mu F]}{(\clause{q}{\tilde{t}})}{\hat{P}\;\Fold{F}{\alpha}\;\nu}{t}{P'}{\tau}{Subderivation}
      \judgeunrollPf{\Xi, a:\tau}{\Theta, a:\tau}{\nu:\hat{P}[\mu F]}{(\clause{q}{\tilde{t}})}{\hat{P}\;\Fold{F}{\alpha}\;\nu}{t'}{Q'}{\tau}{By \ih}
      \judgeequivPf[+]{\Theta,a:\tau}{P'}{Q'}{\ditto}
    \end{llproof}
    \\
    By \DeclUnrollId,
    \[
      \judgeunroll*{\Xi}{\Theta}{\nu : (\Id\otimes\hat{P}')[\mu F']}
      {(\clause{(a,q)}{\tilde{t}})}
      {(\Id\otimes\hat{P}')\;\Fold{F'}{\alpha}\;\nu}{t'}
      {\extype{a:\tau}
        {\comprehend{\nu:\mu F'}{ \Fold{F'}{\alpha}\,{\nu} =_\tau a } \times Q'}}
      {\tau}
    \]
    as desired.

    Finally, we show the desired type equivalence:
    
    \begin{llproof}
      \judgeequivPf[]{\Theta}{F}{F'}{Given}
      \judgeequivPf[]{\Theta, a:\tau}{F}{F'}{By \Lemmaref{lem:ix-level-weakening}}
      \judgeentailPf{\Theta, a:\tau}{a=a}{By \Lemmaref{lem:equivassert}}
    \end{llproof}

    By \TpEquivPosFix,
    \[
      \judgeequiv{\Theta, a:\tau}{\comprehend{\nu:\mu F}{ \Fold{F}{\alpha}\,{\nu} =_\tau a }}{\comprehend{\nu:\mu F'}{ \Fold{F'}{\alpha}\,{\nu} =_\tau a }}
    \]
    By \TpEquivPosProd,
    \[
      \judgeequiv{\Theta, a:\tau}{\comprehend{\nu:\mu F}{ \Fold{F}{\alpha}\,{\nu} =_\tau a } \times P'}{\comprehend{\nu:\mu F'}{ \Fold{F'}{\alpha}\,{\nu} =_\tau a } \times Q'}
    \]
    By \TpEquivPosEx,
    \[
      \judgeequiv{\Theta}{\extype{a:\tau}\comprehend{\nu:\mu F}{ \Fold{F}{\alpha}\,{\nu} =_\tau a } \times P'}{\extype{a:\tau}\comprehend{\nu:\mu F'}{ \Fold{F'}{\alpha}\,{\nu} =_\tau a } \times Q'}
    \]

    \ProofCaseRule{\DeclUnrollConstEx}
    \[
      \Infer{}{
        \judgeunroll{\Xi,a:\tau'}{\Theta,a:\tau'}{\nu:(\Const{Q_0}\otimes\hat{P})[\mu F]}{(\clause{(\bap,q)}{\tilde{t}})}{(\Const{Q_0}\otimes\hat{P})\;\Fold{F}{\alpha}\;\nu}{t}{P'}{\tau} }
      { \judgeunroll*{\Xi}{\Theta}{\nu :(\Const{\extype{a:\tau'}{Q_0}}\otimes\hat{P})[\mu F]}{(\clause{(\pack{a}{\bap}, q)}{\tilde{t}})}{(\Const{\extype{a:\tau'}{Q_0}}\otimes\hat{P})\;\Fold{F}{\alpha}\;\nu}{t}{\extype{a:\tau'}{P'}}{\tau} }
    \]
    \begin{llproof}
      \judgeentailPf{\Theta}{t=t'}{Given}
      \judgeentailPf{\Theta,a:\tau'}{t=t'}{By \Lemmaref{lem:prop-truth-weakening}}
      \proofsep
      \judgeequivPf{\Theta}{\Const{\extype{a:\tau'}{Q_0}}\otimes\hat{P}}{G'}{Given}
      \Pf{G'}{=}{\Const{\extype{a:\tau'} Q_0'} \otimes \hat{P}'}{By inversion}
      \judgeequivPf{\Theta, a:\tau'}{Q_0}{Q_0'}{\ditto}
      \judgeequivPf{\Theta}{\hat{P}}{\hat{P}'}{\ditto}
      \judgeequivPf{\Theta, a:\tau'}{\hat{P}}{\hat{P}'}{By \Lemmaref{lem:ix-level-weakening}}
      \judgeequivPf{\Theta, a:\tau'}{\Const{Q_0}}{\Const{Q_0'}}{By \FunEquivConst}
      \judgeequivPf{\Theta, a:\tau'}{\Const{Q_0} \otimes \hat{P}}{\Const{Q_0'} \otimes \hat{P}'}{By \FunEquivProd}
    \end{llproof} 
    \\
    Applying the \ih to the subderivation
    \[
      \judgeunroll{\Xi,a:\tau'}{\Theta,a:\tau'}{\nu:(\Const{Q_0}\otimes\hat{P})[\mu F]}{(\clause{(\bap,q)}{\tilde{t}})}{(\Const{Q_0}\otimes\hat{P})\;\Fold{F}{\alpha}\;\nu}{t}{P'}{\tau}
    \]
    yields a $Q'$ such that
    \[
      \judgeunroll{\Xi,a:\tau'}{\Theta,a:\tau'}{\nu:(\Const{Q_0'}\otimes\hat{P}')[\mu F']}{(\clause{(\bap,q)}{\tilde{t}})}{(\Const{Q_0'}\otimes\hat{P}')\;\Fold{F'}{\alpha}\;\nu}{t'}{Q'}{\tau}
    \]
    and $\judgeequiv[+]{\Theta,a:\tau'}{P'}{Q'}$.
    By \DeclUnrollConstEx,
    \[
      \judgeunroll*{\Xi}{\Theta}{\nu :(\Const{\extype{a:\tau'}{Q_0'}}\otimes\hat{P}')[\mu F']}{(\clause{(\pack{a}{\bap}, q)}{\tilde{t}})}{(\Const{\extype{a:\tau'}{Q_0'}}\otimes\hat{P}')\;\Fold{F'}{\alpha}\;\nu}{t'}{\extype{a:\tau'}{Q'}}{\tau}
    \]
    By \TpEquivPosEx, $\judgeequiv[+]{\Theta}{\extype{a:\tau'}P'}{\extype{a:\tau'}Q'}$.

    \ProofCaseRule{\DeclUnrollConst}
    \[
      \Infer{}{
        \judgeunroll{\Xi}{\Theta}{\nu:\hat{P}[\mu F]}{(\clause{q}{\tilde{t}})}{\hat{P}\;\Fold{F}{\alpha}\;\nu}{t}{P'}{\tau} }
      { \judgeunroll{\Xi}{\Theta}{\nu :(\Const{P_0}\otimes\hat{P})[\mu F]}{(\clause{(\wild,q)}{\tilde{t}})}{(\Const{P_0}\otimes\hat{P})\;\Fold{F}{\alpha}\;\nu}{t}{P_0 \times P'}{\tau} }
    \]
    \begin{llproof}
      \judgeentailPf{\Theta}{t=t'}{Given}
      \judgeequivPf{\Theta}{\Const{P_0}\otimes\hat{P}}{G'}{Given}
      \Pf{G'}{=}{\Const{P_0'}\otimes\hat{P}'}{By inversion}
      \judgeequivPf[+]{\Theta}{P_0}{P_0'}{\ditto}
      \judgeequivPf{\Theta}{\hat{P}}{\hat{P}'}{\ditto}
    \end{llproof}

    Applying the \ih to the subderivation
    \[
      \judgeunroll{\Xi}{\Theta}{\nu:\hat{P}[\mu F]}{(\clause{q}{\tilde{t}})}{\hat{P}\;\Fold{F}{\alpha}\;\nu}{t}{P'}{\tau}
    \]
    yields a $Q'$ such that
    \[
      \judgeunroll{\Xi}{\Theta}{\nu:\hat{P}'[\mu F']}{(\clause{q}{\tilde{t}})}{\hat{P}'\;\Fold{F'}{\alpha}\;\nu}{t'}{Q'}{\tau}
    \]
    and $\judgeequiv[+]{\Theta}{P'}{Q'}$.
    By \DeclUnrollConst,
    \[
      \judgeunroll{\Xi}{\Theta}{\nu :(\Const{P_0'}\otimes\hat{P}')[\mu F']}{(\clause{(\wild,q)}{\tilde{t}})}{(\Const{P_0'}\otimes\hat{P}')\;\Fold{F'}{\alpha}\;\nu}{t'}{P_0' \times Q'}{\tau}
    \]
    By \TpEquivPosProd, $\judgeequiv[+]{\Theta}{P_0 \times P'}{P_0' \times Q'}$.

    \DerivationProofCase{\DeclUnrollUnit}
    { }
    { \judgeunroll{\Xi}{\Theta}{\nu:I[\mu F]}{(\clause{\unitexp}{\tilde{t}})}{I\;\Fold{F}{\alpha}\;\nu}{t}{1 \land (t = \tilde{t})}{\tau} }
    \begin{llproof}
      \judgeequivPf{\Theta}{I}{G'}{Given}
      \Pf{G'}{=}{I}{By inversion}
      \Hand\judgeunrollPf{\Xi}{\Theta}{\nu:I[\mu F]}{(\clause{\unitexp}{\tilde{t}})}{I\;\Fold{F}{\alpha}\;\nu}{t'}{1 \land (t' = \tilde{t})}{\tau}{By \DeclUnrollUnit}
      \decolumnizePf
      \judgeequivPf[+]{\Theta}{1}{1}{By \TpEquivPosUnit}
      \judgeentailPf{\Theta}{t=t'}{Given}
      \judgeentailPf{\Theta}{\tilde{t} = \tilde{t}}{By \Lemmaref{lem:equivassert}}
      \judgeequivPf{\Theta}{t = \tilde{t}}{t' = \tilde{t}}{By \PrpEquivEq}
      \Hand\judgeequivPf[+]{\Theta}{1 \land (t = \tilde{t})}{1 \land (t' = \tilde{t})}{By \TpEquivPosWith \qedhere}
    \end{llproof}
  \end{itemize}
\end{proof}

\begin{lemma}[Prp.\ Equiv.\ Symmetric]
  \label{lem:symmetric-equiv-prop}
  If $\judgeequiv[]{\Theta}{\phi}{\psi}$, then $\judgeequiv[]{\Theta}{\psi}{\phi}$.
\end{lemma}
\begin{proof}
  By structural induction on the given derivation,
  using \Lemmaref{lem:equivassert} as needed.
\end{proof}

\begin{lemma}[Tp./Fun.\ Equiv.\ Symmetric]
  \label{lem:symmetric-equiv-tp-fun}
  ~
  \begin{enumerate}
  \item If $\judgeequiv[\pm]{\Theta}{A}{B}$,
    then $\judgeequiv[\pm]{\Theta}{B}{A}$
    by a derivation of lesser or equal height.
  \item If $\judgeequiv[]{\Theta}{\mathcal{F}}{\mathcal{F}'}$,
    then $\judgeequiv[]{\Theta}{\mathcal{F}'}{\mathcal{F}}$
    by a derivation of lesser or equal height.
  \end{enumerate}
\end{lemma}
\begin{proof}
  By mutual induction
  on the structure of the given type or functor equivalence derivation,
  using \Lemmaref{lem:symmetric-equiv-prop},
  \Lemmaref{lem:equivassert} as needed.
\end{proof}

\begin{lemma}[Equiv.\ Prop.\ Meaning]
  \label{lem:equiv-prop-meaning}
  If $\judgeequiv{\Theta}{\phi}{\psi}$ and $|- \delta : \Theta$
  then $\sem{\delta}{\phi} = \sem{\delta}{\psi}$.
\end{lemma}
\begin{proof}
  By structural induction on the proposition equivalence derivation.
\end{proof}

\begin{lemma}[Equiv.\ Resp.\ Prp.\ Truth]
  \label{lem:equiv-respects-prop-truth}
  If $\judgeentail{\Theta}{\phi}$ and $\judgeequiv{\Theta}{\phi}{\psi}$,
  then $\judgeentail{\Theta}{\psi}$.
\end{lemma}
\begin{proof}
  Follows from \Lemmaref{lem:equiv-prop-meaning}.
\end{proof}

\begin{lemma}[Ctx.\ Equiv.\ Compat.]
  \label{lem:ctx-equiv-compat}
  Assume $\judgeequiv{\Theta_1}{\Theta}{\Theta'}$.
  \begin{enumerate}
  \item If $\judgeterm{\Theta_1, \Theta, \Theta_2}{t}{\tau}$,
    then $\judgeterm{\Theta_1, \Theta', \Theta_2}{t}{\tau}$.
  \item If $|- \delta : \Theta_1, \Theta, \Theta_2$,
    then $|- \delta : \Theta_1, \Theta', \Theta_2$.
  \item If $\judgeentail{\Theta_1, \Theta, \Theta_2}{\phi}$,
    then $\judgeentail{\Theta_1, \Theta', \Theta_2}{\phi}$.
  \item If $\judgeequiv{\Theta_1, \Theta, \Theta_2}{\phi}{\psi}$,
    then $\judgeequiv{\Theta_1, \Theta', \Theta_2}{\phi}{\psi}$.
  \item If $\judgeextract[\pm]{\Theta_1, \Theta, \Theta_2}{A}{A'}{\Theta'}$,
    then $\judgeextract[\pm]{\Theta_1, \Theta', \Theta_2}{A}{A'}{\Theta'}$.
  \item If $\judgeequiv{\Theta_1, \Theta, \Theta_2}{\mathcal{F}}{\mathcal{G}}$,
    then $\judgeequiv{\Theta_1, \Theta', \Theta_2}{\mathcal{F}}{\mathcal{G}}$.
  \item If $\judgeequiv[\pm]{\Theta_1, \Theta, \Theta_2}{A}{B}$,
    then $\judgeequiv[\pm]{\Theta_1, \Theta', \Theta_2}{A}{B}$.
  \item If $\judgesub[\pm]{\Theta_1, \Theta, \Theta_2}{A}{B}$,
    then $\judgesub[\pm]{\Theta_1, \Theta', \Theta_2}{A}{B}$.
  \item If $\judgeunroll{\Xi}{\Theta_1, \Theta, \Theta_2}{\nu:G[\mu F]}{\beta}{G\;\Fold{F}{\alpha}\;\nu}{t}{P}{\tau}$,\\
    then $\judgeunroll{\Xi}{\Theta_1, \Theta', \Theta_2}{\nu:G[\mu F]}{\beta}{G\;\Fold{F}{\alpha}\;\nu}{t}{P}{\tau}$.
  \item If $\judgesynhead{\Theta_1, \Theta, \Theta_2}{\Gamma}{h}{P}$,
    then $\judgesynhead{\Theta_1, \Theta', \Theta_2}{\Gamma}{h}{P}$.
  \item If $\judgesynexp{\Theta_1, \Theta, \Theta_2}{\Gamma}{\be}{\upshift{P}}$,
    then $\judgesynexp{\Theta_1, \Theta', \Theta_2}{\Gamma}{\be}{\upshift{P}}$.
  \item If $\judgechkval{\Theta_1, \Theta, \Theta_2}{\Gamma}{v}{P}$,
    then $\judgechkval{\Theta_1, \Theta', \Theta_2}{\Gamma}{v}{P}$.
  \item If $\judgechkexp{\Theta_1, \Theta, \Theta_2}{\Gamma}{e}{N}$,
    then $\judgechkexp{\Theta_1, \Theta', \Theta_2}{\Gamma}{e}{N}$.
  \item If $\judgechkmatch{\Theta_1, \Theta, \Theta_2}{\Gamma}{P}{\clauses{\pa}{e}{i}{I}}{N}$,
    then
    $\judgechkmatch{\Theta_1, \Theta', \Theta_2}{\Gamma}{P}{\clauses{\pa}{e}{i}{I}}{N}$.
  \item If $\judgespine{\Theta_1, \Theta, \Theta_2}{\Gamma}{s}{N}{\upshift{P}}$
    then $\judgespine{\Theta_1, \Theta', \Theta_2}{\Gamma}{s}{N}{\upshift{P}}$.
  \end{enumerate}
  Further, none of the consequent derivations increase in height.
\end{lemma}
\begin{proof}
  ~
  All necessary presupposed well-formedness derivations with $\Theta'$
  instead of $\Theta$ are proved by structural induction on them.
  \begin{enumerate}
  \item By structural induction on the index sorting derivation.
  \item By lexicographic induction on, first, the structure of $\Theta_2$, and,
    second, the structure of $\Theta$.
  \item
    ~\\
    \begin{llproof}
      \judgsubsPf{}{\delta}{\Theta_1, \Theta', \Theta_2}{Assume}
      \judgeequivPf{\Theta_1}{\Theta'}{\Theta}{By repeated \Lemmaref{lem:symmetric-equiv-prop}}
      \judgsubsPf{}{\delta}{\Theta_1, \Theta, \Theta_2}{By part (2)}
      \judgeentailPf{\Theta_1, \Theta, \Theta_2}{\phi}{Given}
      \Pf{\sem{\delta}{\phi}}{=}{1}{By inversion on \PropTrue}
      \judgeentailPf{\Theta_1, \Theta', \Theta_2}{\phi}{By \PropTrue}
    \end{llproof} 
  \item By structural induction on the proposition equivalence derivation,
    using part (3) as needed.
  \item By structural induction on the extraction derivation.
  \item By structural induction on the functor equivalence derivation.
    This part is proved mutually with part (7).
  \item By structural induction on the type equivalence derivation,
    using parts (3) and (4) as needed.
    This part is proved mutually with part (6).
  \item By structural induction on the subtyping derivation,
    using parts (1), (3), (5), (6), and (7) as needed.
  \item By structural induction the unrolling derivation.
  \end{enumerate}
  Parts (10) through (15) are proved by mutual induction
  on the structure of the program typing derivation,
  using previous parts as needed. \qedhere
\end{proof}

\begin{lemma}[Equiv.\ Implies Subtyping]
  \label{lem:equiv-implies-sub}
  If $\judgeequiv[\pm]{\Theta}{A}{B}$, then $\judgesub[\pm]{\Theta}{A}{B}$.
\end{lemma}
\begin{proof}
  By induction on the height of the given equivalence derivation.
  \begin{itemize}
    \DerivationProofCase{\TpEquivPosUnit}
    { }
    {
      \judgeequiv[+]{\Theta}{1}{1}
    }
    \begin{llproof}
      \judgesubPf[+]{\Theta}{1}{1}{By \DeclSubPosUnit}
    \end{llproof} 

    \ProofCaseRule{\TpEquivPosVoid}
    Similarly to \TpEquivPosUnit case, simply apply \DeclSubPosVoid.

    \DerivationProofCase{\TpEquivPosProd}
    {
      \judgeequiv[+]{\Theta}{P_1}{Q_1}
      \\
      \judgeequiv[+]{\Theta}{P_2}{Q_2}
    }
    {
      \judgeequiv[+]{\Theta}{P_1 \times P_2}{Q_1 \times Q_2}
    }
    \begin{llproof}
      \judgeequivPf[+]{\Theta}{P_k}{Q_k}{Subderivations}
      \judgesubPf[+]{\Theta}{P_k}{Q_k}{By \ih (twice)}
      \judgesubPf[+]{\Theta}{P_1 \times P_2}{Q_1 \times Q_2}{By \DeclSubPosProd}
    \end{llproof}

    \DerivationProofCase{\TpEquivPosSum}
    {
      \judgeequiv[+]{\Theta}{P_1}{Q_1}
      \\
      \judgeequiv[+]{\Theta}{P_2}{Q_2}
    }
    {
      \judgeequiv[+]{\Theta}{P_1 + P_2}{Q_1 + Q_2}
    }
    \begin{llproof}
      \judgeequivPf[+]{\Theta}{P_k}{Q_k}{Subderivations}
      \judgesubPf[+]{\Theta}{P_1 + P_2}{Q_1 + Q_2}{By \DeclSubPosSum}
    \end{llproof}

    \DerivationProofCase{\TpEquivPosWith}
    {
      \judgeequiv[+]{\Theta}{P}{Q}
      \\
      \judgeequiv[]{\Theta}{\phi}{\psi}
    }
    {
      \judgeequiv[+]{\Theta}{P \land \phi}{Q \land \psi}
    }
    \begin{llproof}
      \judgeequivPf[+]{\Theta}{P}{Q}{Subderivation}
      \judgesubPf[+]{\Theta}{P}{Q}{By \ih}
      \judgeextractPf[+]{\Theta}{P}{P'}{\Theta_P}{By \Lemmaref{lem:sub-extract-inversion}}
      \judgesubPf[+]{\Theta, \Theta_P}{P'}{Q}{\ditto}
      \judgesubPf[+]{\Theta, \psi, \Theta_P}{P'}{Q}{By \Lemmaref{lem:ix-level-weakening}}
      \judgeentailPf{\Theta, \psi, \Theta_P}{\psi}{By \Lemmaref{lem:propvar}}
      \judgesubPf[+]{\Theta, \psi, \Theta_P}{P'}{Q \land \psi}{By \DeclSubPosWithR}
      \judgeequivPf[]{\Theta}{\phi}{\psi}{Subderivation}
      \judgeequivPf[]{\Theta}{\psi}{\phi}{By \Lemmaref{lem:symmetric-equiv-prop}}
      \judgesubPf[+]{\Theta, \phi, \Theta_P}{P'}{Q \land \psi}{By \Lemmaref{lem:ctx-equiv-compat}}
      \judgeextractPf[+]{\Theta}{P \land \phi}{P'}{\phi, \Theta_P}{By \ExtractWith}
      \judgesubPf[+]{\Theta}{P \land \phi}{Q \land \psi}{By \DeclSubPosL}
    \end{llproof}

    \DerivationProofCase{\TpEquivPosEx}
    {
      \judgeequiv[+]{\Theta, a:\tau}{P}{Q}
    }
    {
      \judgeequiv[+]{\Theta}{\extype{a:\tau}{P}}{\extype{a:\tau}{Q}}
    }
    \begin{llproof}
      \judgeequivPf[+]{\Theta, a:\tau}{P}{Q}{Subderivation}
      \judgesubPf[+]{\Theta, a:\tau}{P}{Q}{By \ih}
      \judgeextractPf[+]{\Theta, a:\tau}{P}{P'}{\Theta_P}{By \Lemmaref{lem:sub-extract-inversion}}
      \judgesubPf[+]{\Theta, a:\tau, \Theta_P}{P'}{\underbrace{Q}_{[a/a]Q}}{\ditto}
      \judgetermPf{\Theta, a:\tau, \Theta_P}{a}{\tau}{By \IxVar}
      \judgesubPf[+]{\Theta, a:\tau, \Theta_P}{P'}{\extype{a:\tau}Q}{By \DeclSubPosExR}
      \judgeextractPf[+]{\Theta}{\extype{a:\tau}P}{P'}{a:\tau,\Theta_P}{By \ExtractEx}
      \judgesubPf[+]{\Theta}{\extype{a:\tau}P}{\extype{a:\tau}Q}{By \DeclSubPosL}
    \end{llproof}

    \DerivationProofCase{\TpEquivPosFix}
    {
      \judgeequiv[]{\Theta}{F}{G}
      \\
      \judgeentail{\Theta}{t = t'}
    }
    {
      \judgeequiv[+]{\Theta}{\comprehend{\nu:\mu F}{\Fold{F}{\alpha}\,\nu =_\tau t}}{\comprehend{\nu:\mu G}{\Fold{G}{\alpha}\,\nu =_\tau t'}}
    }
    \begin{llproof}
      \judgeequivPf[]{\Theta}{F}{G}{Subderivation}
      \judgeentailPf{\Theta}{t = t'}{Subderivation}
      \judgesubPf[+]{\Theta}{\comprehend{\nu:\mu F}{\Fold{F}{\alpha}\,\nu =_\tau t}}{\comprehend{\nu:\mu G}{\Fold{G}{\alpha}\,\nu =_\tau t'}}{By \DeclSubPosFix}
    \end{llproof}

    \DerivationProofCase{\TpEquivPosDownshift}
    {
      \judgeequiv[-]{\Theta}{N}{M}
    }
    {
      \judgeequiv[+]{\Theta}{\downshift{N}}{\downshift{M}}
    }
    \begin{llproof}
      \judgeequivPf[-]{\Theta}{N}{M}{Subderivation}
      \judgesubPf[-]{\Theta}{N}{M}{By \ih}
      \judgesubPf[-]{\Theta}{\downshift{N}}{\downshift{M}}{By \DeclSubPosDownshift}
    \end{llproof}

    \ProofCaseRule{\TpEquivNegUpshift}
    Similar to \TpEquivPosDownshift case.

    \ProofCaseRule{\TpEquivNegImp}
    Similar to \TpEquivPosWith case.

    \ProofCaseRule{\TpEquivNegAll}
    Similar to \TpEquivPosEx case.

    \DerivationProofCase{\TpEquivNegArrow}
    {
      \Dee_1 \derives \judgeequiv[+]{\Theta}{P}{Q}
      \\
      \judgeequiv[-]{\Theta}{N}{M}
    }
    {
      \judgeequiv[-]{\Theta}{P \to N}{Q \to M}
    }  
    \begin{llproof}
      \judgeequivPf[+]{\Theta}{P}{Q}{Subderivation}
      \judgeequivPf[+]{\Dee_1' \derives \Theta}{Q}{P}{By \Lemmaref{lem:symmetric-equiv-tp-fun}}
      \Pf{\hgt{\Dee_1'}}{\leq}{\hgt{\Dee_1}}{\ditto}
      \judgesubPf[+]{\Theta}{Q}{P}{By \ih}
      \judgeequivPf[-]{\Theta}{N}{M}{Subderivation}
      \judgesubPf[-]{\Theta}{N}{M}{By \ih}
      \judgesubPf[-]{\Theta}{P \to N}{Q \to M}{By \DeclSubNegArrow}
    \end{llproof}
    \qedhere
  \end{itemize}
\end{proof}

\begin{lemma}[Unroll to Mutual Subtypes]
  \label{lem:unroll-to-mutual-sub}
  ~\\
  If $\judgeunroll{\Xi}{\Theta}{\nu:G[\mu F]}{\beta}{G\;\Fold{F}{\alpha}\;\nu}{t}{P}{\tau}$\\
  and $\judgeequiv[]{\Theta}{G}{G'}$
  and $\judgeequiv[]{\Theta}{F}{F'}$
  and $\judgeentail{\Theta}{t=t'}$,\\
  then there exists $Q$
  such that $\judgeunroll{\Xi}{\Theta}{\nu:G'[\mu F']}{\beta}{G'\;\Fold{F'}{\alpha}\;\nu}{t'}{Q}{\tau}$\\
  and $\judgesub[+]{\Theta}{P}{Q}$ and $\judgesub[+]{\Theta}{Q}{P}$.
\end{lemma}
\begin{proof}
  ~\\
  \begin{llproof}
    \Hand\judgeunrollPf{\Xi}{\Theta}{\nu:G'[\mu F']}{\beta}{G'\;\Fold{F'}{\alpha}\;\nu}{t'}{Q}{\tau}{By \Lemmaref{lem:unroll-to-equiv}}
    \decolumnizePf
    \judgeequivPf[+]{\Theta}{P}{Q}{\ditto}
    \Hand\judgesubPf[+]{\Theta}{P}{Q}{By \Lemmaref{lem:equiv-implies-sub}}
    \judgeequivPf[+]{\Theta}{Q}{P}{By \Lemmaref{lem:symmetric-equiv-tp-fun}}
    \Hand\judgesubPf[+]{\Theta}{Q}{P}{By \Lemmaref{lem:equiv-implies-sub}}
  \end{llproof} 
  \qedhere
\end{proof}

\begin{lemma}[Sub.\ and Equiv.\ Compatible]
  \label{lem:sub-equiv-compatible}
  If $\judgesub[\pm]{\Theta}{A}{B}$
  and $\judgeequiv[\pm]{\Theta}{A}{A'}$
  and $\judgeequiv[\pm]{\Theta}{B}{B'}$,\\
  then $\judgesub[\pm]{\Theta}{A'}{B'}$.
\end{lemma}
\begin{proof}
  ~\\
  \begin{llproof}
    \judgeequivPf[\pm]{\Theta}{A}{A'}{Given}
    \judgeequivPf[\pm]{\Theta}{A'}{A}{By \Lemmaref{lem:symmetric-equiv-tp-fun}}
    \judgesubPf[\pm]{\Theta}{A'}{A}{By \Lemmaref{lem:equiv-implies-sub}}
    \judgesubPf[\pm]{\Theta}{A}{B}{Given}
    \judgesubPf[\pm]{\Theta}{A'}{B}{By \Lemmaref{lem:trans-sub}}
    \judgeequivPf[\pm]{\Theta}{B}{B'}{Given}
    \judgesubPf[\pm]{\Theta}{B}{B'}{By \Lemmaref{lem:equiv-implies-sub}}
    \judgesubPf[\pm]{\Theta}{A'}{B'}{By \Lemmaref{lem:trans-sub}}
  \end{llproof}
\end{proof}

\begin{lemma}[Ctx.\ Equiv.\ Var.\ Push]
  \label{lem:ctx-equiv-var-push}
  If $\judgeequiv{\Theta_0, a:\tau}{\Theta_1}{\Theta_2}$,
  then $\judgeequiv{\Theta_0}{a:\tau, \Theta_1}{a:\tau, \Theta_2}$.
\end{lemma}
\begin{proof}
  By structural induction on the given context equivalence derivation.
\end{proof}

\begin{lemma}[Ctx.\ Equiv.\ Append]
  \label{lem:ctx-equiv-append}
  If $\judgctx{(\Theta_0, \Theta_1)}$
  and $\judgctx{(\Theta_0', \Theta_2)}$
  and $\judgeequiv{\Theta}{\Theta_0}{\Theta_0'}$
  and $\judgeequiv{\Theta}{\Theta_1}{\Theta_2}$,
  then $\judgeequiv{\Theta}{\Theta_0, \Theta_1}{\Theta_0', \Theta_2}$.
\end{lemma}
\begin{proof}
  By structural induction on the derivation of $\judgeequiv{\Theta}{\Theta_1}{\Theta_2}$,
  using \Lemmaref{lem:ix-level-weakening} in the \CtxEquivProp case.
\end{proof}

\begin{lemma}[Equiv.\ Extract]
  \label{lem:equiv-extract}
  If $\judgeextract[\pm]{\Theta}{A}{A'}{\Theta_A}$
  and $\judgeequiv[\pm]{\Theta}{A}{B}$,\\
  then $\judgeextract[\pm]{\Theta}{B}{B'}{\Theta_B}$
  and $\judgeequiv[\pm]{\Theta, \Theta_A}{A'}{B'}$
  and $\judgeequiv{\Theta}{\Theta_A}{\Theta_B}$.
\end{lemma}
\begin{proof}
  By structural induction on the given extraction derivation.
  We will show only the negative part; the positive part is similar.
  \begin{itemize}
    \DerivationProofCase{\ExtractStopNeg}
    {
    }
    {
      \judgeextract[-]{\Theta}{\underbrace{\upshift{P}}_A}{\underbrace{\upshift{P}}_{A'}}{\underbrace{\cdot}_{\Theta_A}}
    }
    \begin{llproof}
      \Pf{B}{=}{\upshift{Q}}{By inversion on \TpEquivNegUpshift}
      \judgeextractPf[-]{\Theta}{B}{B}{\underbrace{\cdot}_{\Theta_B}}{By \ExtractStopNeg}
      \Hand\judgeequivPf[-]{\Theta, \underbrace{\cdot}_{\Theta_A}}{\underbrace{A}_{A'}}{\underbrace{B}_{B'}}{Given}
      \Hand\judgeequivPf{\Theta}{\underbrace{\cdot}_{\Theta_A}}{\underbrace{\cdot}_{\Theta_B}}{By \CtxEquivEmpty}
    \end{llproof} 

    \DerivationProofCase{\ExtractImp}
    {
      \judgeextract[-]{\Theta}{N}{N'}{\Theta_A'}
    }
    {
      \judgeextract[-]{\Theta}{\phi \implies N}{N'}{\phi, \Theta_A'}
    }
    \begin{llproof}
      \judgeequivPf[-]{\Theta}{\phi \implies N}{B}{Given}
      \Pf{B}{=}{\psi \implies M}{By inversion on \TpEquivNegImp}
      \judgeequivPf[-]{\Theta}{N}{M}{\ditto}
      \judgeequivPf[]{\Theta}{\phi}{\psi}{\ditto}
      \judgeextractPf[-]{\Theta}{N}{N'}{\Theta_A'}{Subderivation}
      \judgeextractPf[-]{\Theta}{\psi \implies M}{B'}{\Theta_B}{Given}
      \Pf{\Theta_B}{=}{\psi, \Theta_B'}{By inversion on \ExtractImp}
      \judgeextractPf[-]{\Theta}{M}{B'}{\Theta_B'}{\ditto}
      \judgeequivPf[-]{\Theta, \Theta_A'}{N'}{B'}{By \ih}
      \judgeequivPf[]{\Theta}{\Theta_A'}{\Theta_B'}{\ditto}
      \Hand\judgeequivPf[-]{\Theta, \phi, \Theta_A'}{N'}{B'}{By \Lemmaref{lem:ix-level-weakening}}
      \Hand\judgeequivPf[]{\Theta}{\phi, \Theta_A'}{\psi, \Theta_B'}{Add entry}
    \end{llproof}

    \DerivationProofCase{\ExtractAll}
    {
      \judgeextract[-]{\Theta, a:\tau}{N}{N'}{\Theta_A'}
    }
    {
      \judgeextract[-]{\Theta}{\alltype{a:\tau}{N}}{N'}{a:\tau, \Theta_A'}
    }
    \begin{llproof}
      \judgeequivPf[-]{\Theta}{\alltype{a:\tau} N}{B}{Given}
      \Pf{B}{=}{\alltype{a:\tau} M}{By inversion on \TpEquivNegAll}
      \judgeequivPf[-]{\Theta, a:\tau}{N}{M}{\ditto}
      \judgeextractPf[-]{\Theta, a:\tau}{N}{N'}{\Theta_A'}{Subderivation}
      \judgeextractPf[-]{\Theta}{\alltype{a:\tau} M}{M'}{\Theta_B}{Given}
      \Pf{\Theta_B}{=}{a:\tau, \Theta_B'}{By inversion on \ExtractAll}
      \judgeextractPf[-]{\Theta, a:\tau}{M}{M'}{\Theta_B'}{\ditto}
      \Hand\judgeequivPf[-]{\Theta, a:\tau, \Theta_A'}{N'}{M'}{By \ih}
      \judgeequivPf[]{\Theta, a:\tau}{\Theta_A'}{\Theta_B'}{\ditto}
      \Hand\judgeequivPf[]{\Theta}{a:\tau, \Theta_A'}{a:\tau, \Theta_B'}{By \Lemmaref{lem:ctx-equiv-var-push}}
    \end{llproof}

    \DerivationProofCase{\ExtractArrow}
    {
      \judgeextract[+]{\Theta}{P}{P'}{\Theta_P}
      \\
      \judgeextract[-]{\Theta}{N}{N'}{\Theta_N}
    }
    {
      \judgeextract[-]{\Theta}{P \to N}{P' \to N'}{\Theta_P, \Theta_N}
    }
    \begin{llproof}
      \judgeequivPf[-]{\Theta}{P \to N}{B}{Given}
      \Pf{B}{=}{Q \to M}{By inversion on \TpEquivNegArrow}
      \judgeequivPf[+]{\Theta}{P}{Q}{\ditto}
      \judgeequivPf[-]{\Theta}{N}{M}{\ditto}
      \proofsep
      \judgeextractPf[-]{\Theta}{Q \to M}{B'}{\Theta_B}{Given}
      \Pf{B'}{=}{Q' \to M'}{By inversion on \ExtractArrow}
      \Pf{\Theta_B}{=}{\Theta_Q, \Theta_M}{\ditto}
      \judgeextractPf[+]{\Theta}{Q}{Q'}{\Theta_Q}{\ditto}
      \judgeextractPf[-]{\Theta}{M}{M'}{\Theta_M}{\ditto}
      \proofsep
      \judgeextractPf[+]{\Theta}{P}{P'}{\Theta_P}{Subderivation}
      \judgeequivPf[+]{\Theta, \Theta_P}{P'}{Q'}{By positive part of this lemma}
      \judgeequivPf{\Theta}{\Theta_P}{\Theta_Q}{\ditto}
      \proofsep
      \judgeextractPf[-]{\Theta}{N}{N'}{\Theta_N}{Subderivation}
      \judgeequivPf[-]{\Theta, \Theta_N}{N'}{M'}{By \ih}
      \judgeequivPf{\Theta}{\Theta_N}{\Theta_M}{\ditto}
      \proofsep
      \judgeequivPf[+]{\Theta, \Theta_P, \Theta_N}{P'}{Q'}{By \Lemmaref{lem:ix-level-weakening}}
      \judgeequivPf[-]{\Theta, \Theta_P, \Theta_N}{N'}{M'}{By \Lemmaref{lem:ix-level-weakening}}
      \Hand\judgeequivPf[-]{\Theta, \Theta_P, \Theta_N}{P' \to N'}{Q' \to M'}{By \TpEquivNegArrow}
      \decolumnizePf
      \Hand\judgeequivPf{\Theta}{\Theta_P, \Theta_N}{\Theta_Q, \Theta_M}{By \Lemmaref{lem:ctx-equiv-append}}
    \end{llproof} 
    \qedhere
  \end{itemize}
\end{proof}

\begin{lemma}
  \label{lem:extracted-subtype-of-downshift}
  If $\judgesub[+]{\Theta}{P}{\downshift{M}}$ and $\judgeextract{\Theta}{P}{P}{\cdot}$,
  then there exists $N$ such that $P = \downshift{N}$.
\end{lemma}
\begin{proof}
  Because $\judgeextract{\Theta}{P}{P}{\cdot}$,
  the given subtyping derivation cannot be concluded by \DeclSubPosL;
  by rule inspection, the only possible concluding rule is \DeclSubPosDownshift.
  By inversion on \DeclSubPosDownshift, there exists $N$ such that $P = \downshift{N}$.
\end{proof}

\begin{lemma}[Value Instantiate]
  \label{lem:value-instantiate}
  If $\Dee :: \judgechkval{\Theta}{\Gamma}{v}{P}$
  and $\judgeextract[+]{\Theta}{P}{P'}{\Theta'}$,\\
  then there exist $\Theta |- \sigma : \Theta'$ and $\Dee'$
  such that $\Dee' :: \judgechkval{\Theta}{\Gamma}{v}{[\sigma]P'}$
  and $\hgt{\Dee'} \leq \hgt{\Dee}$.
\end{lemma}
\begin{proof}
  By structural induction on $\Dee$.
  We case analyze rules concluding $\Dee$:
  \begin{itemize}
    \DerivationProofCase{\DeclChkValVar}
    {
      P \neq \exists, \land 
      \\
      (x:Q) \in \Gamma
      \\
      \arrayenvb{\Dee_1 \\ \judgesub[+]{\Theta}{Q}{P}}
    }
    {
      \Dee \derives~~ \judgechkval{\Theta}{\Gamma}{x}{P}
    }
    \begin{llproof}
      \Pf{(x:Q)}{\in}{\Gamma}{Subderivation}
      \judgeextractPf[+]{\Theta}{Q}{Q}{\cdot}{By inversion on $\Gamma$ WF}
      \judgeextractPf[+]{\Theta}{P}{P'}{\Theta'}{Given}
      \judgesubPf[+]{\Dee_1 \derives~~ \Theta}{Q}{P}{Subderivation}
      \Hand\judgsubsPf{\underbrace{\Theta, \cdot}_\Theta}{\sigma}{\Theta'}{By \Lemmaref{lem:sub-instantiate}}
      \judgesubPf[+]{\Dee_1' \derives~~ \underbrace{\Theta, \cdot}_\Theta}{Q}{[\sigma]P'}{\ditto}
      \Pf{\hgt{\Dee_1'}}{\leq}{\hgt{\Dee_1}}{\ditto}
      \Pf{P}{\neq}{\exists, \land}{Premise}
      \Pf{[\sigma]P}{\neq}{\exists, \land}{Straightforward}
      \Hand\judgechkvalPf{\Dee' \derives~~ \Theta}{\Gamma}{x}{[\sigma]P'}{By \DeclChkValVar}
      \Pf{\hgt{\Dee'}}{=}{\hgt{\Dee_1'}+1}{By \defn of $\hgt{-}$}
      \Pf{}{\leq}{\hgt{\Dee_1}+1}{By above inequality}
      \Pf{}{=}{\hgt{\Dee}}{By \defn of $\hgt{-}$}
    \end{llproof}

    \DerivationProofCase{\DeclChkValPair}
    { \arrayenvb{\Dee_1 \derives \judgechkval{\Theta}{\Gamma}{v_1}{P_1}} \\
      \arrayenvb{\Dee_2 \derives \judgechkval{\Theta}{\Gamma}{v_2}{P_2}} } 
    { \Dee \derives \judgechkval{\Theta}{\Gamma}{\pair{v_1}{v_2}}{P_1 \times P_2} }
    \begin{llproof}
      \judgeextractPf[+]{\Theta}{P_1 \times P_2}{P'}{\Theta'}{Given}
      \Pf{P'}{=}{P_1' \times P_2'}{By inversion}
      \Pf{\Theta'}{=}{\Theta_1, \Theta_2}{\ditto}
      \judgeextractPf[+]{\Theta}{P_1}{P_1'}{\Theta_1}{\ditto}
      \judgeextractPf[+]{\Theta}{P_2}{P_2'}{\Theta_2}{\ditto}
      \judgechkvalPf{\Dee_1 \derives~~ \Theta}{\Gamma}{v_1}{P_1}{Subderivation}
      \judgsubsPf{\Theta}{\sigma_1}{\Theta_1}{By \ih}
      \judgechkvalPf{\Dee_1' \derives~~ \Theta}{\Gamma}{v_1}{[\sigma_1]P_1'}{\ditto}
      \Pf{\hgt{\Dee_1'}}{\leq}{\hgt{\Dee_1}}{\ditto}
      \judgechkvalPf{\Dee_2 \derives~~ \Theta}{\Gamma}{v_2}{P_2}{Subderivation}
      \judgsubsPf{\Theta}{\sigma_2}{\Theta_2}{By \ih}
      \judgechkvalPf{\Dee_2' \derives~~ \Theta}{\Gamma}{v_2}{[\sigma_2]P_2'}{\ditto}
      \Pf{\hgt{\Dee_2'}}{\leq}{\hgt{\Dee_2}}{\ditto}
      \proofsep
      \Hand\judgsubsPf{\Theta}{\sigma_1, \sigma_2}{\Theta_1, \Theta_2}{By \Lemmaref{lem:subs-append}}
      \proofsep
      \judgetpPf{\Theta, \Theta_1}{P_1'}{\dontcare}{By \Lemmaref{lem:extract-to-type-wf}}
      \judgetpPf{\Theta, \Theta_2}{P_2'}{\dontcare}{By \Lemmaref{lem:extract-to-type-wf}}
      \judgetpPf{\Theta}{[\id_\Theta, \sigma_2]P_2'}{\dontcare}{By \Lemmaref{lem:syn-subs-tp-fun-alg}}
      \trailingjust{(after prepending $\id_\Theta$ to $\sigma$)}
      \judgetpPf{\Theta}{[\sigma_2]P_2'}{\dontcare}{By \Lemmaref{lem:id-subs-id}}
      \judgctxPf{(\Theta, \Theta_1, \Theta_2)}{By \Lemmaref{lem:extract-to-ctx-wf}}
      \Pf{\emptyset}{=}{\dom{\Theta_2} \sect \FV{P_1'}}{Follows from above}
      \Pf{\emptyset}{=}{\dom{\Theta_1} \sect \FV{[\sigma_2]P_2'}}{\ditto}
      \proofsep
      \judgechkvalPf{\Dee_1' \derives~~ \Theta}{\Gamma}{v_1}{[\sigma_1, \sigma_2]P_1'}{$\because \emptyset = \dom{\Theta_2} \sect \FV{P_1'}$, $[\sigma_2]P_1' = P_1'$}
      \judgechkvalPf{\Dee_2' \derives~~ \Theta}{\Gamma}{v_2}{[\sigma_1, \sigma_2]P_2'}{$\because \emptyset = \dom{\Theta_1} \sect \FV{[\sigma_2]P_2'}$,}
      \trailingjust{$[\sigma_1]([\sigma_2]P_2') = [\sigma_2]P_2'$}
      \judgechkvalPf{\Theta}{\Gamma}{\pair{v_1}{v_2}}{[\sigma_1, \sigma_2]P_2' \times [\sigma_1, \sigma_2]P_2'}{By \DeclChkValPair}
      \Hand\judgechkvalPf{\Dee' \derives~~ \Theta}{\Gamma}{\pair{v_1}{v_2}}{[\sigma_1, \sigma_2](P_2' \times P_2')}{By \defn of substitution}
      \Pf{\hgt{\Dee'}}{=}{\max(\hgt{\Dee_1'},\hgt{\Dee_2'}) + 1}{By \defn of $\hgt{-}$}
      \Pf{}{\leq}{\max(\hgt{\Dee_1},\hgt{\Dee_2}) + 1}{By above inequalities}
      \Pf{}{=}{\hgt{\Dee}}{By \defn of $\hgt{-}$}
    \end{llproof} 

    \DerivationProofCase{\DeclChkValExists}
    {
      \arrayenvb{\Dee_1 \\ \judgechkval{\Theta}{\Gamma}{v}{[t/a]P_0}}
      \\
      \judgeterm{\Theta}{t}{\tau}
    }
    { \Dee \derives \judgechkval{\Theta}{\Gamma}{v}{(\extype{a:\tau} P_0)} }
    \begin{llproof}
      \judgechkvalPf{\Dee_1 \derives~~ \Theta}{\Gamma}{v}{[t/a]P_0}{Subderivation}
      \judgeextractPf[+]{\Theta}{\extype{a:\tau} P_0}{P'}{\Theta'}{Given}
      \Pf{\Theta'}{=}{a:\tau, \Theta_0'}{By inversion on extraction}
      \judgeextractPf[+]{\Theta, a:\tau}{P_0}{P'}{\Theta_0'}{\ditto}
      \judgsubsPf{\Theta}{\id_\Theta}{\Theta}{By \Lemref{lem:id-subs-typing}}
      \judgetermPf{\Theta}{t}{\tau}{Subderivation}
      \judgetermPf{\Theta}{[\id_\Theta]t}{\tau}{Identity substitution}
      \judgsubsPf{\Theta}{\id_\Theta, t/a}{\Theta, a:\tau}{By \IxSyn}
      \judgeextractPf[+]{\Theta}{[\id_\Theta, t/a]P_0}{[\id_\Theta, t/a]P'}{[\id_\Theta, t/a]\Theta_0'}{By \Lemref{lem:syn-subs-extract}}
      \judgeextractPf[+]{\Theta}{[t/a]P_0}{[t/a]P'}{[t/a]\Theta_0'}{By \Lemref{lem:id-subs-id}}
      \judgsubsPf{\Theta}{\sigma}{[t/a]\Theta_0'}{By \ih}
      \judgechkvalPf{\Dee_1' \derives~~ \Theta}{\Gamma}{v}{[\sigma]([t/a]P')}{\ditto}
      \Pf{\hgt{\Dee_1'}}{\leq}{\hgt{\Dee_1}}{\ditto}
      \Pf{}{<}{\hgt{\Dee}}{By \defn of $\hgt{-}$}
      \Pf{\hgt{\Dee_1'}}{\leq}{\hgt{\Dee_1}}{\ditto}
      \proofsep
      \judgctxPf{(\Theta, a:\tau, \Theta_0')}{By \Lemref{lem:extract-to-ctx-wf}}
      \Pf{\emptyset}{=}{\dom{\Theta} \sect \dom{\Theta_0'}}{By inversion on ctx.\ WF}
      \Pf{}{=}{\dom{\Theta} \sect \dom{[t/a]\Theta_0'}}{By \defn of $[-]-$}
      \Hand\judgsubsPf{\Theta}{t/a, \sigma}{a:\tau, \Theta_0'}{By \Lemref{lem:subs-typing-undo-subs}}
      \proofsep
      \judgechkvalPf{\Dee_1' \derives~~ \Theta}{\Gamma}{v}{[[\sigma]t/a]([\sigma]P')}{By \Lemref{lem:barendregt}}
      \Pf{[\sigma]t}{=}{t}{$\emptyset = \dom{\Theta} \sect \dom{[t/a]\Theta_0'}$}
      \judgechkvalPf{\Dee_1' \derives~~ \Theta}{\Gamma}{v}{[t/a]([\sigma]P')}{$[\sigma]t=t$}
      \Hand\judgechkvalPf{\Dee_1' \derives~~ \Theta}{\Gamma}{v}{[t/a, \sigma]P'}{By \defn of $[-]-$}
    \end{llproof} 

    \DerivationProofCase{\DeclChkValWith}
    {
      \arrayenvb{\Dee_1 \derives \judgechkval{\Theta}{\Gamma}{v}{P_0}}
      \\
      \judgeentail{\Theta}{\phi}
    }
    { \Dee \derives \judgechkval{\Theta}{\Gamma}{v}{P_0 \land \phi} }
    \begin{llproof}
      \judgechkvalPf{\Dee_1 \derives~~ \Theta}{\Gamma}{v}{P_0}{Subderivation}
      \judgeextractPf[+]{\Theta}{P_0 \land \phi}{P'}{\Theta'}{Given}
      \Pf{\Theta'}{=}{\phi, \Theta_0'}{By inversion}
      \judgeextractPf[+]{\Theta}{P_0}{P'}{\Theta_0'}{\ditto}
      \judgsubsPf{\Theta}{\sigma}{\Theta_0'}{By \ih}
      \Hand\judgechkvalPf{\Dee_1' \derives~~ \Theta}{\Gamma}{v}{[\sigma]P'}{\ditto}
      \Pf{\hgt{\Dee_1'}}{\leq}{\hgt{\Dee_1}}{\ditto}
      \Pf{}{<}{\hgt{\Dee}}{By \defn of $\hgt{-}$}
      \judgeentailPf{\Theta}{\phi}{Subderivation}
      \Hand\judgsubsPf{\Theta}{\sigma}{\phi, \Theta_0'}{By \Lemmaref{lem:syn-subs-deep-entry}}
    \end{llproof}

    \item \textbf{Cases} \DeclChkValUnit, \DeclChkValIn{k}, \DeclChkValFix,
      \DeclChkValDownshift:\\
      In these cases, $P' = P$ and $\Theta' = \cdot$, by inversion on extraction.
      Choose $\sigma = \cdot$ (and the same given value typing derivation). \qedhere
  \end{itemize}
\end{proof}

\begin{lemma}
  \label{lem:match-apply-ex-and-with}
  If $\judgeextract[+]{\Theta}{P}{P'}{\Theta'}$
  and $\judgechkmatch{\Theta, \Theta'}{\Gamma}{P'}{\clauses{\pa}{e}{i}{I}}{N}$,\\
  then $\judgechkmatch{\Theta}{\Gamma}{P}{\clauses{\pa}{e}{i}{I}}{N}$.
\end{lemma}
\begin{proof}
  By structural induction on the given extraction derivation.
\end{proof}

\begin{lemma}[Spine Instantiate]
  \label{lem:spine-instantiate}
  If $\Dee \derives \judgespine{\Theta}{\Gamma}{s}{N}{\upshift{P}}$
  and $\judgeextract[-]{\Theta}{N}{N'}{\Theta_N}$,\\
  then there exist $\Theta |- \sigma : \Theta_N$
  and $\Dee' \derives \judgespine{\Theta}{\Gamma}{s}{[\sigma]N'}{\upshift{P}}$
  such that $\hgt{\Dee'} \leq \hgt{\Dee}$.
\end{lemma}
\begin{proof}
  By induction on the height of the extraction derivation.
  \begin{itemize}
    \DerivationProofCase{\ExtractStopNeg}
    {
    }
    {
      \judgeextract[-]{\Theta}{\upshift{P}}{\upshift{P}}{\cdot}
    }
    Choose $\sigma = \cdot$ and $\Dee' = \Dee$.

    \DerivationProofCase{\ExtractImp}
    {
      \judgeextract[-]{\Theta}{N_0}{N'}{\Theta_0'}
    }
    {
      \judgeextract[-]{\Theta}{\phi \implies N_0}{N'}{\phi, \Theta_0'}
    }
    \begin{llproof}
      \judgespinePf{\Dee \derives \Theta}{\Gamma}{s}{\phi \implies N_0}{\upshift{P}}{Given}
      \judgeentailPf{\Theta}{\phi}{By inversion on \DeclSpineImplies}
      \judgespinePf{\Dee_1 \derives \Theta}{\Gamma}{s}{N_0}{\upshift{P}}{\ditto}
      \judgeextractPf[-]{\Theta}{N_0}{N'}{\Theta_0'}{Subderivation}
      \judgsubsPf{\Theta}{\sigma}{\Theta_0'}{By \ih}
      \judgespinePf{\Dee_1' \derives \Theta}{\Gamma}{s}{[\sigma]N'}{\upshift{P}}{\ditto}
      \Pf{\hgt{\Dee_1'}}{\leq}{\hgt{\Dee_1}}{\ditto}
      \Pf{}{<}{\hgt{\Dee}}{By \defn of $\hgt{-}$}
      \judgsubsPf{\Theta}{\sigma}{\phi, \Theta_0'}{By \Lemmaref{lem:syn-subs-deep-entry}}
    \end{llproof}

    \DerivationProofCase{\ExtractAll}
    {
      \judgeextract[-]{\Theta, a:\tau}{N_0}{N'}{\Theta_0'}
    }
    {
      \judgeextract[-]{\Theta}{\alltype{a:\tau}{N_0}}{N'}{a:\tau, \Theta_0'}
    }
    \begin{llproof}
      \judgespinePf{\Dee \derives \Theta}{\Gamma}{s}{\alltype{a:\tau} N_0}{\upshift{P}}{Given}
      \judgetermPf{\Theta}{t}{\tau}{By inversion (\DeclSpineAll)}
      \judgespinePf{\Dee_1 \derives \Theta}{\Gamma}{s}{[t/a]N_0}{\upshift{P}}{\ditto}
      \judgeextractPf[-]{\Theta, a:\tau}{N_0}{N'}{\Theta_0'}{Subderivation}
      \judgeextractPf[-]{\Theta}{[t/a]N_0}{[t/a]N'}{[t/a]\Theta_0'}{By \Lemref{lem:syn-subs-extract}}
      \Pf{}{}{\text{(lesser or equal hgt.)}}{}
      \judgsubsPf{\Theta}{\sigma_0}{\Theta_0'}{By \ih}
      \judgespinePf{\Dee_1' \derives \Theta}{\Gamma}{s}{[\sigma_0]([t/a]N')}{\upshift{P}}{\ditto}
      \Pf{\hgt{\Dee_1'}}{\leq}{\hgt{\Dee_1}}{\ditto}
      \Pf{}{<}{\hgt{\Dee}}{By \defn of $\hgt{-}$}
      \judgespinePf{\Dee_1' \derives \Theta}{\Gamma}{s}{[[\sigma_0]t/a]([\sigma_0]N')}{\upshift{P}}{By \Lemref{lem:barendregt}}
      \Pf{\emptyset}{=}{\dom{\Theta}\sect\dom{\Theta_0'}}{By \Lemref{lem:extract-to-ctx-wf}}
      \judgespinePf{\Dee_1' \derives \Theta}{\Gamma}{s}{[t/a]([\sigma_0]N')}{\upshift{P}}{$\emptyset = \dom{\Theta}\sect\dom{\Theta_0'}$}
      \trailingjust{(so $[\sigma_0]t = t$)}
      \judgespinePf{\Dee_1' \derives \Theta}{\Gamma}{s}{[t/a, \sigma_0]N'}{\upshift{P}}{By \defn of $[-]-$}
      \judgsubsPf{\Theta}{t/a, \sigma_0}{a:\tau, \Theta_0'}{By \Lemref{lem:syn-subs-deep-entry}}
    \end{llproof}

    \DerivationProofCase{\ExtractArrow}
    {
      \judgeextract[+]{\Theta}{Q}{Q'}{\Theta_1}
      \\
      \judgeextract[-]{\Theta}{N_0}{N_0'}{\Theta_2}
    }
    {
      \judgeextract[-]{\Theta}{Q \to N_0}{Q' \to N_0'}{\Theta_1, \Theta_2}
    }
    \begin{llproof}
      \judgespinePf{\Dee \derives \Theta}{\Gamma}{s}{Q \to N_0}{\upshift{P}}{Given}
      \Pf{s}{=}{v, s_0}{By inversion on \DeclSpineApp}
      \judgechkvalPf{\Dee_1 \derives \Theta}{\Gamma}{v}{Q}{\ditto}
      \judgespinePf{\Dee_2 \derives \Theta}{\Gamma}{s_0}{N_0}{\upshift{P}}{\ditto}
      \judgeextractPf[+]{\Theta}{Q}{Q'}{\Theta_1}{Subderivation}
      \judgsubsPf{\Theta}{\sigma_1}{\Theta_1}{By \Lemmaref{lem:value-instantiate}}
      \judgechkvalPf{\Dee_1' \derives \Theta}{\Gamma}{v}{[\sigma_1]Q'}{\ditto}
      \Pf{\hgt{\Dee_1'}}{\leq}{\hgt{\Dee_1}}{\ditto}
      \judgeextractPf[-]{\Theta}{N_0}{N_0'}{\Theta_2}{Subderivation}
      \judgsubsPf{\Theta}{\sigma_2}{\Theta_2}{By \ih}
      \judgespinePf{\Dee_2' \derives \Theta}{\Gamma}{s_0}{[\sigma_2]N_0'}{\upshift{P}}{\ditto}
      \Pf{\hgt{\Dee_2'}}{\leq}{\hgt{\Dee_2}}{\ditto}
      \Hand\judgsubsPf{\Theta}{\sigma_1, \sigma_2}{\Theta_1, \Theta_2}{By \Lemmaref{lem:subs-append}}
      \decolumnizePf
      \Pf{\emptyset}{=}{\dom{\Theta}\sect\dom{\Theta_1}\sect\dom{\Theta_2}}{By \Lemmaref{lem:extract-to-ctx-wf}}
      \decolumnizePf
      \judgechkvalPf{\Dee_1' \derives \Theta}{\Gamma}{v}{[\sigma_1, \sigma_2]Q'}{$\emptyset = \dom{\Theta}\sect\dom{\Theta_1}\sect\dom{\Theta_2}$}
      \judgespinePf{\Dee_2' \derives \Theta}{\Gamma}{s_0}{[\sigma_1, \sigma_2]N_0'}{\upshift{P}}{$\emptyset = \dom{\Theta}\sect\dom{\Theta_1}\sect\dom{\Theta_2}$}
      \decolumnizePf
      \judgespinePf{\Dee' \derives \Theta}{\Gamma}{v, s_0}{[\sigma_1, \sigma_2]Q' \to [\sigma_1, \sigma_2]N_0'}{\upshift{P}}{By \DeclSpineApp}
      \Hand\judgespinePf{\Dee' \derives \Theta}{\Gamma}{v, s_0}{[\sigma_1, \sigma_2](Q' \to N_0')}{\upshift{P}}{By \defn of subst.}
      \decolumnizePf
      \Pf{\hgt{\Dee'}}{=}{\max(\hgt{\Dee_1'}, \hgt{\Dee_2'}) + 1}{By \defn of $\hgt{-}$}
      \Pf{}{\leq}{\max(\hgt{\Dee_1}, \hgt{\Dee_2}) + 1}{By above inequalities}
      \Pf{}{=}{\hgt{\Dee}}{By \defn of $\hgt{-}$}
    \end{llproof}  \qedhere
  \end{itemize}
\end{proof}

\begin{lemma}[Subsumption Admissibility]
  \label{lem:subsumption-admissibility}
  Assume $\judgesub[+]{\Theta}{\Gamma'}{\Gamma}$. Then:
  \begin{enumerate}
  \item If $\judgesynhead{\Theta}{\Gamma}{h}{P}$,
    then either:
    \begin{enumerate}
    \item $\judgesynhead{\Theta}{\Gamma'}{h}{P}$; or
    \item there exists $P'$ 
      such that $\judgeextract[+]{\Theta}{P'}{P'}{\cdot}$
      and $\judgesub[+]{\Theta}{P'}{P}$
      and $\judgesynhead{\Theta}{\Gamma'}{h}{P'}$.
    \end{enumerate}
  \item If $\judgesynexp{\Theta}{\Gamma}{\be}{\upshift{P}}$,
    then there exists $P'$ such that $\judgesub[-]{\Theta}{\upshift{P'}}{\upshift{P}}$
    and $\judgesynexp{\Theta}{\Gamma'}{\be}{\upshift{P'}}$.
  \item If $\judgechkval{\Theta}{\Gamma}{v}{P}$
    and $\judgesub[+]{\Theta}{P}{Q}$,
    then $\judgechkval{\Theta}{\Gamma'}{v}{Q}$.
  \item If $\judgechkexp{\Theta}{\Gamma}{e}{N}$
    and $\judgesub[-]{\Theta}{N}{M}$,
    then $\judgechkexp{\Theta}{\Gamma'}{e}{M}$.
  \item If $\judgechkmatch{\Theta}{\Gamma}{P}{\clauses{\pa}{e}{i}{I}}{N}$
    and $\judgesub[-]{\Theta}{N}{M}$
    and $\judgesub[+]{\Theta}{Q}{P}$,
    then\\
    $\judgechkmatch{\Theta}{\Gamma'}{Q}{\clauses{\pa}{e}{i}{I}}{M}$.
  \item If $\judgespine{\Theta}{\Gamma}{s}{N}{\upshift{P}}$
    and $\judgesub[-]{\Theta}{M}{N}$,
    then there exists $P'$ such that
    $\judgesub[-]{\Theta}{\upshift{P'}}{\upshift{P}}$
    and $\judgespine{\Theta}{\Gamma'}{s}{M}{\upshift{P'}}$.
  \end{enumerate}
\end{lemma}
\begin{proof}
  By lexicographic induction on,
  first, the height of the given typing derivation;
  second, the size (\Figureref{fig:size}) of the type in verification position
  of the given (non-contextual) subtyping derivation, \ie
  $\size{Q}$ for part (3),
  $\size{N}$ for part (4),
  $\size{P}$ for part (5), and
  $\size{M}$ for part (6); and
  third, the height of the (non-contextual) subtyping derivation, \ie (the hgt.\ of)
  $\judgesub[+]{\Theta}{P}{Q}$ for part (3),
  $\judgesub[-]{\Theta}{N}{M}$ for part (4),
  $\judgesub[+]{\Theta}{Q}{P}$ for part (5), and
  $\judgesub[-]{\Theta}{M}{N}$ for part (6).
  \begin{enumerate}
  \item %
    \begin{itemize}
      \DerivationProofCase{\DeclSynHeadVar}
      {
        (x : P) \in \Gamma
      }
      {
        \judgesynhead{\Theta}{\Gamma}{x}{P}
      }
      \begin{llproof}
        \Pf{(x : P)}{\in}{\Gamma}{Subderivation}
        \judgesubPf[+]{\Theta}{\Gamma'}{\Gamma}{Given}
        \Pf{(x:P')}{\in}{\Gamma'}{By inversion}
        \Hand\judgesubPf[+]{\Theta}{P'}{P}{\ditto}
        \Hand\judgeextractPf[+]{\Theta}{P'}{P'}{\cdot}{By inversion on $\Gamma'$ WF}
        \Hand\judgesynheadPf{\Theta}{\Gamma'}{x}{P'}{By \DeclSynHeadVar}
      \end{llproof}

      \DerivationProofCase{\DeclSynValAnnot}
      {
        \judgetp{\Theta}{P}{\Xi}
        \\
        \judgechkval{\Theta}{\Gamma}{v}{P}
      }
      {
        \judgesynhead{\Theta}{\Gamma}{\annoexp{v}{P}}{P}
      }
      \begin{llproof}
        \judgesubPf[+]{\Theta}{\Gamma'}{\Gamma}{Given}
        \judgesubPf[+]{\Theta}{P}{P}{By \Lemmaref{lem:refl-sub}}
        \judgechkvalPf{\Theta}{\Gamma}{v}{P}{Subderivation}
        \judgechkvalPf{\Theta}{\Gamma'}{v}{P}{By i.h.}
        \Hand\judgesynheadPf{\Theta}{\Gamma'}{\annoexp{v}{P}}{P}{By \DeclSynValAnnot}
      \end{llproof}
    \end{itemize}

  \item %
    \begin{itemize}
      \DerivationProofCase{\DeclSynSpineApp}
      { \judgesynhead{\Theta}{\Gamma}{h}{\downshift{N}} \\
        \judgespine{\Theta}{\Gamma}{s}{N}{\upshift{P}} }
      { \judgesynexp{\Theta}{\Gamma}{h(s)}{\upshift{P}} }
      \begin{llproof}
        \judgesubPf[+]{\Theta}{\Gamma'}{\Gamma}{Given}
        \judgesynheadPf{\Theta}{\Gamma}{h}{\downshift{N}}{Subderivation}
      \end{llproof}

      By \ih, either
      (a) $\judgesynhead{\Theta}{\Gamma'}{h}{\downshift{N}}$; or\\
      (b) there exists $Q$ 
      such that $\judgeextract[+]{\Theta}{Q}{Q}{\cdot}$
      and $\judgesub[+]{\Theta}{Q}{\downshift{N}}$
      and $\judgesynhead{\Theta}{\Gamma'}{h}{Q}$.

      Consider subcases (a) and (b):

      \begin{itemize}
      \item \textbf{Case} (a):\\
        \begin{llproof}
          \judgespinePf{\Theta}{\Gamma}{s}{N}{\upshift{P}}{Subderivation}
          \judgesubPf[-]{\Theta}{N}{N}{By \Lemmaref{lem:refl-sub}}
          \judgespinePf{\Theta}{\Gamma'}{s}{N}{\upshift{P'}}{By \ih}
          \Hand\judgesubPf[-]{\Theta}{\upshift{P'}}{\upshift{P}}{\ditto}
          \Hand\judgesynexpPf{\Theta}{\Gamma'}{h(s)}{\upshift{P'}}{By \DeclSynSpineApp}
        \end{llproof} 
      \item \textbf{Case} (b):\\
        \begin{llproof}
          \Pf{Q}{=}{\downshift{M}}{By \Lemref{lem:extracted-subtype-of-downshift}}
          \judgesubPf[+]{\Theta}{\downshift{M}}{\downshift{N}}{Rewrite above}
          \judgesubPf[-]{\Theta}{M}{N}{By inversion on \DeclSubPosDownshift}
          \judgespinePf{\Theta}{\Gamma}{s}{N}{\upshift{P}}{Subderivation}
          \judgespinePf{\Theta}{\Gamma'}{s}{M}{\upshift{P'}}{By \ih}
          \Hand\judgesubPf[-]{\Theta}{\upshift{P'}}{\upshift{P}}{\ditto}
          \Hand\judgesynexpPf{\Theta}{\Gamma'}{h(s)}{\upshift{P'}}{By \DeclSynSpineApp}
        \end{llproof} 
      \end{itemize}

      \DerivationProofCase{\DeclSynExpAnnot}
      {
        \judgetp{\Theta}{P}{\Xi}
        \\
        \judgechkexp{\Theta}{\Gamma}{e}{\upshift{P}}
      }
      { \judgesynexp{\Theta}{\Gamma}{\annoexp{e}{\upshift{P}}}{\upshift{P}} }
      \begin{llproof}
        \judgesubPf[+]{\Theta}{\Gamma'}{\Gamma}{Given}
        \Hand\judgesubPf[-]{\Theta}{\upshift{P}}{\upshift{P}}{By \Lemmaref{lem:refl-sub}}
        \judgechkexpPf{\Theta}{\Gamma}{e}{\upshift{P}}{Subderivation}
        \judgechkexpPf{\Theta}{\Gamma'}{e}{\upshift{P}}{By \ih}
        \Hand\judgesynexpPf{\Theta}{\Gamma'}{\annoexp{e}{\upshift{P}}}{\upshift{P}}{By \DeclSynExpAnnot}
      \end{llproof}
    \end{itemize}

  \item %
    The \DeclChkValVar case is independent of
    the structure of the given subtyping derivation, and we consider it first.
    Then we consider the three subcases for the final rule of the subtyping derivation
    that are independent of the structure of $P$.
    
    \begin{itemize}
      \DerivationProofCase{\DeclChkValVar}
      { P \neq \exists, \land \\ (x:P') \in \Gamma \\ \judgesub[+]{\Theta}{P'}{P} }
      { \judgechkval{\Dee :: \Theta}{\Gamma}{x}{P} }
      \begin{llproof}
        \judgesubPf[+]{\Theta}{P'}{P}{Subderivation}
        \judgesubPf[+]{\Theta}{P}{Q}{Given}
        \judgesubPf[+]{\Theta}{P'}{Q}{By \Lemmaref{lem:trans-sub}}
        \Pf{(x:P')}{\in}{\Gamma}{Subderivation}
        \simplePf{\Theta}{P'}{By inversion on $\Gamma$ WF}
        \judgesubPf[+]{\Theta}{\Gamma'}{\Gamma}{Given}
      \end{llproof} 
      \begin{itemize}
      \item \textbf{Case} $Q \neq \exists, \land$\\
        \begin{llproof}
          \Pf{(x:P'')}{\in}{\Gamma'}{By inversion}
          \judgesubPf[+]{\Theta}{P''}{P'}{\ditto}
          \judgesubPf[+]{\Theta}{P''}{Q}{By \Lemmaref{lem:trans-sub}}
          \Hand\judgechkvalPf{\Theta}{\Gamma'}{x}{Q}{By \DeclChkValVar}
        \end{llproof} 
      \item \textbf{Case} $Q = \extype{a:\tau} Q_0$\\
        \begin{llproof}
          \Pf{(x:P')}{\in}{\Gamma}{Above}
          \simplePf{\Theta}{P'}{By inversion on $\Gamma$ WF}
          \neqPf{P'}{\exists}{By inversion on extraction}
          \judgesubPf[+]{\Theta}{P'}{P'}{By \Lemmaref{lem:refl-sub}}
          \judgechkvalPf{\Dee' :: \Theta}{\Gamma}{x}{P'}{By \DeclChkValVar}
          \eqPf{\hgt{\Dee'}}{\hgt{\Dee}}{By \defn of $\hgt{-}$}
          \proofsep
          \judgesubPf[+]{\Theta}{P'}{P}{Subderivation}
          \judgesubPf[+]{\Dee :: \Theta}{P}{\extype{a:\tau} Q_0}{Rewrite given}
          \judgesubPf[+]{\Theta}{P'}{\extype{a:\tau} Q_0}{By \Lemmaref{lem:trans-sub}}
          \judgesubPf[+]{\Theta}{P'}{[t/a]Q_0}{By inversion on \DeclSubPosExR}
          \judgetermPf{\Theta}{t}{\tau}{\ditto}
          \Pf{\size{[t/a]Q_0}}{<}{\size{\extype{a:\tau} Q_0}}{By \defn of size}
          \judgechkvalPf{\Theta}{\Gamma'}{x}{[t/a]Q_0}{By \ih (same typing hgt., but smaller $Q$ size)}
          \Hand\judgechkvalPf{\Theta}{\Gamma'}{x}{\underbrace{\extype{a:\tau} Q_0}_Q}{By \DeclChkValExists}
        \end{llproof}
      \item \textbf{Case} $Q = Q_0 \land \phi$\\
        \begin{llproof}
          \judgechkvalPf{\Dee' :: \Theta}{\Gamma}{x}{P'}{See beginning of subcase $Q = \extype{a:\tau} Q_0$}
          \eqPf{\hgt{\Dee'}}{\hgt{\Dee}}{\ditto}
          \proofsep
          \judgesubPf[+]{\Theta}{P'}{P}{Subderivation}
          \judgesubPf[+]{\Theta}{P}{Q_0 \land \phi}{Rewrite given}
          \judgesubPf[+]{\Theta}{P'}{Q_0 \land \phi}{By \Lemmaref{lem:trans-sub}}
          \judgeentailPf{\Theta}{\phi}{By inversion on \DeclSubPosWithR}
          \judgesubPf[+]{\Theta}{P'}{Q_0}{\ditto}
          \judgechkvalPf{\Theta}{\Gamma'}{x}{Q_0}{By \ih (same typing height, but smaller $Q$ size)}
          \Hand\judgechkvalPf{\Theta}{\Gamma'}{x}{\underbrace{Q_0 \land \phi}_Q}{By \DeclChkValWith}
        \end{llproof}
      \end{itemize}

      \DerivationProofCase{\DeclSubPosWithR}
      {\judgesub[+]{\Theta}{P}{Q'} \\ \judgeentail{\Theta}{\phi} }
      {\judgesub[+]{\Theta}{P}{Q' \land \phi} }
      \begin{llproof}
        \judgesubPf[+]{\Theta}{\Gamma'}{\Gamma}{Given}
        \judgechkvalPf{\Theta}{\Gamma}{v}{P}{Given}
        \judgesubPf[+]{\Theta}{P}{Q'}{Subderivation}
        \judgechkvalPf{\Theta}{\Gamma'}{v}{Q'}{By \ih (same typing height, but smaller $Q$ size)}
        \judgeentailPf{\Theta}{\phi}{Subderivation}
        \judgechkvalPf{\Theta}{\Gamma'}{v}{Q' \land \phi}{By \DeclChkValWith}
      \end{llproof}

      \DerivationProofCase{\DeclSubPosExR}
      { 
        \judgesub[+]{\Theta}{P}{[t/a]Q'} 
        \\
        \judgeterm{\Theta}{t}{\tau} 
      }
      {\judgesub[+]{\Theta}{P}{\extype{a:\tau}{Q'}} }
      \begin{llproof}
        \judgesubPf[+]{\Theta}{\Gamma'}{\Gamma}{Given}
        \judgechkvalPf{\Theta}{\Gamma}{v}{P}{Given}
        \judgesubPf[+]{\Theta}{P}{[t/a]Q'}{Subderivation}
        \judgechkvalPf{\Theta}{\Gamma'}{v}{[t/a]Q'}{By \ih (same typing height, but smaller $Q$ size)}
        \judgetermPf{\Theta}{t}{\tau}{Subderivation}
        \judgechkvalPf{\Theta}{\Gamma'}{v}{\extype{a:\tau}{Q'}}{By \DeclChkValExists}
      \end{llproof}

      \DerivationProofCase{\DeclSubPosL}
      {
        \judgeextract[+]{\Theta}{P}{P'}{\Theta'}
        \\
        \Theta' \neq \cdot
        \\
        \arrayenvb{\E_1 \\ \judgesub[+]{\Theta, \Theta'}{P'}{Q}}
      }
      {
        \judgesub[+]{\E \derives ~~\Theta}{P}{Q}
      }
      \begin{llproof}
        \judgechkvalPf{\Dee \derives~~ \Theta}{\Gamma}{v}{P}{Given}
        \judgeextractPf[+]{\Theta}{P}{P'}{\Theta'}{Subderivation}
        \judgsubsPf{\Theta}{\sigma}{\Theta'}{By \Lemmaref{lem:value-instantiate}}
        \judgechkvalPf{\Dee' \derives~~ \Theta}{\Gamma}{v}{[\sigma]P'}{\ditto}
        \Pf{\hgt{\Dee'}}{\leq}{\hgt{\Dee}}{\ditto}
        \judgsubsPf{\Theta}{\id_\Theta}{\Theta}{By \Lemmaref{lem:id-subs-typing}}
        \judgsubsPf{\Theta}{\id_\Theta, \sigma}{\Theta, \Theta'}{By \Lemmaref{lem:id-prepend}}
        \judgesubPf[+]{\Theta, \Theta'}{P'}{Q}{Subderivation}
        \judgesubPf[+]{\E_1' \derives~~ \Theta}{[\id_\Theta, \sigma]P'}{Q}{By \Lemmaref{lem:syn-subs-sub}}
        \Pf{\hgt{\E_1'}}{\leq}{\hgt{\E_1}}{\ditto}
        \Pf{}{<}{\hgt{\E}}{By \defn of $\hgt{-}$}
        \judgesubPf[+]{\Theta}{[\sigma]P'}{Q}{By \Lemmaref{lem:id-subs-id}}
        \judgechkvalPf{\Theta}{\Gamma'}{v}{Q}{By \ih}
        \trailingjust{(same typing height; \dots}
        \trailingjust{\dots same $Q$ size; \dots}
        \trailingjust{\dots but smaller subtyping height)}
      \end{llproof} 
    \end{itemize}

    Next, we consider the remaining (i.e., non-variable) cases for the final rule
    of the derivation of the value typing judgement.
    Because we already covered the cases where the rule concluding
    the subtyping derivation has general $P$ on the left hand side,
    for cases
    \DeclChkValUnit, \DeclChkValFix, \DeclChkValDownshift, \DeclChkValPair,
    \DeclChkValIn{k},
    we need only consider one subcase (for the final rule of the subtyping derivation)
    in which $P$ has the specific form of the introduced type of the value typing case.

    \begin{itemize}
      \DerivationProofCase{\DeclChkValExists}
      {
        \judgechkval{\Theta}{\Gamma}{v}{[t/a]P'}
        \\
        \judgeterm{\Theta}{t}{\tau}
      }
      { \judgechkval{\Theta}{\Gamma}{v}{(\extype{a:\tau} P')} }
      The only three possible subcases for rules concluding
      $\judgesub[+]{\Theta}{\extype{a:\tau} P'}{Q}$
      (\ie, \DeclSubPosWithR, \DeclSubPosExR, and \DeclSubPosL)
      were already covered above.

      \DerivationProofCase{\DeclChkValWith}
      {
        \judgechkval{\Theta}{\Gamma}{v}{P'}
        \\
        \judgeentail{\Theta}{\phi}
      }
      { \judgechkval{\Theta}{\Gamma}{v}{P' \land \phi} }
      The only three possible subcases for rules concluding
      $\judgesub[+]{\Theta}{P' \land \phi}{Q}$
      (\ie, \DeclSubPosWithR, \DeclSubPosExR, and \DeclSubPosL)
      were already covered above.

      \DerivationProofCase{\DeclChkValFix}
      { \judgeunroll{\cdot}{\Theta}{\nu:F[\mu F]}{\alpha}{F\;\Fold{F}{\alpha}\;\nu}{t}{P'}{\tau}
        \\ 
        \judgechkval{\Theta}{\Gamma}{v}{P'} }
      { \judgechkval{\Theta}{\Gamma}{\roll{v}}
        {\comprehend{\nu:\mu F}{\Fold{F}{\alpha}\,{\nu} =_\tau t}} }
      \begin{itemize}
        \DerivationProofCase{\DeclSubPosFix}
        { \judgeequiv{\Theta}{F}{G} \\ \judgeentail{\Theta}{t = t'} }
        { \judgesub[+]{\Theta}{\comprehend{\nu:\mu F}{\Fold{F}{\alpha}\,\nu =_\tau t}}{\comprehend{\nu:\mu G}{\Fold{G}{\alpha}\,\nu =_\tau t'}} }
        \begin{llproof}
          \judgeunrollPf{\cdot}{\Theta}{\nu:F[\mu F]}{\alpha}{F\;\Fold{F}{\alpha}\;\nu}{t}{P'}{\tau}{Subderivation}
          \judgeequivPf{\Theta}{F}{G}{Subderivation}
          \judgeentailPf{\Theta}{t = t'}{Subderivation}
          \judgeunrollPf{\cdot}{\Theta}{\nu:G[\mu G]}{\alpha}{G\;\Fold{G}{\alpha}\;\nu}{t'}{Q'}{\tau}{By \Lemref{lem:unroll-to-mutual-sub}}
          \judgesubPf[+]{\Theta}{P'}{Q'}{\ditto}
          \judgechkvalPf{\Theta}{\Gamma}{v}{P'}{Subderivation}
          \judgesubPf[+]{\Theta}{\Gamma'}{\Gamma}{Given}
          \judgechkvalPf{\Theta}{\Gamma'}{v}{Q'}{By \ih (smaller typing hgt.)}
          \judgechkvalPf{\Theta}{\Gamma'}{\roll{v}}{\comprehend{\nu:\mu G}{\Fold{G}{\alpha}\,{\nu} =_\tau t'}}{By \DeclChkValFix}
        \end{llproof}
      \end{itemize}

      \DerivationProofCase{\DeclChkValDownshift}
      { \judgechkexp{\Theta}{\Gamma}{e}{N} }
      { \judgechkval{\Theta}{\Gamma}{\thunk{e}}{\downshift{N}} }
      \begin{itemize}
        \DerivationProofCase{\DeclSubPosDownshift}
        { \judgesub[-]{\Theta}{N}{M} }
        {\judgesub[+]{\Theta}{\downshift{N}}{\downshift{M}}}
        \begin{llproof}
          \judgesubPf[+]{\Theta}{\Gamma'}{\Gamma}{Given}
          \judgesubPf[-]{\Theta}{N}{M}{Subderivation}
          \judgechkexpPf{\Theta}{\Gamma}{e}{N}{Subderivation}
          \judgechkexpPf{\Theta}{\Gamma'}{e}{M}{By \ih (smaller typing hgt.)}
          \judgechkvalPf{\Theta}{\Gamma'}{\thunk{e}}{\downshift{M}}{By \DeclChkValDownshift}
        \end{llproof} 
      \end{itemize}

      \DerivationProofCase{\DeclChkValIn{k}}
      { \judgechkval{\Theta}{\Gamma}{v}{P_k} }
      { \judgechkval{\Theta}{\Gamma}{\inj{k}{v}}{P_1 + P_2} }
      \begin{itemize}
        \DerivationProofCase{\DeclSubPosSum}
        {
          \judgeequiv[+]{\Theta}{P_1}{Q_1}
          \\
          \judgeequiv[+]{\Theta}{P_2}{Q_2}
        }
        {
          \judgesub[+]{\Theta}{P_1 + P_2}{Q_1 + Q_2}
        }
        \begin{llproof}
          \judgesubPf[+]{\Theta}{\Gamma'}{\Gamma}{Given}
          \judgeequivPf[+]{\Theta}{P_k}{Q_k}{Subderivation}
          \judgesubPf[+]{\Theta}{P_k}{Q_k}{By \Lemmaref{lem:equiv-implies-sub}}
          \judgechkvalPf{\Theta}{\Gamma}{v}{P_k}{Subderivation}
          \judgechkvalPf{\Theta}{\Gamma'}{v}{Q_k}{By \ih (smaller typing hgt.)}
          \judgechkvalPf{\Theta}{\Gamma'}{\inj{k}{v}}{Q_1 + Q_2}{By \DeclChkValIn{k}}
        \end{llproof} 
      \end{itemize}

      \ProofCaseRule{\DeclChkValPair}
      Straightforward. Use \ih for each corresponding value typing and subtyping
      subderivation, then reapply \DeclChkValPair.

      \ProofCaseRule{\DeclChkValUnit}\\
      \begin{llproof}
        \Pf{Q}{=}{1}{By inversion on \DeclSubPosUnit}
        \judgechkvalPf{\Theta}{\Gamma'}{\unit}{1}{By \DeclChkValUnit}
      \end{llproof} 
    \end{itemize}

  \item %
    We first consider the
    three
    cases for the concluding rule
    of the expression typing derivation that are independent of
    the structure of the subtyping derivation,
    \ie, \DeclChkExpLet, \DeclChkExpMatch, and \DeclChkExpUnreachable.
    (After these
    three
    cases, we assume
    neither \DeclChkExpLet nor \DeclChkExpMatch
    nor \DeclChkExpUnreachable
    conclude $\Dee$.)
    Next, we consider the case of rule concluding the subtyping derivation
    that is independent of the program typing derivation,
    namely, \DeclSubNegR.
    Finally, we case analyze the remaining rules
    that can conclude the program typing derivation;
    each such case has exactly one corresponding subcase
    for the final rule of the subtyping derivation
    because the \DeclSubNegR subcase is already independently covered.
    \begin{itemize}
      \ProofCaseRule{\DeclChkExpUnreachable}
      By \DeclChkExpUnreachable.

      \DerivationProofCase{\DeclChkExpLet}
      { \simple{\Theta}{N} \\
        \judgesynexp{\Theta}{\Gamma}{\be}{\upshift{P}} \\
        \judgeextract[+]{\Theta}{P}{P'}{\Theta_P} \\
        \judgechkexp{\Theta, \Theta_P}{\Gamma, x:P'}{e'}{N} }
      { \Dee \derives~~ \judgechkexp{\Theta}{\Gamma}{\Let{x}{\be}{e'}}{N} }
      \begin{llproof}
        \judgesubPf[+]{\Theta}{\Gamma'}{\Gamma}{Given}
        \judgesynexpPf{\Theta}{\Gamma}{\be}{\upshift{P}}{Subderivation}
        \judgesynexpPf{\Theta}{\Gamma'}{\be}{\upshift{Q}}{By \ih, there exists such a $Q$}
        \judgesubPf[-]{\Theta}{\upshift{Q}}{\upshift{P}}{\ditto}
        \judgesubPf[+]{\Theta}{Q}{P}{By inversion on \DeclSubNegUpshift}
        \judgeextractPf[+]{\Theta}{Q}{Q'}{\Theta_Q}{By \Lemref{lem:extract-determinism}}
        \judgeextractPf{\Theta}{P}{P'}{\Theta_P}{Subderivation}
        \judgsubsPf{\Theta, \Theta_Q}{\sigma}{\Theta_P}{By \Lemref{lem:sub-instantiate}}
        \judgesubPf[+]{\Theta, \Theta_Q}{Q'}{[\sigma]P'}{\ditto}
        \judgesubPf[+]{\Theta, \Theta_Q}{\Gamma'}{\Gamma}{By \Lemref{lem:ix-level-weakening}}
        \judgesubPf[+]{\Theta, \Theta_Q}{\Gamma', x:Q'}{\Gamma, x:[\sigma]P'}{Add entry}
        \proofsep
        \judgechkexpPf{\Dee_2 \derives~~ \Theta, \Theta_P}{\Gamma, x:P'}{e'}{N}{Subderivation}
        \judgechkexpPf{\Dee_2' \derives~~ \Theta, \Theta_Q}{[\sigma](\Gamma, x:P')}{[\sigma]e'}{[\sigma]N}{By \Corref{cor:ix-syn-subs}}
        \Pf{\hgt{\Dee_2'}}{\leq}{\hgt{\Dee_2}}{\ditto}
        \Pf{}{<}{\hgt{\Dee}}{By \defn of $\hgt{-}$}
        \judgechkexpPf{\Dee_2' \derives~~ \Theta, \Theta_Q}{[\sigma]\Gamma, x:[\sigma]P'}{[\sigma]e'}{[\sigma]N}{By \defn of $[-]-$}
        \judgechkexpPf{\Dee_2' \derives~~ \Theta, \Theta_Q}{\Gamma, x:[\sigma]P'}{e'}{N}{$\dom{\Theta} \sect \dom{\Theta_P} = \emptyset$}
        \proofsep
        \judgesubPf[-]{\Theta}{N}{M}{Given}
        \judgeextractPf[-]{\Theta}{M}{M'}{\Theta_M}{By \Lemref{lem:extract-determinism}}
      \end{llproof}
      By \Lemmaref{lem:extract-disjunction},
      either
      (a) $\simple{\Theta}{M}$; or
      (b) $\Theta_M \neq \cdot$.
      \begin{itemize}
      \item \textbf{Case }(a)\\
        \begin{llproof}
          \judgechkexpPf{\Theta, \Theta_Q}{\Gamma', x:Q'}{e'}{M}{By \ih (smaller typing hgt.)}
          \judgechkexpPf{\Theta}{\Gamma'}{\Let{x}{\be}{e'}}{M}{By \DeclChkExpLet}
        \end{llproof}
      \item \textbf{Case }(b)\\
        \begin{llproof}
          \judgesubPf[-]{\Theta, \Theta_M}{N}{M'}{By \Lemref{lem:sub-extract-inversion}}
          \trailingjust{\dots and \Lemref{lem:extract-determinism}}
          \judgesubPf[-]{\Theta, \Theta_M, \Theta_Q}{N}{M'}{By \Lemref{lem:ix-level-weakening}}
          \judgechkexpPf{\Dee_2'' \derives~~ \Theta, \Theta_M, \Theta_Q}{\Gamma, x:[\sigma]P'}{e'}{N}{By \Lemref{lem:ix-level-weakening}}
          \Pf{\hgt{\Dee_2''}}{=}{\hgt{\Dee_2'}}{\ditto}
          \judgechkexpPf{\Theta, \Theta_M, \Theta_Q}{\Gamma', x:Q'}{e'}{M'}{By \ih (smaller typing hgt.)}
          \simplePf{\Theta}{M'}{By \Lemref{lem:extract-terminates}}
          \judgechkexpPf{\Theta, \Theta_M}{\Gamma'}{\Let{x}{\be}{e'}}{M'}{By \DeclChkExpLet}
          \judgechkexpPf{\Theta}{\Gamma'}{\Let{x}{\be}{e'}}{M}{By \DeclChkExpExtract}
        \end{llproof}
      \end{itemize}

      \DerivationProofCase{\DeclChkExpMatch}
      {
        \simple{\Theta}{N}
        \\
        \judgesynhead{\Theta}{\Gamma}{h}{P}
        \\
        \judgechkmatch{\Theta}{\Gamma}{P}{\clauses{\pa}{e}{i}{I}}{N}
      }
      {
        \judgechkexp{\Theta}{\Gamma}{\match{h}{\clauses{\pa}{e}{i}{I}}}{N}
      }
      \begin{llproof}
        \judgesubPf[+]{\Theta}{\Gamma'}{\Gamma}{Given}
        \judgesynheadPf{\Theta}{\Gamma}{h}{P}{Subderivation}
        \judgesynheadPf{\Theta}{\Gamma'}{h}{P'}{By \ih and \Lemmaref{lem:refl-sub}}
        \judgesubPf[+]{\Theta}{P'}{P}{\ditto}
        \judgechkmatchPf{\Dee_2 \derives \Theta}{\Gamma}{P}{\clauses{\pa}{e}{i}{I}}{N}{Subderivation}
        \judgesubPf[-]{\Theta}{N}{M}{Given}
        \judgeextractPf[-]{\Theta}{M}{M'}{\Theta_M}{By \Lemref{lem:extract-determinism}}
      \end{llproof}
      ~\\
      By \Lemmaref{lem:extract-disjunction},
      either
      (a) $\simple{\Theta}{M}$; or
      (b) $\Theta_M \neq \cdot$.
      \begin{itemize}
      \item \textbf{Case }(a)\\
        \begin{llproof}
          \judgechkmatchPf{\Theta}{\Gamma'}{P'}{\clauses{\pa}{e}{i}{I}}{M}{By \ih (smaller typing hgt.)}
          \judgechkexpPf{\Theta}{\Gamma'}{\match{h}{\clauses{\pa}{e}{i}{I}}}{M}{By \DeclChkExpMatch}
        \end{llproof}
      \item \textbf{Case }(b)\\
        \begin{llproof}
          \judgesubPf[-]{\Theta, \Theta_M}{N}{M'}{By \Lemref{lem:sub-extract-inversion}}
          \trailingjust{\dots and \Lemref{lem:extract-determinism}}
          \judgechkmatchPf{\Dee_2' \derives~~ \Theta, \Theta_M}{\Gamma}{P}{{\clauses{\pa}{e}{i}{I}}}{N}{By \Lemref{lem:ix-level-weakening}}
          \Pf{\hgt{\Dee_2'}}{=}{\hgt{\Dee_2}}{\ditto}
          \judgechkmatchPf{\Theta, \Theta_M}{\Gamma'}{P}{{\clauses{\pa}{e}{i}{I}}}{M'}{By \ih (smaller typing hgt.)}
          \simplePf{\Theta}{M'}{By \Lemref{lem:extract-terminates}}
          \judgechkexpPf{\Theta, \Theta_M}{\Gamma'}{\match{h}{\clauses{\pa}{e}{i}{I}}}{M'}{By \DeclChkExpMatch}
          \judgechkexpPf{\Theta}{\Gamma'}{\match{h}{\clauses{\pa}{e}{i}{I}}}{M}{By \DeclChkExpExtract}
        \end{llproof}
      \end{itemize}

      \DerivationProofCase{\DeclChkExpRec}
      {
        \arrayenvb{
          \simple{\Theta}{N}
          \\
          \judgesub[-]{\Theta}{\alltype{a:\kindnat} N'}{N}
        }
        \\
        \judgechkexp{\Theta, a:\kindnat}{\Gamma, x:\downshift{\alltype{a':\kindnat} a' < a \implies [a'/a]N'}}{e_0}{N'}
      }
      {
        \judgechkexp{\Theta}{\Gamma}{\rec{x : (\alltype{a:\kindnat} N')}{e_0}}{N}
      }
      \begin{llproof}
        \judgesubPf[+]{\Theta}{\Gamma'}{\Gamma}{Given}
        \decolumnizePf
        \judgesubPf[+]{\Theta, a:\kindnat}{\downshift{\alltype{a':\kindnat} a' < a \implies [a'/a]N'}}{\downshift{\alltype{a':\kindnat} a' < a \implies [a'/a]N'}}{By \Lemref{lem:refl-sub}}
        \decolumnizePf
        \judgesubPf[+]{\Theta, a:\kindnat}{\Gamma', x:\downshift{\alltype{a':\kindnat} a' < a \implies [a'/a]N'}}{\Gamma, x:\downshift{\alltype{a':\kindnat} a' < a \implies [a'/a]N'}}{Add entry}
        \decolumnizePf
        \judgesubPf[-]{\Theta, a:\kindnat}{N'}{N'}{By \Lemmaref{lem:refl-sub}}
        \decolumnizePf
        \judgechkexpPf{\Theta, a:\kindnat}{\Gamma, x:\downshift{\alltype{a':\kindnat} a' < a \implies [a'/a]N'}}{e_0}{N'}{Subderivation}
        \judgechkexpPf{\Theta, a:\kindnat}{\Gamma', x:\downshift{\alltype{a':\kindnat} a' < a \implies [a'/a]N'}}{e_0}{N'}{By \ih (smaller typing hgt.)}
        \decolumnizePf
        \judgesubPf[-]{\Theta}{\alltype{a:\kindnat} N'}{N}{Premise}
        \judgesubPf[-]{\Theta}{N}{M}{Given}
        \judgesubPf[-]{\Theta}{\alltype{a:\kindnat} N'}{M}{By \Lemmaref{lem:trans-sub}}
        \decolumnizePf
        \judgeextractPf[-]{\Theta}{M}{M'}{\Theta_M}{By \Lemref{lem:extract-determinism}}
      \end{llproof}
      ~\\
      By \Lemmaref{lem:extract-disjunction},
      either
      (a) $\simple{\Theta}{M}$; or
      (b) $\Theta_M \neq \cdot$.
      \begin{itemize}
      \item \textbf{Case }(a)\\
        \begin{llproof}
          \simplePf{\Theta}{M}{Current subcase}
          \judgesubPf[-]{\Theta}{\alltype{a:\kindnat} N'}{M}{Above}
          \judgechkexpPf{\Theta, a:\kindnat}{\Gamma', x:\downshift{\alltype{a':\kindnat} a' < a \implies [a'/a]N'}}{e_0}{N'}{Above}
          \decolumnizePf
          \judgechkexpPf{\Theta}{\Gamma'}{\rec{x : (\alltype{a:\kindnat} N')}{e_0}}{M}{By \DeclChkExpRec}
        \end{llproof} 
      \item \textbf{Case }(b)\\
        \begin{llproof}
          \judgesubPf[-]{\Theta}{\alltype{a:\kindnat} N'}{M}{Above}
          \judgesubPf[-]{\Theta, \Theta_M}{\alltype{a:\kindnat} N'}{M'}{By \Lemmaref{lem:sub-extract-inversion}}
          \trailingjust{\dots and \Lemmaref{lem:extract-determinism}}
          \simplePf{\Theta, \Theta_M}{M'}{By \Lemmaref{lem:extract-terminates}}
          \decolumnizePf
          \judgechkexpPf{\Theta, a:\kindnat}{\Gamma', x:\downshift{\alltype{a':\kindnat} a' < a \implies [a'/a]N'}}{e_0}{N'}{Above}
          \judgechkexpPf{\Theta, \Theta_M, a:\kindnat}{\Gamma', x:\downshift{\alltype{a':\kindnat} a' < a \implies [a'/a]N'}}{e_0}{N'}{By \Lemref{lem:ix-level-weakening}}
          \decolumnizePf
          \judgechkexpPf{\Theta, \Theta_M}{\Gamma'}{\rec{x : (\alltype{a:\kindnat} N')}{e_0}}{M'}{By \DeclChkExpRec}
          \judgeextractPf[-]{\Theta}{M}{M'}{\Theta_M}{Above}
          \judgechkexpPf{\Theta}{\Gamma'}{\rec{x : (\alltype{a:\kindnat} N')}{e_0}}{M}{By \DeclChkExpExtract}
        \end{llproof} 
      \end{itemize}

      \DerivationProofCase{\DeclSubNegR}
      {
        \judgeextract[-]{\Theta}{M}{M'}{\Theta'}
        \\
        \Theta' \neq \cdot
        \\
        \judgesub[+]{\Theta, \Theta'}{N}{M'}
      }
      {
        \judgesub[-]{\Theta}{N}{M}
      }
      \begin{llproof}
        \judgesubPf[+]{\Theta}{\Gamma'}{\Gamma}{Given}
        \judgesubPf[+]{\Theta, \Theta'}{\Gamma'}{\Gamma}{By \Lemmaref{lem:ix-level-weakening}}
        \judgechkexpPf{\Theta}{\Gamma}{e}{N}{Given}
        \judgechkexpPf{\Theta, \Theta'}{\Gamma}{e}{N}{By \Lemmaref{lem:ix-level-weakening}}
        \Pf{}{}{\text{(equal height)}}{\ditto}
        \judgesubPf[-]{\Theta, \Theta'}{N}{M'}{Subderivation}
        \judgechkexpPf{\Theta, \Theta'}{\Gamma'}{e}{M'}{By \ih}
        \trailingjust{(same typing height; \dots}
        \trailingjust{\dots same $N$ size; \dots}
        \trailingjust{\dots but smaller subtyping height)}
        \judgeextractPf[-]{\Theta}{M}{M'}{\Theta'}{Subderivation}
        \judgechkexpPf{\Theta}{\Gamma'}{e}{M}{By \DeclChkExpExtract}
      \end{llproof}

      \DerivationProofCase{\DeclChkExpLam}
      { \simple{\Theta}{P \to N_0} \\ \judgechkexp{\Theta}{\Gamma, x:P}{e_0}{N_0} }
      { \judgechkexp{\Theta}{\Gamma}{\fun{x}{e_0}}{P \to N_0} }
      \begin{itemize}
        \DerivationProofCase{\DeclSubNegArrow}
        { \judgesub[+]{\Theta}{Q}{P} \\ \judgesub[-]{\Theta}{N_0}{M_0} }
        { \judgesub[-]{\Theta}{P \to N_0}{Q \to M_0} }
        \begin{llproof}
          \judgeextractPf[+]{\Theta}{Q}{Q'}{\Theta_Q}{By \Lemref{lem:extract-determinism}}
          \judgeextractPf[-]{\Theta}{M_0}{M_0'}{\Theta_{M_0}}{By \Lemref{lem:extract-determinism}}
          \judgesubPf[+]{\Theta}{Q}{P}{Subderivation}
          \judgesubPf[+]{\Theta, \Theta_Q}{Q'}{P}{By \Lemref{lem:sub-extract-inversion}}
          \judgesubPf[+]{\Theta, \Theta_Q, \Theta_{M_0}}{Q'}{P}{By \Lemref{lem:ix-level-weakening}}
          \judgesubPf[+]{\Theta}{\Gamma'}{\Gamma}{Given}
          \judgesubPf[+]{\Theta, \Theta_Q, \Theta_{M_0}}{\Gamma'}{\Gamma}{By \Lemref{lem:ix-level-weakening}}
          \judgesubPf[+]{\Theta, \Theta_Q, \Theta_{M_0}}{\Gamma', x:Q'}{\Gamma, x:P}{Add entry}
          \proofsep
          \judgesubPf[-]{\Theta}{N_0}{M_0}{Subderivation}
          \judgesubPf[-]{\Theta, \Theta_{M_0}}{N_0}{M_0'}{By \Lemref{lem:sub-extract-inversion}}
          \judgesubPf[-]{\Theta, \Theta_Q, \Theta_{M_0}}{N_0}{M_0}{By \Lemref{lem:ix-level-weakening}}
          \proofsep
          \judgechkexpPf{\Dee_1 \derives \Theta}{\Gamma, x:P}{e_0}{N_0}{Subderivation}
          \judgechkexpPf{\Dee_1' \derives \Theta, \Theta_Q, \Theta_{M_0}}{\Gamma, x:P}{e_0}{N_0}{By \Lemref{lem:ix-level-weakening}}
          \Pf{\hgt{\Dee_1'}}{=}{\hgt{\Dee_1}}{\ditto}
          \Pf{}{<}{\Dee}{By \defn of $\hgt{-}$}
          \judgechkexpPf{\Theta, \Theta_Q, \Theta_{M_0}}{\Gamma', x:Q'}{e_0}{M_0'}{By \ih}
          \trailingjust{(smaller typing height)}
          \proofsep
          \judgeextractPf[-]{\Theta}{Q \to M_0}{Q' \to M_0'}{\Theta_Q, \Theta_{M_0}}{By \ExtractArrow}
        \end{llproof}
        ~\\
        By \Lemmaref{lem:extract-disjunction},
        either
        (a) $Q' \to M_0' = Q \to M_0$ and $(\Theta_Q, \Theta_{M_0}) = \cdot$; or
        (b) $(\Theta_Q, \Theta_{M_0}) \neq \cdot$.
        Similarly to cases \DeclChkExpLet and \DeclChkExpMatch,
        if (a), apply rule \DeclChkExpLam;
        if (b), first apply rule \DeclChkExpLam, and then rule \DeclChkExpExtract.
      \end{itemize}

      \DerivationProofCase{\DeclChkExpUpshift}
      { \judgechkval{\Theta}{\Gamma}{v}{P'} }
      { \judgechkexp{\Theta}{\Gamma}{\Return{v}}{\upshift{P'}} }
      \begin{itemize}
        \DerivationProofCase{\DeclSubNegUpshift}
        {
          \judgesub[+]{\Theta}{P'}{Q}
        }
        {\judgesub[-]{\Theta}{\upshift{P'}}{\upshift{Q}} }
        \begin{llproof}
          \judgesubPf[+]{\Theta}{\Gamma'}{\Gamma}{Given}
          \judgesubPf[+]{\Theta}{P'}{Q}{Subderivation}
          \judgechkvalPf{\Theta}{\Gamma}{v}{P'}{Subderivation}
          \judgechkvalPf{\Theta}{\Gamma'}{v}{Q}{By \ih (smaller typing height)}
          \judgechkexpPf{\Theta}{\Gamma'}{\Return{v}}{\upshift{Q}}{By \DeclChkExpUpshift}
        \end{llproof} 
      \end{itemize}

      \DerivationProofCase{\DeclChkExpExtract}
      {
        \judgeextract{\Theta}{N}{N'}{\Theta_N}
        \\
        \Theta_N \neq \cdot
        \\
        \judgechkexp{\Theta, \Theta_N}{\Gamma}{e}{N'}
      }
      {
        \Dee \derives \judgechkexp{\Theta}{\Gamma}{e}{N}
      }
      \begin{llproof}
        \judgesubPf[+]{\Theta}{\Gamma'}{\Gamma}{Given}
        \judgesubPf[+]{\Theta}{N}{M}{Given}
        \judgeextractPf{\Theta}{N}{N'}{\Theta_N}{Subderivation}
        \judgeextractPf{\Theta}{M}{M'}{\Theta_M}{By \Lemmaref{lem:extract-determinism}}
        \judgsubsPf{\Theta, \Theta_M}{\sigma}{\Theta_N}{By \Lemmaref{lem:sub-instantiate}}
        \judgesubPf[+]{\Theta, \Theta_M}{[\sigma]N'}{M'}{\ditto}
        \judgechkexpPf{\Dee_1 \derives \Theta, \Theta_N}{\Gamma}{e}{N'}{Subderivation}
        \judgechkexpPf{\Dee_1' \derives \Theta, \Theta_M}{[\sigma]\Gamma}{[\sigma]e}{[\sigma]N'}{By \Corollaryref{cor:ix-syn-subs}}
        \Pf{\hgt{\Dee_1'}}{\leq}{\hgt{\Dee_1}}{\ditto}
        \Pf{}{<}{\hgt{\Dee}}{By \defn of $\hgt{-}$}
        \Pf{\emptyset}{=}{\dom{\Theta} \sect \dom{\Theta_N}}{By \Lemmaref{lem:extract-to-ctx-wf}}
        \judgechkexpPf{\Dee_1' \derives \Theta, \Theta_M}{\Gamma}{e}{[\sigma]N'}{$\emptyset = \dom{\Theta} \sect \dom{\Theta_N}$}
        \judgechkexpPf{\Theta, \Theta_M}{\Gamma'}{e}{M'}{By \ih (smaller typing height)}
        \judgechkexpPf{\Theta}{\Gamma'}{e}{M}{By \DeclChkExpExtract}
      \end{llproof}
    \end{itemize}

  \item %
    We first consider the one case for the final rule of the
    positive subtyping derivation that is independent of the structure of $P$,
    \ie \DeclSubPosL.
    Then we consider cases for the final rule of the typing derivation;
    each such case has exactly one corresponding subcase
    for the final rule of the subtyping derivation that's
    determined by the structure of $P$ for the given case
    (the \DeclSubPosL case already being covered).
    \begin{itemize}
      \DerivationProofCase{\DeclSubPosL}
      {
        \judgeextract[+]{\Theta}{Q}{Q'}{\Theta_Q}
        \\
        \Theta' \neq \cdot
        \\
        \judgesub[+]{\Theta, \Theta_Q}{Q'}{P}
      }
      {
        \judgesub[+]{\Theta}{Q}{P}
      }
      \begin{llproof}
        \judgesubPf[+]{\Theta, \Theta_Q}{Q'}{P}{Subderivation}
        \judgesubPf[+]{\Theta}{\Gamma'}{\Gamma}{Given}
        \judgesubPf[+]{\Theta, \Theta_Q}{\Gamma'}{\Gamma}{By \Lemmaref{lem:ix-level-weakening}}
        \judgesubPf[-]{\Theta}{N}{M}{Given}
        \judgesubPf[-]{\Theta, \Theta_Q}{N}{M}{By \Lemmaref{lem:ix-level-weakening}}
        \judgechkmatchPf{\Theta}{\Gamma}{P}{\clauses{\pa}{e}{i}{I}}{N}{Given}
        \judgechkmatchPf{\Theta, \Theta_Q}{\Gamma}{P}{\clauses{\pa}{e}{i}{I}}{N}{By \Lemmaref{lem:ix-level-weakening}}
        \Pf{}{}{\text{(equal height)}}{\ditto}
        \judgechkmatchPf{\Theta, \Theta_Q}{\Gamma'}{Q'}{\clauses{\pa}{e}{i}{I}}{M}{By \ih}
        \trailingjust{(same typing height; \dots}
        \trailingjust{\dots same $P$ size; \dots}
        \trailingjust{\dots but smaller subtyping hgt.)}
        \judgeextractPf[+]{\Theta}{Q}{Q'}{\Theta_Q}{Subderivation}
        \judgechkmatchPf{\Theta}{\Gamma'}{Q}{\clauses{\pa}{e}{i}{I}}{M}{By \Lemref{lem:match-apply-ex-and-with}}
      \end{llproof} 

      \DerivationProofCase{\DeclChkMatchEx}
      {\judgechkmatch{\Theta, a:\tau}{\Gamma}{P'}{\clauses{\pa}{e}{i}{I}}{N}}
      {\judgechkmatch{\Theta}{\Gamma}{\extype{a:\tau}{P'}}{\clauses{\pa}{e}{i}{I}}{N}}
      \begin{itemize}
        \DerivationProofCase{\DeclSubPosExR}
        { 
          \judgesub[+]{\Theta}{Q}{[t/a]P'} 
          \\
          \judgeterm{\Theta}{t}{\tau} 
        }
        {\judgesub[+]{\Theta}{Q}{\extype{a:\tau}{P'}} }
        \begin{llproof}
          \judgsubsPf{\Theta;\cdot}{\idsubs{\Theta}{\cdot}}{\Theta;\cdot}{By \Lemmaref{lem:id-subs-typing}}
          \judgetermPf{\Theta}{\underbrace{t}_{[\cdot]t}}{\tau}{Subderivation}
          \judgsubsPf{\Theta;\cdot}{t/a}{a:\tau;\cdot}{By \IxSyn (after \EmptySyn)}
          \judgechkmatchPf{\Theta, a:\tau}{\Gamma}{P'}{\clauses{\pa}{e}{i}{I}}{N}{Subderivation}
          \judgechkmatchPf{\Theta}{[t/a]\Gamma}{[t/a]P'}{\clauses{\pa}{e}{i}{I}}{[t/a]N}{By \Corref{cor:ix-syn-subs}}
          \Pf{}{}{\text{(equal or lesser height)}}{\ditto}
          \judgechkmatchPf{\Theta}{\Gamma}{[t/a]P'}{\clauses{\pa}{e}{i}{I}}{N}{Variable $a$ not free in $\Gamma$ or $N$}
          \judgesubPf[+]{\Theta}{\Gamma'}{\Gamma}{Given}
          \judgesubPf[-]{\Theta}{N}{M}{Given}
          \judgesubPf[+]{\Theta}{Q}{[t/a]P'}{Subderivation}
          \judgechkmatchPf{\Theta}{\Gamma'}{Q}{\clauses{\pa}{e}{i}{I}}{M}{By \ih (smaller typing hgt.)}
        \end{llproof} 
      \end{itemize}

      \DerivationProofCase{\DeclChkMatchWith}
      {\judgechkmatch{\Theta, \phi}{\Gamma}{P'}{\clauses{\pa}{e}{i}{I}}{N}}
      {\judgechkmatch{\Theta}{\Gamma}{P' \land \phi}{\clauses{\pa}{e}{i}{I}}{N}}
      \begin{itemize}
        \DerivationProofCase{\DeclSubPosWithR}
        {\judgesub[+]{\Theta}{Q}{P'} \\ \judgeentail{\Theta}{\phi} }
        {\judgesub[+]{\Theta}{Q}{P' \land \phi} }
        \begin{llproof}
          \judgesubPf[+]{\Theta}{\Gamma'}{\Gamma}{Given}
          \judgesubPf[+]{\Theta,\phi}{\Gamma'}{\Gamma}{By \Lemmaref{lem:ix-level-weakening}}
          \judgesubPf[-]{\Theta}{N}{M}{Given}
          \judgesubPf[-]{\Theta,\phi}{N}{M}{By \Lemmaref{lem:ix-level-weakening}}
          \judgesubPf[+]{\Theta}{Q}{P'}{Subderivation}
          \judgesubPf[+]{\Theta,\phi}{Q}{P'}{By \Lemmaref{lem:ix-level-weakening}}
          \judgechkmatchPf{\Theta,\phi}{\Gamma}{P'}{\clauses{\pa}{e}{i}{I}}{N}{Subderivation}
          \judgechkmatchPf{\Theta,\phi}{\Gamma'}{Q}{\clauses{\pa}{e}{i}{I}}{M}{By \ih (smaller typing hgt.)}
          \judgeentailPf{\Theta}{\phi}{Subderivation}
          \judgechkmatchPf{\Theta}{\Gamma'}{Q}{\clauses{\pa}{e}{i}{I}}{M}{By \Lemmaref{lem:typing-consequence}}
        \end{llproof} 
      \end{itemize}

      \DerivationProofCase{\DeclChkMatchUnit}
      { \judgechkexp{\Theta}{\Gamma}{e}{N} }
      { \judgechkmatch{\Theta}{\Gamma}{1}{\setof{\clause{\unit}{e}}}{N} }
      \begin{itemize}
        \DerivationProofCase{\DeclSubPosUnit}
        { }
        {\judgesub[+]{\Theta}{1}{1}}
        \begin{llproof}
          \judgesubPf[+]{\Theta}{\Gamma'}{\Gamma}{Given}
          \judgesubPf[-]{\Theta}{N}{M}{Given}
          \judgechkexpPf{\Theta}{\Gamma}{e}{N}{Subderivation}
          \judgechkexpPf{\Theta}{\Gamma'}{e}{M}{By \ih (smaller typing hgt.)}
          \judgechkmatchPf{\Theta}{\Gamma'}{1}{\setof{\clause{\unit}{e}}}{M}{By \DeclChkMatchUnit}
        \end{llproof} 
      \end{itemize}

      \DerivationProofCase{\DeclChkMatchPair}
      { \arrayenvb{\judgeextract{\Theta}{P_1}{P_1'}{\Theta_{P_1}} \\
          \judgeextract{\Theta}{P_2}{P_2'}{\Theta_{P_2}}} \\
        \judgechkexp{\Theta, \Theta_{P_1}, \Theta_{P_2}}{\Gamma, x_1:P_1', x_2:P_2'}{e}{N} }
      {
        \judgechkmatch{\Theta}{\Gamma}{P_1 \times P_2}
        {\setof{\clause{\pair{x_1}{x_2}}{e}}}{N}
      }
      \begin{itemize}
        \DerivationProofCase{\DeclSubPosProd}
        {
          \judgesub[+]{\Theta}{Q_1}{P_1}
          \\
          \judgesub[+]{\Theta}{Q_2}{P_2}
        }
        {\judgesub[+]{\Theta}{Q_1 \times Q_2}{P_1 \times P_2}}
        \begin{llproof}
          \judgesubPf[+]{\Theta}{\Gamma'}{\Gamma}{Given}
          \judgeextractPf[+]{\Theta}{Q_1}{Q_1'}{\Theta_{Q_1}}{By \Lemref{lem:extract-determinism}}
          \judgeextractPf[+]{\Theta}{Q_2}{Q_2'}{\Theta_{Q_2}}{By \Lemref{lem:extract-determinism}}
          \judgesubPf[+]{\Theta, \Theta_{Q_1}, \Theta_{Q_2}}{\Gamma'}{\Gamma}{By (repeated) \Lemref{lem:ix-level-weakening}}
          \judgesubPf[+]{\Theta}{Q_1}{P_1}{Subderivation}
          \judgesubPf[+]{\Theta}{Q_2}{P_2}{Subderivation}
          \judgsubsPf{\Theta, \Theta_{Q_1}}{\sigma_1}{\Theta_{P_1}}{By \Lemref{lem:sub-instantiate}}
          \judgesubPf[+]{\Theta, \Theta_{Q_1}}{Q_1'}{[\sigma_1]P_1'}{\ditto}
          \judgsubsPf{\Theta, \Theta_{Q_2}}{\sigma_2}{\Theta_{P_2}}{By \Lemref{lem:sub-instantiate}}
          \judgesubPf[+]{\Theta, \Theta_{Q_2}}{Q_2'}{[\sigma_2]P_2'}{\ditto}
          \judgsubsPf{\Theta, \Theta_{Q_1}, \Theta_{Q_2}}{\sigma_1}{\Theta_{P_1}}{By \Lemref{lem:syn-subs-weakening}}
          \judgsubsPf{\Theta, \Theta_{Q_1}, \Theta_{Q_2}}{\sigma_2}{\Theta_{P_2}}{By \Lemref{lem:syn-subs-weakening}}
          \judgsubsPf{\Theta, \Theta_{Q_1}, \Theta_{Q_2}}{\sigma_1, \sigma_2}{\Theta_{P_1}, \Theta_{P_2}}{By \Lemref{lem:subs-append}}
          \judgesubPf[+]{\Theta, \Theta_{Q_1}, \Theta_{Q_2}}{Q_1'}{[\sigma_1]P_1'}{By \Lemref{lem:syn-subs-weakening}}
          \judgesubPf[+]{\Theta, \Theta_{Q_1}, \Theta_{Q_2}}{Q_2'}{[\sigma_2]P_2'}{By \Lemref{lem:syn-subs-weakening}}
        \end{llproof}
        ~\\
        By reasoning similar to that found in the \DeclChkValPair case
        of \Lemmaref{lem:value-instantiate},
        \[
          \judgesub[+]{\Theta, \Theta_{Q_1}, \Theta_{Q_2}}{Q_k'}{[\sigma_1, \sigma_2]P_k'}
        \]
        for $k=1,2$.\\
        \begin{llproof}
          \judgesubPf[+]{\Theta, \Theta_{Q_1}, \Theta_{Q_2}}{\Gamma',x_1:Q_1',x_2:Q_2'}{\Gamma,x_1:[\sigma_1, \sigma_2]P_1',x_2:[\sigma_1, \sigma_2]P_2'}{By above}
          \decolumnizePf
          \judgechkexpPf{\Theta, \Theta_{P_1}, \Theta_{P_2}}{\Gamma, x_1:P_1', x_2:P_2'}{e}{N}{Subderivation}
          \judgechkexpPf{\Theta, \Theta_{Q_1}, \Theta_{Q_2}}{[\sigma_1, \sigma_2](\Gamma, x_1:P_1', x_2:P_2')}{[\sigma_1, \sigma_2]e}{[\sigma_1, \sigma_2]N}{By \Corref{cor:ix-syn-subs}}
          \Pf{}{}{\text{(lesser or equal hgt.)}}{}
          \decolumnizePf
          \judgechkexpPf{\Theta, \Theta_{Q_1}, \Theta_{Q_2}}{\Gamma, x_1:[\sigma_1, \sigma_2]P_1', x_2:[\sigma_1, \sigma_2]P_2'}{e}{N}{By \Lemref{lem:extract-to-ctx-wf}}
          \trailingjust{(and \defn of subst.)}
          \judgesubPf[-]{\Theta}{N}{M}{Given}
          \judgesubPf[-]{\Theta, \Theta_{Q_1}, \Theta_{Q_2}}{N}{M}{By \Lemref{lem:ix-level-weakening}}
          \judgechkexpPf{\Theta, \Theta_{Q_1}, \Theta_{Q_2}}{\Gamma', x_1:Q_1', x_2:Q_2'}{e}{M}{By \ih}
          \trailingjust{(smaller typing hgt.)}
          \decolumnizePf
          \judgechkmatchPf{\Theta}{\Gamma'}{Q_1 \times Q_2}{\setof{\clause{\pair{x_1}{x_2}}{e}}}{M}{By \DeclChkMatchPair}
        \end{llproof}
      \end{itemize}

      \ProofCaseRule{\DeclChkMatchSum}
      Similar to case for rule \DeclChkMatchPair.

      \DerivationProofCase{\DeclChkMatchVoid}
      { }
      { \judgechkmatch{\Theta}{\Gamma}{0}{\setof{}}{N} }
      \begin{itemize}
        \DerivationProofCase{\DeclSubPosVoid}
        { }
        {\judgesub[+]{\Theta}{0}{0}}
        \begin{llproof}
          \judgechkmatchPf{\Theta}{\Gamma'}{0}{\setof{}}{M}{By \DeclChkMatchVoid}
        \end{llproof} 
      \end{itemize}

      \DerivationProofCase{\DeclChkMatchFix}
      {
        \arrayenvb{\judgeunroll{\cdot}{\Theta}{\nu:F[\mu F]}{\alpha}{F\;\Fold{F}{\alpha}\;\nu}{t}{P'}{\tau} \\ \judgeextract[+]{\Theta}{P'}{P''}{\Theta_{P'}}}
        \\
        \judgechkexp{\Theta, \Theta_{P'}}{\Gamma, x:P''}{e}{N}
      }
      {
        \judgechkmatch{\Theta}{\Gamma}{\comprehend{\nu:\mu F}
          {\Fold{F}{\alpha}\,{\nu} =_\tau t}}{\setof{\clause{\roll{x}}{e}}}{N}
      }
      \begin{itemize}
        \DerivationProofCase{\DeclSubPosFix}
        { \judgeequiv{\Theta}{G}{F} \\ \judgeentail{\Theta}{t' = t} }
        { \judgesub[+]{\Theta}{\comprehend{\nu:\mu G}{\Fold{G}{\alpha}\,\nu =_\tau t'}}{\comprehend{\nu:\mu F}{\Fold{F}{\alpha}\,\nu =_\tau t}} }
        \begin{llproof}
          \judgeunrollPf{\cdot}{\Theta}{\nu:F[\mu F]}{\alpha}{F\;\Fold{F}{\alpha}\;\nu}{t}{P'}{\tau}{Subderivation}
          \judgeentailPf{\Theta}{t' = t}{Subderivation}
          \judgeentailPf{\Theta}{t = t'}{By \Lemref{lem:equivassert}}
          \judgeequivPf{\Theta}{G}{F}{Subderivation}
          \judgeequivPf{\Theta}{F}{G}{By \Lemref{lem:symmetric-equiv-tp-fun}}
          \judgeunrollPf{\cdot}{\Theta}{\nu:G[\mu G]}{\alpha}{G\;\Fold{G}{\alpha}\;\nu}{t'}{Q'}{\tau}{By \Lemref{lem:unroll-to-mutual-sub}}
          \decolumnizePf
          \judgesubPf[+]{\Theta}{Q'}{P'}{\ditto}
          \judgeextractPf[+]{\Theta}{Q'}{Q''}{\Theta_{Q'}}{By \Lemref{lem:extract-determinism}}
          \judgsubsPf{\Theta, \Theta_{Q'}}{\sigma}{\Theta_{P'}}{By \Lemref{lem:sub-instantiate}}
          \judgesubPf[+]{\Theta, \Theta_{Q'}}{Q''}{[\sigma]P''}{\ditto}
          \proofsep
          \judgesubPf[+]{\Theta}{\Gamma'}{\Gamma}{Given}
          \judgesubPf[+]{\Theta, \Theta_{Q'}}{\Gamma'}{\Gamma}{By \Lemref{lem:ix-level-weakening}}
          \judgesubPf[+]{\Theta, \Theta_{Q'}}{\Gamma',x:Q''}{\Gamma,x:[\sigma]P''}{Add entry}
          \judgechkexpPf{\Theta, \Theta_{P'}}{\Gamma, x:P''}{e}{N}{Subderivation}
          \judgechkexpPf{\Theta, \Theta_{Q'}}{[\sigma](\Gamma, x:P'')}{[\sigma]e}{[\sigma]N}{By \Corref{cor:ix-syn-subs}}
          \Pf{}{}{\text{(less or equal hgt.)}}{}
          \judgechkexpPf{\Theta, \Theta_{Q'}}{\Gamma, x:[\sigma]P''}{e}{N}{By \Lemref{lem:extract-to-ctx-wf}}
          \trailingjust{(and \defn of subst.)}
          \judgesubPf[-]{\Theta}{N}{M}{Given}
          \judgesubPf[-]{\Theta, \Theta_{Q'}}{N}{M}{By \Lemref{lem:ix-level-weakening}}
          \judgechkexpPf{\Theta, \Theta_Q'}{\Gamma', x:Q''}{e}{M}{By \ih}
          \trailingjust{(smaller typing height)}
          \decolumnizePf
          \judgechkmatchPf{\Theta}{\Gamma'}{\comprehend{\nu:\mu G}                {\Fold{G}{\alpha}\,{\nu} =_\tau t'}}{\setof{\clause{\roll{x}}{e}}}{M}{By \DeclChkMatchFix}
        \end{llproof}
      \end{itemize}
    \end{itemize}

  \item %
    We case analyze the rule concluding the subtyping derivation.
    \begin{itemize}
      \DerivationProofCase{\DeclSubNegUpshift}
      {
        \judgesub[+]{\Theta}{P'}{P}
      }
      {\judgesub[-]{\Theta}{\upshift{P'}}{\upshift{P}} }
      \begin{itemize}
        \DerivationProofCase{\DeclSpineNil}
        { }
        {\judgespine{\Theta}{\Gamma}{\cdot}{\upshift{P}}{\upshift{P}}}
        \begin{llproof}
          \Hand\judgesubPf[-]{\Theta}{\upshift{P'}}{\upshift{P}}{Given}
          \judgetpPf{\Theta}{\upshift{P'}}{\Xi}{Presupposed derivation}
          \Hand\judgespinePf{\Theta}{\Gamma'}{\cdot}{\upshift{P'}}{\upshift{P'}}{By \DeclSpineNil}
        \end{llproof}
      \end{itemize}

      \DerivationProofCase{\DeclSubNegImpL}
      {\judgesub[-]{\Theta}{M'}{N} \\ \judgeentail{\Theta}{\phi} }
      {\judgesub[-]{\Theta}{\phi \implies M'}{N} }
      \begin{llproof}
        \judgespinePf{\Theta}{\Gamma}{s}{N}{\upshift{P}}{Given}
        \judgesubPf[+]{\Theta}{\Gamma'}{\Gamma}{Given}
        \judgesubPf[-]{\Theta}{M'}{N}{Subderivation}
        \judgespinePf{\Theta}{\Gamma'}{s}{M'}{\upshift{P'}}{By \ih (same typing height, but smaller $M$ size)}
        \Hand\judgesubPf[-]{\Theta}{\upshift{P'}}{\upshift{P}}{\ditto}
        \judgeentailPf{\Theta}{\phi}{Subderivation}
        \Hand\judgespinePf{\Theta}{\Gamma'}{s}{\phi \implies M'}{\upshift{P'}}{By \DeclSpineImplies}
      \end{llproof}

      \DerivationProofCase{\DeclSubNegAllL}
      {
        \judgesub[-]{\Theta}{[t/a]M'}{N} 
        \\ 
        \judgeterm{\Theta}{t}{\tau} 
      }
      {\judgesub[-]{\Theta}{\alltype{a:\tau}{M'}}{N} }
      \begin{llproof}
        \judgespinePf{\Theta}{\Gamma}{s}{N}{\upshift{P}}{Given}
        \judgesubPf[+]{\Theta}{\Gamma'}{\Gamma}{Given}
        \judgesubPf[-]{\Theta}{[t/a]M'}{N}{Subderivation}
        \judgespinePf{\Theta}{\Gamma'}{s}{[t/a]M'}{\upshift{P'}}{By \ih (same typing height, but smaller $M$ size)}
        \Hand\judgesubPf[-]{\Theta}{\upshift{P'}}{\upshift{P}}{\ditto}
        \judgetermPf{\Theta}{t}{\tau}{Subderivation}
        \Hand\judgespinePf{\Theta}{\Gamma'}{s}{\alltype{a:\tau}{M'}}{\upshift{P'}}{By \DeclSpineAll}
      \end{llproof}

      \DerivationProofCase{\DeclSubNegR}
      {
        \judgeextract[-]{\Theta}{N}{N'}{\Theta_N}
        \\
        \Theta_N \neq \cdot
        \\
        \E_0 \derives \judgesub[-]{\Theta, \Theta_N}{M}{N'}
      }
      {
        \E \derives \judgesub[-]{\Theta}{M}{N}
      }
      \begin{llproof}
        \judgespinePf{\Dee \derives \Theta}{\Gamma}{s}{N}{\upshift{P}}{Given}
        \judgeextractPf[-]{\Theta}{N}{N'}{\Theta_N}{Subderivation}
        \judgsubsPf{\Theta}{\sigma}{\Theta_N}{By \Lemmaref{lem:spine-instantiate}}
        \judgespinePf{\Dee' \derives \Theta}{\Gamma}{s}{[\sigma]N'}{\upshift{P}}{\ditto}
        \Pf{\hgt{\Dee'}}{\leq}{\hgt{\Dee}}{\ditto}
        \judgesubPf[-]{\E_0 \derives \Theta, \Theta_N}{M}{N'}{Subderivation}
        \judgesubPf[-]{\E_0' \derives \Theta}{[\sigma]M}{[\sigma]N'}{By \Lemmaref{lem:syn-subs-sub}}
        \Pf{\hgt{\E_0'}}{\leq}{\hgt{\E_0}}{\ditto}
        \Pf{}{<}{\hgt{\E}}{By \defn of $\hgt{-}$}
        \Pf{\emptyset}{=}{\dom{\Theta}\sect\dom{\Theta_N}}{By \Lemmaref{lem:extract-to-ctx-wf}}
        \judgesubPf[-]{\E_0' \derives \Theta}{M}{[\sigma]N'}{$\emptyset = \dom{\Theta}\sect\dom{\Theta_N}$}
        \judgespinePf{\Theta}{\Gamma'}{s}{M}{\upshift{P'}}{By \ih}
        \trailingjust{(same typing height; \dots}
        \trailingjust{\dots same $M$ size; \dots}
        \trailingjust{\dots but smaller subtyping height)}
        \judgesubPf[-]{\Theta}{\upshift{P'}}{\upshift{P}}{\ditto}
      \end{llproof}

      \DerivationProofCase{\DeclSubNegArrow}
      { \judgesub[+]{\Theta}{Q}{Q'} \\ \judgesub[-]{\Theta}{M'}{N'} }
      { \judgesub[-]{\Theta}{Q' \to M'}{Q \to N'} }  
      \begin{itemize}
        \DerivationProofCase{\DeclSpineApp}
        {\judgechkval{\Theta}{\Gamma}{v}{Q} \\ \judgespine{\Theta}{\Gamma}{s_0}{N'}{\upshift{P}} }
        {\judgespine{\Theta}{\Gamma}{v,s_0}{Q \to N'}{\upshift{P}}}
        \begin{llproof}
          \judgesubPf[+]{\Theta}{\Gamma'}{\Gamma}{Given}
          \judgechkvalPf{\Theta}{\Gamma}{v}{Q}{Subderivation}
          \judgesubPf[+]{\Theta}{Q}{Q'}{Subderivation}
          \judgechkvalPf{\Theta}{\Gamma'}{v}{Q'}{By \ih (smaller typing height)}
          \judgespinePf{\Theta}{\Gamma}{s_0}{N'}{\upshift{P}}{Subderivation}
          \judgesubPf[-]{\Theta}{M'}{N'}{Subderivation}
          \judgespinePf{\Theta}{\Gamma'}{s_0}{M'}{\upshift{P'}}{By \ih (smaller typing height)}
          \judgesubPf[-]{\Theta}{\upshift{P'}}{\upshift{P}}{\ditto}
          \judgespinePf{\Theta}{\Gamma'}{v,s_0}{Q' \to M'}{\upshift{P'}}{By \DeclSpineApp}
        \end{llproof}
        \qedhere
      \end{itemize}
    \end{itemize}
  \end{enumerate}
\end{proof}

\begin{lemma}[Syntactic Substitution]
  \label{lem:syn-subs}
  Assume $\Theta_0; \Gamma_0 |- \sigma : \Theta; \Gamma$. Then:
  \begin{enumerate}
  \item If $\judgesynhead{\Theta}{\Gamma}{h}{P}$, then either:
    \begin{enumerate}[(a)]
    \item
      $\judgesynhead{\Theta_0}{\Gamma_0}{[\sigma]h}{[\filterprog{\sigma}]P}$; or
    \item
      there exists $Q$
      such that $\judgeextract[+]{\Theta_0}{Q}{Q}{\cdot}$
      and $\judgesub[+]{\Theta_0}{Q}{[\filterprog{\sigma}]P}$ 
      and $\judgesynhead{\Theta_0}{\Gamma_0}{[\sigma]h}{Q}$.
    \end{enumerate}
  \item If $\judgesynexp{\Theta}{\Gamma}{\be}{\upshift{P}}$,
    then there exists $Q$ such that $\judgesub[-]{\Theta_0}{\upshift{Q}}{[\filterprog{\sigma}]\upshift{P}}$
    and $\judgesynexp{\Theta_0}{\Gamma_0}{[\sigma]\be}{\upshift{Q}}$.
  \item If $\judgechkval{\Theta}{\Gamma}{v}{P}$,
    then $\judgechkval{\Theta_0}{\Gamma_0}{[\sigma]v}{[\filterprog{\sigma}]P}$.
  \item If $\judgechkexp{\Theta}{\Gamma}{e}{N}$,
    then $\judgechkexp{\Theta_0}{\Gamma_0}{[\sigma]e}{[\filterprog{\sigma}]N}$.
  \item If $\judgechkmatch{\Theta}{\Gamma}{P}{\clauses{\pa}{e}{i}{I}}{N}$,
    then
    $\judgechkmatch{\Theta_0}{\Gamma_0}{[\filterprog{\sigma}]P}
    {[\sigma]\clauses{\pa}{e}{i}{I}}{[\filterprog{\sigma}]N}$.
  \item If $\judgespine{\Theta}{\Gamma}{s}{N}{\upshift{P}}$,
    then 
    $\judgespine{\Theta_0}{\Gamma_0}{[\sigma]s}{[\filterprog{\sigma}]N}{[\filterprog{\sigma}]\upshift{P}}$.
  \end{enumerate}
\end{lemma}
\begin{proof}
  By mutual induction on the structure of the program typing derivation.
  \begin{enumerate}
  \item
    \begin{itemize}
      \DerivationProofCase{\DeclSynHeadVar}
      {
        (x : P) \in \Gamma
      }
      {
        \judgesynhead{\Theta}{\Gamma}{x}{P}
      }
      \begin{llproof}
        \Pf{(x:P)}{\in}{\Gamma}{By inversion on \DeclSynHeadVar}
        \judgsubsPf{\Theta_0;\Gamma_0}{\sigma}{\Theta;\Gamma}{Given}
        \eqPf{\Theta}{\Theta_1, \Theta_2}{By inversion}
        \eqPf{\Gamma}{\Gamma_1, x:P, \Gamma_2}{\ditto}
        \eqPf{\sigma}{\sigma_1, \subs{v_1}{P}{x}, \sigma_2}{\ditto}
        \judgsubsPf{\Theta_0;\Gamma_0}{\sigma_1}{\Theta_1;\Gamma_1}{\ditto}
        \judgechkvalPf{\Theta_0}{\Gamma_0}{[\sigma_1]v_1}{[\filterprog{\sigma_1}]P}{\ditto}
        \judgsubsPf{\Theta_0;\Gamma_0}{\sigma_1, \subs{v_1}{P}{x}}{\Theta_1;\Gamma_1,x:P}{\ditto}
        \judgectxPf{\Theta_1}{\Gamma_1,x:P}{Presupposed derivation}
        \judgetpPf{\Theta_1}{P}{\dontcare}{By inversion}
        \eqPf{[\filterprog{\sigma_1}]P}{[\filterprog{\sigma}]P}{By inversion, $\FV{P} \cap \dom{\Theta_2} = \emptyset$}
      \end{llproof}
      \begin{itemize}
      \item \textbf{Case} $v_1 = x$:\\
        Because $\dom{\sigma_1} \subseteq \dom{\Theta_1} \cup \dom{\Gamma_1}$
        and $x \notin \dom{\Theta_1} \cup \dom{\Gamma_1}$,\\
        we have $[\sigma_1]v_1 = [\sigma_1]x = x$.\\
        \begin{llproof}
          \Pf{\judgechkval{\Theta_0}{\Gamma_0}{x}{[\filterprog{\sigma_1}]P}}{}{}{Rewrite above}
          \Pf{\judgesub{\Theta_0}{Q}{[\filterprog{\sigma_1}]P}}{}{}{By inversion on \DeclChkValVar, there is such a $Q$}
          \Pf{(x : Q) \in \Gamma_0}{}{}{\ditto}
          \Hand\Pf{\judgesub{\Theta_0}{Q}{[\filterprog{\sigma}]P}}{}{}{By equality}
          \Pf{\judgectx{\Theta_0}{\Gamma_0}}{}{}{Presupposed derivation}
          \Hand\Pf{\judgeextract{\Theta_0}{Q}{Q}{\cdot}}{}{}{By inversion on ctx.\ WF}
          \Pf{\judgesynhead{\Theta_0}{\Gamma_0}{x}{Q}}{}{}{By \DeclSynHeadVar}
          \Pf{\judgesynhead{\Theta_0}{\Gamma_0}{[\sigma_1]v_1}{Q}}{}{}{By equality}
          \Hand\Pf{\judgesynhead{\Theta_0}{\Gamma_0}{[\sigma]x}{Q}}{}{}{By \defn of \defsubst at heads ($v_1 = x$)}
        \end{llproof} 
      \item \textbf{Case} $v_1 \neq x$:\\
        \begin{llproof}
          \Pf{\judgetp{\Theta}{P}{\Xi}}{}{}{By \Lemmaref{lem:ix-level-weakening}}
          \Pf{\judgetp{\Theta_0}{[\filterprog{\sigma}]P}{\dontcare}}{}{}{By \Lemmaref{lem:syn-subs-tp-fun-alg}}
          \Pf{\judgechkval{\Theta_0}{\Gamma_0}{[\sigma_1]v_1}{[\filterprog{\sigma_1}]P}}{}{}{Above}
          \Pf{\judgesynhead{\Theta_0}{\Gamma_0}{([\sigma_1]v_1:[\sigma_1]P)}{[\filterprog{\sigma_1}]{P}}}{}{}{By \DeclSynValAnnot}
          \Pf{\judgesynhead{\Theta_0}{\Gamma_0}{[\sigma_1](v_1:P)}{[\filterprog{\sigma_1}]{P}}}{}{}{By \defn of \defsubst}
          \Pf{\judgesynhead{\Theta_0}{\Gamma_0}{[\sigma]x}{[\filterprog{\sigma_1}]{P}}}{}{}{By \defn of \defsubst ($x$ is a head and $v_1 \neq x$)}
          \Pf{\judgesynhead{\Theta_0}{\Gamma_0}{[\sigma]x}{[\filterprog{\sigma}]{P}}}{}{}{By equality}
        \end{llproof}
      \end{itemize}
      
      \DerivationProofCase{\DeclSynValAnnot}
      {
        \judgetp{\Theta}{P}{\Xi}
        \\
        \judgechkval{\Theta}{\Gamma}{v}{P}
      }
      {
        \judgesynhead{\Theta}{\Gamma}{\annoexp{v}{P}}{P}
      }
      \begin{llproof}
        \Pf{\judgetp{\Theta_0}{[\filterprog{\sigma}]P}{\dontcare}}{}{}{By \Lemmaref{lem:syn-subs-tp-fun-alg}}
        \Pf{\judgechkval{\Theta_0}{\Gamma_0}{[\sigma]v}{[\filterprog{\sigma}]P}}{}{}{By i.h.}
        \Pf{\judgesynhead{\Theta_0}{\Gamma_0}{\annoexp{[\sigma]v}{[\filterprog{\sigma}]P}}{[\filterprog{\sigma}]P}}{}{}{By \DeclSynValAnnot}
        \Hand\Pf{\judgesynhead{\Theta_0}{\Gamma_0}{[\sigma]\annoexp{v}{P}}{[\filterprog{\sigma}]P}}{}{}{By def. of subst.}
      \end{llproof} 

    \end{itemize}

  \item
    \begin{itemize}
      \DerivationProofCase{\DeclSynSpineApp}
      { \judgesynhead{\Theta}{\Gamma}{h}{\downshift{N}} \\
        \judgespine{\Theta}{\Gamma}{s}{N}{\upshift{P}} }
      { \judgesynexp{\Theta}{\Gamma}{h(s)}{\upshift{P}} }
      \begin{llproof}
        \Pf{\Theta_0;\Gamma_0|-\sigma:\Theta;\Gamma}{}{}{Given}
        \Pf{\judgespine{\Theta}{\Gamma}{s}{N}{\upshift{P}}}{}{}{Subderivation}
        \Pf{\judgespine{\Theta_0}{\Gamma_0}{[\filterprog{\sigma}]s}{[\sigma]N}{[\filterprog{\sigma}]\upshift{P}}}{}{}{By \ih}
        \Pf{\judgespine{\Theta_0}{\Gamma_0}{[\sigma]s}{[\filterprog{\sigma}]N}{\upshift{[\filterprog{\sigma}]P}}}{}{}{By \defn of subst.}
        \proofsep
        \Pf{\judgesynhead{\Theta}{\Gamma}{h}{\downshift{N}}}{}{}{Subderivation}
      \end{llproof}
     
      By \ih, either
      (a) $\judgesynhead{\Theta_0}{\Gamma_0}{[\sigma]h}{[\filterprog{\sigma}]\downshift{N}}$; or
      (b) there exists $Q$ such that\\
          $\judgeextract[+]{\Theta_0}{Q}{Q}{\cdot}$
          and $\judgesub[+]{\Theta_0}{Q}{[\filterprog{\sigma}]\downshift{N}}$
          and $\judgesynhead{\Theta_0}{\Gamma_0}{[\sigma]h}{Q}$.

      Consider subcases (a) and (b):

      \begin{itemize}
      \item \textbf{Case} (a):

        \begin{llproof}
          \Pf{\judgesynhead{\Theta_0}{\Gamma_0}{[\sigma]h}{\downshift{[\filterprog{\sigma}]N}}}{}{}{By \defn of subst.}
          \Pf{\judgesynexp{\Theta_0}{\Gamma_0}{([\sigma]h)([\sigma]s)}{\upshift{[\filterprog{\sigma}]P}}}{}{}{By \DeclSynSpineApp}
          \Hand\Pf{\judgesynexp{\Theta_0}{\Gamma_0}{[\sigma](h(s))}{[\filterprog{\sigma}]\upshift{P}}}{}{}{By \defn of subst.}
          \Hand\Pf{\judgesub{\Theta_0}{[\filterprog{\sigma}]\upshift{P}}{[\filterprog{\sigma}]\upshift{P}}}{}{}{By \Lemmaref{lem:refl-sub}}
        \end{llproof} 

      \item \textbf{Case} (b):
      \end{itemize}
      \begin{llproof}
        \Pf{\judgeextract[+]{\Theta_0}{Q}{Q}{\cdot}}{}{}{Current subcase}
        \Pf{\judgesub[+]{\Theta_0}{Q}{\downshift{[\filterprog{\sigma}]N}}}{}{}{By \defn of subst.}
        \Pf{Q = \downshift{M}}{}{}{By \Lemref{lem:extracted-subtype-of-downshift}}
        \Pf{\judgesub[+]{\Theta_0}{M}{[\filterprog{\sigma}]N}}{}{}{By inversion on \DeclSubPosDownshift}
        \Pf{\judgesub[+]{\Theta_0}{\Gamma_0}{\Gamma_0}}{}{}{By \Lemref{lem:refl-sub}}
        \Pf{\judgespine{\Theta_0}{\Gamma_0}{[\sigma]s}{M}{\upshift{P'}}}{}{}{By \Lemref{lem:subsumption-admissibility}, $\exists$ such a $P'$}
        \Hand\Pf{\judgesub[+]{\Theta_0}{\upshift{P'}}{\upshift{[\filterprog{\sigma}]P}}}{}{}{\ditto}
        \Hand\Pf{\judgesynexp{\Theta_0}{\Gamma_0}{[\sigma](h(s))}{\upshift{P'}}}{}{}{By \DeclSynSpineApp and \defn of subst.}
      \end{llproof}

      \DerivationProofCase{\DeclSynExpAnnot}
      {
        \judgetp{\Theta}{P}{\Xi}
        \\
        \judgechkexp{\Theta}{\Gamma}{e}{\upshift{P}}
      }
      { \judgesynexp{\Theta}{\Gamma}{\annoexp{e}{\upshift{P}}}{\upshift{P}} }
      \begin{llproof}
        \Pf{\Theta_0;\Gamma_0|-\sigma:\Theta;\Gamma}{}{}{Given}
        \Pf{\judgechkexp{\Theta}{\Gamma}{e}{\upshift{P}}}{}{}{Subderivation}
        \Pf{\judgechkexp{\Theta_0}{\Gamma_0}{[\sigma]e}{[\filterprog{\sigma}]\upshift{P}}}{}{}{By i.h.}
        \Hand\Pf{\judgesynexp{\Theta_0}{\Gamma_0}{[\sigma]\annoexp{e}{\upshift{P}}}{[\filterprog{\sigma}]\upshift{P}}}{}{}{By \DeclSynExpAnnot and \defn of subst.}
        \Hand\Pf{\judgesub{\Theta_0}{[\filterprog{\sigma}]\upshift{P}}{[\filterprog{\sigma}]\upshift{P}}}{}{}{By \Lemmaref{lem:refl-sub}}
      \end{llproof} 
    \end{itemize}

  \item
    \begin{itemize}
      \DerivationProofCase{\DeclChkValVar}
      { P \neq \exists, \land \\ (x:Q) \in \Gamma \\ \judgesub[+]{\Theta}{Q}{P} }
      { \judgechkval{\Theta}{\Gamma}{x}{P} }
      \begin{llproof}
        \inPf{(x:Q)}{\Gamma}{Subderivation}
        \judgsubsPf{\Theta_0;\Gamma_0}{\sigma}{\Theta;\Gamma}{Given}
        \eqPf{\Theta}{\Theta_1,\Theta_2}{By inversion}
        \eqPf{\Gamma}{\Gamma_1,x:Q,\Gamma_2}{\ditto}
        \eqPf{\sigma}{\sigma_1,\subs{v_1}{Q}{x},\sigma_2}{\ditto}
        \judgsubsPf{\Theta_0; \Gamma_0}{\sigma_1}{\Theta_1; \Gamma_1}{\ditto}
        \judgechkvalPf{\Theta_0}{\Gamma_0}{[\sigma_1]v_1}{[\filterprog{\sigma_1}]Q}{\ditto}
        \judgsubsPf{\Theta_0;\Gamma_0}{\sigma_1, \subs{v_1}{Q}{x}}{\Theta_1;\Gamma_1,x:Q}{\ditto}
        \judgectxPf{\Theta_1}{\Gamma_1, x:Q}{Presupposed derivation}
        \judgetpPf{\Theta_1}{Q}{\dontcare}{By inversion}
        \eqPf{[\filterprog{\sigma_1}]Q}{[\filterprog{\sigma}]Q}{By inversion, $\FV{Q} \cap \dom{\Theta_2} = \emptyset$}
        \judgechkvalPf{\Theta_0}{\Gamma_0}{[\sigma_1]v_1}{[\filterprog{\sigma}]Q}{By equality}
        \judgesubPf[+]{\Theta}{Q}{P}{Subderivation}
        \judgesubPf[+]{\Theta_0}{[\filterprog{\sigma}]Q}{[\filterprog{\sigma}]P}{By \Lemmaref{lem:syn-subs-sub}}
        \judgesubPf[+]{\Theta_0}{\Gamma_0}{\Gamma_0}{By \Lemmaref{lem:refl-sub}}
        \judgechkvalPf{\Theta_0}{\Gamma_0}{[\sigma_1]v_1}{[\filterprog{\sigma}]P}{By \Lemmaref{lem:subsumption-admissibility}}
        \judgechkvalPf{\Theta_0}{\Gamma_0}{[\sigma]x}{[\filterprog{\sigma}]P}{By \defn of \defsubst (here, $x$ is a value)}
      \end{llproof} 

      \item \textbf{Cases} \DeclChkValUnit, \DeclChkValPair, \DeclChkValIn{k}:
        Straightforward.

      \DerivationProofCase{\DeclChkValExists}
      {
        \judgechkval{\Theta}{\Gamma}{v}{[t/a]P'}
        \\
        \judgeterm{\Theta}{t}{\tau}
      }
      { \judgechkval{\Theta}{\Gamma}{v}{(\extype{a:\tau} P')} }
      \begin{llproof}
        \Pf{\judgechkval{\Theta_0}{\Gamma_0}{[\sigma]v}{[\filterprog{\sigma}]([t/a]P')}}{}{}{By i.h.}
        \Pf{\judgechkval{\Theta_0}{\Gamma_0}{[\sigma]v}{[[\filterprog{\sigma}]t/a]([\filterprog{\sigma}]P')}}{}{}{By \Lemmaref{lem:barendregt}}
        \Pf{\judgeterm{\Theta_0}{[\filterprog{\sigma}]t}{\tau}}{}{}{By \Lemmaref{lem:syn-subs-ix}}
        \Pf{\judgechkval{\Theta_0}{\Gamma_0}{[\sigma]v}{[\filterprog{\sigma}](\extype{a:\tau} P')}}{}{}{By \DeclChkValExists and def. of subst.}
      \end{llproof}

      \DerivationProofCase{\DeclChkValWith}
      {
        \judgechkval{\Theta}{\Gamma}{v}{P'}
        \\
        \judgeentail{\Theta}{\phi}
      }
      { \judgechkval{\Theta}{\Gamma}{v}{P' \land \phi} }
      \begin{llproof}
        \Pf{\Theta_0;\Gamma_0|-\sigma:\Theta;\Gamma}{}{}{Given}
        \Pf{\judgechkval{\Theta}{\Gamma}{v}{P'}}{}{}{Subderivation}
        \Pf{\judgechkval{\Theta_0}{\Gamma_0}{[\sigma]v}{[\filterprog{\sigma}]P'}}{}{}{By i.h.}
        \Pf{\judgeentail{\Theta}{\phi}}{}{}{Subderivation}
        \Pf{\judgeentail{\Theta_0}{[\filterprog{\sigma}]\phi}}{}{}{By \Lemmaref{lem:syn-subs-prop-true}}
        \Pf{\judgechkval{\Theta_0}{\Gamma_0}{[\sigma]v}{[\filterprog{\sigma}](P' \land \phi)}}{}{}{By \DeclChkValWith and def. of subst.}
      \end{llproof}

      \DerivationProofCase{\DeclChkValFix}
      { \judgeunroll{\cdot}{\Theta}{\nu:F[\mu F]}{\alpha}{F\;\Fold{F}{\alpha}\;\nu}{t}{P'}{\tau}
        \\ 
        \judgechkval{\Theta}{\Gamma}{v'}{P'} }
      { \judgechkval{\Theta}{\Gamma}{\roll{v'}}
        {\comprehend{\nu:\mu F}{\Fold{F}{\alpha}\,{\nu} =_\tau t}} }
      By \Lemmaref{lem:syn-subs-unroll},
      \[
        \judgeunroll{\cdot}{\Theta_0}{\nu:([\filterprog{\sigma}]F)[\mu [\filterprog{\sigma}]F]}{[\filterprog{\sigma}]\alpha}{([\filterprog{\sigma}]F)\;\Fold{[\filterprog{\sigma}]F}{[\filterprog{\sigma}]\alpha}\;\nu}{[\filterprog{\sigma}]t}{[\filterprog{\sigma}]P'}{\tau}
      \]
      By the induction hypothesis,
      \[
        \judgechkval{\Theta_0}{\Gamma_0}{[\sigma]v'}{[\filterprog{\sigma}]P'}
      \]
      By \DeclChkValFix and def. of subst.
      \[
        \judgechkval{\Theta_0}{\Gamma_0}{[\sigma]\roll{v'}}
        {[\filterprog{\sigma}]\comprehend{\nu:\mu F}{\Fold{F}{\alpha}\,{\nu} =_\tau t}}
      \]

      \ProofCaseRule{\DeclChkValDownshift}
      Straightforward.
    \end{itemize}

  \item
    \begin{itemize}
      \ProofCaseRule{\DeclChkExpUpshift}:
      Straightforward.

      \DerivationProofCase{\DeclChkExpLet}
      { \simple{\Theta}{N} \\
        \judgesynexp{\Theta}{\Gamma}{\be}{\upshift{P}} \\
        \judgeextract[+]{\Theta}{P}{P'}{\Theta_P} \\
        \judgechkexp{\Theta, \Theta_P}{\Gamma, x:P'}{e'}{N} }
      { \judgechkexp{\Theta}{\Gamma}{\Let{x}{\be}{e'}}{N} }
      \begin{llproof}
        \judgsubsPf{\Theta_0;\Gamma_0}{\sigma}{\Theta;\Gamma}{Given}
        \judgesynexpPf{\Theta}{\Gamma}{\be}{\upshift{P}}{Subderivation}
        \judgesynexpPf{\Theta_0}{\Gamma_0}{[\sigma]\be}{M}{By \ih}
        \judgesubPf[+]{\Theta_0}{M}{[\filterprog{\sigma}]\upshift{P}}{\ditto}
        \judgesubPf[+]{\Theta_0}{M}{\upshift{[\filterprog{\sigma}]P}}{By \defn of subst.}
        \eqPf{M}{\upshift{Q}}{By inversion on \DeclSubNegUpshift}
        \judgesubPf[+]{\Theta_0}{Q}{[\filterprog{\sigma}]P}{\ditto}
        \decolumnizePf
        \judgeextractPf[+]{\Theta}{P}{P'}{\Theta_P}{Subderivation}
        \judgeextractPf[+]{\Theta_0}{[\filterprog{\sigma}]P}{[\filterprog{\sigma}]P'}{[\filterprog{\sigma}]\Theta_P}{By \Lemref{lem:syn-subs-extract}}
        \judgsubsPf{\Theta_0,[\filterprog{\sigma}]\Theta_P;\Gamma_0}{\sigma,\id_{\Theta_P}}{\Theta,\Theta_P;\Gamma}{By \Lemref{lem:id-subs-ext-ix-level}}
        \judgsubsPf{\Theta_0,[\filterprog{\sigma}]\Theta_P;\Gamma_0,x:[\filterprog{\sigma}]P'}{\sigma,\id_{\Theta_P},\subs{x}{P'}{x}}{\Theta,\Theta_P;\Gamma,x:P'}{By \Lemref{lem:id-subs-ext-prog}}
        \trailingjust{and \defn of $\id_{\Theta_P}$}
        \decolumnizePf
        \judgechkexpPf{\Theta, \Theta_P}{\Gamma, x:P'}{e'}{N}{Subderivation}
        \judgechkexpPf{\Theta_0, [\filterprog{\sigma}]\Theta_P}{\Gamma_0, x:[\filterprog{\sigma}]P'}{[\sigma]e'}{[\filterprog{\sigma}]N}{By \ih}
        \trailingjust{(and \defn of id.\ subst.)}
        \proofsep
        \judgeextractPf[+]{\Theta_0}{Q}{Q'}{\Theta_Q}{By \Lemref{lem:extract-determinism}}
        \judgsubsPf{\Theta_0, \Theta_Q}{\sigma'}{[\filterprog{\sigma}]\Theta_P}{By \Lemref{lem:sub-instantiate}}
        \judgesubPf[+]{\Theta_0, \Theta_Q}{Q'}{[\sigma']([\filterprog{\sigma}]P')}{\ditto}
        \decolumnizePf
        \judgechkexpPf{\Theta_0, \Theta_Q}{[\sigma'](\Gamma_0, x:[\filterprog{\sigma}]P')}{[\sigma']([\filterprog{\sigma}]e')}{[\sigma']([\filterprog{\sigma}]N)}{By \Corref{cor:ix-syn-subs}}
        \judgechkexpPf{\Theta_0, \Theta_Q}{[\sigma']\Gamma_0, x:[\sigma']([\filterprog{\sigma}]P')}{[\sigma']([\filterprog{\sigma}]e')}{[\sigma']([\filterprog{\sigma}]N)}{By \defn of subst.}
        \judgctxPf{(\Theta_0, [\filterprog{\sigma}]\Theta_P)}{By \Lemref{lem:extract-to-ctx-wf}}
        \eqPf{\emptyset}{\dom{\Theta_0}\sect\dom{[\filterprog{\sigma}]\Theta_P}}{By inversion on ctx. WF}
        \judgechkexpPf{\Theta_0, \Theta_Q}{\Gamma_0, x:[\sigma']([\filterprog{\sigma}]P')}{[\filterprog{\sigma}]e'}{[\filterprog{\sigma}]N}{By line above}
        \decolumnizePf
        \Pf{\judgesub[+]{\Theta_0}{\Gamma_0}{\Gamma_0}}{}{}{By repeated \Lemref{lem:refl-sub}}
        \Pf{\judgesub[+]{\Theta_0, \Theta_Q}{\Gamma_0}{\Gamma_0}}{}{}{By \Lemref{lem:ix-level-weakening}}
        \Pf{\judgesub[+]{\Theta_0, \Theta_Q}{\Gamma_0,x:Q'}{\Gamma_0,x:[\sigma']([\filterprog{\sigma}]P')}}{}{}{Add entry}
        \Pf{\judgesub[-]{\Theta_0}{[\filterprog{\sigma}]N}{[\filterprog{\sigma}]N}}{}{}{By \Lemref{lem:refl-sub}}
        \Pf{\judgesub[-]{\Theta_0, \Theta_Q}{[\filterprog{\sigma}]N}{[\filterprog{\sigma}]N}}{}{}{By \Lemref{lem:ix-level-weakening}}
        \Pf{\judgechkexp{\Theta_0, \Theta_Q}{\Gamma_0, x:Q'}{[\sigma]e'}{[\filterprog{\sigma}]N}}{}{}{By \Lemref{lem:subsumption-admissibility}}
        \Pf{\simple{\Theta}{N}}{}{}{Subderivation}
        \Pf{\simple{\Theta_0}{[\filterprog{\sigma}]N}}{}{}{By \Lemref{lem:syn-subs-extract}}
        \Pf{\judgechkexp{\Theta_0}{\Gamma_0}{[\sigma](\Let{x}{\be}{e'})}{[\filterprog{\sigma}]N}}{}{}{By \DeclChkExpLet and def. of subst.}
      \end{llproof}

      \DerivationProofCase{\DeclChkExpMatch}
      {
        \simple{\Theta}{N}
        \\
        \judgesynhead{\Theta}{\Gamma}{h}{P}
        \\
        \judgechkmatch{\Theta}{\Gamma}{P}{\clauses{\pa}{e}{i}{I}}{N}
      }
      {
        \judgechkexp{\Theta}{\Gamma}{\match{h}{\clauses{\pa}{e}{i}{I}}}{N}
      }
      \begin{llproof}
        \Pf{\Theta_0;\Gamma_0|-\sigma:\Theta;\Gamma}{}{}{Given}
        \Pf{\simple{\Theta}{N}}{}{}{Subderivation}
        \Pf{\simple{\Theta_0}{[\filterprog{\sigma}]N}}{}{}{By \Lemref{lem:syn-subs-extract}}
        \Pf{\judgechkmatch{\Theta}{\Gamma}{P}{\clauses{\pa}{e}{i}{I}}{N}}{}{}{Subderivation}
        \Pf{\judgechkmatch{\Theta_0}{\Gamma_0}{[\filterprog{\sigma}]P}{[\sigma]\clauses{\pa}{e}{i}{I}}{[\filterprog{\sigma}]N}}{}{}{By i.h.}
        \proofsep
        \Pf{\judgesynhead{\Theta}{\Gamma}{h}{P}}{}{}{Subderivation}
      \end{llproof}

      By \ih, either
      (a) $\judgesynhead{\Theta_0}{\Gamma_0}{[\sigma]h}{[\filterprog{\sigma}]P}$; or
      (b) there exists $Q$ such that\\
      $\judgeextract[+]{\Theta_0}{Q}{Q}{\cdot}$
      and $\judgesub[+]{\Theta_0}{Q}{[\filterprog{\sigma}]P}$
      and $\judgesynhead{\Theta_0}{\Gamma_0}{[\sigma]h}{Q}$.

      Consider subcases (a) and (b):
      \begin{itemize}
      \item \textbf{Case} (a): Apply \DeclChkExpMatch, and use \defn of substitution.
      \item \textbf{Case} (b):
        
        \begin{llproof}
          \Pf{\judgesub[+]{\Theta_0}{\Gamma_0}{\Gamma_0}}{}{}{By \Lemref{lem:refl-sub}}
          \Pf{\judgesub[+]{\Theta_0}{[\filterprog{\sigma}]N}{[\filterprog{\sigma}]N}}{}{}{By Lemma~\ref{lem:refl-sub}}
          \Pf{\judgechkmatch{\Theta_0}{\Gamma_0}{Q}{[\sigma]\clauses{\pa}{e}{i}{I}}{[\filterprog{\sigma}]N}}{}{}{By Lemma~\ref{lem:subsumption-admissibility}}
          \Pf{\judgechkexp{\Theta_0}{\Gamma_0}{[\sigma](\match{h}{\clauses{\pa}{e}{i}{I}})}{[\filterprog{\sigma}]N}}{}{}{By \DeclChkExpMatch and def. of subst.}
        \end{llproof} 
      \end{itemize}

      \ProofCaseRule{\DeclChkExpLam}
      Similar to \DeclChkExpLet case, but simpler.

      \DerivationProofCase{\DeclChkExpRec}
      {
        \arrayenvb{
          \simple{\Theta}{N}
          \\
          \judgesub[-]{\Theta}{\alltype{a:\kindnat} M}{N}
        }
        \\
        \judgechkexp{\Theta, a:\kindnat}{\Gamma, x:\overbrace{\downshift{\alltype{a':\kindnat} a' < a \implies [a'/a]M}}^{Q}}{e_0}{M}
      }
      {
        \judgechkexp{\Theta}{\Gamma}{\rec{x : (\alltype{a:\kindnat} M)}{e_0}}{N}
      }
      \begin{llproof}
        \judgsubsPf{\Theta_0;\Gamma_0}{\sigma}{\Theta;\Gamma}{Given}
        \simplePf{\Theta}{N}{Premise}
        \simplePf{\Theta_0}{[\filterprog{\sigma}]N}{By \Lemmaref{lem:syn-subs-extract}}
        \proofsep
        \judgesubPf[-]{\Theta}{\alltype{a:\kindnat} M}{N}{Subderivation}
        \judgesubPf[-]{\Theta_0}{[\filterprog{\sigma}](\alltype{a:\kindnat} M)}{[\filterprog{\sigma}]N}{By \Lemmaref{lem:syn-subs-sub}}
        \judgesubPf[-]{\Theta_0}{\alltype{a:\kindnat} [\filterprog{\sigma}]M}{[\filterprog{\sigma}]N}{By \defn of \defsubst}
        \decolumnizePf
        \judgechkexpPf{\Theta, a:\kindnat}{\Gamma, x:\downshift{\alltype{a':\kindnat} a' < a \implies [a'/a]M}}{e_0}{M}{Subderivation}
        \decolumnizePf
        \judgsubsPf{\Theta_0,a:\kindnat;\Gamma_0}{\sigma,a/a}{\Theta,a:\kindnat;\Gamma}{By \Lemref{lem:id-subs-ext-ix}}
        \judgsubsPf{\Theta_0,a:\kindnat;\Gamma_0, x:[\filterprog{\sigma},a/a]Q}{\sigma,a/a,\subs{x}{Q}{x}}{\Theta,a:\kindnat;\Gamma, x:Q}{By \Lemref{lem:id-subs-ext-prog}}
        \decolumnizePf
        \judgsubsPf{\Theta_0,a:\kindnat;\Gamma_0, x:\downshift{\alltype{a':\kindnat} a' < a \implies [\filterprog{\sigma}]([a'/a]M)}}{\sigma,a/a,\subs{x}{Q}{x}}{\Theta,a:\kindnat;\Gamma, x:Q}{By \defn}
      \end{llproof}

      By properties of substitution,
      \[
        \Theta_0,a:\kindnat;\Gamma_0, x:\downshift{\alltype{a':\kindnat} a' < a \implies [a'/a]([\filterprog{\sigma}]M)} |- \sigma,a/a,\subs{x}{Q}{x} : \Theta,a:\kindnat;\Gamma, x:Q
      \]

      \begin{llproof}
        \judgechkexpPf{\Theta_0, a:\kindnat}{\Gamma_0, x:\downshift{\alltype{a':\kindnat} a' < a \implies [a'/a]([\filterprog{\sigma}]M)}}{[\sigma]e_0}{[\filterprog{\sigma}]M}{By \ih\dots}
        \trailingjust{\dots and id.\ subst.}
        \decolumnizePf
        \judgechkexpPf{\Theta_0}{\Gamma_0}{\rec{x : (\alltype{a:\kindnat} [\filterprog{\sigma}]M)}{[\sigma]e_0}}{[\filterprog{\sigma}]N}{By \DeclChkExpRec}
        \judgechkexpPf{\Theta_0}{\Gamma_0}{[\sigma](\rec{x : (\alltype{a:\kindnat} M)}{e_0})}{[\filterprog{\sigma}]N}{By \defn of \defsubst}
      \end{llproof} 

      \ProofCaseRule{\DeclChkExpExtract}
      Similar to \DeclChkExpLam case.

      \ProofCaseRule{\DeclChkExpUnreachable}
      Use \Lemmaref{lem:filter-out-prog-vars-syn} \Lemmaref{lem:subst-inconsistent}.
    \end{itemize}
  \item
    \begin{itemize}
      \DerivationProofCase{\DeclChkMatchEx}
      {\judgechkmatch{\Theta, a:\tau}{\Gamma}{P'}{\clauses{\pa}{e}{i}{I}}{N}}
      {\judgechkmatch{\Theta}{\Gamma}{\extype{a:\tau}{P'}}{\clauses{\pa}{e}{i}{I}}{N}}
      \begin{llproof}
        \Pf{\Theta_0;\Gamma_0 |- \sigma : \Theta;\Gamma}{}{}{Given}
        \Pf{\Theta_0,a:\tau;\Gamma_0 |- \sigma,a/a : \Theta,a:\tau;\Gamma}{}{}{By Lemma~\ref{lem:id-subs-ext-ix}}
        \Pf{\judgechkmatch{\Theta, a:\tau}{\Gamma}{P'}{\clauses{\pa}{e}{i}{I}}{N}}{}{}{Subderivation}
        \Pf{\judgechkmatch{\Theta_0, a:\tau}{\Gamma_0}{[\filterprog{\sigma},a/a]P'}{[\sigma,a/a]\clauses{\pa}{e}{i}{I}}{[\filterprog{\sigma},a/a]N}}{}{}{By i.h.}
        \Pf{\judgechkmatch{\Theta_0, a:\tau}{\Gamma_0}{[\filterprog{\sigma}]P'}{[\sigma]\clauses{\pa}{e}{i}{I}}{[\filterprog{\sigma}]N}}{}{}{Apply id.\ subst.}
        \Pf{\judgechkmatch{\Theta_0}{\Gamma_0}{[\filterprog{\sigma}]\extype{a:\tau}{P'}}{[\sigma]\clauses{\pa}{e}{i}{I}}{[\filterprog{\sigma}]N}}{}{}{By \DeclChkMatchEx \dots}
        \trailingjust{\dots and \defn of \defsubst}
      \end{llproof}

      \ProofCaseRule{\DeclChkMatchWith}
      Similar to case for \DeclChkMatchEx.

      \DerivationProofCase{\DeclChkMatchUnit}
      { \judgechkexp{\Theta}{\Gamma}{e}{N} }
      { \judgechkmatch{\Theta}{\Gamma}{1}{\setof{\clause{\unit}{e}}}{N} }
      \begin{llproof}
        \Pf{\Theta_0;\Gamma_0 |- \sigma : \Theta;\Gamma}{}{}{Given}
        \Pf{\judgechkexp{\Theta}{\Gamma}{e}{N}}{}{}{Subderivation}
        \Pf{\judgechkexp{\Theta_0}{\Gamma_0}{[\sigma]e}{[\filterprog{\sigma}]N}}{}{}{By i.h.}
        \Pf{\judgechkmatch{\Theta_0}{\Gamma_0}{1}{\setof{\clause{\unit}{[\sigma]e}}}{[\filterprog{\sigma}]N}}{}{}{By \DeclChkMatchUnit}
        \Pf{\judgechkmatch{\Theta_0}{\Gamma_0}{[\filterprog{\sigma}]1}{[\sigma]\setof{\clause{\unit}{e}}}{[\filterprog{\sigma}]N}}{}{}{By def. of subst.}
      \end{llproof}

      \DerivationProofCase{\DeclChkMatchPair}
      { \arrayenvb{\judgeextract{\Theta}{P_1}{P_1'}{\Theta_1} \\
          \judgeextract{\Theta}{P_2}{P_2'}{\Theta_2}} \\
        \judgechkexp{\Theta, \Theta_1, \Theta_2}{\Gamma, x_1:P_1', x_2:P_2'}{e}{N} }
      {
        \judgechkmatch{\Theta}{\Gamma}{P_1 \times P_2}
        {\setof{\clause{\pair{x_1}{x_2}}{e}}}{N}
      }
      \begin{llproof}
        \Pf{\Theta_0;\Gamma_0 |- \sigma : \Theta;\Gamma}{}{}{Given}
      \end{llproof}
      \\
      By \Lemmaref{lem:id-subs-ext-ix-level} (twice),
      \Lemmaref{lem:id-subs-id},
      \Lemmaref{lem:id-subs-ext-prog} (twice),
      and the fact that program variables $x$ cannot appear free in types,
      \begin{align*}
        \Theta_0, [\filterprog{\sigma}]\Theta_1, [\filterprog{\sigma}]\Theta_2;\Gamma_0,x_1:[\filterprog{\sigma}]P_1',x_2:[\filterprog{\sigma}]P_2'
        &|- \sigma,\id_{\Theta_1},\id_{\Theta_2},\subs{x_1}{P_1'}{x_1},\subs{x_2}{P_2'}{x_2} \\
        &: \Theta, \Theta_1, \Theta_2;\Gamma,x_1:P_1',x_2:P_2'
      \end{align*}
      \begin{llproof}
        \Pf{\sigma' = \sigma,\id_{\Theta_1},\id_{\Theta_2},\subs{x_1}{P_1'}{x_1},\subs{x_2}{P_2'}{x_2}}{}{}{Define}
        \Pf{\judgechkexp{\Theta, \Theta_1, \Theta_2}{\Gamma, x_1:P_1', x_2:P_2'}{e}{N}}{}{}{Subderivation}
        \Pf{\judgechkexp{\Theta_0, [\filterprog{\sigma}]\Theta_1, [\filterprog{\sigma}]\Theta_2}{\Gamma_0, x_1:[\filterprog{\sigma}]P_1', x_2:[\filterprog{\sigma}]P_2'}{[\sigma']e}{[\filterprog{\sigma'}]N}}{}{}{By \ih}
        \Pf{\judgechkexp{\Theta_0, [\filterprog{\sigma}]\Theta_1, [\filterprog{\sigma}]\Theta_2}{\Gamma_0, x_1:[\filterprog{\sigma}]P_1', x_2:[\filterprog{\sigma}]P_2'}{[\sigma]e}{[\filterprog{\sigma}]N}}{}{}{By \Lemref{lem:id-subs-id}}
        \Pf{\judgeextract{\Theta}{P_k}{P_k'}{\Theta_k}}{}{}{Subderivations}
        \Pf{\judgeextract{\Theta_0}{[\filterprog{\sigma}]P_k}{[\filterprog{\sigma}]P_k'}{[\filterprog{\sigma}]\Theta_k}}{}{}{By \Lemref{lem:syn-subs-extract}}
        \Pf{\judgechkmatch{\Theta_0}{\Gamma_0}{[\filterprog{\sigma}]P_1 \times [\filterprog{\sigma}]P_2}{\setof{\clause{\pair{x_1}{x_2}}{[\sigma]e}}}{[\filterprog{\sigma}]N}}{}{}{By \DeclChkMatchPair}
        \Pf{\judgechkmatch{\Theta_0}{\Gamma_0}{[\filterprog{\sigma}](P_1 \times P_2)}{[\sigma]\setof{\clause{\pair{x_1}{x_2}}{e}}}{[\filterprog{\sigma}]N}}{}{}{By def. of subst.}
      \end{llproof}

      \ProofCaseRule{\DeclChkMatchSum}
      Similar to case for \DeclChkMatchPair.

      \DerivationProofCase{\DeclChkMatchVoid}
      { }
      { \judgechkmatch{\Theta}{\Gamma}{0}{\setof{}}{N} }
      \begin{llproof}
        \Pf{\judgechkmatch{\Theta_0}{\Gamma_0}{0}{\setof{}}{[\filterprog{\sigma}]N}}{}{}{By \DeclChkMatchVoid}
        \Pf{\judgechkmatch{\Theta_0}{\Gamma_0}{[\filterprog{\sigma}]0}{[\sigma]\setof{}}{[\filterprog{\sigma}]N}}{}{}{By def. of subst.}
      \end{llproof}

      \DerivationProofCase{\DeclChkMatchFix}
      {
        \arrayenvb{\judgeunroll{\cdot}{\Theta}{\nu:F[\mu F]}{\alpha}{F\;\Fold{F}{\alpha}\;\nu}{t}{Q}{\tau} \\ \judgeextract[+]{\Theta}{Q}{Q'}{\Theta_Q}}
        \\
        \judgechkexp{\Theta, \Theta_Q}{\Gamma, x:Q'}{e}{N}
      }
      {
        \judgechkmatch{\Theta}{\Gamma}{\comprehend{\nu:\mu F}
          {\Fold{F}{\alpha}\,{\nu} =_\tau t}}{\setof{\clause{\roll{x}}{e}}}{N}
      }
        We are given $\Theta_0;\Gamma_0 |- \sigma : \Theta;\Gamma$.
        Consider the subderivation:
        \[
          \judgeunroll{\cdot}{\Theta}{\nu:F[\mu F]}{\alpha}{F\;\Fold{F}{\alpha}\;\nu}{t}{Q}{\tau}
        \]
        By \Lemmaref{lem:syn-subs-unroll},
        \[
          \judgeunroll{\cdot}{\Theta_0}{\nu:[\filterprog{\sigma}]F[\mu [\filterprog{\sigma}]F]}{[\filterprog{\sigma}]\alpha}{[\filterprog{\sigma}]F\;\Fold{[\filterprog{\sigma}]F}{[\filterprog{\sigma}]\alpha}\;\nu}{[\filterprog{\sigma}]t}{[\filterprog{\sigma}]Q}{\tau}
        \]
        By \Lemmaref{lem:id-subs-ext-ix-level},
        \Lemmaref{lem:id-subs-ext-prog},
        \Lemmaref{lem:id-subs-id},
        and the fact that program variables (such as $x$) cannot appear in types,
        \[
          \Theta_0, [\filterprog{\sigma}]\Theta_Q; \Gamma_0, x:[\filterprog{\sigma}]Q' |- \sigma, \id_{\Theta_Q}, \subs{x}{Q'}{x} : \Theta, \Theta_Q; \Gamma, x:Q'
        \]
        Consider the typing subderivation $\judgechkexp{\Theta, \Theta_Q}{\Gamma, x:Q'}{e}{N}$.
        By the induction hypothesis and \Lemref{lem:id-subs-id},
        \[
          \judgechkexp{\Theta_0, [\filterprog{\sigma}]Q}{\Gamma_0, x:[\filterprog{\sigma}]Q'}{[\sigma]e}{[\filterprog{\sigma}]N}
        \]
        By \Lemmaref{lem:syn-subs-extract} on the extraction subderivation,
        \[
          \judgeextract[+]{\Theta_0}{[\filterprog{\sigma}]Q}{[\filterprog{\sigma}]Q'}{[\filterprog{\sigma}]\Theta_Q}
        \]
        By \DeclChkMatchFix,
        \[
          \judgechkmatch{\Theta_0}{\Gamma_0}{\comprehend{\nu:\mu [\filterprog{\sigma}]F}{\Fold{[\filterprog{\sigma}]F}{[\filterprog{\sigma}]\alpha}\,{\nu} =_\tau [\filterprog{\sigma}]t}}{\setof{\clause{\roll{x}}{[\sigma]e}}}{[\filterprog{\sigma}]N}
        \]
        By definition of substitution,
        \[
          \judgechkmatch{\Theta_0}{\Gamma_0}{[\filterprog{\sigma}]\comprehend{\nu:\mu F}{\Fold{F}{\alpha}\,{\nu} =_\tau t}}{[\sigma]\setof{\clause{\roll{x}}{e}}}{[\filterprog{\sigma}]N}
        \]
      \end{itemize}
  \item
    \begin{itemize}
      \DerivationProofCase{\DeclSpineAll}
      {
        \judgeterm{\Theta}{t}{\tau}
        \\
        \judgespine{\Theta}{\Gamma}{s}{[t/a]N'}{\upshift{P}}
      }
      {\judgespine{\Theta}{\Gamma}{s}{(\alltype{a:\tau}{N'})}{\upshift{P}}}
      \begin{llproof}
        \Pf{\Theta_0;\Gamma_0 |- \sigma : \Theta;\Gamma}{}{}{Given}
        \Pf{\judgespine{\Theta}{\Gamma}{s}{[t/a]N'}{\upshift{P}}}{}{}{Subderivation}
        \Pf{\judgespine{\Theta_0}{\Gamma_0}{[\sigma]s}{[\filterprog{\sigma}]([t/a]N')}{[\filterprog{\sigma}]\upshift{P}}}{}{}{By i.h.}
        \Pf{\judgespine{\Theta_0}{\Gamma_0}{[\sigma]s}{[[\filterprog{\sigma}]t/a]([\filterprog{\sigma}]N')}{[\filterprog{\sigma}]\upshift{P}}}{}{}{By \Lemmaref{lem:barendregt}}
        \Pf{\judgeterm{\Theta}{t}{\tau}}{}{}{Subderivation}
        \Pf{\judgeterm{\Theta_0}{[\filterprog{\sigma}]t}{\tau}}{}{}{By Lemma~\ref{lem:syn-subs-ix}}
        \Pf{\judgespine{\Theta_0}{\Gamma_0}{[\sigma]s}{[\filterprog{\sigma}](\alltype{a:\tau}{N'})}{[\filterprog{\sigma}]\upshift{P}}}{}{}{By \DeclSpineAll and def. of subst.}
      \end{llproof}

      \DerivationProofCase{\DeclSpineImplies}
      {\judgeentail{\Theta}{\phi} \\ \judgespine{\Theta}{\Gamma}{s}{N'}{\upshift{P}}  }
      {\judgespine{\Theta}{\Gamma}{s}{(\phi \implies N')}{\upshift{P}}}
      \begin{llproof}
        \Pf{\judgespine{\Theta}{\Gamma}{s}{N'}{\upshift{P}}}{}{}{Subderivation}
        \Pf{\judgespine{\Theta_0}{\Gamma_0}{[\sigma]s}{[\filterprog{\sigma}]N'}{[\filterprog{\sigma}]\upshift{P}}}{}{}{By i.h.}
        \Pf{\judgeentail{\Theta}{\phi}}{}{}{Subderivation}
        \Pf{\judgeentail{\Theta_0}{[\filterprog{\sigma}]\phi}}{}{}{By \Lemmaref{lem:syn-subs-prop-true}}
        \Pf{\judgespine{\Theta_0}{\Gamma_0}{[\sigma]s}{[\filterprog{\sigma}](\phi \implies N')}{[\filterprog{\sigma}]\upshift{P}}}{}{}{By \DeclSpineImplies and def. of subst.}
      \end{llproof}

      \item \textbf{Cases} \DeclSpineApp and \DeclSpineNil:
        Straightforward.
        \qedhere
    \end{itemize}
  \end{enumerate}
\end{proof}

\begin{lemma}[Prog.\ Typing Resp.\ Equiv.]
  \label{lem:prog-typing-respects-equiv}
  Assume $\judgeequiv[+]{\Theta}{\Gamma'}{\Gamma}$.
  \begin{enumerate}
  \item If $\judgesynhead{\Theta}{\Gamma}{h}{P}$
    then there exists $Q$
    such that $\judgesynhead{\Theta}{\Gamma'}{h}{Q}$
    and $\judgeequiv[+]{\Theta}{P}{Q}$.
  \item If $\judgesynexp{\Theta}{\Gamma}{\be}{\upshift{P}}$
    then there exists $Q$
    such that $\judgesynexp{\Theta}{\Gamma'}{\be}{\upshift{Q}}$
    and $\judgeequiv[-]{\Theta}{\upshift{P}}{\upshift{Q}}$.
  \item If $\judgechkval{\Theta}{\Gamma}{v}{P}$
    and $\judgeequiv[+]{\Theta}{P}{Q}$,
    then $\judgechkval{\Theta}{\Gamma'}{v}{Q}$.
  \item If $\judgechkexp{\Theta}{\Gamma}{e}{N}$
    and $\judgeequiv[-]{\Theta}{N}{M}$,
    then $\judgechkexp{\Theta}{\Gamma'}{e}{M}$.
  \item If $\judgechkmatch{\Theta}{\Gamma}{P}{\clauses{\pa}{e}{i}{I}}{N}$
    and $\judgeequiv[+]{\Theta}{P}{Q}$
    and $\judgeequiv[-]{\Theta}{N}{M}$,\\
    then $\judgechkmatch{\Theta}{\Gamma'}{Q}{\clauses{\pa}{e}{i}{I}}{M}$.
  \item If $\judgespine{\Theta}{\Gamma}{s}{N}{\upshift{P}}$
    and $\judgeequiv[-]{\Theta}{N}{M}$,\\
    then there exists $Q$
    such that $\judgespine{\Theta}{\Gamma'}{s}{M}{\upshift{Q}}$
    and $\judgeequiv[-]{\Theta}{\upshift{P}}{\upshift{Q}}$.
  \end{enumerate}
\end{lemma}
\begin{proof}
  By mutual induction on the structure of the given program typing derivation.
  \begin{enumerate}
  \item
    \begin{itemize}
      \DerivationProofCase{\DeclSynHeadVar}
      {
        (x : P) \in \Gamma
      }
      {
        \judgesynhead{\Theta}{\Gamma}{x}{P}
      }
      \begin{llproof}
        \judgeequivPf[+]{\Theta}{\Gamma'}{\Gamma}{Given}
        \Pf{(x:P)}{\in}{\Gamma}{Subderivation}
        \Pf{(x:Q)}{\in}{\Gamma'}{By inversion}
        \judgeequivPf[+]{\Theta}{P}{Q}{\ditto}
        \judgesynheadPf{\Theta}{\Gamma'}{x}{Q}{By \DeclSynHeadVar}
      \end{llproof}

      \DerivationProofCase{\DeclSynValAnnot}
      {
        \judgetp{\Theta}{P}{\Xi}
        \\
        \judgechkval{\Theta}{\Gamma}{v}{P}
      }
      {
        \judgesynhead{\Theta}{\Gamma}{\annoexp{v}{P}}{P}
      }
      \begin{llproof}
        \judgeequivPf[+]{\Theta}{\Gamma'}{\Gamma}{Given}
        \judgeequivPf[+]{\Theta}{P}{P}{By \Lemmaref{lem:refl-equiv-tp-fun}}
        \judgechkvalPf{\Theta}{\Gamma}{v}{P}{Subderivation}
        \judgechkvalPf{\Theta}{\Gamma'}{v}{P}{By \ih}
        \judgesynheadPf{\Theta}{\Gamma'}{\annoexp{v}{P}}{P}{By \DeclSynValAnnot}
      \end{llproof} 
    \end{itemize}
    
  \item
    \begin{itemize}
      \DerivationProofCase{\DeclSynSpineApp}
      { \judgesynhead{\Theta}{\Gamma}{h}{\downshift{N}} \\
        \judgespine{\Theta}{\Gamma}{s}{N}{\upshift{P}} }
      { \judgesynexp{\Theta}{\Gamma}{h(s)}{\upshift{P}} }
      \begin{llproof}
        \judgeequivPf[+]{\Theta}{\Gamma}{\Gamma'}{Given}
        \judgesynheadPf{\Theta}{\Gamma}{h}{\downshift{N}}{Subderivation}
        \judgesynheadPf{\Theta}{\Gamma'}{h}{Q}{By \ih}
        \judgeequivPf[+]{\Theta}{\downshift{N}}{Q}{\ditto}
        \Pf{Q}{=}{\downshift{M}}{By inversion on \TpEquivPosDownshift}
        \judgeequivPf[-]{\Theta}{N}{M}{\ditto}
        \judgespinePf{\Theta}{\Gamma}{s}{N}{\upshift{P}}{Subderivation}
        \judgespinePf{\Theta}{\Gamma'}{s}{M}{\upshift{P'}}{By \ih}
        \judgeequivPf[-]{\Theta}{\upshift{P}}{\upshift{P'}}{\ditto}
        \judgesynexpPf{\Theta}{\Gamma'}{h(s)}{\upshift{P'}}{By \DeclSynSpineApp}
      \end{llproof}

      \DerivationProofCase{\DeclSynExpAnnot}
      {
        \judgetp{\Theta}{P}{\Xi}
        \\
        \judgechkexp{\Theta}{\Gamma}{e}{\upshift{P}}
      }
      { \judgesynexp{\Theta}{\Gamma}{\annoexp{e}{\upshift{P}}}{\upshift{P}} }
      \begin{llproof}
        \judgeequivPf[+]{\Theta}{\Gamma'}{\Gamma}{Given}
        \judgechkexpPf{\Theta}{\Gamma}{e}{\upshift{P}}{Subderivation}
        \judgeequivPf[+]{\Theta}{\upshift{P}}{\upshift{P}}{By \Lemmaref{lem:refl-equiv-tp-fun}}
        \judgechkexpPf{\Theta}{\Gamma'}{e}{\upshift{P}}{By \ih}
        \judgesynexpPf{\Theta}{\Gamma'}{\annoexp{e}{\upshift{P}}}{\upshift{P}}{By \DeclSynExpAnnot}
      \end{llproof}
    \end{itemize}

  \item Follows from \Lemmaref{lem:equiv-implies-sub}
    and \Lemmaref{lem:subsumption-admissibility}.

  \item Follows from \Lemmaref{lem:equiv-implies-sub}
    and \Lemmaref{lem:subsumption-admissibility}.

  \item Follows from \Lemmaref{lem:equiv-implies-sub}
    and \Lemmaref{lem:subsumption-admissibility}
    and \Lemmaref{lem:symmetric-equiv-tp-fun}.

  \item
    \begin{itemize}
      \DerivationProofCase{\DeclSpineNil}
      { }
      {\judgespine{\Theta}{\Gamma}{\cdot}{\upshift{P}}{\upshift{P}}}
      \begin{llproof}
        \judgeequivPf[-]{\Theta}{\upshift{P}}{M}{Given}
        \Pf{M}{=}{\upshift{Q}}{By inversion}
        \judgeequivPf[+]{\Theta}{P}{Q}{\ditto}
        \Hand\judgeequivPf[-]{\Theta}{\upshift{P}}{\upshift{Q}}{By \TpEquivNegUpshift}
        \Hand\judgespinePf{\Theta}{\Gamma'}{\cdot}{\upshift{Q}}{\upshift{Q}}{By \DeclSpineNil}
      \end{llproof} 

      \DerivationProofCase{\DeclSpineAll}
      {
        \judgeterm{\Theta}{t}{\tau}
        \\
        \judgespine{\Theta}{\Gamma}{s}{[t/a]N_0}{\upshift{P}}
      }
      {
        \judgespine{\Theta}{\Gamma}{s}{\alltype{a:\tau}{N_0}}{\upshift{P}}
      }
      \begin{llproof}
        \judgeequivPf[+]{\Theta}{\Gamma'}{\Gamma}{Given}
        \judgeequivPf[-]{\Theta}{\alltype{a:\tau}{N_0}}{M}{Given}
        \eqPf{M}{\alltype{a:\tau} M_0}{By inversion}
        \judgeequivPf[-]{\Theta, a:\tau}{N_0}{M_0}{\ditto}
        \judgetermPf{\Theta}{t}{\tau}{Premise}
        \judgeequivPf[-]{\Theta}{[t/a]N_0}{[t/a]M_0}{By \Lemmaref{lem:syn-subs-equiv-tp-fun}}
        \judgespinePf{\Theta}{\Gamma}{s}{[t/a]N_0}{\upshift{P}}{Subderivation}
        \judgespinePf{\Theta}{\Gamma'}{s}{[t/a]M_0}{\upshift{Q}}{By \ih}
        \Hand\judgeequivPf[-]{\Theta}{\upshift{P}}{\upshift{Q}}{\ditto}
        \Hand\judgespinePf{\Theta}{\Gamma'}{s}{M}{\upshift{Q}}{By \DeclSpineAll}
      \end{llproof} 

      \DerivationProofCase{\DeclSpineImplies}
      {
        \judgeentail{\Theta}{\phi}
        \\
        \judgespine{\Theta}{\Gamma}{s}{N_0}{\upshift{P}}
      }
      {
        \judgespine{\Theta}{\Gamma}{s}{\phi \implies N_0}{\upshift{P}}
      }
      \begin{llproof}
        \judgeequivPf[+]{\Theta}{\Gamma'}{\Gamma}{Given}
        \judgeequivPf[-]{\Theta}{\phi \implies N_0}{M}{Given}
        \eqPf{M}{\psi \implies M_0}{By inversion}
        \judgeequivPf{\Theta}{\phi}{\psi}{\ditto}
        \judgeequivPf[-]{\Theta}{N_0}{M_0}{\ditto}
        \judgespinePf{\Theta}{\Gamma}{s}{N_0}{\upshift{P}}{Subderivation}
        \judgespinePf{\Theta}{\Gamma'}{s}{M_0}{\upshift{Q}}{By \ih}
        \Hand\judgeequivPf[-]{\Theta}{\upshift{P}}{\upshift{Q}}{\ditto}
        \judgeentailPf{\Theta}{\phi}{Premise}
        \judgeentailPf{\Theta}{\psi}{By \Lemmaref{lem:equiv-respects-prop-truth}}
        \Hand\judgespinePf{\Theta}{\Gamma'}{s}{M}{\upshift{Q}}{By \DeclSpineImplies}
      \end{llproof} 

      \DerivationProofCase{\DeclSpineApp}
      {
        \judgechkval{\Theta}{\Gamma}{v}{Q}
        \\
        \judgespine{\Theta}{\Gamma}{s_0}{N_0}{\upshift{P}}
      }
      {
        \judgespine{\Theta}{\Gamma}{v,s_0}{Q \to N_0}{\upshift{P}}
      }
      \begin{llproof}
        \judgeequivPf[+]{\Theta}{\Gamma'}{\Gamma}{Given}
        \judgeequivPf[-]{\Theta}{Q \to N_0}{M}{Given}
        \eqPf{M}{Q' \to M_0}{By inversion}
        \judgeequivPf[-]{\Theta}{Q}{Q'}{\ditto}
        \judgeequivPf[-]{\Theta}{N_0}{M_0}{\ditto}
        \judgechkvalPf{\Theta}{\Gamma}{v}{Q}{Subderivation}
        \judgechkvalPf{\Theta}{\Gamma'}{v}{Q'}{By \ih}
        \judgespinePf{\Theta}{\Gamma}{s_0}{N_0}{\upshift{P}}{Subderivation}
        \judgespinePf{\Theta}{\Gamma'}{s_0}{M_0}{\upshift{P'}}{By \ih}
        \Hand\judgeequivPf[-]{\Theta}{\upshift{P}}{\upshift{P'}}{\ditto}
        \judgespinePf{\Theta}{\Gamma'}{s}{M}{\upshift{P'}}{By \DeclSpineApp}
      \end{llproof} 
      \qedhere
    \end{itemize}
  \end{enumerate}

\end{proof}

\section{Unrefined System and Erasure}

\begin{lemma}[Unrefined Syntactic Substitution]
  \label{lem:unref-syn-subs}
  Assume $\Gamma_0 |- \sigma : \Gamma$. Then:
  \begin{enumerate}
  \item If $\unrefsynhead{\Gamma}{h}{P}$,
    then $\unrefsynhead{\Gamma_0}{[\sigma]h}{P}$.
  \item If $\unrefsynexp{\Gamma}{\be}{\upshift{P}}$,
    then $\unrefsynexp{\Gamma_0}{[\sigma]\be}{\upshift{P}}$.
  \item If $\unrefchkval{\Gamma}{v}{P}$,
    then $\unrefchkval{\Gamma_0}{[\sigma]v}{P}$.
  \item If $\unrefchkexp{\Gamma}{e}{N}$,
    then $\unrefchkexp{\Gamma_0}{[\sigma]e}{N}$.
  \item If $\unrefchkmatch{\Gamma}{P}{\clauses{\pa}{e}{i}{I}}{N}$,
    then
    $\unrefchkmatch{\Gamma_0}{P}{[\sigma]\clauses{\pa}{e}{i}{I}}{N}$.
  \item If $\unrefspine{\Gamma}{s}{N}{\upshift{P}}$,
    then 
    $\unrefspine{\Gamma_0}{[\sigma]s}{N}{\upshift{P}}$.
  \end{enumerate}
\end{lemma}
\begin{proof}
  Similar to \Lemmaref{lem:syn-subs}, but simpler.
\end{proof}

\begin{lemma}[Functor Monotone]
  \label{lem:functor-monotone}
  If $\unreffunctor{F}$ and $X, Y \in \Set$ and $X \subseteq Y$,
  then $\sem{}{F} X \subseteq \sem{}{F} Y$
  (interpreting $F$ as a $\Set$ endofunctor).
\end{lemma}
\begin{proof}
  By structural induction on $F$.
\end{proof}

\begin{lemma}
  \label{lem:aux-repeated-monotonicity}
  If $\unreffunctor{F}$ and $n \in \kindnat$,
  then $\sem{}{F}^n \emptyset \subseteq \sem{}{F}^{n+1} \emptyset$
  (interpreting $F$ as a $\Set$ endofunctor).
\end{lemma}
\begin{proof}
  By induction on $n$, using \Lemmaref{lem:functor-monotone} in the inductive step.
\end{proof}

\begin{lemma}[Repeated Monotonicity]
  \label{lem:repeated-monotonicity}
  If $m, n \in \kindnat$ and $m \leq n$ and $\unreffunctor{F}$,\\
  then $\sem{}{F}^m \emptyset \subseteq \sem{}{F}^n \emptyset$
  (interpreting $F$ as a $\Set$ endofunctor).
\end{lemma}
\begin{proof}
  Follows from \Lemref{lem:aux-repeated-monotonicity}.
\end{proof}

\begin{lemma}[Mu Superset]
  \label{lem:mu-superset}
  If $\unreffunctor{F}$ and $n \in \kindnat$,
  then $\sem{}{F}^n \emptyset \subseteq \mu\sem{}{F}$.
\end{lemma}
\begin{proof}
  ~\\
  \begin{llproof}
    \subseteqPf{\sem{}{F}^n \emptyset}{\cup_{k \in \kindnat} \sem{}{F}^k \emptyset}{Set theory}
    \eqPf{}{\mu\sem{}{F}}{By \defn}
  \end{llproof} 
\end{proof}

\begin{lemma}[Reverse Mu Superset]
  \label{lem:reverse-mu-superset}
  If $\unreffunctor{\mathcal{F}}$ and $\unreffunctor{F}$
  and $V \in \sem{}{\mathcal{F}} (\mu \sem{}{F})$,\\
  then there exists $n \in \kindnat$
  such that $V \in \sem{}{\mathcal{F}} (\sem{}{F}^n \emptyset)$.
\end{lemma}
\begin{proof}
  ~
  \begin{itemize}
    \ProofCaseThing{\mathcal{F} = I}\\
    By \defn, $\sem{}{I} (\mu\sem{}{F}) = \one$, so $V = \bullet$,
    which is in $\one = \sem{}{I} (\sem{}{F}^n \emptyset)$ for all $n \in \kindnat$.

    \ProofCaseThing{\mathcal{F} = \Const{P}}\\
    By \defn, $\sem{}{\Const{P}} (\mu\sem{}{F}) = \sem{}{P}$, so $V \in \sem{}{P}$.
    But $\sem{}{P} = \sem{}{\Const{P}} (\sem{}{F}^n \emptyset)$
    for all $n \in \kindnat$.
   
    \ProofCaseThing{\mathcal{F} = \Id}\\
    By \defn, $\sem{}{\Id} (\mu\sem{}{F}) = \mu\sem{}{F}$, so $V \in \mu\sem{}{F}$.
    By \defn, $\mu\sem{}{F} = \cup_{k \in \kindnat} \sem{}{F}\emptyset$,
    so there exists $n \in \kindnat$
    such that $V \in \sem{}{F}^n \emptyset = \sem{}{\Id} (\sem{}{F}^n \emptyset)$.

    \ProofCaseThing{\mathcal{F} = B \otimes \hat{P}}\\
    By \defn, $\sem{}{B \otimes \hat{P}} (\mu\sem{}{F}) = \sem{}{B} (\mu\sem{}{F}) \times \sem{}{\hat{P}} (\mu\sem{}{F})$,
    so there exist $V_1 \in \sem{}{B} (\mu\sem{}{F})$
    and $V_2 \in \sem{}{\hat{P}} (\mu\sem{}{F})$
    such that $V = (V_1, V_2)$.
    By \ih, there exists $n_1$ such that $V_1 \in \sem{}{B} (\sem{}{F}^{n_1} \emptyset)$.
    By \ih, there exists $n_2$ such that $V_2 \in \sem{}{\hat{P}} (\sem{}{F}^{n_2} \emptyset)$.
    Let $n = \max\{n_1, n_2\}$.
    By \Lemmaref{lem:repeated-monotonicity},
    $V_1 \in \sem{}{B} (\sem{}{F}^n \emptyset)$
    and $V_2 \in \sem{}{\hat{P}} (\sem{}{F}^n \emptyset)$.
    Therefore,
    $V = (V_1, V_2) \in \sem{}{B} (\sem{}{F}^n \emptyset) \times \sem{}{\hat{P}} (\sem{}{F}^n \emptyset) = \sem{}{B \otimes \hat{P}} (\sem{}{F}^n \emptyset)$,
    as desired.

    \ProofCaseThing{\mathcal{F} = F_1 \oplus F_2}\\
    By \defn, $\sem{}{F_1 \oplus F_2} (\mu\sem{}{F}) = \sem{}{F_1} (\mu\sem{}{F}) \uplus \sem{}{F_2} (\mu\sem{}{F})$,
    so there exist $j \in \{1,2\}$ and $V' \in \sem{}{F_j} (\mu\sem{}{F})$
    such that $V = (j, V')$.
    By \ih, there exists $n$ such that $V' \in \sem{}{F_j} (\sem{}{F}^n \emptyset)$.
    Therefore $V = (j, V') \in \sem{}{F_1} (\sem{}{F}^n \emptyset) \uplus \sem{}{F_2} (\sem{}{F}^n \emptyset) = \sem{}{F_1 \oplus F_2} (\sem{}{F}^n \emptyset)$. \qedhere
  \end{itemize}
\end{proof}

\begin{lemma}[Mu is Fixed Point]
  \label{lem:unref-mu-unroll-equal}
  If $\unreffunctor{F}$, then $\mu \sem{}{F} = \sem{}{F} (\mu \sem{}{F})$.
\end{lemma}
\begin{proof}
  ~
  \begin{itemize}
    \item ``$\subseteq$''\\
      Suppose $V \in \mu \sem{}{F}$.
      Then there exists $n > 0$ such that $V \in \sem{}{F}^n \emptyset$.
      By \Lemmaref{lem:mu-superset}, $\sem{}{F}^{n-1} \emptyset \subseteq \mu\sem{}{F}$.
      Now,

      \begin{llproof}
        \eqPf{\sem{}{F}^n \emptyset}{\sem{}{F} (\sem{}{F}^{n-1} \emptyset)}{By \defn}
        \subseteqPf{}{\sem{}{F} (\mu \sem{}{F})}{By \Lemmaref{lem:functor-monotone}}
      \end{llproof} 
    \item ``$\supseteq$''\\
      Suppose $V \in \sem{}{F} \mu\sem{}{F}$.
      By \Lemmaref{lem:reverse-mu-superset},
      there exists $n \in \kindnat$
      such that $V \in \sem{}{F} (\sem{}{F}^n \emptyset)$.
      But

      \begin{llproof}
        \eqPf{\sem{}{F} (\sem{}{F}^n \emptyset)}{\sem{}{F}^{n+1} \emptyset}{By \defn}
        \subseteqPf{}{\mu \sem{}{F}}{By \Lemmaref{lem:mu-superset}}
      \end{llproof} 
      \qedhere
  \end{itemize}
\end{proof}

\begin{definition}[Predomain]
  \label{def:predomain}
  A \emph{complete partial order (cpo)} or \emph{predomain}
  is a poset $(D, \ordsym)$ that is chain-complete,
  \ie, such that every chain $d_0 \sqsubseteq d_1 \sqsubseteq \cdots$ in $D$
  has a least upper bound (\emph{lub}) $\sqcup_{k \in \kindnat} d_k \in D$.
  We often write $\ordsym[D]$ for the partial order of a predomain with set $D$.
\end{definition}

\begin{definition}[Domain]
  \label{def:domain}
  A \emph{complete pointed partial order (cppo)} or \emph{domain}
  is a triple $(D, \ordsym, \bott)$
  where $(D, \ordsym)$ is a predomain,
  $\bott \in D$,
  and $\bott \sqsubseteq d$ for all $d \in D$.
  We often write $\bott[D]$ for the bottom element $\bott$ of a domain with set $D$.
\end{definition}

\begin{definition}[Continuous]
  \label{def:continuous}
  A function $f : D \to E$ of predomains (or domains) is \emph{continuous}
  if $f$ is monotone
  (if $d \sqsubseteq_D d'$, then $f(d) \sqsubseteq_E f(d')$)
  and respects least upper bounds
  (if $d_0 \sqsubseteq d_1 \sqsubseteq \cdots$ is a chain in $D$,
  then $\sqcup_{k \in \kindnat} f(d_k) = f(\sqcup_{k \in \kindnat} d_k)$).
\end{definition}

\begin{lemma}[Predomains and Domains are Cats]
  \label{lem:predomains-and-domains-are-cats}
  ~
  \begin{enumerate}
  \item Predomains and continuous functions form a category $\Cpo$.
  \item Domains and continuous functions form a category $\Cppo$.
  \end{enumerate}
\end{lemma}
\begin{proof}
  The composition of continuous functions is continuous,
  the identity function is continuous,
  and function composition is associative.
\end{proof}

\begin{definition}[Cpo]
  \label{def:Cpo}
  Define the category of predomains $\Cpo$
  to have cpos as objects and continuous functions as morphisms.
\end{definition}

\begin{definition}[Cppo]
  \label{def:Cppo}
  Define the category of domains $\Cppo$
  to have cppos as objects and continuous functions as morphisms.
\end{definition}

\begin{lemma}[Predomain Constructions]
  \label{lem:predomain-constructions}
  ~
  \begin{enumerate}
  \item The empty set $\emptyset$ with empty ordering $\emptyset$ is a predomain.
  \item A singleton $\{d\}$ with ordering $d \sqsubseteq d$
    is a predomain.
  \item If $D$ and $E$ are predomains,
    then $D \times E$ with component-wise ordering $\ordsym[D \times E]$ is a predomain.
  \item If $D$ and $E$ are predomains,
    then $D \uplus E$ with injection-wise ordering $\ordsym[D \uplus E]$ is a predomain.
  \item If $D$ and $E$ are predomains,
    then $D \Rightarrow E = \comprehend{f : D \to E}{f\text{ is continuous}}$
    with pointwise ordering $\ordsym[D \Rightarrow E]$ is a predomain.
  \end{enumerate}
\end{lemma}
\begin{proof}
  Each part is straightforward.
\end{proof}

\begin{lemma}[Least Fixed Point]
  \label{lem:least-fixed-point}
  If $D \in \Cppo$ and $f : D \to D$ is continuous,
  then $f$ has a least fixed point $\sqcup_{k \in \kindnat} f^k(\bott[D])$ in $D$.
\end{lemma}
\begin{proof}
  See proof of Theorem 4.12 in the textbook of \citet{GunterTextbook}.
\end{proof}

\begin{lemma}
  \label{lem:aux-unref-type-denotations}
  If $\unreffunctor{\mathcal{F}}$ and $\unreffunctor{F}$
  and $\ord[\sem{}{\mathcal{F}} (\sem{}{F}^m \emptyset)]{d}{d'}$,
  then for all $n > m$,
  we have $\ord[\sem{}{\mathcal{F}} (\sem{}{F}^n \emptyset)]{d}{d'}$.
\end{lemma}
\begin{proof}
  By lexicographic induction on,
  first, $m$ and,
  second, the structure of $\mathcal{F}$.
\end{proof}

\begin{lemma}[Mu Chain-Complete]
  \label{lem:mu-chain-complete}
  If $\unreffunctor{\mathcal{F}}$ and $\unreffunctor{F}$\\
  and $\sem{}{\mathcal{F}} : \Cpo \to \Cpo$ and $\sem{}{F} : \Cpo \to \Cpo$\\
  and $d_0 \sqsubseteq_{\sem{}{\mathcal{F}} (\sem{}{F}^{n_0} \emptyset)} d_1 \sqsubseteq_{\sem{}{\mathcal{F}} (\sem{}{F}^{n_1} \emptyset)} \cdots d_k \sqsubset_{\sem{}{\mathcal{F}} (\sem{}{F}^{n_k} \emptyset)} \cdots$,\\
  then there exists a least upper bound $\sqcup_{k \in \kindnat} d_k$
  in $\sem{}{\mathcal{F}}(\mu\sem{}{F})$.
\end{lemma}
\begin{proof}
  By lexicographic induction on,
  first, $\min\comprehend{n_k}{k \in \kindnat}$ and,
  second, the structure of $\mathcal{F}$.
  
  For all $k$, let $D_k = \sem{}{F}^{n_k} \emptyset$.
  \begin{itemize}
    \ProofCaseThing{\mathcal{F} = I}\\
    By \defn, for all $k \in \kindnat$, we have $\sem{}{I} D_k = \one$;
    so $d_k = \bullet$ for all $k$.
    then $\sqcup_k d_k = \bullet$.
    
    \ProofCaseThing{\mathcal{F} = \Const{Q}}\\
    For all $k$, we have $\sem{}{\Const{Q}} D_k = \sem{}{Q}$;
    so $d_k \in \sem{}{Q}$ for all $k$.
    By \Lemmaref{lem:predomain-constructions}, $\emptyset \in \Cpo$.
    Therefore, by \defn of \defsem and because $\sem{}{\mathcal{F}} : \Cpo \to \Cpo$,
    we have $\sem{}{Q} = \sem{}{\Const{Q}} \emptyset \in \Cpo$.
    Therefore, the chain $d_0 \sqsubseteq \cdots$
    has a least upper bound $\sqcup_k d_k$
    in $\sem{}{Q} = \sem{}{\Const{Q}}(\mu\sem{}{F})$.

    \ProofCaseThing{\mathcal{F} = \Id}\\
    By \defn, for all $k$, we have
    $\sem{}{\Id} D_k = D_k = \sem{}{F}^{n_k} \emptyset = \sem{}{F} (\sem{}{F}^{n_k - 1} \emptyset)$.
    For all $k$, we have $n_k > 0$ (otherwise $d_k$ would not exist),
    so $n_k - 1 \in \kindnat$ for all $k$.
    Now, $\min\comprehend{n_k - 1}{k \in \kindnat} < \min\comprehend{n_k}{k \in \kindnat}$,
    and we are given $\sem{}{F} : \Cpo \to \Cpo$,
    so by the \ih, the chain $d_0 \sqsubseteq \cdots$
    has a least upper bound $\sqcup_k d_k$ in $\sem{}{F} (\mu \sem{}{F})$;
    and the latter equals $\mu\sem{}{F}$ by \Lemmaref{lem:unref-mu-unroll-equal},
    which equals $\sem{}{\Id} (\mu\sem{}{F})$ by definition.

    \ProofCaseThing{\mathcal{F} = B \otimes \hat{P}}\\
    By \defn, for all $k$, we have
    $\sem{}{B \otimes \hat{P}} D_k = \sem{}{B} D_k \times \sem{}{\hat{P}} D_k$;
    so for all $k$,
    there exist $d_{k1} \in \sem{}{B} D_k$ and $d_{k2} \in \sem{}{\hat{P}} D_k$
    such that $d_k = (d_{k1}, d_{k2})$.
    By \ih, chain $d_{01} \sqsubseteq \cdots$ has a least upper bound $\sqcup_k d_{k1}$
    in $\sem{}{B} (\mu\sem{}{F})$.
    By \ih, chain $d_{02} \sqsubseteq \cdots$ has a least upper bound $\sqcup_k d_{k2}$
    in $\sem{}{\hat{P}} (\mu\sem{}{F})$.
    Then $\sqcup_k d_k = (\sqcup_k d_{k1}, \sqcup_k d_{k2})$
    is the least upper bound for chain $d_k \sqsubseteq \cdots$ in
    $\sem{}{B} (\mu\sem{}{F}) \times \sem{}{\hat{P}} (\mu\sem{}{F}) = \sem{}{B\otimes\hat{P}}(\mu\sem{}{F})$.

    \ProofCaseThing{\mathcal{F} = F_1 \oplus F_2}\\
    By \defn, for all $k$, we have
    $\sem{}{F_1 \oplus F_2} D_k = \sem{}{F_1} D_k \uplus \sem{}{F_2} D_k$;
    so for all $k$, there exists $d_{k}' \in \sem{}{F_j} D_k$
    such that $d_k = (j, d_{k}')$.
    By \ih, the chain $d_{0}' \sqsubseteq \cdots$ has a least upper bound $\sqcup_k d_{k}'$
    in $\sem{}{F_j} (\mu\sem{}{F})$.
    Then $\sqcup_k d_k = (j, \sqcup_k d_{k}')$
    is the least upper bound for chain $d_k \sqsubseteq \cdots$
    in $\sem{}{F_1} (\mu\sem{}{F}) \uplus \sem{}{F_2} (\mu\sem{}{F}) = \sem{}{F_1 \oplus F_2} (\mu\sem{}{F})$. \qedhere
  \end{itemize} 
\end{proof}

\begin{lemma}[Repeated Cpo Functor]
  \label{lem:repeated-cpo-functor}
  If $F : \Cpo \to \Cpo$ and $X \in \Cpo$,
  then for all $k \in \kindnat$, we have $F^k(X) \in \Cpo$.
\end{lemma}
\begin{proof}
  By induction on $k$.
\end{proof}

\begin{lemma}[Unref.\ Type Denotations]
  \label{lem:unref-type-denotations}
  ~
  \begin{enumerate}
  \item If $\unreftp{P}$, then $\sem{}{P} \in \Cpo$.
  \item If $\unreftp{N}$, then $\sem{}{N} \in \Cppo$.
  \item If $\unreffunctor{\mathcal{F}}$,
    then $\sem{}{\mathcal{F}}$ is a $\Cpo$ endofunctor.
  \end{enumerate}
\end{lemma}
\begin{proof}
  By mutual induction on the structure of $P$, $N$ or $\mathcal{F}$.
  \begin{enumerate}
  \item
    \begin{itemize}
      \ProofCaseThing{P = 0}\\
      \begin{llproof}
        \eqPf{\sem{}{0}}{\emptyset}{By \defn}
        \inPf{}{\Cpo}{By \Lemmaref{lem:predomain-constructions}}
      \end{llproof} 

      \ProofCaseThing{P = P_1 + P_2}\\
      \begin{llproof}
        \inPf{\sem{}{P_1}}{\Cpo}{By \ih}
        \inPf{\sem{}{P_2}}{\Cpo}{By \ih}
        \eqPf{\sem{}{P_1 + P_2}}{(\sem{}{P_1} \uplus \sem{}{P_2}, \ordsym[\sem{}{P_1} \uplus \sem{}{P_2}])}{By \defn}
        \inPf{}{\Cpo}{By \Lemmaref{lem:predomain-constructions}}
      \end{llproof} 

      \ProofCaseThing{P = 1}\\
      Similar to $P = 0$ case.

      \ProofCaseThing{P = P_1 \times P_2}\\
      Similar to $P = P_1 + P_2$ case.

      \ProofCaseThing{P = \downshift{N}}\\
      \begin{llproof}
        \eqPf{\sem{}{\downshift{N}}}{(\sem{}{N}, \ordsym[\sem{}{N}])}{By \defn}
        \inPf{}{\Cpo}{By \ih}
      \end{llproof} 

      \ProofCaseThing{P = \mu F}\\
      We first show that $(\sem{}{\mu F}, \ordsym[\sem{}{\mu F}])$ is a poset.

      Suppose $V \in \sem{}{\mu F}$.
      Then there exists $L \in \kindnat$
      such that $V \in \sem{}{F}^L \emptyset$.
      By \Lemmaref{lem:predomain-constructions}, $\emptyset \in \Cpo$.
      By \ih (part (3)), $\sem{}{F} : \Cpo \to \Cpo$.
      By \Lemmaref{lem:repeated-cpo-functor},
      $\sem{}{F}^L \emptyset \in \Cpo$;
      therefore (by reflexivity), $\ord[\sem{}{F}^{L} \emptyset]{V}{V}$.
      By \defn, $\ord[\sem{}{\mu F}]{V}{V}$, so $\ordsym[\sem{}{\mu F}]$ is reflexive.

      Suppose $\ord[\sem{}{\mu F}]{V_1}{V_2}$ and $\ord[\sem{}{\mu F}]{V_2}{V_3}$.
      Then there exist $L_1$ and $L_2$
      such that $\ord[\sem{}{F}^{L_1}\emptyset]{V_1}{V_2}$
      and $\ord[\sem{}{F}^{L_2}\emptyset]{V_2}{V_3}$.
      Let $L = \max(L_1, L_2)$.
      By \Lemref{lem:aux-unref-type-denotations},
      $\ord[\sem{}{F}^{L}\emptyset]{V_1}{V_2}$
      and $\ord[\sem{}{F}^{L}\emptyset]{V_2}{V_3}$.
      By \Lemmaref{lem:predomain-constructions}, $\emptyset \in \Cpo$.
      By \ih (part (3)), $\sem{}{F} : \Cpo \to \Cpo$.
      By \Lemmaref{lem:repeated-cpo-functor},
      $\sem{}{F}^{L} \emptyset \in \Cpo$;
      therefore (by transitivity of $\ordsym[\sem{}{F}^L \emptyset]$),
      $\ord[\sem{}{F}^{L} \emptyset]{V_1}{V_3}$.
      By \defn, $\ord[\sem{}{\mu F}]{V_1}{V_3}$,
      so $\ordsym[\sem{}{\mu F}]$ is transitive.

      Similarly, $\ordsym[\sem{}{\mu F}]$ is antisymmetric,
      so $\sem{}{\mu F}$ is a poset.

      To conclude, we show that $\sem{}{\mu F}$ is chain-complete.
      Suppose $d_0 \sqsubseteq \cdots$ is a chain in $\sem{}{\mu F}$.
      By \defn, for all $k$, there exists $n_k$
      such that $\ord[\sem{}{F}^{n_k+1}\emptyset]{d_k}{d_{k+1}}$.
      Above, we have $\sem{}{F} : \Cpo \to \Cpo$.
      By \Lemmaref{lem:mu-chain-complete},
      the chain $d_0 \sqsubseteq \cdots$ has a least fixed point $\sqcup_k d_k$
      in $\sem{}{F} (\mu \sem{}{F})$;
      and the latter equals $\mu\sem{}{F}$ by \Lemmaref{lem:unref-mu-unroll-equal},
      which equals $\sem{}{\mu F}$ by definition.
    \end{itemize}
  \item
    \begin{itemize}
      \ProofCaseThing{N = \upshift{P}}\\
      By \defn,
      \[
        \sem{}{\upshift{P}} = (\sem{}{P} \uplus \{\bott[\uparrow]\}, \comprehend{((1,d),(1,d'))}{\ord[\sem{}{P}]{d}{d'}} \cup \comprehend{((2, \bott[\uparrow]), d)}{d \in \sem{}{\upshift{P}}}, (2, \bott[\uparrow]))
      \]
      By \ih (part (1)), $\sem{}{P} \in \Cpo$.
      This together with the definition of $\ordsym[\sem{}{\upshift{P}}]$ above,
      that is,
      \[
        \comprehend{((1,d),(1,d'))}{\ord[\sem{}{P}]{d}{d'}} \cup \comprehend{((2, \bott[\uparrow]), d)}{d \in \sem{}{\upshift{P}}}
      \]
      implies that $\sem{}{\upshift{P}}$ is a poset.
      The definition also implies that
      $\bott[\sem{}{\upshift{P}}] = (2, \bott[\uparrow])$
      is the bottom element of $\sem{}{\upshift{P}}$.

      To conclude this case, we show that $\sem{}{\upshift{P}}$ is chain-complete.
      Suppose $d_0 \sqsubseteq \cdots$ is a chain in $\sem{}{\upshift{P}}$.
      If for all $k$, $d_k = (2, \bott[\uparrow])$,
      then $\sqcup_k d_k = (2, \bott[\uparrow])$.
      Else, if there exist $j$ and $d_j'$ such that $d_j = (1, d_j')$,
      then, by definition of $\ordsym[\sem{}{\upshift{P}}]$ (above),
      for all $m > j$, there exists $d_m' \in \sem{}{P}$
      such that $d_m = (1, d_m')$ and $\ord[\sem{}{P}]{d_m'}{d_{m+1}'}$;
      by the \ih, $\sem{}{P} \in \Cpo$, hence chain-complete,
      so the chain $d_j' \sqsubseteq \cdots$ has a least upper bound
      $\sqcup_{k \geq j} d_k'$ in $\sem{}{P}$;
      then $\sqcup_k d_k = (1, \sqcup_{k \geq j} d_k')$
      is the least upper bound of chain $d_0 \sqsubseteq \cdots$
      in $\sem{}{P} \uplus \{\bott[\uparrow]\} = \sem{}{\upshift{P}}$.

      \ProofCaseThing{N = P \to N_0}\\
      By \defn, $\sem{}{P \to N_0}$ is the set $\sem{}{P} \Rightarrow \sem{}{N_0}$
      of continuous functions from $\sem{}{P}$ to $\sem{}{N_0}$.
      By \ih, $\sem{}{P} \in \Cpo$.
      By \ih, $\sem{}{N_0} \in \Cppo$; therefore $\sem{}{N_0} \in \Cpo$.
      By \Lemmaref{lem:predomain-constructions},
      $\sem{}{P} \Rightarrow \sem{}{N_0}$ (with pointwise ordering) is a $\Cpo$.

      To conclude this case,
      we show $\ord[]{\bott[\sem{}{P \to N_0}]}{f}$ for all $f \in \sem{}{P \to N_0}$.
      To this end, suppose $f \in \sem{}{P \to N_0}$ and $x \in \sem{}{P}$.
      Then:\\
      \begin{llproof}
        \eqPf{\bott[\sem{}{P \to N_0}] \; x}{\bott[\sem{}{N_0}]}{By \defn}
        \ordPf{}{f(x)}{By \ih, $\sem{}{N_0} \in \Cppo$}
      \end{llproof} 
    \end{itemize}
  \item
    For each case below, $X$ is an arbitrary predomain ($\Cpo$),
    and $f \in \Hom_\Cpo(X, Y)$ is an arbitrary continuous function.
    \begin{itemize}
      \ProofCaseThing{\mathcal{F} = I}\\
      \begin{llproof}
        \eqPf{\sem{}{I} X}{\one}{By \defn}
        \inPf{}{\Cpo}{By \Lemmaref{lem:predomain-constructions}}
      \end{llproof} 
     
      Further, $\sem{}{I} f = \id_\one$ is continuous
      because identity functions are continuous.

      \ProofCaseThing{\mathcal{F} = \Const{Q}}\\
      \begin{llproof}
        \eqPf{\sem{}{\Const{Q}} X}{\sem{}{Q}}{By \defn}
        \inPf{}{\Cpo}{By \ih}
      \end{llproof} 

      Further, $\sem{}{\Const{Q}} f = \id_{\sem{}{Q}}$ is continuous
      because identity functions are continuous.

      \ProofCaseThing{\mathcal{F} = \Id}\\
      \begin{llproof}
        \eqPf{\sem{}{\Id} X}{X}{By \defn}
        \inPf{}{\Cpo}{Given}
      \end{llproof} 

      Further, $\sem{}{\Id} f = f$ is continuous.

      \ProofCaseThing{\mathcal{F} = B \otimes \hat{P}}\\
      \begin{llproof}
        \inPf{X}{\Cpo}{Given}
        \inPf{\sem{}{B} X}{\Cpo}{By \ih}
        \inPf{\sem{}{\hat{P}} X}{\Cpo}{By \ih}
        \eqPf{\sem{}{B \otimes \hat{P}} X}{\sem{}{B} X \times \sem{}{\hat{P}} X}{By \defn}
        \inPf{}{\Cpo}{By \Lemmaref{lem:predomain-constructions}}
      \end{llproof}

      Further, $\sem{}{B \otimes \hat{P}} f = ((\sem{}{B} f) \circ \pi_1, (\sem{}{\hat{P}} f) \circ \pi_2)$
      is continuous by two uses of the \ih,
      by the fact that projections $\pi_k$ are continuous,
      by the fact that the composition of continuous functions is continuous,
      and by the fact that the universal pair $(g_1, g_2)$
      of continuous functions $g_1$ and $g_2$ is continuous.

      \ProofCaseThing{\mathcal{F} = F_1 \oplus F_2}\\
      \begin{llproof}
        \inPf{X}{\Cpo}{Given}
        \inPf{\sem{}{F_1} X}{\Cpo}{By \ih}
        \inPf{\sem{}{F_2} X}{\Cpo}{By \ih}
        \eqPf{\sem{}{F_1 \oplus F_2} X}{\sem{}{F_1} X \uplus \sem{}{F_2} X}{By \defn}
        \inPf{}{\Cpo}{By \Lemmaref{lem:predomain-constructions}}
      \end{llproof}

      Further, $\sem{}{F_1 \oplus F_2} f = [\mathit{inj}_1 \circ (\sem{}{F_1} f), \mathit{inj}_2 \circ (\sem{}{F_2} f)]$
      is continuous by two uses of the \ih,
      by the fact that injections $\mathit{inj}_k$ are continuous,
      by the fact that the composition of continuous functions is continuous,
      and by the fact that the universal copair $[g_1, g_2]$
      of continuous functions $g_1$ and $g_2$ is continuous.
      \qedhere
    \end{itemize}
  \end{enumerate}
\end{proof}

\begin{lemma}[Unref.\ Unroll Sound]
  \label{lem:unref-unroll-soundness}
  If $\unrefunroll{G}{F}{P}$, then $\sem{}{G} (\mu \sem{}{F}) = \sem{}{P}$.
\end{lemma}
\begin{proof}
  By structural induction on the derivation of $\unrefunroll{G}{F}{P}$.
\end{proof}

\begin{definition}[Directed Poset]
  A subset $D \subseteq P$ of a poset $P$ is \emph{directed}
  if every finite subset $E \subseteq D$ of it has an upper bound in $D$.
\end{definition}

\begin{lemma}[Exchange]
  \label{lem:exchange}
  If $D_1$ and $D_2$ are directed posets, $E$ is a predomain,
  and $f : D_1 \times D_2 \to E$ is a monotone function,
  then $\sqcup_{x_1 \in D_1} \sqcup_{x_2 \in D_2} f(x_1, x_2) = \sqcup_{x_2 \in D_2} \sqcup_{x_1 \in D_1} f(x_1, x_2)$
\end{lemma}
\begin{proof}
  See Lemma 4.9 (Exchange) in the textbook of \citet{GunterTextbook}.
\end{proof}

\begin{lemma}[Diagonal]
  \label{lem:diagonal}
  If $D$ is a directed poset and $E$ is a predomain and $f : D \times D \to E$
  is a monotone function,
  then $\sqcup_{x \in D} \sqcup_{y \in D} f(x, y) = \sqcup_{x \in D} f(x, x)$.
\end{lemma}
\begin{proof}
  This is Lemma 4.16 (Diagonal) in the textbook of \citet{GunterTextbook}.
\end{proof}

\begin{lemma}[Application Diagonal]
  \label{lem:application-diagonal}
  If $D$ and $E$ are predomains\\
  and $f_0 \sqsubseteq f_1 \sqsubseteq \cdots$ is a chain in $D \Rightarrow E$
  and $x_0 \sqsubseteq x_1 \sqsubseteq \cdots$ is a chain in $D$,\\
  then $\sqcup_{k \in \kindnat} \sqcup_{j \in \kindnat} f_k x_j = \sqcup_{k \in \kindnat} f_k x_k$.
\end{lemma}
\begin{proof}
  Define the function $g : \kindnat \times \kindnat \to E$
  by $(k, j) \mapsto f_k x_j$.
  The poset $\kindnat \times \kindnat$
  (with componentwise ordering where $\kindnat$ has usual order $\leq$)
  is directed and the function $g$ is monotone,
  so the goal follows by \Lemmaref{lem:diagonal}. \qedhere
\end{proof}

\begin{lemma}[App.\ lub Distributes]
  \label{lem:app-lub-distributes}
  If $D$ and $E$ are predomains\\
  and $f_0 \sqsubseteq f_1 \sqsubseteq \cdots$ is a chain in $D \Rightarrow E$
  and $x_0 \sqsubseteq x_1 \sqsubseteq \cdots$ is a chain in $D$,\\
  then
  $\sqcup_{k \in \kindnat} f_k x_k = (\sqcup_{k \in \kindnat} f_k) (\sqcup_{k \in \kindnat} x_k)$.
\end{lemma}
\begin{proof}
  ~\\
  \begin{llproof}
    \eqPf{\sqcup_{k \in \kindnat} f_k x_k}{\sqcup_{k \in \kindnat} \sqcup_{j \in \kindnat} f_k x_j}{By \Lemmaref{lem:application-diagonal}}
    \eqPf{}{\sqcup_{k \in \kindnat} f_k (\sqcup_{j \in \kindnat} x_j)}{For all $k$, we have $f_k$ continuous}
    \eqPf{}{(\sqcup_{k \in \kindnat} f_k) (\sqcup_{k \in \kindnat} x_k)}{By \defn}
  \end{llproof} 
\end{proof}

\begin{lemma}[Cut lub]
  \label{lem:cut-lub}
  If $D$ is a predomain
  and $d_0 \sqsubseteq d_1 \sqsubseteq \cdots$ is a chain in $D$,
  then for all $m \in \kindnat$,
  we have $\sqcup_{k \in \kindnat} d_k = \sqcup_{k \geq m} d_k$.
\end{lemma}
\begin{proof}
  Straightforward.
\end{proof}

The following two lemmas
(\Lemmaref{lem:continuous-maps} and \Lemmaref{lem:unrefined-type-soundness})
are mutually recursive.

\begin{lemma}[Continuous Maps]
  \label{lem:continuous-maps}
  Suppose $|- \delta_1 : \Gamma_1$ and $|- \delta_2 : \Gamma_2$
  and $\judgctx{(\Gamma_1, y:Q, \Gamma_2)}$.
  \begin{enumerate}
  \item If $\unrefsynhead{\Gamma_1, y:Q, \Gamma_2}{h}{P}$,
    then the function
    $\sem{}{Q} \to \sem{}{P}$
    defined by $d \mapsto \sem{\delta_1, d/y, \delta_2}{h}$
    is continuous.
  \item If $\unrefsynexp{\Gamma_1, y:Q, \Gamma_2}{g}{\upshift{P}}$,
    then the function
    $\sem{}{Q} \to \sem{}{\upshift{P}}$
    defined by $d \mapsto \sem{\delta_1, d/y, \delta_2}{g}$
    is continuous.
  \item If $\unrefchkval{\Gamma_1, y:Q, \Gamma_2}{v}{P}$,
    then the function
    $\sem{}{Q} \to \sem{}{P}$
    defined by $d \mapsto \sem{\delta_1, d/y, \delta_2}{v}$
    is continuous.
  \item If $\unrefchkexp{\Gamma_1, y:Q, \Gamma_2}{e}{N}$,
    then the function
    $\sem{}{Q} \to \sem{}{N}$
    defined by $d \mapsto \sem{\delta_1, d/y, \delta_2}{e}$
    is continuous.
  \item If $\unrefchkmatch{\Gamma_1, y:Q, \Gamma_2}{P}{\clauses{\pa}{e}{i}{I}}{N}$,
    then the function
    $\sem{}{Q} \to \sem{}{N}$
    defined by $d \mapsto \sem{\delta_1, d/y, \delta_2}{\clauses{\pa}{e}{i}{I}}$
    is continuous.
  \item If $\unrefspine{\Gamma_1, y:Q, \Gamma_2}{s}{N}{\upshift{P}}$,
    then the function
    $\sem{}{Q} \to \sem{}{N} \to \sem{}{\upshift{P}}$
    defined by $d \mapsto \sem{\delta_1, d/y, \delta_2}{s}$
    is continuous.
  \end{enumerate}
\end{lemma}
\begin{proof}
  By mutual induction on structure of the given typing derivation.
  Note that we implicitly use \Lemmaref{lem:ix-meaning-weakening-invariant}
  when extending semantic substitutions.

  We use the mutually recursive \Lemmaref{lem:unrefined-type-soundness} below
  to obtain the fact that subderivations (program subterms) denote
  continuous and hence monotone functions
  (\eg, in the \UnrefSynSpineApp of part (2)).

  \begin{enumerate}
  \item Straightforward.
  \item Use \Lemmaref{lem:app-lub-distributes} in the \UnrefSynSpineApp case.
    Otherwise straightforward.
  \item All cases are straightforward.
  \item
    \begin{itemize}
      \DerivationProofCase{\UnrefChkExpLet}
      { \unrefsynexp{\Gamma_1, y:Q, \Gamma_2}{\be}{\upshift{P}} \\
        \unrefchkexp{\Gamma_1, y:Q, \Gamma_2, x:P}{e_0}{N} }
      { \unrefchkexp{\Gamma_1, y:Q, \Gamma_2}{\Let{x}{\be}{e_0}}{N} }
      Consider cases of whether or not $\sqcup_{k \in \kindnat} \sem{\delta_1, d_k/y, \delta_2}{\be} = (2, \bott[\uparrow])$.
      The former case is easy.
      In the latter case, there exist a minimal $m$
      and $V_m \in \sem{\delta_1, d_m/y, \delta_2}{P}$
      such that $\sem{\delta_1, d_m/y, \delta_2}{\be} = (1, V_m)$
      and, for all $j \geq m$, there are $V_j \in \sem{\delta_1, d_j/y, \delta_2}{P}$
      such that $\sem{\delta_1, d_j/y, \delta_2}{\be} = (1, V_j)$;
      use \Lemmaref{lem:diagonal} and \Lemmaref{lem:cut-lub}.

      \ProofCaseRule{\UnrefChkExpMatch}
      Use \Lemmaref{lem:app-lub-distributes}.

      \ProofCaseRule{\UnrefChkExpRec}
      Use \Lemmaref{lem:exchange}.

      \item The remaining cases are straightforward.
    \end{itemize}
  \item
    Straightforward.
  \item Use \Lemmaref{lem:app-lub-distributes} in the \UnrefSpineApp case.
    Otherwise straightforward.
  \end{enumerate}
\end{proof}

\begin{lemma}[Unrefined Type Soundness]
  \label{lem:unrefined-type-soundness}
  Assume $|- \delta : \Gamma$.
  \begin{enumerate}
  \item If $\unrefsynhead{\Gamma}{h}{P}$,
    then $\sem{\delta}{\unrefsynhead{\Gamma}{h}{P}} \in \sem{}{P}$.
  \item If $\unrefsynexp{\Gamma}{\be}{\upshift{P}}$,
    then $\sem{\delta}{\unrefsynexp{\Gamma}{\be}{\upshift{P}}} \in \sem{}{\upshift{P}}$.
  \item If $\unrefchkval{\Gamma}{v}{P}$,
    then $\sem{\delta}{\unrefchkval{\Gamma}{v}{P}} \in \sem{}{P}$.
  \item If $\unrefchkexp{\Gamma}{e}{N}$,
    then $\sem{\delta}{\unrefchkexp{\Gamma}{e}{N}} \in \sem{}{N}$.
  \item If $\unrefchkmatch{\Gamma}{P}{\clauses{\pa}{e}{i}{I}}{N}$,
    then $\sem{\delta}{\unrefchkmatch{\Gamma}{P}{\clauses{\pa}{e}{i}{I}}{N}} \in \sem{}{P} \Rightarrow \sem{}{N}$.
  \item If $\unrefspine{\Gamma}{s}{N}{\upshift{P}}$,
    then $\sem{\delta}{\unrefspine{\Gamma}{s}{N}{\upshift{P}}} \in \sem{}{N} \Rightarrow \sem{}{\upshift{P}}$.
  \end{enumerate}
\end{lemma}
\begin{proof}
  Straightforward.
  Uses the mutually recursive \Lemmaref{lem:continuous-maps} above
  to obtain the continuity of maps denoted by program subterms
  (\eg, in the \UnrefChkExpLam case of part (4)).
  Also uses \Lemmaref{lem:unref-type-denotations}.

  The \UnrefChkValFix and \UnrefChkMatchFix cases use
  \Lemmaref{lem:unref-mu-unroll-equal}
  and \Lemmaref{lem:unref-unroll-soundness}.

  The \UnrefChkExpRec case uses \Lemmaref{lem:least-fixed-point}.
\end{proof}

\begin{lemma}[Unrefined Substitution Type Soundness]
  \label{lem:unref-type-soundness-subs}
  If $\Gamma_0 |- \sigma : \Gamma$,
  then $|- \sem{\delta}{\sigma} : \Gamma$ for all $|- \delta : \Gamma_0$.
\end{lemma}
\begin{proof}
  Similar to \Lemmaref{lem:type-soundness-subs}, but simpler:
  there's no corresponding \IxSyn or \PropSyn case,
  and the \UnrefValSyn case uses \Lemmaref{lem:unrefined-type-soundness}
  and has no need for an unrefined version of \Lemmaref{lem:subs-soundness-wf}.
\end{proof}

\begin{lemma}[Unrefined Substitution Soundness]
  \label{lem:unref-subs-soundness}
  Assume $\Gamma_0 |- \sigma : \Gamma$ and $|- \delta : \Gamma_0$.
  \begin{enumerate}
  \item If $\Dee :: \unrefsynhead{\Gamma}{h}{P}$,
    then $\sem{\delta}{\unrefsynhead{\Gamma_0}{[\sigma]h}{P}} = \sem{\sem{\delta}{\sigma}}{\unrefsynhead{\Gamma}{h}{P}}$.
  \item If $\Dee :: \unrefsynexp{\Gamma}{\be}{\upshift{P}}$,
    then $\sem{\delta}{\unrefsynexp{\Gamma_0}{[\sigma]\be}{\upshift{P}}} = \sem{\sem{\delta}{\sigma}}{\unrefsynexp{\Gamma}{\be}{\upshift{P}}}$.
  \item If $\Dee :: \unrefchkval{\Gamma}{v}{P}$,
    then $\sem{\delta}{\unrefchkval{\Gamma_0}{[\sigma]v}{P}} = \sem{\sem{\delta}{\sigma}}{\unrefchkval{\Gamma}{v}{P}}$.
  \item If $\Dee :: \unrefchkexp{\Gamma}{e}{N}$,
    then $\sem{\delta}{\unrefchkexp{\Gamma_0}{[\sigma]e}{N}} = \sem{\sem{\delta}{\sigma}}{\unrefchkexp{\Gamma}{e}{N}}$.
  \item If $\Dee :: \unrefchkmatch{\Gamma}{P}{\clauses{\pa}{e}{i}{I}}{N}$,\\
    then $\sem{\delta}{\unrefchkmatch{\Gamma_0}{P}{[\sigma]\clauses{\pa}{e}{i}{I}}{N}} = \sem{\sem{\delta}{\sigma}}{\unrefchkmatch{\Gamma}{P}{\clauses{\pa}{e}{i}{I}}{N}}$.
  \item If $\Dee :: \unrefspine{\Gamma}{s}{N}{\upshift{P}}$,
    then $\sem{\delta}{\unrefspine{\Gamma_0}{[\sigma]s}{N}{\upshift{P}}} = \sem{\sem{\delta}{\sigma}}{\unrefspine{\Gamma}{s}{N}{\upshift{P}}}$.
  \end{enumerate}
\end{lemma}
\begin{proof}
  Note that by \Lemmaref{lem:unref-type-soundness-subs}, we know that
  $|- \sem{\delta}{\sigma} : \Gamma$.

  For each part in the lemma statement, we are universally quantifying over
  ``substituted'' derivations;
  at the beginning of each part we assume such a derivation is given.

  Proceed by mutual induction on the structure of the program term.
  For each part, we consider cases for the rule concluding $\Dee$.
  While the denotations of program terms are only defined
  when those terms are well-typed (\ie, have a typing derivation),
  the denotations themselves are defined merely on the syntax of the program terms.
  Keeping this in mind,
  whenever the \ih is used, the necessary ``substituted'' subderivations
  are obtained by inversion on the given ``substituted'' derivation
  (the unrefined program typing rules are syntax directed).
  \begin{enumerate}
  \item
    \begin{itemize}
      \DerivationProofCase{\UnrefSynHeadVar}
      {
        (x : P) \in \Gamma
      }
      {
        \unrefsynhead{\Gamma}{x}{P}
      }
      \begin{llproof}
        \Pf{(x : P) \in \Gamma}{}{}{Subderivation}
        \Pf{\Gamma_0 |- \sigma : \Gamma}{}{}{Given}
        \Pf{\unrefchkval{\Gamma_0}{[\sigma_1]v}{P}}{}{}{By inversion}
        \Pf{\sigma = \sigma_1, \subs{v}{P}{x}, \sigma_2}{}{}{\ditto}
        \Pf{\Gamma_0 |- \sigma_1 : \Gamma_1}{}{}{\ditto}
      \end{llproof}
      \begin{itemize}
      \item \textbf{Case} $v = x$ \\
        \begin{align*}
          \hspace{2em}&\hspace{-2em}\sem{\sem{\delta}{\sigma}}{\unrefsynhead{\Gamma}{x}{P}} \\
          &= \sem{\delta}{\sigma}(x) &&\quad\text{By \defn of denotation} \\
          &= \sem{\delta}{\unrefchkval{\Gamma_0}{[\sigma_1]v}{P}} &&\quad\text{By \defn of }\sem{\delta}{\sigma} \\
          &= \sem{\delta}{\unrefchkval{\Gamma_0}{[\sigma_1]x}{P}} &&\quad v=x \\
          &= \sem{\delta}{\unrefchkval{\Gamma_0}{x}{P}} &&\quad x \notin \dom{\sigma_1} \\
          &= \delta(x) &&\quad\text{By \defn of \defsem} \\
          &= \sem{\delta}{\unrefsynhead{\Gamma_0}{x}{P}} &&\quad\text{By \defn of \defsem} \\
          &= \sem{\delta}{\unrefsynhead{\Gamma_0}{[\sigma]x}{P}}  &&\quad\text{By \defn of \defsubst} \\
        \end{align*}
      \item \textbf{Case} $v \neq x$ \\
        \begin{align*}
          \hspace{2em}&\hspace{-2em}\sem{\sem{\delta}{\sigma}}{\unrefsynhead{\Gamma}{x}{P}} \\
          &= \sem{\delta}{\sigma}(x) &&\quad\text{By \defn of denotation} \\
          &= \sem{\delta}{\unrefchkval{\Gamma_0}{[\sigma_1]v}{P}} &&\quad\text{By \defn of }\sem{\delta}{\sigma} \\
          &= \sem{\delta}{\unrefchkval{\Gamma_0}{\annoexp{[\sigma_1]v}{[\sigma_1]P}}{P}} &&\quad\text{By \defn of \defsem} \\
          &= \sem{\delta}{\unrefchkval{\Gamma_0}{[\sigma_1]\annoexp{v}{P}}{P}} &&\quad\text{By \defn of \defsubst} \\
          &= \sem{\delta}{\unrefsynhead{\Gamma_0}{[\sigma]x}{P}}  &&\quad\text{By \defn of \defsubst} \\
        \end{align*}
      \end{itemize}

      \DerivationProofCase{\UnrefSynValAnnot}
      {
        \unrefchkval{\Gamma}{v}{P}
      }
      {
        \unrefsynhead{\Gamma}{\annoexp{v}{P}}{P}
      }
      \begin{align*}
        \hspace{2em}&\hspace{-2em}\sem{\sem{\delta}{\sigma}}{\unrefsynhead{\Gamma}{\annoexp{v}{P}}{P}} \\
        &= \sem{\sem{\delta}{\sigma}}{\unrefchkval{\Gamma}{v}{P}} &&\quad\text{By def.} \\
        &= \sem{\delta}{\unrefchkval{\Gamma_0}{[\sigma]v}{P}} &&\quad\text{By i.h.}  \\
        &= \sem{\delta}{[\sigma]v} &&\quad\text{By def.}  \\
        &= \sem{\delta}{\annoexp{[\sigma]v}{P}} &&\quad\text{By def.} \\
        &= \sem{\delta}{\unrefsynhead{\Gamma_0}{\annoexp{[\sigma]v}{P}}{P}} &&\quad\text{By def.}  \\
        &= \sem{\delta}{\unrefsynhead{\Gamma_0}{[\sigma]\annoexp{v}{P}}{P}} &&\quad\text{By def.}
      \end{align*}
    \end{itemize}

  \item
    \begin{itemize}
      \DerivationProofCase{\UnrefSynSpineApp}
      { \unrefsynhead{\Gamma}{h}{\downshift{N}} \\
        \unrefspine{\Gamma}{s}{N}{\upshift{P}} }
      { \unrefsynexp{\Gamma}{h(s)}{\upshift{P}} }
      Straightforward:
      Suppose $\unrefsynexp{\Gamma_0}{[\sigma](h(s))}{\upshift{P}}$. Then
      \begin{align*}
        \sem{\sem{\delta}{\sigma}}{\unrefsynexp{\Gamma}{h(s)}{\upshift{P}}}
        &= \sem{\sem{\delta}{\sigma}}{h(s)} &&\quad\text{By \defn} \\
        &= \sem{\sem{\delta}{\sigma}}{s} \sem{\sem{\delta}{\sigma}}{h} &&\quad\text{By \defn} \\
        &= \sem{\delta}{[\sigma]s} \sem{\delta}{[\sigma]h} &&\quad\text{By \ih} \\
        &= \sem{\delta}{([\sigma]h)([\sigma]s)} &&\quad\text{By \defn} \\
        &= \sem{\delta}{[\sigma](h(s))} &&\quad\text{By \defn} \\
        &= \sem{\delta}{\unrefsynexp{\Gamma_0}{[\sigma](h(s))}{\upshift{P}}} &&\quad\text{By \defn}
      \end{align*}

      \ProofCaseRule{\UnrefSynExpAnnot}
      Straightforward.
    \end{itemize}

  \item
    \begin{itemize}
      \DerivationProofCase{\UnrefChkValVar}
      { (x:P) \in \Gamma }
      { \unrefchkval{\Gamma}{x}{P} }
      \begin{llproof}
        \Pf{(x : P) \in \Gamma}{}{}{Premise}
        \Pf{\Gamma_0 |- \sigma : \Gamma}{}{}{Given}
        \Pf{\unrefchkval{\Gamma_0}{[\sigma_1]v}{P}}{}{}{By inversion}
        \Pf{\sigma = \sigma_1, \subs{v}{P}{x}, \sigma_2}{}{}{\ditto}
      \end{llproof} 
      ~\\
      \begin{align*}
        \sem{\sem{\delta}{\sigma}}{\unrefchkval{\Gamma}{x}{P}}
        &= \sem{\delta}{\sigma}(x) &&\quad\text{By \defn of \defsem} \\
        &= \sem{\delta}{\unrefchkval{\Gamma_0}{[\sigma_1]v}{P}} &&\quad\text{By \defn of }\sem{\delta}{\sigma} \\
        &= \sem{\delta}{\unrefchkval{\Gamma_0}{[\sigma]x}{P}} &&\quad\text{By \defn of \defsubst}
      \end{align*}
      (Note that $x$ in the last line is a \emph{value} variable.)

      \ProofCaseRule{\UnrefChkValUnit}
      Straightforward.
      
      \ProofCaseRule{\UnrefChkValPair}
      Straightforward.

      \ProofCaseRule{\UnrefChkValIn{k}}
      Straightforward.

      \DerivationProofCase{\UnrefChkValFix}
      { \unrefunroll{F}{F}{P}
        \\ 
        \unrefchkval{\Gamma}{v_0}{P} }
      { \unrefchkval{\Gamma}{\roll{v_0}}{\mu F} }
      Straightforward:
      \begin{align*}
        \sem{\sem{\delta}{\sigma}}{\roll{v_0}}
        &= \sem{\sem{\delta}{\sigma}}{v_0} &&\quad\text{By \defn} \\
        &= \sem{\delta}{[\sigma]v_0} &&\quad\text{By \ih} \\
        &= \sem{\delta}{\roll{[\sigma]v_0}} &&\quad\text{By \defn} \\
        &= \sem{\delta}{[\sigma]\roll{v_0}} &&\quad\text{By \defn} \\
      \end{align*}

      \ProofCaseRule{\UnrefChkValDownshift}
      Straightforward.
    \end{itemize}

  \item
    \begin{itemize}
      \ProofCaseRule{\UnrefChkExpUpshift}
      Straightforward.

      \DerivationProofCase{\UnrefChkExpLet}
      { \unrefsynexp{\Gamma}{\be}{\upshift{P}} \\
        \unrefchkexp{\Gamma, x:P}{e_0}{N} }
      { \unrefchkexp{\Gamma}{\Let{x}{\be}{e_0}}{N} }
      \begin{llproof}
        \Pf{\Gamma_0 |- \sigma : \Gamma}{}{}{Given}
        \Pf{\unrefsynexp{\Gamma}{\be}{\upshift{P}}}{}{}{Subderivation}
        \Pf{|- \delta : \Gamma}{}{}{Given}
      \end{llproof}
      \\
      But
      \begin{align*}
        \sem{\sem{\delta}{\sigma}}{g}
        &= \sem{\delta}{[\sigma]\be} &&\quad\text{By \ih} \\
        &\in \sem{}{\upshift{P}} &&\quad\text{By \Lemmaref{lem:unrefined-type-soundness}} \\
        &= \sem{}{P} \uplus \{\bott[\uparrow]\} &&\quad\text{By \defn}
      \end{align*}
      
      \begin{itemize}
        \ProofCaseThing{\sem{\sem{\delta}{\sigma}}{g} = (2, \bott[\uparrow])}
        Then, by definition of \defsem
        and by $\sem{\sem{\delta}{\sigma}}{g} = \sem{\delta}{[\sigma]\be}$,
        \[
          \sem{\sem{\delta}{\sigma}}{\Let{x}{\be}{e_0}} = \bott[\sem{}{N}]
          = \sem{\delta}{[\sigma](\Let{x}{\be}{e_0})}
        \]
        
        \ProofCaseThing{\sem{\sem{\delta}{\sigma}}{g} = (1, V)}\\
        \begin{llproof}
          \Pf{|- \delta, \sem{\delta}{[\sigma]\be}/x : \Gamma_0,x:P}{}{}{By above and \UnrefValSem}
          \Pf{|- \sem{\delta}{\sigma}, \sem{\sem{\delta}{\sigma}}{g}/x : \Gamma, x:P}{}{}{By above}
        \end{llproof} 
        \\
        Further,\\
        \begin{llproof}
          \Pf{\Gamma_0,x:P |- \sigma, \subs{x}{P}{x} : \Gamma,x:P}{}{}{By (unrefined version of) \Lemref{lem:id-subs-ext-prog}}
        \end{llproof} 
        \\
        Therefore,
        \begin{align*}
          \sem{\delta}{[\sigma](\Let{x}{\be}{e_0})}
          &= \sem{\delta}{\Let{x}{[\sigma]\be}{[\sigma]e_0}} &&\quad\text{By \defn} \\
          &= \sem{\delta,\sem{\delta}{[\sigma]\be}/x}{[\sigma]e_0} &&\quad\text{By \defn} \\
          &= \sem{\delta,\sem{\sem{\delta}{\sigma}}{\be}/x}{[\sigma]e_0} &&\quad\text{By \ih (as seen above)} \\
          &= \sem{\delta,\sem{\sem{\delta}{\sigma}}{\be}/x}{[\sigma, \subs{x}{P}{x}]e_0} &&\quad\text{By (unrefined version of) \Lemref{lem:id-subs-id}} \\
          &= \sem{\sem{\delta, \sem{\sem{\delta}{\sigma}}{\be}/x}{\sigma,\subs{x}{P}{x}}}{e_0} &&\quad\text{By \ih} \\
          &= \sem{\left(\sem{\delta,\sem{\sem{\delta}{\sigma}}{\be}/x}{\sigma}\right), \sem{\sem{\delta}{\sigma}}{\be}/x}{e_0} &&\quad\text{By \defn} \\
          &= \sem{\sem{\delta}{\sigma}, \sem{\sem{\delta}{\sigma}}{\be}/x}{e_0} &&\quad\text{Because } x \notin \dom{\Gamma_0} \\
          &= \sem{\sem{\delta}{\sigma}}{\Let{x}{\be}{e_0}} &&\quad\text{By \defn}
        \end{align*}
      \end{itemize}

      \ProofCaseRule{\UnrefChkExpMatch}
      Straightforward.

      \ProofCaseRule{\UnrefChkExpLam}
      Similar to case for \UnrefChkExpLet, but simpler.

      \DerivationProofCase{\UnrefChkExpRec}
      {
        \unrefchkexp{\Gamma, x:\downshift{N}}{e_0}{N}
      }
      {
        \unrefchkexp{\Gamma}{\rec{x}{e_0}}{N}
      }
      By definition,
      $\sem{\delta}{\rec{x}{[\sigma]e_0}} = \sqcup_{k \in \kindnat} d_k$
      where $d_0 = \bott[\sem{}{N}]$ and $d_{n+1} = \sem{\delta, d_n/x}{[\sigma]e_0}$.
      By definition,
      $\sem{\sem{\delta}{\sigma}}{\rec{x}{e_0}} = \sqcup_{k \in \kindnat} d_k'$
      where $d_0' = \bott[\sem{}{N}]$
      and $d_{n+1}' = \sem{\sem{\delta}{\sigma}, d_n'/x}{e_0}$.
      Therefore, it suffices to show $d_k = d_k'$ for all $k \in \kindnat$,
      which we will do by induction on $k$.
      Clearly, $d_0 = \bott = d_0'$.
      Suppose $d_n = d_n'$. It now suffices to show $d_{n+1} = d_{n+1}'$.

      \begin{llproof}
        \Pf{\Gamma_0, x:\downshift{N} |- \sigma, \subs{x}{\downshift{N}}{x} : \Gamma, x:\downshift{N}}{}{}{By \Lemref{lem:id-subs-ext-prog}}
        \trailingjust{(unrefined version)}
        \eqPf{d_{n+1}}{\sem{\delta, d_n/x}{[\sigma]e_0}}{By \defn}
        \eqPf{}{\sem{\delta, d_n/x}{[\sigma, \subs{x}{\downshift{N}}{x}]e_0}}{Identity substitution}
        \eqPf{}{\sem{\sem{\delta, d_n/x}{\sigma, \subs{x}{\downshift{N}}{x}}}{e_0}}{By \ih}
        \eqPf{}{\sem{\sem{\delta}{\sigma}, d_n/x}{e_0}}{By \defn}
        \eqPf{}{\sem{\sem{\delta}{\sigma}, d_n'/x}{e_0}}{Above supposition}
        \eqPf{}{d_{n+1}'}{By \defn}
      \end{llproof} 

      \ProofCaseRule{\UnrefChkExpDiverge}
      Straightforward.
    \end{itemize}

  \item
    \begin{itemize}
      \ProofCaseRule{\UnrefChkMatchUnit}
      Straightforward.

      \ProofCaseRule{\UnrefChkMatchPair}
      Similar to case for \UnrefChkExpLam.

      \ProofCaseRule{\UnrefChkMatchSum}
      Similar to case for \UnrefChkExpLam.

      \ProofCaseRule{\UnrefChkMatchVoid}
      Both sides are the empty function.

      \DerivationProofCase{\UnrefChkMatchFix}
      {
        \unrefunroll{F}{F}{P}
        \\
        \unrefchkexp{\Gamma, x:P}{e}{N}
      }
      {
        \unrefchkmatch{\Gamma}{\mu F}{\setof{\clause{\roll{x}}{e}}}{N}
      }
     \begin{llproof}
       \Pf{V \in \mu\sem{}{F}}{}{}{Suppose}
       \Pf{\mu\sem{}{F} = \sem{}{F} (\mu\sem{}{F})}{}{}{By \Lemmaref{lem:unref-mu-unroll-equal}}
       \Pf{\sem{}{F} (\mu\sem{}{F}) = \sem{}{P}}{}{}{By \Lemmaref{lem:unref-unroll-soundness}}
       \Pf{V \in \sem{}{P}}{}{}{Follows from above}
       \Pf{|- \delta : \Gamma_0}{}{}{Given}
       \Pf{|- \delta, V/x : \Gamma_0, x:P}{}{}{By \UnrefValSem}
       \Pf{|- \sem{\delta}{\sigma} : \Gamma}{}{}{Above}
       \Pf{|- \sem{\delta}{\sigma}, V/x : \Gamma, x:P}{}{}{By \UnrefValSem}
       \Pf{\Gamma_0 |- \sigma : \Gamma}{}{}{Given}
       \Pf{\Gamma_0, x:P |- \sigma, \subs{x}{P}{x} : \Gamma, x:P}{}{}{By (unrefined version of) \Lemref{lem:id-subs-ext-prog}}
     \end{llproof}
     \\
     Therefore,
     \begin{align*}
       \hspace{1em}&\hspace{-1em}\sem{\delta}{\unrefchkmatch{\Gamma_0}{\mu F}{[\sigma]\setof{\clause{\roll{x}}{e}}}{N}} V \\
       &= \sem{\delta}{\unrefchkmatch{\Gamma_0}{\mu F}{\setof{\clause{\roll{x}}{[\sigma]e}}}{N}} V &&\quad\text{By \defn} \\
       &= \sem{\delta, V/x}{\unrefchkexp{\Gamma_0, x:P}{[\sigma]e}{N}} &&\quad\text{By \defn} \\
       &= \sem{\delta, V/x}{\unrefchkexp{\Gamma_0, x:P}{[\sigma, \subs{x}{P}{x}]e}{N}} &&\quad\text{Identity substitution} \\
       &= \sem{\sem{\delta, V/x}{\sigma, \subs{x}{P}{x}}}{\unrefchkexp{\Gamma, x:P}{e}{N}} &&\quad\text{By \ih} \\
       &= \sem{\sem{\delta}{\sigma}, V/x}{\unrefchkexp{\Gamma, x:P}{e}{N}} &&\quad\text{By \defn} \\
       &= \sem{\sem{\delta}{\sigma}, V/x}{\unrefchkmatch{\Gamma}{\mu F}{\setof{\clause{\roll{x}}{e}}}{N}} &&\quad\text{By \defn}
     \end{align*}
    \end{itemize}

  \item
    \begin{itemize}
      \ProofCaseRule{\UnrefSpineApp}
      Straightforward.

      \DerivationProofCase{\UnrefSpineNil}
      { }
      {\unrefspine{\Gamma}{\cdot}{\upshift{P}}{\upshift{P}}}
      Both sides are $\id_{\sem{}{\upshift{P}}}$. \qedhere
    \end{itemize}
  \end{enumerate}
\end{proof}

\begin{lemma}[Type Subset of Erasure]
  \label{lem:type-subset-erasure}
  If $\judgetp{\Theta}{A}{\Xi}$, then $\sem{\delta}{A} \subseteq \sem{}{\erase{A}}$
  for all $|- \delta : \Theta$.
\end{lemma}
\begin{proof}
  By structural induction on the derivation of $\judgetp{\Theta}{A}{\Xi}$.
\end{proof}

\begin{lemma}[Functor Application Subset of Erasure]
  \label{lem:functor-subset-erasure}
  If $\judgefunctor{\Theta}{\mathcal{F}}{\Xi}$ and $X \in \Set$,
  then $\sem{\delta}{\mathcal{F}} X \subseteq \sem{}{\erase{\mathcal{F}}} X$
  for all $|- \delta : \Theta$.
\end{lemma}
\begin{proof}
  By structural induction on the derivation of $\judgefunctor{\Theta}{\mathcal{F}}{\Xi}$,
  and case analysis on its concluding rule.
  The \DeclFunctorConst case uses \Lemmaref{lem:type-subset-erasure}.
\end{proof}

\begin{lemma}[Mu Subset of Erasure]
  \label{lem:mu-subset-erasure}
  If $\judgefunctor{\Theta}{\mathcal{F}}{\Xi}$ and $|- \delta : \Theta$,
  then $\mu\sem{\delta}{\mathcal{F}} \subseteq \mu\sem{}{\erase{\mathcal{F}}}$.
\end{lemma}
\begin{proof}
  Suppose $V \in \mu\sem{\delta}{\mathcal{F}}$.
  By definition,
  $\mu\sem{\delta}{\mathcal{F}} = \bigcup_{k=0}^{\infty} \sem{\delta}{\mathcal{F}}^k \emptyset$.
  Therefore, there exists $n \in \mathbb{N}$ such that $V \in \sem{\delta}{\mathcal{F}}^n \emptyset$.
  We have $\sem{\delta}{\mathcal{F}}^n \emptyset \subseteq \sem{}{\erase{\mathcal{F}}}^n \emptyset$ by induction on $n$,
  using \Lemmaref{lem:functor-subset-erasure}.
  But $\sem{}{\erase{\mathcal{F}}}^n \emptyset \subseteq \mu\sem{}{\erase{\mathcal{F}}}$
  by set theory and definition of $\mu$,
  so $V \in \mu\sem{}{\erase{\mathcal{F}}}$, as desired.
\end{proof}

\begin{lemma}[Refined and Unrefined $\mathrm{fmap}$ Agree]
  \label{lem:ref-unref-fmap-agree}
  If $\judgefunctor{\Theta}{\mathcal{F}}{\Xi}$
  and $f$ is a function from set $X$ to set $Y$
  and $|- \delta : \Theta$,
  then $\sem{\delta}{\mathcal{F}} f = \sem{}{\erase{\mathcal{F}}} f$
  on $\sem{\delta}{\mathcal{F}} X$.
\end{lemma}
\begin{proof}
  By structural induction on the derivation of
  $\judgefunctor{\Theta}{\mathcal{F}}{\Xi}$.
\end{proof}

\begin{lemma}[Erasure Subst.\ Invariant]
  \label{lem:erasure-subs}
  Assume $\Theta_0 |- \sigma : \Theta$.
  \begin{enumerate}
  \item If $\judgetp{\Theta}{A}{\Xi}$,
    then $\erase{[\sigma]A} = \erase{A}$.
  \item If $\judgefunctor{\Theta}{\mathcal{F}}{\Xi}$,
    then $\erase{[\sigma]\mathcal{F}} = \erase{\mathcal{F}}$.
  \end{enumerate}
\end{lemma}
\begin{proof}
  By mutual induction on the structure of $A$ and $\mathcal{F}$.
\end{proof}

\begin{lemma}[Extraction Erases to Equality]
  \label{lem:extract-erases-to-equality}
  If $\judgeextract[]{\Theta}{A}{A'}{\Theta'}$, then $\erase{A} = \erase{A'}$.
\end{lemma}
\begin{proof}
  By structural induction on the extraction derivation,
  using the definition of erasure.
\end{proof}

\begin{lemma}[Equivalence Erases to Equality]
  \label{lem:equivalence-erases-to-equality}
  ~
  \begin{enumerate}
  \item If $\judgeequiv[\pm]{\Theta}{A}{B}$, then $\erase{A} = \erase{B}$.
  \item If $\judgeequiv[]{\Theta}{\mathcal{F}}{\mathcal{F}'}$, then $\erase{\mathcal{F}} = \erase{\mathcal{F}'}$.
  \end{enumerate}
\end{lemma}
\begin{proof}
  By mutual induction on the structure of the equivalence derivation,
  using the definition of erasure.
\end{proof}

\begin{lemma}[Subtyping Erases to Equality]
  \label{lem:subtyping-erases-to-equality}
  If $\judgesub[\pm]{\Theta}{A}{B}$, then $\erase{A} = \erase{B}$.
\end{lemma}
\begin{proof}
  By induction on the structure of the given subtyping derivation,
  using the definition of erasure.
  The \DeclSubPosSum and \DeclSubPosFix cases
  use \Lemmaref{lem:equivalence-erases-to-equality}.
  The \DeclSubPosL and \DeclSubNegR cases use \Lemmaref{lem:extract-erases-to-equality}.
  The \DeclSubPosExR and \DeclSubNegAllL cases use \Lemmaref{lem:erasure-subs}.
\end{proof}

\begin{lemma}[Erasure Respects Unrolling]
  \label{lem:unroll-erasure}
  ~\\
  If
  $\judgeunroll{\Xi}{\Theta}{ \nu:G[\mu F] }{\beta}{G\;\Fold{F}{\alpha}\;\nu}{t}{P}{\tau}$,
  then
  $\unrefunroll{\erase{G}}{\erase{F}}{\erase{P}}$.
\end{lemma}
\begin{proof}
  By structural induction on the derivation of the given refined unrolling judgement,
  using the definition of erasure.
\end{proof}

\begin{lemma}[Erasure Respects Typing]
  \label{lem:erasure-respects-typing}
  ~
  \begin{enumerate}
    \item If $\judgesynhead{\Theta}{\Gamma}{h}{P}$,
      then $\unrefsynhead{\erase{\Gamma}}{\erase{h}}{\erase{P}}$.
    \item If $\judgesynexp{\Theta}{\Gamma}{\be}{\upshift{P}}$,
      then $\unrefsynexp{\erase{\Gamma}}{\erase{\be}}{\erase{\upshift{P}}}$.
    \item If $\judgechkval{\Theta}{\Gamma}{v}{P}$,
      then $\unrefchkval{\erase{\Gamma}}{\erase{v}}{\erase{P}}$.
    \item If $\judgechkexp{\Theta}{\Gamma}{e}{N}$,
      then $\unrefchkexp{\erase{\Gamma}}{\erase{e}}{\erase{N}}$.
    \item If $\judgechkmatch{\Theta}{\Gamma}{P}{\clauses{\pa}{e}{i}{I}}{N}$,
      then $\unrefchkmatch{\erase{\Gamma}}{\erase{P}}{\erase{\clauses{\pa}{e}{i}{I}}}{\erase{N}}$.
    \item If $\judgespine{\Theta}{\Gamma}{s}{N}{\upshift{P}}$,
      then $\unrefspine{\erase{\Gamma}}{\erase{s}}{\erase{N}}{\erase{\upshift{P}}}$.
  \end{enumerate}
\end{lemma}
\begin{proof}
  By mutual structural induction on the given refined typing derivation,
  considering cases for its concluding rule.
  The \DeclChkValVar case of part (3)
  and the \DeclChkExpRec case of part (4)
  use \Lemmaref{lem:subtyping-erases-to-equality}.
  The \DeclChkExpRec case of part (4)
  and the \DeclChkValExists (\DeclSpineAll) case of part (3) (part (6)) uses \Lemmaref{lem:erasure-subs}.
  The \DeclChkValFix (\DeclChkMatchFix) case of part (3) (part (5)) uses \Lemmaref{lem:unroll-erasure}.
  The \DeclChkExpExtract case of part (4)
  and the \DeclChkMatchPair, \DeclChkMatchSum, and \DeclChkMatchFix cases of part (5),
  use \Lemmaref{lem:extract-erases-to-equality}.
\end{proof}

\begin{lemma}[Erasure Respects Substitution]
  \label{lem:erasure-respects-subs}
  Assume $\Theta_0; \Gamma_0 |- \sigma : \Theta; \Gamma$.
  \begin{enumerate}
  \item If $\judgesynhead{\Theta}{\Gamma}{h}{P}$,
    then $\erase{[\sigma]h} = [\erase{\sigma}]\erase{h}$.
  \item If $\judgesynexp{\Theta}{\Gamma}{\be}{\upshift{P}}$,
    then $\erase{[\sigma]\be} = [\erase{\sigma}]\erase{\be}$.
  \item If $\judgechkval{\Theta}{\Gamma}{v}{P}$,
    then $\erase{[\sigma]v} = [\erase{\sigma}]\erase{v}$.
  \item If $\judgechkexp{\Theta}{\Gamma}{e}{N}$,
    then $\erase{[\sigma]e} = [\erase{\sigma}]\erase{e}$.
  \item If $\judgechkmatch{\Theta}{\Gamma}{P}{\clauses{\pa}{e}{i}{I}}{N}$,
    then $\erase{[\sigma]\clauses{\pa}{e}{i}{I}} = [\erase{\sigma}]\erase{\clauses{\pa}{e}{i}{I}}$.
  \item If $\judgespine{\Theta}{\Gamma}{s}{N}{\upshift{P}}$,
    then $\erase{[\sigma]s} = [\erase{\sigma}]\erase{s}$.
  \end{enumerate}
\end{lemma}
\begin{proof}
  By lexicographic induction on,
  first, $\size[\sigma]{\sigma}$ (defined in \Figref{fig:size-prog-chi}), and,
  second, the structure of the given program term typing derivation.
  The proof is straightforward:
  apply the \ih as needed,
  follow the definitions of substitution and erasure,
  and use \Lemmaref{lem:erasure-subs},
  \Lemmaref{lem:id-subs-ext-ix},
  and \Lemmaref{lem:id-subs-ext-prog} as needed
  (note that the latter two lemmas do not increase the size of $\sigma$,
  because $\size[\sigma]{-}$ does not count identity substitution entries).
  We only show the first part.
  \begin{itemize}
    \DerivationProofCase{\DeclSynHeadVar}
    {
      (x : P) \in \Gamma
    }
    {
      \judgesynhead{\Theta}{\Gamma}{x}{P}
    }
    \begin{llproof}
      \Pf{\Theta_0; \Gamma_0 |- \sigma : \Theta; \Gamma}{}{}{Given}
      \Pf{(x : P) \in \Gamma}{}{}{By inversion}
      \Pf{\sigma = \sigma_1, \subs{v}{P}{x}, \sigma_2}{}{}{By inversion}
      \Pf{\erase{\sigma} = \erase{\sigma_1}, \subs{\erase{v}}{\erase{P}}{x}, \erase{\sigma_2}}{}{}{By \defn of erasure}
    \end{llproof} 
    \begin{itemize}
    \item \textbf{Case} $v = x$ \\
      \begin{align*}
        \erase{[\sigma]x}
        &= \erase{[\sigma_1]v} &&\quad\text{By \defn of substitution} \\
        &= \erase{[\sigma_1]x} &&\quad v=x \\
        &= x &&\quad x \notin \dom{\sigma_1} \\
        &= [\erase{\sigma_1}]x &&\quad x \notin \dom{\erase{\sigma_1}} \\
        &= [\erase{\sigma_1}]\erase{x} &&\quad\text{By \defn of erasure} \\
        &= [\erase{\sigma_1}]\erase{v} &&\quad v=x \\
        &= [\erase{\sigma}]\erase{x} &&\quad\text{By \defn of erasure}
      \end{align*}
    \item \textbf{Case} $v \neq x$\\
      \begin{align*}
        \erase{[\sigma]x}
        &= \erase{[\sigma_1]\annoexp{v}{P}} &&\quad\text{By \defn of substitution} \\
        &= \erase{\annoexp{[\sigma_1]v}{[\sigma_1]P}} &&\quad\text{By \defn of substitution} \\
        &= \annoexp{\erase{[\sigma_1]v}}{\erase{[\sigma_1]P}} &&\quad\text{By \defn of erasure} \\
        &= \annoexp{\erase{[\sigma_1]v}}{\erase{P}} &&\quad\text{By \Lemmaref{lem:erasure-subs}} \\
        &= \annoexp{[\erase{\sigma_1}]\erase{v}}{\erase{P}} &&\quad\text{By \ih} \\
        &= \annoexp{[\erase{\sigma_1}]\erase{v}}{[\erase{\sigma_1}]\erase{P}} &&\quad\text{No program terms in }\erase{P} \\
        &= [\erase{\sigma_1}]\annoexp{\erase{v}}{\erase{P}} &&\quad\text{By definition of \defsubst} \\
        &= [\erase{\sigma}]x &&\quad\text{By \defn of substitution} \\
        &= [\erase{\sigma}]\erase{x} &&\quad\text{By \defn of erasure}
      \end{align*}
    \end{itemize}

    \DerivationProofCase{\DeclSynValAnnot}
    {
      \judgetp{\Theta}{P}{\Xi}
      \\
      \judgechkval{\Theta}{\Gamma}{v}{P}
    }
    {
      \judgesynhead{\Theta}{\Gamma}{\annoexp{v}{P}}{P}
    }
    \begin{align*}
      \erase{[\sigma]\annoexp{v}{P}}
      &= \erase{\annoexp{[\sigma]v}{[\sigma]P}} &&\quad\text{By \defn of substitution} \\
      &= \annoexp{\erase{[\sigma]v}}{\erase{[\sigma]P}} &&\quad\text{By \defn of erasure} \\
      &= \annoexp{[\erase{\sigma}]\erase{v}}{\erase{[\sigma]P}} &&\quad\text{By \ih} \\
      &= \annoexp{[\erase{\sigma}]\erase{v}}{\erase{P}} &&\quad\text{By \Lemmaref{lem:erasure-subs}} \\
      &= \annoexp{[\erase{\sigma}]\erase{v}}{[\erase{\sigma}]\erase{P}} &&\quad\text{No variables in }\erase{P} \\
      &= [\erase{\sigma}]\annoexp{\erase{v}}{\erase{P}} &&\quad\text{By \defn of substitution} \\
      &= [\erase{\sigma}]\erase{\annoexp{v}{P}} &&\quad\text{By \defn of erasure} \qedhere
    \end{align*}
  \end{itemize}
\end{proof}

\begin{lemma}[Erasure Respects Substitution Typing]
  \label{lem:erasure-respects-subs-typing}
  ~
  \begin{enumerate}
    \item If $\Theta_0; \Gamma_0 |- \sigma : \Theta; \Gamma$,
      then $\erase{\Gamma_0} |- \erase{\sigma} : \erase{\Gamma}$.
    \item If $|- \delta : \Theta; \Gamma$, then $|- \erase{\delta} : \erase{\Gamma}$.
  \end{enumerate}
\end{lemma}
\begin{proof}
  ~
  \begin{enumerate}
  \item By structural induction on the derivation of 
    $\Theta_0; \Gamma_0 |- \sigma : \Theta; \Gamma$, 
     considering cases for its concluding rule,
     and using the definition of erasure.
     The \ValSyn case uses \Lemmaref{lem:erasure-respects-typing}
     and \Lemmaref{lem:erasure-subs}
     and \Lemmaref{lem:erasure-respects-subs}.
  \item By structural induction on the derivation of 
    $|- \delta : \Theta; \Gamma$, 
    considering cases for its concluding rule,
    and using the definition of erasure.
    The \ValSem case uses \Lemmaref{lem:type-subset-erasure}. \qedhere
  \end{enumerate}
\end{proof}

\begin{lemma}[Erasure Respects Semantic Substitution]
  \label{lem:erasure-respects-sem-subs}
  If $\Theta_0; \Gamma_0 |- \sigma : \Theta; \Gamma$
  and $|- \delta : \Theta_0; \Gamma_0$,
  then $\erase{\sem{\delta}{\sigma}} = \sem{\erase{\delta}}{\erase{\sigma}}$.
\end{lemma}
\begin{proof}
  By structural induction on the derivation of
  $\Theta_0; \Gamma_0 |- \sigma : \Theta; \Gamma$,
  considering cases for its concluding rule.
\end{proof}

\section{Semantic Metatheory of Refined System}

\begin{definition}[Fold]
  \label{def:fold}
  In \Figureref{fig:denotation-types1}, we informally define $\mathit{fold}$.
  Given $\judgefunctor{\Theta}{F}{\dontcare}$ and $\judgealgebra{\Xi}{\Theta}{\alpha}{F}{\tau}$
  and $|- \delta : \Theta$,
  we define the function
  $\fold{\sem{\delta}{F}}{\sem{\delta}{\alpha}} : \mu \sem{\delta}{F} \to \sem{}{\tau}$ as follows.
  If $\mu\sem{\delta}{F}$ is empty, then define it to be the empty function.
  If $\mu\sem{\delta}{F}$ is not empty,
  then for each $V \in \mu \sem{\delta}{F}$,
  choose $n > 0$ such that $V \in \sem{\delta}{F}^n \emptyset$
  (such an $n$ exists because $\mu \sem{\delta}{F} = \cup_{k \in \kindnat} \sem{\delta}{F}^k \emptyset$ by definition)
  and define $(\fold{\sem{\delta}{F}}{\sem{\delta}{\alpha}}) \; V$
  by $(\foldn{n}{\sem{\delta}{F}}{\sem{\delta}{\alpha}}) \; V$
  where $\foldn{0}{\sem{\delta}{F}}{\sem{\delta}{\alpha}} : \emptyset \to \sem{}{\tau}$ is the empty function
  and $\foldn{k+1}{\sem{\delta}{F}}{\sem{\delta}{\alpha}} : \sem{\delta}{F}^{k+1} \emptyset \to \sem{}{\tau}$
  is defined by $\foldn{k+1}{\sem{\delta}{F}}{\sem{\delta}{\alpha}} = \sem{\delta}{\alpha} \circ (\sem{\delta}{F} (\foldn{k}{\sem{\delta}{F}}{\sem{\delta}{\alpha}}))$.
  The function $\textit{fold}$ is well-defined
  because the choice of $n$ does not matter.
\end{definition}

For any sort $\tau$,
we give it the discrete order $d \sqsubseteq d$ for all $d \in \sem{}{\tau}$.

\begin{lemma}[Index Cpo]
  \label{lem:index-cpo}
  For any sort $\tau$, the pair $(\sem{}{\tau}, \ordsym[\sem{}{\tau}])$,
  where $\ordsym[\sem{}{\tau}]$ is the discrete order
  $d \sqsubseteq_{\sem{}{\tau}} d$ for all $d \in \sem{}{\tau}$,
  is a predomain.
\end{lemma}
\begin{proof}
  Straightforward.
\end{proof}

\begin{lemma}[Equal Functor Mu]
  \label{lem:mu-cong}
  Assume $F$ and $G$ are functors from $\Set$ to $\Set$ such that $F = G$.
  Then $\mu F = \mu G$.
\end{lemma}
\begin{proof}
  Since $F = G$, we know $F X = G X$ for all $X \in \Set$.
  If $k = 0$, then $F^k\emptyset = \emptyset = G^k\emptyset$.
  Assume $F^k\emptyset = G^k\emptyset$ for fixed $k \geq 0$;
  then $F^{k+1}\emptyset = F (F^k \emptyset) = G (G^k \emptyset) = G^{k+1}\emptyset$.
  Therefore, $\mu F = \cup_{k\in\kindnat} F^k \emptyset = \cup_{k\in\kindnat}G^k \emptyset = \mu G$.
\end{proof}

\begin{lemma}[Equal Func.\ Alg.\ Mu]
  \label{lem:fold-cong}
  Assume $F$ and $G$ are endofunctors on $\Set$.
  Let $X \in \Set$.
  Assume $\alpha : F X \to X$ and $\beta : G X \to X$ are algebras.
  If $F = G$ and $\alpha = \beta$, then $\fold{F}{\alpha} = \fold{G}{\beta}$.
\end{lemma}
\begin{proof}
  Note that $\fold{F}{\alpha} : \mu F \to X$ and $\fold{G}{\beta} : \mu G \to X$.
  By \Lemmaref{lem:mu-cong}, $\mu F = \mu G$, \ie, these functions have equal domain.
  By definition, $\mu F = \cup_{k\in\kindnat} F^k \emptyset$.
  Denote by $f_k$ and $g_k$ the $k$th level
  functions $\fold{F}{\alpha}$ and $\fold{G}{\beta}$
  from $F^k \emptyset$ to $X$, respectively.
  It suffices to prove that $f_k = g_k$ for all $k \in \kindnat$.
  If $k = 0$, then both $f_k$ and $g_k$ equal the unique empty function.
  Assume $f_n = g_n$ for fixed $n \geq 0$;
  then:
  \begin{align*}
    f_{n+1}
    &= \alpha \circ F f_n &&\quad\text{By def.} \\
    &= \beta \circ G f_n &&\quad\text{Because $F = G$ and $\alpha = \beta$} \\
    &= \beta \circ G g_n &&\quad\text{By i.h.} \\
    &= g_{n+1} &&\quad\text{By def.} \qedhere
  \end{align*}
\end{proof}

\begin{lemma}[Filter Out Program Vars.]
  \label{lem:filter-out-prog-vars}
  If $|- \delta : \Theta; \Gamma$, then $|- \filterprog{\delta} : \Theta$.
\end{lemma}
\begin{proof}
  By structural induction on the given semantic substitution derivation.
\end{proof}

\begin{lemma}[Meaning Weak.\ Invariant]
  \label{lem:tp-fun-alg-meaning-weakening-invariant}
  ~\\
  Assume $|- \delta : \Theta_1, \Theta_2$ and $|- \delta' : \Theta_1, \Theta, \Theta_2$
  and $\delta'\restriction_{\Theta_1, \Theta_2} = \delta$.
  \begin{enumerate}
    \item If $\judgetp{\Theta_1, \Theta_2}{A}{\Xi}$,
      then
      $\sem{\delta}{\judgetp{\Theta_1, \Theta_2}{A}{\Xi}}
      = \sem{\delta'}{\judgetp{\Theta_1, \Theta, \Theta_2}{A}{\Xi}}$
      (for any such weakened derivation).
    \item If $\judgefunctor{\Theta_1, \Theta_2}{\mathcal{F}}{\Xi}$,
      then
      $\sem{\delta}{\judgefunctor{\Theta_1, \Theta_2}{\mathcal{F}}{\Xi}}
      = \sem{\delta'}{\judgefunctor{\Theta_1, \Theta, \Theta_2}{\mathcal{F}}{\Xi}}$
      (for any such weakened derivation).
    \item If $\judgealgebra{\Xi}{\Theta_1,\Theta_2}{\alpha}{F}{\tau}$,
      then
      $\sem{\delta}{\judgealgebra{\Xi}{\Theta_1,\Theta_2}{\alpha}{F}{\tau}}
      = \sem{\delta'}{\judgealgebra{\Xi}{\Theta_1,\Theta,\Theta_2}{\alpha}{F}{\tau}}$
      (for any such weakened derivation).
  \end{enumerate}
\end{lemma}
\begin{proof}
  By mutual structural induction on the 
  given type, functor or algebra well-formedness derivation.
  Uses
  \Lemmaref{lem:ix-meaning-weakening-invariant},
  \Lemref{lem:mu-cong}, and
  \Lemref{lem:fold-cong}.
\end{proof}

\begin{lemma}[Strong Index Substitution Soundness]
  \label{lem:subs-soundness-ix-strong}
  If $\Theta_0; \cdot |- \sigma : \Theta; \cdot$ and $|- \delta : \Theta_0$
  and $\Xi \subseteq \Theta$ and $\judgeterm{\Xi}{t}{\tau}$,
  then $\sem{(\sem{\delta}{\sigma})\restriction_\Xi}{\judgeterm{\Xi}{t}{\tau}} = \sem{\delta\restriction_{\fvimctx{\sigma}{\Xi}}}{\judgeterm{\fvimctx{\sigma}{\Xi}}{[\sigma]t}{\tau}}$.
\end{lemma}
\begin{proof}
  The proof is similar to that of \Lemmaref{lem:subs-soundness-ix},
  keeping track of restrictions.
\end{proof}

\begin{lemma}[Functor Type Soundness]
  \label{lem:type-soundness-functor}
  If $\judgefunctor{\Theta}{\mathcal{F}}{\Xi}$ and $|- \delta : \Theta$,
  then $\sem{\delta}{\mathcal{F}}$ is a functor from $\rCpo$ to $\rCpo$.
\end{lemma}
\begin{proof}
  By structural induction on $\mathcal{F}$.
  Use \Lemmaref{lem:unref-type-denotations}
  and \Lemmaref{lem:type-subset-erasure}.
\end{proof}

\begin{lemma}[Algebra Type Soundness]
  \label{lem:type-soundness-algebra}
  If $\judgealgebra{\Xi}{\Theta}{\alpha}{F}{\tau}$
  and $\Xi \subseteq \Theta$
  and $|- \delta : \Theta$,\\
  then $\sem{\delta}{\alpha} : \sem{\delta}{F} \sem{}{\tau} \to \sem{}{\tau}$.
\end{lemma}
\begin{proof}
By induction on the structure of $\alpha$, considering cases for the rule
concluding $\judgealgebra{\Xi}{\Theta}{\alpha}{F}{\tau}$:
\begin{itemize}
  \DerivationProofCase{\DeclAlgUnit}
  { \judgeterm{\Xi}{t}{\tau} }
  { \judgealgebra{\Xi}{\Theta}{\clause{\unitexp}{t}}{I}{\tau} }
  Assume $V \in \sem{\delta}{I} \sem{}{\tau}$.
  The latter equals $\one$ by definition, so $V = \bullet$.\\
  \begin{llproof}
    \Pf{\judgeterm{\Xi}{t}{\tau}}{}{}{Subderivation}
    \Pf{|- \delta : \Theta}{}{}{Given}
    \Pf{\Xi \subseteq \Theta}{}{}{Given}
    \Pf{|- \delta\restriction_\Xi : \Xi}{}{}{Straightforward}
  \end{llproof}
  ~\\
  Now,
  \begin{align*}
    \sem{\delta}{\clause{\unitexp}{t}} V
    &= \sem{\delta}{\clause{\unitexp}{t}} \bullet \\
    &= \sem{\delta\restriction_\Xi}{t} &&\quad\text{By \defn of denotation} \\
    &\in \sem{}{\tau} &&\quad\text{By \Lemmaref{lem:type-soundness-ix}}
  \end{align*}

  \DerivationProofCase{\DeclAlgSum}
  {
    \arrayenvb{
      \composeinj{1}{\alpha}{\alpha_1}
      \\
      \composeinj{2}{\alpha}{\alpha_2}
    }
    \\
    \arrayenvb{
      \Dee_1 :: \judgealgebra{\Xi}{\Theta}{\alpha_1}{F_1}{\tau}
      \\
      \Dee_2 :: \judgealgebra{\Xi}{\Theta}{\alpha_2}{F_2}{\tau}
    }
  }
  {
    \judgealgebra{\Xi}{\Theta}{\alpha}{(F_1 \oplus F_2)}{\tau}
  }
  Assume $V \in \sem{\delta}{F_1 \oplus F_2}\sem{}{\tau}$.
  By definition, the latter equals
  $\sem{\delta}{F_1}\sem{}{\tau} \uplus \sem{\delta}{F_2}\sem{}{\tau}$,
  so there exists $j \in \{1,2\}$ such that $V = (j, V_j)$
  and $V_j \in \sem{\delta}{F_j}\sem{}{\tau}$.
  By the induction hypothesis for subderivation $\Dee_j$,
  $\sem{\delta}{\alpha_j} : \sem{\delta}{F_j}\sem{}{\tau} \to \sem{}{\tau}$.
  Therefore,
  $\sem{\delta}{\alpha} V
  = \sem{\delta}{\alpha} (j, V_j)
  = \sem{\delta}{\alpha_j} V_j
  \in \sem{}{\tau}$.

  \DerivationProofCase{\DeclAlgIdProd}
  { \Dee_1 :: \judgealgebra{\Xi, a:\tau}{\Theta, a:\tau}{\clause{q}{t}}{\hat{P}}{\tau} }
  { \judgealgebra{\Xi}{\Theta}{\clause{(a, q)}{t}}{(\Id \otimes \hat{P})}{\tau} }
  Assume $V \in \sem{\delta}{\Id \otimes \hat{P}}\sem{}{\tau}$.
  By definition, the latter equals
  $\sem{}{\tau} \times \sem{\delta}{\hat{P}}\sem{}{\tau}$,
  so there exist $d \in \sem{}{\tau}$
  and $V' \in \sem{\delta}{\hat{P}}\sem{}{\tau}$
  such that $V = (d, V')$.
  By \IxSem, $|- \delta, d/a : \Theta, a:\tau$
  By the \ih for subderivation $\Dee_1$,
  $\sem{\delta, d/a}{\clause{q}{t}} : \sem{\delta, d/a}{\hat{P}}\sem{}{\tau} \to \sem{}{\tau}$.
  Therefore,
  $\sem{\delta}{\clause{(a,q)}{t}} (d, V')
  = \sem{\delta,d/a}{\clause{q}{t}} V'
  \in \sem{}{\tau}$.

  \DerivationProofCase{\DeclAlgConstProd}
  { \Dee_1 :: \judgealgebra{\Xi}{\Theta}{\clause{q}{t}}{\hat{P}}{\tau} \\ \judgetp{\Theta}{Q}{\dontcare} }
  { \judgealgebra{\Xi}{\Theta}{\clause{(\wild, q)}{t}}{(\Const{Q} \otimes \hat{P})}{\tau} }
  Assume $V \in \sem{\delta}{\Const{Q} \otimes \hat{P}}\sem{}{\tau}$.
  By definition, the latter equals
  $\sem{\delta}{Q} \times \sem{\delta}{\hat{P}} \sem{}{\tau}$.
  Therefore, there exist $V_1 \in \sem{\delta}{Q}$
  and $V_2 \in \sem{\delta}{\hat{P}} \sem{}{\tau}$
  such that $V = (V_1, V_2)$.
  By the \ih for subderivation $\Dee_1$,
  $\sem{\delta}{\clause{q}{t}} : \sem{\delta}{\hat{P}}\sem{}{\tau} \to \sem{}{\tau}$.
  Therefore,
  $\sem{\delta}{\clause{(\wild, q)}{t}} V
  = \sem{\delta}{\clause{(\wild, q)}{t}} (V_1, V_2)
  = \sem{\delta}{\clause{q}{t}} V_2
  \in \sem{}{\tau}$.

  \DerivationProofCase{\DeclAlgExConstProd}
  { \Dee_1 :: \judgealgebra{\Xi, a:\tau'}{\Theta, a:\tau'}{\clause{(\bap, q)}{t}}{(\Const{Q} \otimes \hat{P})}{\tau} }
  { \judgealgebra{\Xi}{\Theta}
    {\clause{(\pack{a}{\bap}, q)}{t}}
    {(\Const{\extype{a:\tau'}{Q}} \otimes \hat{P})}{\tau} }
  We want to show that
  \[
    \sem{\delta}{\clause{(\pack{a}{\bap}, q)}{t}} :
      \sem{\delta}{\Const{\extype{a:\tau'}{Q}} \otimes \hat{P}} \sem{}{\tau}
        \to \sem{}{\tau}
  \]
  Suppose $V$ is an element of the domain.
  Then, by definition of denotation,
  there exist $V_1 \in \comprehend{V \in \sem{}{\erase{Q}}}{\extype{d\in\sem{}{\tau'}}{V\in\sem{\delta,d/a}{Q}}}$
  and $V_2 \in \sem{\delta}{\hat{P}}\sem{}{\tau}$
  such that $V = (V_1, V_2)$.
  Now, by definition, 
  \[
    \sem{\delta}{\clause{(\pack{a}{\bap}, q)}{t}} (V_1, V_2)
    = \sem{\delta, d/a}{\clause{(\bap, q)}{t}} (V_1, V_2)
  \]
  where $d \in \sem{}{\tau'}$ is fixed such that $V_1 \in \sem{\delta, d/a}{Q}$
  (that $V_1$ inhabits the set above implies such a $d$ exists).
  By the \ih for subderivation $\Dee_1$ (and by definition of denotation), 
  $\sem{\delta,d/a}{\clause{(\bap, q)}{t}} :
     \sem{\delta, d/a}{Q} \times \sem{\delta, d/a}{\hat{P}}\sem{}{\tau}
       \to \sem{}{\tau}$.
  Because $a$ does not appear free in $\hat{P}$, we have
  $\sem{\delta, d/a}{\hat{P}} \sem{}{\tau} = \sem{\delta}{\hat{P}} \sem{}{\tau}$.
  Therefore,
  $\sem{\delta,d/a}{\clause{(\bap, q)}{t}} (V_1, V_2) \in \sem{}{\tau}$,
  as desired. \qedhere
\end{itemize}
\end{proof}

\begin{lemma}[Type Substitution Soundness]
  \label{lem:subs-soundness-wf}
  Assume $\E :: \Theta_0;\cdot |- \sigma : \Theta; \cdot$. Then:
\begin{enumerate}
\item If $\Dee :: \judgetp{\Theta}{A}{\Xi}$,
  then $\sem{\delta}{[\sigma]A} = \sem{\sem{\delta}{\E}}{A}$
  for all $|- \delta : \Theta_0$.
\item If $\Dee :: \judgefunctor{\Theta}{\mathcal{F}}{\Xi}$,
  then $\sem{\delta}{[\sigma]\mathcal{F}} = \sem{\sem{\delta}{\E}}{\mathcal{F}}$
  for all $|- \delta : \Theta_0$.
\item If $\Dee :: \judgealgebra{\Xi}{\Theta}{\alpha}{F}{\tau}$,
  then $\sem{\delta}{[\sigma]\alpha} = \sem{\sem{\delta}{\E}}{\alpha}$
  for all $|- \delta : \Theta_0$.
\end{enumerate}
\end{lemma}
\begin{proof}
  By mutual induction on the structure of the
  type, functor, and algebra well-formedness derivation $\Dee$;
  in each part, we consider cases for $\Dee$'s concluding rule.
  Note that $|- \sem{\delta}{\sigma} : \Theta; \cdot$
  by \Lemmaref{lem:type-soundness-subs-ix-prop}.
  In each part, we assume a derivation of
  $\judgetp{\Theta_0}{[\sigma]A}{\dontcare}$,
  $\judgefunctor{\Theta_0}{[\sigma]\mathcal{F}}{\dontcare}$, or
  $\judgealgebra{\fvimctx{\sigma}{\Xi}}{\Theta_0}{[\sigma]\alpha}{([\sigma]F)}{\tau}$.
  \begin{enumerate}
  \item
    \begin{itemize}
      \DerivationProofCase{\DeclTpVoid}
      { }
      { \judgetp{\Theta}{0}{\cdot}  }
      Both denotations are $\emptyset$.

      \DerivationProofCase{\DeclTpUnit}
      { }
      { \judgetp{\Theta}{1}{\cdot}  }
      Both denotations are $\{\bullet\}$.

      \DerivationProofCase{\DeclTpSum}
      { \judgetp{\Theta}{P_1}{\Xi_1} \\ \judgetp{\Theta}{P_2}{\Xi_2} }
      { \judgetp{\Theta}{P_1 + P_2}{\Xi_1 \sect \Xi_2}  }
      \begin{align*}
        \hspace{2em}&\hspace{-2em}\sem{\delta}{\judgetp{\Theta_0}{[\sigma](P_1 + P_2)}{\dontcare}} \\
          &= \sem{\delta}{\judgetp{\Theta_0}{[\sigma]P_1 + [\sigma]P_2}{\dontcare}} &&\quad\text{By def., $[\sigma](P_1+P_2) = [\sigma]P_1 + [\sigma]P_2$} \\
          &= \sem{\delta}{[\sigma]P_1} \uplus \sem{\delta}{[\sigma]P_2} &&\quad\text{By definition} \\
          &= \sem{\sem{\delta}{\sigma}}{P_1} \uplus \sem{\sem{\delta}{\sigma}}{P_2} &&\quad\text{By i.h.} \\
          &= \sem{\sem{\delta}{\sigma}}{\judgetp{\Theta}{P_1+P_2}{\dontcare}} &&\quad\text{By definition}
      \end{align*}

      \ProofCaseRule{\DeclTpProd}
      Similar to case for \DeclTpSum.

      \DerivationProofCase{\DeclTpEx}
      { \judgetp{\Theta, a:\tau}{P}{\Xi, a:\tau} }
      { \judgetp{\Theta}{\extype{a:\tau}{P}}{\Xi} }
      \begin{llproof}
        \Pf{\Theta_0;\Gamma_0 |- \sigma : \Theta; \Gamma}{}{}{Given}
        \Pf{\Theta_0,a:\tau;\Gamma_0 |- \sigma,a/a : \Theta,a:\tau; \Gamma}{}{}{By \Lemmaref{lem:id-subs-ext-ix}}
      \end{llproof}
      \\
      \begin{align*}
        \hspace{2em}&\hspace{-2em}\sem{\delta}{\judgetp{\Theta_0}{[\sigma](\extype{a:\tau}{P})}{\dontcare}} \\
          &= \sem{\delta}{\judgetp{\Theta_0}{\extype{a:\tau}{[\sigma]P}}{\dontcare}} &&\quad\text{By \defn}\\
          &= \sem{\delta}{\extype{a:\tau}{[\sigma,a/a]P}} &&\quad\text{Subst. $a/a$ is identity}\\
          &= \comprehend{V \in \sem{}{\erase{ [\sigma,a/a]P}}}{\extype{d \in \sem{}{\tau}}{V \in \sem{\delta,d/a}{[\sigma,a/a]P}}} &&\quad\text{By \defn} \\
          &= \comprehend{V \in \sem{}{\erase{P}}}{\extype{d \in \sem{}{\tau}}{V \in \sem{\delta,d/a}{[\sigma,a/a]P}}} &&\quad\text{By \Lemref{lem:erasure-subs}}\\
          &= \comprehend{V \in \sem{}{\erase{P}}}{\extype{d \in \sem{}{\tau}}{V \in \sem{\sem{\delta,d/a}{\sigma,a/a}}{P}}} &&\quad\text{By \ih}\\
          &= \comprehend{V \in \sem{}{\erase{P}}}{\extype{d \in \sem{}{\tau}}{V \in \sem{\sem{\delta}{\sigma}, d/a}{P}}} &&\quad\text{By \defn}\\
          &= \sem{\sem{\delta}{\sigma}}{\judgetp{\Theta}{\extype{a:\tau}{P}}{\dontcare}} &&\quad\text{By def.}
      \end{align*}

      \ProofCaseRule{\DeclTpDown}
      Similar to case for \DeclTpSum, but simpler.

      \ProofCaseRule{\DeclTpWith}
      Similar to case for \DeclTpEx, but using \Lemmaref{lem:subs-soundness-ix}.

      \DerivationProofCase{\DeclTpFixVar}
      {
        \judgefunctor{\Theta}{F}{\Xi'}
        \\
        \judgealgebra{\cdot}{\Theta}{\alpha}{F}{\tau}
        \\
        (b : \tau) \in \Theta
      }
      {
        \judgetp{\Theta}{\comprehend{\nu:\mu F}{\Fold{F}{\alpha}\,{\nu} =_\tau b}}{\Xi' \union b:\tau}
      }
      \begin{llproof}
        \Pf{\judgefunctor{\Theta}{F}{\Xi'}}{}{}{Subderivation}
        \Pf{\sem{\delta}{[\sigma]F} = \sem{\sem{\delta}{\sigma}}{F}}{}{}{By i.h.}
        \Pf{\mu\sem{\delta}{[\sigma]F} = \mu\sem{\sem{\delta}{\sigma}}{F}}{}{}{By \Lemmaref{lem:mu-cong}}
        \Pf{\judgealgebra{\cdot}{\Theta}{\alpha}{F}{\tau}}{}{}{Subderivation}
        \Pf{\sem{\delta}{[\sigma]\alpha} = \sem{\sem{\delta}{\sigma}}{\alpha}}{}{}{By i.h.}
        \Pf{(b : \tau) \in \Theta}{}{}{By inversion}
        \Pf{\judgeterm{\Theta}{b}{\tau}}{}{}{By \IxVar}
        \Pf{\sem{\delta}{[\sigma]b} = \sem{\sem{\delta}{\sigma}}{b}}{}{}{By \Lemmaref{lem:subs-soundness-ix}}
      \end{llproof} 
      \\
      By \Lemmaref{lem:fold-cong}, for every
      $d \in \mu\sem{\delta}{[\sigma]F} = \mu\sem{\sem{\delta}{\sigma}}{F}$,
      we have
      \[
        (\fold{\sem{\delta}{[\sigma]F}}{\sem{\delta}{[\sigma]\alpha}})\;d
        = (\fold{\sem{\sem{\delta}{\sigma}}{F}}{\sem{\sem{\delta}{\sigma}}{\alpha}})\;d
      \]
      By \Lemmaref{lem:subs-soundness-ix},
      $\sem{\delta}{[\sigma]t} = \sem{\sem{\delta}{\sigma}}{t}$.
      Therefore, for every
      $d \in \mu\sem{\delta}{[\sigma]F} = \mu\sem{\sem{\delta}{\sigma}}{F}$,
      $(\fold{\sem{\delta}{[\sigma]F}}{\sem{\delta}{[\sigma]\alpha}})\;d
      = \sem{\delta}{[\sigma]t}$
      if and only if
      $(\fold{\sem{\sem{\delta}{\sigma}}{F}}{\sem{\sem{\delta}{\sigma}}{\alpha}})\;d
      = \sem{\sem{\delta}{\sigma}}{t}$,
      and we have:
      \begin{align*}
        \hspace{2em}&\hspace{-2em}\sem{\delta}{\judgetp{\Theta_0}{[\sigma]\comprehend{\nu:\mu F}{\Fold{F}{\alpha}\,{\nu} =_\tau b}}{\dontcare}}\\
        &= \sem{\delta}{\judgetp{\Theta_0}{\comprehend{\nu:\mu [\sigma]F}{\Fold{[\sigma]F}{[\sigma]\alpha}\,{\nu} =_\tau [\sigma]b}}{\dontcare}} &&\quad\text{By \defn}\\
        &= \comprehend{d \in \mu\sem{}{\erase{ [\sigma]F}}}{d \in \mu\sem{\delta}{[\sigma]F} \land (\fold{\sem{\delta}{[\sigma]F}}{\sem{\delta}{[\sigma]\alpha}})\;d = \sem{\delta}{[\sigma]b}} &&\quad\text{By \defn} \\
        &= \comprehend{d \in \mu\sem{}{\erase{F}}}{d \in \mu\sem{\delta}{[\sigma]F} \land (\fold{\sem{\delta}{[\sigma]F}}{\sem{\delta}{[\sigma]\alpha}})\;d = \sem{\delta}{[\sigma]b}} &&\quad\text{By \Lemref{lem:erasure-subs}} \\
        &= \comprehend{d \in \mu\sem{}{\erase{F}}}{d \in \mu\sem{\sem{\delta}{\sigma}}{F} \land (\fold{\sem{\sem{\delta}{\sigma}}{F}}{\sem{\sem{\delta}{\sigma}}{\alpha}})\;d = \sem{\sem{\delta}{\sigma}}{b}} &&\quad\text{Above} \\
        &= \sem{\sem{\delta}{\sigma}}{\judgetp{\Theta}{\comprehend{\nu:\mu F}{\Fold{F}{\alpha}\,{\nu} =_\tau b}}{\dontcare}} &&\quad\text{By \defn}
      \end{align*}

      \ProofCaseRule{\DeclTpFix}
      Similar to case for \DeclTpFixVar.

      \ProofCaseRule{\DeclTpAll}
      Similar to case for \DeclTpEx.

      \ProofCaseRule{\DeclTpImplies}
      Similar to case for \DeclTpWith.

      \ProofCaseRule{\DeclTpArrow}
      Similar to case for \DeclTpEx.

      \ProofCaseRule{\DeclTpUp}
      Similar to case for \DeclTpDown.
    \end{itemize}

  \item
    \begin{itemize}
      \DerivationProofCase{\DeclFunctorConst}
      { \judgetp{\Theta}{P}{\Xi} } 
      { \judgefunctor{\Theta}{\Const{P}}{\Xi} } 
      Let $X \in \Set$. Then:
      \begin{align*}
        \sem{\delta}{{\judgefunctor{\Theta_0}{[\sigma]\Const{P}}{\dontcare}} X} 
        &= \sem{\delta}{\judgefunctor{\Theta_0}{\Const{[\sigma]P}}{\dontcare}} X &&\quad\text{By def.} \\
        &= \sem{\delta}{\judgetp{\Theta_0}{[\sigma]P}{\dontcare}} &&\quad\text{By def.} \\
        &= \sem{\sem{\delta}{\sigma}}{\judgetp{\Theta}{P}{\dontcare}} &&\quad\text{By i.h.} \\
        &= \sem{\sem{\delta}{\sigma}}{\judgefunctor{\Theta}{\Const{P}}{\dontcare}} X &&\quad\text{By def.}
      \end{align*}

      \DerivationProofCase{\DeclFunctorId}
      {  } 
      { \judgefunctor{\Theta}{\Id}{\cdot} } 
      Let $X \in \Set$. Then:
      \begin{align*}
        \sem{\delta}{\judgefunctor{\Theta_0}{[\sigma]\Id}{\dontcare}} X
        &= \sem{\delta}{\judgefunctor{\Theta_0}{\Id}{\dontcare}} X &&\quad\text{By def.} \\
        &= X &&\quad\text{By def.} \\
        &= \sem{\sem{\delta}{\sigma}}{\judgefunctor{\Theta}{\Id}{\dontcare}} X &&\quad\text{By def.}
      \end{align*}

      \ProofCaseRule{\DeclFunctorUnit}
      Similar to case for \DeclFunctorId
      (there, both sides equal $X$; here, both sides equal $\{\bullet\}$).

      \ProofCaseRule{\DeclFunctorProd}
      Similar to case for \DeclTpProd of previous part.

      \ProofCaseRule{\DeclFunctorSum}
      Similar to case for \DeclTpSum of previous part.
    \end{itemize}

  \item
    \begin{itemize}
      \DerivationProofCase{\DeclAlgSum}
      {
        \arrayenvb{
          \composeinj{1}{\alpha}{\alpha_1}
          \\
          \composeinj{2}{\alpha}{\alpha_2}
        }
        \\
        \arrayenvb{
          \judgealgebra{\Xi}{\Theta}{\alpha_1}{F_1}{\tau}
          \\
          \judgealgebra{\Xi}{\Theta}{\alpha_2}{F_2}{\tau}
        }
      }
      {
        \judgealgebra{\Xi}{\Theta}{\alpha}{(F_1 \oplus F_2)}{\tau}
      }
      By \Lemref{lem:alg-pat-subs}, $\composeinj{k}{([\sigma]\alpha)}{[\sigma]\alpha_k}$.\\
      Let $d' \in \sem{\delta}{[\sigma]F_1 \oplus [\sigma]F_2}\sem{}{\tau}$.
      Then $d' = (k, d)$ for some $k\in\{1,2\}$, and:
      \begin{align*}
        d' = (k, d) &\in \sem{\delta}{[\sigma]F_1 \oplus [\sigma]F_2}\sem{}{\tau} \\
        &= (\sem{\delta}{[\sigma]F_1}\sem{}{\tau}) \uplus (\sem{\delta}{[\sigma]F_2}\sem{}{\tau}) &&\quad\text{By def.} \\
        &= (\sem{\sem{\delta}{\sigma}}{F_1}\sem{}{\tau}) \uplus (\sem{\sem{\delta}{\sigma}}{F_2}\sem{}{\tau}) &&\quad\text{By i.h.} \\
        &= \sem{\sem{\delta}{\sigma}}{F_1 \oplus F_2} \sem{}{\tau} &&\quad\text{By def.}
      \end{align*}
      Then:
      \begin{align*}
        \hspace{2em}&\hspace{-2em}\sem{\delta}{\judgealgebra{\fvimctx{\sigma}{\Xi}}{\Theta_0}{[\sigma]\alpha}{([\sigma]F_1 \oplus [\sigma]F_2)}{\tau}} (k,d) \\
        &= \sem{\delta}{\judgealgebra{\fvimctx{\sigma}{\Xi}}{\Theta_0}{[\sigma]\alpha_k}{[\sigma]F_k}{\tau}} d &&\quad\text{By def.} \\ 
        &= \sem{\sem{\delta}{\sigma}}{\judgealgebra{\Xi}{\Theta}{\alpha_k}{F_k}{\tau}} d &&\quad\text{By i.h.} \\ 
        &= \sem{\sem{\delta}{\sigma}}{\judgealgebra{\Xi}{\Theta}{\alpha}{(F_1 \oplus F_2)}{\tau}} (k,d) &&\quad\text{By def.}
      \end{align*}

      \DerivationProofCase{\DeclAlgUnit}
      { \judgeterm{\Xi}{t}{\tau} }
      { \judgealgebra{\Xi}{\Theta}{\clause{\unitexp}{t}}{I}{\tau} }
      Note that $\fvimctx{\sigma}{\Xi} \subseteq \Theta_0$
      by \Lemmaref{lem:fvimctx-subset}.
      \begin{align*}
        \hspace{2em}&\hspace{-2em}\sem{\delta}{\judgealgebra{\fvimctx{\sigma}{\Xi}}{\Theta_0}{[\sigma](\clause{\unitexp}{t})}{([\sigma]I)}{\tau}} \bullet \\
        &= \sem{\delta}{\judgealgebra{\fvimctx{\sigma}{\Xi}}{\Theta_0}{\clause{\unitexp}{[\sigma]t}}{I}{\tau}} \bullet &&\quad\text{By def.} \\
        &= \sem{\delta\restriction_{\fvimctx{\sigma}{\Xi}}}{\judgeterm{\fvimctx{\sigma}{\Xi}}{[\sigma]t}{\tau}}  &&\quad\text{By \defn} \\
        &= \sem{(\sem{\delta}{\sigma})\restriction_\Xi}{\judgeterm{\Xi}{t}{\tau}} &&\quad\text{By \Lemref{lem:subs-soundness-ix-strong}} \\
        &= \sem{\sem{\delta}{\sigma}}{\judgealgebra{\Xi}{\Theta}{\clause{\unitexp}{t}}{I}{\tau}}\bullet &&\quad\text{By def.}
      \end{align*}

      \DerivationProofCase{\DeclAlgIdProd}
      { \judgealgebra{\Xi,a:\tau}{\Theta, a:\tau}{\clause{q}{t}}{\hat{P}}{\tau} }
      { \judgealgebra{\Xi}{\Theta}{\clause{(a, q)}{t}}{(\Id \otimes \hat{P})}{\tau} }
      Let $(d,d') \in \sem{}{\tau}\times\sem{\delta}{[\sigma]\hat{P}}\sem{}{\tau}$.
      Then by i.h.,
      $(d,d') \in \sem{}{\tau}\times\sem{\sem{\delta}{\sigma}}{\hat{P}}\sem{}{\tau}$.
      Because $d \in \sem{}{\tau}$, $a \notin \Theta_0$ (by $\alpha$-renaming),
      and $|- \delta : \Theta_0; \Gamma_0$,
      we have $|- \delta,d/a : \Theta_0,a:\tau; \Gamma_0$.
      The derivation for the present case presupposes $\judgeterm{\Theta}{t}{\tau}$.
      By \Lemmaref{lem:id-subs-ext-ix}, $\Theta_0,a:\tau;\Gamma |- \sigma,a/a : \Theta,a:\tau;\Gamma$.
      By \Lemmaref{lem:syn-subs-ix}, $\judgeterm{\Theta_0}{[\sigma]t}{\tau}$.
      Now,
      \begin{align*}
        \hspace{2em}&\hspace{-2em}\sem{\delta}{\judgealgebra{\fvimctx{\sigma}{\Xi}}{\Theta_0}{\clause{(a, q)}{[\sigma]t}}{(\Id \otimes [\sigma]\hat{P})}{\tau}} (d,d') \\
        &= \sem{\delta,d/a}{\judgealgebra{\fvimctx{\sigma}{\Xi},a:\tau}{\Theta_0, a:\tau}{\clause{q}{[\sigma,a/a]t}}{[\sigma,a/a]\hat{P}}{\tau}} d' &&\quad\text{By def.}\\
        &= \sem{\delta,d/a}{\judgealgebra{\fvimctx{\sigma,a/a}{(\Xi,a:\tau)}}{\Theta_0, a:\tau}{\clause{q}{[\sigma,a/a]t}}{[\sigma,a/a]\hat{P}}{\tau}} d' &&\quad\text{By \Lemref{lem:fvimctx-id-convenient}}\\
        &= \sem{\sem{\delta,d/a}{\sigma,a/a}}{\judgealgebra{\Xi,a:\tau}{\Theta, a:\tau}{\clause{q}{t}}{\hat{P}}{\tau}} d' &&\quad\text{By i.h.}\\
        &= \sem{\sem{\delta}{\sigma},d/a}{\judgealgebra{\Xi,a:\tau}{\Theta, a:\tau}{\clause{q}{t}}{\hat{P}}{\tau}} d' &&\quad\text{By def.}\\
        &= \sem{\sem{\delta}{\sigma}}{\judgealgebra{\Xi}{\Theta}{\clause{(a, q)}{t}}{(\Id \otimes \hat{P})}{\tau}} (d,d')  &&\quad\text{By def.}
      \end{align*}

      \DerivationProofCase{\DeclAlgConstProd}
      { \judgealgebra{\Xi}{\Theta}{\clause{q}{t}}{\hat{P}}{\tau} \\ \judgetp{\Theta}{Q}{\dontcare} }
      { \judgealgebra{\Xi}{\Theta}{\clause{(\wild, q)}{t}}{(\Const{Q} \otimes \hat{P})}{\tau} }
      \begin{llproof}
        \Pf{\judgetp{\Theta}{Q}{\Xi}}{}{}{Subderivation}
        \Pf{\judgetp{\Theta_0}{[\sigma]Q}{\dontcare}}{}{}{By \Lemmaref{lem:syn-subs-tp-fun-alg}}
        \Pf{\sem{\delta}{[\sigma]Q}=\sem{\sem{\delta}{\sigma}}{Q}}{}{}{By i.h.}
      \end{llproof} 
      \begin{align*}
        \hspace{2em}&\hspace{-2em}\sem{\delta}{\judgealgebra{\fvimctx{\sigma}{\Xi}}{\Theta_0}{\clause{(\wild, q)}{[\sigma]t}}{(\Const{[\sigma]Q} \otimes [\sigma]\hat{P})}{\tau}} \\
        &= \sem{\delta}{\judgealgebra{\fvimctx{\sigma}{\Xi}}{\Theta_0}{\clause{q}{[\sigma]t}}{([\sigma]\hat{P})}{\tau}} &&\quad\text{By def.} \\
        &= \sem{\sem{\delta}{\sigma}}{\judgealgebra{\Xi}{\Theta}{\clause{q}{t}}{\hat{P}}{\tau}} &&\quad\text{By i.h.} \\
        &= \sem{\sem{\delta}{\sigma}}{\judgealgebra{\Xi}{\Theta}{\clause{(\wild, q)}{t}}{(\Const{Q} \otimes \hat{P})}{\tau}}  &&\quad\text{By def.}
      \end{align*}

      \DerivationProofCase{\DeclAlgExConstProd}
      { \judgealgebra{\Xi, a:\tau'}{\Theta, a:\tau'}{\clause{(\bap, q)}{t}}{(\Const{Q} \otimes \hat{P})}{\tau} }
      { \judgealgebra{\Xi}{\Theta}
        {\clause{(\pack{a}{\bap}, q)}{t}}
        {(\Const{\extype{a:\tau'}{Q}} \otimes \hat{P})}{\tau} }
      Note that by \Lemmaref{lem:id-subs-ext-ix},
      $\Theta_0,a:\tau';\Gamma |- \sigma,a/a : \Theta,a:\tau';\Gamma$.
      Let
      \begin{align*}
        (V_1, V_2) &\in \sem{\delta}{\Const{\extype{a:\tau'}{[\sigma]Q}} \otimes [\sigma]\hat{P}} \sem{}{\tau} \\
        &= \sem{\delta}{\extype{a:\tau'}{[\sigma]Q}} \times \sem{\delta}{[\sigma]\hat{P}}\sem{}{\tau} &&\quad\text{By \defn} \\
        &= \comprehend{V \in \sem{}{\erase{[\sigma]Q}}}{\extype{d \in \sem{}{\tau'}}{V \in \sem{\delta, d/a}{[\sigma]Q}}} \times \sem{\delta}{[\sigma]\hat{P}}\sem{}{\tau} &&\quad\text{By \defn} \\
        &= \comprehend{V \in \sem{}{\erase{Q}}}{\extype{d \in \sem{}{\tau'}}{V \in \sem{\delta, d/a}{[\sigma]Q}}} \times \sem{\delta}{[\sigma]\hat{P}}\sem{}{\tau} &&\quad\text{By \Lemref{lem:erasure-subs}} \\
        &= \comprehend{V \in \sem{}{\erase{Q}}}{\extype{d \in \sem{}{\tau'}}{V \in \sem{\delta, d/a}{[\sigma,a/a]Q}}} \times \sem{\delta}{[\sigma]\hat{P}}\sem{}{\tau} &&\quad\text{By \Lemref{lem:id-subs-id}} \\
        &= \comprehend{V \in \sem{}{\erase{Q}}}{\extype{d \in \sem{}{\tau'}}{V \in \sem{\sem{\delta}{\sigma},d/a}{Q}}} \times \sem{\sem{\delta}{\sigma}}{\hat{P}}\sem{}{\tau} &&\quad\text{By \ih} \\
      \end{align*}
      Define $V = (V_1, V_2)$.
      Below, $d$ is arbitrarily chosen and fixed 
      such that $V_1 \in \sem{\sem{\delta}{\sigma},d/a}{Q}$,
      as per the definition of the denotational semantics of algebras;
      it exists by above.
      Now,
      \begin{align*}
        &\sem{\delta}{\judgealgebra{\fvimctx{\sigma}{\Xi}}{\Theta_0}{[\sigma](\clause{(\pack{a}{\bap}, q)}{t})}{[\sigma](\Const{\extype{a:\tau'}{Q}} \otimes \hat{P})}{\tau}} V \\
        &= \sem{\delta,d/a}{\judgealgebra{\fvimctx{\sigma}{\Xi},a:\tau'}{\Theta_0, a:\tau'}{\clause{(\bap, q)}{[\sigma]t}}{(\Const{[\sigma]Q} \otimes [\sigma]\hat{P})}{\tau}} V \quad\text{By \defn}\\
        &= \sem{\delta,d/a}{\judgealgebra{\fvimctx{\sigma}{\Xi},a:\tau'}{\Theta_0, a:\tau'}{\clause{(\bap, q)}{[\sigma,a/a]t}}{(\Const{[\sigma,a/a]Q} \otimes [\sigma,a/a]\hat{P})}{\tau}} V \\
        &= \sem{\delta,d/a}{\judgealgebra{\fvimctx{\sigma,a/a}{(\Xi,a:\tau')}}{\Theta_0, a:\tau'}{\clause{(\bap, q)}{[\sigma,a/a]t}}{(\Const{[\sigma,a/a]Q} \otimes [\sigma,a/a]\hat{P})}{\tau}} V \\
        &= \sem{\sem{\delta,d/a}{\sigma,a/a}}{\judgealgebra{\Xi,a:\tau'}{\Theta, a:\tau'}{\clause{(\bap, q)}{t}}{(\Const{Q} \otimes \hat{P})}{\tau}} V \quad\text{By i.h.}\\
        &= \sem{\sem{\delta}{\sigma},d/a}{\judgealgebra{\Xi,a:\tau'}{\Theta, a:\tau'}{\clause{(\bap, q)}{t}}{(\Const{Q} \otimes \hat{P})}{\tau}} V \quad\text{By def.}\\
        &= \sem{\sem{\delta}{\sigma}}{\judgealgebra{\Xi}{\Theta}{\clause{(\pack{a}{\bap}, q)}{t}}{(\Const{\extype{a:\tau'}{Q}} \otimes \hat{P})}{\tau}} V \quad\text{By def.}
      \end{align*}
      where the justifications for the two unjustified lines above
      are \Lemmaref{lem:id-subs-id} and \Lemref{lem:fvimctx-id-convenient},
      respectively. \qedhere
    \end{itemize}
  \end{enumerate}
\end{proof}

\begin{lemma}[Ref.\ Equal Functor Mu]
  \label{lem:mu-unroll-equal}
  If $\judgefunctor{\Theta}{F}{\Xi}$,\\
  then $\mu \sem{\delta}{F} = \sem{\delta}{F} (\mu \sem{\delta}{F})$
  for all $|- \delta : \Theta$.
\end{lemma}
\begin{proof}
  Similar to \Lemmaref{lem:unref-mu-unroll-equal}.
\end{proof}

\begin{lemma}[Semantic Unroll]
  \label{lem:refined-mu-unroll-equal}
  If $\judgetp{\Theta}{\comprehend{\nu:\mu F}{\Fold{F}{\alpha}\,{\nu} =_\tau t}}{\Xi}$,
  then 
  \[
    \comprehend{V \in \sem{\delta}{F} (\mu\sem{\delta}{F})}{\sem{\delta}{\alpha} (\sem{\delta}{F}(\fold{\sem{\delta}{F}}{\sem{\delta}{\alpha}})\;V) = \sem{\delta}{t}}
    = \comprehend{V \in \mu\sem{\delta}{F}}{(\fold{\sem{\delta}{F}}{\sem{\delta}{\alpha}})\;V = \sem{\delta}{t}}
  \]
  for all $|- \delta : \Theta$.
\end{lemma}
\begin{proof}
  This follows from \Lemmaref{lem:mu-unroll-equal} and the \defn of $\mathit{fold}$ (\Defnref{def:fold}).
\end{proof}

The following is an auxiliary lemma for \Lemmaref{lem:value-determined-soundness}:

\begin{lemma}[Aux.\ Xi Sound]
  \label{lem:aux-fold-equal}
  If $\judgefunctor{\Theta}{F}{\Xi}$
  and $\judgealgebra{\cdot}{\Theta}{\alpha}{F}{\tau}$
  and $|- \delta_1 : \Theta$ and $|- \delta_2 : \Theta$
  and $n \in \mathbb{N}$
  and $V \in \sem{\delta_1}{F}^{n+1} \emptyset$
  and $V \in \sem{\delta_2}{F}^{n+1} \emptyset$
  and $\sem{\delta_1}{\alpha} = \sem{\delta_2}{\alpha}$
  on $\sem{\delta_1}{F}^{n+1} \sem{}{\tau} \sect \sem{\delta_2}{F}^{n+1} \sem{}{\tau}$,\\
  then $(\foldn{n+1}{\sem{\delta_1}{F}}{\sem{\delta_1}{\alpha}})\;V = (\foldn{n+1}{\sem{\delta_1}{F}}{\sem{\delta_2}{\alpha}})\;V$.
\end{lemma}
\begin{proof}
  By induction on $n$.

  By \Defnref{def:fold}, for all $k \in \{1,2\}$,
  \[
    (\foldn{n+1}{\sem{\delta_k}{F}}{\sem{\delta_k}{\alpha}})\;V
    = \sem{\delta_k}{\alpha} (\sem{\delta_k}{F} (\foldn{n}{\sem{\delta_k}{F}}{\sem{\delta_k}{\alpha}})\;V)
  \]
  By definition,
  \[
    \sem{\delta_k}{F} (\foldn{n}{\sem{\delta_k}{F}}{\sem{\delta_k}{\alpha}})
    : \sem{\delta_k}{F}^{n+1} \emptyset \to \sem{\delta_k}{F} \sem{}{\tau}
  \]
  Therefore, 
  \[
    \sem{\delta_k}{F} (\foldn{n}{\sem{\delta_k}{F}}{\sem{\delta_k}{\alpha}})\;V
    \in \sem{\delta_1}{F} \sem{}{\tau} \sect \sem{\delta_2}{F} \sem{}{\tau}
  \]
  for all $k$.
  But we are given that $\sem{\delta_1}{\alpha} = \sem{\delta_2}{\alpha}$
  on $\sem{\delta_1}{F} \sem{}{\tau} \sect \sem{\delta_2}{F} \sem{}{\tau}$,
  so it suffices to show
  \[
    \sem{\delta_1}{F} (\foldn{n}{\sem{\delta_1}{F}}{\sem{\delta_1}{\alpha}})\;V
    = \sem{\delta_2}{F} (\foldn{n}{\sem{\delta_2}{F}}{\sem{\delta_2}{\alpha}})\;V
  \]

  First, consider the case $n = 0$.
  Then for $k=1,2$,
  $\foldn{0}{\sem{\delta_k}{F}}{\sem{\delta_k}{\alpha}} : \emptyset \to \sem{}{\tau}$
  is the empty function, independently of $F$.
  Then, noting that $V \in \sem{\delta_k}{F} \emptyset \subseteq \sem{}{\erase{F}} \emptyset$ by \Lemmaref{lem:functor-subset-erasure},
  we have
  \begin{align*}
    \hspace{1em}&\hspace{-1em}\sem{\delta_1}{F} (\foldn{0}{\sem{\delta_1}{F}}{\sem{\delta_1}{\alpha}})\;V \\
    &= \sem{}{\erase{F}} (\foldn{0}{\sem{\delta_1}{F}}{\sem{\delta_1}{\alpha}})\;V  &&\quad\text{By \Lemmaref{lem:ref-unref-fmap-agree}} \\
    &= \sem{}{\erase{F}} (\foldn{0}{\sem{\delta_2}{F}}{\sem{\delta_2}{\alpha}})\;V  &&\quad\text{Empty function is unique} \\
    &= \sem{\delta_2}{F} (\foldn{0}{\sem{\delta_2}{F}}{\sem{\delta_2}{\alpha}})\;V &&\quad\text{By \Lemmaref{lem:ref-unref-fmap-agree}}
  \end{align*}

  Finally, given the \ih for $n = m-1$, \ie,
  $(\foldn{m}{\sem{\delta_1}{F}}{\sem{\delta_2}{\alpha}})\;V = (\foldn{m}{\sem{\delta_1}{F}}{\sem{\delta_2}{\alpha}})\;V$,
  prove the equation holds for $n = m$, \ie
  $(\foldn{m+1}{\sem{\delta_1}{F}}{\sem{\delta_2}{\alpha}})\;V = (\foldn{m+1}{\sem{\delta_1}{F}}{\sem{\delta_2}{\alpha}})\;V$.
  As argued at the beginning of the proof, to show the latter, the following suffices:
  \begin{align*}
    \hspace{1em}&\hspace{-1em}\sem{\delta_1}{F} (\foldn{m}{\sem{\delta_1}{F}}{\sem{\delta_1}{\alpha}})\;V \\
    &= \sem{}{\erase{F}} (\foldn{m}{\sem{\delta_1}{F}}{\sem{\delta_1}{\alpha}})\;V &&\quad\text{By \Lemmaref{lem:ref-unref-fmap-agree}} \\
    &= \sem{}{\erase{F}} (\foldn{m}{\sem{\delta_2}{F}}{\sem{\delta_2}{\alpha}})\;V &&\quad\text{By \ih} \\
    &= \sem{\delta_2}{F} (\foldn{m}{\sem{\delta_2}{F}}{\sem{\delta_2}{\alpha}})\;V &&\quad\text{By \Lemmaref{lem:ref-unref-fmap-agree}}
  \end{align*}
  Note that (for $k=1,2$)
  \begin{align*}
    V
    &\in \sem{\delta_k}{F}^{m+1} \emptyset &&\quad\text{Above} \\
    &= \sem{\delta_k}{F} (\sem{\delta_k}{F}^{m} \emptyset) &&\quad\text{By \defn} \\
    &\subseteq \sem{}{\erase{F}} (\sem{\delta_k}{F}^{m} \emptyset) &&\quad\text{By \Lemmaref{lem:functor-subset-erasure}}
  \end{align*}
  to see that the function applications involving $\sem{}{\erase{F}}$ make sense.
\end{proof}

\begin{lemma}[Soundness of Value-Determined Indexes]
  \label{lem:value-determined-soundness}
  Assume $|- \delta_1 : \Theta$ and $\delta_2 : \Theta$.
  \begin{enumerate}
    \item If $\Dee :: \judgetp{\Theta}{P}{\Xi}$
      and $V \in \sem{\delta_1}{P}$ and $V \in \sem{\delta_2}{P}$,
      then $\delta_1\restriction_\Xi = \delta_2\restriction_\Xi$.
    \item If $\Dee :: \judgefunctor{\Theta}{\mathcal{F}}{\Xi}$
      and $X_1$ and $X_2$ are sets
      and $V \in \sem{\delta_1}{\mathcal{F}} X_1$
      and $V \in \sem{\delta_2}{\mathcal{F}} X_2$,\\
      then $\delta_1\restriction_\Xi = \delta_2\restriction_\Xi$.
    \item If $\Dee :: \judgealgebra{\Xi}{\Theta}{\alpha}{F}{\tau}$
      and $\Xi \subseteq \Theta$
      and $\delta_1\restriction_\Xi = \delta_2\restriction_\Xi$,\\
      then $\sem{\delta_1}{\alpha} = \sem{\delta_2}{\alpha}$
      on $\sem{\delta_1}{F} \sem{}{\tau} \sect \sem{\delta_2}{F} \sem{}{\tau}$.
  \end{enumerate}
\end{lemma}
\begin{proof}
  By mutual induction on structure of 
  the given type or functor well-formedness derivation $\Dee$.
  In each part, we consider cases for the rule concluding $\Dee$.
  \begin{enumerate}
  \item
    \begin{itemize}
      \DerivationProofCase{\DeclTpVoid}
      { }
      { \judgetp{\Theta}{0}{\cdot}  }
      $\delta_1\restriction_\cdot = \text{empty function} = \delta_2\restriction_\cdot$

      \item \textbf{Cases }\DeclTpUnit and \DeclTpDown:
      Similar to case for \DeclTpVoid.

      \DerivationProofCase{\DeclTpSum}
      { \Dee_1 :: \judgetp{\Theta}{P_1}{\Xi_1} \\ \Dee_2 :: \judgetp{\Theta}{P_2}{\Xi_2} }
      { \judgetp{\Theta}{P_1 + P_2}{\Xi_1 \sect \Xi_2}  }
      Assume $V \in \sem{\delta_k}{P_1 + P_2}$ for $k=1,2$.
      By definition of denotation,
      $\sem{\delta_k}{P_1 + P_2} = \sem{\delta_k}{P_1} \uplus \sem{\delta_k}{P_2}$
      for $k=1,2$,
      so there exists $j \in \{1, 2\}$ such that $V = (j, V_j)$
      and (for $k=1,2$) $V_j \in \sem{\delta_k}{P_j}$.
      By the \ih for subderivation $\Dee_j$, we have
      $\delta_1\restriction_{\Xi_j} = \delta_2\restriction_{\Xi_j}$.
      But $\Xi_1 \cap \Xi_2 \subseteq \Xi_j$, so
      $\delta_1\restriction_{\Xi_1 \cap \Xi_2} = \delta_2\restriction_{\Xi_1 \cap \Xi_2}$.

      \DerivationProofCase{\DeclTpProd}
      { \Dee_1 :: \judgetp{\Theta}{P_1}{\Xi_1} \\ \Dee_2 :: \judgetp{\Theta}{P_2}{\Xi_2} }
      { \judgetp{\Theta}{P_1 \times P_2}{\Xi_1 \union \Xi_2}  }
      Assume $V \in \sem{\delta_k}{P_1 \times P_2}$ for $k=1,2$.
      By definition of denotation,
      $\sem{\delta_k}{P_1 \times P_2} = \sem{\delta_k}{P_1} \times \sem{\delta_k}{P_2}$ for $k=1,2$,
      so there exist $V_1$ and $V_2$ such that $V_j \in \sem{\delta_k}{P_j}$
      for all $j\in\{1,2\}$ and $k\in\{1,2\}$.
      By the \ih for $\Dee_1$ and the \ih for $\Dee_2$, we have
      $\delta_1\restriction_{\Xi_k} = \delta_2\restriction_{\Xi_k}$
      for $k=1,2$.
      Therefore,
      $\delta_1\restriction_{\Xi_1 \cup \Xi_2} = \delta_2\restriction_{\Xi_1 \cup \Xi_2}$.

      \DerivationProofCase{\DeclTpEx}
      { \Dee_1 :: \judgetp{\Theta, a:\tau}{P'}{\Xi, a:\tau} }
      { \judgetp{\Theta}{\extype{a:\tau}{P'}}{\Xi} }
      Assume $V \in \sem{\delta_k}{\extype{a:\tau}{P'}}$ for $k=1,2$.
      By definition of denotation,
      there exist $d_1 \in \sem{}{\tau}$ and $d_2 \in \sem{}{\tau}$
      such that $V \in \sem{\delta_k,d_k/a}{P'}$ for $k=1,2$.
      By the \ih for subderivation $\Dee_1$,
      $(\delta_1, d_1/a)\restriction_{\Xi, a:\tau}
      = (\delta_2, d_2/a)\restriction_{\Xi, a:\tau}$.
      Therefore,
      $\delta_1\restriction_\Xi = \delta_2\restriction_\Xi$.

      \DerivationProofCase{\DeclTpWith}
      {
        \judgetp{\Theta}{P'}{\Xi}
        \\
        \judgeterm{\Theta}{\phi}{\Booltype}
      }
      { \judgetp{\Theta}{P' \land \phi}{\Xi} }
      Assume $V \in \sem{\delta_k}{P' \land \phi}$ for $k=1,2$.
      By definition of denotation, we have
      $V \in \sem{\delta_k}{P'}$ for $k=1,2$.
      By the \ih, $\delta_1\restriction_\Xi = \delta_2\restriction_\Xi$.

      \DerivationProofCase{\DeclTpFixVar}
      {
        \Dee_1 :: \judgefunctor{\Theta}{F}{\Xi'}
        \\
        \Dee_2 :: \judgealgebra{\cdot}{\Theta}{\alpha}{F}{\tau}
        \\
        (b : \tau) \in \Theta
      }
      {
        \judgetp{\Theta}{\comprehend{\nu:\mu F}{\Fold{F}{\alpha}\,{\nu} =_\tau b}}{\Xi' \cup b:\tau}
      }
      Assume $V \in \sem{\delta_k}{\comprehend{\nu:\mu F}{\Fold{F}{\alpha}\,{\nu} =_\tau b}}$ for $k=1,2$.
      By definition of denotation,
      \[
        \sem{\delta_k}{\comprehend{\nu:\mu F}{\Fold{F}{\alpha}\,{\nu} =_\tau b}}
          = \comprehend{V \in \mu\sem{}{\erase{F}}}{V \in \mu\sem{\delta_k}{F} \land (\fold{\sem{\delta_k}{F}}{\sem{\delta_k}{\alpha}})\;V = \delta_k(b)}
      \]
      for $k=1,2$.
      Therefore, for $k=1,2$, we have $V \in \mu\sem{\delta_k}{F}$
      and
      \begin{equation}
        \delta_k(b) = (\fold{\sem{\delta_k}{F}}{\sem{\delta_k}{\alpha}})\;V \tag{A}
      \end{equation}

      By \Lemmaref{lem:mu-unroll-equal},
      $\mu\sem{\delta_k}{F} = \sem{\delta_k}{F}(\mu\sem{\delta_k}{F})$
      for $k=1,2$.
      By the \ih (part (2)) for subderivation $\Dee_1$, we have
      $\delta_1\restriction_{\Xi'} = \delta_2\restriction_{\Xi'}$.
      If $(b:\tau) \in \Xi'$, then we are done; we will assume otherwise.

      To complete this case, it suffices to show $\delta_1(b) = \delta_2(b)$.
      To this end, by (A), it suffices to show
      \[
        (\fold{\sem{\delta_1}{F}}{\sem{\delta_1}{\alpha}})\;V
        = (\fold{\sem{\delta_2}{F}}{\sem{\delta_2}{\alpha}})\;V
      \]

      By definition of $\mu$,
      there exist $n_1, n_2 \in \mathbb{N}$ such that
      $V \in \sem{\delta_k}{F}^{n_k+1} \emptyset$ for $k=1,2$.
      Let $n = \max{\{n_1,n_2\}}$.
      Then $V \in \sem{\delta_k}{F}^{n+1} \emptyset$,
      and, by definition of $\mathit{fold}$ (\Defnref{def:fold}),
      \[
        (\fold{\sem{\delta_k}{F}}{\sem{\delta_k}{\alpha}})\;V
        = (\foldn{n+1}{\sem{\delta_k}{F}}{\sem{\delta_k}{\alpha}})\;V
      \]
      for $k=1,2$.

      We have $\delta_1\restriction_\cdot = \cdot = \delta_2\restriction_\cdot$.
      By the \ih (part (3)), $\sem{\delta_1}{\alpha} = \sem{\delta_2}{\alpha}$
      on $\sem{\delta_1}{F}\sem{}{\tau} \sect \sem{\delta_2}{F}\sem{}{\tau}$.\\
      Since
      $\sem{\delta_k}{F}^{n+1} \sem{}{\tau} \subseteq \sem{\delta_k}{F} \sem{}{\tau}$,
      we know $\sem{\delta_1}{\alpha} = \sem{\delta_2}{\alpha}$
      on $\sem{\delta_1}{F}^{n+1}\sem{}{\tau} \sect \sem{\delta_2}{F}^{n+1}\sem{}{\tau}$.
      Therefore, by \Lemref{lem:aux-fold-equal},
      \[
        (\foldn{n+1}{\sem{\delta_1}{F}}{\sem{\delta_1}{\alpha}})\;V
        = (\foldn{n+1}{\sem{\delta_2}{F}}{\sem{\delta_2}{\alpha}})\;V
      \]
      completing this case.

      \ProofCaseRule{\DeclTpFix}
      Similar to the case for \DeclTpFixVar, but simpler:
      there's no need to show anything like the equation
      $\delta_1(b) = \delta_2(b)$
      that's needed in that case.
    \end{itemize}

  \item
    \begin{itemize}
      \DerivationProofCase{\DeclFunctorConst}
      { \judgetp{\Theta}{P}{\Xi} } 
      { \judgefunctor{\Theta}{\Const{P}}{\Xi} } 
      Assume $V \in \sem{\delta_k}{\Const{P}} X_k$ for $k=1,2$.
      By definition of denotation,
      $\sem{\delta_k}{\Const{P}} X_k = \sem{\delta_k}{P}$
      for $k=1,2$.
      Therefore, $V \in \sem{\delta_k}{P}$ for $k=1,2$.
      By the \ih (part (1)) for subderivation $\judgetp{\Theta}{P}{\Xi}$,
      we have $\delta_1\restriction_\Xi = \delta_2\restriction_\Xi$.

      \DerivationProofCase{\DeclFunctorId}
      {  } 
      { \judgefunctor{\Theta}{\Id}{\underbrace{\cdot}_\Xi} } 
      $\delta_1\restriction_\cdot = \cdot = \delta_2\restriction_\cdot$

      \ProofCaseRule{\DeclFunctorUnit}
      Similar to case for \DeclFunctorId ($\Xi = \cdot$).

      \DerivationProofCase{\DeclFunctorProd}
      { \judgefunctor{\Theta}{B}{\Xi_1} \\ \judgefunctor{\Theta}{\hat{P}}{\Xi_2} } 
      { \judgefunctor{\Theta}{B \otimes \hat{P}}{\underbrace{\Xi_1 \union \Xi_2}_\Xi} } 
      Assume $V \in \sem{\delta_k}{B \otimes \hat{P}} X_k$ for $k=1,2$.
      By definition of denotation,
      $\sem{\delta_k}{B \otimes \hat{P}} X_k = \sem{\delta_k}{B} X_k \times \sem{\delta_k}{\hat{P}} X_k$
      for $k=1,2$.
      By the \ih for subderivation $\judgefunctor{\Theta}{B}{\Xi_1}$,
      we have $\delta_1\restriction_{\Xi_1} = \delta_2\restriction_{\Xi_1}$.
      By the \ih for subderivation $\judgefunctor{\Theta}{\hat{P}}{\Xi_2}$,
      we have $\delta_1\restriction_{\Xi_2} = \delta_2\restriction_{\Xi_2}$.
      By these two equations and set theory, we have
      $\delta_1\restriction_{\Xi_1 \cup \Xi_2} = \delta_2\restriction_{\Xi_1 \cup \Xi_2}$.

      \DerivationProofCase{\DeclFunctorSum}
      { \judgefunctor{\Theta}{F_1}{\Xi_1} \\ \judgefunctor{\Theta}{F_2}{\Xi_2} } 
      { \judgefunctor{\Theta}{F_1 \oplus F_2}{\underbrace{\Xi_1 \sect \Xi_2}_\Xi} } 
      Assume $V \in \sem{\delta_k}{F_1 \oplus F_2} X_k$ for $k=1,2$.
      By definition of denotation,
      $\sem{\delta_k}{F_1 \oplus F_2} X_k
      = \sem{\delta_k}{F_1} X_k \uplus \sem{\delta_k}{F_2} X_k$
      for $k=1,2$.
      Therefore, there exists $j \in \{1,2\}$
      such that $V = (j, V_j)$
      and $V_j \in \sem{\delta_k}{F_j} X_k$
      for $k=1,2$.
      By the \ih for subderivation $\judgefunctor{\Theta}{F_j}{\Xi_j}$,
      we have $\delta_1\restriction_{\Xi_j} = \delta_2\restriction_{\Xi_j}$.
      But $\Xi_1 \cap \Xi_2 \subseteq \Xi_j$, so
      $\delta_1\restriction_{\Xi_1 \cap \Xi_2} = \delta_2\restriction_{\Xi_1 \cap \Xi_2}$
      by set theory.

    \end{itemize}

  \item
    Note that, by \Lemmaref{lem:type-soundness-algebra}, we have
    $\sem{\delta_k}{\alpha} : \sem{\delta_k}{F}\sem{}{\tau} \to \sem{}{\tau}$.
    for $k=1,2$.
    We will prove the desired equality
    $\sem{\delta_1}{\alpha} = \sem{\delta_2}{\alpha}$
    on $\sem{\delta_1}{F} \sem{}{\tau} \sect \sem{\delta_2}{F} \sem{}{\tau}$
    by function extensionality, namely
    by showing these functions are equal at an arbitrary argument.
    To that end, suppose
    \[
      V \in \sem{\delta_1}{F} \sem{}{\tau} \sect \sem{\delta_2}{F} \sem{}{\tau}
    \]
    \begin{itemize}
      \DerivationProofCase{\DeclAlgUnit}
      { \judgeterm{\Xi}{t}{\tau} }
      { \judgealgebra{\Xi}{\Theta}{\clause{\unitexp}{t}}{I}{\tau} }
      By definition of denotation, $V = \bullet$.
      Now,
      \begin{align*}
        \sem{\delta_1}{\clause{\unitexp}{t}} \bullet
        &= \sem{\delta_1\restriction_\Xi}{t} &&\quad\text{By \defn of denotation} \\
        &= \sem{\delta_2\restriction_\Xi}{t} &&\quad\text{Because } \delta_1\restriction_\Xi = \delta_2\restriction_\Xi \text{, as given} \\
        &= \sem{\delta_2}{\clause{\unitexp}{t}} \bullet &&\quad\text{By \defn of denotation}
      \end{align*}

      \DerivationProofCase{\DeclAlgSum}
      {
        \arrayenvb{
          \composeinj{1}{\alpha}{\alpha_1}
          \\
          \composeinj{2}{\alpha}{\alpha_2}
        }
        \\
        \arrayenvb{
          \Dee_1 :: \judgealgebra{\Xi}{\Theta}{\alpha_1}{F_1}{\tau}
          \\
          \Dee_2 :: \judgealgebra{\Xi}{\Theta}{\alpha_2}{F_2}{\tau}
        }
      }
      {
        \judgealgebra{\Xi}{\Theta}{\alpha}{(F_1 \oplus F_2)}{\tau}
      }
      By definition of denotation, $V = (j, V_j)$
      where $V_j \in \sem{\delta_k}{F_j} \sem{}{\tau}$ for all $k$.
      Then
      \begin{align*}
        \sem{\delta_1}{\alpha} (j, V_j)
        &= \sem{\delta_1}{\alpha_j} V_j &&\quad\text{By \defn of denotation} \\
        &= \sem{\delta_2}{\alpha_j} V_j &&\quad\text{By \ih} \\
        &= \sem{\delta_2}{\alpha} (j, V_j) &&\quad\text{By \defn of denotation}
      \end{align*}

      \DerivationProofCase{\DeclAlgIdProd}
      { \Dee_1 :: \judgealgebra{\Xi, a:\tau}{\Theta, a:\tau}{\clause{q}{t}}{\hat{P}}{\tau} }
      { \judgealgebra{\Xi}{\Theta}{\clause{(a, q)}{t}}{(\Id \otimes \hat{P})}{\tau} }
      By definition of denotation, $V = (d, V')$
      where $d \in \sem{}{\tau}$
      and $V' \in \sem{\delta_k}{\hat{P}} \sem{}{\tau}$.
      Then
      \begin{align*}
        \sem{\delta_1}{\clause{(a, q)}{t}} (d, V')
        &= \sem{\delta_1,d/a}{\clause{q}{t}} V' &&\quad\text{By \defn of denotation} \\
        &= \sem{\delta_2,d/a}{\clause{q}{t}} V' &&\quad\text{By \ih for subderivation }\Dee_1 \\
        &= \sem{\delta_2}{\clause{(a, q)}{t}} (d, V')  &&\quad\text{By \defn of denotation}
      \end{align*}

      \DerivationProofCase{\DeclAlgConstProd}
      { \Dee_1 :: \judgealgebra{\Xi}{\Theta}{\clause{q}{t}}{\hat{P}}{\tau} \\ \judgetp{\Theta}{Q}{\dontcare} }
      { \judgealgebra{\Xi}{\Theta}{\clause{(\wild, q)}{t}}{(\Const{Q} \otimes \hat{P})}{\tau} }
      By definition of denotation, $V = (V_1, V_2)$
      where $V_1 \in \sem{\delta_k}{Q}$ for all $k$
      and $V_2 \in \sem{\delta_k}{\hat{P}} \sem{}{\tau}$ for all $k$.
      Then
      \begin{align*}
        \sem{\delta_1}{\clause{(\wild, q)}{t}} (V_1, V_2)
        &= \sem{\delta_1}{\clause{q}{t}} V_2 &&\quad\text{By \defn of denotation} \\
        &= \sem{\delta_2}{\clause{q}{t}} V_2 &&\quad\text{By \ih for subderivation }\Dee_1 \\
        &= \sem{\delta_2}{\clause{(\wild, q)}{t}} (V_1, V_2) &&\quad\text{By \defn of denotation}
      \end{align*}

      \DerivationProofCase{\DeclAlgExConstProd}
      { \Dee_1 :: \judgealgebra{\Xi, a:\tau'}{\Theta, a:\tau'}{\clause{(\bap, q)}{t}}{(\Const{Q} \otimes \hat{P})}{\tau} }
      { \judgealgebra{\Xi}{\Theta}
        {\clause{(\pack{a}{\bap}, q)}{t}}
        {(\Const{\extype{a:\tau'}{Q}} \otimes \hat{P})}{\tau} }
      By definition of denotation, $V = (V_1, V_2)$
      where $V_1 \in \sem{\delta_k}{\extype{a:\tau'}{Q}}$ for all $k$
      and $V_2 \in \sem{\delta_k}{\hat{P}} \sem{}{\tau}$ for all $k$.
      On the one hand,
      \begin{align*}
        \hspace{1em}&\hspace{-1em}\sem{\delta_1}{\clause{(\pack{a}{\bap}, q)}{t}} (V_1, V_2) \\
                    &= \sem{\delta_1, d_1/a}{\clause{(\bap, q)}{t}} (V_1, V_2) &&\quad\text{By \defn of denotation} \\
                    &= \sem{\delta_2, d_1/a}{\clause{(\bap, q)}{t}} (V_1, V_2) &&\quad\text{By \ih on subderivation }\Dee_1
      \end{align*}
      where $d_1 \in \sem{}{\tau'}$ is fixed such that
      \begin{equation}
        V_1 \in \sem{\delta_1,d_1/a}{Q} \tag{A}
      \end{equation}
      On the other hand, by definition,
      \[
        \sem{\delta_2}{\clause{(\pack{a}{\bap}, q)}{t}} (V_1, V_2)
        = \sem{\delta_2, d_2/a}{\clause{(\bap, q)}{t}} (V_1, V_2)
      \]
      where $d_2 \in \sem{}{\tau'}$ is fixed such that
      \begin{equation}
        V_1 \in \sem{\delta_2,d_2/a}{Q} \tag{B}
      \end{equation}
      To complete the proof, it suffices to show $d_1 = d_2$.
      Consider the presupposed derivation
      $\judgefunctor{\Theta}{\Const{\extype{a:\tau'}{Q}} \otimes \hat{P}}{\Xi'}$.
      By inversion on the functor and type well-formedness rules,
      there exists a subderivation
      $\Dee' :: \judgetp{\Theta, a:\tau'}{Q}{\Xi'', a:\tau'}$.
      By the \ih (part (1)) for subderivation $\Dee'$, (A), (B),
      $|- \delta_1, d_1/a : \Theta, a:\tau'$, and
      $|- \delta_2, d_2/a : \Theta, a:\tau'$,
      we have
      $(\delta_1, d_1/a)\restriction_{\Xi'', a:\tau'}
      = (\delta_2, d_2/a)\restriction_{\Xi'', a:\tau'}$.
      This implies $d_1 = d_2$, as desired. \qedhere
    \end{itemize}
  \end{enumerate}
\end{proof}

\begin{lemma}[Unrolling Soundness]
  \label{lem:unroll-soundness}
  Assume $|- \delta : \Theta$ and $\Xi \subseteq \Theta$.
  If
  \[
    \judgeunroll{\Xi}{\Theta}{ \nu:G[\mu F] }{\beta}{G\;\Fold{F}{\alpha}\;\nu}{t}{P}{\tau}
  \]
  then
  \[
    \comprehend
    {V \in \sem{\delta}{G} (\mu\sem{\delta}{F})}
    {\sem{\delta}{\beta} (\sem{\delta}{G}(\fold{\sem{\delta}{F}}{\sem{\delta}{\alpha}})\;V) = \sem{\delta}{t}} = \sem{\delta}{P}
  \]
\end{lemma}
\begin{proof}
  By structural induction on the given unrolling derivation.
  \begin{itemize}
    \DerivationProofCase{\DeclUnrollUnit}
    { }
    { \judgeunroll{\Xi}{\Theta}{\nu:I[\mu F]}{(\clause{\unitexp}{t'})}{I\;\Fold{F}{\alpha}\;\nu}{t}{1 \land (t = t')}{\tau} }
    The equations that are unjustified below hold by definition of denotation:
    \begin{align*}
      \hspace{1em}&\hspace{-1em}
      \comprehend{V \in \sem{\delta}{I}(\mu\sem{\delta}{F})}{\sem{\delta}{\clause{\unitexp}{t'}}(\sem{\delta}{I}\;\Fold{\sem{\delta}{F}}{\sem{\delta}{\alpha}}\;V) = \sem{\delta}{t}} \\
      &= \comprehend{V \in \one}{\sem{\delta}{\clause{\unitexp}{t'}}(\sem{\delta}{I}\;\Fold{\sem{\delta}{F}}{\sem{\delta}{\alpha}}\;V) = \sem{\delta}{t}} \\
      &= \comprehend{V \in \one}{\sem{\delta}{\clause{\unitexp}{t'}} \bullet  = \sem{\delta}{t}} \\
      &= \comprehend{V \in \one}{\sem{\delta}{t'} = \sem{\delta}{t}} \\
      &= \comprehend{V \in \one}{\sem{\delta}{t} = \sem{\delta}{t'}} &&\quad\text{By symmetry of equality} \\
      &= \comprehend{V \in \one}{\sem{\delta}{t = t'} = \one} \\
      &= \comprehend{V \in \one}{V \in \one \text{ and }\sem{\delta}{t = t'} = \one} \\
      &= \comprehend{V \in \sem{}{\unref{1}}}{V \in \sem{\delta}{1} \text{ and }\sem{\delta}{t = t'} = \one} \\
      &= \comprehend{V \in \sem{}{\erase{1}}}{V \in \sem{\delta}{1} \text{ and }\sem{\delta}{t = t'} = \one} &&\quad\text{By \defn of erasure}\\
      &= \sem{\delta}{1 \land (t = t')}
    \end{align*}

    \DerivationProofCase{\DeclUnrollSum}
    {\arrayenvbl{
        \composeinj{1}{\beta}{\beta_1}
        \\
        \composeinj{2}{\beta}{\beta_2}
      }
      \\
      \arrayenvbl{
        \judgeunroll{\Xi}{\Theta}{\nu:G_1[\mu F]}{\beta_1}{G_1\;\Fold{F}{\alpha}\;\nu}{t}{P_1}{\tau}
        \\
        \judgeunroll{\Xi}{\Theta}{\nu:G_2[\mu F]}{\beta_2}{G_2\;\Fold{F}{\alpha}\;\nu}{t}{P_2}{\tau}
      }
    }
    { \judgeunroll{\Xi}{\Theta}{\nu:(G_1 \oplus G_2)[\mu F]}{\beta}{(G_1 \oplus G_2)\;\Fold{F}{\alpha}\;\nu}{t}{P_1 + P_2}{\tau} }
    By the \ih for the given (unrolling) subderivations,
    \begin{equation}
      \comprehend
      {V \in \sem{\delta}{G_k}(\mu\sem{\delta}{F})}
      {\sem{\delta}{\beta_k} (\sem{\delta}{G_k}(\fold{\sem{\delta}{F}}{\sem{\delta}{\alpha}})\;V) = \sem{\delta}{t}}
      = \sem{\delta}{P_k} \tag{*}
    \end{equation}
    for $k = 1, 2$.
    \begin{itemize}
    \item ($\subseteq$)
      Suppose
      $V \in \sem{\delta}{G_1 \oplus G_2} (\mu\sem{\delta}{F})$
      and
      ${\sem{\delta}{\beta} (\sem{\delta}{G_1 \oplus G_2}(\fold{\sem{\delta}{F}}{\sem{\delta}{\alpha}})\;V) = \sem{\delta}{t}}$.
      By definition,
      $\sem{\delta}{G_1 \oplus G_2} (\mu\sem{\delta}{F})
      = \sem{\delta}{G_1} (\mu\sem{\delta}{F}) \uplus \sem{\delta}{G_2}(\mu\sem{\delta}{F})$,
      so, for some $k \in \{1, 2\}$,
      we have $V = (k, V_k)$ where $V_k \in \sem{\delta}{G_k} (\mu\sem{\delta}{F})$.
      Therefore,
      \begin{align*}
        \sem{\delta}{\beta_k} (\sem{\delta}{G_k}(\fold{\sem{\delta}{F}}{\sem{\delta}{\alpha}})\;V_k)
        &= \sem{\delta}{\beta} (\sem{\delta}{G_1 \oplus G_2}(\fold{\sem{\delta}{F}}{\sem{\delta}{\alpha}})\;(k, V_k)) \\
        &= \sem{\delta}{\beta} (\sem{\delta}{G_1 \oplus G_2}(\fold{\sem{\delta}{F}}{\sem{\delta}{\alpha}})\;V) \\
        &= \sem{\delta}{t}
      \end{align*}
      By (*), $V_k \in \sem{\delta}{P_k}$,
      so
      \begin{align*}
        V
        &= (k, V_k) \\
        &\in \sem{\delta}{P_1} \uplus \sem{\delta}{P_2} \\
        &= \sem{\delta}{P_1+P_2}
      \end{align*}
      as desired.
    \item ($\supseteq$)
      Suppose $V \in \sem{\delta}{P_1 + P_2}$.
      By definition,
      $\sem{\delta}{P_1 + P_2} = \sem{\delta}{P_1} \uplus \sem{\delta}{P_2}$.
      Hence, for some $k \in \{1, 2\}$, we have $V = (k, V_k)$
      where $V_k \in \sem{\delta}{P_k}$.
      By (*), 
      $V_k \in \sem{\delta}{G_k}(\mu\sem{\delta}{F})$
      and
      $\sem{\delta}{\beta_k} (\sem{\delta}{G_k}(\fold{\sem{\delta}{F}}{\sem{\delta}{\alpha}})\;V_k) = \sem{\delta}{t}$.
      Therefore,
      \begin{align*}
        V
        &= (k, V_k) \\
        &\in \sem{\delta}{G_1}(\mu\sem{\delta}{F}) \uplus\sem{\delta}{G_2} (\mu\sem{\delta}{F}) \\
        &= \sem{\delta}{G_1 \oplus G_2}(\mu\sem{\delta}{F})
      \end{align*}
      and
      \begin{align*}
        \sem{\delta}{\beta} (\sem{\delta}{G_1 \oplus G_2}(\fold{\sem{\delta}{F}}{\sem{\delta}{\alpha}})\;V)
        &= \sem{\delta}{\beta} (\sem{\delta}{G_1 \oplus G_2}(\fold{\sem{\delta}{F}}{\sem{\delta}{\alpha}})\;(k, V_k)) \\
        &= \sem{\delta}{\beta_k} (\sem{\delta}{G_k}(\fold{\sem{\delta}{F}}{\sem{\delta}{\alpha}})\;V_k) \\
        &= \sem{\delta}{t}
      \end{align*}
      as desired.
    \end{itemize}

    \DerivationProofCase{\DeclUnrollId}
    {
      \Dee :: \judgeunroll{\Xi,a:\tau}{\Theta, a:\tau}{\nu:\hat{P}[\mu F]}
      {(\clause{q}{t'})}{\hat{P}\;\Fold{F}{\alpha}\;\nu}{t}{Q}{\tau} 
    }
    {
      \judgeunroll*{\Xi}{\Theta}{\nu : (\Id\otimes\hat{P})[\mu F]}
      {(\clause{(a,q)}{t'})}
      {(\Id\otimes\hat{P})\;\Fold{F}{\alpha}\;\nu}{t}
      {\extype{a:\tau}
        {\comprehend{\nu:\mu F}{ \Fold{F}{\alpha}\,{\nu} =_\tau a } \times Q}}
      {\tau}
    }
    \begin{itemize}
      \item ($\subseteq$)
        Assume
        \[
          V
          \in
          \comprehend{V \in \sem{\delta}{\Id \otimes \hat{P}}(\mu\sem{\delta}{F})}
          {\sem{\delta}{\clause{(a,q)}{t'}} \left( \sem{\delta}{\Id \otimes \hat{P}} \;\Fold{\sem{\delta}{F}}{\sem{\delta}{\alpha}}\;V \right) = \sem{\delta}{t}}
        \]
        Then, by definition of denotation, there exist $V_1$ and $V_2$
        such that $V = (V_1, V_2)$
        and $V_1 \in \mu\sem{\delta}{F}$
        and $V_2 \in \sem{\delta}{\hat{P}} (\mu \sem{\delta}{F})$
        and $\sem{\delta,\Fold{\sem{\delta}{F}}{\sem{\delta}{\alpha}}\;V_1/a}{\clause{q}{t'}} (\sem{\delta}{\hat{P}}\;\Fold{\sem{\delta}{F}}{\sem{\delta}{\alpha}}\;V_2) = \sem{\delta}{t}$.

        By definition of denotation and \Lemmaref{lem:type-subset-erasure},
        it suffices to show there exists $d \in \sem{}{\tau}$ such that
        $V \in \sem{\delta,d/a}{\comprehend{\nu:\mu F}{ \Fold{F}{\alpha}\,{\nu} =_\tau a } \times Q}$.
        Put $d = \Fold{\sem{\delta}{F}}{\sem{\delta}{\alpha}}\;V_1$.
        By definition of denotation (noting $V = (V_1, V_2)$),
        it suffices to show
        \[
          V_1 \in \sem{\delta, d/a}{\comprehend{\nu:\mu F}{ \Fold{F}{\alpha}\,{\nu} =_\tau a }}
        \]
        and
        $V_2 \in \sem{\delta,d/a}{Q}$.
        The former follows from the facts that
        $V_1 \in \mu\sem{\delta}{F}$,
        $a$ is not free in $F$,
        and $d = \Fold{\sem{\delta}{F}}{\sem{\delta}{\alpha}}\;V_1$.
        The latter follows from the induction hypothesis
        for subderivation $\Dee$
        and semantic substitution $|- \delta, d/a : \Theta, a:\tau$,
        and from the facts that
        $V_2 \in \sem{\delta}{\hat{P}} (\mu \sem{\delta}{F})$,
        $a$ is not free in $F$, $\hat{P}$, $\alpha$, or $t$,
        and $\sem{\delta,d/a}{\clause{q}{t'}} (\sem{\delta}{\hat{P}}\;\Fold{\sem{\delta}{F}}{\sem{\delta}{\alpha}}\;V_2) = \sem{\delta}{t}$.
      \item ($\supseteq$)
        Assume
        $V \in \sem{\delta}{\extype{a:\tau}{\comprehend{\nu:\mu F}{ \Fold{F}{\alpha}\,{\nu} =_\tau a } \times Q}}$.
        By definition of denotation (and erasure)
        and the fact that $a$ is not free in $F$ or $\alpha$,
        \begin{align*}
          \hspace{1em}&\hspace{-1em}\sem{\delta}{\extype{a:\tau}{\comprehend{\nu:\mu F}{ \Fold{F}{\alpha}\,{\nu} =_\tau a } \times Q}} \\
          &= \comprehend{(V_1, V_2) \in \sem{}{\mu\erase{F}} \times \sem{}{\erase{Q}}}{\extype{d\in\sem{}{\tau}} V_1 \in \sem{\delta,d/a}{\comprehend{\nu:\mu F}{ \Fold{F}{\alpha}\,{\nu} =_\tau a }} \text{ and } V_2 \in \sem{\delta,d/a}{Q}}  \\
          &= \comprehend{(V_1, V_2) \in \sem{}{\mu\erase{F}} \times \sem{}{\erase{Q}}}{\extype{d\in\sem{}{\tau}} V_1 \in \comprehend{V \in \mu\sem{\delta}{F}}{\Fold{\sem{\delta}{F}}{\sem{\delta}{\alpha}}\;V = d} \text{ and } V_2 \in \sem{\delta,d/a}{Q}} 
        \end{align*}
        Therefore,
        there exist $d \in \sem{}{\tau}$, $V_1$, and $V_2$
        such that $V = (V_1, V_2)$
        and
        \begin{equation}
          V_1 \in \comprehend{V \in \mu\sem{\delta}{F}}{\Fold{\sem{\delta}{F}}{\sem{\delta}{\alpha}}\;V = d} \tag{A}
        \end{equation}
        and $V_2 \in \sem{\delta,d/a}{Q}$.

        Now, it suffices to show that
        $V_1 \in \mu\sem{\delta}{F}$
        and $V_2 \in \sem{\delta}{\hat{P}}(\mu\sem{\delta}{F})$
        and 
        \[
          \sem{\delta, \Fold{\sem{\delta}{F}}{\sem{\delta}{\alpha}}\;V_1/a}{\clause{q}{t'}} (\sem{\delta}{\hat{P}}\;\Fold{\sem{\delta}{F}}{\sem{\delta}{\alpha}}\;V_2) = \sem{\delta}{t}
        \]
        That
        $V_1 \in \mu\sem{\delta}{F}$
        follows from (A).
        By the induction hypothesis,
        the fact that $V_2 \in \sem{\delta,d/a}{Q}$,
        and the fact that $a$ is not free in $F$, $\hat{P}$, $\alpha$, or $t$,
        we have
        \begin{equation}
          V_2 \in \comprehend{V \in \sem{\delta}{\hat{P}}(\mu\sem{\delta}{F})}{\sem{\delta,d/a}{\clause{q}{t'}} (\sem{\delta}{\hat{P}}\;\Fold{\sem{\delta}{F}}{\sem{\delta}{\alpha}}\;V) = \sem{\delta}{t}} \tag{B}
        \end{equation}
        Therefore, $V_2 \in \sem{\delta}{\hat{P}}(\mu\sem{\delta}{F})$.
        Finally, by (A) we have
        $d = \Fold{\sem{\delta}{F}}{\sem{\delta}{\alpha}}\;V_1$,
        and so the equation we need follows from (B).
    \end{itemize}

  \item \textbf{Case} \DeclUnrollConstEx:
    \[
      \Infer{}
      {
        \Dee :: \judgeunroll{\Xi,a:\tau'}{\Theta,a:\tau'}{\nu:(\Const{Q}\otimes\hat{P})[\mu F]}{(\clause{(\bap,q)}{t'})}{(\Const{Q}\otimes\hat{P})\;\Fold{F}{\alpha}\;\nu}{t}{Q'}{\tau} }
      {
        \judgeunroll*{\Xi}{\Theta}{\nu :(\Const{\extype{a:\tau'}{Q}}\otimes\hat{P})[\mu F]}{(\clause{(\pack{a}{\bap}, q)}{t'})}{(\Const{\extype{a:\tau'}{Q}}\otimes\hat{P})\;\Fold{F}{\alpha}\;\nu}{t}{\extype{a:\tau'}{Q'}}{\tau}
      }
    \]
    \begin{itemize}
      \item ($\subseteq$)
        Assume that $V$ is in the left-hand side
        of the set equality we are proving, i.e. that
        \[
          V \in \sem{\delta}{\Const{\extype{a:\tau'}{Q}}\otimes\hat{P}} (\mu\sem{\delta}{F})
        \]
        and
        \[
          \sem{\delta}{\clause{(\pack{a}{\bap}, q)}{t'}} \left(\sem{\delta}{\Const{\extype{a:\tau'}{Q}}\otimes\hat{P}}(\fold{\sem{\delta}{F}}{\sem{\delta}{\alpha}})\;V\right) = \sem{\delta}{t}
        \]
        We want to show that $V \in \sem{\delta}{\extype{a:\tau'}{Q'}}$.
        By definition of denotation,
        $\sem{\delta}{\extype{a:\tau'}{Q'}}
        = \comprehend{V \in \sem{}{\erase{Q'}}}{\extype{d\in\sem{}{\tau'}}{V \in \sem{\delta,d/a}{Q'}}}$.
        So, by \Lemmaref{lem:type-subset-erasure},
        it suffices to show that there exists a $d \in \sem{}{\tau'}$
        such that $V \in \sem{\delta,d/a}{Q'}$.
        
        By definition of denotation,
        \begin{align*}
          \sem{\delta}{\Const{\extype{a:\tau'}{Q}}\otimes\hat{P}} (\mu\sem{\delta}{F})
          &= \sem{\delta}{\extype{a:\tau'}{Q}} \times \sem{\delta}{\hat{P}} (\mu\sem{\delta}{F}) \\
          &= \comprehend{V \in \sem{}{\erase{Q}}}{\extype{d\in\sem{}{\tau'}} \sem{\delta,d/a}{Q}} \times \sem{\delta}{\hat{P}} (\mu\sem{\delta}{F})
        \end{align*}
        Therefore, there exist $d \in \sem{}{\tau'}$, $V_1$, and $V_2$
        such that $V = (V_1, V_2)$
        and $V_1 \in \sem{\delta, d/a}{Q}$
        and $V_2 \in \sem{\delta}{\hat{P}} (\mu\sem{\delta}{F})$, and we have
        \begin{align*}
          V 
          &= (V_1, V_2) \\
          &\in \sem{\delta, d/a}{Q} \times \sem{\delta}{\hat{P}} (\mu\sem{\delta}{F}) \\
          &= \sem{\delta, d/a}{Q} \times \sem{\delta, d/a}{\hat{P}} (\mu\sem{\delta, d/a}{F}) &&\quad a \text{ not free in } F \text{ or } \hat{P} \\
          &= \sem{\delta, d/a}{\Const{Q} \otimes \hat{P}} (\mu\sem{\delta,d/a}{F})
        \end{align*}
        Further,
        \begin{align*}
          \sem{\delta}{t}
          &= \sem{\delta}{\clause{(\pack{a}{\bap}, q)}{t'}} \left(\sem{\delta}{\Const{\extype{a:\tau'}{Q}}\otimes\hat{P}}(\fold{\sem{\delta}{F}}{\sem{\delta}{\alpha}})\;V\right) &&\quad\text{Above} \\
          &= \sem{\delta}{\clause{(\pack{a}{\bap}, q)}{t'}} \left( \sem{\delta}{\Const{\extype{a:\tau'}{Q}}} \Fold{\sem{\delta}{F}}{\sem{\delta}{\alpha}}\;V_1, \sem{\delta}{\hat{P}} \Fold{\sem{\delta}{F}}{\sem{\delta}{\alpha}}\;V_2 \right) &&\quad\text{By \defn} \\
          &= \sem{\delta}{\clause{(\pack{a}{\bap}, q)}{t'}} \left( V_1, \sem{\delta}{\hat{P}} \Fold{\sem{\delta}{F}}{\sem{\delta}{\alpha}}\;V_2 \right) &&\quad\text{By \defn} \\
          &= \sem{\delta, d'/a}{\clause{(\bap, q)}{t'}} \left( V_1, \sem{\delta}{\hat{P}} \Fold{\sem{\delta}{F}}{\sem{\delta}{\alpha}}\;V_2 \right) &&\quad\text{By \defn}
        \end{align*}
        where $d' \in \sem{}{\tau'}$
        is such that $V_1 \in \sem{\delta, d'/a}{Q}$
        (such a $d'$ exists because $V_1 \in \sem{\delta}{\extype{a:\tau'}{Q}}$).
        By inversion on \DeclTpEx, there exists $\Xi'$
        such that $\judgetp{\Theta, a:\tau'}{Q}{\Xi', a:\tau'}$.
        By \Lemmaref{lem:value-determined-soundness}, $d' = d$.
        So, continuing, we have
        \begin{align*}
          \hspace{1em}&\hspace{-1em}\sem{\delta, d'/a}{\clause{(\bap, q)}{t'}} \left( V_1, \sem{\delta}{\hat{P}} \Fold{\sem{\delta}{F}}{\sem{\delta}{\alpha}}\;V_2 \right) \\
          &= \sem{\delta, d/a}{\clause{(\bap, q)}{t'}} \left( V_1, \sem{\delta}{\hat{P}} \Fold{\sem{\delta}{F}}{\sem{\delta}{\alpha}}\;V_2 \right) &&\quad\text{By }d' = d\text{ (above)} \\
          &= \sem{\delta, d/a}{\clause{(\bap, q)}{t'}} \left( \sem{\delta, d/a}{\Const{Q}} \Fold{\sem{\delta}{F}}{\sem{\delta}{\alpha}}\;V_1, \sem{\delta}{\hat{P}} \Fold{\sem{\delta}{F}}{\sem{\delta}{\alpha}}\;V_2 \right) &&\quad\text{By \defn of \defsem} \\
          &= \sem{\delta, d/a}{\clause{(\bap, q)}{t'}} \left( \sem{\delta, d/a}{\Const{Q} \otimes \hat{P}} \Fold{\sem{\delta}{F}}{\sem{\delta}{\alpha}}\;V \right) &&\quad\text{\defn and }a\notin\FV{\hat{P}} \\
          &= \sem{\delta, d/a}{\clause{(\bap, q)}{t'}} \left( \sem{\delta, d/a}{\Const{Q} \otimes \hat{P}} \Fold{\sem{\delta,d/a}{F}}{\sem{\delta,d/a}{\alpha}}\;V \right) &&\quad a \text{ not  free in }F\text{ or }\alpha
        \end{align*}
        Therefore (by above equations, and since $a$ is not free in $t$),
        \[
          \sem{\delta, d/a}{\clause{(\bap, q)}{t'}} \left( \sem{\delta, d/a}{\Const{Q} \otimes \hat{P}} \Fold{\sem{\delta,d/a}{F}}{\sem{\delta,d/a}{\alpha}}\;V \right)
          = \sem{\delta, d/a}{t}
        \]
        We have just shown that $V$ is in the left-hand side of the set equation
        we get by the induction hypothesis for subderivation $\Dee$
        and semantic substitution $|- \delta, d/a : \Theta, a:\tau'$;
        by the same set equation, $V \in \sem{\delta, d/a}{Q'}$,
        which is what we wanted to show.
      \item ($\supseteq$)
        Similar to the ``$\subseteq$'' part;
        also uses \Lemmaref{lem:value-determined-soundness}.
    \end{itemize}

    \DerivationProofCase{\DeclUnrollConst}
    {
      \judgeunroll{\Xi}{\Theta}{\nu:\hat{P}[\mu F]}{(\clause{q}{t'})}{\hat{P}\;\Fold{F}{\alpha}\;\nu}{t}{Q'}{\tau} }
    { \judgeunroll{\Xi}{\Theta}{\nu :(\Const{Q}\otimes\hat{P})[\mu F]}{(\clause{(\wild,q)}{t'})}{(\Const{Q}\otimes\hat{P})\;\Fold{F}{\alpha}\;\nu}{t}{Q \times Q'}{\tau} }
    The equations lacking justification below hold by definition of denotation:
    \begin{align*}
      \hspace{1em}&\hspace{-1em}
      \comprehend{V \in \sem{\delta}{\Const{Q}\otimes\hat{P}}(\mu\sem{\delta}{F})}{\sem{\delta}{\clause{(\wild,q)}{t'}} \left(\sem{\delta}{\Const{Q}\otimes\hat{P}}\;\Fold{\sem{\delta}{F}}{\sem{\delta}{\alpha}}\;V\right) = \sem{\delta}{t}} \\
      &= \comprehend{(V_1, V_2) \in \sem{\delta}{Q} \times \sem{\delta}{\hat{P}}(\mu\sem{\delta}{F})}{\sem{\delta}{\clause{(\wild,q)}{t'}} \left(\sem{\delta}{\Const{Q}\otimes\hat{P}}\;\Fold{\sem{\delta}{F}}{\sem{\delta}{\alpha}}\;(V_1, V_2)\right) = \sem{\delta}{t}} \\
      &= \comprehend{(V_1, V_2) \in \sem{\delta}{Q} \times \sem{\delta}{\hat{P}}(\mu\sem{\delta}{F})}{\sem{\delta}{\clause{q}{t'}} \left(\sem{\delta}{\hat{P}}\;\Fold{\sem{\delta}{F}}{\sem{\delta}{\alpha}}\;V_2\right) = \sem{\delta}{t}} \\
      &= \sem{\delta}{Q} \times \comprehend{V \in \sem{\delta}{\hat{P}}(\mu\sem{\delta}{F})}{\sem{\delta}{\clause{q}{t'}} \left(\sem{\delta}{\hat{P}}\;\Fold{\sem{\delta}{F}}{\sem{\delta}{\alpha}}\;V\right) = \sem{\delta}{t}} \qquad\text{By set theory} \\
      &= \sem{\delta}{Q} \times \sem{\delta}{Q'} \qquad\text{By \ih} \\
      &= \sem{\delta}{Q \times Q'} \qedhere
    \end{align*}
  \end{itemize}
\end{proof}

\begin{lemma}[Soundness of Type/Functor Equiv.]
  \label{lem:type-soundness-equiv}
  Assume $|- \delta : \Theta$.
  \begin{enumerate}
    \item If $\judgeequiv[\pm]{\Theta}{A}{B}$,
      then $\sem{\delta}{A} = \sem{\delta}{B}$.
    \item If $\judgeequiv[]{\Theta}{\mathcal{F}}{\mathcal{G}}$,
      then $\sem{\delta}{\mathcal{F}} = \sem{\delta}{\mathcal{G}}$.
  \end{enumerate}
\end{lemma}
\begin{proof}
  By mutual induction on the structure of the equivalence derivation,
  following the definition of the denotational semantics.
  \begin{enumerate}
  \item
    The \TpEquivPosWith and \TpEquivNegImp cases use \Lemmaref{lem:equiv-prop-meaning}.
    We show the hardest case (remaining cases are straightforward):
    \begin{itemize}
      \DerivationProofCase{\TpEquivPosFix}
      {
        \judgeequiv[]{\Theta}{F}{G}
        \\
        \judgeentail{\Theta}{t = t'}
      }
      {
        \judgeequiv[+]{\Theta}{\comprehend{\nu:\mu F}{\Fold{F}{\alpha}\,\nu =_\tau t}}{\comprehend{\nu:\mu G}{\Fold{G}{\alpha}\,\nu =_\tau t'}}
      }
      \begin{llproof}
        \judgeequivPf[]{\Theta}{F}{G}{Subderivation}
        \Pf{\sem{\delta}{F}}{=}{\sem{\delta}{G}}{By \ih (part (2))}
        \Pf{\mu\sem{\delta}{F}}{=}{\mu\sem{\delta}{G}}{By \Lemmaref{lem:mu-cong}}
        \Pf{\erase{F}}{=}{\erase{G}}{By \Lemmaref{lem:equivalence-erases-to-equality}}
        \judgeentailPf{\Theta}{t = t'}{Subderivation}
        \Pf{\sem{\delta}{t}}{=}{\sem{\delta}{t'}}{By inversion on \PropTrue and \defn}
      \end{llproof}
      ~\\
      Now,
      \begin{align*}
        \hspace{1em}&\hspace{-1em}\sem{\delta}{\comprehend{\nu:\mu F}{\Fold{F}{\alpha}\,\nu =_\tau t}}\\
        &= \comprehend{ V \in \mu\sem{}{\erase{F}} }{V \in \mu\sem{\delta}{F} \land (\fold{\sem{\delta}{F}}{\sem{\delta}{\alpha}})\;V = \sem{\delta}{t} } &&\quad\text{By \defn} \\
        &= \comprehend{ V \in \mu\sem{}{\erase{G}} }{V \in \mu\sem{\delta}{F} \land (\fold{\sem{\delta}{F}}{\sem{\delta}{\alpha}})\;V = \sem{\delta}{t} } &&\quad\erase{F}=\erase{G} \\
        &= \comprehend{ V \in \mu\sem{}{\erase{G}} }{V \in \mu\sem{\delta}{G} \land (\fold{\sem{\delta}{F}}{\sem{\delta}{\alpha}})\;V = \sem{\delta}{t} } &&\quad \mu\sem{\delta}{F}=\mu\sem{\delta}{G} \\
        &= \comprehend{ V \in \mu\sem{}{\erase{G}} }{V \in \mu\sem{\delta}{G} \land (\fold{\sem{\delta}{G}}{\sem{\delta}{\alpha}})\;V = \sem{\delta}{t} } &&\quad \text{\Lemref{lem:fold-cong}} \\
        &= \comprehend{ V \in \mu\sem{}{\erase{G}} }{V \in \mu\sem{\delta}{G} \land (\fold{\sem{\delta}{G}}{\sem{\delta}{\alpha}})\;V = \sem{\delta}{t'} } &&\quad\sem{\delta}{t}=\sem{\delta}{t'} \\
        &= \sem{\delta}{\comprehend{\nu:\mu G}{\Fold{G}{\alpha}\,\nu =_\tau t'}} &&\quad\text{By \defn}
      \end{align*}
    \end{itemize}
  \item Straightforward. \qedhere
  \end{enumerate}
\end{proof}

\begin{lemma}[Positive Value Extract]
  \label{lem:positive-value-extract}
  If $\judgetp{\Theta}{P}{\dontcare}$
  and $|- \delta : \Theta$
  and $V \in \sem{\delta}{P}$
  and $\judgeextract[+]{\Theta}{P}{P'}{\Theta'}$,
  then there exists $\delta'$
  such that $|- \delta, \delta' : \Theta, \Theta'$
  and $V \in \sem{\delta, \delta'}{P'}$.
\end{lemma}
\begin{proof}
  By structural induction on the given extraction derivation.
  We case analyze rules concluding it:
  \begin{itemize}
    \DerivationProofCase{\ExtractStopPos}
    {
      P \neq \exists, \land, \text{ or } \times
    }
    {
      \judgeextract[+]{\Theta}{P}{P}{\cdot}
    }
    Choose $\delta' = \cdot$.

    \DerivationProofCase{\ExtractWith}
    {
      \judgeextract[+]{\Theta}{P_0}{P'}{\Theta_0'}
    }
    {
      \judgeextract[+]{\Theta}{P_0 \land \phi}{P'}{\phi, \Theta_0'}
    }
    \begin{llproof}
      \Pf{V}{\in}{\sem{\delta}{P_0 \land \phi}}{Given}
      \Pf{V}{\in}{\sem{}{\erase{P_0}}}{By \defn of denotation}
      \Pf{V}{\in}{\sem{\delta}{P_0}}{\ditto}
      \Pf{\sem{\delta}{\phi}}{=}{\one}{\ditto}
      \judgeextractPf[+]{\Theta}{P_0}{P'}{\Theta_0'}{Subderivation}
      \judgsubsPf{}{\delta, \delta'}{\Theta, \Theta_0'}{By \ih}
      \Hand\Pf{V}{\in}{\sem{\delta, \delta'}{P'}}{\ditto}
      \Hand\judgsubsPf{}{\delta, \delta'}{\Theta, \phi, \Theta_0'}{By \Lemmaref{lem:sem-subs-deep-entry}}
    \end{llproof}

    \DerivationProofCase{\ExtractEx}
    {
      \judgeextract[+]{\Theta, a:\tau}{P_0}{P'}{\Theta_0'}
    }
    {
      \judgeextract[+]{\Theta}{\extype{a:\tau}{P_0}}{P'}{a:\tau, \Theta_0'}
    }
    \begin{llproof}
      \Pf{V}{\in}{\sem{\delta}{\extype{a:\tau}{P_0}}}{Given}
      \Pf{V}{\in}{\sem{}{\erase{P_0}}}{By \defn of denotation}
      \Pf{d}{\in}{\sem{}{\tau}}{\ditto}
      \Pf{V}{\in}{\sem{\delta, d/a}{P_0}}{\ditto}
      \judgsubsPf{}{\delta, d/a}{\Theta, a:\tau}{By \IxSem}
      \judgeextractPf[+]{\Theta, a:\tau}{P_0}{P'}{\Theta_0'}{Subderivation}
      \Hand\judgsubsPf{}{\delta, \underbrace{d/a, \delta_0'}_{\delta'}}{\Theta, a:\tau, \Theta_0'}{By \ih}
      \Hand\Pf{V}{\in}{\sem{\delta, \delta'}{P'}}{\ditto}
    \end{llproof}

    \DerivationProofCase{\ExtractProd}
    {
      \judgeextract[+]{\Theta}{P_1}{P_1'}{\Theta_1}
      \\
      \judgeextract[+]{\Theta}{P_2}{P_2'}{\Theta_2}
    }
    {
      \judgeextract[+]{\Theta}{P_1 \times P_2}{P_1' \times P_2'}{\Theta_1, \Theta_2}
    }
    \begin{llproof}
      \Pf{V}{\in}{\sem{\delta}{P_1 \times P_2}}{Given}
      \Pf{V}{=}{(V_1, V_2)}{By \defn of denotation}
      \Pf{V_1}{\in}{\sem{\delta}{P_1}}{\ditto}
      \Pf{V_2}{\in}{\sem{\delta}{P_2}}{\ditto}
      \judgeextractPf[+]{\Theta}{P_1}{P_1'}{\Theta_1}{Subderivation}
      \judgsubsPf{}{\delta, \delta_1}{\Theta, \Theta_1}{By \ih}
      \Pf{V_1}{\in}{\sem{\delta, \delta_1}{P_1'}}{\ditto}
      \judgeextractPf[+]{\Theta}{P_2}{P_2'}{\Theta_2}{Subderivation}
      \judgsubsPf{}{\delta, \delta_2}{\Theta, \Theta_2}{By \ih}
      \Pf{V_2}{\in}{\sem{\delta, \delta_2}{P_2'}}{\ditto}
      \Hand\judgsubsPf{}{\delta, \delta_1, \delta_2}{\Theta, \Theta_1, \Theta_2}{By \Lemmaref{lem:sem-subs-append}}
      \judgetpPf{\Theta, \Theta_1}{P_1'}{\dontcare}{By \Lemmaref{lem:extract-to-type-wf}}
      \judgetpPf{\Theta, \Theta_1, \Theta_2}{P_1'}{\dontcare}{By \Lemmaref{lem:ix-level-weakening}}
      \Pf{\sem{\delta, \delta_1}{P_1'}}{=}{\sem{\delta, \delta_1, \delta_2}{P_1'}}{By \Lemmaref{lem:tp-fun-alg-meaning-weakening-invariant}}
      \Pf{V_1}{\in}{\sem{\delta, \delta_1, \delta_2}{P_1'}}{By above equation}
      \Pf{V_2}{\in}{\sem{\delta, \delta_1, \delta_2}{P_2'}}{Similarly}
      \Pf{V}{=}{(V_1, V_2)}{Above}
      \Pf{}{\in}{\sem{\delta, \delta_1, \delta_2}{P_1'} \times \sem{\delta, \delta_1, \delta_2}{P_2'}}{By above}
      \Hand\Pf{}{=}{\sem{\delta, \delta_1, \delta_2}{P_1' \times P_2'}}{By \defn of denotation}
    \end{llproof}
    \qedhere
  \end{itemize}
\end{proof}

\begin{lemma}[Negative Value Extract]
  \label{lem:negative-value-extract}
  If $\judgetp{\Theta}{N}{\dontcare}$
  and $f \in \sem{}{\erase{N}}$\\
  and $\judgeextract[-]{\Theta}{N}{N'}{\Theta'}$,
  and $|- \delta : \Theta$\\
  and $f \in \sem{\delta, \delta'}{N'}$
        for all $\delta'$ such that $|- \delta, \delta' : \Theta, \Theta'$,\\
  then $f \in \sem{\delta}{N}$.
\end{lemma}
\begin{proof}
  By structural induction on $N$.
  \begin{itemize}
    \DerivationProofCase{\ExtractStopNeg}
    {
    }
    {
      \judgeextract[-]{\Theta}{\upshift{P}}{\upshift{P}}{\cdot}
    }
    We are given (in this case, $\Theta' = \cdot$ and $N' = N$):
    for all $\delta'$
    such that $|- \delta, \delta' : \Theta, \cdot$,
    we have $f \in \sem{\delta, \delta'}{N}$.
    Applying this to $\delta' = \cdot$, we get the desired result.

    \DerivationProofCase{\ExtractImp}
    {
      \judgeextract[-]{\Theta}{N_0}{N'}{\Theta_0'}
    }
    {
      \judgeextract[-]{\Theta}{\phi \implies N_0}{N'}{\phi, \Theta_0'}
    }
    By definition of denotation,
    $\sem{\delta}{\phi \implies N_0} = \comprehend{f \in \sem{}{\erase{N_0}}}{\sem{\delta}{\phi} = \one \implies f \in \sem{\delta}{N_0}}$.
    So, it suffices to show that
    $f \in \sem{}{\erase{N_0}}$
    and that if $\sem{\delta}{\phi} = \one$, then $f \in \sem{\delta}{N_0}$.

    Given $f \in \sem{}{\erase{\phi \implies N_0}}$,
    we have $f \in \sem{}{\erase{N_0}}$
    by definition of erasure.

    Assume $\sem{\delta}{\phi} = \one$.
    We are given $|- \delta : \Theta$.
    By \PropSem, $|- \delta : \Theta, \phi$.
    By \Lemmaref{lem:ix-level-weakening} on the given extraction subderivation,
    $\judgeextract[-]{\Theta, \phi}{N_0}{N'}{\Theta_0'}$.
    We are given:
    for all $\delta''$ such that $|- \delta, \delta'' : \Theta, \phi, \Theta_0'$,
    we have $f \in \sem{\delta, \delta''}{N'}$.
    By the induction hypothesis, $f \in \sem{\delta}{N_0}$, as desired.

    \DerivationProofCase{\ExtractAll}
    {
      \judgeextract[-]{\Theta, a:\tau}{N_0}{N'}{\Theta_0'}
    }
    {
      \judgeextract[-]{\Theta}{\alltype{a:\tau}{N_0}}{N'}{a:\tau, \Theta_0'}
    }
    By definition of denotation, it suffices to show that $f \in \sem{}{\erase{N_0}}$
    and that for all $d \in \sem{}{\tau}$, we have $f \in \sem{\delta, d/a}{N_0}$.

    Given $f \in \sem{}{\erase{\alltype{a:\tau} N_0}}$,
    we have $f \in \sem{}{\erase{N_0}}$
    by definition of erasure.

    Assume $d \in \sem{}{\tau}$.
    We are given $|- \delta : \Theta$.
    By \IxSem, $|- \delta, d/a : \Theta, a:\tau$.
    We are given:
    for all $\delta''$ such that $|- \delta, \delta'' : \Theta, a:\tau, \Theta_0'$,
    we have $f \in \sem{\delta, \delta''}{N'}$.
    Assume $\delta'$ is such that $|- \delta, d/a, \delta' : \Theta, a:\tau, \Theta_0'$.
    Then $\delta'' = d/a, \delta'$ is such that $|- \delta, \delta'' : \Theta, a:\tau, \Theta_0'$;
    by universal quantification elimination, we have $f \in \sem{\delta, d/a, \delta'}{N'}$.
    Therefore,
    for all $\delta'$ such that $|- \delta, d/a, \delta' : \Theta, a:\tau, \Theta_0'$,
    we have $f \in \sem{\delta, d/a, \delta'}{N'}$.
    We have the subderivation $\judgeextract[-]{\Theta, a:\tau}{N_0}{N'}{\Theta_0'}$.
    By the induction hypothesis, $f \in \sem{\delta, d/a}{N_0}$, as desired.

    \DerivationProofCase{\ExtractArrow}
    {
      \judgeextract[+]{\Theta}{P}{P'}{\Theta_1}
      \\
      \judgeextract[-]{\Theta}{N_0}{N_0'}{\Theta_2}
    }
    {
      \judgeextract[-]{\Theta}{P \to N_0}{P' \to N_0'}{\Theta_1, \Theta_2}
    }
    By definition of denotation, it suffices to show that $f \in \sem{}{\erase{P \to N_0}}$
    and that for all $V \in \sem{\delta}{P}$, we have $f(V) \in \sem{\delta}{N_0}$.

    We are already given the former.

    Assume $V \in \sem{\delta}{P}$.

    By \Lemmaref{lem:ix-level-weakening} on the well-formedness judgment
    $\judgetp{\Theta}{N_0}{\dontcare}$
    obtained by inversion on the presupposed derivation
    $\judgetp{\Theta}{P \to N_0}{\dontcare}$,
    we have $\judgetp{\Theta, \Theta_1}{N_0}{\dontcare}$.

    We have
    \begin{align*}
      f &\in \sem{}{\erase{P \to N_0}} &&\quad\text{Given} \\
        &= \sem{}{\erase{P} \to \erase{N_0}} &&\quad\text{By \defn of erasure} \\
        &= \sem{}{\erase{P}} \Rightarrow \sem{}{\erase{N_0}} &&\quad\text{By \defn of denotation}
    \end{align*}
    By \Lemmaref{lem:type-subset-erasure}, $V \in \sem{}{\erase{P}}$.
    Thus, $f(V) \in \sem{}{\erase{N_0}}$.

    By \Lemmaref{lem:ix-level-weakening} on the negative extraction subderivation,
    we have $\judgeextract[-]{\Theta, \Theta_1}{N_0}{N_0'}{\Theta_2}$.

    Consider the subderivation $\judgeextract[+]{\Theta}{P}{P'}{\Theta_1}$.
    By \Lemmaref{lem:positive-value-extract}, there exists $\delta_1$
    such that $|- \delta, \delta_1 : \Theta, \Theta_1$
    and $V \in \sem{\delta, \delta_1}{P'}$.
    Assume $\delta_2$ is such that $|- \delta, \delta_1, \delta_2 : \Theta, \Theta_1, \Theta_2$.
    We are given:
    for all $\delta'$ such that $|- \delta, \delta' : \Theta, \Theta_1, \Theta_2$,
    we have $f \in \sem{\delta, \delta'}{P' \to N_0'}$.
    By universal quantification elimination with $\delta' = \delta_1, \delta_2$,
    we have $f \in \sem{\delta, \delta_1, \delta_2}{P' \to N_0'}$.
    By definition of denotation, for all $V' \in \sem{\delta, \delta_1, \delta_2}{P'}$,
    we have $f(V') \in \sem{\delta, \delta_1, \delta_2}{N_0'}$.
    By \Lemmaref{lem:tp-fun-alg-meaning-weakening-invariant},
    $V \in \sem{\delta, \delta_1, \delta_2}{P'}$.
    By universal quantification elimination, $f(V) \in \sem{\delta, \delta_1, \delta_2}{N_0'}$.
    Therefore, for all $\delta_2$ such that $|- \delta, \delta_1, \delta_2 : \Theta, \Theta_1, \Theta_2$,
    we have $f(V) \in \sem{\delta, \delta_1, \delta_2}{N_0'}$.
    
    By the induction hypothesis, $f(V) \in \sem{\delta, \delta_1}{N_0}$.
    By \Lemmaref{lem:tp-fun-alg-meaning-weakening-invariant},
    $f(V) \in \sem{\delta}{N_0}$, as desired. \qedhere
  \end{itemize}
\end{proof}

\begin{lemma}[Soundness of Subtyping]
  \label{lem:type-soundness-sub}
  Assume $|- \delta : \Theta$.
  If $\judgesub[\pm]{\Theta}{A}{B}$,
  then $\sem{\delta}{A} \subseteq \sem{\delta}{B}$.
\end{lemma}
\begin{proof}
  By structural induction on the given subtyping derivation,
  considering cases for its concluding rule:
  \begin{itemize}
    \DerivationProofCase{\DeclSubPosVoid}
    { }
    {\judgesub[+]{\Theta}{0}{0}}
    $\sem{\delta}{0} = \emptyset \subseteq \emptyset = \sem{\delta}{0}$

    \DerivationProofCase{\DeclSubPosUnit}
    { }
    {\judgesub[+]{\Theta}{1}{1}}
    $\sem{\delta}{1} = \one \subseteq \one = \sem{\delta}{1}$

    \DerivationProofCase{\DeclSubPosProd}
    {
      \judgesub[+]{\Theta}{P_1}{Q_1}
      \\
      \judgesub[+]{\Theta}{P_2}{Q_2}
    }
    {\judgesub[+]{\Theta}{P_1 \times P_2}{Q_1 \times Q_2}}
    \begin{align*}
      \sem{\delta}{P_1 \times P_2}
      &= \sem{\delta}{P_1} \times \sem{\delta}{P_2} &&\quad\text{By \defn of denotation} \\
      &\subseteq \sem{\delta}{Q_1} \times \sem{\delta}{Q_2} &&\quad\text{By \ih and set theory} \\
      &= \sem{\delta}{Q_1 \times Q_2} &&\quad\text{By \defn of denotation}
    \end{align*}

    \DerivationProofCase{\DeclSubPosSum}
    {
      \judgeequiv[+]{\Theta}{P_1}{Q_1}
      \\
      \judgeequiv[+]{\Theta}{P_2}{Q_2}
    }
    {
      \judgesub[+]{\Theta}{P_1 + P_2}{Q_1 + Q_2}
    }
    \begin{llproof}
      \judgeequivPf[+]{\Theta}{P_k}{Q_k}{Subderivations}
      \Pf{\sem{\delta}{P_k}}{=}{\sem{\delta}{Q_k}}{By \Lemmaref{lem:type-soundness-equiv}}
      \Pf{\sem{\delta}{P_1 + P_2}}{=}{\sem{\delta}{P_1} \uplus \sem{\delta}{P_2}}{By \defn of denotation}
      \Pf{}{=}{\sem{\delta}{Q_1} \uplus \sem{\delta}{Q_2}}{By above equations}
      \Pf{}{=}{\sem{\delta}{Q_1 + Q_2}}{By \defn of denotation}
    \end{llproof}

    \DerivationProofCase{\DeclSubPosL}
    {
      \judgeextract[+]{\Theta}{P}{P'}{\Theta'}
      \\
      \Theta' \neq \cdot
      \\
      \judgesub[+]{\Theta, \Theta'}{P'}{Q}
    }
    {
      \judgesub[+]{\Theta}{P}{Q}
    }
    \begin{llproof}
      \Pf{V}{\in}{\sem{\delta}{P}}{Suppose}
      \judgeextractPf[+]{\Theta}{P}{P'}{\Theta'}{Subderivation}
      \judgsubsPf{}{\delta, \delta'}{\Theta, \Theta'}{By \Lemmaref{lem:positive-value-extract}}
      \Pf{V}{\in}{\sem{\delta, \delta'}{P'}}{\ditto}
      \judgesubPf[+]{\Theta, \Theta'}{P'}{Q}{Subderivation}
      \Pf{V}{\in}{\sem{\delta, \delta'}{Q}}{By \ih}
      \Pf{}{=}{\sem{\delta}{Q}}{By \Lemmaref{lem:tp-fun-alg-meaning-weakening-invariant}}
    \end{llproof} 

    \DerivationProofCase{\DeclSubPosWithR}
    {\judgesub[+]{\Theta}{P}{Q} \\ \judgeentail{\Theta}{\phi} }
    {\judgesub[+]{\Theta}{P}{Q \land \phi} }
    \begin{llproof}
      \Pf{V \in \sem{\delta}{\judgetp{\Theta}{P}{\Xi}}}{}{}{Suppose}
      \Pf{\judgesub[+]{\Theta}{P}{Q}}{}{}{Subderivation}
      \Pf{\sem{\delta}{\judgetp{\Theta}{P}{\Xi}}
        \subseteq \sem{\delta}{\judgetp{\Theta}{Q}{\Xi'}}}{}{}{By \ih}
      \Pf{V \in \sem{\delta}{\judgetp{\Theta}{Q}{\Xi'}}}{}{}{Follows from above}
      \Pf{\sem{\delta}{\judgetp{\Theta}{Q}{\Xi'}} \subseteq \sem{}{\erase{Q}}}{}{}{By \Lemmaref{lem:type-subset-erasure}}
      \Pf{V \in \sem{}{\erase{Q}}}{}{}{Follows from above}
      \Pf{\judgeentail{\Theta}{\phi}}{}{}{Subderivation}
      \Pf{\sem{\delta}{\phi} = \one}{}{}{By inversion on \PropTrue}
    \end{llproof} 
    \\
    Therefore,
    \begin{align*}
      V
      &\in \comprehend{V \in \sem{}{\erase{Q}}}{V \in \sem{\delta}{Q} \land \sem{\delta}{\phi} = \one} &&\quad\text{Follows from above} \\
      &= \sem{\delta}{Q \land \phi} &&\quad\text{By \defn of denotation}
    \end{align*}

    \DerivationProofCase{\DeclSubPosExR}
    { 
      \judgesub[+]{\Theta}{P}{[t/a]Q} 
      \\
      \judgeterm{\Theta}{t}{\tau} 
    }
    {\judgesub[+]{\Theta}{P}{\extype{a:\tau}{Q}} }
    \begin{llproof}
      \Pf{V \in \sem{\delta}{P}}{}{}{Suppose}
      \Pf{\judgesub[+]{\Theta}{P}{[t/a]Q}}{}{}{Subderivation}
      \Pf{\sem{\delta}{P} \subseteq \sem{\delta}{[t/a]Q}}{}{}{By \ih}
      \Pf{\judgeterm{\Theta}{t}{\tau}}{}{}{Subderivation}
      \Pf{\sem{\delta}{t} \in \sem{}{\tau}}{}{}{By \Lemmaref{lem:type-soundness-ix}}
    \end{llproof}
    \\
    But 
    \begin{align*}
      \sem{\delta}{[t/a]Q}
      &= \sem{\delta}{[\id_\Theta, t/a]Q} &&\quad\text{Identity substitution} \\
      &= \sem{\sem{\delta}{\id_\Theta, t/a}}{Q} &&\quad\text{\Lemmaref{lem:subs-soundness-wf}} \\
      &= \sem{\delta, \sem{\delta}{t}/a}{Q} &&\quad\text{By \defn of denotation}
    \end{align*}
    Therefore, $V \in \sem{\delta,\sem{\delta}{t}/a}{Q}$. Further:
    \\
    \begin{llproof}
      \Pf{\sem{\delta, \sem{\delta}{t}/a}{Q} \subseteq \sem{}{\erase{Q}}}{}{}{By \Lemmaref{lem:type-subset-erasure}}
      \Pf{V \in \sem{}{\erase{Q}}}{}{}{Follows from above}
    \end{llproof} 
    \\
    Therefore,
    \begin{align*}
      V
      &\in \comprehend{V \in \sem{}{\erase{Q}}}{\extype{d \in \sem{}{\tau}}{V \in \sem{\delta, d/a}{Q}}} &&\quad\text{Follows from above} \\
      &= \sem{\delta}{\extype{a:\tau}{Q}} &&\quad\text{By \defn of denotation}
    \end{align*}

    \DerivationProofCase{\DeclSubPosFix}
    { \judgeequiv{\Theta}{F}{G} \\ \judgeentail{\Theta}{t = t'} }
    { \judgesub[+]{\Theta}{\comprehend{\nu:\mu F}{\Fold{F}{\alpha}\,\nu =_\tau t}}{\comprehend{\nu:\mu G}{\Fold{G}{\alpha}\,\nu =_\tau t'}} }
    Similar to the \TpEquivPosFix case of \Lemmaref{lem:type-soundness-equiv}.

    \DerivationProofCase{\DeclSubPosDownshift}
    {
      \judgesub[-]{\Theta}{N}{M}
    }
    {\judgesub[+]{\Theta}{\downshift{N}}{\downshift{M}}}
    Straightforward (use \ih for subderivation $\judgesub[-]{\Theta}{N}{M}$
    and the definition of denotation).

    \ProofCaseRule{\DeclSubNegUpshift}
    Similar to case for dual rule \DeclSubPosDownshift.

    \DerivationProofCase{\DeclSubNegR}
    {
      \judgeextract[-]{\Theta}{M}{M'}{\Theta'}
      \\
      \Theta' \neq \cdot
      \\
      \judgesub[-]{\Theta, \Theta'}{N}{M'}
    }
    {
      \judgesub[-]{\Theta}{N}{M}
    }
    Assume $f \in \sem{\delta}{N}$. We want to show $f \in \sem{\delta}{M}$.

    We have:
    \begin{align*}
      f 
      &\in \sem{\delta}{N} &&\quad\text{Above} \\
      &\subseteq \sem{}{\erase{N}} &&\quad\text{By \Lemmaref{lem:type-subset-erasure}}\\
      &= \sem{}{\erase{M}} &&\quad\text{By \Lemmaref{lem:subtyping-erases-to-equality}}
    \end{align*}
    
    We are given $|- \delta : \Theta$,
    and have the subderivation $\judgeextract[-]{\Theta}{M}{M'}{\Theta'}$.

    Assume $\delta'$ is such that $|- \delta, \delta' : \Theta, \Theta'$.
    By the \ih on the subtyping subderivation,
    $\sem{\delta, \delta'}{N} \subseteq \sem{\delta, \delta'}{M'}$.
    By \Lemmaref{lem:tp-fun-alg-meaning-weakening-invariant},
    $\sem{\delta, \delta'}{N} = \sem{\delta}{N}$.
    Since $f \in \sem{\delta}{N}$, we have $f \in \sem{\delta, \delta'}{M'}$.
    Therefore, for all $\delta'$ such that $|- \delta, \delta' : \Theta, \Theta'$,
    we have $f \sem{\delta, \delta'}{M'}$.

    By \Lemmaref{lem:negative-value-extract},
    $f \in \sem{\delta}{M}$,
    as desired.

    \DerivationProofCase{\DeclSubNegImpL}
    {\judgesub[-]{\Theta}{N}{M} \\ \judgeentail{\Theta}{\phi} }
    {\judgesub[-]{\Theta}{\phi \implies N}{M} }
    \begin{llproof}
      \Pf{V \in \sem{\delta}{\phi \implies N}}{}{}{Suppose}
      \Pf{\judgeentail{\Theta}{\phi}}{}{}{Subderivation}
      \Pf{\sem{\delta}{\phi} = \one}{}{}{By inversion on \PropTrue}
      \Pf{V \in \sem{\delta}{N}}{}{}{By \defn of $\sem{\delta}{\phi \implies N}$}
      \Pf{\judgesub[-]{\Theta}{N}{M}}{}{}{Subderivation}
      \Pf{\sem{\delta}{N} \subseteq \sem{\delta}{M}}{}{}{By \ih}
      \Pf{V \in \sem{\delta}{M}}{}{}{Follows from above}
    \end{llproof} 

    \ProofCaseRule{\DeclSubNegAllL}
    Similar to case for dual rule \DeclSubPosExR.

    \DerivationProofCase{\DeclSubNegArrow}
    { \judgesub[+]{\Theta}{Q}{P} \\ \judgesub[-]{\Theta}{N}{M} }
    { \judgesub[-]{\Theta}{P \to N}{Q \to M} }  
    Suppose $f \in \sem{\delta}{P \to N}$.
    We want to show that $f \in \sem{\delta}{Q \to M}$.
 
    By \defn of denotation, $f \in \sem{}{\erase{P \to N}}$
    and for all $V \in \sem{\delta}{P}$, we have $f(V) \in \sem{\delta}{N}$.

    By \defn of denotation, it suffices to show that $f \in \sem{}{\erase{Q \to M}}$
    and that for all $V \in \sem{\delta}{Q}$, we have $f(V) \in \sem{\delta}{M}$.
    The former follows form \Lemmaref{lem:subtyping-erases-to-equality}.
    To prove the latter, suppose $V \in \sem{\delta}{Q}$.
    By the \ih for the positive subtyping subderivation, $V \in \sem{\delta}{P}$.
    Therefore, $f(V) \in \sem{\delta}{N}$.
    By the \ih for the negative subtyping subderivation,
    $f(V) \in \sem{\delta}{M}$. \qedhere
  \end{itemize}
\end{proof}

\begin{lemma}[Fold Continuous]
  \label{lem:fold-continuous}
  Assume $|- \delta : \Theta$ and $\judgealgebra{\cdot}{\Theta}{\alpha}{F}{\tau}$.
  \begin{enumerate}
  \item If $\sem{\delta}{\alpha}$ is monotone,
    then $\fold{\sem{\delta}{F}}{\sem{\delta}{\alpha}}$ is monotone.
  \item If $\sem{\delta}{\alpha}$ respects least upper bounds,
    then $\fold{\sem{\delta}{F}}{\sem{\delta}{\alpha}}$ respects least upper bounds.
  \item If $\sem{\delta}{\alpha}$ is continuous,
    then $\fold{\sem{\delta}{F}}{\sem{\delta}{\alpha}}$ is continuous.
  \end{enumerate}
\end{lemma}
\begin{proof}
  We only show part (1).
  Part (2) is similar.
  Part (3) follows from parts (1) and (2) and the definition of continuous.

  By definition of $\textit{fold}$ (\Defnref{def:fold}),
  it suffices to show that $\foldn{k}{\sem{\delta}{F}}{\sem{\delta}{\alpha}}$
  is monotone for all $k \in \kindnat$.
  
  The empty function $\foldn{0}{\sem{\delta}{F}}{\sem{\delta}{\alpha}}$
  vacuously is monotone.

  For the inductive step,
  we assume $\foldn{n}{\sem{\delta}{F}}{\sem{\delta}{\alpha}}$ is monotone
  and prove $\foldn{n+1}{\sem{\delta}{F}}{\sem{\delta}{\alpha}}$ is monotone.
  By definition,
  \[
    \foldn{n+1}{\sem{\delta}{F}}{\sem{\delta}{\alpha}}
    = \sem{\delta}{\alpha} \circ (\sem{\delta}{F} (\foldn{n}{\sem{\delta}{F}}{\sem{\delta}{\alpha}}))
  \]
  By \ih, $\foldn{n}{\sem{\delta}{F}}{\sem{\delta}{\alpha}}$ is monotone.
  It is straightforward to check that
  $\sem{\delta}{F}$ takes monotone functions to monotone functions.
  Therefore, $\sem{\delta}{F} (\foldn{n}{\sem{\delta}{F}}{\sem{\delta}{\alpha}})$
  is monotone.
  We are given that $\sem{\delta}{\alpha}$ is monotone.
  Because the composition of monotone functions is monotone,
  $\sem{\delta}{\alpha} \circ (\sem{\delta}{F} (\foldn{n}{\sem{\delta}{F}}{\sem{\delta}{\alpha}}))$ is monotone,
  which concludes the proof.
\end{proof}

\begin{lemma}
  \label{lem:aux-upward-closure}
  If $\judgefunctor{\Theta}{\mathcal{F}}{\dontcare}$
  and $\judgefunctor{\Theta}{F}{\dontcare}$
  and $k \in \kindnat$
  and $V \in \sem{}{\erase{\mathcal{F}}} (\sem{}{\erase{F}}^k \emptyset)$
  and $|- \delta : \Theta$
  and $V \in \sem{\delta}{\mathcal{F}} (\mu\sem{\delta}{F})$,
  then $V \in \sem{\delta}{\mathcal{F}} (\sem{\delta}{F}^k \emptyset)$.
\end{lemma}
\begin{proof}
  By lexicographic induction on,
  first, $k$, and,
  second, the structure of $\mathcal{F}$.
\end{proof}

\begin{lemma}[Upward Closure]
  \label{lem:upward-closure}
  Assume $|- \delta : \Theta$.
  \begin{enumerate}
  \item If $\judgealgebra{\Xi}{\Theta}{\alpha}{F}{\tau}$
    and $\Xi \subseteq \Theta$
    then $\sem{\delta}{\alpha}$ is monotone.
  \item If $\judgefunctor{\Theta}{\mathcal{F}}{\dontcare}$
    and $\judgefunctor{\Theta}{F}{\dontcare}$
    and $k \in \kindnat$\\
    and $V \in \sem{\delta}{\mathcal{F}} (\sem{\delta}{F}^k \emptyset)$
    and $\ord[\sem{}{\erase{\mathcal{F}}} (\sem{}{\erase{F}}^k \emptyset)]{V}{V'}$,\\
    then $V' \in \sem{\delta}{\mathcal{F}} (\sem{\delta}{F}^k \emptyset)$.
  \item If $\judgetp{\Theta}{A}{\dontcare}$
    and $V \in \sem{\delta}{A}$ and $\ord[\sem{}{\erase{A}}]{V}{V'}$,
    then $V' \in \sem{\delta}{A}$.
  \end{enumerate}
\end{lemma}
\begin{proof}
  By lexicographic induction on,
  first, $\size{A}$/$\size{F}$ (parts (1), (2) and (3), mutually), and,
  second, $\langle k, \mathcal{F} \text{ structure} \rangle$ (part (2)),
  where $\langle \dots \rangle$ denotes lexicographic order.
  We define $\size{-}$ in \Figureref{fig:size}.
  \begin{enumerate}
  \item
    By \Lemmaref{lem:type-soundness-algebra},
    $\sem{\delta}{\alpha} : \sem{\delta}{F} \sem{}{\tau} \to \sem{}{\tau}$.
    We case analyze the given algebra well-formedness derivation.

    \begin{itemize} 
      \DerivationProofCase{\DeclAlgExConstProd}
      {
        \judgealgebra{\Xi, a:\tau'}{\Theta, a:\tau'}{\clause{(\bap, q)}{t}}{(\Const{Q} \otimes \hat{P})}{\tau}
      }
      {
        \judgealgebra{\Xi}{\Theta}{\clause{(\pack{a}{\bap}, q)}{t}}{(\Const{\extype{a:\tau'}{Q}} \otimes \hat{P})}{\tau}
      }
      Suppose $\ord[\sem{\delta}{\Const{\extype{a:\tau'}{Q}} \otimes \hat{P}} \sem{}{\tau}]{V}{V'}$.
      By \defn of $\sem{}{-}$,
      $\sem{\delta}{\Const{\extype{a:\tau'}{Q}} \otimes \hat{P}} \sem{}{\tau} = \sem{\delta}{\extype{a:\tau'} Q} \times \sem{\delta}{\hat{P}} \sem{}{\tau}$.
      Therefore, there exist $V_1$ and $V_1'$ in $\sem{\delta}{\extype{a:\tau'} Q}$
      and $V_2$ and $V_2'$ in $\sem{\delta}{\hat{P}} \sem{}{\tau}$
      such that $V = (V_1, V_2)$ and $V' = (V_1', V_2')$
      and $\ord[\sem{\delta}{\extype{a:\tau'} Q}]{V_1}{V_1'}$
      and $\ord[\sem{\delta}{\hat{P}} \sem{}{\tau}]{V_2}{V_2'}$.

      On one hand,

      \begin{llproof}
        \judgealgebraPf{\Xi, a:\tau'}{\Theta, a:\tau'}{\clause{(\bap, q)}{t}}{(\Const{Q} \otimes \hat{P})}{\tau}{Subderivation}
        \proofsep
        \eqPf{\sem{\delta}{\clause{(\pack{a}{\bap}, q)}{t}} \; (V_1, V_2)}{\sem{\delta, d/a}{\clause{(\bap, q)}{t}} \; (V_1, V_2)}{By \defn (for some $d \in \sem{}{\tau}$\dots}
        \trailingjust{\dots such that $V_1 \in \sem{\delta, d/a}{Q}$)}
        \ordPf{}{\sem{\delta, d/a}{\clause{(\bap, q)}{t}} \; (V_1', V_2')}{By \ih}
      \end{llproof}

      On the other hand,

      \begin{llproof}
        \eqPf{\sem{\delta}{\clause{(\pack{a}{\bap}, q)}{t}} \; (V_1', V_2')}{\sem{\delta, d'/a}{\clause{(\bap, q)}{t}} \; (V_1', V_2')}{By \defn (for some $d' \in \sem{}{\tau}$\dots}
        \trailingjust{\dots such that $V_1' \in \sem{\delta, d'/a}{Q}$)}
      \end{llproof}

      By \defn of $\ord[\sem{\delta}{\extype{a:\tau'} Q}]{V_1}{V_1'}$,
      we have $\ord[\sem{}{\erase{Q}}]{V_1}{V_1'}$.
      Above, we have $V_1 \in \sem{\delta, d/a}{Q}$.
      By \ih (part (3)), $V_1' \in \sem{\delta, d/a}{Q}$.
      But (above) $V_1' \in \sem{\delta, d'/a}{Q}$,
      so, by \Lemmaref{lem:value-determined-soundness}, $d = d'$,
      which concludes this case.

      \item The remaining cases are straightforward.
    \end{itemize} 

  \item We case analyze $\mathcal{F}$.
    \begin{itemize}
      \ProofCaseThing{\mathcal{F} = I}

      \begin{llproof}
        \eqPf{\sem{}{I} (\sem{}{\erase{F}}^k \emptyset)}{\one}{By \defn}
        \ordPf[\one]{V}{V'}{Rewrite given}
        \inPf{V'}{\one}{By \defn of $\sqsubseteq$}
        \eqPf{}{\sem{\delta}{I} (\sem{\delta}{F}^k \emptyset)}{By \defn}
      \end{llproof} 

      \ProofCaseThing{\mathcal{F} = \Id}

      \begin{llproof}
        \eqPf{\sem{}{\Id} (\sem{}{\erase{F}}^k \emptyset)}{\sem{}{\erase{F}}^k \emptyset}{By \defn}
        \ordPf[\sem{}{\erase{F}} (\sem{}{\erase{F}}^{k-1} \emptyset)]{V}{V'}{Rewrite given}
        \inPf{V}{\sem{\delta}{F} (\sem{\delta}{F}^{k-1} \emptyset)}{Rewrite given similarly}
        \inPf{V'}{\sem{\delta}{F} (\sem{\delta}{F}^{k-1} \emptyset)}{By \ih}
        \eqPf{}{\sem{\delta}{\Id} (\sem{\delta}{F}^k \emptyset)}{By \defn}
      \end{llproof} 

      \ProofCaseThing{\mathcal{F} = \Const{Q}}

      \begin{llproof}
        \eqPf{\sem{}{\Const{Q}} (\sem{}{\erase{F}}^k \emptyset)}{\sem{}{Q}}{By \defn}
        \ordPf[\sem{}{Q}]{V}{V'}{Rewrite given}
        \inPf{V}{\sem{\delta}{Q}}{Rewrite given similarly}
        \inPf{V'}{\sem{\delta}{Q}}{By \ih}
        \eqPf{}{\sem{\delta}{\Const{Q}} (\sem{\delta}{F}^k \emptyset)}{By \defn}
      \end{llproof}

      \ProofCaseThing{\mathcal{F} = B \otimes \hat{P}}

      \begin{llproof}
        \eqPf{\sem{}{\erase{B} \otimes \erase{\hat{P}}} (\sem{}{\erase{F}}^k \emptyset)}{\sem{}{\erase{B}} (\sem{}{\erase{F}}^k \emptyset) \times \sem{}{\erase{\hat{P}}} (\sem{}{\erase{F}}^k \emptyset)}{By \defn}
        \ordPf[\sem{}{\erase{B}} (\sem{}{\erase{F}}^k \emptyset) \times \sem{}{\erase{\hat{P}}} (\sem{}{\erase{F}}^k \emptyset)]{V}{V'}{Rewrite given}
        \inPf{V}{\sem{\delta}{B} (\sem{\delta}{F}^k \emptyset) \times \sem{\delta}{\hat{P}} (\sem{\delta}{F}^k \emptyset)}{Rewrite given similarly}
        \eqPf{V}{(V_1, V_2)}{By inversion}
        \inPf{V_1}{\sem{\delta}{B} (\sem{\delta}{F}^k \emptyset) }{\ditto}
        \inPf{V_2}{\sem{\delta}{\hat{P}} (\sem{\delta}{F}^k \emptyset) }{\ditto}
        \inPf{V'}{\sem{}{\erase{B}} (\sem{}{\erase{F}}^k \emptyset) \times \sem{}{\erase{\hat{P}}} (\sem{}{\erase{F}}^k \emptyset)}{By \defn of $\sqsubseteq_\times$}
        \eqPf{V'}{(V_1', V_2')}{\ditto}
        \inPf{V_1'}{\sem{}{\erase{B}} (\sem{}{\erase{F}}^k \emptyset)}{\ditto}
        \inPf{V_2'}{\sem{}{\erase{\hat{P}}} (\sem{}{\erase{F}}^k \emptyset)}{\ditto}
        \ordPf[\sem{}{\erase{B}} (\sem{}{\erase{F}}^k \emptyset)]{V_1}{V_1'}{\ditto}
        \ordPf[\sem{}{\erase{\hat{P}}} (\sem{}{\erase{F}}^k \emptyset)]{V_2}{V_2'}{\ditto}
        \inPf{V_1'}{\sem{\delta}{B} (\sem{\delta}{F}^{k} \emptyset)}{By \ih}
        \inPf{V_2'}{\sem{\delta}{\hat{P}} (\sem{\delta}{F}^{k} \emptyset)}{By \ih}
        \eqPf{V'}{(V_1', V_2')}{Above}
        \inPf{}{\sem{\delta}{B} (\sem{\delta}{F}^{k} \emptyset) \times \sem{\delta}{\hat{P}} (\sem{\delta}{F}^{k} \emptyset)}{By set theory}
        \eqPf{}{\sem{\delta}{B \otimes \hat{P}} (\sem{\delta}{F}^{k} \emptyset)}{By \defn}
      \end{llproof} 

      \ProofCaseThing{\mathcal{F} = F_1 \oplus F_2}
      Similar to $\mathcal{F} = B \otimes \hat{P}$ case.
    \end{itemize}

  \item We case analyze $A$.
    \begin{itemize}
      \ProofCaseThing{A = \comprehend{\nu : \mu F}{\Fold{F}{\alpha}\,{\nu} =_\tau t}}

      \begin{llproof}
        \inPf{V}{\sem{\delta}{\comprehend{\nu : \mu F}{\Fold{F}{\alpha}\,{\nu} =_\tau t}}}{Given}
        \eqPf{}{\comprehend{ V \in \mu\sem{}{\erase{F}} }{V \in \mu\sem{\delta}{F} \text{ and } (\fold{\sem{\delta}{F}}{\sem{\delta}{\alpha}})\;V = \sem{\delta}{t} }}{By \defn}
        \decolumnizePf
        \inPf{V}{\mu\sem{\delta}{F}}{Follows from above}
        \eqPf{(\fold{\sem{\delta}{F}}{\sem{\delta}{\alpha}})\;V}{\sem{\delta}{t}}{\ditto}
        \proofsep
        \ordPf[\sem{}{\mu \erase{F}}]{V}{V'}{Given}
        \ordPf[\sem{}{\erase{F}} (\sem{}{\erase{F}}^n \emptyset)]{V}{V'}{By \defn of $\sqsubseteq$, there exists such an $n$}
        \inPf{V}{\sem{}{\erase{F}} (\sem{}{\erase{F}}^n \emptyset)}{\ditto}
        \inPf{V}{\mu\sem{\delta}{F}}{Above}
        \eqPf{}{\sem{\delta}{F} (\mu\sem{\delta}{F})}{By \Lemmaref{lem:mu-unroll-equal}}
        \inPf{V}{\sem{\delta}{F} (\sem{\delta}{F}^n \emptyset)}{By \Lemref{lem:aux-upward-closure}}
        \proofsep
        \inPf{V'}{\sem{\delta}{F} (\sem{\delta}{F}^n \emptyset)}{By \ih}
        \subseteqPf{}{\cup_{k\in\kindnat}\sem{\delta}{F}^k \emptyset}{By set theory}
        \eqPf{}{\mu\sem{\delta}{F}}{By \defn}
        \subseteqPf{}{\mu\sem{}{\erase{F}}}{By \Lemmaref{lem:mu-subset-erasure}}
        \decolumnizePf
        \judgetpPf{\Theta}{\comprehend{\nu : \mu F}{\Fold{F}{\alpha}\,{\nu} =_\tau t}}{\dontcare}{Given}
        \judgealgebraPf{\cdot}{\Theta}{\alpha}{F}{\tau}{By inversion}
        \Pf{}{}{\sem{\delta}{\alpha}\text{ monotone}}{By \ih}
        \Pf{\sem{\delta}{\alpha}}{:}{\sem{\delta}{F} \sem{}{\tau} \to \sem{}{\tau}}{By \Lemmaref{lem:type-soundness-algebra}}
        \Pf{}{}{(\fold{\sem{\delta}{F}}{\sem{\delta}{\alpha}})\text{ monotone}}{By \Lemmaref{lem:fold-continuous}}
        \decolumnizePf
        \ordPf[\sem{}{\mu \erase{F}}]{V}{V'}{Above}
        \ordPf[\sem{}{\tau}]{(\fold{\sem{\delta}{F}}{\sem{\delta}{\alpha}})\;V}{(\fold{\sem{\delta}{F}}{\sem{\delta}{\alpha}})\;V'}{By \defn of monotone}
        \eqPf{\sem{\delta}{t}}{(\fold{\sem{\delta}{F}}{\sem{\delta}{\alpha}})\;V}{Above}
        \eqPf{}{(\fold{\sem{\delta}{F}}{\sem{\delta}{\alpha}})\;V'}{$\sem{}{\tau}$ has discrete order}
        \decolumnizePf
        \inPf{V'}{\sem{\delta}{\comprehend{\nu : \mu F}{\Fold{F}{\alpha}\,{\nu} =_\tau t}}}{Follows from above}
      \end{llproof} 

      \item The remaining cases are straightforward. \qedhere
    \end{itemize}
  \end{enumerate}
\end{proof}

\subsection{Refined Type and Substitution Soundness}

\begin{lemma}[Sem.\ Subs.\ Prog.\ Entry]
  \label{lem:sem-subs-deep-entry-prog}
  If $|- \delta, V/x : \Theta; \Gamma, x:P$
  and $|- \delta, \delta' : \Theta, \Theta'; \Gamma$,\\
  then $|- \delta, V/x, \delta' : \Theta, \Theta'; \Gamma, x:P$.
\end{lemma}
\begin{proof}
  By structural induction on the derivation of
  $|- \delta, \delta' : \Theta, \Theta'; \Gamma$.
  The \ValSem case is impossible by the requirement that there is exactly
  one entry for each variable in $\delta, \delta'$.
\end{proof}

\begin{lemma}[Sem.\ Subs.\ Ix.\ Extend]
  \label{lem:sem-subs-ix-extend}
  If $|- \delta : \Theta; \Gamma$
  and $|- \filterprog{\delta}, \delta' : \Theta, \Theta'$,\\
  then $|- \delta, \delta' : \Theta, \Theta'; \Gamma$.
\end{lemma}
\begin{proof}
  By structural induction on the derivation of $|- \delta : \Theta, \Gamma$.
  The \ValSem case uses \Lemmaref{lem:sem-subs-deep-entry-prog}.
\end{proof}

\begin{theorem}[Type Soundness]
  \label{thm:type-soundness}
  Assume $|- \delta : \Theta ; \Gamma$. Then:
  \begin{enumerate}
    \item If $\Dee :: \judgesynhead{\Theta}{\Gamma}{h}{P}$,
      then $\sem{\delta}{h} \in \sem{\filterprog{\delta}}{P}$.
    \item If $\Dee :: \judgesynexp{\Theta}{\Gamma}{\be}{\upshift{P}}$,
      then $\sem{\delta}{\be} \in \sem{\filterprog{\delta}}{\upshift{P}}$.
    \item If $\Dee :: \judgechkval{\Theta}{\Gamma}{v}{P}$,
      then $\sem{\delta}{v} \in \sem{\filterprog{\delta}}{P}$.
    \item If $\Dee :: \judgechkexp{\Theta}{\Gamma}{e}{N}$,
      then $\sem{\delta}{e} \in \sem{\filterprog{\delta}}{N}$.
    \item If $\Dee :: \judgechkmatch{\Theta}{\Gamma}{P}{\clauses{\pa}{e}{i}{I}}{N}$,
      then $\sem{\delta}{\clauses{\pa}{e}{i}{I}} \in \sem{\filterprog{\delta}}{P} \Rightarrow \sem{\filterprog{\delta}}{N}$.
    \item If $\Dee :: \judgespine{\Theta}{\Gamma}{s}{N}{\upshift{P}}$,
      then $\sem{\delta}{s} \in \sem{\filterprog{\delta}}{N} \Rightarrow \sem{\filterprog{\delta}}{\upshift{P}}$.
  \end{enumerate}
\end{theorem}
\begin{proof}
  By mutual induction on the structure of $\Dee$.
  Note that, throughout the proof,
  we implicitly use
  \Lemmaref{lem:filter-out-prog-vars},
  \Lemmaref{lem:erasure-respects-subs},
  and \Lemmaref{lem:filter-out-props}.
  By \Lemmaref{lem:unrefined-type-soundness},
  we only need to show that the denotation of the program term is in the refined set
  (as stated in the theorem).
  \begin{enumerate}
  \item
    \begin{itemize}
      \DerivationProofCase{\DeclSynHeadVar}
      {
        (x : P) \in \Gamma
      }
      {
        \judgesynhead{\Theta}{\Gamma}{x}{P}
      }
      \begin{llproof}
        \Pf{\sem{\delta}{x}}{=}{\sem{\erase{\delta}}{\erase{x}}}{By \defn of \defsem}
        \Pf{}{=}{\sem{\erase{\delta}}{x}}{By \defn of \deferase}
        \Pf{}{=}{(\erase{\delta})(x)}{By \defn of \defsem}
        \Pf{}{=}{\delta(x)}{By \defn of \deferase}
        \Pf{}{\in}{\sem{\filterprog{\delta}}{P}}{By inversion on $|- \delta : \Theta; \Gamma$ and $(x : P) \in \Gamma$,}
        \trailingjust{using \Lemmaref{lem:tp-fun-alg-meaning-weakening-invariant} if needed}
      \end{llproof} 

      \DerivationProofCase{\DeclSynValAnnot}
      {
        \judgetp{\Theta}{P}{\Xi}
        \\
        \judgechkval{\Theta}{\Gamma}{v}{P}
      }
      {
        \judgesynhead{\Theta}{\Gamma}{\annoexp{v}{P}}{P}
      }
      \begin{llproof}
        \Pf{|- \delta : \Theta;\Gamma}{}{}{Given}
        \Pf{\judgechkval{\Theta}{\Gamma}{v}{P}}{}{}{Subderivation}
        \proofsep
        \Pf{\sem{\delta}{\annoexp{v}{P}}}{=}{\sem{\erase{\delta}}{\erase{\annoexp{v}{P}}}}{By \defn of \defsem}
        \Pf{}{=}{\sem{\erase{\delta}}{\annoexp{\erase{v}}{\erase{P}}}}{By \defn of \deferase}
        \Pf{}{=}{\sem{\erase{\delta}}{\erase{v}}}{By \defn of \defsem}
        \Pf{}{=}{\sem{\delta}{v}}{By \defn of \defsem}
        \Pf{}{\in}{\sem{\filterprog{\delta}}{P}}{By \ih}
      \end{llproof}
    \end{itemize}

  \item
    \begin{itemize}
      \DerivationProofCase{\DeclSynSpineApp}
      { \judgesynhead{\Theta}{\Gamma}{h}{\downshift{N}} \\
        \judgespine{\Theta}{\Gamma}{s}{N}{\upshift{P}} }
      { \judgesynexp{\Theta}{\Gamma}{h(s)}{\upshift{P}} }
      \begin{llproof}
        \judgsubsPf{}{\delta}{\Theta;\Gamma}{Given}
        \judgesynheadPf{\Theta}{\Gamma}{h}{\downshift{N}}{Subderivation}
        \Pf{\sem{\delta}{h}}{\in}{\sem{\filterprog{\delta}}{\downshift{N}}}{By \ih}
        \Pf{}{=}{\sem{\filterprog{\delta}}{N}}{By \defn of \defsem}
        \judgespinePf{\Theta}{\Gamma}{s}{N}{\upshift{P}}{Subderivation}
        \Pf{\sem{\delta}{s}}{\in}{\sem{\filterprog{\delta}}{N} \Rightarrow \sem{\filterprog{\delta}}{\upshift{P}}}{By \ih}
        \proofsep
        \Pf{\sem{\delta}{h(s)}}{=}{\sem{\erase{\delta}}{\erase{h(s)}}}{By \defn of \defsem}
        \Pf{}{=}{\sem{\erase{\delta}}{\erase{h} (\erase{s})}}{By \defn of \deferase}
        \Pf{}{=}{\sem{\erase{\delta}}{\erase{s}} \sem{\erase{\delta}}{\erase{h}}}{By \defn of \defsem}
        \Pf{}{=}{\sem{\delta}{s} \sem{\delta}{h}}{By \defn of \defsem}
        \Pf{}{\in}{\sem{\filterprog{\delta}}{\upshift{P}}}{Function application}
      \end{llproof}

      \DerivationProofCase{\DeclSynExpAnnot}
      {
        \judgetp{\Theta}{P}{\Xi}
        \\
        \judgechkexp{\Theta}{\Gamma}{e}{\upshift{P}}
      }
      { \judgesynexp{\Theta}{\Gamma}{\annoexp{e}{\upshift{P}}}{\upshift{P}} }
      Similar to the \DeclSynValAnnot case of part (1).
    \end{itemize}

  \item
    \begin{itemize}
      \DerivationProofCase{\DeclChkValVar}
      { P \neq \exists, \land \\ (x:Q) \in \Gamma \\ \judgesub[+]{\Theta}{Q}{P} }
      { \judgechkval{\Theta}{\Gamma}{x}{P} }
      \begin{llproof}
        \judgsubsPf{}{\delta}{\Theta;\Gamma}{Given}
        \Pf{\sem{\delta}{x}}{=}{\sem{\erase{\delta}}{\erase{x}}}{By \defn of \defsem}
        \Pf{}{=}{\sem{\erase{\delta}}{x}}{By \defn of \deferase}
        \Pf{}{=}{(\erase{\delta})(x)}{By \defn of \defsem}
        \Pf{}{=}{\delta(x)}{By \defn of \deferase}
        \Pf{}{\in}{\sem{\filterprog{\delta}}{Q}}{By inversion on $|- \delta : \Theta; \Gamma$ and subderivation $(x:Q) \in \Gamma$,}
        \trailingjust{using \Lemmaref{lem:tp-fun-alg-meaning-weakening-invariant} if needed}
        \Pf{}{\subseteq}{\sem{\filterprog{\delta}}{P}}{By \Lemmaref{lem:type-soundness-sub}}
        \trailingjust{for subderivation $\judgesub[+]{\Theta}{Q}{P}$}
      \end{llproof} 

      \ProofCaseRule{\DeclChkValUnit}
      Straightforward.

      \ProofCaseRule{\DeclChkValPair}
      Straightforward.

      \ProofCaseRule{\DeclChkValIn{k}}
      Straightforward.

      \DerivationProofCase{\DeclChkValExists}
      {
        \judgechkval{\Theta}{\Gamma}{v}{[t/a]P_0}
        \\
        \judgeterm{\Theta}{t}{\tau}
      }
      { \judgechkval{\Theta}{\Gamma}{v}{(\extype{a:\tau} P_0)} }
      \begin{llproof}
        \judgetermPf{\Theta}{t}{\tau}{Subderivation}
        \Pf{\sem{\filterprog{\delta}}{t}}{\in}{\sem{}{\tau}}{By \Lemmaref{lem:type-soundness-ix}}
        \proofsep
        \Pf{\sem{\delta}{v}}{=}{\sem{\erase{\delta}}{\erase{v}}}{By \defn of \defsem}
        \Pf{}{=}{\sem{\delta}{\judgechkval{\Theta}{\Gamma}{v}{[t/a]P_0}}}{By \defn of \defsem}
        \Pf{}{\in}{\sem{\filterprog{\delta}}{[t/a]P_0}}{By \ih}
        \Pf{}{=}{\sem{\filterprog{\delta}}{[\id_\Theta, t/a]P_0}}{By \Lemref{lem:id-subs-id}}
        \Pf{}{=}{\sem{\sem{\filterprog{\delta}}{\id_\Theta, t/a}}{P_0}}{By \Lemref{lem:subs-soundness-wf}}
        \Pf{}{=}{\sem{\filterprog{\delta}, \sem{\filterprog{\delta}}{t}}{P_0}}{By \defn of \defsem}
        \Pf{}{\subseteq}{\comprehend{V \in \sem{}{\erase{P_0}}}{\extype{d\in\sem{}{\tau}}{V \in \sem{\filterprog{\delta},d/a}{P_0}}}}{By \Lemmaref{lem:type-subset-erasure}}
        \trailingjust{(and because $\sem{\filterprog{\delta}}{t} \in \sem{}{\tau}$, above)}
        \Pf{}{=}{\sem{\filterprog{\delta}}{\extype{a:\tau}P_0}}{By \defn of \defsem}
      \end{llproof} 

      \ProofCaseRule{\DeclChkValWith}
      Similar to case for \DeclChkValExists.

      \DerivationProofCase{\DeclChkValFix}
      { \judgeunroll{\cdot}{\Theta}{\nu:F[\mu F]}{\alpha}{F\;\Fold{F}{\alpha}\;\nu}{t}{Q}{\tau}
        \\ 
        \judgechkval{\Theta}{\Gamma}{v_0}{Q} }
      { \judgechkval{\Theta}{\Gamma}{\roll{v_0}}
        {\comprehend{\nu:\mu F}{\Fold{F}{\alpha}\,{\nu} =_\tau t}} }
      \begin{align*}
        \hspace{1em}&\hspace{-1em}\sem{\delta}{\roll{v_0}} \\
        &= \sem{\erase{\delta}}{\erase{\roll{v_0}}} &&\quad\text{By \defn of \defsem} \\
        &= \sem{\erase{\delta}}{\roll{\erase{v_0}}} &&\quad\text{By \defn of \deferase} \\
        &={\sem{\erase{\delta}}{\erase{v_0}}} &&\quad\text{By \defn of \defsem}\\
        &={\sem{\delta}{\judgechkval{\Theta}{\Gamma}{v_0}{Q}}} &&\quad\text{By \defn of \defsem} \\
        &\in{\sem{\filterprog{\delta}}{Q}} &&\quad \text{By \ih} \\
        &={\comprehend{V \in \sem{\filterprog{\delta}}{F} (\mu\sem{\filterprog{\delta}}{F})}{\sem{\filterprog{\delta}}{\alpha} (\sem{\filterprog{\delta}}{F}(\fold{\sem{\filterprog{\delta}}{F}}{\sem{\filterprog{\delta}}{\alpha}})\;V) = \sem{\filterprog{\delta}}{t}}} &&\quad\text{By \Lemref{lem:unroll-soundness}} \\
        &= \comprehend{V \in \mu\sem{\filterprog{\delta}}{F}}{(\fold{\sem{\filterprog{\delta}}{F}}{\sem{\filterprog{\delta}}{\alpha}})\;V = \sem{\filterprog{\delta}}{t}} &&\quad\text{By \Lemref{lem:refined-mu-unroll-equal}} \\
        &\subseteq \comprehend{V \in \mu\sem{}{\erase{F}}}{V \in \mu\sem{\filterprog{\delta}}{F} \land (\fold{\sem{\filterprog{\delta}}{F}}{\sem{\filterprog{\delta}}{\alpha}})\;V = \sem{\filterprog{\delta}}{t}} &&\quad\text{By \Lemref{lem:mu-subset-erasure}} \\
        &= \sem{\filterprog{\delta}}{\comprehend{\nu:\mu F}{\Fold{F}{\alpha}\,{\nu} =_\tau t}} &&\quad\text{By \defn of \defsem}
      \end{align*} 

      \ProofCaseRule{\DeclChkValDownshift}
      Straightforward.
    \end{itemize}

  \item
    \begin{itemize}
      \ProofCaseRule{\DeclChkExpUpshift}
      Straightforward.

      \DerivationProofCase{\DeclChkExpLet}
      { \simple{\Theta}{N} \\
        \judgesynexp{\Theta}{\Gamma}{\be}{\upshift{P}} \\
        \judgeextract[+]{\Theta}{P}{P'}{\Theta'} \\
        \judgechkexp{\Theta, \Theta'}{\Gamma, x:P'}{e_0}{N} }
      { \judgechkexp{\Theta}{\Gamma}{\Let{x}{\be}{e_0}}{N} }
      \begin{llproof}
        \judgesynexpPf{\Theta}{\Gamma}{\be}{\upshift{P}}{Subderivation}
        \Pf{\sem{\delta}{\be}}{\in}{\sem{\filterprog{\delta}}{\upshift{P}}}{By \ih}
        \Pf{}{=}{\comprehend{(1, V)}{V \in \sem{\filterprog{\delta}}{P}}}{By \defn of \defsem}
      \end{llproof}

      Therefore, there exists $V \in \sem{\filterprog{\delta}}{P}$
      such that $\sem{\delta}{g} = (1, V)$.
      By definition, $\sem{\delta}{g} = \sem{\erase{\delta}}{\erase{g}}$.
      By transitivity, $\sem{\erase{\delta}}{\erase{g}} = (1, V)$.

      \begin{llproof}
        \judgeextractPf[+]{\Theta}{P}{P'}{\Theta'}{Subderivation}
        \judgsubsPf{}{\filterprog{\delta}, \delta'}{\Theta, \Theta'}{By \Lemmaref{lem:positive-value-extract}}
        \Pf{V}{\in}{\sem{\underbrace{\filterprog{\delta}, \delta'}_{\filterprog{\delta, \delta'}}}{P'}}{\ditto}
        \judgsubsPf{}{\delta}{\Theta; \Gamma}{Given}
        \judgsubsPf{}{\delta, \delta'}{\Theta, \Theta'; \Gamma}{By \Lemmaref{lem:sem-subs-ix-extend}}
        \judgsubsPf{}{\delta, \delta', V/x}{\Theta, \Theta'; \Gamma, x:P'}{By \ValSem}
        \judgechkexpPf{\Theta, \Theta'}{\Gamma, x:P'}{e_0}{N}{Subderivation}
        \proofsep
        \Pf{\sem{\delta}{\Let{x}{\be}{e_0}}}{=}{\sem{\erase{\delta}}{\erase{\Let{x}{\be}{e_0}}}}{By \defn of \defsem}
        \Pf{}{=}{\sem{\erase{\delta}}{\Let{x}{\erase{\be}}{\erase{e_0}}}}{By \defn of \deferase}
        \Pf{}{=}{\sem{\erase{\delta}, V/x}{\erase{e_0}}}{By \defn of \defsem}
        \Pf{}{=}{\sem{\erase{\delta, \delta', V/x}}{\erase{e_0}}}{By \defn of \deferase}
        \Pf{}{=}{\sem{\delta, \delta', V/x}{e_0}}{By \defn of \defsem}
        \Pf{}{\in}{\sem{\filterprog{\delta, \delta', V/x}}{N}}{By \ih}
        \Pf{}{=}{\sem{\filterprog{\delta}, \delta'}{N}}{By \defn of \deffilterprog}
        \Pf{}{=}{\sem{\filterprog{\delta}}{N}}{$\FV{N} \sect \dom{\Theta'} = \emptyset$}
      \end{llproof}

      \ProofCaseRule{\DeclChkExpMatch}
      Straightforward.

      \ProofCaseRule{\DeclChkExpLam}
      Similar to \DeclChkExpLet case, but simpler.

      \DerivationProofCase{\DeclChkExpRec}
      {
        \arrayenvb{
          \simple{\Theta}{N}
          \\
          \judgesub[-]{\Theta}{\alltype{a:\kindnat} M}{N}
        }
        \\
        \judgechkexp{\Theta, a:\kindnat}{\Gamma, x:\downshift{\alltype{a':\kindnat} a' < a \implies [a'/a]M}}{e_0}{M}
      }
      {
        \judgechkexp{\Theta}{\Gamma}{\rec{x : (\alltype{a:\kindnat} M)}{e_0}}{N}
      }
      For all $k \in \kindnat$,
      define $X_k = \sem{\filterprog{\delta}, k/a}{\downshift{\alltype{a':\kindnat} a' < a \implies [a'/a]M}}$.
      For each $k$,

      \begin{llproof}
        \eqPf{X_k}{\sem{\filterprog{\delta}, k/a}{\downshift{\alltype{a':\kindnat} a' < a \implies [a'/a]M}}}{By \defn of $X_k$}
        \eqPf{}{\sem{\filterprog{\delta}, k/a}{\alltype{a':\kindnat} a' < a \implies [a'/a]M}}{By \defn of \defsem}
        \eqPf{}{\comprehend{f \in \sem{}{\erase{[a'/a]M}}}{\alltype{n < k} f \in \sem{\filterprog{\delta}, k/a, n/a'}{[a'/a]M}}}{By \defn of \defsem}
        \eqPf{}{\comprehend{f \in \sem{}{\erase{M}}}{\alltype{n < k} f \in \sem{\filterprog{\delta}, k/a, n/a'}{[a'/a]M}}}{By \Lemref{lem:erasure-subs}}
        \eqPf{}{\comprehend{f \in \sem{}{\erase{M}}}{\alltype{n < k} f \in \sem{\filterprog{\delta}, n/a'}{[a'/a]M}}}{By \Lemref{lem:tp-fun-alg-meaning-weakening-invariant}}
        \eqPf{}{\comprehend{f \in \sem{}{\erase{M}}}{\alltype{n < k} f \in \sem{\filterprog{\delta}, n/a}{M}}}{By \Lemref{lem:subs-soundness-wf}}
      \end{llproof}

      Define $g : \sem{}{\erase{M}} \to \sem{}{\erase{M}}$
      by $V \mapsto \sem{\erase{\delta}, V/x}{\erase{e_0}}$
      (well-defined by \Lemmaref{lem:unrefined-type-soundness}).
      By \Lemmaref{lem:continuous-maps}, $g$ is continuous
      (\ie, $g$ is monotonic and respects lubs).

      For each $k \in \kindnat$, define $g_k$ as follows:
      \begin{align*}
        g_0 &= \bott[\sem{}{\erase{M}}] \\
        g_{j+1} &= g(g_j)
      \end{align*}

      Because $g_0$ is the bottom element of $\sem{}{\erase{M}}$
      and $g : \sem{}{\erase{M}} \to \sem{}{\erase{M}}$ is monotone,
      we know $\ord[\sem{}{\erase{M}}]{g_k}{g_{k+1}}$ for all $k \in \kindnat$.

      By definition of $\sem{}{-}$ and $\erase{-}$,
      \begin{align*}
        \sem{\delta}{\rec{x : (\alltype{a:\kindnat} M)}{e_0}} &= \sem{\erase{\delta}}{\rec{x}{\erase{e_0}}} \\
        &= \bigsqcup_{k \in \kindnat} g_k
      \end{align*}
      Therefore, to complete this case,
      it suffices to show that
      $\sqcup_{k \in \kindnat} g_k \in \sem{\filterprog{\delta}}{N}$.

      We now show a lemma that $g_k \in X_k$ for all $k \in \kindnat$.
      We have $g_0 = \bott[\sem{}{\erase{M}}] \in \sem{}{\erase{M}}$
      because $\sem{}{\erase{M}} \in \Cppo$ by \Lemmaref{lem:unref-type-denotations}.
      For the inductive step, we assume $g_m \in X_m$
      and will prove $g_{m+1} \in X_{m+1}$.
      Now,

      \begin{llproof}
        \eqPf{g_{m+1}}{g(g_m)}{By \defn of $g_k$}
        \eqPf{}{\sem{\erase{\delta}, g_m/x}{\erase{e_0}}}{By \defn of $g$}
        \eqPf{}{\sem{\delta, m/a, g_m/x}{e_0}}{By \defn of $\sem{}{-}$}
        \inPf{}{\sem{\filterprog{\delta}, m/a}{M}}{By \ih and \defn of $\filterprog{-}$}
        \subseteqPf{}{\sem{}{\erase{M}}}{By \Lemmaref{lem:type-subset-erasure}}
        \decolumnizePf
        \ordPf[\sem{}{\erase{M}}]{g_m}{g_{m+1}}{Above}
        \inPf{g_m}{X_m}{Above}
        \inPf{g_{m+1}}{X_m}{By \Lemmaref{lem:upward-closure}}
        \decolumnizePf
        \inPf{g_{m+1}}{X_m \cap \sem{\filterprog{\delta}, m/a}{M}}{By set theory}
        \eqPf{}{\comprehend{f \in \sem{}{\erase{M}}}{\alltype{n < m} f \in \sem{\filterprog{\delta}, n/a}{M}} \cap \sem{\filterprog{\delta}, m/a}{M}}{Above}
        \eqPf{}{\comprehend{f \in \sem{}{\erase{M}}}{\alltype{n < m+1} f \in \sem{\filterprog{\delta}, n/a}{M}}}{By set theory}
        \eqPf{}{X_{m+1}}{Above}
      \end{llproof}

      We have just proved that $g_k \in X_k$ for all $k \in \kindnat$.

      Because $\sem{}{\erase{M}} \in \Cppo$, it is chain-complete.
      But $g_0 \sqsubseteq g_1 \sqsubseteq \cdots$ is a chain in $\sem{}{\erase{M}}$,
      so it has a lub $\sqcup_{k\in\kindnat} g_k$ in $\sem{}{\erase{M}}$
      (by chain-completeness).

      Next, we show that 
      $\sqcup_{k\in\kindnat} g_k \in \sem{\filterprog{\delta}, j/a}{M}$
      for all $j \in \kindnat$.
      Suppose $j \in \kindnat$. Then:

      \begin{llproof}
        \ordPf[\sem{}{\erase{M}}]{g_{j+1}}{\sqcup_{k\in\kindnat} g_k}{By \defn of lub}
        \inPf{g_{j+1}}{X_{j+1}}{}
        \eqPf{}{\comprehend{f \in \sem{}{\erase{M}}}{\alltype{n < j+1} f \in \sem{\filterprog{\delta}, n/a}{M}}}{Above}
        \subseteqPf{}{\sem{\filterprog{\delta}, j/a}{M}}{By set theory}
        \inPf{\sqcup_{k\in\kindnat} g_k}{\sem{\filterprog{\delta}, j/a}{M}}{By \Lemmaref{lem:upward-closure}}
      \end{llproof} 

      Therefore,
      $\sqcup_{k\in\kindnat} g_k \in \sem{\filterprog{\delta}, j/a}{M}$
      for all $j \in \kindnat$, and we have:

      \begin{llproof}
        \inPf{\sqcup_{k\in\kindnat} g_k}{\bigcap_{j \in \kindnat} \sem{\filterprog{\delta}, j/a}{M}}{By set theory}
        \eqPf{}{\sem{\filterprog{\delta}}{\alltype{a:\kindnat} M}}{By \defn of $\sem{}{-}$ and \Lemmaref{lem:type-subset-erasure}}
        \subseteqPf{}{\sem{\filterprog{\delta}}{N}}{By \Lemmaref{lem:type-soundness-sub}}
        \trailingjust{(with premise $\judgesub[-]{\Theta}{\alltype{a:\kindnat} M}{N}$)}
      \end{llproof}

      That completes this case.

      \DerivationProofCase{\DeclChkExpExtract}
      {
        \judgeextract{\Theta}{N}{N'}{\Theta'}
        \\
        \Theta' \neq \cdot
        \\
        \judgechkexp{\Theta, \Theta'}{\Gamma}{e}{N'}
      }
      {
        \judgechkexp{\Theta}{\Gamma}{e}{N}
      }
      \begin{llproof}
        \judgeextractPf{\Theta}{N}{N'}{\Theta'}{Subderivation}
        \judgctxPf{(\Theta, \Theta')}{By \Lemmaref{lem:extract-to-ctx-wf}}
        \judgsubsPf{}{\filterprog{\delta}, \delta'}{\Theta, \Theta'}{Assume such a $\delta'$}
        \judgsubsPf{}{\delta, \delta'}{\Theta, \Theta'; \Gamma}{By \Lemmaref{lem:sem-subs-ix-extend}}
        \judgechkexpPf{\Theta, \Theta'}{\Gamma}{e}{N'}{Subderivation}
        \Pf{\sem{\delta, \delta'}{e}}{\in}{\sem{\filterprog{\delta, \delta'}}{N'}}{By \ih}
        \Pf{}{=}{\sem{\filterprog{\delta}, \delta'}{N'}}{By \defn of \deffilterprog}
        \Pf{\forall\delta'\text{ s.t. }}{|-}{\filterprog{\delta}, \delta' : \Theta, \Theta', \sem{\delta}{e} \in \sem{\filterprog{\delta}, \delta'}{N'}}{Universal quantification intro.}
        \trailingjust{(and $\sem{\delta, \delta'}{e} = \sem{\delta}{e}$)}
        \proofsep
        \judgechkexpPf{\Theta}{\Gamma}{e}{N}{Given}
        \unrefchkexpPf{\erase{\Gamma}}{\erase{e}}{\erase{N}}{By \Lemmaref{lem:erasure-respects-typing}}
        \Pf{\sem{\delta, \delta'}{e}}{=}{\sem{\erase{\delta, \delta'}}{\erase{e}}}{By \defn of \defsem}
        \Pf{}{=}{\sem{\erase{\delta}}{\erase{e}}}{By \defn of \deferase}
        \Pf{}{\in}{\sem{}{\erase{N}}}{By \Lemmaref{lem:unrefined-type-soundness}}
        \proofsep
        \Pf{\sem{\delta}{e}}{\in}{\sem{\filterprog{\delta}}{N}}{By \Lemmaref{lem:negative-value-extract}}
      \end{llproof} 

      \ProofCaseRule{\DeclChkExpUnreachable}
      Straightforward. Use \Lemmaref{lem:unref-type-denotations}.
    \end{itemize}

  \item
    \begin{itemize}
      \DerivationProofCase{\DeclChkMatchEx}
      {\judgechkmatch{\Theta, a:\tau}{\Gamma}{P_0}{\clauses{\pa}{e}{i}{I}}{N}}
      {\judgechkmatch{\Theta}{\Gamma}{\extype{a:\tau}{P_0}}{\clauses{\pa}{e}{i}{I}}{N}}
      Assume $V \in \sem{\filterprog{\delta}}{\extype{a:\tau}{P_0}}$.
      It suffices to show
      $\sem{\delta}{\clauses{\pa}{e}{i}{I}} V \in \sem{\filterprog{\delta}}{N}$.

      By definition of denotation, there exists $d \in \sem{}{\tau}$
      such that $V \in \sem{\filterprog{\delta},d/a}{P_0}$.

      \begin{llproof}
        \judgsubsPf{}{\delta, d/a}{\Theta, a:\tau; \Gamma}{By \IxSem}
        \judgechkmatchPf{\Theta, a:\tau}{\Gamma}{P_0}{\clauses{\pa}{e}{i}{I}}{N}{Subderivation}
        \Pf{\sem{\delta, d/a}{\clauses{\pa}{e}{i}{I}}}{\in}{\sem{\filterprog{\delta, d/a}}{P_0} \Rightarrow \sem{\filterprog{\delta, d/a}}{N}}{By \ih}
        \Pf{\sem{\delta, d/a}{\clauses{\pa}{e}{i}{I}}}{\in}{\sem{\filterprog{\delta}, d/a}{P_0} \Rightarrow \sem{\filterprog{\delta}, d/a}{N}}{By \defn of \deffilterprog}
        \Pf{\sem{\delta}{\clauses{\pa}{e}{i}{I}}}{\in}{\sem{\filterprog{\delta}, d/a}{P_0} \Rightarrow \sem{\filterprog{\delta}, d/a}{N}}{$a \notin \FV{\clauses{\pa}{e}{i}{I}}$}
        \Pf{\sem{\delta}{\clauses{\pa}{e}{i}{I}}}{\in}{\sem{\filterprog{\delta}, d/a}{P_0} \Rightarrow \sem{\filterprog{\delta}}{N}}{$a \notin \FV{N}$}
        \Pf{\sem{\delta}{\clauses{\pa}{e}{i}{I}} V}{\in}{\sem{\filterprog{\delta}}{N}}{Function application}
      \end{llproof} 

      \ProofCaseRule{\DeclChkMatchWith}
      Similar to case for \DeclChkMatchEx.

      \DerivationProofCase{\DeclChkMatchFix}
      {
        \judgeunroll{\cdot}{\Theta}{\nu:F[\mu F]}{\alpha}{F\;\Fold{F}{\alpha}\;\nu}{t}{Q}{\tau}
        \\
        \judgeextract{\Theta}{Q}{Q'}{\Theta'}
        \\
        \judgechkexp{\Theta, \Theta'}{\Gamma, x:Q'}{e}{N}
      }
      {
        \judgechkmatch{\Theta}{\Gamma}{\comprehend{\nu:\mu F}
          {\Fold{F}{\alpha}\,{\nu} =_\tau t}}{\setof{\clause{\roll{x}}{e}}}{N}
      }
      \begin{llproof}
        \Pf{V}{\in}{\sem{\filterprog{\delta}}{\comprehend{\nu:\mu F}{\Fold{F}{\alpha}\,{\nu} =_\tau t}}}{Suppose}
        \Pf{V}{\in}{\mu\sem{}{\erase{F}}}{By \defn of \defsem}
        \Pf{V}{\in}{\mu\sem{\filterprog{\delta}}{F}}{\ditto}
        \Pf{(\fold{\sem{\filterprog{\delta}}{F}}{\sem{\filterprog{\delta}}{\alpha}})\;V}{=}{\sem{\filterprog{\delta}}{t}}{\ditto}
        \Pf{V}{\in}{\comprehend{V \in \mu\sem{\filterprog{\delta}}{F}}{(\fold{\sem{\filterprog{\delta}}{F}}{\sem{\filterprog{\delta}}{\alpha}})\;V = \sem{\filterprog{\delta}}{t}}}{By above}
      \end{llproof}

      But
      \begin{align*}
        \hspace{1em}&\hspace{-1em}\comprehend{V \in \mu\sem{\filterprog{\delta}}{F}}{(\fold{\sem{\filterprog{\delta}}{F}}{\sem{\filterprog{\delta}}{\alpha}})\;V = \sem{\filterprog{\delta}}{t}} \\
        &= \comprehend{V \in \sem{\filterprog{\delta}}{F} (\mu\sem{\filterprog{\delta}}{F})}{\sem{\filterprog{\delta}}{\alpha} (\sem{\filterprog{\delta}}{F}(\fold{\sem{\filterprog{\delta}}{F}}{\sem{\filterprog{\delta}}{\alpha}})\;V) = \sem{\filterprog{\delta}}{t}} &&\quad\text{By \Lemref{lem:refined-mu-unroll-equal}} \\
        &= \sem{\filterprog{\delta}}{Q} &&\quad\text{By \Lemref{lem:unroll-soundness}}
      \end{align*}
      so $V \in \sem{\filterprog{\delta}}{Q}$.
      This together with the extraction subderivation implies
      yields a $\delta'$ such that $|- \filterprog{\delta}, \delta' : \Theta, \Theta'$
      and $V \in \sem{\filterprog{\delta}, \delta'}{Q'}$
      by \Lemmaref{lem:positive-value-extract}.
      By \defn of \deffilterprog, $V \in \sem{\filterprog{\delta, \delta'}}{Q'}$.
      By \ValSem, $|- \delta, \delta', V/x : \Theta, \Theta', \Gamma,x:Q'$. Now,

      \begin{llproof}
        \Pf{\sem{\delta}{\setof{\clause{\roll{x}}{e}}} V}{=}{\sem{\erase{\delta}}{\erase{\setof{\clause{\roll{x}}{e}}}} V}{By \defn of \defsem}
        \Pf{}{=}{\sem{\erase{\delta}}{\setof{\clause{\roll{x}}{\erase{e}}}} V}{By \defn of \deferase}
        \Pf{}{=}{\sem{\erase{\delta}, V/x}{\erase{e}}}{By \defn of \defsem}
        \Pf{}{=}{\sem{\erase{\delta, \delta', V/x}}{\erase{e}}}{By \defn of \deferase}
        \Pf{}{=}{\sem{\delta, \delta', V/x}{e}}{By \defn of \defsem}
        \Pf{}{\in}{\sem{\filterprog{\delta, \delta', V/x}}{N}}{By \ih}
        \Pf{}{=}{\sem{\filterprog{\delta}, \delta'}{N}}{By \defn of \deffilterprog}
        \Pf{}{=}{\sem{\filterprog{\delta}}{N}}{$\FV{N} \sect \dom{\Theta'} = \emptyset$}
      \end{llproof} 

      \item The remaining cases are straightforward
        or similar to previous cases.
    \end{itemize}

  \item
    \begin{itemize}
      \DerivationProofCase{\DeclSpineAll}
      {
        \judgeterm{\Theta}{t}{\tau}
        \\
        \judgespine{\Theta}{\Gamma}{s}{[t/a]N_0}{\upshift{P}}
      }
      {\judgespine{\Theta}{\Gamma}{s}{(\alltype{a:\tau}{N_0})}{\upshift{P}}}
      Suppose $f \in \sem{\filterprog{\delta}}{\alltype{a:\tau}{N_0}}$.
      By definition of denotation,
      $f \in \sem{\filterprog{\delta}, d/a}{N_0}$ for all $d \in \sem{}{\tau}$.

      By \Lemmaref{lem:type-soundness-ix}, $\sem{\filterprog{\delta}}{t} \in \sem{}{\tau}$.
      So, $f \in \sem{\filterprog{\delta},\sem{\filterprog{\delta}}{t}/a}{N_0}$.

      By \ih,
      $\sem{\delta}{\judgespine{\Theta}{\Gamma}{s}{[t/a]N}{\upshift{P}}}
      \in \sem{\filterprog{\delta}}{[t/a]N} \Rightarrow \sem{\filterprog{\delta}}{\upshift{P}}$.
      Now,
      \begin{align*}
        \sem{\filterprog{\delta}}{[t/a]N}
        &= \sem{\filterprog{\delta}}{[\id_{\Theta;\cdot},t/a]N} &&\quad\text{By \Lemmaref{lem:id-subs-id}} \\
        &= \sem{\sem{\filterprog{\delta}}{\id_{\Theta;\cdot},t/a}}{N} &&\quad\text{By \Lemmaref{lem:subs-soundness-wf}} \\
        &= \sem{\filterprog{\delta},\sem{\filterprog{\delta}}{t}/a}{N} &&\quad\text{By \defn of \defsem}
      \end{align*}
      so $\sem{\delta}{s} \in \sem{\filterprog{\delta},\sem{\filterprog{\delta}}{t}/a}{N} \Rightarrow \sem{\delta}{\upshift{P}}$.
      By function application, $\sem{\delta}{s} f \in \sem{\filterprog{\delta}}{\upshift{P}}$,
      as desired.

      \ProofCaseRule{\DeclSpineImplies}
      Similar to case for \DeclSpineAll.

      \ProofCaseRule{\DeclSpineApp}
      Straightforward.

      \ProofCaseRule{\DeclSpineNil}
      Straightforward. \qedhere
    \end{itemize}
  \end{enumerate}
\end{proof}

\begin{lemma}[Filter Resp.\ Sem.\ Subst.]
  \label{lem:filter-respects-sem-subs}
  If $\Theta_0; \Gamma_0 |- \sigma : \Theta; \Gamma$
  and $|- \delta : \Theta_0; \Gamma_0$,\\
  then $\filterprog{\sem{\delta}{\sigma}} = \sem{\filterprog{\delta}}{\filterprog{\sigma}}$.
\end{lemma}
\begin{proof}
  By structural induction on the derivation of
  $\Theta_0; \Gamma_0 |- \sigma : \Theta; \Gamma$.
\end{proof}

\begin{lemma}[Substitution Type Soundness]
  \label{lem:type-soundness-subs}
  ~\\
  If $\E :: \Theta_0;\Gamma_0 |- \sigma : \Theta; \Gamma$,
  then $|- \sem{\delta}{\E} : \Theta; \Gamma$
  for all $|- \delta : \Theta_0; \Gamma_0$,
\end{lemma}
\begin{proof}
  By induction on the height of the derivation $\E$.
  Consider cases for the rule concluding $\E$:
  \begin{itemize}
    \DerivationProofCase{\EmptySyn}
    {
    }
    {
      \Theta_0; \Gamma_0 |- \cdot : \cdot; \cdot
    }
    \begin{llproof}
      \Pf{|- \cdot : \cdot;\cdot}{}{}{By \EmptySem}
    \end{llproof}
    \DerivationProofCase{\IxSyn}
    {
      \E' :: \Theta_0; \Gamma_0 |- \sigma' : \Theta'; \Gamma
      \\
      \Theta_0 |- [\filterprog{\sigma'}]t : \tau
    }
    {
      \E :: \Theta_0; \Gamma_0 |- \sigma', t/a : \Theta', a:\tau; \Gamma
    }
    \begin{llproof}
      \Pf{\E' :: \Theta_0; \Gamma_0 |- \sigma' : \Theta'; \Gamma}{}{}{Subderivation}
      \Pf{|- \sem{\delta}{\E'} : \Theta'; \Gamma}{}{}{By i.h.}
      \Pf{\Theta_0 |- [\filterprog{\sigma'}]t : \tau}{}{}{Subderivation}
      \Pf{\sem{\filterprog{\delta}}{[\filterprog{\sigma'}]t} \in \sem{}{\tau}}{}{}{By~\Lemref{lem:type-soundness-ix}}
      \Pf{|- \underbrace{\sem{\delta}{\E'}, \sem{\filterprog{\delta}}{[\filterprog{\sigma'}]t}/a}_{\sem{\delta}{\E}} : \Theta', a:\tau; \Gamma}{}{}{By \IxSem and \Defnref{def:den-syn-subs}}
    \end{llproof}
    \DerivationProofCase{\PropSyn}
    {
      \E' :: \Theta_0; \Gamma_0 |- \sigma : \Theta'; \Gamma
      \\
      \judgeentail{\Theta_0}{[\filterprog{\sigma}]\phi}
    }
    {
      \E :: \Theta_0; \Gamma_0 |- \sigma : \Theta', \phi; \Gamma
    }
    \begin{llproof}
      \judgsubsPf{\Theta_0;\Gamma_0}{\sigma}{\Theta';\Gamma}{Subderivation}
      \judgsubsPf{}{\sem{\delta}{\sigma}}{\Theta'; \Gamma}{By \ih}
      \judgsubsPf{}{\filterprog{\delta}}{\Theta_0}{By \Lemmaref{lem:filter-out-prog-vars}}
      \judgsubsPf{\Theta_0}{\filterprog{\sigma}}{\Theta'}{By \Lemmaref{lem:filter-out-prog-vars-syn}}
      \judgeentailPf{\Theta_0}{[\filterprog{\sigma}]\phi}{Subderivation}
      \Pf{1}{=}{\sem{\filterprog{\delta}}{[\filterprog{\sigma}]\phi}}{By inversion on \PropTrue}
      \Pf{}{=}{\sem{\sem{\filterprog{\delta}}{\filterprog{\sigma}}}{\phi}}{By \Lemmaref{lem:subs-soundness-ix}}
      \Pf{}{=}{\sem{\filterprog{\sem{\delta}{\sigma}}}{\phi}}{By \Lemmaref{lem:filter-respects-sem-subs}}
      \judgsubsPf{}{\sem{\delta}{\sigma}}{\Theta', \phi; \Gamma}{By \PropSem}
    \end{llproof}
    \DerivationProofCase{\ValSyn}
    {
      \E' :: \Theta_0; \Gamma_0 |- \sigma' : \Theta; \Gamma'
      \\
      \Dee :: \judgechkval{\Theta_0}{\Gamma_0}{[\sigma']v}{[\filterprog{\sigma'}]P}
    }
    {
      \E :: \Theta_0; \Gamma_0 |- \sigma', \subs{v}{P}{x} : \Theta; \Gamma', x:P
    }
    \begin{llproof}
      \judgsubsPf{\Theta_0; \Gamma_0}{\sigma'}{\Theta; \Gamma'}{Subderivation}
      \judgsubsPf{}{\sem{\delta}{\sigma'}}{\Theta;\Gamma'}{By \ih}
      \Pf{\sem{\delta}{[\sigma']v}}{\in}{\sem{\filterprog{\delta}}{[\filterprog{\sigma'}]P}}{By~\Theoremref{thm:type-soundness}}
      \Pf{}{=}{\sem{\sem{\filterprog{\delta}}{\filterprog{\sigma'}}}{P}}{By~\Theoremref{lem:subs-soundness-wf}}
      \Pf{}{=}{\sem{\filterprog{\sem{\delta}{\sigma'}}}{P}}{By \Lemmaref{lem:filter-respects-sem-subs}}
      \judgsubsPf{}{\sem{\delta}{\sigma'}, \sem{\delta}{[\sigma']v}/x}{\Theta;\Gamma',x:P}{By \ValSem}
      \judgsubsPf{}{\sem{\delta}{\sigma', v/x}}{\Theta;\Gamma',x:P}{By \defn of \defsem}
    \end{llproof}
    \qedhere
  \end{itemize}
\end{proof}

\begin{theorem}[Soundness of Substitution]
  \label{thm:subs-soundness}
  Assume $\E :: \Theta_0;\Gamma_0 |- \sigma : \Theta; \Gamma$. Then:
  \begin{enumerate}
  \item If
    $\Dee :: \judgesynhead{\Theta}{\Gamma}{h}{P}$,
    then for all $Q$
    such that $\judgesub[+]{\Theta_0}{Q}{[\filterprog{\sigma}]P}$
    and $\judgesynhead{\Theta_0}{\Gamma_0}{[\sigma]h}{Q}$,
    we have
    $\sem{\delta}{\judgesynhead{\Theta_0}{\Gamma_0}{[\sigma]h}{Q}}
    = \sem{\sem{\delta}{\sigma}}{\judgesynhead{\Theta}{\Gamma}{h}{P}}$
    for all $|- \delta : \Theta_0; \Gamma_0$.
  \item If
    $\Dee :: \judgesynexp{\Theta}{\Gamma}{\be}{\upshift{P}}$,
    then for all $Q$
    such that $\judgesub[-]{\Theta_0}{\upshift{Q}}{[\filterprog{\sigma}]\upshift{P}}$
    and $\judgesynexp{\Theta_0}{\Gamma_0}{[\sigma]\be}{\upshift{Q}}$,
    we have
    $\sem{\delta}{\judgesynexp{\Theta_0}{\Gamma_0}{[\sigma]\be}{\upshift{Q}}}
    = \sem{\sem{\delta}{\sigma}}{\judgesynexp{\Theta}{\Gamma}{\be}{\upshift{P}}}$
    for all $|- \delta : \Theta_0; \Gamma_0$.
  \item If
    $\Dee :: \judgechkval{\Theta}{\Gamma}{v}{P}$,
    then
    $\sem{\delta}{\judgechkval{\Theta_0}{\Gamma_0}{[\sigma]v}{[\filterprog{\sigma}]P}}
    = \sem{\sem{\delta}{\sigma}}{\judgechkval{\Theta}{\Gamma}{v}{P}}$
    for all $|- \delta : \Theta_0; \Gamma_0$.
  \item If
    $\Dee :: \judgechkexp{\Theta}{\Gamma}{e}{N}$,
    then
    $\sem{\delta}{\judgechkexp{\Theta_0}{\Gamma_0}{[\sigma]e}{[\filterprog{\sigma}]N}}
    = \sem{\sem{\delta}{\sigma}}{\judgechkexp{\Theta}{\Gamma}{e}{N}}$
    for all $|- \delta : \Theta_0; \Gamma_0$.
  \item If
    $\Dee :: \judgechkmatch{\Theta}{\Gamma}{P}{\clauses{\pa}{e}{i}{I}}{N}$,
    then
    $\sem{\delta}{\judgechkmatch{\Theta_0}{\Gamma_0}{[\filterprog{\sigma}]P}
      {\clauses{\pa}{[\sigma]e}{i}{I}}{[\filterprog{\sigma}]N}}
    =\\ \sem{\sem{\delta}{\sigma}}{\judgechkmatch{\Theta}{\Gamma}{P}{\clauses{\pa}{e}{i}{I}}{N}}$
    for all $|- \delta : \Theta_0; \Gamma_0$.
  \item If
    $\Dee :: \judgespine{\Theta}{\Gamma}{s}{N}{\upshift{P}}$,
    then 
    $\sem{\delta}{\judgespine{\Theta_0}{\Gamma_0}
      {[\sigma]s}{[\filterprog{\sigma}]N}{\upshift{[\filterprog{\sigma}]P}}}
    = \sem{\sem{\delta}{\sigma}}{\judgespine{\Theta}{\Gamma}{s}{N}{\upshift{P}}}$
    for all $|- \delta : \Theta_0; \Gamma_0$.
  \end{enumerate}
\end{theorem}
\begin{proof}
  Note that by \Lemmaref{lem:type-soundness-subs}, we know that
  $|- \sem{\delta}{\E} : \Theta; \Gamma$.

  The proof of each part is similar.
  The first two parts are the most complicated, but also similar to each other.
  So, we will show only part (1):

  We are given a derivation $\judgesynhead{\Theta}{\Gamma}{h}{P}$.
  Suppose $\judgesub[+]{\Theta_0}{Q}{[\filterprog{\sigma}]P}$
  and $\judgesynhead{\Theta_0}{\Gamma_0}{[\sigma]h}{Q}$.
  By \Lemmaref{lem:erasure-respects-typing},
  $\unrefsynhead{\erase{\Gamma}}{\erase{h}}{\erase{P}}$
  and
  $\unrefsynhead{\erase{\Gamma_0}}{\erase{[\sigma]h}}{\erase{Q}}$.
  Further,
  \begin{align*}
    \erase{Q}
    &= \erase{[\filterprog{\sigma}]P} &&\quad\text{By \Lemmaref{lem:subtyping-erases-to-equality}} \\
    &= \erase{P} &&\quad\text{By \Lemmaref{lem:erasure-subs}}
  \end{align*}

  We are also given derivations
  $\Theta_0; \Gamma_0 |- \sigma : \Theta; \Gamma$
  and
  $|- \delta : \Theta_0; \Gamma_0$.
  By \Lemmaref{lem:erasure-respects-subs-typing},
  $\erase{\Gamma_0} |- \erase{\sigma} : \erase{\Gamma}$
  and
  $|- \erase{\delta} : \erase{\Gamma_0}$.
  
  Therefore,
  \begin{align*}
    \hspace{1em}&\hspace{-1em}\sem{\delta}{\judgesynhead{\Theta_0}{\Gamma_0}{[\sigma]h}{Q}} \\
    &= \sem{\erase{\delta}}{\unrefsynhead{\erase{\Gamma_0}}{\erase{[\sigma]h}}{\erase{Q}}} &&\quad\text{By \defn of \defsem} \\
    &= \sem{\erase{\delta}}{\unrefsynhead{\erase{\Gamma_0}}{[\erase{\sigma}]\erase{h}}{\erase{Q}}} &&\quad\text{By \Lemmaref{lem:erasure-respects-subs}} \\
    &= \sem{\sem{\erase{\delta}}{\erase{\sigma}}}{\unrefsynhead{\erase{\Gamma}}{{\erase{h}}}{\erase{Q}}} &&\quad\text{By \Lemmaref{lem:unref-subs-soundness}} \\
    &= \sem{\erase{\sem{\delta}{\sigma}}}{\unrefsynhead{\erase{\Gamma}}{{\erase{h}}}{\erase{Q}}} &&\quad\text{By \Lemmaref{lem:erasure-respects-sem-subs}} \\
    &= \sem{\erase{\sem{\delta}{\sigma}}}{\unrefsynhead{\erase{\Gamma}}{{\erase{h}}}{\erase{P}}} &&\quad\text{Above }(\erase{Q} = \erase{P}) \\
    &= \sem{\sem{\delta}{\sigma}}{\judgesynhead{\Theta}{\Gamma}{h}{P}} &&\quad\text{By \defn of \defsem} \qedhere
  \end{align*}
\end{proof}

\section{Algorithmic Extension}

\begin{lemma}[Weaken Extension]
  \label{lem:weaken-ext}
  ~
  \begin{enumerate}
  \item If $\extend{\Theta_1, \Theta_2}{\Delta}{\Delta'}$
    and $\judgctx{(\Theta_1, \Theta_0, \Theta_2)}$,
    then $\extend{\Theta_1, \Theta_0, \Theta_2}{\Delta}{\Delta'}$.
  \item If $\rextend{\Theta_1, \Theta_2}{\Delta}{\Delta'}$
    and $\judgctx{(\Theta_1, \Theta_0, \Theta_2)}$,
    then $\rextend{\Theta_1, \Theta_0, \Theta_2}{\Delta}{\Delta'}$.
  \end{enumerate}
\end{lemma}
\begin{proof}
  Each part by structural induction on the given extension derivation,
  using \Lemmaref{lem:ix-level-weakening} as needed.
\end{proof}

\begin{lemma}[Extension Soundness]
  \label{lem:extension-sound}
  If $\extend{\Theta}{\Delta}{\Delta'}$,
  then $\rextend{\Theta}{\Delta}{\Delta'}$.
\end{lemma}
\begin{proof}
  Straightforward. Use \Lemmaref{lem:equivassert}.
\end{proof}

\subsection{Reflexivity and Transitivity}

\begin{lemma}[Ext.\ Reflexivity]
  \label{lem:ext-refl}
  ~
  \begin{enumerate}
  \item If $\judgectx{\Theta}{\Delta}$, then $\extend{\Theta}{\Delta}{\Delta}$.
  \item If $\judgectx{\Theta}{\Delta}$, then $\rextend{\Theta}{\Delta}{\Delta}$.
  \end{enumerate}
\end{lemma}
\begin{proof}
  Part (1) is straightforward.
  For part (2), use part (1) and \Lemmaref{lem:extension-sound}.
\end{proof}

\begin{lemma}[Ext.\ Transitivity]
  \label{lem:ext-trans}
  ~
  \begin{enumerate}
  \item If $\extend{\Theta}{\Delta_1}{\Delta_2}$
    and $\extend{\Theta}{\Delta_2}{\Delta_3}$,
    then $\extend{\Theta}{\Delta_1}{\Delta_3}$.
  \item If $\rextend{\Theta}{\Delta_1}{\Delta_2}$
    and $\rextend{\Theta}{\Delta_2}{\Delta_3}$,
    then $\rextend{\Theta}{\Delta_1}{\Delta_3}$.
  \end{enumerate}
\end{lemma}
\begin{proof}
  Straightforward.
  Part (2) uses \Lemmaref{lem:equivassert}.
\end{proof}

\subsection{Extension Weakening}

\begin{lemma}[Ext.\ Weakening (Ixs.)]
  \label{lem:ext-weak-ix}
  ~
  \begin{enumerate}
  \item If $\judgeterm{\Theta; \Delta}{t}{\tau}$
    and $\extend{\Theta}{\Delta}{\Delta'}$,
    then $\judgeterm{\Theta; \Delta'}{t}{\tau}$.
  \item If $\judgeterm{\Theta; \Delta}{t}{\tau}$
    and $\rextend{\Theta}{\Delta}{\Delta'}$,
    then $\judgeterm{\Theta; \Delta'}{t}{\tau}$.
  \end{enumerate}
\end{lemma}
\begin{proof}
  Straightforward.
\end{proof}

\begin{lemma}[Ext.\ Weakening (Types)]
  \label{lem:ext-weak-tp}
  ~
  \begin{enumerate}
  \item
    If $\judgetp{\Theta; \Delta}{A}{\Xi}$ and $\extend{\Theta}{\Delta}{\Delta'}$,
    then $\judgetp{\Theta; \Delta'}{A}{\Xi}$.
  \item
    If $\judgefunctor{\Theta; \Delta}{\mathcal{F}}{\Xi}$ and $\extend{\Theta}{\Delta}{\Delta'}$,
    then $\judgefunctor{\Theta; \Delta'}{\mathcal{F}}{\Xi}$.
  \item
    If $\judgealgebra{\Xi}{\Theta; \Delta}{\alpha}{F}{\tau}$ and $\extend{\Theta}{\Delta}{\Delta'}$,
    then $\judgealgebra{\Xi}{\Theta; \Delta'}{\alpha}{F}{\tau}$.
  \item
    If $\judgetp{\Theta; \Delta}{A}{\Xi}$ and $\rextend{\Theta}{\Delta}{\Delta'}$,
    then $\judgetp{\Theta; \Delta'}{A}{\Xi}$.
  \item
    If $\judgefunctor{\Theta; \Delta}{\mathcal{F}}{\Xi}$ and $\rextend{\Theta}{\Delta}{\Delta'}$,
    then $\judgefunctor{\Theta; \Delta'}{\mathcal{F}}{\Xi}$.
  \item
    If $\judgealgebra{\Xi}{\Theta; \Delta}{\alpha}{F}{\tau}$ and $\rextend{\Theta}{\Delta}{\Delta'}$,
    then $\judgealgebra{\Xi}{\Theta; \Delta'}{\alpha}{F}{\tau}$.
  \end{enumerate}
\end{lemma}
\begin{proof}
  By structural induction on the given well-formedness derivation.
  Parts (1), (2), and (3) are mutually recursive;
  and so are parts (4), (5), and (6).
  Use \Lemmaref{lem:ext-weak-ix}
  and \Lemmaref{lem:weaken-ext} as needed.
\end{proof}

\begin{lemma}[Ext.\ Weak.\ (Unroll)]
  \label{lem:ext-weak-unroll}
  ~\\
  If $\judgeunroll{\Xi}{\Theta; \Delta}{\nu:G[\mu F]}{\beta}{G\; \Fold{F}{\alpha}\;\nu}{\ahat}{Q}{\tau}$
  and $\rextend{\Theta}{\Delta}{\Delta'}$,\\
  then $\judgeunroll{\Xi}{\Theta; \Delta'}{\nu:G[\mu F]}{\beta}{G\; \Fold{F}{\alpha}\;\nu}{\ahat}{Q}{\tau}$.
\end{lemma}
\begin{proof}
  By structural induction on the given unrolling derivation
\end{proof}

\subsection{Extending Judgments}

\begin{lemma}[Ext.\ Solve Entry]
  \label{lem:deep-solve-extend}
  ~
  \begin{enumerate}
  \item If $\judgectx{\Theta}{\Delta_1, \ahat:\tau, \Delta_2}$
    and $\judgeterm{\Theta}{t}{\tau}$,
    then $\extend{\Theta}{\Delta_1, \ahat:\tau, \Delta_2}{\Delta_1, \hypeq{\ahat}{\tau}{t}, \Delta_2}$.
  \item If $\judgectx{\Theta}{\Delta_1, \ahat:\tau, \Delta_2}$
    and $\judgeterm{\Theta}{t}{\tau}$,
    then $\rextend{\Theta}{\Delta_1, \ahat:\tau, \Delta_2}{\Delta_1, \hypeq{\ahat}{\tau}{t}, \Delta_2}$.
  \end{enumerate}
\end{lemma}
\begin{proof}
  We show part (1) by structural induction on $\Delta_2$:
  \begin{itemize}
  \item \textbf{Case} $\Delta_2 = \cdot$\\
    By \Lemmaref{lem:ext-refl}, $\extend{\Theta}{\Delta_1}{\Delta_1}$.
    By \ExtSolve, $\extend{\Theta}{\Delta_1, \ahat:\tau}{\Delta_1, \hypeq{\ahat}{\tau}{t}}$.
  \item \textbf{Case} $\Delta_2 = \Delta_2', \bhat:\tau$\\
    By \ih, $\extend{\Theta}{\Delta_1, \ahat:\tau, \Delta_2'}{\Delta_1, \hypeq{\ahat}{\tau}{t}, \Delta_2'}$.
    Apply \ExtUnsolved.
  \end{itemize}
  Part (2) follows from part (1) and \Lemmaref{lem:extension-sound}.
\end{proof}

\begin{lemma}[Prop.\ Equiv.\ Inst.\ Extends]
  \label{lem:propequivinst-extends}
  If $\algpropequivinst{\Theta; \Delta}{\phi}{\psi}{\Delta'}$,
  then $\extend{\Theta}{\Delta}{\Delta'}$.
\end{lemma}
\begin{proof}
  Either \PropEquivInst or \PropEquivNoInst concludes
  $\algpropequivinst{\Theta; \Delta}{\phi}{\psi}{\Delta'}$.
  If \PropEquivNoInst, use \Lemmaref{lem:ext-refl}.
  If \PropEquivInst, use \Lemmaref{lem:deep-solve-extend}.
\end{proof}

\begin{lemma}[Alg.\ Equiv.\ Extends]
  \label{lem:alg-equiv-extends}
  ~
  \begin{enumerate}
  \item If $\algequiv[\pm]{\Theta; \Delta}{A}{B}{W}{\Delta'}$,
    then $\extend{\Theta}{\Delta}{\Delta'}$.
  \item If $\algequiv[]{\Theta; \Delta}{\F}{\Gee}{W}{\Delta'}$,
    then $\extend{\Theta}{\Delta}{\Delta'}$.
  \end{enumerate}
\end{lemma}
\begin{proof}
  By structural induction on the given equivalence derivation.
  Both parts are mutually recursive.
  Use \Lemmaref{lem:ext-refl} and \Lemmaref{lem:ext-trans} as needed.
  In the \AlgTpEquivPosFixInst case, use \Lemmaref{lem:deep-solve-extend}.
  In the \AlgTpEquivPosWith case, use \Lemmaref{lem:propequivinst-extends}.
\end{proof}

\begin{lemma}[Inst.\ Extends]
  \label{lem:inst-extends}
  If $\alginst{\Theta; \Delta}{\phi}{\Delta'}$,
  then $\extend{\Theta}{\Delta}{\Delta'}$.
\end{lemma}
\begin{proof}
  Either \RuleInst or \RuleNoInst concludes $\alginst{\Theta; \Delta}{\phi}{\Delta'}$.
  If \RuleNoInst, use \Lemmaref{lem:ext-refl}.
  If \RuleInst, use \Lemmaref{lem:deep-solve-extend}.
\end{proof}

\begin{lemma}[Alg.\ Sub.\ Extends]
  \label{lem:alg-sub-extends}
  If $\algsub[\pm]{\Theta; \Delta}{A}{B}{W}{\Delta'}$,
  then $\extend{\Theta}{\Delta}{\Delta'}$.
\end{lemma}
\begin{proof}
  By structural induction on the given subtyping derivation.
  Use \Lemmaref{lem:alg-equiv-extends},
  \Lemmaref{lem:ext-refl} and \Lemmaref{lem:ext-trans} as needed.
  In the \AlgSubPosFixInst case, use \Lemmaref{lem:deep-solve-extend}.
  In the \AlgSubPosWithR case, use \Lemmaref{lem:inst-extends}.
\end{proof}

\begin{lemma}[Typing Extends]
  \label{lem:typing-extends}
  ~
  \begin{enumerate}
    \item
      If $\algchk{\Theta; \Delta}{\Gamma}{v}{P}{\chi}{\Delta'}$
      then $\extend{\Theta}{\Delta}{\Delta'}$.
    \item
      If $\algspine{\Theta; \Delta}{\Gamma}{s}{N}{\upshift{P}}{\chi}{\Delta'}$
      then $\extend{\Theta}{\Delta}{\Delta'}$.
  \end{enumerate}
\end{lemma}
\begin{proof}
  Each part is proved by structural induction on the given type-checking derivation.
  Use \Lemmaref{lem:ext-refl}, \Lemmaref{lem:ext-trans},
  \Lemmaref{lem:alg-sub-extends}, and \Lemmaref{lem:inst-extends} as needed.
  Part (2) uses part (1) (but not the other way around).
\end{proof}

\subsection{Relaxed Extension}

\begin{lemma}[Relaxed Solve Entry]
  \label{lem:relaxed-deep-entry}
  If $\rextend{\Theta}{\Delta_1, \ahat:\tau, \Delta_2}{\Omega}$
  and $\judgeentail{\Theta}{t = [\Omega]\ahat}$,\\
  then $\rextend{\Theta}{\Delta_1, \hypeq{\ahat}{\tau}{t}, \Delta_2}{\Omega}$.
\end{lemma}
\begin{proof}
  By structural induction on $\Delta_2$.
\end{proof}

\begin{lemma}[Extend Relaxed Reflexively]
  \label{lem:extend-relaxed-reflexively}
  ~\\
  If $\judgectx{\Theta}{\Delta, \Delta_0}$
  and $\judgectx{\Theta}{\Delta', \Delta_0}$
  and $\rextend{\Theta}{\Delta}{\Delta'}$,
  then $\rextend{\Theta}{\Delta, \Delta_0}{\Delta', \Delta_0}$.
\end{lemma}
\begin{proof}
  By structural induction on $\Delta_0$,
  using \Lemmaref{lem:equivassert} as needed.
\end{proof}

\section{Algorithmic Substitution and Well-Formedness Properties}

\begin{lemma}[Alg.\ to Decl.\ WF]
  \label{lem:alg-to-decl-wf}
  Assume $\extend{\Theta}{\Delta}{\Omega}$
  or $\rextend{\Theta}{\Delta}{\Omega}$.
  \begin{enumerate}
  \item If $\judgeterm{\Theta; \Delta}{t}{\tau}$,
    then $\judgeterm{\Theta}{[\Omega]t}{\tau}$.
  \item If $\judgetp{\Theta; \Delta}{A}{\Xi}$,
    then there exists $\Xi' \supseteq [\Omega]\Xi$
    such that $\judgetp{\Theta}{[\Omega]A}{\Xi'}$.
  \item If $\judgefunctor{\Theta; \Delta}{\mathcal{F}}{\Xi}$,
    then there exists $\Xi' \supseteq [\Omega]\Xi$
    such that $\judgefunctor{\Theta}{[\Omega]\mathcal{F}}{\Xi'}$.
  \item If $\judgealgebra{\Xi}{\Theta; \Delta}{\alpha}{F}{\tau}$,
    then $\judgealgebra{\Xi}{\Theta}{\alpha}{([\Omega]F)}{\tau}$.
  \end{enumerate}
\end{lemma}
\begin{proof}
  Assume $\rextend{\Theta}{\Delta}{\Omega}$.
  Each part is proved by structural induction on the given well-formedness derivation.
  Parts (2), (3), and (4) are mutually recursive.
  The \AlgTpEx and \AlgTpAll cases of part (2)
  and the \AlgAlgExConstProd case of part (4)
  use \Lemmaref{lem:weaken-ext}.
  Cases of parts (2) and (3) use algorithmic variants of
  \Lemref{lem:union-subs} and \Lemref{lem:intersection-subs}.

  If we instead assume $\extend{\Theta}{\Delta}{\Omega}$,
  the parts follow by the preceding with \Lemmaref{lem:extension-sound}.
\end{proof}

\begin{lemma}[Decl.\ WF is Ground]
  \label{lem:decl-wf-ground}
  ~
  \begin{enumerate}
  \item If $\judgeterm{\Theta}{t}{\tau}$,
    then $\ground{t}$.
  \item If $\judgetp{\Theta}{A}{\Xi}$,
    then $\ground{A}$.
  \item If $\judgefunctor{\Theta}{\mathcal{F}}{\Xi}$,
    then $\ground{\mathcal{F}}$.
  \item If $\judgealgebra{\Xi}{\Theta}{\alpha}{F}{\tau}$,
    then $\ground{\alpha}$.
  \end{enumerate}
\end{lemma}
\begin{proof}
  Straightforward.
\end{proof}

\begin{lemma}[Ground Alg.\ to Decl.\ WF]
  \label{lem:ground-to-decl}
  ~
  \begin{enumerate}
  \item
    If $\judgeterm{\Theta; \Delta}{t}{\tau}$
    and $\ground{t}$,
    then $\judgeterm{\Theta}{t}{\tau}$.
  \item
    If $\judgetp{\Theta; \Delta}{A}{\Xi}$
    and $\ground{A}$,
    then $\judgetp{\Theta}{A}{\Xi}$.
  \item
    If $\judgefunctor{\Theta; \Delta}{\mathcal{F}}{\Xi}$
    and $\ground{\mathcal{F}}$,
    then $\judgefunctor{\Theta}{\mathcal{F}}{\Xi}$.
  \end{enumerate}
\end{lemma}
\begin{proof}
  Straightforward.
\end{proof}

\begin{lemma}[Algebras are Ground]
  \label{lem:algebras-are-ground}
  If $\judgealgebra{\Xi}{\Theta; \Delta}{\alpha}{F}{\tau}$,
  then $\ground{\alpha}$.
\end{lemma}
\begin{proof}
  By structural induction on the given algebra well-formedness derivation.
\end{proof}

\begin{lemma}[Ground Alg.\ WF]
  \label{lem:ground-alg-wf}
  Assume $\extend{\Theta}{\Delta}{\Omega}$
  or $\rextend{\Theta}{\Delta}{\Omega}$.
  \begin{enumerate}
  \item If $\judgeterm{\Theta; \Delta}{t}{\tau}$,
    then $\ground{[\Omega]t}$.
  \item If $\judgetp{\Theta; \Delta}{A}{\Xi}$,
    then $\ground{[\Omega]A}$.
  \item If $\judgefunctor{\Theta; \Delta}{\mathcal{F}}{\Xi}$,
    then $\ground{[\Omega]\mathcal{F}}$.
  \end{enumerate}
\end{lemma}
\begin{proof}
  Follows from \Lemmaref{lem:alg-to-decl-wf} and \Lemmaref{lem:decl-wf-ground}.
\end{proof}

\begin{lemma}[Decl.\ to Ground Alg.\ WF]
  \label{lem:decl-to-ground-alg-wf}
  ~
  \begin{enumerate}
  \item If $\judgeterm{\Theta}{t}{\tau}$, then $\judgeterm{\Theta; \cdot}{t}{\tau}$.
  \item If $\judgetp{\Theta}{A}{\Xi}$, then $\judgetp{\Theta; \cdot}{A}{\Xi}$.
  \item If $\judgefunctor{\Theta}{\mathcal{F}}{\Xi}$,
    then $\judgefunctor{\Theta; \cdot}{\mathcal{F}}{\Xi}$.
  \item If $\judgealgebra{\Xi}{\Theta}{\alpha}{F}{\tau}$,
    then $\judgealgebra{\Xi}{\Theta; \cdot}{\alpha}{F}{\tau}$.
  \end{enumerate}
\end{lemma}
\begin{proof}
  Straightforward.
\end{proof}

\begin{lemma}[Evar Rename]
  \label{lem:evar-rename}
  Assume $\judgctx{(\Theta_1, \Theta_2)}$.
  \begin{enumerate}
  \item If $\judgeterm{\Theta_1, a:\tau, \Theta_2; \Delta}{t}{\tau}$
    and $\ahat \notin \dom{\Delta}$,
    then $\judgeterm{\Theta_1, \Theta_2; \Delta, \ahat:\tau}{[\ahat/a]t}{\tau}$.
  \item If $\judgetp{\Theta_1, a:\tau, \Theta_2; \Delta}{A}{\Xi}$
    and $(a:\tau) \notin \Xi$
    and $\ahat \notin \dom{\Delta}$,\\
    then $\judgetp{\Theta_1, \Theta_2; \Delta, \ahat:\tau}{[\ahat/a]A}{\Xi}$.
  \item If $\judgefunctor{\Theta_1, a:\tau, \Theta_2; \Delta}{\mathcal{F}}{\Xi}$
    and $(a:\tau) \notin \Xi$
    and $\ahat \notin \dom{\Delta}$,\\
    then $\judgefunctor{\Theta_1, \Theta_2; \Delta, \ahat:\tau}{[\ahat/a]\mathcal{F}}{\Xi}$.
  \item If $\judgealgebra{\Xi_i}{\Theta_1, a:\tau, \Theta_2; \Delta}{\alpha}{F}{\tau}$
    then $\judgealgebra{\Xi_i}{\Theta_1, \Theta_2; \Delta, \ahat:\tau}{[\ahat/a]\alpha}{([\ahat/a]F)}{\tau}$.
  \end{enumerate}
\end{lemma}
\begin{proof}
  Each part is proved by structural induction on the given index sorting
  or well-formedness derivation.
  Parts (2), (3), and (4) are mutually recursive.
\end{proof}

\begin{lemma}[Value-det.\ Evar Rename]
  \label{lem:value-determined-evar-rename}
  Assume $\judgctx{(\Theta_1, \Theta_2)}$.
  \begin{enumerate}
  \item If $\judgetp{\Theta_1, a:\tau, \Theta_2; \Delta}{A}{\Xi}$
    and $(a:\tau) \in \Xi$
    and $\ahat \notin \dom{\Delta} \cup \dom{\Xi}$,\\
    then $\judgetp{\Theta_1, \Theta_2; \Delta, \ahat:\tau}{[\ahat/a]A}{(\Xi-a), \ahat:\tau}$.
  \item If $\judgefunctor{\Theta_1, a:\tau, \Theta_2; \Delta}{\mathcal{F}}{\Xi}$
    and $(a:\tau) \in \Xi$
    and $\ahat \notin \dom{\Delta} \cup \dom{\Xi}$,\\
    then $\judgefunctor{\Theta_1, \Theta_2; \Delta, \ahat:\tau}{[\ahat/a]\mathcal{F}}{(\Xi-a), \ahat:\tau}$.
  \end{enumerate}
\end{lemma}
\begin{proof}
  By mutual induction on the structure of the given type or functor
  well-formedness derivation.
  \begin{enumerate}
  \item
    \begin{itemize}
      \ProofCasesRules{\AlgTpVoid, \AlgTpUnit, \AlgTpDown, \AlgTpUp}
      Impossible because they conclude with $\Xi = \cdot$.
      
      \DerivationProofCase{\AlgTpSum}
      { \judgetp{\Theta_1, a:\tau, \Theta_2; \Delta}{P_1}{\Xi_1} \\ \judgetp{\Theta_1, a:\tau, \Theta_2; \Delta}{P_2}{\Xi_2} }
      { \judgetp{\Theta_1, a:\tau, \Theta_2; \Delta}{P_1 + P_2}{\Xi_1 \sect \Xi_2}  }
      \begin{llproof}
        \Pf{(a:\tau)}{\in}{\Xi_1 \sect \Xi_2}{Given}
        \Pf{(a:\tau)}{\in}{\Xi_1}{Set theory}
        \Pf{(a:\tau)}{\in}{\Xi_2}{\ditto}
        \judgetpPf{\Theta_1, a:\tau, \Theta_2; \Delta}{P_1}{\Xi_1}{Subderivation}
        \judgetpPf{\Theta_1, \Theta_2; \Delta, \ahat:\tau}{[\ahat/a]P_1}{(\Xi_1 - a), \ahat:\tau}{By \ih}
        \judgetpPf{\Theta_1, \Theta_2; \Delta, \ahat:\tau}{[\ahat/a]P_2}{(\Xi_2 - a), \ahat:\tau}{Similarly}
        \decolumnizePf
        \judgetpPf{\Theta_1, \Theta_2; \Delta, \ahat:\tau}{[\ahat/a]P_1 + [\ahat/a]P_2}{((\Xi_1 - a), \ahat:\tau) \sect ((\Xi_2 - a), \ahat:\tau)}{By \AlgTpSum}
        \judgetpPf{\Theta_1, \Theta_2; \Delta, \ahat:\tau}{[\ahat/a](P_1 + P_2)}{((\Xi_1 - a), \ahat:\tau) \sect ((\Xi_2 - a), \ahat:\tau)}{By \defsubst \defn}
        \decolumnizePf
        \Pf{((\Xi_1 - a), \ahat:\tau) \sect ((\Xi_2 - a), \ahat:\tau)}{=}{((\Xi_1 \sect \Xi_2) - a), \ahat:\tau}{Set theory}
        \Pf{}{=}{(\Xi - a), \ahat:\tau}{Current case}
      \end{llproof} 

      \ProofCaseRule{\AlgTpProd}
      Similar to \AlgTpSum case.

      \DerivationProofCase{\AlgTpEx}
      { \judgetp{\Theta_1, a:\tau, \Theta_2, b:\tau'; \Delta}{P}{\Xi, b:\tau'} }
      { \judgetp{\Theta_1, a:\tau, \Theta_2; \Delta}{\extype{b:\tau'}{P}}{\Xi} }
      \begin{llproof}
        \Pf{(a:\tau)}{\in}{\Xi}{Given}
        \Pf{(a:\tau)}{\in}{\Xi, b:\tau'}{Set theory}
        \judgetpPf{\Theta_1, a:\tau, \Theta_2, b:\tau'; \Delta}{P}{\Xi, b:\tau'}{Subderivation}
        \judgetpPf{\Theta_1, \Theta_2, b:\tau'; \Delta, \ahat:\tau}{[\ahat/a]P}{((\Xi, b:\tau')-a),\ahat:\tau}{By \ih}
        \Pf{((\Xi, b:\tau')-a),\ahat:\tau}{=}{((\Xi - a), b:\tau'),\ahat:\tau}{$a \neq b$}
        \Pf{}{=}{(\Xi - a), \ahat:\tau, b:\tau'}{Unordered}
        \judgetpPf{\Theta_1, \Theta_2, b:\tau'; \Delta, \ahat:\tau}{[\ahat/a]P}{(\Xi - a), \ahat:\tau, b:\tau'}{By equation}
        \judgetpPf{\Theta_1, \Theta_2; \Delta, \ahat:\tau}{\extype{b:\tau'}[\ahat/a]P}{(\Xi - a), \ahat:\tau}{By \AlgTpEx}
        \judgetpPf{\Theta_1, \Theta_2; \Delta, \ahat:\tau}{[\ahat/a]\extype{b:\tau'}P}{(\Xi - a), \ahat:\tau}{By \defn of \defsubst}
      \end{llproof}

      \DerivationProofCase{\AlgTpFixVar}
      {
        \judgefunctor{\Theta; \Delta}{F}{\Xi_F}
        \\
        \judgealgebra{\cdot}{\Theta; \Delta}{\alpha}{F}{\tau'}
        \\
        (b : \tau') \in \Theta
      }
      {
        \judgetp{\underbrace{\Theta}_{\Theta_1, a:\tau, \Theta_2}; \Delta}{\comprehend{\nu:\mu F}{\Fold{F}{\alpha}\,{\nu} =_{\tau'} b}}{\Xi_F \union b:\tau'}
      }
      \begin{llproof}
        \Pf{(a:\tau)}{\in}{\Xi_F \union b:\tau'}{Given}
      \end{llproof} 
      \begin{itemize}
        \item \textbf{Subcase} $(a:\tau) \notin \Xi_F$:\\
          \begin{llproof}
            \Pf{a}{=}{b}{Current subcase}
            \Pf{\tau}{=}{\tau'}{\ditto}
            \judgefunctorPf{\Theta; \Delta}{F}{\Xi_F}{Subderivation}
            \judgefunctorPf{\Theta_1, \Theta_2; \Delta, \ahat:\tau}{[\ahat/a]F}{\Xi_F}{By \Lemmaref{lem:evar-rename}}
            \judgealgebraPf{\cdot}{\Theta; \Delta}{\alpha}{F}{\tau'}{Subderivation}
            \judgealgebraPf{\cdot}{\Theta_1, \Theta_2; \Delta, \ahat:\tau}{[\ahat/a]\alpha}{([\ahat/a]F)}{\tau'}{By \Lemmaref{lem:evar-rename}}
            \decolumnizePf
            \judgetpPf{\Theta_1, \Theta_2; \Delta, \ahat:\tau}{\comprehend{\nu:\mu [\ahat/a]F}{\Fold{}{[\ahat/a]\alpha}\,{\nu} =_{\tau} \ahat}}{\Xi_F \union \ahat:\tau}{By \AlgTpFixEVar}
            \judgetpPf{\Theta_1, \Theta_2; \Delta, \ahat:\tau}{\comprehend{\nu:\mu [\ahat/a]F}{\Fold{}{[\ahat/a]\alpha}\,{\nu} =_{\tau} \ahat}}{\Xi_F \union \ahat:\tau}{By \AlgTpFixEVar}
            \judgetpPf{\Theta_1, \Theta_2; \Delta, \ahat:\tau}{[\ahat/a]\comprehend{\nu:\mu F}{\Fold{F}{\alpha}\,{\nu} =_{\tau'} b}}{\Xi_F \union \ahat:\tau}{By \defn of subst.}
            \decolumnizePf
            \Pf{\Xi_F \union \ahat:\tau}{=}{\Xi_F, \ahat:\tau}{$\ahat \notin \dom{\Xi_F}$}
            \Pf{}{=}{\Xi_F - a, \ahat:\tau}{$a \notin \dom{\Xi_F}$}
            \Pf{}{=}{(\Xi_F \union b:\tau') - a, \ahat:\tau}{Set theory, $b=a$}
            \Pf{}{=}{\Xi - a, \ahat:\tau}{Current case}
          \end{llproof} 
        \item \textbf{Subcase} $(a:\tau) \in \Xi_F$:\\
          Similar to preceding subcase, 
          but using \ih rather than \Lemmaref{lem:evar-rename}
          on the functor well-formedness subderivation.
          If $a = b$, use \AlgTpFixEVar.
          If $a \neq b$, use \AlgTpFixVar.
      \end{itemize}

      \ProofCasesRules{\AlgTpFixEVar, \AlgTpFixSolvedEVar, \AlgTpFix}
      Each is similar to the \AlgTpFixVar case, but simpler.
      The \AlgTpFix case uses \Lemmaref{lem:evar-rename}
      on its index sorting subderivation.

      \ProofCasesRules{\AlgTpWith, \AlgTpImplies}
      Straightforward. Uses \Lemmaref{lem:evar-rename}.

      \ProofCaseRule{\AlgTpAll}
      Similar to \AlgTpEx case.

      \ProofCaseRule{\AlgTpArrow}
      Similar to \AlgTpProd case.
    \end{itemize}
  \item Similar to part (1).
    \qedhere
  \end{enumerate}
\end{proof}

\begin{lemma}[Ext.\ Subst.\ Invariance]
  \label{lem:ext-subst-invariance}
  Assume $\extend{\Theta}{\Delta}{\Delta'}$.
  \begin{enumerate}
  \item
    If $\judgeterm{\Theta; \Delta}{t}{\tau}$,
    then $[\Delta']([\Delta]t) = [\Delta']t$.
  \item
    If $\judgetp{\Theta; \Delta}{A}{\Xi}$,
    then $[\Delta']([\Delta]A) = [\Delta']A$.
  \item
    If $\judgefunctor{\Theta; \Delta}{\mathcal{F}}{\Xi}$,
    then $[\Delta']([\Delta]\mathcal{F}) = [\Delta']\mathcal{F}$.
  \item
    If $\judgewf{\Theta; \Delta}{W}$, then $[\Delta']([\Delta]W) = [\Delta']W$.
  \item
    If $\judgewf{\Theta; \Delta}{\chi}$, then $[\Delta']([\Delta]\chi) = [\Delta']\chi$.
  \end{enumerate}
\end{lemma}
\begin{proof}
  Part (1) is proved by structural induction on $t$.
  Parts (2) and (3) are proved by mutual induction on the structure of the derivation.
  Part (4) is proved by structural induction on $W$.
  Part (5) is proved by structural induction on $\chi$.
\end{proof}

\begin{lemma}[Alg.\ Subst.\ on Xi]
  \label{lem:alg-subst-Xi}
  Assume $\judgectx{\Theta}{\Delta}$.
  For all $\hat{\Xi}_1$ and $\hat{\Xi}_2$, the set
  \[
    \left( [\Delta]\hat{\Xi}_1 \cap [\Delta]\hat{\Xi}_2 \right) \setminus [\Delta](\hat{\Xi}_1 \cap \hat{\Xi}_2)
  \]
  is equal to the set containing elements $(a:\tau)$
  such that $(a:\tau) \notin [\Delta](\hat{\Xi}_1 \cap \hat{\Xi}_2)$\\
  and such that there exist
  $(\ahat_1:\tau)\in\hat{\Xi}_1$ and $(\ahat_2:\tau)\in\hat{\Xi}_2$\\
  such that
  $\ahat_1 \neq \ahat_2$
  and
  $[\Delta](\ahat_1 : \tau) = (a:\tau) = [\Delta](\ahat_2 : \tau)$.
\end{lemma}
\begin{proof}
  The ``$\supseteq$'' direction is straightforward.
  
  Assume
  $\xi \in \left( [\Delta]\hat{\Xi}_1 \cap [\Delta]\hat{\Xi}_2 \right) \setminus [\Delta](\hat{\Xi}_1 \cap \hat{\Xi}_2)$.
  
  \begin{enumerate}
  \item \textbf{Case} $\xi = (a:\tau)$
    
    Either $(a:\tau) \in \hat{\Xi}_1 \cap \hat{\Xi}_2$,\\
    or there exist $(\ahat_1 : \tau) \in \hat{\Xi}_1$
    and $(\ahat_2 : \tau) \in \hat{\Xi}_2$\\
    such that $\ahat_1 \in \dom{\Delta}$
    and $\ahat_2 \in \dom{\Delta}$
    and $[\Delta](\ahat_1 : \tau) = (a:\tau) = [\Delta](\ahat_2 : \tau)$.\\
    The first subcase is impossible
    because $(a:\tau) \in [\Delta](\hat{\Xi}_1 \cap \hat{\Xi}_2)$ is
    a contradiction.\\
    To complete the proof for the second subcase,
    it suffices to show that $\ahat_1 \neq \ahat_2$:\\
    if $\ahat_1 = \ahat_2$ then $(a:\tau) \in [\Delta](\hat{\Xi}_1 \cap \hat{\Xi}_2)$,
    a contradiction.

  \item \textbf{Case} $\xi = (\ahat:\tau)$

    We show this case is impossible.\\
    Either $(\ahat:\tau) \in \hat{\Xi}_1 \cap \hat{\Xi}_2$
    and $\ahat \notin \dom{\Delta}$,\\
    or there exist $(\ahat_1 : \tau) \in \hat{\Xi}_1$
    and $(\ahat_2 : \tau) \in \hat{\Xi}_2$\\
    such that $\ahat_1 \in \dom{\Delta}$
    and $\ahat_2 \in \dom{\Delta}$
    and $[\Delta](\ahat_1 : \tau) = (\ahat:\tau) = [\Delta](\ahat_2 : \tau)$.\\
    The first subcase is impossible
    because $(\ahat:\tau) \in [\Delta](\hat{\Xi}_1 \cap \hat{\Xi}_2)$ is
    a contradiction.\\
    The second subcase is also impossible:\\
    by inversion on the given $\judgectx{\Theta}{\Delta}$,
    all solutions in $\Delta$ are ground.
    \qedhere
  \end{enumerate}
\end{proof}

\begin{lemma}[Right-hand Subst.]
  \label{lem:right-hand-subst}
  ~
  \begin{enumerate}
  \item
    If $\judgeterm{\Theta; \Delta}{t}{\tau}$,
    then $\judgeterm{\Theta; \Delta}{[\Delta]t}{\tau}$.
  \item
    If $\judgetp{\Theta; \Delta}{A}{\Xi}$,
    then there exists $\Xi' \supseteq [\Delta]\Xi$
    such that $\judgetp{\Theta; \Delta}{[\Delta]A}{\Xi'}$;\\
    moreover, $\dom{\Xi' \setminus [\Delta]\Xi}$
    does not contain existential variables.
  \item
    If $\judgefunctor{\Theta; \Delta}{\mathcal{F}}{\Xi}$,
    then there exists $\Xi' \supseteq [\Delta]\Xi$
    such that $\judgefunctor{\Theta; \Delta}{[\Delta]\mathcal{F}}{\Xi'}$;\\
    moreover, $\dom{\Xi' \setminus [\Delta]\Xi}$
    does not contain existential variables.
  \item
    If $\judgealgebra{\Xi}{\Theta; \Delta}{\alpha}{F}{\tau}$,
    then $\judgealgebra{\Xi}{\Theta; \Delta}{\alpha}{([\Delta]F)}{\tau}$.
  \end{enumerate}
\end{lemma}
\begin{proof}
  Part (1) is proved by structural induction on the index sorting derivation.
  Parts (2), (3), and (4) are proved by mutual induction on the given derivation.
  The proof is largely similar to \Lemmaref{lem:syn-subs-tp-fun-alg},
  the main difference being the use of \Lemmaref{lem:alg-subst-Xi}
  for rule cases concluding a judgment outputting a set intersection,
  and the only other differences being some more basic set theory.
\end{proof}

\begin{lemma}[Right-hand Subst. (Unroll)]
  \label{lem:right-hand-subst-unroll}
  ~\\
  If $\judgeunroll{\Xi}{\Theta; \Delta}{\nu:G[\mu F]}{\beta}{G\; \Fold{F}{\alpha}\;\nu}{t}{P}{\tau}$,\\
  then $\judgeunroll{\Xi}{\Theta; \Delta}{\nu:([\Delta]G)[\mu ([\Delta]F)]}{\beta}{([\Delta]G)\; \Fold{[\Delta]F}{\alpha}\;\nu}{[\Delta]t}{[\Delta]P}{\tau}$.
\end{lemma}
\begin{proof}
  By structural induction on the given unrolling derivation.
  Use \Lemmaref{lem:algebras-are-ground}.
\end{proof}

\begin{lemma}[Complete Extraction]
  \label{lem:complete-alg-extract}
  ~
  \begin{enumerate}
  \item If $\judgeextract[\pm]{\Theta; \Delta}{A}{A'}{\Thetahat}$
    and $\extend{\Theta}{\Delta}{\Omega}$,
    then $\judgeextract[\pm]{\Theta}{[\Omega]A}{[\Omega]A'}{[\Omega]\Thetahat}$.
  \item If $\judgeextract[\pm]{\Theta; \Delta}{A}{A'}{\Thetahat}$
    and $\rextend{\Theta}{\Delta}{\Omega}$,
    then $\judgeextract[\pm]{\Theta}{[\Omega]A}{[\Omega]A'}{[\Omega]\Thetahat}$.
  \end{enumerate}
\end{lemma}
\begin{proof}
  Part (2) is proved by structural induction on the given extraction derivation.
  Part (1) follows from part (2) and \Lemmaref{lem:extension-sound}.
\end{proof}

\begin{lemma}[Alg.\ Unrolling Output WF]
  \label{lem:alg-unroll-output-wf}
  ~\\
  If $\judgeunroll{\Xi}{\Theta; \Delta}{\nu:G[\mu F]}{\beta}{G\; \Fold{F}{\alpha}\;\nu}{t}{P}{\tau}$
  and $\judgefunctor{\Theta; \Delta}{G}{\Xi_G}$,\\
  then there exists $\Xi'$ such that $\judgetp{\Theta; \Delta}{P}{\Xi'}$
  and $\Xi_G \subseteq \Xi'$.
\end{lemma}
\begin{proof}
  By structural induction on the given unrolling derivation.
\end{proof}

\begin{lemma}[Complete Alg.\ Unrolling]
  \label{lem:complete-alg-unroll}
  ~\\
  If $\extend{\Theta}{\Delta}{\Omega}$
  and $\judgeunroll{\Xi}{\Theta; \Delta}{\nu:G[\mu F]}{\beta}{G\; \Fold{F}{\alpha}\;\nu}{t}{P}{\tau}$,\\
  then $\judgeunroll{\Xi}{\Theta}{\nu:([\Omega]G)[\mu [\Omega]F]}{\beta}{([\Omega]G)\; \Fold{[\Omega]F}{\alpha}\;\nu}{[\Omega]t}{[\Omega]P}{\tau}$.
\end{lemma}
\begin{proof}
  By structural induction on the given unrolling derivation.
\end{proof}

\begin{lemma}[Extract Applied]
  \label{lem:extract-applied}
  If $\judgeextract{\Theta; \Delta}{A}{A'}{\Thetahat}$
  and $[\Delta]A = A$,\\
  then $[\Delta]\Thetahat = \Thetahat$
  and $[\Delta]A' = A'$.
\end{lemma}
\begin{proof}
  By structural induction on the given extraction derivation.
\end{proof}

\begin{lemma}[Unroll Applied]
  \label{lem:unroll-applied}
  If $\judgeunroll{\Xi}{\Theta; \Delta}{\nu:G[\mu F]}{\beta}{G\; \Fold{F}{\alpha}\;\nu}{t}{P}{\tau}$
  and $[\Delta]G = G$
  and $[\Delta]F = F$
  and $[\Delta]t = t$,
  then $[\Delta]P = P$.
\end{lemma}
\begin{proof}
  By structural induction on the given unrolling derivation,
  case analyzing its concluding rule.
  Each case is straightforward.
  The \DeclUnrollUnit case uses \Lemmaref{lem:algebras-are-ground}.
\end{proof}

\begin{lemma}[Apply Idempotent]
  \label{lem:apply-idempotent}
  For any object $\mathcal{O}$ that's well-typed under $\judgectx{\Theta}{\Delta}$
  among perhaps other contexts,
  we have $[\Delta]([\Delta]\mathcal{O}) = [\Delta]\mathcal{O}$.
\end{lemma}
\begin{proof}
  By structural induction on $\mathcal{O}$.
  Follows from the fact that index term solutions in $\Delta$ are ground.
\end{proof}

\begin{lemma}[Output Applied]
  \label{lem:output-applied}
  ~
  \begin{enumerate}
  \item If $\algequiv[]{\Theta; \Delta}{\F}{\Gee}{W}{\Delta'}$
    and $\ground{\F}$ and $[\Delta]\Gee = \Gee$,
    then $[\Delta']W = W$.
  \item If $\algequiv[+]{\Theta; \Delta}{P}{Q}{W}{\Delta'}$
    and $\ground{P}$ and $[\Delta]Q = Q$,
    then $[\Delta']W = W$.
  \item If $\algequiv[-]{\Theta; \Delta}{N}{M}{W}{\Delta'}$
    and $\ground{M}$ and $[\Delta]N = N$,
    then $[\Delta']W = W$.
  \item If $\algsub[+]{\Theta; \Delta}{P}{Q}{W}{\Delta'}$
    and $\ground{P}$ and $[\Delta]Q = Q$,
    then $[\Delta']W = W$.
  \item If $\algsub[-]{\Theta; \Delta}{N}{M}{W}{\Delta'}$
    and $\ground{M}$ and $[\Delta]N = N$,
    then $[\Delta']W = W$.
  \item If $\algchk{\Theta; \Delta}{\Gamma}{v}{P}{\chi}{\Delta'}$
    and $[\Delta]P = P$,
    then $[\Delta']\chi = \chi$.
  \item If $\algspine{\Theta; \Delta}{\Gamma}{s}{N}{\upshift{P}}{\chi}{\Delta'}$
    and $[\Delta]N = N$,
    then $[\Delta']\chi = \chi$
    and $[\Delta']P = P$.
  \end{enumerate}
\end{lemma}
\begin{proof}
  By structural induction on the given derivation,
  using
  \Lemmaref{lem:ext-subst-invariance},
  \Lemmaref{lem:ext-refl},
  \Lemmaref{lem:extract-applied},
  and
  \Lemmaref{lem:apply-idempotent}
  as needed.
  Parts (1) and (2) are mutually recursive.
  Part (4) uses parts (1) and (2).
  Part (6) uses part (4).
  Part (7) uses part (6).
\end{proof}

\begin{lemma}[Uncomplete Extract]
  \label{lem:uncomplete-extract}
  If $\rextend{\Theta}{\Delta}{\Omega}$
  and $\judgeextract[\pm]{\Theta}{[\Omega]A}{A_\Omega}{\Theta_\Omega}$,\\
  then there exist $A'$ and $\Thetahat$
  such that $\judgeextract[\pm]{\Theta; \Delta}{A}{A'}{\Thetahat}$
  and $[\Omega]A' = A_\Omega$
  and $[\Omega]\Thetahat = \Theta_\Omega$.
\end{lemma}
\begin{proof}
  By structural induction on the given extraction derivation,
  using \Lemmaref{lem:weaken-ext} as needed.
\end{proof}

\begin{lemma}[Uncomplete Unrolling]
  \label{lem:uncomplete-alg-unroll}
  ~\\
  If $\rextend{\Theta}{\Delta}{\Omega}$
  and $\judgeunroll{\Xi}{\Theta}{\nu:([\Omega]G)[\mu [\Omega]F]}{\beta}{([\Omega]G)\; \Fold{[\Omega]F}{\alpha}\;\nu}{[\Omega]t}{P}{\tau}$\\
  and $\judgefunctor{\Theta; \Delta}{G}{\dontcare}$
  and $\judgefunctor{\Theta; \Delta}{F}{\dontcare}$
  and $\judgeterm{\Theta; \Delta}{t}{\tau}$,\\
  then there exists $P'$
  such that $\judgeunroll{\Xi}{\Theta; \Delta}{\nu:G[\mu F]}{\beta}{G\; \Fold{F}{\alpha}\;\nu}{t}{P'}{\tau}$
  and $[\Omega]P' = P$.
\end{lemma}
\begin{proof}
  By structural induction on the given unrolling derivation.
\end{proof}

\section{Algorithmic Decidability}
\label{sec:apx-decidability}

It is straightforward to check that all the well-formedness (including index sorting)
judgments are decidable (and we often don't mention this in subsequent proofs).

When measuring judgments, we only take their inputs into account,
and often put ``$\dontcare$'' for their outputs to reflect this.

\begin{lemma}[Decidability of Algebra Pattern Match]
  \label{lem:decidable-algebra-pattern}
  ~
  \begin{enumerate}
  \item Given algebra $\alpha$,
    it is decidable whether there exists $\alpha_1$
    such that $\composeinj{1}{\alpha}{\alpha_1}$.
  \item Given algebra $\alpha$,
    it is decidable whether there exists $\alpha_2$
    such that $\composeinj{2}{\alpha}{\alpha_2}$.
  \end{enumerate}
\end{lemma}
\begin{proof}
  Straightforward
  (measure $\composeinj{k}{\alpha}{\dontcare}$ by the number of clauses in $\alpha$).
\end{proof}

\begin{lemma}[Decidability of Instantiation]
  \label{lem:decidable-inst}
  ~
  \begin{enumerate}
  \item Given $\judgeterm{\Theta; \Delta}{\phi}{\Booltype}$,\\
    it is decidable whether there exists $\Delta'$
    such that $\alginst{\Theta; \Delta}{\phi}{\Delta'}$.
  \item Given $\judgeterm{\Theta; \Delta}{\phi}{\Booltype}$
    and $\judgeterm{\Theta; \Delta}{\psi}{\Booltype}$,\\
    it is decidable whether there exists $\Delta'$
    such that $\algpropequivinst{\Theta; \Delta}{\phi}{\psi}{\Delta'}$.
  \end{enumerate}
\end{lemma}
\begin{proof}
  Each part is straightforward.
\end{proof}

\begin{lemma}[Decidability of Equivalence]
  \label{lem:decidable-equiv}
  ~
  \begin{enumerate}
  \item
    Given $\judgefunctor{\Theta; \Delta}{\mathcal{F}}{\dontcare}$
    and $\judgefunctor{\Theta; \Delta}{\mathcal{G}}{\dontcare}$,\\
    it is decidable whether there exist $W$ and $\Delta'$ such that
    $\algequiv[]{\Theta; \Delta}{\mathcal{F}}{\mathcal{G}}{W}{\Delta'}$.
  \item
    Given $\judgetp{\Theta; \Delta}{A}{\dontcare}$
    and $\judgetp{\Theta; \Delta}{B}{\dontcare}$,\\
    it is decidable whether there exist $W$ and $\Delta'$ such that
    $\algequiv[\pm]{\Theta; \Delta}{A}{B}{W}{\Delta'}$.
  \end{enumerate}
\end{lemma}
\begin{proof}
  For rules deriving judgments of the form
  \[
    \begin{array}[t]{lll}
      \algequiv[]{\Theta; \Delta}{\mathcal{F}}{\mathcal{G}}{\dontcare}{\dontcare} \\
      \algequiv[\pm]{\Theta; \Delta}{A}{B}{\dontcare}{\dontcare}
    \end{array}
  \]
  (where we write ``$\dontcare$'' for parts of the judgments that are outputs),
  the following induction measure on such judgments is adequate to prove decidability:
  $\size{\mathcal{F}} + \size{\mathcal{G}}$
  or $\size{A} + \size{B}$
  (\defsize is defined in Figures~\ref{fig:size} and~\ref{fig:size-prog-chi}).

  In each rule deriving one of the above judgments,
  every premise of such judgmental form
  is smaller than the conclusion (according to the above measure);
  note that size is independent of index variables, so, for example,
  in rule \AlgTpEquivPosSum,
  $\size{[\Delta'']Q_2} = \size{Q_2} < \size{Q_1 + Q_2}$.
  Further, it is easy to show that every premise not of the above judgmental forms
  (those found in \AlgTpEquivPosWith, \AlgTpEquivPosFix, \AlgTpEquivPosFixInst, and \AlgTpEquivPosFixInstFun)
  is decidable.
\end{proof}

\begin{lemma}[Decidability of Extraction]
  \label{lem:decidable-extract}
  Given $\judgetp{\Theta; \Delta}{A}{\dontcare}$,\\
  it is decidable whether there exist $A'$ and $\Theta'$
  such that $\judgeextract[\pm]{\Theta; \Delta}{A}{A'}{\Theta'}$.
\end{lemma}
\begin{proof}
  Measuring the judgment $\judgeextract[\pm]{\Theta; \Delta}{A}{\dontcare}{\dontcare}$
  (where we put ``$\dontcare$'' for outputs)
  by $\size{A}$, it is straightforward to show that,
  in each rule deriving the above judgment, every premise gets smaller.
\end{proof}

\begin{theorem}[Decidability of Subtyping]
  \label{thm:decidable-sub}
  Given $\judgetp{\Theta; \Delta}{A}{\dontcare}$
  and $\judgetp{\Theta; \Delta}{B}{\dontcare}$,\\
  it is decidable whether there exist $W$ and $\Delta'$ such that
  $\algsub[\pm]{\Theta; \Delta}{A}{B}{W}{\Delta'}$.
\end{theorem}
\begin{proof}
  Similar to \Lemmaref{lem:decidable-equiv},
  but cases \AlgSubPosSum, \AlgSubPosFix, \AlgSubPosFixInst, and \AlgSubPosFixInstFun
  use \Lemmaref{lem:decidable-equiv},
  cases \AlgSubPosDownshift and \AlgSubNegUpshift
  use \Lemmaref{lem:decidable-extract},
  and case \AlgSubPosWithR
  uses \Lemmaref{lem:decidable-inst}.
\end{proof}

\begin{lemma}[Decidability of Unrolling]
  \label{lem:decidable-unroll}
  ~\\
  Given $\judgealgebra{\Xi}{\Theta; \Delta}{\beta}{G}{\tau}$
  and $\judgealgebra{\cdot}{\Theta; \Delta}{\alpha}{F}{\tau}$
  and $\judgeterm{\Theta; \Delta}{t}{\tau}$,\\
  it is decidable whether there exists $P$ such that
  $\judgeunroll{\Xi}{\Theta; \Delta}{ \nu:G[\mu F] }{\beta}{G\;\Fold{F}{\alpha}\;\nu}{t}{P}{\tau}$.
\end{lemma}
\begin{proof}
  Measuring the judgment
  $\judgeunroll{\Xi}{\Theta; \Delta}{ \nu:G[\mu F] }{\beta}{G\;\Fold{F}{\alpha}\;\nu}{t}{\dontcare}{\tau}$
  (putting ``$\dontcare$'' for the output)
  by $\size{G}$,
  it is easy to show that,
  in each rule deriving said judgment,
  every premise gets smaller.
  The algebra pattern matching premises $\composeinj{k}{\alpha}{\alpha_k}$
  in \AlgUnrollSum are decidable by \Lemmaref{lem:decidable-algebra-pattern}.
\end{proof}

\begin{lemma}[Decidability of Prop.\ Truth]
  \label{lem:decidable-prop-truth}
  Given $\judgeterm{\Theta}{\phi}{\Booltype}$,
  it is decidable whether $\judgeentail{\Theta}{\phi}$.
\end{lemma}
\begin{proof}
  See \citet{BarrettSMT}.
\end{proof}

\begin{lemma}[Decidability of Prop.\ Equiv.]
  \label{lem:decidable-prop-equiv}
  Given $\judgeterm{\Theta}{\phi}{\Booltype}$
  and $\judgeterm{\Theta}{\psi}{\Booltype}$,\\
  it is decidable whether $\judgeequiv{\Theta}{\phi}{\psi}$.
\end{lemma}
\begin{proof}
  In each rule concluding $\judgeequiv{\Theta}{\phi}{\psi}$,
  for every premise of this form,
  the structure of both $\phi$ and $\psi$ get smaller.
  The \PrpEquivEq and \PrpEquivLeq cases use \Lemmaref{lem:decidable-prop-truth}.
\end{proof}

\begin{lemma}[Alg.\ Equiv.\ Shrinks Constraints]
  \label{lem:alg-equiv-shrinks-problem}
  ~
  \begin{enumerate}
  \item If $\algequiv[]{\Theta; \Delta}{\mathcal{F}}{\mathcal{G}}{W}{\Delta'}$, 
    then $\size{W} < \size{\mathcal{F}} + \size{\mathcal{G}} + 1$.
  \item If $\algequiv[+]{\Theta; \Delta}{P}{Q}{W}{\Delta'}$,
    then $\size{W} < \size{\poseqprob{P}{Q}}$.
  \item If $\algequiv[-]{\Theta; \Delta}{N}{M}{W}{\Delta'}$,
    then $\size{W} < \size{\negeqprob{N}{M}}$.
  \end{enumerate}
\end{lemma}
\begin{proof}
  Similar to proof of \Lemmaref{lem:semidecl-equiv-shrinks-problem}.
\end{proof}

\begin{lemma}[Alg.\ Sub.\ Shrinks Constraints]
  \label{lem:alg-sub-shrinks-problem}
  ~
  \begin{enumerate}
  \item If $\algsub[+]{\Theta; \Delta}{P}{Q}{W}{\Delta'}$,
    then $\size{W} < \size{\possubprob{P}{Q}}$.
  \item If $\algsub[-]{\Theta; \Delta}{N}{M}{W}{\Delta'}$,
    then $\size{W} < \size{\negsubprob{N}{M}}$.
  \end{enumerate}
\end{lemma}
\begin{proof}
  Similar to proof of \Lemmaref{lem:semidecl-sub-shrinks-problem}.
\end{proof}

\begin{lemma}[Decidability of Constraint Checking]
  \label{lem:decidable-problem-checking}
  Given $\judgewf{\Theta}{W}$,\\
  it is decidable whether $\entailwah{\Theta}{W}$.
\end{lemma}
\begin{proof}
  We measure the judgment $\entailwah{\Theta}{W}$ by $\size{W}$
  (see \Figureref{fig:size}).
  We will show that, in each rule deriving said judgment,
  every premise is either smaller than the conclusion (according to this measure)
  or already known to be decidable.

  \begin{itemize}
    \ProofCaseRule{\WTrueProp}
    The premise is decidable by \Lemmaref{lem:decidable-prop-truth}.

    \ProofCaseRule{\WTruePrpEquiv}
    The premise is decidable by \Lemmaref{lem:decidable-prop-equiv}.

    \ProofCaseRule{\WTrueAnd, \WTrueImpl, \WTrueAll}
    For every premise, $\size{W}$ decreases.

    \ProofCaseRule{\WTruePosSub, \WTrueNegSub}
    The first premise is decidable by \Theoremref{thm:decidable-sub}.
    By \Lemmaref{lem:alg-sub-shrinks-problem} on the first premise,
    the second premise decreases in size.

    \ProofCaseRule{\WTruePosEquiv, \WTrueNegEquiv}
    Similar to the \WTruePosSub and \WTrueNegSub cases,
    but using \Lemmaref{lem:decidable-equiv} and \Lemmaref{lem:alg-equiv-shrinks-problem}
    rather than the theorem and lemma listed in those cases.
    \qedhere
  \end{itemize}
\end{proof}

\begin{lemma}[Program Typing Shrinks Constraints]
  \label{lem:program-typing-shrinks-problems}
  ~
  \begin{enumerate}
  \item If $\algchk{\Theta; \Delta}{\Gamma}{v}{P}{\chi}{\Delta'}$,
    then $\size{\chi} \leq \size{v}$.
  \item If $\algspine{\Theta; \Delta}{\Gamma}{s}{N}{\upshift{P}}{\chi}{\Delta'}$,
    then $\size{\chi} \leq \size{s}$.
  \end{enumerate}
\end{lemma}
\begin{proof}
  Each part is proved by structural induction on the given typing derivation.
  \begin{enumerate}
  \item
    \begin{itemize}
      \DerivationProofCase{\AlgChkValVar}
      {
        P \neq \exists, \land 
        \\
        (x:Q)\in\Gamma
        \\
        \algsub[+]{\Theta; \Delta}{Q}{P}{\Wah}{\Delta'}
      }
      {
        \algchk{\Theta; \Delta}{\Gamma}{x}{P}{\Wah}{\Delta'}
      }
      \begin{llproof}
        \Pf{\size{\underbrace{\Wah}_{\Wah, \cdot}}}{=}{\size{\cdot}}{By \defn of \defsize[\chi]}
        \Pf{}{=}{0}{By \defn of \defsize[\chi]}
        \Pf{}{=}{\size{x}}{By \defn of \defsize[v]}
      \end{llproof} 

      \DerivationProofCase{\AlgChkValUnit}
      {
      }
      {
        \algchk{\Theta; \Delta}{\Gamma}{\unit}{\unitty}{\cdot}{\Delta}
      }
      \begin{llproof}
        \Pf{\size{\cdot}}{=}{0}{By \defn of \defsize[\chi]}
        \Pf{}{=}{\size{\unit}}{By \defn of \defsize[v]}
      \end{llproof}

      \DerivationProofCase{\AlgChkValPair}
      {
        \algchk{\Theta; \Delta}{\Gamma}{v_1}{P_1}{\chi_1}{\Delta''}
        \\
        \algchk{\Theta; \Delta''}{\Gamma}{v_2}{[\Delta'']P_2}{\chi_2}{\Delta'}
      }
      {
        \algchk{\Theta; \Delta}{\Gamma}{\pair{v_1}{v_2}}
        {(P_1 \times P_2)}{[\Delta']\chi_1, \chi_2}{\Delta'}
      }
      \begin{llproof}
        \Pf{\size{[\Delta']\chi_1, \chi_2}}{=}{\size{\chi_1, \chi_2}}{Size independent of index terms}
        \Pf{}{=}{\size{\chi_1} + \size{\chi_2}}{Straightforward}
        \Pf{}{\leq}{\size{v_1} + \size{v_2}}{By \ih (twice)}
        \Pf{}{<}{\size{v_1} + \size{v_2} + 1}{}
        \Pf{}{=}{\size{\pair{v_1}{v_2}}}{By \defn of \defsize[v]}
      \end{llproof}

      \DerivationProofCase{\AlgChkValIn{k}}
      {
        \algchk{\Theta; \Delta}{\Gamma}{v_k}{P_k}{\chi}{\Delta'}
      }
      {
        \algchk{\Theta; \Delta}{\Gamma}{\inj{k}{v_k}}{(P_1 + P_2)}{\chi}{\Delta'}
      }
      \begin{llproof}
        \Pf{\size{\chi}}{\leq}{\size{v_k}}{By \ih}
        \Pf{}{<}{\size{v_k} + 1}{}
        \Pf{}{=}{\size{\inj{k}{v_k}}}{By \defn of \defsize[v]}
      \end{llproof} 

      \DerivationProofCase{\AlgChkValExists}
      {
        \algchk{\Theta; \Delta, \ahat:\tau}{\Gamma}{v}{[\ahat / a]P_0}{\chi}{\Delta', \hypeq{\ahat}{\tau}{t}}
      }
      {
        \algchk{\Theta; \Delta}{\Gamma}{v}{(\extype{a : \tau} P_0)}{\chi}{\Delta'}
      }
      \begin{llproof}
        \Pf{\size{\chi}}{\leq}{\size{v}}{By \ih}
      \end{llproof}

      \DerivationProofCase{\AlgChkValWith}
      {
        \algchk{\Theta; \Delta}{\Gamma}{v}{P_0}{\chi_0}{\Delta''}
        \\
        \alginst{\Theta; \Delta''}{[\Delta'']\phi}{\Delta'}
      }
      {
        \algchk{\Theta; \Delta}{\Gamma}{v}{(P_0 \andty \phi)}{([\Delta']\phi, [\Delta']\chi_0)}{\Delta'}
      }
      \begin{llproof}
        \Pf{\size{[\Delta']\phi, [\Delta']\chi_0}}{=}{\size{\phi, \chi_0}}{Size unaffected by index substitution}
        \Pf{}{=}{\size{\chi_0}}{By \defn of \defsize}
        \Pf{}{\leq}{\size{v}}{By \ih}
      \end{llproof}

      \DerivationProofCase{\AlgChkValFix}
      {
        \judgeunroll{\cdot}{\Theta; \Delta}{\nu:F[\mu F]}{\alpha}{F\; \Fold{F}{\alpha}\;\nu}{t}{Q}{\tau}
        \\
        \algchk{\Theta; \Delta}{\Gamma}{v_0}{Q}{\chi}{\Delta'}
      }
      {
        \algchk{\Theta; \Delta}{\Gamma}{\into{v_0}}{\comprehend{\nu : \mu F}{\Fold{F}{\alpha}\,{\nu} =_\tau t}}{\chi}{\Delta'}
      }
      \begin{llproof}
        \Pf{\size{\chi}}{\leq}{\size{v_0}}{By \ih}
        \Pf{}{<}{\size{v_0} + 1}{}
        \Pf{}{=}{\size{\into{v_0}}}{By \defn of \defsize}
      \end{llproof}

      \DerivationProofCase{\AlgChkValDownshift}
      {
      }
      {
        \algchk{\Theta; \Delta}{\Gamma}{\thunk{e}}{\downshift{N}}{(e <= N)}{\Delta}
      }
      \begin{llproof}
        \Pf{\size{e <= N}}{=}{\size{e} + \size{\cdot} + 1}{By \defn of \defsize[\chi]}
        \Pf{}{=}{\size{e} + 1}{By \defn of \defsize[\chi]}
        \Pf{}{=}{\size{\thunk{e}}}{By \defn of \defsize[e]}
      \end{llproof}
    \end{itemize}
    
  \item
    \begin{itemize}
      \DerivationProofCase{\AlgSpineAll}
      {
        \algspine{\Theta; \Delta, \ahat : \tau}{\Gamma}{s}{[\ahat/a]{N_0}}{\upshift{P}}{\chi}{\Delta', \hypeq{\ahat}{\tau}{t}}
      }
      {
        \algspine{\Theta; \Delta}{\Gamma}{s}{(\alltype{a:\tau}N_0)}{\upshift{P}}{\chi}{\Delta'}
      }
      \begin{llproof}
        \Pf{\size{\chi}}{\leq}{\size{s}}{By \ih}
      \end{llproof}

      \ProofCaseRule{\AlgSpineImplies}
      Similar to \AlgChkValWith case of part (3).

      \DerivationProofCase{\AlgSpineApp}
      {
        \algchk{\Theta; \Delta}{\Gamma}{v}{Q}{\chi_1}{\Delta''}
        \\
        \algspine{\Theta; \Delta''}{\Gamma}{s_0}{[\Delta'']N_0}{\upshift{P}}{\chi_2}{\Delta'}
      }
      {
        \algspine{\Theta; \Delta}{\Gamma}{v, s_0}{Q -> N_0}{\upshift{P}}{[\Delta']\chi_1, \chi_2}{\Delta'}
      }
      \begin{llproof}
        \Pf{\size{[\Delta']\chi_1, \chi_2}}{=}{\size{\chi_1, \chi_2}}{Size independent of index terms}
        \Pf{}{=}{\size{\chi_1} + \size{\chi_2}}{Straightforward}
        \Pf{}{\leq}{\size{v} + \size{\chi_2}}{By part (1)}
        \Pf{}{\leq}{\size{v} + \size{s_0}}{By \ih (second subderivation)}
        \Pf{}{<}{\size{v} + \size{s_0} + 1}{}
        \Pf{}{=}{\size{v, s_0}}{By \defn of \defsize[s]}
      \end{llproof}

      \DerivationProofCase{\AlgSpineNil}
      {}
      {
        \algspine{\Theta; \Delta}{\Gamma}{\cdot}{\upshift{P}}{\upshift{P}}{\True}{\Delta}
      }
      \begin{llproof}
        \Pf{\size{\underbrace{\True}_{\True, \cdot}}}{=}{\size{\cdot}}{By \defn of \defsize[\chi]}
        \Pf{}{=}{0}{By \defn of \defsize[\chi]}
        \Pf{}{=}{\size{\cdot}}{By \defn of \defsize[s]}
      \end{llproof}
      \qedhere
    \end{itemize}
  \end{enumerate}
\end{proof}

\begin{theorem}[Decidability of Typing]
  \label{thm:decidable-typing}
  ~
  \begin{enumerate}
  \item Given $\judgewf{\Theta; \Gamma}{\chi}$,
    it is decidable whether $\algneg{\Theta}{\Gamma}{\chi}$.
  \item Given $\judgectx{\Theta}{\Gamma}$
    and head $h$,
    it is decidable whether there exists $P$ such that
    $\algsynhead{\Theta}{\Gamma}{h}{P}$.
  \item Given $\judgectx{\Theta}{\Gamma}$
    and bound expression $\be$,\\
    it is decidable whether there exists $P$ such that
    $\algsynexp{\Theta}{\Gamma}{\be}{\upshift{P}}$.
  \item Given $\judgectx{\Theta}{\Gamma}$
    and $\judgetp{\Theta; \Delta}{P}{\dontcare}$
    and value $v$,\\
    it is decidable whether there exists $\chi$ and $\Delta'$ such that
    $\algchk{\Theta; \Delta}{\Gamma}{v}{P}{\chi}{\Delta'}$.
  \item Given $\judgectx{\Theta}{\Gamma}$
    and $\judgetp{\Theta}{N}{\dontcare}$
    and expression $e$,\\
    it is decidable whether $\algchkneg{\Theta}{\Gamma}{e}{N}$.
  \item Given $\judgectx{\Theta}{\Gamma}$
    and $\judgetp{\Theta}{P}{\dontcare}$
    and $\judgetp{\Theta}{N}{\dontcare}$
    and match expression $\clauses{\pa}{e}{i}{I}$,\\
    it is decidable whether
    $\algchkmatch{\Theta}{\Gamma}{P}{\clauses{\pa}{e}{i}{I}}{N}$.
  \item Given $\judgectx{\Theta}{\Gamma}$
    and $\judgetp{\Theta; \Delta}{N}{\Xi}$
    and spine $s$,\\
    it is decidable whether there exists $\chi$ and $\Delta'$ such that
    $\algspine{\Theta; \Delta}{\Gamma}{s}{N}{\upshift{P}}{\chi}{\Delta'}$.
  \end{enumerate}
\end{theorem}
\begin{proof}
  For rules deriving judgments of the form
  \[
    \begin{array}[t]{lll}
      \algneg{\Theta}{\Gamma}{\chi} \\
      \algsynhead{\Theta}{\Gamma}{h}{\dontcare} \\
      \algsynexp{\Theta}{\Gamma}{\be}{\dontcare} \\
      \algchk{\Theta; \Delta}{\Gamma}{v}{A}{\dontcare}{\dontcare} \\
      \algchkneg{\Theta}{\Gamma}{e}{A} \\
      \algchkmatch{\Theta}{\Gamma}{A}{\clauses{\pa}{e}{i}{I}}{N} \\
      \algspine{\Theta; \Delta}{\Gamma}{s}{A}{\dontcare}{\dontcare}{\dontcare}
    \end{array}
  \]
  (where we write ``$\dontcare$'' for parts of the judgments that are outputs),
  the following induction measure on such judgments is adequate to prove decidability:
  \[
    \left\langle
      \size{\chi} / \size{h} / \size{\be} / \size{v} / \size{e} / \size{\clauses{\pa}{e}{i}{I}} / \size{s},
      \size{A}/\#_{W}(\chi)
    \right\rangle
  \]
  where $\langle \dots \rangle$ denotes lexicographic order,
  \ie, first, the size of constraints ($\chi$) or the program term ($h$ etc.),
  and, second, the size of the type $A$
  or the number of $W$s listed in $\chi$.

  We will show that in each rule deriving one of the above judgments,
  every premise is either smaller than the conclusion (according to the above measure)
  or already known to be decidable.

  Note that whenever we mention the size of constraints $\chi$, a program term or type,
  we are referring to $\size{-}$,
  defined in Figures~\ref{fig:size} and~\ref{fig:size-prog-chi}.

  \begin{itemize}
    \ProofCaseRule{\ChkProblemsEmpty}
    No premises.

    \ProofCaseRule{\ChkProblemsNegChk}
    ~\\
    First premise:

    \begin{llproof}
      \Pf{\size{e}}{<}{\size{e} + \size{\chi} + 1}{}
      \Pf{}{=}{\size{(e <= N), \chi}}{By \defn of \defsize[\chi]}
    \end{llproof} 

    Second premise:

    \begin{llproof}
      \Pf{\size{\chi}}{<}{\size{e} + \size{\chi} + 1}{}
      \Pf{}{=}{\size{(e <= N), \chi}}{By \defn of \defsize[\chi]}
    \end{llproof}

    \ProofCaseRule{\ChkProblemsWah}

    The first premise is decidable by \Lemmaref{lem:decidable-problem-checking}.
    In the second premise, the size of $\chi$ stays the same, but it loses a $W$.
  \end{itemize}

  Next, we consider the synthesis judgments:

  \begin{itemize}
    \ProofCaseRule{\AlgSynHeadVar}
    No nontrivial premises.

    \ProofCaseRule{\AlgSynValAnnot}
    It is easy to show that the first premise
    (of form $\judgetp{\Theta; \Delta}{P}{\Xi}$)
    is decidable.
    The second premise is smaller because the program term decreases in size.
    As for the last premise:
    ~\\
    \begin{llproof}
      \Pf{\size{\chi}}{\leq}{\size{v}}{By \Lemmaref{lem:program-typing-shrinks-problems}}
      \Pf{}{<}{\size{v} + 1}{}
      \Pf{}{=}{\size{\annoexp{v}{P}}}{By \defn of \defsize}
    \end{llproof} 

    \ProofCaseRule{\AlgSynSpineApp}
    The first and second premises are smaller
    because the program term decreases in size.
    As for the last premise:
    ~\\
    \begin{llproof}
      \Pf{\size{\chi}}{\leq}{\size{s}}{By \Lemmaref{lem:program-typing-shrinks-problems}}
      \Pf{}{<}{\size{h} + \size{s} + 1}{}
      \Pf{}{=}{\size{h(s)}}{By \defn of \defsize}
    \end{llproof} 

    \ProofCaseRule{\AlgSynExpAnnot}
    It is easy to show that the first premise
    (of form $\judgetp{\Theta; \Delta}{P}{\Xi}$)
    is decidable.
    The second premise is smaller because the program term decreases in size.
  \end{itemize}

  Next, we consider the (non-spine or -match) type checking judgments:

  \begin{itemize}
    \ProofCaseRule{\AlgChkValVar}
    The first two premises are trivial.
    The third premise is decidable by \Theoremref{thm:decidable-sub}.

    \ProofCasesRules{\AlgChkValUnit, \AlgChkValDownshift}
    No premises.

    \ProofCasesRules{\AlgChkValPair, \AlgChkValIn{k}, \AlgChkExpMatch}
    For all premises, the size of the program term or constraints decreases.

    \ProofCaseRule{\AlgChkValExists}
    The size of the term stays the same,
    but the size of $A$ decreases (variable renaming does not affect size).

    \ProofCaseRule{\AlgChkValWith}
    In the first premise, the size of the term stays the same,
    but the size of $A$ decreases.
    The second premise is decidable by \Lemmaref{lem:decidable-inst}.

    \ProofCaseRule{\AlgChkValFix}
    The first premise is decidable by \Lemmaref{lem:decidable-unroll}.
    In the second premise, the program term (value) size decreases
    (note that we use \Lemmaref{lem:alg-unroll-output-wf}
    to know the premise's input type is well-formed).

    \ProofCaseRule{\AlgChkExpUpshift}
    In the first premise, the size of the program term decreases.
    As for the second premise:\\
    \begin{llproof}
      \Pf{\size{\chi}}{\leq}{\size{v}}{By \Lemmaref{lem:program-typing-shrinks-problems}}
      \Pf{}{<}{\size{v} + 1}{}
      \Pf{}{=}{\size{\Return{v}}}{By \defn of \defsize}
    \end{llproof} 

    \ProofCasesRules{\AlgChkExpLet, \AlgChkExpLam}
    In the non-extraction premises, the program term size decreases.
    The extraction premise is decidable by \Lemmaref{lem:decidable-extract}.

    \ProofCaseRule{\AlgChkExpRec}
    The extraction premise is decidable by \Lemmaref{lem:decidable-extract}.
    The subtyping premise is decidable by \Theoremref{thm:decidable-sub}.
    The constraint checking premise $\entailwah{\Theta}{W}$
    is decidable by \Lemmaref{lem:decidable-problem-checking}.
    In the expression type checking premise,
    the program term (expression) size decreases.

    \ProofCaseRule{\AlgChkExpExtract}
    The extraction premise is decidable by \Lemmaref{lem:decidable-extract}.
    It is easy to show that it is decidable whether $\Theta' \neq \cdot$.
    In the expression type-checking premise,
    the program term (expression) stays the same, but since $\Theta' \neq \cdot$,
    the type $A$ loses at least one $\exists$, $\forall$, $\land$, or $\implies$,
    decreasing the size of $A$.

    \ProofCaseRule{\AlgChkExpUnreachable}
    Only has one non-extraction premise which is decidable by \Lemmaref{lem:decidable-prop-truth}.
  \end{itemize}

  Next, we consider the match judgment:

  \begin{itemize}
    \ProofCasesRules{\AlgChkMatchEx, \AlgChkMatchWith}
    Program term stays the same, but the size of $A$ decreases.
   
    \ProofCaseRule{\AlgChkMatchVoid}
    No premises.

    \ProofCasesRules{\AlgChkMatchUnit, \AlgChkMatchPair, \AlgChkMatchSum}
    The extraction premises are decidable by \Lemmaref{lem:decidable-extract}.
    In all the non-extraction premises, the program term size decreases.

    \ProofCaseRule{\AlgChkMatchFix}
    The unrolling premise is decidable by \Lemmaref{lem:decidable-unroll}.
    The extraction premise is decidable by \Lemmaref{lem:decidable-extract}.
    In the remaining premise, the program term size decreases
    (note that we use \Lemmaref{lem:alg-unroll-output-wf}
    to know the premise's input type is well-formed).
  \end{itemize}

  Finally, we consider the spine judgment:

  \begin{itemize}
    \ProofCasesRules{\AlgSpineAll, \AlgSpineImplies}
    Program term (spine) stays the same,
    but the size of $A$ decreases (size is unaffected by variable renaming).

    \ProofCaseRule{\AlgSpineApp}
    The size of the program term (spine) decreases.
    
    \ProofCaseRule{\AlgSpineNil}
    No premises.
    \qedhere
  \end{itemize}
\end{proof}

\section{Semideclarative System}

\subsection{Miscellaneous Properties of Semideclarative System}

\begin{lemma}[Semidecl.\ Equiv.\ Shrinks Constraints]
  \label{lem:semidecl-equiv-shrinks-problem}
  ~
  \begin{enumerate}
  \item If $\semideclequiv[]{\Theta}{\mathcal{F}}{\mathcal{G}}{W}$, 
    then $\size{W} < \size{\mathcal{F}} + \size{\mathcal{G}} + 1$.
  \item If $\semideclequiv[+]{\Theta}{P}{Q}{W}$,
    then $\size{W} < \size{\poseqprob{P}{Q}}$.
  \item If $\semideclequiv[-]{\Theta}{N}{M}{W}$,
    then $\size{W} < \size{\negeqprob{N}{M}}$.
  \end{enumerate}
\end{lemma}
\begin{proof}
  Each part is proved by structural induction on the given equivalence derivation;
  parts (1) and (2) are mutually recursive.
  We follow the definition of size given in \Figureref{fig:size}.
\end{proof}

\begin{lemma}[Semidecl.\ Sub.\ Shrinks Constraint]
  \label{lem:semidecl-sub-shrinks-problem}
  ~
  \begin{enumerate}
  \item If $\semideclsub[+]{\Theta}{P}{Q}{W}$,
    then $\size{W} < \size{\possubprob{P}{Q}}$.
  \item If $\semideclsub[-]{\Theta}{N}{M}{W}$,
    then $\size{W} < \size{\negsubprob{N}{M}}$.
  \end{enumerate}
\end{lemma}
\begin{proof}
  Each part is proved by structural induction on the given equivalence derivation.
  We follow the definition of size given in \Figureref{fig:size}.
  We show only the trickiest cases, each of part (1).
  Part (2) cases are similar to part (1) cases of their dual rules
  (e.g., rule \SemiDeclSubNegUpshift is dual to \SemiDeclSubPosDownshift
  and its case proof is similar to the latter's).
  \begin{itemize}
    \DerivationProofCase{\SemiDeclSubPosFix}
    {
      \semideclequiv{\Theta}{F}{G}{W_0}
    }
    {
      \semideclsub[+]{\Theta}{\underbrace{\comprehend{\nu:\mu F}{\Fold{F}{\alpha}\,\nu =_\tau t}}_P}{\underbrace{\comprehend{\nu:\mu G}{\Fold{G}{\alpha}\,\nu =_\tau t'}}_Q}{W_0 \land t = t'}
    }
    \begin{llproof}
      \Pf{\size{W_0 \land t = t'}}{=}{\size{W_0} + \size{t = t'} + 1}{By \defn of \defsize}
      \Pf{}{=}{\size{W_0} + 1}{By \defn of \defsize}
      \Pf{}{<}{\size{F} + \size{G} + 1 + 1}{By \Lemmaref{lem:semidecl-equiv-shrinks-problem}}
      \Pf{}{<}{\size{F} + 1 + \size{G} + 1 + 1}{Add $1$}
      \Pf{}{=}{\size{P} + \size{Q} + 1}{By \defn of \defsize}
      \Pf{}{=}{\size{\possubprob{P}{Q}}}{By \defn of \defsize}
    \end{llproof} 

    \DerivationProofCase{\SemiDeclSubPosDownshift}
    {
      \judgeextract[-]{\Theta}{M}{M'}{\Theta'}
    }
    {
      \semideclsub[+]{\Theta}{\downshift{N}}{\downshift{M}}{\Theta' \implies^\ast \negsubprob{N}{M'}}
    }
    \begin{llproof}
      \Pf{\size{\Theta' \implies^\ast \negsubprob{N}{M'}}}{=}{\size{\Theta'} + \size{\negsubprob{N}{M'}}}{By \defn of $\implies^\ast$ and of \defsize}
      \Pf{}{=}{\size{\Theta'} + \size{N} + \size{M'} + 1}{By \defn of \defsize}
      \Pf{}{=}{\size{N} + \size{M} + 1}{By \Lemmaref{lem:extract-size-eqn}}
      \Pf{}{<}{\size{N} + 1 + \size{M} + 1 + 1}{Add $2$}
      \Pf{}{=}{\size{\downshift{N}} + \size{\downshift{M}} + 1}{By \defn of \defsize}
      \Pf{}{=}{\size{\possubprob{\downshift{N}}{\downshift{M}}}}{By \defn of \defsize}
    \end{llproof}
    \qedhere
  \end{itemize}
\end{proof}

\begin{lemma}[Semidecl.\ Typing Shrinks]
  \label{lem:semidecl-typing-shrinks}
  ~
  \begin{enumerate}
  \item
    If $\semideclchkval{\Theta}{\Gamma}{v}{P}{\chi}$
    then $\size{\chi} \leq \size{v}$.
  \item
    If $\semideclspine{\Theta}{\Gamma}{s}{N}{\upshift{P}}{\chi}$,
    then $\size{\chi} \leq \size{s}$.
  \end{enumerate}
\end{lemma}
\begin{proof}
  Similar to \Lemmaref{lem:program-typing-shrinks-problems}, but slightly simpler.
\end{proof}

\begin{lemma}[Equiv.\ Respects Size]
  \label{lem:equiv-respects-size}
  ~
  \begin{enumerate}
  \item If $\judgeequiv[]{\Theta}{\F}{\Gee}$, then $\size{\F} = \size{\Gee}$.
  \item If $\judgeequiv[\pm]{\Theta}{A}{B}$, then $\size{A} = \size{B}$.
  \item If $\wahequiv{\Theta}{W}{W'}$, then $\size{W} = \size{W'}$.
  \end{enumerate}
\end{lemma}
\begin{proof}
  Each part is proved by structural induction on the given equivalence derivation.
  Parts (1) and (2) are mutually recursive.
  Part (3) uses part (2).
\end{proof}

\begin{lemma}[Ix.\ Sandwich]
  \label{lem:ix-sandwich}
  If $\rextend{\Theta}{\Delta}{\Omega}$
  and $\judgeterm{\Theta; \Delta}{t}{\tau}$,\\
  then $\judgeentail{\Theta}{[\Omega]t = [\Omega]([\Delta]t)}$.
\end{lemma}
\begin{proof}
  Either $t$ is not an existential variable or it is.

  If it is not, then by definition $[\Omega]t = t = [\Omega]([\Delta]t)$
  and the goal follows by \Lemmaref{lem:equivassert}.

  Now suppose it is, \ie $t = \ahat$ for some $\ahat$.
  Either $\Delta$ does not solve $\ahat$ or it does.
  If it does not, then $[\Omega]([\Delta]t) = [\Omega]t$
  and the goal follows by \Lemmaref{lem:equivassert}.
  If it does, then the goal follows
  by inversion on $\rextend{\Theta}{\Delta}{\Omega}$
  and by \Lemmaref{lem:equivassert}.
\end{proof}

\begin{lemma}[Prop.\ Sandwich]
  \label{lem:prop-sandwich}
  If $\rextend{\Theta}{\Delta}{\Omega}$
  and $\judgeterm{\Theta; \Delta}{\phi}{\Booltype}$,\\
  then $\judgeequiv[]{\Theta}{[\Omega]\phi}{[\Omega]([\Delta]\phi)}$.
\end{lemma}
\begin{proof}
  By structural induction $\phi$, using \Lemmaref{lem:ix-sandwich} as needed.
\end{proof}

\begin{lemma}[Tp.\ Sandwich]
  \label{lem:tp-sandwich}
  If $\rextend{\Theta}{\Delta}{\Omega}$
  and $\judgetp{\Theta; \Delta}{A}{\dontcare}$,\\
  then $\judgeequiv[\pm]{\Theta}{[\Omega]A}{[\Omega]([\Delta]A)}$.
\end{lemma}
\begin{proof}
  By structural induction on $A$,
  using \Lemmaref{lem:ix-sandwich} and \Lemmaref{lem:prop-sandwich} as needed.
\end{proof}

\begin{lemma}[Prob.\ Sandwich]
  \label{lem:wah-sandwich}
  If $\rextend{\Theta}{\Delta}{\Omega}$
  and $\judgewf{\Theta; \Delta}{W}$,\\
  then $\wahequiv{\Theta}{[\Omega]W}{[\Omega]([\Delta]W)}$.
\end{lemma}
\begin{proof}
  By structural induction on $W$,
  using \Lemmaref{lem:ix-sandwich},
  \Lemmaref{lem:prop-sandwich},
  and \Lemmaref{lem:tp-sandwich}
  as needed.
\end{proof}

\begin{lemma}[Equiv.\ Sandwich]
  \label{lem:equiv-sandwich}
  Assume $\rextend{\Theta}{\Delta}{\Omega}$.
  \begin{enumerate}
    \item If $\Dee :: \semideclequiv{\Theta}{\F}{[\Omega]\Gee}{W}$
      and $\judgefunctor{\Theta; \Delta}{\Gee}{\dontcare}$,\\
      then there exists
      $\Dee' :: \semideclequiv{\Theta}{\F}{[\Omega]([\Delta]\Gee)}{W'}$\\
      such that $\wahequiv{\Theta}{W}{W'}$
      and $\hgt{\Dee} = \hgt{\Dee'}$.
    \item If $\Dee :: \semideclequiv{\Theta}{P}{[\Omega]Q}{W}$
      and $\judgetp{\Theta; \Delta}{Q}{\dontcare}$,\\
      then there exists $\Dee' :: \semideclsub{\Theta}{P}{[\Omega]([\Delta]Q)}{W'}$\\
      such that $\wahequiv{\Theta}{W}{W'}$
      and $\hgt{\Dee} = \hgt{\Dee'}$.
    \item If $\Dee :: \semideclequiv{\Theta}{[\Omega]M}{N}{W}$
      and $\judgetp{\Theta; \Delta}{M}{\dontcare}$,\\
      then there exists $\Dee' :: \semideclsub{\Theta}{[\Omega]([\Delta]M)}{N}{W'}$\\
      such that $\wahequiv{\Theta}{W}{W'}$
      and $\hgt{\Dee} = \hgt{\Dee'}$.
  \end{enumerate}
\end{lemma}
\begin{proof}
  By structural induction on the semideclarative equivalence derivation,
  using \Lemmaref{lem:ix-sandwich},
  \Lemmaref{lem:prop-sandwich},
  and \Lemmaref{lem:tp-sandwich}
  as needed.
  Parts (1) and (2) are mutually recursive.
\end{proof}

\begin{lemma}[Sub.\ Sandwich]
  \label{lem:sub-sandwich}
  Assume $\rextend{\Theta}{\Delta}{\Omega}$.
  \begin{enumerate}
    \item If $\Dee :: \semideclsub{\Theta}{P}{[\Omega]Q}{W}$
      and $\judgetp{\Theta; \Delta}{Q}{\dontcare}$,\\
      then there exists $\Dee' :: \semideclsub{\Theta}{P}{[\Omega]([\Delta]Q)}{W'}$\\
      such that $\wahequiv{\Theta}{W}{W'}$
      and $\hgt{\Dee} = \hgt{\Dee'}$.
    \item If $\Dee :: \semideclsub{\Theta}{[\Omega]M}{N}{W}$
      and $\judgetp{\Theta; \Delta}{M}{\dontcare}$,\\
      then there exists $\Dee' :: \semideclsub{\Theta}{[\Omega]([\Delta]M)}{N}{W'}$\\
      such that $\wahequiv{\Theta}{W}{W'}$
      and $\hgt{\Dee} = \hgt{\Dee'}$.
  \end{enumerate}
\end{lemma}
\begin{proof}
  By structural induction on the semideclarative subtyping derivation,
  using \Lemmaref{lem:ix-sandwich},
  \Lemmaref{lem:prop-sandwich},
  \Lemmaref{lem:tp-sandwich}
  and \Lemmaref{lem:equiv-sandwich}
  as needed.
\end{proof}

\begin{lemma}[Probs.\ Sandwich]
  \label{lem:probs-sandwich}
  If $\rextend{\Theta}{\Delta}{\Omega}$
  and $\judgewf{\Theta; \Delta; \Gamma}{\chi}$,\\
  then $\chiequiv{\Theta}{\Gamma}{[\Omega]\chi}{[\Omega]([\Delta]\chi)}$.
\end{lemma}
\begin{proof}
  By structural induction on $\chi$,
  using \Lemmaref{lem:tp-sandwich}
  and \Lemmaref{lem:equiv-sandwich}
  as needed.
\end{proof}

\begin{lemma}[Typing Sandwich]
  \label{lem:typing-sandwich}
  Assume $\rextend{\Theta}{\Delta}{\Omega}$.
  \begin{enumerate}
  \item If $\Dee :: \semideclchkval{\Theta}{\Gamma}{v}{[\Omega]P}{\chi}$
    and $\judgetp{\Theta; \Delta}{P}{\dontcare}$,\\
    then there exists
    $\Dee' :: \semideclchkval{\Theta}{\Gamma}{P}{[\Omega]([\Delta]Q)}{\chi'}$\\
    such that $\chiequiv{\Theta}{\Gamma}{\chi}{\chi'}$
    and $\hgt{\Dee} = \hgt{\Dee'}$.
  \item If $\Dee :: \semideclspine{\Theta}{\Gamma}{s}{[\Omega]N}{\upshift{P}}{\chi}$
    and $\judgetp{\Theta; \Delta}{N}{\dontcare}$,\\
    then there exists
    $\Dee' :: \semideclspine{\Theta}{\Gamma}{s}{[\Omega]([\Delta]N)}{\upshift{P'}}{\chi'}$\\
    such that $\judgeequiv[+]{\Theta}{P}{P'}$
    and $\chiequiv{\Theta}{\Gamma}{\chi}{\chi'}$
    and $\hgt{\Dee} = \hgt{\Dee'}$.
  \end{enumerate}
\end{lemma}
\begin{proof}
  By structural induction on the semideclarative typing derivation,
  using \Lemmaref{lem:ix-sandwich},
  \Lemmaref{lem:prop-sandwich},
  \Lemmaref{lem:tp-sandwich},
  \Lemmaref{lem:equiv-sandwich}
  and \Lemmaref{lem:sub-sandwich}
  as needed.
\end{proof}

\subsection{Semideclarative Functor and Type Equivalence}

\begin{lemma}[Semidecl.\ Equiv.\ Sound]
  \label{lem:semidecl-equiv-sound}
  ~
  \begin{enumerate}
  \item If $\semideclequiv[\pm]{\Theta}{A}{B}{W}$ and $\semideclentailwah{\Theta}{W}$,
    then $\judgeequiv[\pm]{\Theta}{A}{B}$.
  \item If $\semideclequiv[]{\Theta}{\mathcal{F}}{\mathcal{G}}{W}$
    and $\semideclentailwah{\Theta}{W}$,
    then $\judgeequiv[]{\Theta}{\mathcal{F}}{\mathcal{G}}$.
  \end{enumerate}
\end{lemma}
\begin{proof}
  By lexicographic induction on,
  first, $\size{W}$,
  and, second, the structure of the given equivalence derivation.
  Parts (1) and (2) are mutually recursive.
  Case analyze the given equivalence derivation;
  each case uniquely determines the rule concluding $\semideclentailwah{\Theta}{W}$.
  Use \Lemmaref{lem:semidecl-equiv-shrinks-problem}
  and \Lemmaref{lem:semidecl-sub-shrinks-problem} as needed.
\end{proof}

\begin{lemma}[Semidecl.\ Equiv.\ Complete]
  \label{lem:semidecl-equiv-complete}
  ~
  \begin{enumerate}
  \item
    If $\judgeequiv[\pm]{\Theta}{A}{B}$,
    then $\semideclequiv[\pm]{\Theta}{A}{B}{W}$ and $\semideclentailwah{\Theta}{W}$.
  \item
    If $\judgeequiv[]{\Theta}{\mathcal{F}}{\mathcal{G}}$,
    then $\semideclequiv[]{\Theta}{\mathcal{F}}{\mathcal{G}}{W}$
    and $\semideclentailwah{\Theta}{W}$.
  \end{enumerate}
\end{lemma}
\begin{proof}
  By structural induction on the given type or functor equivalence derivation.
  Parts (1) and (2) are mutually recursive.
\end{proof}

\subsection{Restricted Declarative Subtyping}

\begin{lemma}[Res.\ Decl.\ Sub.\ Sound]
  \label{lem:res-decl-sub-sound}
  If $\resjudgesub[\pm]{\Theta}{A}{B}$, then $\judgesub[\pm]{\Theta}{A}{B}$.
\end{lemma}
\begin{proof}
  By structural induction on $\resjudgesub[\pm]{\Theta}{A}{B}$.
  The positive and negative parts are mutually recursive.
\end{proof}

\begin{lemma}[Res.\ Decl.\ Sub.\ Complete]
  \label{lem:res-decl-sub-complete}
  ~
  \begin{enumerate}
  \item If $\judgesub[+]{\Theta}{P}{Q}$,
    then $\resjudgesub[-]{\Theta}{\upshift{P}}{\upshift{Q}}$.
  \item If $\judgesub[-]{\Theta}{N}{M}$,
    then $\resjudgesub[+]{\Theta}{\downshift{N}}{\downshift{M}}$.
  \end{enumerate}
\end{lemma}
\begin{proof}
  By structural induction on the given subtyping derivation.
  We show one of the relatively interesting cases.
  \begin{itemize}
    \DerivationProofCase{\DeclSubNegArrow}
    {
      \judgesub[+]{\Theta}{Q}{P}
      \\
      \judgesub[-]{\Theta}{N_0}{M_0}
    }
    {
      \judgesub[-]{\Theta}{P \to N_0}{Q \to M_0}
    }  
    \begin{llproof}
      \judgesubPf[+]{\Theta}{Q}{P}{Subderivation}
      \resjudgesubPf[-]{\Theta}{\upshift{Q}}{\upshift{P}}{By \ih}
      \judgeextractPf[+]{\Theta}{Q}{Q'}{\Theta_Q}{By inversion}
      \resjudgesubPf[+]{\Theta, \Theta_Q}{Q'}{P}{\ditto}
      \judgesubPf[-]{\Theta}{N_0}{M_0}{Subderivation}
      \resjudgesubPf[+]{\Theta}{\downshift{N_0}}{\downshift{M_0}}{By \ih}
      \judgeextractPf[-]{\Theta}{M_0}{M_0'}{\Theta_{M_0}}{By inversion}
      \resjudgesubPf[-]{\Theta, \Theta_{M_0}}{N_0}{M_0'}{\ditto}
      \resjudgesubPf[+]{\Theta, \Theta_Q, \Theta_{M_0}}{Q'}{P}{By \Lemmaref{lem:ix-level-weakening}}
      \trailingjust{(extended)}
      \resjudgesubPf[-]{\Theta, \Theta_Q, \Theta_{M_0}}{N_0}{M_0'}{By \Lemmaref{lem:ix-level-weakening}}
      \trailingjust{(extended)}
      \resjudgesubPf[-]{\Theta, \Theta_Q, \Theta_{M_0}}{P \to N_0}{Q' \to M_0'}{By \ResDeclSubNegArrow}
      \judgeextractPf[-]{\Theta}{Q \to M_0}{Q' \to M_0'}{\Theta_Q, \Theta_{M_0}}{By \ExtractArrow}
      \resjudgesubPf[+]{\Theta}{\downshift{(P \to N_0)}}{\downshift{(Q \to M_0)}}{By \ResDeclSubPosDownshift}
    \end{llproof}
    \qedhere
  \end{itemize}
\end{proof}

\subsection{Semideclarative Subtyping}

\begin{lemma}[Aux.\ Semidecl.\ Sub.\ Sound]
  \label{lem:aux-semidecl-sub-sound}
  ~
  \begin{enumerate}
  \item If $\semideclsub[+]{\Theta}{P}{Q}{W}$ and $\semideclentailwah{\Theta}{W}$,
    then $\resjudgesub[+]{\Theta}{P}{Q}$.
  \item If $\semideclsub[-]{\Theta}{N}{M}{W}$ and $\semideclentailwah{\Theta}{W}$,
    then $\resjudgesub[-]{\Theta}{N}{M}$.
  \end{enumerate}
\end{lemma}
\begin{proof}
  By lexicographic induction on,
  first, $\size{W}$,
  and, second, the height of the given subtyping derivation.
  Use \Lemmaref{lem:semidecl-equiv-sound}
  and \Lemmaref{lem:semidecl-sub-shrinks-problem}
  as needed.
\end{proof}

\begin{lemma}[Semidecl.\ Sub.\ Sound]
  \label{lem:semidecl-sub-sound}
  ~
  \begin{enumerate}
  \item If $\semideclsub[+]{\Theta}{P}{Q}{W}$ and $\semideclentailwah{\Theta}{W}$,
    then $\judgesub[+]{\Theta}{P}{Q}$.
  \item If $\semideclsub[-]{\Theta}{N}{M}{W}$ and $\semideclentailwah{\Theta}{W}$,
    then $\judgesub[-]{\Theta}{N}{M}$.
  \end{enumerate}
\end{lemma}
\begin{proof}
  Follows from \Lemmaref{lem:aux-semidecl-sub-sound}
  and \Lemmaref{lem:res-decl-sub-sound}.
\end{proof}

\begin{lemma}[Aux.\ Semidecl.\ Sub.\ Complete]
  \label{lem:aux-semidecl-sub-complete}
  ~
  \begin{enumerate}
  \item
    If $\resjudgesub[+]{\Theta}{P}{Q}$,
    then $\semideclsub[+]{\Theta}{P}{Q}{W}$ and $\semideclentailwah{\Theta}{W}$.
  \item
    If $\resjudgesub[-]{\Theta}{N}{M}$,
    then $\semideclsub[-]{\Theta}{N}{M}{W}$ and $\semideclentailwah{\Theta}{W}$.
  \end{enumerate}
\end{lemma}
\begin{proof}
  By structural induction on the given subtyping derivation,
  using \Lemmaref{lem:semidecl-equiv-complete} as needed.
\end{proof}

\begin{lemma}[Semidecl.\ Sub.\ Complete]
  \label{lem:semidecl-sub-complete}
  ~
  \begin{enumerate}
  \item
    If $\judgesub[+]{\Theta}{P}{Q}$,
    then $\semideclsub[-]{\Theta}{\upshift{P}}{\upshift{Q}}{W}$
    and $\semideclentailwah{\Theta}{W}$
  \item
    If $\judgesub[-]{\Theta}{N}{M}$,
    then $\semideclsub[+]{\Theta}{\downshift{N}}{\downshift{M}}{W}$
    and $\semideclentailwah{\Theta}{W}$.
  \end{enumerate}
\end{lemma}
\begin{proof}
  Follows from \Lemmaref{lem:res-decl-sub-complete}
  and \Lemmaref{lem:aux-semidecl-sub-complete}.
\end{proof}

\subsection{Constraint Equivalence}

\begin{lemma}[Ctx.\ Equiv. Compat.\ (Semidecl.)]
  \label{lem:ctx-equiv-compat-semidecl}
  Assume $\judgeequiv{\Theta_1}{\Theta}{\Theta'}$.
  \begin{enumerate}
  \item If $\semideclequiv{\Theta_1, \Theta, \Theta_2}{\F}{\Gee}{W}$,
    then $\semideclequiv{\Theta_1, \Theta', \Theta_2}{\F}{\Gee}{W}$.
  \item If $\semideclequiv[\pm]{\Theta_1, \Theta, \Theta_2}{A}{B}{W}$,
    then $\semideclequiv[\pm]{\Theta_1, \Theta', \Theta_2}{A}{B}{W}$.
  \item If $\semideclsub[\pm]{\Theta_1, \Theta, \Theta_2}{A}{B}{W}$,
    then $\semideclsub[\pm]{\Theta_1, \Theta', \Theta_2}{A}{B}{W}$.
  \end{enumerate}
\end{lemma}
\begin{proof}
  Parts (1) and (2) follow from
  \Lemmaref{lem:semidecl-equiv-sound}, \Lemmaref{lem:semidecl-equiv-complete}, and \Lemmaref{lem:ctx-equiv-compat}.
  Part (3) follows from
  \Lemmaref{lem:semidecl-sub-sound}, \Lemmaref{lem:semidecl-sub-complete}, and \Lemmaref{lem:ctx-equiv-compat}.
\end{proof}

\begin{lemma}[Ctx.\ Equiv. Compat.\ (Prob.)]
  \label{lem:ctx-equiv-compat-wah}
  If $\judgeequiv{\Theta_1}{\Theta}{\Theta'}$
  and $\semideclentailwah{\Theta_1, \Theta, \Theta_2}{W}$,\\
  then $\semideclentailwah{\Theta_1, \Theta', \Theta_2}{W}$.
\end{lemma}
\begin{proof}
  By induction on $\size{W}$ (\Figref{fig:size})
  and case analysis on the given semideclarative $W$-verification derivation.
  Use
  \Lemmaref{lem:ctx-equiv-compat},
  \Lemmaref{lem:ctx-equiv-compat-semidecl},
  \Lemmaref{lem:semidecl-equiv-shrinks-problem}, and
  \Lemmaref{lem:semidecl-sub-shrinks-problem}
  as needed.
\end{proof}

\begin{lemma}[Equiv.\ Weakening]
  \label{lem:equiv-weakening}
  If $\wahequiv{\Theta_1, \Theta_2}{W}{W'}$,
  then $\wahequiv{\Theta_1, \Theta_0, \Theta_2}{W}{W'}$.
\end{lemma}
\begin{proof}
  By structural induction on the given constraint equivalence derivation,
  using \Lemmaref{lem:ix-level-weakening} as needed.
\end{proof}

\begin{lemma}[Equiv.\ Respects Entailment]
  \label{lem:equiv-respects-entail}
  If $\semideclentailwah{\Theta}{W}$ and $\wahequiv{\Theta}{W}{W'}$,
  then $\semideclentailwah{\Theta}{W'}$.
\end{lemma}
\begin{proof}
  By structural induction on $W$.
  We case analyze rules concluding $\wahequiv{\Theta}{W}{W'}$.
  The \WahEquivPrp case uses \Lemmaref{lem:equiv-respects-prop-truth}.
  The \WahEquivPrpEq case uses \Lemmaref{lem:trans-equiv} (proposition part).
  The \WahEquivImplies case uses \Lemmaref{lem:ctx-equiv-compat-wah}
  after using the \ih with the equivalence subderivation
  weakened by \Lemmaref{lem:equiv-weakening}.
  The \SemiDeclWTrueAll case is straightforward.
  \begin{itemize}
    \DerivationProofCase{\WahEquivPosSub}
    {
      \judgeequiv[+]{\Theta}{P}{P'}
      \\
      \judgeequiv[+]{\Theta}{Q}{Q'}
    }
    {
      \wahequiv{\Theta}{\possubprob{P}{Q}}{\possubprob{P'}{Q'}}
    }
    The only possible rule concluding
    $\semideclentailwah{\Theta}{\possubprob{P}{Q}}$ is \SemiDeclWTruePosSub.
    \begin{itemize}
      \DerivationProofCase{\SemiDeclWTruePosSub}
      {
        \semideclsub[+]{\Theta}{P}{Q}{\widetilde{W}}
        \\
        \semideclentailwah{\Theta}{\widetilde{W}}
      }
      {
        \semideclentailwah{\Theta}{\possubprob{P}{Q}}
      }
      \begin{llproof}
        \semideclsubPf[+]{\Theta}{P}{Q}{\widetilde{W}}{Subderivation}
        \semideclentailwahPf{\Theta}{\widetilde{W}}{Subderivation}
        \judgesubPf[+]{\Theta}{P}{Q}{By \Lemmaref{lem:semidecl-sub-sound}}
        \judgeequivPf[+]{\Theta}{P}{P'}{Subderivation}
        \judgeequivPf[+]{\Theta}{Q}{Q'}{Subderivation}
        \judgesubPf[+]{\Theta}{P'}{Q'}{By \Lemmaref{lem:sub-equiv-compatible}}
        \semideclsubPf[+]{\Theta}{P'}{Q'}{\widetilde{W}'}{By \Lemmaref{lem:semidecl-sub-complete}}
        \semideclentailwahPf{\Theta}{\widetilde{W}'}{\ditto}
        \semideclentailwahPf{\Theta}{\possubprob{P'}{Q'}}{By \SemiDeclWTruePosSub}
      \end{llproof} 
    \end{itemize}
    
    \ProofCaseRule{\SemiDeclWTrueNegSub}
    Similar to \SemiDeclWTruePosSub case.

    \ProofCaseRule{\SemiDeclWTruePosEquiv}
    Similar to \SemiDeclWTruePosSub case, but using
    \Lemmaref{lem:semidecl-equiv-sound}
    rather than \Lemmaref{lem:semidecl-sub-sound},
    \Lemmaref{lem:semidecl-equiv-complete}
    rather than \Lemmaref{lem:semidecl-sub-complete},
    and both \Lemmaref{lem:trans-equiv} and \Lemmaref{lem:symmetric-equiv-tp-fun}
    rather than \Lemmaref{lem:sub-equiv-compatible}.

    \ProofCaseRule{\SemiDeclWTrueNegEquiv}
    Similar to \SemiDeclWTruePosEquiv case. \qedhere
  \end{itemize}
\end{proof}

\begin{lemma}[Prob.\ Equiv.\ Reflexive]
  \label{lem:wah-equiv-refl}
  If $\judgewf{\Theta}{W}$, then $\wahequiv{\Theta}{W}{W}$.
\end{lemma}
\begin{proof}
  By structural induction on $W$,
  using \Lemmaref{lem:refl-equiv-prop} and \Lemmaref{lem:refl-equiv-tp-fun}
  as needed.
\end{proof}

\begin{lemma}[Prob.\ Equiv.\ Symmetric]
  \label{lem:wah-equiv-symmetric}
  If $\wahequiv{\Theta}{W_1}{W_2}$,
  then $\wahequiv{\Theta}{W_2}{W_1}$.
\end{lemma}
\begin{proof}
  By structural induction on $W_1$,
  using \Lemmaref{lem:symmetric-equiv-prop} and \Lemmaref{lem:symmetric-equiv-tp-fun}
  as needed.
\end{proof}

\begin{lemma}[Prob.\ Equiv.\ Transitive]
  \label{lem:wah-equiv-trans}
  If $\wahequiv{\Theta}{W}{\widetilde{W}}$
  and $\wahequiv{\Theta}{\widetilde{W}}{W'}$,
  then $\wahequiv{\Theta}{W}{W'}$.
\end{lemma}
\begin{proof}
  By structural induction on $W$, using \Lemmaref{lem:trans-equiv} as needed.
\end{proof}

\begin{lemma}[Probs.\ Equiv.\ Reflexive]
  \label{lem:chi-equiv-refl}
  If $\judgewf{\Theta; \Gamma}{\chi}$, then $\chiequiv{\Theta}{\Gamma}{\chi}{\chi}$.
\end{lemma}
\begin{proof}
  By structural induction on $\chi$,
  using \Lemmaref{lem:wah-equiv-refl} and \Lemmaref{lem:refl-equiv-tp-fun}
  as needed.
\end{proof}

\begin{lemma}[Probs.\ Equiv.\ Symmetric]
  \label{lem:chi-equiv-symmetric}
  If $\chiequiv{\Theta}{\Gamma}{\chi_1}{\chi_2}$,
  then $\chiequiv{\Theta}{\Gamma}{\chi_2}{\chi_1}$.
\end{lemma}
\begin{proof}
  By structural induction on $\chi_1$,
  using \Lemmaref{lem:wah-equiv-symmetric} and \Lemmaref{lem:symmetric-equiv-tp-fun}
  as needed.
\end{proof}

\begin{lemma}[Probs.\ Equiv.\ Transitive]
  \label{lem:chi-equiv-trans}
  If $\chiequiv{\Theta}{\Gamma}{\chi}{\widetilde{\chi}}$
  and $\chiequiv{\Theta}{\Gamma}{\widetilde{\chi}}{\chi'}$,
  then $\chiequiv{\Theta}{\Gamma}{\chi}{\chi'}$.
\end{lemma}
\begin{proof}
  By structural induction on $\chi$,
  using
  \Lemmaref{lem:trans-equiv}
  and
  \Lemmaref{lem:wah-equiv-trans}
  as needed.
\end{proof}

\section{Algorithmic Soundness}

\subsection{Algorithmic Functor and Type Equivalence}

\begin{lemma}[Aux.\ Alg.\ Equiv.\ Sound]
  \label{lem:aux-alg-equiv-sound}
  ~
  \begin{enumerate}
  \item If $\algequiv[+]{\Theta; \Delta}{P}{Q}{W}{\Delta'}$
    and $\judgetp{\Theta; \Delta}{Q}{\Xi}$
    and $\ground{P}$
    and $[\Delta]Q = Q$
    and $\extend{\Theta}{\Delta'}{\Omega}$,
    then there exists $W'$
    such that $\semideclequiv[+]{\Theta}{P}{[\Omega]Q}{W'}$
    and $\wahequiv{\Theta}{[\Omega]W}{W'}$.
  \item If $\algequiv[-]{\Theta; \Delta}{N}{M}{W}{\Delta'}$
    and $\judgetp{\Theta; \Delta}{N}{\Xi}$
    and $\ground{M}$
    and $[\Delta]N = N$
    and $\extend{\Theta}{\Delta'}{\Omega}$,
    then there exists $W'$
    such that $\semideclequiv[-]{\Theta}{[\Omega]N}{M}{W'}$
    and $\wahequiv{\Theta}{[\Omega]W}{W'}$.
  \item If $\algequiv[]{\Theta; \Delta}{\mathcal{F}}{\mathcal{G}}{W}{\Delta'}$
    and $\judgefunctor{\Theta; \Delta}{\mathcal{G}}{\Xi}$
    and $\ground{\mathcal{F}}$
    and $[\Delta]\Gee = \Gee$
    and $\extend{\Theta}{\Delta'}{\Omega}$,
    then there exists $W'$
    such that $\semideclequiv[]{\Theta}{\mathcal{F}}{[\Omega]\mathcal{G}}{W'}$
    and $\wahequiv{\Theta}{[\Omega]W}{W'}$.
  \end{enumerate}
\end{lemma}
\begin{proof}
  By structural induction on the given algorithmic subtyping derivation.
  Similar to the proof of \Lemmaref{lem:aux-alg-sub-sound}, but simpler,
  and using \Lemmaref{lem:propequivinst-extends} rather than \Lemmaref{lem:inst-extends}
  (in the \AlgTpEquivPosWith case).
\end{proof}

\subsection{Algorithmic Subtyping}

\begin{lemma}[Aux.\ Alg.\ Sub.\ Sound]
  \label{lem:aux-alg-sub-sound}
  ~
  \begin{enumerate}
    \item
      If $\Dee :: \algsub[+]{\Theta; \Delta}{P}{Q}{W}{\Delta'}$
      and $\judgetp{\Theta; \Delta}{Q}{\Xi}$
      and $\ground{P}$
      and $[\Delta]Q = Q$
      and $\extend{\Theta}{\Delta'}{\Omega}$,
      then there exists $\Dee' :: \semideclsub[+]{\Theta}{P}{[\Omega]Q}{W'}$
      such that $\wahequiv{\Theta}{[\Omega]W}{W'}$
      and $\hgt{\Dee} = \hgt{\Dee'}$.
    \item
      If $\Dee :: \algsub[-]{\Theta; \Delta}{N}{M}{W}{\Delta'}$
      and $\judgetp{\Theta; \Delta}{N}{\Xi}$
      and $\ground{M}$
      and $[\Delta]N = N$
      and $\extend{\Theta}{\Delta'}{\Omega}$,
      then there exists $\Dee' :: \semideclsub[-]{\Theta}{[\Omega]N}{M}{W'}$
      such that $\wahequiv{\Theta}{[\Omega]W}{W'}$
      and $\hgt{\Dee} = \hgt{\Dee'}$.
  \end{enumerate}
\end{lemma}
\begin{proof}
  By structural induction on the given algorithmic subtyping derivation.
  We do not explicitly show the height equality,
  as it can be easily seen from context
  (\eg, when using \ih, subderivations have same height by \ih,
  then we apply the corresponding semideclarative rule).
  We also elide the easy well-formedness, groundness, and $[\Delta]A = A$ checks,
  \eg when using the \ih.
  \begin{enumerate}
  \item
    \begin{itemize}
      \DerivationProofCase{\AlgSubPosVoid}
      { }
      {
        \algsub[+]{\Theta; \Delta}{0}{0}{\True}{\Delta}
      }
      \begin{llproof}
        \Hand\semideclsubPf[+]{\Theta}{0}{\underbrace{0}_{[\Omega]0}}{\True}{By \SemiDeclSubPosVoid}
        \judgeequivPf[]{\Theta}{\True}{\True}{By \PrpEquivT}
        \Hand\wahequivPf{\Theta}{\underbrace{\True}_{[\Omega]\True}}{\True}{By \WahEquivPrp}
      \end{llproof} 

      \ProofCaseRule{\AlgSubPosUnit}
      Similar to \AlgSubPosVoid case.

      \DerivationProofCase{\AlgSubPosProd}
      {
        \algsub[+]{\Theta; \Delta}{P_1}{Q_1}{\Wah_1}{\Delta''}
        \\
        \algsub[+]{\Theta; \Delta''}{P_2}{[\Delta'']Q_2}{\Wah_2}{\Delta'}
      }
      {
        \algsub[+]{\Theta; \Delta}{P_1 \times P_2}{Q_1 \times Q_2}{[\Delta']\Wah_1 \land \Wah_2}{\Delta'}
      }
      \begin{llproof}
        \algsubPf[+]{\Theta; \Delta''}{P_2}{[\Delta'']Q_2}{\Wah_2}{\Delta'}{Subderivation}
        \extendPf{\Theta}{\Delta''}{\Delta'}{By \Lemmaref{lem:alg-sub-extends}}
        \extendPf{\Theta}{\Delta'}{\Omega}{Given}
        \extendPf{\Theta}{\Delta''}{\Omega}{By \Lemmaref{lem:ext-trans}}
        \algsubPf[+]{\Theta; \Delta}{P_1}{Q_1}{\Wah_1}{\Delta''}{Subderivation}
        \semideclsubPf[+]{\Theta}{P_1}{[\Omega]Q_1}{\Wah_1'}{By \ih}
        \wahequivPf{\Theta}{[\Omega]\Wah_1}{\Wah_1'}{\ditto}
        \wahequivPf{\Theta}{[\Omega]([\Delta']\Wah_1)}{\Wah_1'}{By \Lemmaref{lem:ext-subst-invariance}}
        \decolumnizePf
        \judgetpPf{\Theta; \Delta}{Q_1 \times Q_2}{\dontcare}{Given}
        \judgetpPf{\Theta; \Delta}{Q_2}{\dontcare}{By inversion}
        \extendPf{\Theta}{\Delta}{\Delta''}{By \Lemmaref{lem:typing-extends}}
        \judgetpPf{\Theta; \Delta''}{Q_2}{\dontcare}{By \Lemmaref{lem:ext-weak-tp}}
        \judgetpPf{\Theta; \Delta''}{[\Delta'']Q_2}{\dontcare}{By \Lemmaref{lem:right-hand-subst}}
        \Pf{[\Delta'']([\Delta'']Q_2)}{=}{[\Delta'']Q_2}{By \Lemmaref{lem:apply-idempotent}}
        \decolumnizePf
        \semideclsubPf[+]{\Theta}{P_2}{[\Omega]([\Delta'']Q_2)}{\Wah_2'}{By \ih}
        \wahequivPf{\Theta}{[\Omega]\Wah_2}{\Wah_2'}{\ditto}
        \semideclsubPf[+]{\Theta}{P_2}{[\Omega]Q_2}{\Wah_2'}{By \Lemmaref{lem:ext-subst-invariance}}
          \semideclsubPf[+]{\Theta}{P_1 \times P_2}{[\Omega]Q_1 \times [\Omega]Q_2}{\Wah_1' \land \Wah_2'}{By \SemiDeclSubPosProd}
          \Hand\semideclsubPf[+]{\Theta}{P_1 \times P_2}{[\Omega](Q_1 \times Q_2)}{\Wah_1' \land \Wah_2'}{By \defn of \defsubst}
          \wahequivPf{\Theta}{[\Omega]([\Delta']\Wah_1) \land [\Omega]\Wah_2}{\Wah_1' \land \Wah_2'}{By \WahEquivAnd}
          \Hand\wahequivPf{\Theta}{[\Omega]([\Delta'](\Wah_1 \land \Wah_2))}{\Wah_1' \land \Wah_2'}{By \defn of \defsubst}
      \end{llproof}

      \ProofCaseRule{\AlgSubPosSum}
      Similar to \AlgSubPosProd case.

      \DerivationProofCase{\AlgSubPosWithR}
      {
        \algsub[+]{\Theta; \Delta}{P}{Q_0}{\Wah_0}{\Delta''}
        \\
        \alginst{\Theta; \Delta''}{[\Delta'']\phi}{\Delta'}
      }
      {
        \algsub[+]{\Theta; \Delta}{P}{Q_0 \land \phi}{[\Delta']\Wah_0 \land [\Delta']\phi}{\Delta'}
      }
      \begin{llproof}
        \algsubPf[+]{\Theta; \Delta}{P}{Q_0}{\Wah_0}{\Delta''}{Subderivation}
        \alginstPf{\Theta; \Delta''}{[\Delta'']\phi}{\Delta'}{Subderivation}
        \extendPf{\Theta}{\Delta''}{\Delta'}{By \Lemmaref{lem:inst-extends}}
        \extendPf{\Theta}{\Delta'}{\Omega}{Given}
        \extendPf{\Theta}{\Delta''}{\Omega}{By \Lemmaref{lem:ext-trans}}
        \semideclsubPf[+]{\Theta}{P}{[\Omega]Q_0}{\Wah_0'}{By \ih}
        \wahequivPf{\Theta}{[\Omega]\Wah_0}{\Wah_0'}{\ditto}
        \semideclsubPf[+]{\Theta}{P}{[\Omega]Q_0 \land [\Omega]\phi}{\Wah_0' \land [\Omega]\phi}{By \SemiDeclSubPosWithR}
        \Hand\semideclsubPf[+]{\Theta}{P}{[\Omega](Q_0 \land \phi)}{\Wah_0' \land [\Omega]\phi}{By \defn of \defsubst}
        \judgetpPf{\Theta; \Delta}{Q_0 \land \phi}{\dontcare}{Given}
        \judgetermPf{\Theta; \Delta}{\phi}{\Booltype}{By inversion}
        \extendPf{\Theta}{\Delta}{\Delta'}{By \Lemmaref{lem:alg-sub-extends}}
        \judgetermPf{\Theta; \Delta'}{\phi}{\Booltype}{By \Lemmaref{lem:ext-weak-tp}}
        \judgetermPf{\Theta}{[\Omega]\phi}{\Booltype}{By \Lemmaref{lem:alg-to-decl-wf}}
        \judgeequivPf[]{\Theta}{[\Omega]\phi}{[\Omega]\phi}{By \Lemmaref{lem:refl-equiv-prop}}
        \wahequivPf{\Theta}{[\Omega]\phi}{[\Omega]\phi}{By \WahEquivPrp}
        \wahequivPf{\Theta}{[\Omega]([\Delta']\phi)}{[\Omega]\phi}{By \Lemmaref{lem:ext-subst-invariance}}
        \wahequivPf{\Theta}{[\Omega]([\Delta']\Wah_0)}{\Wah_0'}{By \Lemmaref{lem:ext-subst-invariance}}
        \wahequivPf{\Theta}{[\Omega]\Wah_0 \land [\Omega]([\Delta']\phi)}{\Wah_0' \land [\Omega]\phi}{By \WahEquivAnd}
        \Hand\wahequivPf{\Theta}{[\Omega](\Wah_0 \land [\Delta']\phi)}{\Wah_0' \land [\Omega]\phi}{By \defn of \defsubst}
      \end{llproof}

      \DerivationProofCase{\AlgSubPosExR}
      { 
        \algsub[+]{\Theta; \Delta, \ahat:\tau}{P}{[\ahat/a] Q_0}{\Wah}{\Delta', \hypeq{\ahat}{\tau}{t}}
      }
      {
        \algsub[+]{\Theta; \Delta}{P}{\extype{a:\tau} Q_0}{\Wah}{\Delta'}
      }
      \begin{llproof}
        \algsubPf[+]{\Theta; \Delta, \ahat:\tau}{P}{[\ahat/a] Q_0}{\Wah}{\Delta', \hypeq{\ahat}{\tau}{t}}{Subderivation}
        \extendPf{\Theta}{\Delta'}{\Omega}{Given}
        \extendPf{\Theta}{\Delta', \hypeq{\ahat}{\tau}{t}}{\Omega, \hypeq{\ahat}{\tau}{t}}{By \ExtSolved}
        \semideclsubPf[+]{\Theta}{P}{[\Omega, \hypeq{\ahat}{\tau}{t}]([\ahat/a] Q_0)}{\Wah'}{By \ih}
        \wahequivPf{\Theta}{[\Omega, \hypeq{\ahat}{\tau}{t}]\Wah}{\Wah'}{\ditto}
        \Pf{[\Delta', \hypeq{\ahat}{\tau}{t}]W}{=}{W}{By \Lemmaref{lem:output-applied}}
        \Pf{[t/\ahat]W}{=}{W}{Follows from line above}
        \Pf{[\Omega, \hypeq{\ahat}{\tau}{t}]\Wah}{=}{[\Omega]([t/\ahat]\Wah)}{By \defn of \defsubst}
        \Pf{}{=}{[\Omega]\Wah}{By equation}
        \Hand\wahequivPf{\Theta}{[\Omega]\Wah}{\Wah'}{By equation}
        \semideclsubPf[+]{\Theta}{P}{[\Omega]([t/\ahat]([\ahat/a] Q_0))}{\Wah'}{By \defn of \defsubst}
        \semideclsubPf[+]{\Theta}{P}{[\Omega]([t/a] Q_0)}{\Wah'}{By property of subst.}
        \semideclsubPf[+]{\Theta}{P}{[[\Omega]t/a]([\Omega] Q_0)}{\Wah'}{By property of subst.}
        \semideclsubPf[+]{\Theta}{P}{[t/a]([\Omega] Q_0)}{\Wah'}{$\because\ground{t}$}
        \semideclsubPf[+]{\Theta}{P}{\extype{a:\tau}[\Omega] Q_0}{\Wah'}{By \SemiDeclSubPosExR}
        \Hand\semideclsubPf[+]{\Theta}{P}{[\Omega]\extype{a:\tau} Q_0}{\Wah'}{By \defn of \defsubst}
      \end{llproof}

      \item \textbf{Case }
      \[
        \Infer{\AlgSubPosFix}
        {
          \alltype{\ahat \in \dom{\Delta}}{[\Delta]t' \neq \ahat}
          \\
          \algequiv[]{\Theta; \Delta}{F}{G}{W_0}{\Delta'}
        }
        {
          \algsub[+]
          {\Theta; \Delta}
          {\comprehend{\nu:\mu F}{\Fold{F}{\alpha}\,\nu =_\tau t}}
          {\comprehend{\nu:\mu G}{\Fold{G}{\alpha}\,\nu =_\tau t'}}
          {W_0 \land (t = [\Delta']t')}
          {\Delta'}
        }
      \]
      \begin{llproof}
        \algequivPf[]{\Theta; \Delta}{F}{G}{W_0}{\Delta'}{Subderivation}
        \semideclequivPf[]{\Theta}{F}{[\Omega]G}{\Wah_0'}{By \Lemref{lem:aux-alg-equiv-sound}}
        \wahequivPf{\Theta}{[\Omega]\Wah_0}{\Wah_0'}{\ditto}
        \judgeentailPf{\Theta}{t=t}{By \Lemref{lem:equivassert}}
        \judgeentailPf{\Theta}{[\Omega]t'=[\Omega]t'}{By \Lemref{lem:equivassert}}
        \judgeentailPf{\Theta}{[\Omega]([\Delta']t')=[\Omega]t'}{By \Lemref{lem:ext-subst-invariance}}
        \judgeequivPf[]{\Theta}{t = [\Omega]([\Delta']t')}{t = [\Omega]t'}{By \PrpEquivEq}
        \wahequivPf{\Theta}{t = [\Omega]([\Delta']t')}{t = [\Omega]t'}{By \WahEquivPrp}
        \wahequivPf{\Theta}{[\Omega]\Wah_0 \land (t = [\Omega]([\Delta']t'))}{\Wah_0' \land (t = [\Omega]t')}{By \WahEquivAnd}
        \Hand\wahequivPf{\Theta}{[\Omega](\Wah_0 \land (t = [\Delta']t'))}{\Wah_0' \land (t = [\Omega]t')}{By \defn of \defsubst and $\because \ground{t}$}
      \end{llproof}
      ~\\
      By \SemiDeclSubPosFix,
      \[
        \semideclsub[]{\Theta}{\comprehend{\nu:\mu F}{\Fold{F}{\alpha}\,\nu =_\tau t}}{\comprehend{\nu:\mu [\Omega]G}{\Fold{[\Omega]G}{\alpha}\,\nu =_\tau [\Omega]t'}}{\Wah_0' \land (t = [\Omega]t')}
      \]
      By \defn of \defsubst and \Lemmaref{lem:algebras-are-ground},
      \[
        \semideclsub[]{\Theta}{\comprehend{\nu:\mu F}{\Fold{F}{\alpha}\,\nu =_\tau t}}{[\Omega]\comprehend{\nu:\mu G}{\Fold{G}{\alpha}\,\nu =_\tau t'}}{\Wah_0' \land (t = [\Omega]t')}
      \]

      \item \textbf{Case }
      \[
        \Infer{\AlgSubPosFixInst}
        {
          \ground{t}
          \\
          \algequiv[]{\Theta; \Delta}{F}{G}{W_0}{\Delta_1', \ahat:\tau, \Delta_2'}
          \\
          \Delta' = \Delta_1', \hypeq{\ahat}{\tau}{t}, \Delta_2'
        }
        {
          \algsub[+]
          {\Theta; \Delta}
          {\comprehend{\nu:\mu F}{\Fold{F}{\alpha}\,\nu =_\tau t}}
          {\comprehend{\nu:\mu G}{\Fold{G}{\alpha}\,\nu =_\tau \ahat}}
          {
            [\Delta']W_0 \land (t = t)
          }
          {\Delta'}
        }
      \]
      ~\\
      \begin{llproof}
        \extendPf{\Theta}{\Delta_1', \ahat:\tau, \Delta_2'}{\Delta'}{By \Lemmaref{lem:deep-solve-extend}}
        \extendPf{\Theta}{\Delta_1', \ahat:\tau, \Delta_2'}{\Omega}{By \Lemmaref{lem:ext-trans}}
        \decolumnizePf
        \algequivPf[]{\Theta; \Delta}{F}{G}{W_0}{\Delta_1', \ahat:\tau, \Delta_2'}{Subderivation}
        \semideclequivPf[]{\Theta}{F}{[\Omega]G}{W_0'}{By \Lemmaref{lem:aux-alg-equiv-sound}}
        \wahequivPf{\Theta}{[\Omega]W_0}{W_0'}{\ditto}
        \wahequivPf{\Theta}{t=t}{t=t}{By \Lemmaref{lem:wah-equiv-refl}}
        \wahequivPf{\Theta}{[\Omega]W_0 \land (t = t)}{W_0' \land (t = t)}{By \WahEquivAnd}
        \wahequivPf{\Theta}{[\Omega]([\Delta']W_0) \land (t = t)}{W_0' \land (t = t)}{By \Lemmaref{lem:ext-subst-invariance}}
        \Hand\wahequivPf{\Theta}{[\Omega]([\Delta']W_0 \land (t = t))}{W_0' \land (t = t)}{By \defn of \defsubst and $\because\ground{t}$}
      \end{llproof}
      ~\\
      By \SemiDeclSubPosFix,
      \[
        \semideclsub[]{\Theta}{\comprehend{\nu:\mu F}{\Fold{F}{\alpha}\,\nu =_\tau t}}{\comprehend{\nu:\mu [\Omega]G}{\Fold{[\Omega]G}{\alpha}\,\nu =_\tau [\Omega]\ahat}}{W_0' \land (t = [\Omega]\ahat)}
      \]
      By \defn of \defsubst and \Lemmaref{lem:algebras-are-ground},
      \[
        \semideclsub[]{\Theta}{\comprehend{\nu:\mu F}{\Fold{F}{\alpha}\,\nu =_\tau t}}{[\Omega]\comprehend{\nu:\mu G}{\Fold{G}{\alpha}\,\nu =_\tau \ahat}}{W_0' \land (t = [\Omega]\ahat)}
      \]
      By \defn of \defsubst and $\hypeq{\ahat}{\tau}{t} \in \Omega$ (by inversion on extension),
      \[
        \semideclsub[]{\Theta}{\comprehend{\nu:\mu F}{\Fold{F}{\alpha}\,\nu =_\tau t}}{[\Omega]\comprehend{\nu:\mu G}{\Fold{G}{\alpha}\,\nu =_\tau \ahat}}{W_0' \land (t = t)}
      \]

      \ProofCaseRule{\AlgSubPosFixInstFun}
      Similar to \AlgSubPosFixInstFun case, but simpler.

      \DerivationProofCase{\AlgSubPosDownshift}
      {
        \judgeextract{\Theta; \Delta}{M}{M'}{\Thetahat}
      }
      {
        \algsub[+]{\Theta; \Delta}{\downshift{N}}{\downshift{M}}{\Thetahat \implies^\ast \negsubprob{N}{M'}}{\Delta}
      }
      \begin{llproof}
        \judgeextractPf{\Theta; \Delta}{M}{M'}{\Thetahat}{Subderivation}
        \extendPf{\Theta}{\Delta}{\Omega}{Given}
        \judgeextractPf[\pm]{\Theta}{[\Omega]M}{[\Omega]M'}{[\Omega]\Thetahat}{By \Lemmaref{lem:complete-alg-extract}}
        \semideclsubPf[+]{\Theta}{\downshift{N}}{\downshift{[\Omega]M}}{[\Omega]\Thetahat \implies^\ast \negsubprob{N}{[\Omega]M'}}{By \SemiDeclSubPosDownshift}
        \Hand\semideclsubPf[+]{\Theta}{\downshift{N}}{[\Omega]\downshift{M}}{[\Omega]\Thetahat \implies^\ast \negsubprob{N}{[\Omega]M'}}{By \defn of \defsubst}
        \judgewfPf{\Theta}{[\Omega]\Thetahat \implies^\ast \negsubprob{N}{[\Omega]M'}}{Straightforward}
      \end{llproof}
      ~\\
      By \Lemmaref{lem:wah-equiv-refl},
      \[
        \wahequiv{\Theta}{[\Omega]\Thetahat \implies^\ast \negsubprob{N}{[\Omega]M'}}{[\Omega]\Thetahat \implies^\ast \negsubprob{N}{[\Omega]M'}}
      \]
      By \defn of \defsubst and because $\ground{N}$,
      \[
        \wahequiv{\Theta}{[\Omega](\Thetahat \implies^\ast \negsubprob{N}{M'})}{[\Omega]\Thetahat \implies^\ast \negsubprob{N}{[\Omega]M'}}
      \]
    \end{itemize}

  \item
    \begin{itemize}
      \ProofCaseRule{\AlgSubNegUpshift}
      Similar to \AlgSubPosDownshift case of part (1).

      \ProofCaseRule{\AlgSubNegImpL}
      Similar to \AlgSubPosWithR case of part (1), but simpler.

      \ProofCaseRule{\AlgSubNegAllL}
      Similar to \AlgSubPosExR case of part (1).

      \ProofCaseRule{\AlgSubNegArrow}
      Similar to \AlgSubPosProd case of part (1).
      \qedhere
    \end{itemize}
  \end{enumerate}
\end{proof}

\begin{lemma}[Alg.\ Entail.\ Sound]
  \label{lem:alg-entail-sound}
  If $\entailwah{\Theta}{W}$, then $\semideclentailwah{\Theta}{W}$.
\end{lemma}
\begin{proof}
  By induction on $\size{W}$.
  \begin{itemize}
    \DerivationProofCase{\WTruePosSub}
    {
      \algsub[+]{\Theta; \cdot}{P}{Q}{W'}{\cdot}
      \\
      \entailwah{\Theta}{W'}
    }
    {
      \entailwah{\Theta}{\possubprob{P}{Q}}
    }
    \begin{llproof}
      \algsubPf[+]{\Theta; \cdot}{P}{Q}{W'}{\cdot}{Subderivation}
      \judgetpPf{\Theta; \cdot}{Q}{\Xi}{Presupposed derivation}
      \judgetpPf{\Theta}{P}{\dontcare}{Presupposed derivation}
      \Pf{}{}{\ground{P}}{By \Lemmaref{lem:decl-wf-ground}}
      \extendPf{\Theta}{\cdot}{\cdot}{By \ExtEmpty}
      \semideclsubPf[+]{\Theta}{P}{\underbrace{[\cdot]Q}_Q}{W''}{By \Lemmaref{lem:aux-alg-sub-sound}}
      \wahequivPf{\Theta}{\underbrace{[\cdot]W'}_{W'}}{W''}{\ditto}
      \Pf{\size{W'}}{=}{\size{W''}}{By \Lemmaref{lem:equiv-respects-size}}
      \Pf{}{<}{\size{\possubprob{P}{Q}}}{By \Lemmaref{lem:semidecl-sub-shrinks-problem}}
      \entailwahPf{\Theta}{W'}{Subderivation}
      \semideclentailwahPf{\Theta}{W'}{By \ih}
      \semideclentailwahPf{\Theta}{\possubprob{P}{Q}}{By \SemiDeclWTruePosSub}
    \end{llproof}
    
    \ProofCaseRule{\WTrueNegSub}
    Similar to case for \WTruePosSub.

    \ProofCaseRule{\WTruePosEquiv}
    Similar to case for \WTruePosSub,
    but uses \Lemmaref{lem:aux-alg-equiv-sound}
    and \Lemmaref{lem:semidecl-equiv-shrinks-problem}
    instead of \Lemmaref{lem:aux-alg-sub-sound}
    and \Lemmaref{lem:semidecl-sub-shrinks-problem}.

    \ProofCaseRule{\WTrueNegEquiv}
    Similar to case for \WTruePosEquiv.

  \item The remaining cases are straightforward. \qedhere
\end{itemize}
\end{proof}

\begin{theorem}[Alg.\ Sub.\ Sound]
  \label{thm:alg-sub-sound}
  ~
  \begin{enumerate}
  \item If $\algsub[+]{\Theta; \Delta}{P}{Q}{W}{\Delta'}$
    and $\judgetp{\Theta; \Delta}{Q}{\Xi}$
    and $\ground{P}$
    and $[\Delta]Q = Q$
    and $\extend{\Theta}{\Delta'}{\Omega}$
    and $\entailwah{\Theta}{[\Omega]W}$,
    then $\judgesub[+]{\Theta}{P}{[\Omega]Q}$.
  \item If $\algsub[-]{\Theta; \Delta}{N}{M}{W}{\Delta'}$
    and $\judgetp{\Theta; \Delta}{N}{\Xi}$
    and $\ground{M}$
    and $[\Delta]N = N$
    and $\extend{\Theta}{\Delta'}{\Omega}$
    and $\entailwah{\Theta}{[\Omega]W}$,
    then $\judgesub[-]{\Theta}{[\Omega]N}{M}$.
  \end{enumerate}
\end{theorem}
\begin{proof}
  ~
  \begin{enumerate}
  \item ~\\
    \begin{llproof}
      \semideclsubPf[+]{\Theta}{P}{[\Omega]Q}{W'}{By \Lemmaref{lem:aux-alg-sub-sound}}
      \wahequivPf{\Theta}{[\Omega]W}{W'}{\ditto}
      \entailwahPf{\Theta}{[\Omega]W}{Given}
      \semideclentailwahPf{\Theta}{[\Omega]W}{By \Lemmaref{lem:alg-entail-sound}}
      \semideclentailwahPf{\Theta}{W'}{By \Lemmaref{lem:equiv-respects-entail}}
      \judgesubPf[+]{\Theta}{P}{[\Omega]Q}{By \Lemmaref{lem:semidecl-sub-sound}}
    \end{llproof} 
  \item Similar to the first (\ie, positive) part. \qedhere
  \end{enumerate}
\end{proof}

\subsection{Algorithmic Typing}

\begin{theorem}[Alg.\ Typing Sound]
  \label{thm:alg-typing-soundness}
  ~
  \begin{enumerate}
    \item
      If $\algsynhead{\Theta}{\Gamma}{h}{P}$,
      then $\judgesynhead{\Theta}{\Gamma}{h}{P}$.
    \item
      If $\algsynexp{\Theta}{\Gamma}{\be}{\upshift{P}}$,
      then $\judgesynexp{\Theta}{\Gamma}{\be}{\upshift{P}}$.
    \item
      If $\algchk{\Theta; \Delta}{\Gamma}{v}{P}{\chi}{\Delta'}$\\
      and $\judgetp{\Theta; \Delta}{P}{\Xi}$
      and $[\Delta]P = P$
      and $\extend{\Theta}{\Delta'}{\Omega}$
      and $\algneg{\Theta}{\Gamma}{[\Omega]\chi}$,\\
      then $\judgechkval{\Theta}{\Gamma}{[\Omega]v}{[\Omega]P}$.
    \item
      If $\algchkneg{\Theta}{\Gamma}{e}{N}$,
      then $\judgechkexp{\Theta}{\Gamma}{e}{N}$.
    \item
      If $\algchkmatch{\Theta}{\Gamma}{P}{\clauses{\pa}{e}{i}{I}}{N}$,
      then $\judgechkmatch{\Theta}{\Gamma}{P}{\clauses{\pa}{e}{i}{I}}{N}$.
    \item
      If $\algspine{\Theta; \Delta}{\Gamma}{s}{N}{\upshift{P}}{\chi}{\Delta'}$\\
      and $\judgetp{\Theta; \Delta}{N}{\Xi}$
      and $[\Delta]N = N$
      and $\extend{\Theta}{\Delta'}{\Omega}$
      and $\algneg{\Theta}{\Gamma}{[\Omega]\chi}$,\\
      then $\judgespine{\Theta}{\Gamma}{[\Omega]s}{[\Omega]N}{[\Omega]\upshift{P}}$.
  \end{enumerate}
\end{theorem}
\begin{proof}
  By lexicographic induction on,
  first, the structure of the program term 
  ($h$, $\be$, $v$, $e$, $\clauses{\pa}{e}{i}{I}$, or $s$), and,
  second,
  $\size{P}$ for part (3),
  $\size{N}$ for part (4),
  $\size{P}$ for part (5),
  or the size of $N$ for part (6).
  Size $\size{-}$ is defined in \Figureref{fig:size}.
  All parts are mutually recursive.
  \begin{enumerate}
  \item
    \begin{itemize}
      \DerivationProofCase{\AlgSynHeadVar}
      {
        (x : P) \in \Gamma
      }
      {
        \algsynhead{\Theta}{\Gamma}{x}{P}
      }
      \begin{llproof}
        \Pf{(x : P)}{\in}{\Gamma}{Subderivation}
        \judgesynheadPf{\Theta}{\Gamma}{x}{P}{By \DeclSynHeadVar}
      \end{llproof}

      \DerivationProofCase{\AlgSynValAnnot}
      {
        \judgetp{\Theta; \cdot}{P}{\Xi}
        \\
        \algchk{\Theta; \cdot}{\Gamma}{v}{P}{\chi}{\cdot}
        \\
        \algneg{\Theta}{\Gamma}{\chi}
      }
      {
        \algsynhead{\Theta}{\Gamma}{\annoexp{v}{P}}{P}
      }
      \begin{llproof}
        \algchkPf{\Theta; \cdot}{\Gamma}{v}{P}{\chi}{\cdot}{Subderivation}
        \judgetpPf{\Theta; \cdot}{P}{\Xi}{Subderivation}
        \extendPf{\Theta}{\cdot}{\cdot}{By \ExtEmpty}
        \algnegPf{\Theta}{\Gamma}{\underbrace{\chi}_{[\cdot]\chi}}{Subderivation}
        \judgechkvalPf{\Theta}{\Gamma}{v}{[\cdot]P}{By \ih (smaller program term)}
        \judgechkvalPf{\Theta}{\Gamma}{v}{P}{By \defn of $[-]-$}
        \judgesynheadPf{\Theta}{\Gamma}{\annoexp{v}{P}}{P}{By \DeclSynValAnnot}
      \end{llproof} 
    \end{itemize}

  \item
    \begin{itemize}
      \DerivationProofCase{\AlgSynSpineApp}
      { 
        \arrayenvb{
          \algsynhead{\Theta}{\Gamma}{h}{\downshift{N}}
        }
        \\
        \arrayenvb{
          \algspine{\Theta; \cdot}{\Gamma}{s}{N}{\upshift{P}}{\chi}{\cdot}
          \\
          \algneg{\Theta}{\Gamma}{\chi}
        }
      }
      { \algsynexp{\Theta}{\Gamma}{h(s)}{\upshift{P}} }
      \begin{llproof}
        \algsynheadPf{\Theta}{\Gamma}{h}{\downshift{N}}{Subderivation}
        \judgesynheadPf{\Theta}{\Gamma}{h}{\downshift{N}}{By \ih (smaller program term)}
        \algspinePf{\Theta; \cdot}{\Gamma}{s}{N}{\upshift{P}}{\chi}{\cdot}{Subderivation}
        \judgetpPf{\Theta; \cdot}{N}{\Xi'}{Straightforward}
        \extendPf{\Theta}{\cdot}{\cdot}{By \ExtEmpty}
        \algnegPf{\Theta}{\Gamma}{\underbrace{\chi}_{[\cdot]\chi}}{Subderivation}
        \judgespinePf{\Theta}{\Gamma}{s}{\underbrace{[\cdot]N}_N}{\underbrace{[\cdot]\upshift{P}}_{\upshift{P}}}{By \ih (smaller program term)}
        \judgesynexpPf{\Theta}{\Gamma}{h(s)}{\upshift{P}}{By \DeclSynSpineApp}
      \end{llproof} 

      \ProofCaseRule{\AlgSynExpAnnot}
      Straightforward.
    \end{itemize}

  \item
    \begin{itemize}
      \DerivationProofCase{\AlgChkValVar}
      {
        P \neq \exists, \land 
        \\
        (x:Q)\in\Gamma
        \\
        \algsub[+]{\Theta; \Delta}{Q}{P}{\Wah}{\Delta'}
      }
      {
        \algchk{\Theta; \Delta}{\Gamma}{x}{P}{\underbrace{\Wah}_\chi}{\Delta'}
      }
      \begin{llproof}
        \algsubPf[+]{\Theta; \Delta}{Q}{P}{\Wah}{\Delta'}{Subderivation}
        \judgetpPf{\Theta; \Delta}{P}{\Xi}{Given}
        \Pf{[\Delta]P}{=}{P}{Given}
        \Pf{(x:Q)}{\in}{\Gamma}{Subderivation}
        \Pf{}{}{\ground{Q}}{By context well-formedness}
        \extendPf{\Theta}{\Delta'}{\Omega}{Given}
        \algnegPf{\Theta}{\Gamma}{[\Omega]W}{Given}
        \entailwahPf{\Theta}{[\Omega]W}{By inversion}
        \judgesubPf[+]{\Theta}{Q}{[\Omega]P}{By \Theoremref{thm:alg-sub-sound}}
        \Pf{P}{\neq}{\exists, \land}{Premise}
        \Pf{[\Omega]P}{\neq}{\exists, \land}{Straightforward}
        \judgechkvalPf{\Theta}{\Gamma}{x}{[\Omega]P}{By \DeclChkValVar}
        \judgechkvalPf{\Theta}{\Gamma}{[\Omega]x}{[\Omega]P}{By \defn of $[-]$}
      \end{llproof} 

      \DerivationProofCase{\AlgChkValUnit}
      {
      }
      {
        \algchk{\Theta; \Delta}{\Gamma}{\unit}{\unitty}{\cdot}{\Delta}
      }
      \begin{llproof}
        \judgechkvalPf{\Theta}{\Gamma}{\unit}{\unitty}{By \DeclChkValUnit}
        \judgechkvalPf{\Theta}{\Gamma}{[\Omega]\unit}{[\Omega]\unitty}{By \defn of $[-]-$}
      \end{llproof}

      \DerivationProofCase{\AlgChkValPair}
      {
        \algchk{\Theta; \Delta}{\Gamma}{v_1}{P_1}{\chi_1}{\Delta''}
        \\
        \algchk{\Theta; \Delta''}{\Gamma}{v_2}{[\Delta'']P_2}{\chi_2}{\Delta'}
      }
      {
        \algchk{\Theta; \Delta}{\Gamma}{\pair{v_1}{v_2}}
        {(P_1 \times P_2)}{[\Delta']\chi_1, \chi_2}{\Delta'}
      }
      \begin{llproof}
        \algchkPf{\Theta; \Delta}{\Gamma}{v_1}{P_1}{\chi_1}{\Delta''}{Subderivation}
        \judgetpPf{\Theta; \Delta}{P_1 \times P_2}{\Xi}{Given}
        \judgetpPf{\Theta; \Delta}{P_1}{\dontcare}{By inversion on type WF}
        \judgetpPf{\Theta; \Delta}{P_2}{\dontcare}{\ditto}
        \extendPf{\Theta}{\Delta'}{\Omega}{Given}
        \algnegPf{\Theta}{\Gamma}{[\Omega]([\Delta']\chi_1, \chi_2)}{Given}
        \algnegPf{\Theta}{\Gamma}{[\Omega]([\Delta']\chi_1), [\Omega]\chi_2}{By \defn of $[-]-$}
        \algnegPf{\Theta}{\Gamma}{[\Omega]\chi_1, [\Omega]\chi_2}{By \Lemmaref{lem:ext-subst-invariance}}
        \algnegPf{\Theta}{\Gamma}{[\Omega]\chi_1}{By inversion}
        \algnegPf{\Theta}{\Gamma}{[\Omega]\chi_2}{\ditto}
        \algchkPf{\Theta; \Delta''}{\Gamma}{v_2}{[\Delta'']P_2}{\chi_2}{\Delta'}{Subderivation}
        \extendPf{\Theta}{\Delta''}{\Delta'}{By \Lemmaref{lem:typing-extends}}
        \extendPf{\Theta}{\Delta''}{\Omega}{By \Lemmaref{lem:ext-trans}}
        \judgechkvalPf{\Theta}{\Gamma}{[\Omega]v_1}{[\Omega]P_1}{By \ih (smaller program term)}
        \judgetpPf{\Theta; \Delta}{P_2}{\dontcare}{Above}
        \extendPf{\Theta}{\Delta}{\Delta''}{By \Lemmaref{lem:typing-extends}}
        \judgetpPf{\Theta; \Delta''}{P_2}{\dontcare}{By \Lemmaref{lem:ext-weak-tp}}
        \judgetpPf{\Theta; \Delta''}{[\Delta'']P_2}{\dontcare}{By \Lemmaref{lem:right-hand-subst}}
        \Pf{[\Delta'']([\Delta'']P_2)}{=}{[\Delta'']P_2}{By \Lemmaref{lem:apply-idempotent}}
        \judgechkvalPf{\Theta}{\Gamma}{[\Omega]v_2}{[\Omega]P_2}{By \ih (smaller program term)}
        \judgechkvalPf{\Theta}{\Gamma}{\pair{[\Omega]v_1}{[\Omega]v_2}}{[\Omega]P_1 \times [\Omega]P_2}{By \DeclChkValPair}
        \judgechkvalPf{\Theta}{\Gamma}{[\Omega]\pair{v_1}{v_2}}{[\Omega](P_1 \times P_2)}{By \defn of $[-]-$}
      \end{llproof}

      \DerivationProofCase{\AlgChkValIn{1}}
      {
        \algchk{\Theta; \Delta}{\Gamma}{v}{P_1}{\chi}{\Delta'}
      }
      {
        \algchk{\Theta; \Delta}{\Gamma}{\inj{1}{v}}{(P_1 + P_2)}{\chi}{\Delta'}
      }
      \begin{llproof}
        \algchkPf{\Theta; \Delta}{\Gamma}{v}{P_1}{\chi}{\Delta'}{Subderivation}
        \judgechkvalPf{\Theta}{\Gamma}{[\Omega]v}{[\Omega]P_1}{By \ih}
        \judgetpPf{\Theta; \Delta}{P_1 + P_2}{\dontcare}{Presupposed derivation}
        \judgetpPf{\Theta; \Delta}{P_2}{\dontcare}{By inversion on type WF}
        \judgetpPf{\Theta}{[\Omega]P_2}{\dontcare}{By \Lemmaref{lem:alg-to-decl-wf}}
        \judgechkvalPf{\Theta}{\Gamma}{\inj{1}{[\Omega]v}}{[\Omega]P_1 + [\Omega]P_2}{By \DeclChkValIn{1}}
        \judgechkvalPf{\Theta}{\Gamma}{[\Omega]\inj{1}{v}}{[\Omega](P_1 + P_2)}{By \defn of $[-]-$}
      \end{llproof}
      
      \ProofCaseRule{\AlgChkValIn{2}}
      Similar to \AlgChkValIn{1} case.

      \DerivationProofCase{\AlgChkValExists}
      {
        \algchk{\Theta; \Delta, \ahat:\tau}{\Gamma}{v}{[\ahat / a]P_0}{\chi}{\Delta', \hypeq{\ahat}{\tau}{t}}
      }
      {
        \algchk{\Theta; \Delta}{\Gamma}{v}{(\extype{a : \tau} P_0)}{\chi}{\Delta'}
      }
      \begin{llproof}
        \algchkPf{\Theta; \Delta, \ahat:\tau}{\Gamma}{v}{[\ahat / a]P_0}{\chi}{\Delta', \hypeq{\ahat}{\tau}{t}}{Subderivation}
        \judgetpPf{\Theta; \Delta}{\extype{a : \tau} P_0}{\Xi}{Given}
        \judgetpPf{\Theta, a:\tau; \Delta}{P_0}{\Xi', a:\tau}{By inversion on type WF}
        \judgetpPf{\Theta; \Delta, \ahat:\tau}{[\ahat/a]P_0}{\Xi', \ahat:\tau}{By \Lemmaref{lem:value-determined-evar-rename}}
        \Pf{[\Delta, \ahat:\tau]([\ahat/a]P_0)}{=}{[\ahat/a]P_0}{Because $[\Delta]P_0 = P_0$}
        \extendPf{\Theta}{\Delta'}{\Omega}{Given}
        \extendPf{\Theta}{\Delta', \hypeq{\ahat}{\tau}{t}}{\Omega, \hypeq{\ahat}{\tau}{t}}{By \ExtSolved}
        \algnegPf{\Theta}{\Gamma}{[\Omega]\chi}{Given}
        \algnegPf{\Theta}{\Gamma}{[\Omega, \hypeq{\ahat}{\tau}{t}]\chi}{By \Lemmaref{lem:output-applied}}
        \Pf{\size{[\ahat/a]P_0}}{=}{\size{P_0}}{Size independent of variable renaming}
        \Pf{}{<}{\size{P_0} + 1}{Obvious}
        \Pf{}{=}{\size{\extype{a:\tau} P_0}}{By \defn}
        \judgechkvalPf{\Theta}{\Gamma}{[\Omega]v}{[\Omega, \hypeq{\ahat}{\tau}{t}]([\ahat/a]P_0)}{By \ih}
        \judgechkvalPf{\Theta}{\Gamma}{[\Omega]v}{[\Omega]([t/\ahat]([\ahat/a]P_0))}{By \defn of $[-]-$}
        \judgechkvalPf{\Theta}{\Gamma}{[\Omega]v}{[\Omega]([t/a]P_0)}{By composition of subst.}
        \judgechkvalPf{\Theta}{\Gamma}{[\Omega]v}{[[\Omega]t/a]([\Omega]P_0)}{By a property of subst.}
        \judgetermPf{\Theta}{t}{\tau}{By inversion}
        \Pf{}{}{\ground{t}}{By \Lemmaref{lem:decl-wf-ground}}
        \judgechkvalPf{\Theta}{\Gamma}{[\Omega]v}{[t/a]([\Omega]P_0)}{Because $\ground{t}$}
        \judgechkvalPf{\Theta}{\Gamma}{[\Omega]v}{\extype{a:\tau} [\Omega]P_0}{By \DeclChkValExists}
        \judgechkvalPf{\Theta}{\Gamma}{[\Omega]v}{[\Omega]\extype{a:\tau} P_0}{By \defn of $[-]-$}
      \end{llproof} 

      \DerivationProofCase{\AlgChkValWith}
      {
        \algchk{\Theta; \Delta}{\Gamma}{v}{P_0}{\chi_0}{\Delta''}
        \\
        \alginst{\Theta; \Delta''}{[\Delta'']\phi}{\Delta'}
      }
      {
        \algchk{\Theta; \Delta}{\Gamma}{v}{(P_0 \andty \phi)}{([\Delta']\phi, [\Delta']\chi_0)}{\Delta'}
      }
      \begin{llproof}
        \algchkPf{\Theta; \Delta}{\Gamma}{v}{P_0}{\chi_0}{\Delta''}{Subderivation}
        \judgetpPf{\Theta; \Gamma}{P_0 \land \phi}{\dontcare}{Given}
        \judgetpPf{\Theta; \Gamma}{P_0}{\dontcare}{By inversion on type WF}
        \Pf{[\Delta]P_0}{=}{P_0}{Because $[\Delta](P_0 \land \phi) = P_0 \land \phi$}
        \alginstPf{\Theta; \Delta''}{[\Delta'']\phi}{\Delta'}{Subderivation}
        \extendPf{\Theta}{\Delta''}{\Delta'}{By \Lemmaref{lem:inst-extends}}
        \extendPf{\Theta}{\Delta'}{\Omega}{Given}
        \extendPf{\Theta}{\Delta''}{\Omega}{By \Lemmaref{lem:ext-trans}}
        \algnegPf{\Theta}{\Gamma}{[\Omega]([\Delta']\phi, [\Delta']\chi_0)}{Given}
        \algnegPf{\Theta}{\Gamma}{[\Omega]([\Delta']\phi), [\Omega]([\Delta']\chi_0)}{By \defn}
        \entailwahPf{\Theta}{[\Omega]([\Delta']\phi)}{By inversion on \ChkProblemsWah}
        \algnegPf{\Theta}{\Gamma}{[\Omega]([\Delta']\chi_0)}{\ditto}
        \algnegPf{\Theta}{\Gamma}{[\Omega]\chi_0}{By \Lemmaref{lem:ext-subst-invariance}}
        \judgechkvalPf{\Theta}{\Gamma}{[\Omega]v}{[\Omega]P_0}{By \ih}
        \judgeentailPf{\Theta}{[\Omega]([\Delta']\phi)}{By inversion}
        \judgeentailPf{\Theta}{[\Omega]\phi}{By \Lemmaref{lem:ext-subst-invariance}}
        \judgechkvalPf{\Theta}{\Gamma}{[\Omega]v}{[\Omega]P_0 \land [\Omega]\phi}{By \DeclChkValWith}
        \judgechkvalPf{\Theta}{\Gamma}{[\Omega]v}{[\Omega](P_0 \land \phi)}{By \defn}
      \end{llproof} 

      \DerivationProofCase{\AlgChkValFix}
      {
        \judgeunroll{\cdot}{\Theta; \Delta}{\nu:F[\mu F]}{\alpha}{F\; \Fold{F}{\alpha}\;\nu}{t}{Q}{\tau}
        \\
        \algchk{\Theta; \Delta}{\Gamma}{v_0}{Q}{\chi}{\Delta'}
      }
      {
        \algchk{\Theta; \Delta}{\Gamma}{\into{v_0}}{\underbrace{\comprehend{\nu : \mu F}{\Fold{F}{\alpha}\,{\nu} =_\tau t}}_P}{\chi}{\Delta'}
      }
      \begin{llproof}
        \algchkPf{\Theta; \Delta}{\Gamma}{v_0}{Q}{\chi}{\Delta'}{Subderivation}
        \judgeunrollPf{\cdot}{\Theta; \Delta}{\nu:F[\mu F]}{\alpha}{F\; \Fold{F}{\alpha}\;\nu}{t}{Q}{\tau}{Subderivation}
        \judgetpPf{\Theta; \Delta}{Q}{\dontcare}{By \Lemref{lem:alg-unroll-output-wf}}
        \Pf{[\Delta]P}{=}{P}{Given}
        \Pf{[\Delta]F}{=}{F}{By inversion}
        \Pf{[\Delta]t}{=}{t}{\ditto}
        \Pf{[\Delta]Q}{=}{Q}{By \Lemmaref{lem:unroll-applied}}
        \extendPf{\Theta}{\Delta'}{\Omega}{Given}
        \algnegPf{\Theta}{\Gamma}{[\Omega]\chi}{Given}
        \judgechkvalPf{\Theta}{\Gamma}{[\Omega]v_0}{[\Omega]Q}{By \ih}
        \trailingjust{(smaller program term)}
        \decolumnizePf
        \judgeunrollPf{\cdot}{\Theta}{\nu:([\Omega]F)[\mu [\Omega]F]}{\alpha}{[\Omega]F\; \Fold{[\Omega]F}{\alpha}\;\nu}{[\Omega]t}{[\Omega]Q}{\tau}{By \Lemref{lem:complete-alg-unroll}}
        \judgechkvalPf{\Theta}{\Gamma}{\into{[\Omega]v_0}}{\comprehend{\nu : \mu [\Omega]F}{\Fold{[\Omega]F}{\alpha}\,{\nu} =_\tau [\Omega]t}}{By \DeclChkValFix}
        \judgechkvalPf{\Theta}{\Gamma}{[\Omega]\into{v_0}}{[\Omega]\comprehend{\nu : \mu F}{\Fold{F}{\alpha}\,{\nu} =_\tau t}}{By \defn}
      \end{llproof}

      \DerivationProofCase{\AlgChkValDownshift}
      {
      }
      {
        \algchk{\Theta; \Delta}{\Gamma}{\thunk{e}}{\downshift{N}}{(e <= N)}{\Delta}
      }
      \begin{llproof}
        \algnegPf{\Theta}{\Gamma}{[\Omega](e <= N)}{Given}
        \algnegPf{\Theta}{\Gamma}{[\Omega]e <= [\Omega]N}{By \defn}
        \algchknegPf{\Theta}{\Gamma}{[\Omega]e}{[\Omega]N}{By inversion on \ChkProblemsNegChk}
        \Pf{\size{[\Omega]e}}{=}{\size{e}}{Size independent of index terms}
        \Pf{}{<}{\size{e} + 1}{Obvious}
        \Pf{}{=}{\size{\thunk{e}}}{By \defn}
        \judgechkexpPf{\Theta}{\Gamma}{[\Omega]e}{[\Omega]N}{By \ih}
        \judgechkvalPf{\Theta}{\Gamma}{\thunk{[\Omega]e}}{\downshift{[\Omega]N}}{By \DeclChkValDownshift}
        \judgechkvalPf{\Theta}{\Gamma}{[\Omega]\thunk{e}}{[\Omega]\downshift{N}}{By \defn}
      \end{llproof}
    \end{itemize}

  \item
    \begin{itemize}
      \DerivationProofCase{\AlgChkExpUpshift}
      { 
        \algchk{\Theta; \cdot}{\Gamma}{v}{P}{\chi}{\cdot} 
        \\
        \algneg{\Theta}{\Gamma}{\chi}
      }
      {
        \algchkneg{\Theta}{\Gamma}{\Return{v}}{\upshift{P}} 
      }
      \begin{llproof}
        \algchkPf{\Theta; \cdot}{\Gamma}{v}{P}{\chi}{\cdot}{Subderivation}
        \judgetpPf{\Theta; \cdot}{P}{\dontcare}{Presupposed derivation}
        \Pf{[\cdot]P}{=}{P}{By \defn of \defsubst}
        \extendPf{\Theta}{\cdot}{\cdot}{By \ExtEmpty}
        \algnegPf{\Theta}{\Gamma}{\chi}{Subderivation}
        \algnegPf{\Theta}{\Gamma}{[\cdot]\chi}{By \defn of $[-]-$}
        \judgechkvalPf{\Theta}{\Gamma}{[\cdot]v}{[\cdot]P}{By \ih (smaller program term)}
        \judgechkvalPf{\Theta}{\Gamma}{v}{P}{By \defn of $[-]-$}
        \judgechkexpPf{\Theta}{\Gamma}{\Return{v}}{\upshift{P}}{By \DeclChkExpUpshift}
      \end{llproof}

      \DerivationProofCase{\AlgChkExpRec}
      {
        \arrayenv{
          \simple{\Theta}{N}
          \\
          \algsub[-]{\Theta; \cdot}{\alltype{a:\kindnat} M}{N}{W}{\cdot}
          \\
          \entailwah{\Theta}{W}
          \\
          \algchkneg{\Theta, a:\kindnat}{\Gamma, x:\downshift{\alltype{a':\kindnat} a' < a \implies [a'/a]M}}{e_0}{M}
        }
      }
      {
        \algchkneg{\Theta}{\Gamma}{\rec{x : (\alltype{a:\kindnat} M)}{e_0}}{N}
      }
      \begin{llproof}
        \algsubPf[-]{\Theta; \cdot}{\alltype{a:\kindnat} M}{N}{W}{\cdot}{Subderivation}
        \judgetpPf{\Theta; \cdot}{\alltype{a:\kindnat} M}{\dontcare}{Presupposed derivation}
        \judgetpPf{\Theta}{N}{\dontcare}{Presupposed derivation}
        \groundPf{N}{By \Lemmaref{lem:decl-wf-ground}}
        \extendPf{\Theta}{\cdot}{\cdot}{By \ExtEmpty}
        \entailwahPf{\Theta}{W}{Subderivation}
        \judgesubPf[-]{\Theta}{[\cdot](\alltype{a:\kindnat} M)}{N}{By \Theoremref{thm:alg-sub-sound}}
        \judgesubPf[-]{\Theta}{\alltype{a:\kindnat} M}{N}{By \defn of \defsubst}
        \decolumnizePf
        \algchknegPf{\Theta, a:\kindnat}{\Gamma, x:\downshift{\alltype{a':\kindnat} a' < a \implies [a'/a]M}}{e_0}{M}{Subderivation}
        \judgechkexpPf{\Theta, a:\kindnat}{\Gamma, x:\downshift{\alltype{a':\kindnat} a' < a \implies [a'/a]M}}{e_0}{M}{By \ih}
        \simplePf{\Theta}{N}{Subderivation}
        \judgechkexpPf{\Theta}{\Gamma}{\rec{x : (\alltype{a:\kindnat} M)}{e_0}}{N}{By \DeclChkExpRec}
      \end{llproof} 
      
      \item The remaining cases are straightforward.
        For cases \AlgChkExpLet, \AlgChkExpMatch, and \AlgChkExpLam,
        apply the \ih (smaller program term) to the relevant subderivations,
        and then apply the corresponding declarative rule,
        using \Lemmaref{lem:complete-alg-extract} as needed.
        Similarly, for case \AlgChkExpExtract,
        apply the \ih (the program term stays the same,
        but the size of the type that its checked against decreases
        by \Lemmaref{lem:shrinking-extract}
        and the \defn of size in \Figureref{fig:size}),
        and then apply \AlgChkExpExtract
        (after using \Lemmaref{lem:complete-alg-extract}).
    \end{itemize}

  \item Each case is straightforward:
    for each case except for \AlgChkMatchVoid,
    apply the \ih, and then apply the corresponding declarative rule,
    using \Lemmaref{lem:complete-alg-extract} and \Lemmaref{lem:complete-alg-unroll}.
    \begin{itemize}
      \ProofCaseRule{\AlgChkMatchEx}
      The program term stays the same, but $\size{P}$ decreases.

      \ProofCaseRule{\AlgChkMatchWith}
      The program term stays the same, but $\size{P}$ decreases.

      \ProofCaseRule{\AlgChkMatchUnit}
      Smaller program term.

      \ProofCaseRule{\AlgChkMatchPair}
      Smaller program term.

      \ProofCaseRule{\AlgChkMatchSum}
      Smaller program term.

      \ProofCaseRule{\AlgChkMatchVoid}
      Simply apply the corresponding declarative rule, \DeclChkMatchVoid.

      \ProofCaseRule{\AlgChkMatchFix}
      Smaller program term.
    \end{itemize}

  \item
    \begin{itemize}
      \DerivationProofCase{\AlgSpineAll}
      {
        \algspine{\Theta; \Delta, \ahat : \tau}{\Gamma}{s}{[\ahat/a]{N_0}}{\upshift{P}}{\chi}{\Delta', \hypeq{\ahat}{\tau}{t}}
      }
      {
        \algspine{\Theta; \Delta}{\Gamma}{s}{(\alltype{a:\tau}N_0)}{\upshift{P}}{\chi}{\Delta'}
      }
      \begin{llproof}
        \algspinePf{\Theta; \Delta, \ahat : \tau}{\Gamma}{s}{[\ahat/a]{N_0}}{\upshift{P}}{\chi}{\Delta', \hypeq{\ahat}{\tau}{t}}{Subderivation}
        \judgetpPf{\Theta; \Delta}{\alltype{a : \tau} N_0}{\Xi}{Given}
        \judgetpPf{\Theta, a:\tau; \Delta}{N_0}{\Xi', a:\tau}{By inversion on type WF}
        \judgetpPf{\Theta; \Delta, \ahat:\tau}{[\ahat/a]N_0}{\Xi', \ahat:\tau}{By \Lemmaref{lem:value-determined-evar-rename}}
        \Pf{[\ahat/a]N_0}{=}{[\Delta, \ahat:\tau]([\ahat/a]N_0)}{Because $[\Delta]N_0 = N_0$}
        \extendPf{\Theta}{\Delta'}{\Omega}{Given}
        \extendPf{\Theta}{\Delta', \hypeq{\ahat}{\tau}{t}}{\Omega, \hypeq{\ahat}{\tau}{t}}{By \ExtSolved}
        \algnegPf{\Theta}{\Gamma}{[\Omega]\chi}{Given}
        \algnegPf{\Theta}{\Gamma}{[\Omega, \hypeq{\ahat}{\tau}{t}]\chi}{By \Lemmaref{lem:output-applied}}
        \Pf{\size{[\ahat/a]N_0}}{=}{\size{N_0}}{Size independent of variable renaming}
        \Pf{}{<}{\size{N_0} + 1}{Obvious}
        \Pf{}{=}{\size{\alltype{a:\tau} N_0}}{By \defn}
        \decolumnizePf
        \judgespinePf{\Theta}{\Gamma}{[\Omega, \hypeq{\ahat}{\tau}{t}]s}{[\Omega, \hypeq{\ahat}{\tau}{t}]([\ahat/a]{N_0})}{[\Omega, \hypeq{\ahat}{\tau}{t}]\upshift{P}}{By \ih}
        \judgespinePf{\Theta}{\Gamma}{[\Omega]s}{[\Omega, \hypeq{\ahat}{\tau}{t}]([\ahat/a]{N_0})}{[\Omega]\upshift{P}}{$\ahat$ not free in $s$ or $\upshift{P}$}
        \judgespinePf{\Theta}{\Gamma}{[\Omega]s}{[t/a]([\Omega]{N_0})}{[\Omega]\upshift{P}}{By \defn and props.\ of subst.}
        \judgespinePf{\Theta}{\Gamma}{[\Omega]s}{(\alltype{a:\tau}{[\Omega]N_0})}{[\Omega]\upshift{P}}{By \DeclSpineAll}
        \judgespinePf{\Theta}{\Gamma}{[\Omega]s}{[\Omega](\alltype{a:\tau}{N_0})}{[\Omega]\upshift{P}}{By \defn of \defsubst}
      \end{llproof} 

      \DerivationProofCase{\AlgSpineImplies}
      {
        \algspine{\Theta; \Delta}{\Gamma}{s}{N_0}{\upshift{P}}{\chi_0}{\Delta'}
      }
      {
        \algspine{\Theta; \Delta}{\Gamma}{s}{\phi \implies N_0}{\upshift{P}}{[\Delta']\phi,\chi_0}{\Delta'}
      }
      \begin{llproof}
        \algspinePf{\Theta; \Delta}{\Gamma}{s}{N_0}{\upshift{P}}{\chi_0}{\Delta'}{Subderivation}
        \judgetpPf{\Theta; \Delta}{\phi \implies N_0}{\Xi}{Given}
        \judgetpPf{\Theta; \Delta}{N_0}{\dontcare}{By inversion on type WF}
        \extendPf{\Theta}{\Delta'}{\Omega}{Given}
        \algnegPf{\Theta}{\Gamma}{[\Omega]([\Delta']\phi,\chi_0)}{Given}
        \algnegPf{\Theta}{\Gamma}{[\Omega]([\Delta']\phi),[\Omega]\chi_0}{By \defn of $[-]-$}
        \entailwahPf{\Theta}{[\Omega]([\Delta']\phi)}{By inversion on \ChkProblemsWah}
        \algnegPf{\Theta}{\Gamma}{[\Omega]\chi_0}{\ditto}
        \judgespinePf{\Theta}{\Gamma}{[\Omega]s}{[\Omega]N_0}{[\Omega]\upshift{P}}{By \ih (size of $N$ decreased)}
        \entailwahPf{\Theta}{[\Omega]\phi}{By \Lemmaref{lem:ext-subst-invariance}}
        \judgeentailPf{\Theta}{[\Omega]\phi}{By inversion on \WTrueProp}
        \judgespinePf{\Theta}{\Gamma}{[\Omega]s}{[\Omega]\phi \implies [\Omega]N_0}{[\Omega]\upshift{P}}{By \DeclSpineImplies}
        \judgespinePf{\Theta}{\Gamma}{[\Omega]s}{[\Omega](\phi \implies N_0)}{[\Omega]\upshift{P}}{By \defn of $[-]-$}
      \end{llproof} 

      \DerivationProofCase{\AlgSpineApp}
      {
        \algchk{\Theta; \Delta}{\Gamma}{v}{Q}{\chi_1}{\Delta''}
        \\
        \algspine{\Theta; \Delta''}{\Gamma}{s_0}{[\Delta'']N_0}{\upshift{P}}{\chi_2}{\Delta'}
      }
      {
        \algspine{\Theta; \Delta}{\Gamma}{v, s_0}{Q -> N_0}{\upshift{P}}{[\Delta']\chi_1, \chi_2}{\Delta'}
      }
      \begin{llproof}
        \algchkPf{\Theta; \Delta}{\Gamma}{v}{Q}{\chi_1}{\Delta''}{Subderivation}
        \judgetpPf{\Theta; \Delta}{Q -> N_0}{\Xi}{Given}
        \judgetpPf{\Theta; \Delta}{Q}{\dontcare}{By inversion on type WF}
        \judgetpPf{\Theta; \Delta}{N_0}{\dontcare}{\ditto}
        \algspinePf{\Theta; \Delta''}{\Gamma}{s_0}{[\Delta'']N_0}{\upshift{P}}{\chi_2}{\Delta'}{Subderivation}
        \extendPf{\Theta}{\Delta''}{\Delta'}{By \Lemmaref{lem:typing-extends}}
        \extendPf{\Theta}{\Delta'}{\Omega}{Given}
        \extendPf{\Theta}{\Delta''}{\Omega}{By \Lemmaref{lem:ext-trans}}
        \algnegPf{\Theta}{\Gamma}{[\Omega]([\Delta']\chi_1, \chi_2)}{Given}
        \algnegPf{\Theta}{\Gamma}{[\Omega]([\Delta']\chi_1), [\Omega]\chi_2}{By \defn of $[-]-$}
        \algnegPf{\Theta}{\Gamma}{[\Omega]([\Delta']\chi_1)}{By inversion}
        \algnegPf{\Theta}{\Gamma}{[\Omega]\chi_2}{\ditto}
        \algnegPf{\Theta}{\Gamma}{[\Omega]\chi_1}{By \Lemmaref{lem:ext-subst-invariance}}
        \judgechkvalPf{\Theta}{\Gamma}{[\Omega]v}{[\Omega]Q}{By \ih (smaller program term)}
        \extendPf{\Theta}{\Delta}{\Delta''}{By \Lemmaref{lem:typing-extends}}
        \judgetpPf{\Theta; \Delta''}{N_0}{\dontcare}{By \Lemmaref{lem:ext-weak-tp}}
        \judgetpPf{\Theta; \Delta''}{[\Delta'']N_0}{\dontcare}{By \Lemmaref{lem:right-hand-subst}}
        \judgespinePf{\Theta}{\Gamma}{[\Omega]s_0}{[\Omega]N_0}{[\Omega]\upshift{P}}{By \ih (smaller program term)}
        \judgespinePf{\Theta}{\Gamma}{[\Omega]v, [\Omega]s_0}{[\Omega]Q -> [\Omega]N_0}{[\Omega]\upshift{P}}{By \DeclSpineApp}
        \judgespinePf{\Theta}{\Gamma}{[\Omega](v, s_0)}{[\Omega](Q -> N_0)}{[\Omega]\upshift{P}}{By \defn of $[-]-$}
      \end{llproof}

      \DerivationProofCase{\AlgSpineNil}
      {}
      {
        \algspine{\Theta; \Delta}{\Gamma}{\cdot}{\upshift{P}}{\upshift{P}}{\True}{\Delta}
      }
      \begin{llproof}
        \judgespinePf{\Theta}{\Gamma}{\cdot}{\upshift{[\Omega]P}}{\upshift{[\Omega]P}}{By \DeclSpineNil}
        \judgespinePf{\Theta}{\Gamma}{\cdot}{[\Omega]\upshift{P}}{[\Omega]\upshift{P}}{By \defn of $[-]-$}
      \end{llproof}
      \qedhere
    \end{itemize}
  \end{enumerate}
\end{proof}

\section{Algorithmic Completeness}

We often use the following syntactic sugar for the unrolling judgment(s):
\[
  \unroll{\Xi}{\Theta; \Delta}{G}{F}{\beta}{\alpha}{\tau}{t}{P}
\]
is equivalent to
\[
  \judgeunroll{\Xi}{\Theta; \Delta}{ \nu:G[\mu F] }{\beta}{G\;\Fold{F}{\alpha}\;\nu}{t}{P}{\tau}
\]
and similarly for declarative unrolling.

We introduce the following convenient notation for this section.

\begin{definition}[Evar.\ Rename]
  \label{def:evar-rename}
  Given a list of sort declarations for universal index variables\\
  $\Xi = a_1:\tau_1, \dots, a_n:\tau_n$ (where $n \geq 0$),
  define $\Delta_\Xi = \ahat_1:\tau_1, \dots, \ahat_n:\tau_n$,\\
  and define the renaming operation
  $[\Delta_\Xi/\Xi]- = [\ahat_1/a_1, \dots, \ahat_n/a_n]-$.
\end{definition}

\begin{definition}[Solve Evar.\ Rename]
  \label{def:solve-evar-rename}
  Given a list of sort declarations for universal index variables\\
  $\Xi = a_1:\tau_1, \dots, a_n:\tau_n$ (where $n \geq 0$),
  and given $\extend{\Theta}{\Delta_\Xi}{\Omega_\Xi}$
  (where $\Delta_\Xi$ is defined by \Defnref{def:evar-rename}),
  define the substitution operation
  $[\Omega_\Xi/\Xi]- = [([\Omega_\Xi]\ahat_1)/a_1, \dots, ([\Omega_\Xi]\ahat_n)/a_n]-$.
\end{definition}

\begin{lemma}[Alg.\ Ctx.\ Weakening]
  \label{lem:alg-ctx-weakening}
  Assume $\judgectx{\Theta}{\Delta, \Delta_0}$.
  \begin{enumerate}
  \item
    If $\judgeterm{\Theta; \Delta}{t}{\tau}$,
    then $\judgeterm{\Theta; \Delta, \Delta_0}{t}{\tau}$.
  \item
    If $\judgetp{\Theta; \Delta}{A}{\Xi}$,
    then $\judgetp{\Theta; \Delta, \Delta_0}{A}{\Xi}$.
  \item
    If $\judgefunctor{\Theta; \Delta}{\F}{\Xi}$,
    then $\judgefunctor{\Theta; \Delta, \Delta_0}{\F}{\Xi}$.
  \item
    If $\judgealgebra{\Xi}{\Theta; \Delta}{\alpha}{F}{\tau}$,
    then $\judgealgebra{\Xi}{\Theta; \Delta, \Delta_0}{\alpha}{F}{\tau}$.
  \item
    If $\judgeunroll{\Xi}{\Theta; \Delta}{\nu:G[\mu F]}{\beta}{G\; \Fold{F}{\alpha}\;\nu}{t}{Q}{\tau}$,\\
    then $\judgeunroll{\Xi}{\Theta; \Delta, \Delta_0}{\nu:G[\mu F]}{\beta}{G\; \Fold{F}{\alpha}\;\nu}{t}{Q}{\tau}$.
  \end{enumerate}
\end{lemma}
\begin{proof}
  By structural induction on the given derivation.
  Parts (2), (3), and (4) are mutually recursive.
\end{proof}

\begin{lemma}[Equiv.\ Solves Val-det.]
  \label{lem:equiv-solves-val-det}
  ~
  \begin{enumerate}
  \item
    If $\algequiv[]{\Theta; \Delta}{\F}{\Gee}{W}{\Delta'}$
    and $\ground{\F}$
    and $\judgefunctor{\Theta; \Delta}{\Gee}{\Xi}$,\\
    then for all $(\ahat : \tau) \in \Xi$,
    there exists $t$ such that $\judgeterm{\Theta}{t}{\tau}$
    and $(\hypeq{\ahat}{\tau}{t}) \in \Delta'$.
  \item
    If $\algequiv[+]{\Theta; \Delta}{P}{Q}{W}{\Delta'}$
    and $\ground{P}$
    and $\judgetp{\Theta; \Delta}{Q}{\Xi}$,\\
    then for all $(\ahat : \tau) \in \Xi$,
    there exists $t$ such that $\judgeterm{\Theta}{t}{\tau}$
    and $(\hypeq{\ahat}{\tau}{t}) \in \Delta'$.
  \item
    If $\algequiv[-]{\Theta; \Delta}{M}{N}{W}{\Delta'}$
    and $\ground{N}$
    and $\judgetp{\Theta; \Delta}{M}{\Xi}$,\\
    then for all $(\ahat : \tau) \in \Xi$,
    there exists $t$ such that $\judgeterm{\Theta}{t}{\tau}$
    and $(\hypeq{\ahat}{\tau}{t}) \in \Delta'$.
  \end{enumerate}
\end{lemma}
\begin{proof}
  By structural induction on the given algorithmic equivalence derivation.
  Parts (1) and (2) are mutually recursive.
\end{proof}

\begin{lemma}[Sub.\ Solves Val-det.]
  \label{lem:sub-solves-val-det}
  ~
  \begin{enumerate}
  \item
    If $\algsub[+]{\Theta; \Delta}{P}{Q}{W}{\Delta'}$
    and $\ground{P}$
    and $\judgetp{\Theta; \Delta}{Q}{\Xi}$,\\
    then for all $(\ahat : \tau) \in \Xi$,
    there exists $t$ such that $\judgeterm{\Theta}{t}{\tau}$
    and $(\hypeq{\ahat}{\tau}{t}) \in \Delta'$.
  \item
    If $\algsub[-]{\Theta; \Delta}{M}{N}{W}{\Delta'}$
    and $\ground{N}$
    and $\judgetp{\Theta; \Delta}{M}{\Xi}$,\\
    then for all $(\ahat : \tau) \in \Xi$,
    there exists $t$ such that $\judgeterm{\Theta}{t}{\tau}$
    and $(\hypeq{\ahat}{\tau}{t}) \in \Delta'$.
  \end{enumerate}
\end{lemma}
\begin{proof}
  By structural induction on the given algorithmic subtyping derivation,
  using \Lemmaref{lem:equiv-solves-val-det} as needed.
\end{proof}

For the next lemma and wherever it is used, we introduce the following notation.
Given $\Theta$, input-mode $\Xi$ (has no existential variables) and $\Delta$
such that $\judgetp{\Theta, \Xi; \Delta}{P}{\tilde{\Xi}, \Xi}$,
we define $\extype{\Xi}{P}$ inductively on $\Xi$:
\begin{align*}
  \extype{\cdot}{P} &= P \\
  \extype{\Xi, a:\tau}{P} &= \extype{\Xi} (\extype{a:\tau} P)
\end{align*}
Obviously, given $\judgetp{\Theta, \Xi; \Delta}{P}{\tilde{\Xi}, \Xi}$,
by repeated \AlgTpEx, we have $\judgetp{\Theta; \Delta}{\extype{\Xi} P}{\tilde{\Xi}}$.

\begin{lemma}[Pack Unroll Inversion]
  \label{lem:pack-unroll-inversion}
  ~\\
  If
  \[
    \judgeunroll{\Xi}{\Theta; \Delta}{\nu :(\Const{P}\otimes\hat{P})[\mu F]}{(\clause{q}{t'})}{(\Const{P}\otimes\hat{P})\;\Fold{F}{\alpha}\;\nu}{t}{Q}{\tau}
  \]
  then there exist $\Xi_P$, $\Xi'$, $q'$, $P'$, and $Q'$
  such that $\judgetp{\Theta, \Xi'; \Delta}{P'}{\Xi_P, \Xi'}$
  and $Q = \extype{\Xi'} P' \times Q'$
  and
  \[
    \judgeunroll{\Xi, \Xi'}{\Theta, \Xi'; \Delta}{\nu :\hat{P}[\mu F]}{(\clause{q'}{t'})}{\hat{P}\;\Fold{F}{\alpha}\;\nu}{t}{Q'}{\tau}
  \]
\end{lemma}
\begin{proof}
  By structural induction on the given unrolling derivation.
  We case analyze rules concluding it; there are exactly two possibilities:
  \begin{itemize}
    \item \textbf{Case}
    \[
      \Infer{\AlgUnrollConst}
      {
        \judgeunroll{\Xi}{\Theta; \Delta}{\nu:\hat{P}[\mu F]}{(\clause{q'}{t'})}{\hat{P}\;\Fold{F}{\alpha}\;\nu}{t}{Q'}{\tau}
      }
      {
        \judgeunroll{\Xi}{\Theta; \Delta}{\nu :(\Const{P}\otimes\hat{P})[\mu F]}{(\clause{\underbrace{(\wild, q')}_q}{t'})}{(\Const{P}\otimes\hat{P})\;\Fold{F}{\alpha}\;\nu}{t}{\underbrace{P \times Q'}_Q}{\tau}
      }
    \]
    \begin{llproof}
      \LetPf{\Xi'}{\cdot}
      \LetPf{P'}{P}
      \Hand\judgeunrollPf{\Xi}{\Theta; \Delta}{\nu:\hat{P}[\mu F]}{(\clause{q'}{t'})}{\hat{P}\;\Fold{F}{\alpha}\;\nu}{t}{Q'}{\tau}{Subderivation}
      \judgefunctorPf{\Theta; \Delta}{\Const{P} \otimes \hat{P}}{\dontcare}{Presupposed derivation}
      \judgetpPf{\Theta; \Delta}{P}{\Xi_P}{By inversion}
      \Hand\judgetpPf{\Theta, \Xi'; \Delta}{P'}{\Xi_P, \Xi'}{By list append property}
      \trailingjust{(and equations)}
    \end{llproof} 

    \item \textbf{Case} \AlgUnrollConstEx
    \[
      \Infer{}
      {
        \judgeunroll{\Xi,a:\tau'}{\Theta,a:\tau'; \Delta}{\nu:(\Const{P_0}\otimes\hat{P})[\mu F]}{(\clause{(\bap,q_0)}{t'})}{(\Const{P_0}\otimes\hat{P})\;\Fold{F}{\alpha}\;\nu}{t}{Q_0}{\tau}
      }
      {
        \judgeunroll*{\Xi}{\Theta; \Delta}{\nu :(\Const{\overbrace{\extype{a:\tau'}{P_0}}^P}\otimes\hat{P})[\mu F]}{(\clause{\underbrace{(\pack{a}{\bap}, q_0)}_q}{t'})}{(\Const{\extype{a:\tau'}{P_0}}\otimes\hat{P})\;\Fold{F}{\alpha}\;\nu}{t}{\underbrace{\extype{a:\tau'}{Q_0}}_Q}{\tau}
      }
    \]
    \begin{llproof}
      \judgefunctorPf{\Theta; \Delta}{\Const{\extype{a:\tau'} P_0} \otimes \hat{P}}{\dontcare}{Presupposed derivation}
      \judgefunctorPf{\Theta; \Delta}{\Const{\extype{a:\tau'} P_0}}{\tilde{\Xi}}{By inversion on \AlgFunctorProd}
      \judgetpPf{\Theta; \Delta}{\extype{a:\tau'} P_0}{\tilde{\Xi}}{By inversion on \AlgFunctorConst}
      \judgetpPf{\Theta, a:\tau'; \Delta}{P_0}{\tilde{\Xi}, a:\tau'}{By inversion on \AlgTpEx}
    \end{llproof} 

    By the \ih with the above unrolling subderivation,
    there exist $\Xi_0'$, $q'$, $P'$, and $Q'$ such that
    \[
      \judgeunroll{\Xi,a:\tau',\Xi_0'}{\Theta,a:\tau',\Xi_0'; \Delta}{\nu:\hat{P}[\mu F]}{(\clause{q'}{t'})}{\hat{P}\;\Fold{F}{\alpha}\;\nu}{t}{Q'}{\tau}
    \]
    and $\judgetp{\Theta, a:\tau', \Xi_0'; \Delta}{P'}{\tilde{\Xi}, a:\tau', \Xi_0'}$
    and $Q_0 = \extype{\Xi_0'} P' \times Q'$.\\
    \begin{llproof}
      \LetPf{\Xi'}{a:\tau', \Xi_0'}
      \Pf{Q}{=}{\extype{a:\tau'} Q_0}{Current case}
      \Pf{}{=}{\extype{a:\tau'} (\extype{\Xi_0'} P' \times Q')}{By equation}
      \Pf{}{=}{\extype{\Xi'} P' \times Q'}{Straightforward}
    \end{llproof} 
    \qedhere
  \end{itemize}
\end{proof}

\begin{lemma}
  \label{lem:aux-inversion}
  If $\Xi$ is a list of sort declarations
  $a_1:\tau_1, \dots, a_n:\tau_n$ (where $n \geq 0$)\\
  and $\algchk{\Theta; \Delta}{\Gamma}{v}{\extype{\Xi} P}{\chi}{\Delta'}$
  then there exist $\Delta_\Xi$ and $\Omega_\Xi$\\
  such that
  $\algchk{\Theta; \Delta, \Delta_\Xi}{\Gamma}{v}{[\Delta_\Xi/\Xi]P}{\chi}{\Delta', \Omega_{\Xi}}$
\end{lemma}
\begin{proof}
  By induction on the number of entries in $\Xi$.
  If $\Xi = \cdot$, put $\Delta_\Xi = \cdot = \Omega_\Xi$.
  If $\Xi = a_1:\tau_1, \Xi_0$,
  then only \AlgChkValExists can derive the given value typing
  (\AlgChkValVar requires the type not to be $\exists$);
  the goal is attained by inversion on \AlgChkValExists,
  using the \ih on the resulting value typing subderivation
  ($\Xi_0$ is one less in length),
  and using substitution properties,
  following \Defnref{def:evar-rename}.
\end{proof}

\begin{lemma}[Equiv.\ Solves Unrolled Evar.]
  \label{lem:equiv-solves-unrolled-evar}
  Suppose $\Delta = \Delta_1, \ahat:\tau, \Delta_2$.\\
  If $\judgeunroll{\Xi}{\Theta; \Delta}{ \nu:G[\mu F] }{\beta}{G\;\Fold{F}{\alpha}\;\nu}{\ahat}{Q}{\tau}$
  and $\algequiv[+]{\Theta; \Delta}{P}{Q}{W}{\Delta'}$
  and $\ground{P}$
  and $\rextend{\Theta}{\Delta'}{\Omega}$
  and $\semideclentailwah{\Theta}{[\Omega]W}$,
  then there exists $t$ such that $\judgeterm{\Theta}{t}{\tau}$
  and $(\hypeq{\ahat}{\tau}{t}) \in \Delta'$.
\end{lemma}
\begin{proof}
  By structural induction on the given algorithmic type equivalence derivation.
  We elide the easy groundness checks when using the \ih.
  \begin{itemize}
    \DerivationProofCase{\AlgUnrollSum}
    {\arrayenvbl{
        \composeinj{1}{\beta}{\beta_1}
        \\
        \composeinj{2}{\beta}{\beta_2}
      }
      \\
      \arrayenvbl{
        \judgeunroll{\Xi}{\Theta; \Delta}{\nu:G_1[\mu F]}{\beta_1}{G_1\;\Fold{F}{\alpha}\;\nu}{\ahat}{Q_1}{\tau}
        \\
        \judgeunroll{\Xi}{\Theta; \Delta}{\nu:G_2[\mu F]}{\beta_2}{G_2\;\Fold{F}{\alpha}\;\nu}{\ahat}{Q_2}{\tau}
      }
    }
    { \judgeunroll{\Xi}{\Theta; \Delta}{\nu:(G_1 \oplus G_2)[\mu F]}{\beta}{(G_1 \oplus G_2)\;\Fold{F}{\alpha}\;\nu}{\ahat}{Q_1 + Q_2}{\tau} }
    \begin{itemize}
      \DerivationProofCase{\AlgTpEquivPosSum}
      {
        \algequiv[+]{\Theta; \Delta}{P_1}{Q_1}{W_1}{\Delta''}
        \\
        \algequiv[+]{\Theta; \Delta''}{P_2}{[\Delta'']Q_2}{W_2}{\Delta'}
      }
      {
        \algequiv[+]{\Theta; \Delta}{P_1 + P_2}{Q_1 + Q_2}{[\Delta']W_1 \land W_2}{\Delta'}
      }
      \begin{llproof}
        \algequivPf[+]{\Theta; \Delta''}{P_2}{[\Delta'']Q_2}{W_2}{\Delta'}{Subderivation}
        \extendPf{\Theta}{\Delta''}{\Delta'}{By \Lemmaref{lem:alg-equiv-extends}}
        \rextendPf{\Theta}{\Delta''}{\Delta'}{By \Lemmaref{lem:extension-sound}}
        \rextendPf{\Theta}{\Delta'}{\Omega}{Given}
        \rextendPf{\Theta}{\Delta''}{\Omega}{By \Lemmaref{lem:ext-trans}}
        \semideclentailwahPf{\Theta}{[\Omega]([\Delta']W_1 \land W_2)}{Given}
        \semideclentailwahPf{\Theta}{[\Omega]([\Delta']W_1) \land [\Omega]W_2}{By \defn of \defsubst}
        \semideclentailwahPf{\Theta}{[\Omega]([\Delta']W_1)}{By inversion}
        \wahequivPf{\Theta}{[\Omega]W_1}{[\Omega]([\Delta']W_1)}{By \Lemmaref{lem:wah-sandwich}}
        \wahequivPf{\Theta}{[\Omega]([\Delta']W_1)}{[\Omega]W_1}{By \Lemmaref{lem:wah-equiv-symmetric}}
        \semideclentailwahPf{\Theta}{[\Omega]W_1}{By \Lemmaref{lem:equiv-respects-entail}}
        \decolumnizePf
        \algequivPf[+]{\Theta; \Delta}{P_1}{Q_1}{W_1}{\Delta''}{Subderivation}
        \judgeunrollPf{\Xi}{\Theta; \Delta}{\nu:G_1[\mu F]}{\beta_1}{G_1\;\Fold{F}{\alpha}\;\nu}{\ahat}{Q_1}{\tau}{Subderivation}
        \inPf{\hypeq{\ahat}{\tau}{t}}{\Delta''}{By \ih}
        \Hand\judgetermPf{\Theta}{t}{\tau}{\ditto}
        \Hand\inPf{\hypeq{\ahat}{\tau}{t}}{\Delta'}{By inversion on extension}
      \end{llproof} 
    \end{itemize}
    \DerivationProofCase{\AlgUnrollUnit}
    { }
    { \judgeunroll{\Xi}{\Theta; \Delta}{\nu:I[\mu F]}{(\clause{\unitexp}{t_0})}{I\;\Fold{F}{\alpha}\;\nu}{\ahat}{1 \land (\ahat = t_0)}{\tau} }
    \begin{itemize}
      \DerivationProofCase{\AlgTpEquivPosWith}
      {
        \algequiv[+]{\Theta; \Delta}{P_0}{1}{\dontcare}{\Delta''}
        \\
        \algpropequivinst{\Theta; \Delta''}{\phi}{[\Delta''](\ahat = t_0)}{\Delta'}
      }
      {
        \algequiv[+]{\Theta; \Delta}{P_0 \land \phi}{1 \land \ahat = t_0}{\dontcare \land (\phi \equiv [\Delta'](\ahat = t_0))}{\Delta'}
      }
      \begin{llproof}
        \semideclentailwahPf{\Theta}{[\Omega](\dontcare \land (\phi \equiv [\Delta'](\ahat = t_0)))}{Given}
        \semideclentailwahPf{\Theta}{[\Omega]\dontcare \land ([\Omega]\phi \equiv ([\Omega]([\Delta']\ahat) = [\Omega]([\Delta']t_0)))}{By \defn of \defsubst}
        \semideclentailwahPf{\Theta}{[\Omega]\phi \equiv ([\Omega]([\Delta']\ahat) = [\Omega]([\Delta']t_0))}{By inversion on \SemiDeclWTrueAnd}
        \judgeequivPf[]{\Theta}{[\Omega]\phi}{([\Omega]([\Delta']\ahat) = [\Omega]([\Delta']t_0))}{By inversion on \SemiDeclWTruePrpEquiv}
        \judgeequivPf[]{\Theta}{\phi}{([\Omega]([\Delta']\ahat) = [\Omega]([\Delta']t_0))}{Because $\ground{\phi}$}
        \eqPf{\phi}{(t = t')}{By inversion on \PrpEquivEq}
        \algequivPf[+]{\Theta; \Delta}{P_0}{1}{}{\Delta''}{Subderivation}
        \eqPf{\Delta''}{\Delta}{By inversion on \AlgTpEquivPosUnit}
        \eqPf{[\Delta''](\ahat = t_0)}{[\Delta](\ahat = t_0)}{By equality}
        \eqPf{}{(\ahat = [\Delta]t_0)}{By \defn of \defsubst}
        \trailingjust{(and $\because\ahat$ unsolved in $\Delta$)}
        \decolumnizePf
        \algpropequivinstPf{\Theta; \Delta_1, \ahat:\tau, \Delta_2}{(t = t')}{(\ahat = [\Delta]t_0)}{\Delta'}{Rewrite subderivation}
        \Hand\judgetermPf{\Theta}{t}{\tau}{By inversion on \PropEquivInst}
        \Hand\eqPf{\Delta'}{\Delta_1, \hypeq{\ahat}{\tau}{t}, \Delta_2}{\ditto}
      \end{llproof} 
    \end{itemize}
    \item \textbf{Case}
      \[
        \Infer{\AlgUnrollId}
        {
          \unroll{\Xi,a':\tau}{\Theta, a':\tau; \Delta}{\hat{P}}{F}{\clause{q}{t'}}{\alpha}{\tau}{\ahat}{Q_0} 
        }
        {
          \unroll{\Xi}{\Theta; \Delta}{\Id\otimes\hat{P}}{F}{\clause{(a',q)}{t'}}{\alpha}{\tau}{\ahat}{\extype{a':\tau}{\comprehend{\nu:\mu F}{ \Fold{F}{\alpha}\,{\nu} =_\tau a' } \times Q_0}}
        }
      \]
    \begin{itemize}
      \DerivationProofCase{\AlgTpEquivPosEx}
      {
        \algequiv[+]{\Theta, a':\tau; \Delta}{P_0}{\comprehend{\nu:\mu F}{ \Fold{F}{\alpha}\,{\nu} =_\tau a' } \times Q_0}{W_0}{\Delta'}
      }
      {
        \algequiv[+]{\Theta; \Delta}{\extype{a':\tau}{P_0}}{\extype{a':\tau}{\comprehend{\nu:\mu F}{ \Fold{F}{\alpha}\,{\nu} =_\tau a' } \times Q_0}}{\alltype{a':\tau}{W_0}}{\Delta'}
      }
      \begin{llproof}
        \semideclentailwahPf{\Theta}{[\Omega](\alltype{a':\tau}{W_0})}{Given}
        \semideclentailwahPf{\Theta}{(\alltype{a':\tau}{[\Omega]W_0})}{By \defn of \defsubst}
        \semideclentailwahPf{\Theta, a':\tau}{({[\Omega]W_0})}{By inversion}
        \rextendPf{\Theta}{\Delta'}{\Omega}{Given}
        \eqPf{P_0}{P_1 \times P_2}{By inversion on \AlgTpEquivPosProd}
        \eqPf{W_0}{\dontcare \land W_0'}{\ditto}
        \algequivPf[+]{\Theta, a':\tau; \Delta}{P_1}{\comprehend{\nu:\mu F}{ \Fold{F}{\alpha}\,{\nu} =_\tau a' }}{\dontcare}{\Delta''}{\ditto}
        \algequivPf[+]{\Theta, a':\tau; \Delta''}{P_2}{Q_0}{W_0'}{\Delta'}{\ditto}
        \semideclentailwahPf{\Theta, a':\tau}{\dontcare \land [\Omega]W_0'}{By equality and \defn of \defsubst}
        \semideclentailwahPf{\Theta, a':\tau}{[\Omega]W_0'}{By inversion}
      \end{llproof}
      ~\\
      If $\ahat$ is solved in $\Delta''$,
      then we are done by \Lemmaref{lem:alg-equiv-extends} with the second subderivation.
      If $\ahat$ is not solved in $\Delta''$,
      then the goals follow by the \ih with subderivation
      \[
        \unroll{\Xi,a':\tau}{\Theta, a':\tau; \Delta}{\hat{P}}{F}{\clause{q}{t'}}{\alpha}{\tau}{\ahat}{Q_0}
      \]
    \end{itemize}
    \DerivationProofCase{\AlgUnrollConstEx}
    {
      \unroll{\Xi,b:\tau'}{\Theta,b:\tau'; \Delta}{\Const{Q}\otimes\hat{P}}{F}{\clause{(\bap,q)}{t'}}{\alpha}{\tau}{\ahat}{Q_0}
    }
    {
      \unroll{\Xi}{\Theta; \Delta}{\Const{\extype{b:\tau'}{Q}}\otimes\hat{P}}{F}{\clause{(\pack{a}{\bap}, q)}{t'}}{\alpha}{\tau}{\ahat}{\extype{b:\tau'}{Q_0}}
    }
    \begin{itemize}
      \DerivationProofCase{\AlgTpEquivPosEx}
      {
        \algequiv[+]{\Theta, b:\tau'; \Delta}{P_0}{Q_0}{W_0}{\Delta'}
      }
      {
        \algequiv[+]{\Theta; \Delta}{\extype{b:\tau'}{P_0}}{\extype{b:\tau'} Q_0}{\alltype{b:\tau'}{W_0}}{\Delta'}
      }
      \begin{llproof}
        \algequivPf[+]{\Theta, b:\tau'; \Delta}{P_0}{Q_0}{W_0}{\Delta'}{Subderivation}
        \semideclentailwahPf{\Theta}{[\Omega](\alltype{b:\tau'}{W_0})}{Given}
        \semideclentailwahPf{\Theta}{(\alltype{b:\tau'}{[\Omega]W_0})}{By \defn of \defsubst}
        \semideclentailwahPf{\Theta, b:\tau'}{({[\Omega]W_0})}{By inversion}
        \rextendPf{\Theta}{\Delta'}{\Omega}{Given}
        \unrollPf{\Xi,b:\tau';\Theta,b:\tau'; \Delta}{\Const{Q}\otimes\hat{P}}{F}{\clause{(\bap,q)}{t'}}{\alpha}{\tau}{\ahat}{Q_0}{Subderivation}
        \Hand\inPf{\hypeq{\ahat}{\tau}{t}}{\Delta'}{By \ih}
        \Hand\judgetermPf{\Theta}{t}{\tau}{\ditto}
      \end{llproof} 
    \end{itemize}
    \DerivationProofCase{\AlgUnrollConst}
    {
      \unroll{\Xi}{\Theta; \Delta}{\hat{P}}{F}{\clause{q}{t'}}{\alpha}{\tau}{\ahat}{Q_2}
    }
    {
      \unroll{\Xi}{\Theta; \Delta}{\Const{Q_1}\otimes\hat{P}}{F}{\clause{(\wild,q)}{t'}}{\alpha}{\tau}{\ahat}{Q_1 \times Q_2}
    }
    \begin{itemize}
      \DerivationProofCase{\AlgTpEquivPosProd}
      {
        \algequiv[+]{\Theta; \Delta}{P_1}{Q_1}{W_1}{\Delta''}
        \\
        \algequiv[+]{\Theta; \Delta''}{P_2}{[\Delta'']Q_2}{W_2}{\Delta'}
      }
      {
        \algequiv[+]{\Theta; \Delta}{P_1 \times P_2}{Q_1 \times {Q_2}}{[\Delta']W_1 \land W_2}{\Delta'}
      }
    \end{itemize}
    If $\ahat$ is solved in $\Delta''$,
    then we are done by \Lemmaref{lem:alg-equiv-extends} with the second subderivation.
    Suppose $\ahat$ is not solved in $\Delta''$.\\
    \begin{llproof}
      \semideclentailwahPf{\Theta}{[\Omega](\dontcare \land W_2)}{Given}
      \semideclentailwahPf{\Theta}{[\Omega]\dontcare \land [\Omega]W_2}{By \defn of \defsubst}
      \semideclentailwahPf{\Theta}{[\Omega]W_2}{By inversion}
      \rextendPf{\Theta}{\Delta'}{\Omega}{Given}
      \algequivPf[+]{\Theta; \Delta}{P_1}{Q_1}{W_1}{\Delta''}{Subderivation}
      \extendPf{\Theta}{\Delta}{\Delta''}{By \Lemmaref{lem:alg-equiv-extends}}
      \decolumnizePf
      \unrollPf{\Xi;\Theta; \Delta}{\hat{P}}{F}{\clause{q}{t'}}{\alpha}{\tau}{\ahat}{Q_2}{Subderivation}
      \unrollPf{\Xi;\Theta; \Delta''}{\hat{P}}{F}{\clause{q}{t'}}{\alpha}{\tau}{\ahat}{Q_2}{By \Lemmaref{lem:ext-weak-unroll}}
      \decolumnizePf
      \unrollPf{\Xi;\Theta; \Delta''}{[\Delta'']\hat{P}}{[\Delta'']F}{(\clause{q}{t'})}{\alpha}{\tau}{[\Delta'']\ahat}{[\Delta'']Q_2}{By \Lemref{lem:right-hand-subst-unroll}}
      \unrollPf{\Xi;\Theta; \Delta''}{[\Delta'']\hat{P}}{[\Delta'']F}{(\clause{q}{t'})}{\alpha}{\tau}{\ahat}{[\Delta'']Q_2}{$\ahat$ unsolved in $\Delta''$}
      \algequivPf[+]{\Theta; \Delta''}{P_2}{[\Delta'']Q_2}{W_2}{\Delta'}{Subderivation}
      \Hand\inPf{\hypeq{\ahat}{\tau}{t}}{\Delta'}{By \ih}
      \Hand\judgetermPf{\Theta}{t}{\tau}{\ditto}
    \end{llproof} 
    \qedhere
  \end{itemize}
\end{proof}

\begin{lemma}
  \label{lem:aux-inversion-sub}
  If $\Xi$ is a list of sort declarations
  $a_1:\tau_1, \dots, a_n:\tau_n$ (where $n \geq 0$)\\
  and $\algsub[+]{\Theta; \Delta}{P}{\extype{\Xi} Q}{W}{\Delta'}$
  and $\simple{\Theta}{P}$,\\
  then there exist $\Delta_\Xi$ and $\Omega_\Xi$
  such that
  $\algsub{\Theta; \Delta, \Delta_\Xi}{P}{[\Delta_\Xi/\Xi]Q}{W}{\Delta', \Omega_\Xi}$.
\end{lemma}
\begin{proof}
  Because $\simple{\Theta}{P}$, we know $P \neq \exists, \land$,
  and we can get the result by inversion on \AlgSubPosExR (as needed),
  similarly to the proof of \Lemref{lem:aux-inversion},
  which inverts \AlgChkValExists (as needed).
\end{proof}

\begin{lemma}[Sub.\ Solves Unrolled Evar.]
  \label{lem:sub-solves-unrolled-evar}
  Suppose $\Delta = \Delta_1, \ahat:\tau, \Delta_2$.\\
  If $\judgeunroll{\Xi}{\Theta; \Delta}{ \nu:\hat{P}[\mu F] }{\beta}{\hat{P}\;\Fold{F}{\alpha}\;\nu}{\ahat}{Q}{\tau}$
  and $\algsub[+]{\Theta; \Delta}{P}{Q}{W}{\Delta'}$\\
  and $\ground{P}$
  and $\simple{\Theta}{P}$
  then there exists $t$ such that $\judgeterm{\Theta}{t}{\tau}$
  and $(\hypeq{\ahat}{\tau}{t}) \in \Delta'$.
\end{lemma}
\begin{proof}
  By induction on the size (\Figureref{fig:size}) of $Q$.
  We case analyze $\hat{P}$.
  For all cases of $\hat{P}$ except $\Const{P} \otimes \hat{P}$,
  there is exactly one rule that can conclude the given unrolling derivation;
  we write each such case as the given unrolling derivation specialized
  to that case.
  \begin{itemize}
    \item \textbf{Case}
    \[
      \Infer{\AlgUnrollId}
      {
        \unroll{\Xi,a':\tau}{\Theta, a':\tau; \Delta}{\hat{P_0}}{F}{\clause{q}{\tilde{t}}}{\alpha}{\tau}{\ahat}{Q_0} 
      }
      {
        \unroll{\Xi}{\Theta; \Delta}{\Id\otimes\hat{P_0}}{F}{\clause{(a',q)}{\tilde{t}}}{\alpha}{\tau}{\ahat}{\extype{a':\tau}{\comprehend{\nu:\mu F}{ \Fold{F}{\alpha}\,{\nu} =_\tau a' } \times Q_0}}
      }
    \]
    \begin{itemize}
      \DerivationProofCase{\AlgSubPosExR}
      { 
        \algsub[+]{\Theta; \Delta, \ahat':\tau}{P}{{\comprehend{\nu:\mu F}{ \Fold{F}{\alpha}\,{\nu} =_\tau \ahat' } \times [\ahat'/a']Q_0}}{\Wah}{\Delta', \hypeq{\ahat'}{\tau}{t'}}
      }
      {
        \algsub[+]{\Theta; \Delta}{P}{\extype{a':\tau}{\comprehend{\nu:\mu F}{ \Fold{F}{\alpha}\,{\nu} =_\tau a' } \times Q_0}}{\Wah}{\Delta'}
      }
      \begin{llproof}
        \algsubPf[+]{\Theta; \Delta, \ahat':\tau}{P}{{\comprehend{\nu:\mu F}{ \Fold{F}{\alpha}\,{\nu} =_\tau \ahat' } \times [\ahat'/a']Q_0}}{\Wah}{\Delta', \hypeq{\ahat'}{\tau}{t'}}{Subderivation}
        \decolumnizePf
        \simplePf{\Theta}{P}{Given}
        \neqPf{P}{\exists, \land}{Straightforward}
        \eqPf{P}{P' \times P_0}{By inversion on \AlgSubPosProd}
        \algsubPf[+]{\Theta; \Delta, \ahat':\tau}{P'}{\comprehend{\nu:\mu F}{ \Fold{F}{\alpha}\,{\nu} =_\tau \ahat' }}{\dontcare}{\Delta''}{\ditto}
        \algsubPf[+]{\Theta; \Delta''}{P_0}{[\Delta'']([\ahat'/a']Q_0)}{\dontcare}{\Delta', \hypeq{\ahat'}{\tau}{t'}}{\ditto}
        \Pf{}{}{}{}
        \judgefunctorPf{\Theta; \Delta}{F}{\Xi_F}{Presupposed derivation}
        \judgealgebraPf{\Xi}{\Theta; \Delta}{\alpha}{F}{\tau}{Presupposed derivation}
        \judgefunctorPf{\Theta; \Delta, \ahat':\tau}{F}{\Xi_F}{\Lemmaref{lem:alg-ctx-weakening}}
        \judgealgebraPf{\Xi}{\Theta; \Delta, \ahat':\tau}{\alpha}{F}{\tau}{\Lemmaref{lem:alg-ctx-weakening}}
        \decolumnizePf
        \judgetpPf{\Theta; \Delta, \ahat':\tau}{\comprehend{\nu:\mu F}{ \Fold{F}{\alpha}\,{\nu} =_\tau \ahat'}}{\Xi_F \cup \ahat':\tau}{By \AlgTpFixEVar}
        \judgetpPf{\Theta; \Delta, \ahat':\tau}{\comprehend{\nu:\mu F}{ \Fold{F}{\alpha}\,{\nu} =_\tau \ahat'}}{\Xi_F, \ahat':\tau}{$\ahat'$ not free in $F$}
        \groundPf{P'}{Because given $\ground{P' \times P_0}$}
      \end{llproof} 

      By \Lemmaref{lem:sub-solves-val-det},
      $\Delta''$ solves all existential variables in $(\Xi_F, \ahat':\tau)$.
      Therefore, there exist $\Delta_0''$, $\Delta_1''$ and $t_0$
      such that $\Delta'' = \Delta_0'', \hypeq{\ahat'}{\tau}{t_0}, \Delta_1''$.

      \begin{llproof}
        \extendPf{\Theta}{\Delta''}{\Delta', \hypeq{\ahat'}{\tau}{t'}}{By \Lemmaref{lem:alg-sub-extends}}
        \decolumnizePf
        \eqPf{\Delta_1''}{\cdot}{By inversion}
        \eqPf{t_0}{t'}{\ditto}
        \extendPf{\Theta}{\Delta_0''}{\Delta'}{\ditto}
        \algsubPf[+]{\Theta; \Delta''}{P_0}{[\Delta_0'', \hypeq{\ahat'}{\tau}{t'}]([\ahat'/a']Q_0)}{\dontcare}{\Delta', \hypeq{\ahat'}{\tau}{t'}}{Rewrite above}
        \algsubPf[+]{\Theta; \Delta''}{P_0}{[\Delta_0'']([t'/a']Q_0)}{\dontcare}{\Delta', \hypeq{\ahat'}{\tau}{t'}}{By subst.\ properties}
        \decolumnizePf
        \unrollPf{\Xi,a':\tau; \Theta, a':\tau; \Delta}{\hat{P_0}}{F}{\clause{q}{\tilde{t}}}{\alpha}{\tau}{\ahat}{Q_0}{Subderivation}
        \unrollPf{\dontcare; \Theta; \Delta}{\hat{P_0}}{F}{\clause{q}{[t'/a']\tilde{t}}}{\alpha}{\tau}{\ahat}{[t'/a']Q_0}{By \Lemref{lem:syn-subs-unroll}}
        \trailingjust{(alg.\ version)}
        \extendPf{\Theta}{\Delta, \ahat':\tau}{\Delta_0'', \hypeq{\ahat'}{\tau}{t'}}{By \Lemref{lem:alg-sub-extends}}
        \extendPf{\Theta}{\Delta}{\Delta_0''}{By inversion}
        \unrollPf{\dontcare; \Theta; \Delta_0''}{\hat{P_0}}{F}{\clause{q}{[t'/a']\tilde{t}}}{\alpha}{\tau}{\ahat}{[t'/a']Q_0}{By \Lemref{lem:ext-weak-unroll}}
      \end{llproof}

      By \Lemmaref{lem:right-hand-subst-unroll},
      \[
        \unroll{\dontcare}{\Theta; \Delta_0''}{[\Delta_0'']\hat{P_0}}{[\Delta_0'']F}{\clause{q}{[t'/a']\tilde{t}}}{\alpha}{\tau}{[\Delta_0'']\ahat}{[\Delta_0'']([t'/a']Q_0)}
      \]
      By \Lemmaref{lem:alg-ctx-weakening},
      \[
        \unroll{\dontcare}{\Theta; \Delta''}{[\Delta_0'']\hat{P_0}}{[\Delta_0'']F}{\clause{q}{[t'/a']\tilde{t}}}{\alpha}{\tau}{[\Delta_0'']\ahat}{[\Delta_0'']([t'/a']Q_0)}
      \]
      If $\ahat$ is solved in $\Delta''$,
      then we are done by inversion on extension
      and by the fact that $\ahat \neq \ahat'$.
      If $\ahat$ is not solved in $\Delta''$,
      then the goals hold by the \ih with the relevant information above
      (noting that in this case $[\Delta_0'']\ahat$ is $\ahat$).
    \end{itemize}
    \item \textbf{Case} $\hat{P} = \Const{P} \otimes \hat{P_0}$\\
      \begin{llproof}
        \unrollPf{\Xi; \Theta; \Delta}{(\Const{P} \otimes \hat{P_0})}{F}{\beta}{\alpha}{\tau}{\ahat}{Q}{Given}
        \Pf{\beta}{=}{(\clause{q}{t'})}{By inversion}
        \unrollPf{\Xi; \Theta; \Delta}{(\Const{P} \otimes \hat{P_0})}{F}{(\clause{q}{t'})}{\alpha}{\tau}{\ahat}{Q}{By equality}
        \unrollPf{\Xi, \Xi'; \Theta, \Xi'; \Delta}{\hat{P_0}}{F}{(\clause{q'}{t'})}{\alpha}{\tau}{\ahat}{Q'}{By \Lemref{lem:pack-unroll-inversion}}
        \judgetpPf{\Theta, \Xi'; \Delta}{P'}{\dontcare, \Xi'}{\ditto}
        \Pf{Q}{=}{\extype{\Xi'} P' \times Q'}{\ditto}
        \proofsep
        \algsubPf[+]{\Theta; \Delta}{P}{\extype{\Xi'} P' \times Q'}{W}{\Delta'}{Rewrite given}
        \simplePf{\Theta}{P}{Given}
        \algsubPf[+]{\Theta; \Delta, \Delta_{\Xi'}}{P}{[\Delta_{\Xi'}/\Xi'](P' \times Q')}{\dontcare}{\Delta', \Omega_{\Xi'}}{By \Lemref{lem:aux-inversion-sub}}
        \algsubPf[+]{\Theta; \Delta, \Delta_{\Xi'}}{P}{[\Delta_{\Xi'}/\Xi']P' \times [\Delta_{\Xi'}/\Xi']Q'}{\dontcare}{\Delta', \Omega_{\Xi'}}{By \defn of \defsubst}
        \decolumnizePf
        \eqPf{P}{P_1 \times P_2}{By inversion on \AlgSubPosProd}
        \algsubPf[+]{\Theta; \Delta, \Delta_{\Xi'}}{P_1}{[\Delta_{\Xi'}/\Xi']P'}{\dontcare}{\Delta''}{\ditto}
        \algsubPf[+]{\Theta; \Delta''}{P_2}{[\Delta'']([\Delta_{\Xi'}/\Xi']Q')}{\dontcare}{\Delta', \Omega_{\Xi'}}{\ditto}
        \judgetpPf{\Theta, \Xi'; \Delta}{P'}{\dontcare, \Xi'}{Above}
        \groundPf{P_1 \times P_2}{Given}
        \groundPf{P_1}{Straightforward}
        \decolumnizePf
        \judgetpPf{\Theta; \Delta, \Delta_{\Xi'}}{[\Delta_{\Xi'}/\Xi']P'}{\dontcare, \Delta_{\Xi'}}{By \Lemmaref{lem:value-determined-evar-rename}}
        \ForallPf{(\ahat:\tau)}{\Delta_{\Xi'},~\text{there exists}~(\hypeq{\ahat}{\tau}{t}) \in \Delta''}{By \Lemmaref{lem:sub-solves-val-det}}
        \extendPf{\Theta}{\Delta''}{\Delta', \Omega_{\Xi'}}{By \Lemmaref{lem:typing-extends}}
        \eqPf{\Delta''}{\Delta_0'', \Omega_{\Xi'}}{By inversion}
        \extendPf{\Theta}{\Delta_0''}{\Delta'}{\ditto}
        \extendPf{\Theta}{\Delta, \Delta_{\Xi'}}{\Delta_0'', \Omega_{\Xi'}}{By \Lemmaref{lem:typing-extends}}
        \extendPf{\Theta}{\Delta}{\Delta_0''}{By inversion}
        \extendPf{\Theta}{\Delta_{\Xi'}}{\Omega_{\Xi'}}{\ditto}
        \decolumnizePf
        \unrollPf{\Xi, \Xi'; \Theta, \Xi'; \Delta}{\hat{P_0}}{F}{(\clause{q'}{t'})}{\alpha}{\tau}{\ahat}{Q'}{Above}
        \unrollPf{\dontcare; \Theta; \Delta}{\hat{P_0}}{F}{(\clause{q'}{[\Omega_{\Xi'}/\Xi']t'})}{\alpha}{\tau}{\ahat}{[\Omega_{\Xi'}/\Xi']Q'}{By \Lemref{lem:syn-subs-unroll}}
        \trailingjust{(alg.\ version)}
        \unrollPf{\dontcare; \Theta; \Delta_0''}{\hat{P_0}}{F}{(\clause{q'}{[\Omega_{\Xi'}/\Xi']t'})}{\alpha}{\tau}{\ahat}{[\Omega_{\Xi'}/\Xi']Q'}{By \Lemref{lem:ext-weak-unroll}}
      \end{llproof}
      ~\\
      By \Lemmaref{lem:right-hand-subst-unroll} and \defn of \defsubst and $\because \ground{\alpha}$,
      \[
        \unroll{\dontcare}{\Theta; \Delta_0''}{[\Delta_0'']\hat{P_0}}{[\Delta_0'']F}{(\clause{q'}{[\Omega_{\Xi'}/\Xi']t'})}{\alpha}{\tau}{[\Delta_0'']\ahat}{[\Delta_0'']([\Omega_{\Xi'}/\Xi']Q')}
      \]
      By \Lemmaref{lem:alg-ctx-weakening},
      \[
        \unroll{\dontcare}{\Theta; \Delta''}{[\Delta_0'']\hat{P_0}}{[\Delta_0'']F}{(\clause{q'}{[\Omega_{\Xi'}/\Xi']t'})}{\alpha}{\tau}{[\Delta_0'']\ahat}{[\Delta_0'']([\Omega_{\Xi'}/\Xi']Q')}
      \]
      \begin{llproof}
        \algsubPf[+]{\Theta; \Delta''}{P_2}{[\Delta_0'', \Omega_{\Xi'}]([\Delta_{\Xi'}/\Xi']Q')}{\dontcare}{\Delta', \Omega_{\Xi'}}{Rewrite above}
        \algsubPf[+]{\Theta; \Delta''}{P_2}{[\Delta_0'']([\Omega_{\Xi'}/\Xi']Q')}{\dontcare}{\Delta', \Omega_{\Xi'}}{By subst.\ properties (using \Defnref{def:solve-evar-rename})}
      \end{llproof}
      ~\\
      If $\ahat$ is solved in $\Delta''$,
      then by inversion on 
      $\extend{\Theta}{\Delta''}{(\Delta', \Omega_{\Xi'})}$
      it is solved in $\Delta'$ (as desired).
      Suppose $\ahat$ is unsolved in $\Delta''$;
      then it is unsolved in $\Delta_0''$.\\
      \begin{llproof}
        \eqPf{[\Delta_0'']\ahat}{\ahat}{Because $\ahat$ unsolved in $\Delta_0''$}
        \decolumnizePf
        \unrollPf{\dontcare; \Theta; \Delta''}{[\Delta_0'']\hat{P_0}}{[\Delta_0'']F}{(\clause{q'}{[\Omega_{\Xi'}/\Xi']t'})}{\alpha}{\tau}{\ahat}{[\Delta_0'']([\Omega_{\Xi'}/\Xi']Q')}{Above}
        \trailingjust{(rewritten)}
        \decolumnizePf
        \simplePf{\Theta}{P_2}{Straightforward}
        \inPf{(\hypeq{\ahat}{\tau}{t})}{(\Delta', \Omega_{\Xi'})}{By \ih}
        \Hand\judgetermPf{\Theta}{t}{\tau}{\ditto}
        \Hand\inPf{(\hypeq{\ahat}{\tau}{t})}{\Delta'}{$\ahat \notin \dom{\Omega_{\Xi'}}$}
      \end{llproof} 
    \DerivationProofCase{\AlgUnrollUnit}
    { }
    {
      \unroll{\Xi}{\Theta; \Delta}{I}{F}{\clause{\unitexp}{t}}{\alpha}{\tau}{\ahat}{1 \land (\ahat = t)}
    }
    \begin{itemize}
      \DerivationProofCase{\AlgSubPosWithR}
      {
        \algsub[+]{\Theta; \Delta}{P}{1}{\Wah_0}{\Delta''}
        \\
        \alginst{\Theta; \Delta''}{[\Delta''](\ahat = t)}{\Delta'}
      }
      {
        \algsub[+]{\Theta; \Delta}{P}{1 \land (\ahat = t)}{[\Delta']\Wah_0 \land [\Delta'](\ahat = t)}{\Delta'}
      }
    \end{itemize}
    If $\ahat$ is solved in $\Delta''$,
    then we are done by \Lemmaref{lem:alg-equiv-extends}
    with the second subtyping subderivation.
    Suppose $\ahat$ is not solved in $\Delta''$.\\
    \begin{llproof}
      \groundPf{t}{By \Lemmaref{lem:algebras-are-ground}}
      \eqPf{[\Delta''](\ahat = t)}{([\Delta'']\ahat = [\Delta'']t)}{By \defn of \defsubst}
      \eqPf{}{(\ahat = [\Delta'']t)}{$\ahat$ not free in $\Delta''$}
      \eqPf{}{(\ahat = t)}{Because $\ground{t}$}
      \alginstPf{\Theta; \Delta''}{\ahat = t}{\Delta'}{Rewrite second subtyping premise}
      \Hand\inPf{\hypeq{\ahat}{\tau}{t}}{\Delta'}{By inversion on \RuleInst}
      \Hand\judgetermPf{\Theta}{t}{\tau}{\ditto}
    \end{llproof}
    \qedhere
  \end{itemize}
\end{proof}

\subsection{Algorithmic Functor and Type Equivalence}

\begin{lemma}[Aux.\ Alg.\ Equiv.\ Complete]
  \label{lem:aux-alg-equiv-complete}
  ~
  \begin{enumerate}
  \item
    If $\Dee :: \semideclequiv[]{\Theta}{\mathcal{F}}{[\Omega]\mathcal{G}}{W}$
    and $\judgefunctor{\Theta; \Delta}{\mathcal{G}}{\Xi}$
    and $\ground{\mathcal{F}}$
    and $[\Delta]\Gee = \Gee$
    and $\rextend{\Theta}{\Delta}{\Omega}$,
    and $\semideclentailwah{\Theta}{W}$,
    then there exist $W'$ and $\Delta'$
    such that $\algequiv[]{\Theta; \Delta}{\mathcal{F}}{\mathcal{G}}{W'}{\Delta'}$
    and $\rextend{\Theta}{\Delta'}{\Omega}$
    and $\wahequiv{\Theta}{W}{[\Omega]W'}$.
  \item
    If $\Dee :: \semideclequiv[+]{\Theta}{P}{[\Omega]Q}{W}$
    and $\judgetp{\Theta; \Delta}{Q}{\Xi}$
    and $\ground{P}$
    and $[\Delta]Q = Q$
    and $\rextend{\Theta}{\Delta}{\Omega}$,
    and $\semideclentailwah{\Theta}{W}$,
    then there exist $W'$ and $\Delta'$
    such that $\algequiv[+]{\Theta; \Delta}{P}{Q}{W'}{\Delta'}$
    and $\rextend{\Theta}{\Delta'}{\Omega}$
    and $\wahequiv{\Theta}{W}{[\Omega]W'}$.
  \item
    If $\Dee :: \semideclequiv[-]{\Theta}{[\Omega]N}{M}{W}$
    and $\judgetp{\Theta; \Delta}{N}{\Xi}$
    and $\ground{M}$
    and $[\Delta]N = N$
    and $\rextend{\Theta}{\Delta}{\Omega}$,
    and $\semideclentailwah{\Theta}{W}$,
    then there exist $W'$ and $\Delta'$
    such that $\algequiv[-]{\Theta; \Delta}{N}{M}{W'}{\Delta'}$
    and $\rextend{\Theta}{\Delta'}{\Omega}$
    and $\wahequiv{\Theta}{W}{[\Omega]W'}$.
  \end{enumerate}
\end{lemma}
\begin{proof}
  Each part is proved by induction on $\hgt{\Dee}$.
  Parts (1) and (2) are mutually recursive.
  The proof is similar to \Lemmaref{lem:aux-alg-sub-complete}, but simpler.
\end{proof}

\subsection{Algorithmic Subtyping}

\begin{lemma}[Aux.\ Alg.\ Sub.\ Complete]
  \label{lem:aux-alg-sub-complete}
  ~
  \begin{enumerate}
  \item
    If $\Dee :: \semideclsub[+]{\Theta}{P}{[\Omega]Q}{W}$
    and $\judgetp{\Theta; \Delta}{Q}{\Xi}$
    and $\ground{P}$
    and $[\Delta]Q = Q$\\
    and $\rextend{\Theta}{\Delta}{\Omega}$
    and $\semideclentailwah{\Theta}{W}$,\\
    then there exists $\Dee' :: \algsub[+]{\Theta; \Delta}{P}{Q}{W'}{\Delta'}$\\
    such that $\rextend{\Theta}{\Delta'}{\Omega}$
    and $\wahequiv{\Theta}{W}{[\Omega]W'}$
    and $\hgt{\Dee} = \hgt{\Dee'}$.
  \item
    If $\Dee :: \semideclsub[-]{\Theta}{[\Omega]N}{M}{W}$
    and $\judgetp{\Theta; \Delta}{N}{\Xi}$
    and $\ground{M}$
    and $[\Delta]N = N$\\
    and $\rextend{\Theta}{\Delta}{\Omega}$
    and $\semideclentailwah{\Theta}{W}$,\\
    then there exists $\Dee' :: \algsub[-]{\Theta; \Delta}{N}{M}{W'}{\Delta'}$\\
    such that $\rextend{\Theta}{\Delta'}{\Omega}$
    and $\wahequiv{\Theta}{W}{[\Omega]W'}$
    and $\hgt{\Dee} = \hgt{\Dee'}$.
  \end{enumerate}
\end{lemma}
\begin{proof}
  We prove each part by induction on $\hgt{\Dee}$.
  We will elide showing the easy height invariance conclusion,
  as well as easy type well-formedness, groundness, and ``$[\Delta]A = A$''
  checks (\eg when using \ih).
  \begin{enumerate}
  \item
    \begin{itemize}
      \DerivationProofCase{\SemiDeclSubPosVoid}
      {}
      {
        \semideclsub[+]{\Theta}{0}{\underbrace{0}_{[\Omega]0}}{\True}
      }
      \begin{llproof}
        \Hand\algsubPf[+]{\Theta; \Delta}{0}{0}{\True}{\Delta}{By \AlgSubPosVoid}
        \Hand\rextendPf{\Theta}{\Delta}{\Omega}{Given}
        \Hand\wahequivPf{\Theta}{\True}{\underbrace{\True}_{[\Omega]\True}}{By \Lemmaref{lem:wah-equiv-refl}}
      \end{llproof} 

      \ProofCaseRule{\SemiDeclSubPosUnit}
      Similar to \SemiDeclSubPosVoid case.

      \DerivationProofCase{\SemiDeclSubPosProd}
      {
        \semideclsub[+]{\Theta}{P_1}{[\Omega]Q_1}{\Wah_1}
        \\
        \semideclsub[+]{\Dee_2 :: \Theta}{P_2}{[\Omega]Q_2}{\Wah_2}
      }
      {
        \semideclsub[+]{\Theta}{P_1 \times P_2}{[\Omega]Q_1 \times [\Omega]Q_2}{\Wah_1 \land \Wah_2}
      }
      \begin{llproof}
        \semideclsubPf[+]{\Theta}{P_1}{[\Omega]Q_1}{\Wah_1}{Subderivation}
        \rextendPf{\Theta}{\Delta}{\Omega}{Given}
        \semideclentailwahPf{\Theta}{W_1 \land W_2}{Given}
        \semideclentailwahPf{\Theta}{W_1}{By inversion on \WahEquivAnd}
        \semideclentailwahPf{\Theta}{W_2}{\ditto}
        \algsubPf[+]{\Theta; \Delta}{P_1}{Q_1}{\Wah_1'}{\Delta''}{By \ih}
        \rextendPf{\Theta}{\Delta''}{\Omega}{\ditto}
        \wahequivPf{\Theta}{W_1}{[\Omega]W_1'}{\ditto}
        \proofsep
        \judgetpPf{\Theta; \Delta}{Q_1 \times Q_2}{\dontcare}{Given}
        \judgetpPf{\Theta; \Delta}{Q_2}{\dontcare}{By inversion}
        \extendPf{\Theta}{\Delta}{\Delta''}{By \Lemmaref{lem:alg-sub-extends}}
        \judgetpPf{\Theta; \Delta''}{Q_2}{\dontcare}{By \Lemmaref{lem:ext-weak-tp}}
        \judgetpPf{\Theta; \Delta''}{[\Delta'']Q_2}{\dontcare}{By \Lemmaref{lem:right-hand-subst}}
        \semideclsubPf[+]{\Dee_2 :: \Theta}{P_2}{[\Omega]Q_2}{\Wah_2}{Subderivation}
        \semideclsubPf[+]{\Dee_2' :: \Theta}{P_2}{[\Omega]([\Delta'']Q_2)}{\Wah_2'}{By \Lemmaref{lem:sub-sandwich}}
        \wahequivPf{\Theta}{W_2}{W_2'}{\ditto}
        \Pf{\hgt{\Dee_2'}}{=}{\hgt{\Dee_2}}{\ditto}
        \algsubPf[+]{\Theta; \Delta''}{P_2}{[\Delta'']Q_2}{W_2''}{\Delta'}{By \ih}
        \Hand\rextendPf{\Theta}{\Delta'}{\Omega}{\ditto}
        \wahequivPf{\Theta}{W_2'}{[\Omega]W_2''}{\ditto}
        \wahequivPf{\Theta}{W_2}{[\Omega]W_2''}{By \Lemmaref{lem:wah-equiv-trans}}
        \decolumnizePf
        \Hand\algsubPf[+]{\Theta; \Delta}{P_1 \times P_2}{Q_1 \times Q_2}{[\Delta']\Wah_1' \land W_2''}{\Delta'}{By \AlgSubPosProd}
        \decolumnizePf
        \wahequivPf{\Theta}{[\Omega]W_1'}{[\Omega]([\Delta']W_1')}{By \Lemmaref{lem:wah-sandwich}}
        \wahequivPf{\Theta}{W_1}{[\Omega]([\Delta']W_1')}{By \Lemmaref{lem:wah-equiv-trans}}
        \wahequivPf{\Theta}{W_1 \land W_2}{[\Omega]([\Delta']W_1') \land [\Omega]W_2''}{By \WahEquivAnd}
        \Hand\wahequivPf{\Theta}{W_1 \land W_2}{[\Omega]([\Delta']W_1' \land W_2'')}{By \defn of \defsubst}
      \end{llproof}

      \ProofCaseRule{\SemiDeclSubPosSum}
      Similar to \SemiDeclSubPosProd case,
      but using \Lemmaref{lem:aux-alg-equiv-complete}
      rather than the \ih.

      \DerivationProofCase{\SemiDeclSubPosWithR}
      {
        \semideclsub[+]{\Theta}{P}{[\Omega]Q_0}{\Wah_0}
      }
      {
        \semideclsub[+]{\Theta}{P}{[\Omega]Q_0 \land [\Omega]\phi}{\Wah_0 \land [\Omega]\phi}
      }
      \begin{llproof}
        \semideclsubPf[+]{\Theta}{P}{[\Omega]Q_0}{\Wah_0}{Subderivation}
        \rextendPf{\Theta}{\Delta}{\Omega}{Given}
        \semideclentailwahPf{\Theta}{W_0 \land [\Omega]\phi}{Given}
        \semideclentailwahPf{\Theta}{W_0}{By inversion}
        \judgeentailPf{\Theta}{[\Omega]\phi}{\ditto}
        \algsubPf[+]{\Theta; \Delta}{P}{Q_0}{\Wah_0'}{\Delta''}{By \ih}
        \Hand\rextendPf{\Theta}{\Delta''}{\Omega}{\ditto}
        \wahequivPf{\Theta}{W_0}{[\Omega]W_0'}{\ditto}
        \alginstPf{\Theta; \Delta''}{[\Delta'']\phi}{\Delta'}{Inst.\ deterministic}
        \Hand\algsubPf[+]{\Theta; \Delta}{P}{Q_0 \land \phi}{[\Delta']\Wah_0' \land [\Delta']\phi}{\Delta'}{By \AlgSubPosWithR}
      \end{llproof}
      \begin{itemize}
      \item \textbf{Case} $\Delta' = \Delta''$\\
        \begin{llproof}
          \rextendPf{\Theta}{\Delta'}{\Omega}{Rewrite above}
          \wahequivPf{\Theta}{[\Omega]\phi}{[\Omega]([\Delta']\phi)}{By \Lemmaref{lem:wah-sandwich}}
          \proofsep
          \wahequivPf{\Theta}{[\Omega]W_0'}{[\Omega]([\Delta']W_0')}{By \Lemmaref{lem:wah-sandwich}}
          \wahequivPf{\Theta}{W_0}{[\Omega]([\Delta']W_0')}{By \Lemmaref{lem:wah-equiv-trans}}
          \wahequivPf{\Theta}{W_0 \land [\Omega]\phi}{[\Omega]([\Delta']W_0') \land [\Omega]([\Delta']\phi)}{By \WahEquivAnd}
          \Hand\wahequivPf{\Theta}{W_0 \land [\Omega]\phi}{[\Omega]([\Delta']W_0' \land [\Delta']\phi)}{By \defn of \defsubst}
        \end{llproof}
      \item \textbf{Case} $\Delta' \neq \Delta''$\\
        \begin{llproof}
          \eqPf{\phi}{(\ahat = t)}{By inversion on \RuleInst, there exist such $\ahat$ and $t$}
          \eqPf{\Delta''}{\Delta_0'', \ahat:\tau, \Delta_1''}{\ditto}
          \judgetermPf{\Theta}{[\Delta'']t}{\tau}{\ditto}
          \eqPf{\Delta'}{\Delta_0'', \hypeq{\ahat}{\tau}{[\Delta'']t}, \Delta_1''}{\ditto}
          \groundPf{[\Delta'']t}{By \Lemmaref{lem:decl-wf-ground}}
          \judgeentailPf{\Theta}{[\Omega](\ahat = t)}{Rewrite above}
          \judgeentailPf{\Theta}{[\Omega]\ahat = [\Omega]t}{By \defn of \defsubst}
          \judgeentailPf{\Theta}{[\Omega]t = [\Omega]\ahat}{By \Lemmaref{lem:equivassert}}
          \judgeentailPf{\Theta}{[\Omega]t = [\Omega]([\Delta'']t)}{By \Lemmaref{lem:ix-sandwich}}
          \judgeentailPf{\Theta}{[\Omega]t = [\Delta'']t}{Because $\ground{[\Delta'']t}$}
          \judgeentailPf{\Theta}{[\Delta'']t = [\Omega]t}{By \Lemmaref{lem:equivassert}}
          \judgeentailPf{\Theta}{[\Delta'']t = [\Omega]\ahat}{By \Lemmaref{lem:equivassert}}
          \rextendPf{\Theta}{\Delta'}{\Omega}{By \Lemmaref{lem:relaxed-deep-entry}}
          \wahequivPf{\Theta}{[\Omega]\phi}{[\Omega]([\Delta']\phi)}{By \Lemmaref{lem:wah-sandwich}}
          \decolumnizePf
          \Hand\wahequivPf{\Theta}{W_0 \land [\Omega]\phi}{[\Omega]([\Delta']W_0' \land [\Delta']\phi)}{Similar to previous subcase ($\Delta' = \Delta''$)}
        \end{llproof}
      \end{itemize}

      \DerivationProofCase{\SemiDeclSubPosExR}
      { 
        \semideclsub[+]{\Theta}{P}{[t/a]([\Omega]Q_0)}{\Wah}
        \\
        \judgeterm{\Theta}{t}{\tau}
      }
      {
        \semideclsub[+]{\Theta}{P}{\extype{a:\tau} [\Omega]Q_0}{\Wah}
      }
      \begin{llproof}
        \Pf{[t/a]([\Omega]Q_0)}{=}{[[t/a]\Omega]([t/a]Q_0)}{By prop.\ of subst.}
        \Pf{}{=}{[\Omega]([t/a]Q_0)}{$a$ not free in $\Omega$}
        \Pf{}{=}{[\Omega, \hypeq{\ahat}{\tau}{t}]([\ahat/a]Q_0)}{By \defn and prop.\ of subst.}
        \semideclsubPf[+]{\Theta}{P}{[\Omega, \hypeq{\ahat}{\tau}{t}]([\ahat/a]Q_0)}{\Wah}{Rewrite subderivation}
        \judgetpPf{\Theta; \Delta}{\extype{a:\tau} Q_0}{\Xi}{Given}
        \judgetpPf{\Theta, a:\tau; \Delta}{Q_0}{\Xi', a:\tau}{By inversion}
        \judgetpPf{\Theta; \Delta, \ahat:\tau}{[\ahat/a]Q_0}{\Xi', \ahat:\tau}{By \Lemmaref{lem:value-determined-evar-rename}}
        \decolumnizePf
        \rextendPf{\Theta}{\Delta}{\Omega}{Given}
        \rextendPf{\Theta}{\Delta, \ahat:\tau}{\Omega, \hypeq{\ahat}{\tau}{t}}{By \RExtSolve}
        \algsubPf[+]{\Theta; \Delta, \ahat:\tau}{P}{[\ahat/a]Q_0}{\Wah'}{\Delta''}{By \ih}
        \rextendPf{\Theta}{\Delta''}{\Omega, \hypeq{\ahat}{\tau}{t}}{\ditto}
        \wahequivPf{\Theta}{W}{[\Omega, \hypeq{\ahat}{\tau}{t}]W'}{\ditto}
        \Pf{(\hypeq{\ahat}{\tau}{t'})}{\in}{\Delta''}{By \Lemmaref{lem:sub-solves-val-det}}
        \Pf{\Delta''}{=}{\Delta', \hypeq{\ahat}{\tau}{t'}}{By inversion on \RExtSolved}
        \Hand\rextendPf{\Theta}{\Delta'}{\Omega}{\ditto}
        \algsubPf[+]{\Theta; \Delta, \ahat:\tau}{P}{[\ahat/a]Q_0}{\Wah'}{\Delta', \hypeq{\ahat}{\tau}{t'}}{By equation}
        \Hand\algsubPf[+]{\Theta; \Delta}{P}{\extype{a:\tau}Q_0}{\Wah'}{\Delta'}{By \AlgSubPosExR}
        \decolumnizePf
        \Pf{[\Delta', \hypeq{\ahat}{\tau}{t'}]W'}{=}{W'}{By \Lemmaref{lem:output-applied}}
        \Pf{[\Omega, \hypeq{\ahat}{\tau}{t}]W'}{=}{[\Omega]W'}{Follows from line just above}
        \Hand\wahequivPf{\Theta}{W}{[\Omega]W'}{By equation}
      \end{llproof}

      \item \textbf{Case}
      \[
        \Infer{\SemiDeclSubPosFix}
        {
          \semideclequiv{\Theta}{F}{[\Omega]G}{W_0}
        }
        {
          \semideclsub[+]{\Theta}{\comprehend{\nu:\mu F}{\Fold{F}{\alpha}\,\nu =_\tau t}}{\comprehend{\nu:\mu [\Omega]G}{\Fold{[\Omega]G}{\alpha}\,\nu =_\tau [\Omega]t'}}{W_0 \land t = [\Omega]t'}
        }
      \]
      Note that above we implicitly use \Lemmaref{lem:algebras-are-ground}
      (and of course the \defn of \defsubst, as in the other cases).\\
      \begin{llproof}
        \semideclequivPf{\Theta}{F}{[\Omega]G}{W_0}{Subderivation}
        \judgetpPf{\Theta; \Delta}{\comprehend{\nu:\mu G}{\Fold{G}{\alpha}\,\nu =_\tau t'}}{\Xi}{Given}
        \judgefunctorPf{\Theta; \Delta}{G}{\dontcare}{By inversion}
        \Pf{}{}{\ground{F}}{Because $\ground{P}$}
        \rextendPf{\Theta}{\Delta}{\Omega}{Given}
        \semideclentailwahPf{\Theta}{W_0 \land t = [\Omega]t'}{Given}
        \semideclentailwahPf{\Theta}{W_0}{By inversion}
        \semideclentailwahPf{\Theta}{t=[\Omega]t'}{\ditto}
        \algequivPf{\Theta; \Delta}{F}{G}{W_0'}{\Delta'}{By \Lemmaref{lem:aux-alg-equiv-complete}}
        \rextendPf{\Theta}{\Delta'}{\Omega}{\ditto}
        \wahequivPf{\Theta}{W_0}{[\Omega]W_0'}{\ditto}
      \end{llproof}
      ~\\
      \begin{itemize}
      \item \textbf{Case} for all $\ahat \in \dom{\Delta}$, $[\Delta]t' \neq \ahat$:\\
        By \AlgSubPosFix,
        \[
          \algsub[+]{\Theta; \Delta}{\comprehend{\nu:\mu F}{\Fold{F}{\alpha}\,\nu =_\tau t}}{\comprehend{\nu:\mu G}{\Fold{G}{\alpha}\,\nu =_\tau t'}}{W_0' \land t = [\Delta']t'}{\Delta'}
        \]
        \begin{llproof}
          \judgeentailPf{\Theta}{[\Omega]t' = [\Omega]([\Delta']t')}{By \Lemmaref{lem:ix-sandwich}}
          \judgeentailPf{\Theta}{t=t}{By \Lemref{lem:equivassert}}
          \judgeequivPf[]{\Theta}{t = [\Omega]t'}{t = [\Omega]([\Delta']t')}{By \PrpEquivEq}
          \wahequivPf{\Theta}{t = [\Omega]t'}{t = [\Omega]([\Delta']t')}{By \WahEquivPrp}
          \wahequivPf{\Theta}{W_0 \land t = [\Omega]t'}{[\Omega]W_0' \land t = [\Omega]([\Delta']t')}{By \WahEquivAnd}
          \Hand\wahequivPf{\Theta}{W_0 \land t = [\Omega]t'}{[\Omega](W_0' \land t = [\Delta']t')}{By \defn of \defsubst and $\because\ground{t}$}
          \Hand\rextendPf{\Theta}{\Delta'}{\Omega}{Above}
        \end{llproof} 
      \item \textbf{Case} there exists $\ahat \in \dom{\Delta}$ such that $[\Delta]t' = \ahat$, and $\Delta' = \Delta_1', \ahat:\tau, \Delta_2'$:\\
        Because $[\Delta]Q = Q$, we know $[\Delta]t' = t'$, so $t' = \ahat$.
        By \AlgSubPosFixInst,
        \[
          \algsub[+]{\Theta; \Delta}{\comprehend{\nu:\mu F}{\Fold{F}{\alpha}\,\nu =_\tau t}}{\comprehend{\nu:\mu G}{\Fold{G}{\alpha}\,\nu =_\tau t'}}{W_0' \land t = t}{\underbrace{\Delta_1', \hypeq{\ahat}{\tau}{t}, \Delta_2'}_{\Delta''}}
        \]
        \begin{llproof}
          \semideclentailwahPf{\Theta}{t=[\Omega]t'}{Above}
          \judgeentailPf{\Theta}{t=[\Omega]t'}{By inversion}
          \judgeentailPf{\Theta}{t=[\Omega]\ahat}{By equation}
          \Hand\rextendPf{\Theta}{\Delta''}{\Omega}{By \Lemmaref{lem:relaxed-deep-entry}}
          \judgeentailPf{\Theta}{t=t}{By \Lemmaref{lem:equivassert}}
          \judgeentailPf{\Theta}{[\Omega]t'=t}{By \Lemmaref{lem:equivassert}}
          \wahequivPf{\Theta}{t=[\Omega]t'}{t=t}{By \WahEquivPrpEq}
          \wahequivPf{\Theta}{W_0 \land t=[\Omega]t'}{[\Omega]W_0' \land t = t}{By \WahEquivAnd}
          \Hand\wahequivPf{\Theta}{W_0 \land t=[\Omega]t'}{[\Omega](W_0' \land t = t)}{By \defn of \defsubst and $\because\ground{t}$}
        \end{llproof} 
      \item \textbf{Case} there exists $\ahat \in \dom{\Delta}$ such that $[\Delta]t' = \ahat$, and $(\hypeq{\ahat}{\tau}{t''}) \in \Delta'$:\\
        Because $[\Delta]Q = Q$, we know $[\Delta]t' = t'$, so $t' = \ahat$.
        By \AlgSubPosFixInstFun,
        \[
          \algsub[+]{\Theta; \Delta}{\comprehend{\nu:\mu F}{\Fold{F}{\alpha}\,\nu =_\tau t}}{\comprehend{\nu:\mu G}{\Fold{G}{\alpha}\,\nu =_\tau t'}}{W_0' \land t = t''}{\Delta'}
        \]
        \begin{llproof}
          \Hand\rextendPf{\Theta}{\Delta'}{\Omega}{Above}
          \judgeentailPf{\Theta}{t=t}{By \Lemmaref{lem:equivassert}}
          \judgeentailPf{\Theta}{t''=[\Omega]t'}{By inversion on relaxed extension}
          \judgeentailPf{\Theta}{[\Omega]t'=t''}{By \Lemmaref{lem:equivassert}}
          \wahequivPf{\Theta}{t=[\Omega]t'}{t=t''}{By \WahEquivPrpEq}
          \wahequivPf{\Theta}{W_0 \land t=[\Omega]t'}{[\Omega]W_0' \land t = t''}{By \WahEquivAnd}
          \Hand\wahequivPf{\Theta}{W_0 \land t=[\Omega]t'}{[\Omega](W_0' \land t = t'')}{By \defn of \defsubst and $\because\ground{t,t''}$}
        \end{llproof} 
      \end{itemize}
     
      \DerivationProofCase{\SemiDeclSubPosDownshift}
      {
        \judgeextract[-]{\Theta}{[\Omega]M}{M_\Omega}{\Theta_\Omega}
      }
      {
        \semideclsub[+]{\Theta}{\downshift{N}}{\downshift{[\Omega]M}}{\Theta_\Omega \implies^\ast \negsubprob{N}{M_\Omega}}
      }
      \begin{llproof}
        \judgeextractPf[-]{\Theta}{[\Omega]M}{M_\Omega}{\Theta_\Omega}{Subderivation}
        \Hand\rextendPf{\Theta}{\Delta}{\Omega}{Given}
        \judgeextractPf[-]{\Theta; \Delta}{M}{M'}{\Thetahat}{By \Lemmaref{lem:uncomplete-extract}}
        \Pf{[\Omega]M'}{=}{M_\Omega}{\ditto}
        \Pf{[\Omega]\Thetahat}{=}{\Theta_\Omega}{\ditto}
        \Hand\algsubPf[+]{\Theta; \Delta}{\downshift{N}}{\downshift{M}}{\Thetahat \implies^\ast \negsubprob{N}{M'}}{\Delta}{By \AlgSubPosDownshift}
        \decolumnizePf
        \wahequivPf{\Theta}{\Theta_\Omega \implies^\ast \negsubprob{N}{M_\Omega}}{\Theta_\Omega \implies^\ast \negsubprob{N}{M_\Omega}}{By \Lemref{lem:wah-equiv-refl}}
        \wahequivPf{\Theta}{\Theta_\Omega \implies^\ast \negsubprob{N}{M_\Omega}}{[\Omega]\Thetahat \implies^\ast \negsubprob{N}{[\Omega]M'}}{By equations}
        \Hand\wahequivPf{\Theta}{\Theta_\Omega \implies^\ast \negsubprob{N}{M_\Omega}}{[\Omega](\Thetahat \implies^\ast \negsubprob{N}{M'})}{By \defn of \defsubst and $\because\ground{N}$}
      \end{llproof}
    \end{itemize}
  \item
    \begin{itemize}
      \ProofCaseRule{\SemiDeclSubNegUpshift}
      Similar to \SemiDeclSubPosDownshift case of part (1).

      \ProofCaseRule{\SemiDeclSubNegImpL}
      Similar to \SemiDeclSubPosWithR case of part (1).

      \ProofCaseRule{\SemiDeclSubNegAllL}
      Similar to \SemiDeclSubPosExR case of part (1).

      \ProofCaseRule{\SemiDeclSubNegArrow}
      Similar to \SemiDeclSubPosProd case of part (1).
      \qedhere
    \end{itemize}
  \end{enumerate}
\end{proof}

\begin{lemma}[Alg.\ Entail.\ Complete]
  \label{lem:alg-entail-complete}
  If $\semideclentailwah{\Theta}{W}$, then $\entailwah{\Theta}{W}$.
\end{lemma}
\begin{proof}
  By induction on $\size{W}$.
  \begin{itemize}
    \DerivationProofCase{\SemiDeclWTruePosSub}
    {
      \semideclsub[+]{\Theta}{P}{Q}{W'}
      \\
      \semideclentailwah{\Theta}{W'}
    }
    {
      \semideclentailwah{\Theta}{\possubprob{P}{Q}}
    }
    \begin{llproof}
      \Pf{\semideclsub[+]{\Theta}{P}{\underbrace{Q}_{[\cdot]Q}}{W'}}{}{}{Subderivation}
      \Pf{\judgetp{\Theta}{Q}{\Xi}}{}{}{Presupposed derivation}
      \Pf{\judgetp{\Theta; \cdot}{Q}{\Xi}}{}{}{By \Lemmaref{lem:decl-to-ground-alg-wf}}
      \Pf{\judgetp{\Theta}{P}{\Xi}}{}{}{Presupposed derivation}
      \Pf{\ground{P}}{}{}{By \Lemmaref{lem:decl-wf-ground}}
      \Pf{\rextend{\Theta}{\cdot}{\cdot}}{}{}{By \RExtEmpty}
      \Pf{[\cdot]Q = Q}{}{}{By \defn of \defsubst}
      \Pf{\algsub[+]{\Theta; \cdot}{P}{Q}{W''}{\Delta'}}{}{}{By \Lemmaref{lem:aux-alg-sub-complete}}
      \Pf{\rextend{\Theta}{\Delta'}{\cdot}}{}{}{\ditto}
      \Pf{\wahequiv{\Theta}{W'}{\underbrace{[\cdot]W''}_{W''}}}{}{}{\ditto}
      \Pf{\Delta' = \cdot}{}{}{By inversion on \RExtEmpty}
      \Pf{\algsub[+]{\Theta; \cdot}{P}{Q}{W''}{\cdot}}{}{}{Rewrite with $\Delta' = \cdot$}
      \Pf{\semideclentailwah{\Theta}{W'}}{}{}{Subderivation}
      \Pf{\semideclentailwah{\Theta}{W''}}{}{}{By \Lemmaref{lem:equiv-respects-entail}}
      \Pf{\size{W'} = \size{W''}}{}{}{By \Lemmaref{lem:equiv-respects-size}}
      \Pf{\size{W'} < \size{\possubprob{P}{Q}}}{}{}{By \Lemmaref{lem:semidecl-sub-shrinks-problem}}
      \Pf{\size{W''} < \size{\possubprob{P}{Q}}}{}{}{By equation}
      \Pf{\entailwah{\Theta}{W''}}{}{}{By \ih}
      \Pf{\entailwah{\Theta}{\possubprob{P}{Q}}}{}{}{By \WTruePosSub}
    \end{llproof}
    
    \ProofCaseRule{\SemiDeclWTrueNegSub}
    Similar to case for \SemiDeclWTruePosSub.
  
    \ProofCaseRule{\SemiDeclWTruePosEquiv}
    Similar to case for \SemiDeclWTruePosSub,
    but uses \Lemmaref{lem:aux-alg-equiv-complete}
    and \Lemmaref{lem:semidecl-equiv-shrinks-problem}
    instead of \Lemmaref{lem:aux-alg-sub-complete}
    and \Lemmaref{lem:semidecl-sub-shrinks-problem}.
  
    \ProofCaseRule{\SemiDeclWTrueNegEquiv}
    Similar to case for \SemiDeclWTruePosEquiv.
  
    \item The remaining cases are straightforward. \qedhere
  \end{itemize}
\end{proof}

\begin{theorem}[Alg.\ Sub.\ Complete]
  \label{thm:alg-sub-complete}
  ~
  \begin{enumerate}
  \item If $\judgesub[+]{\Theta}{P}{[\Omega]Q}$
    and $\judgetp{\Theta; \Delta}{Q}{\Xi}$
    and $\ground{P}$
    and $[\Delta]Q = Q$
    and $\rextend{\Theta}{\Delta}{\Omega}$,\\
    then there exist $P'$, $\Theta_P$, $W$, and $\Delta'$
    such that $\algsub[+]{\Theta, \Theta_P; \Delta}{P'}{Q}{W}{\Delta'}$\\
    and $\rextend{\Theta, \Theta_P}{\Delta'}{\Omega}$
    and $\entailwah{\Theta, \Theta_P}{[\Omega]W}$
    and $\judgeextract[+]{\Theta}{P}{P'}{\Theta_P}$.
  \item If $\judgesub[-]{\Theta}{[\Omega]N}{M}$
    and $\judgetp{\Theta; \Delta}{N}{\Xi}$
    and $\ground{M}$
    and $[\Delta]N = N$
    and $\rextend{\Theta}{\Delta}{\Omega}$,\\
    then there exist $M'$, $\Theta_M$, $W$, and $\Delta'$
    such that $\algsub[-]{\Theta, \Theta_M; \Delta}{N}{M'}{W}{\Delta'}$\\
    and $\rextend{\Theta, \Theta_M}{\Delta'}{\Omega}$
    and $\entailwah{\Theta, \Theta_M}{[\Omega]W}$
    and $\judgeextract[-]{\Theta}{M}{M'}{\Theta_M}$.
  \end{enumerate}
\end{theorem} 
\begin{proof}
  ~
  \begin{enumerate}
  \item ~\\
    \begin{llproof}
      \Pf{\judgesub[+]{\Theta}{P}{[\Omega]Q}}{}{}{Given}
      \Pf{\semideclsub[-]{\Theta}{\upshift{P}}{\upshift{[\Omega]Q}}{W}}{}{}{By \Lemmaref{lem:semidecl-sub-complete}}
      \Pf{\semideclentailwah{\Theta}{W}}{}{}{\ditto}
      \Hand\Pf{\judgeextract[+]{\Theta}{P}{P'}{\Theta_P}}{}{}{By inversion on \SemiDeclSubNegUpshift}
      \Pf{W = \Theta_P \implies^\ast \possubprob{P'}{[\Omega]Q}}{}{}{\ditto}
      \Pf{\semideclentailwah{\Theta}{\Theta_P \implies^\ast \possubprob{P'}{[\Omega]Q}}}{}{}{Rewrite above entailment}
      \Pf{\semideclentailwah{\Theta, \Theta_P}{\possubprob{P'}{[\Omega]Q}}}{}{}{By inversion on \SemiDeclWTrueImpl and \SemiDeclWTrueAll}
      \Pf{\semideclsub[+]{\Theta, \Theta_P}{P'}{[\Omega]Q}{W'}}{}{}{By inversion on \SemiDeclWTruePosSub}
      \Pf{\semideclentailwah{\Theta, \Theta_P}{W'}}{}{}{\ditto}
      \Pf{\judgetp{\Theta; \Delta}{Q}{\Xi}}{}{}{Given}
      \Pf{\judgetp{\Theta, \Theta_P; \Delta}{Q}{\Xi}}{}{}{By \Lemmaref{lem:ix-level-weakening}}
      \Pf{\rextend{\Theta}{\Delta}{\Omega}}{}{}{Given}
      \Pf{\rextend{\Theta, \Theta_P}{\Delta}{\Omega}}{}{}{By \Lemmaref{lem:weaken-ext}}
      \Pf{[\Delta]Q=Q}{}{}{Given}
      \Pf{\ground{P'}}{}{}{Because $\ground{P}$}
      \Hand\Pf{\algsub[+]{\Theta, \Theta_P; \Delta}{P'}{Q}{W''}{\Delta'}}{}{}{By \Lemmaref{lem:aux-alg-sub-complete}}
      \Hand\Pf{\rextend{\Theta, \Theta_P}{\Delta'}{\Omega}}{}{}{\ditto}
      \Pf{\wahequiv{\Theta, \Theta_P}{W'}{[\Omega]W''}}{}{}{\ditto}
      \Pf{\semideclentailwah{\Theta, \Theta_P}{[\Omega]W''}}{}{}{By \Lemmaref{lem:equiv-respects-entail}}
      \Hand\Pf{\entailwah{\Theta, \Theta_P}{[\Omega]W''}}{}{}{By \Lemmaref{lem:alg-entail-complete}}
    \end{llproof} 
  \item Similar to first part. \qedhere
  \end{enumerate}
\end{proof}

\subsection{Algorithmic Typing}

\begin{lemma}[Append Constraints]
  \label{lem:append-probs}
  We have:
  $\algneg{\Theta}{\Gamma}{\chi}$ and $\algneg{\Theta}{\Gamma}{\chi'}$
  if and only if
  $\algneg{\Theta}{\Gamma}{\chi, \chi'}$.
\end{lemma}
\begin{proof}
  Straightforward.
\end{proof}

\begin{lemma}[Semidecl.\ Typing Sound]
  \label{lem:semidecl-typing-sound}
  ~
  \begin{enumerate}
    \item
      If $\semideclsynhead{\Theta}{\Gamma}{h}{P}$,
      then $\judgesynhead{\Theta}{\Gamma}{h}{P}$.
    \item
      If $\semideclsynexp{\Theta}{\Gamma}{\be}{\upshift{P}}$,
      then $\judgesynexp{\Theta}{\Gamma}{\be}{\upshift{P}}$.
    \item
      If $\semideclchkval{\Theta}{\Gamma}{v}{P}{\chi}$
      and $\semideclneg{\Theta}{\Gamma}{\chi}$,
      then $\judgechkval{\Theta}{\Gamma}{v}{P}$.
    \item
      If $\semideclchkexp{\Theta}{\Gamma}{e}{N}$,
      then $\judgechkexp{\Theta}{\Gamma}{e}{N}$.
    \item
      If $\semideclchkmatch{\Theta}{\Gamma}{P}{\clauses{\pa}{e}{i}{I}}{N}$,
      then $\judgechkmatch{\Theta}{\Gamma}{P}{\clauses{\pa}{e}{i}{I}}{N}$.
    \item
      If $\semideclspine{\Theta}{\Gamma}{s}{N}{\upshift{P}}{\chi}$
      and $\semideclneg{\Theta}{\Gamma}{\chi}$,
      then $\judgespine{\Theta}{\Gamma}{s}{N}{\upshift{P}}$.
  \end{enumerate}
\end{lemma}
\begin{proof}
  By lexicographic induction on,
  first, the size of the program term (\Figureref{fig:size-prog-chi}), and,
  second, the size of the input type (\Figureref{fig:size})
  ($P$ for part (5)).
  All parts are mutually recursive.

  Use \Lemmaref{lem:semidecl-sub-sound} in
  the \SemiDeclChkValVar case of part (3)
  and the \SemiDeclChkExpRec case of part (4).
\end{proof}

\begin{lemma}[Semidecl.\ Typing Complete]
  \label{lem:semidecl-typing-complete}
  ~
  \begin{enumerate}
    \item
      If $\judgesynhead{\Theta}{\Gamma}{h}{P}$,
      then $\semideclsynhead{\Theta}{\Gamma}{h}{P}$.
    \item
      If $\judgesynexp{\Theta}{\Gamma}{\be}{\upshift{P}}$,
      then $\semideclsynexp{\Theta}{\Gamma}{\be}{\upshift{P}}$.
    \item
      If $\judgechkval{\Theta}{\Gamma}{v}{P}$,
      then there exists $\chi$
      such that $\semideclchkval{\Theta}{\Gamma}{v}{P}{\chi}$
      and $\semideclneg{\Theta}{\Gamma}{\chi}$.
    \item
      If $\judgechkexp{\Theta}{\Gamma}{e}{N}$,
      then $\semideclchkexp{\Theta}{\Gamma}{e}{N}$.
    \item
      If $\judgechkmatch{\Theta}{\Gamma}{P}{\clauses{\pa}{e}{i}{I}}{N}$,
      then $\semideclchkmatch{\Theta}{\Gamma}{P}{\clauses{\pa}{e}{i}{I}}{N}$.
    \item
      If $\judgespine{\Theta}{\Gamma}{s}{N}{\upshift{P}}$,
      then there exists $\chi$ such that
      $\semideclspine{\Theta}{\Gamma}{s}{N}{\upshift{P}}{\chi}$
      and $\semideclneg{\Theta}{\Gamma}{\chi}$.
  \end{enumerate}
\end{lemma}
\begin{proof}
  By structural induction on the given typing derivation.
  All parts are mutually recursive.
  \begin{enumerate}
  \item Straightforward.
  \item Straightforward.
  \item 
    \begin{itemize}
      \DerivationProofCase{\DeclChkValVar}
      {
        P \neq \exists, \land 
        \\
        (x:Q) \in \Gamma
        \\
        \judgesub[+]{\Theta}{Q}{P}
      }
      {
        \judgechkval{\Theta}{\Gamma}{x}{P}
      }
      \begin{llproof}
        \judgesubPf[+]{\Theta}{Q}{P}{Subderivation}
        \semideclsubPf[-]{\Theta}{\upshift{Q}}{\upshift{P}}{W}{By \Lemmaref{lem:semidecl-sub-complete}}
        \semideclentailwahPf{\Theta}{W}{\ditto}
        \judgeextractPf[+]{\Theta}{Q}{Q'}{\Theta'}{By inversion on \SemiDeclSubNegUpshift}
        \Pf{W}{=}{\Theta' \implies^\ast \possubprob{Q'}{P}}{\ditto}
        \judgeextractPf[+]{\Theta}{Q}{Q}{\cdot}{By inversion on $\Gamma$ WF}
        \Pf{Q'}{=}{Q}{By \Lemmaref{lem:extract-determinism}}
        \Pf{\Theta'}{=}{\cdot}{\ditto}
        \semideclentailwahPf{\Theta}{\cdot \implies^\ast \possubprob{Q}{P}}{By equations}
        \semideclentailwahPf{\Theta}{\possubprob{Q}{P}}{By \defn of \defimpast}
        \semideclsubPf[+]{\Theta}{Q}{P}{W'}{By inversion on \SemiDeclWTruePosSub}
        \semideclentailwahPf{\Theta}{W'}{\ditto}
        \Hand\semideclnegPf{\Theta}{\Gamma}{W'}{By \SemiChkProblemsEmpty and \SemiChkProblemsWah}
        \Pf{(x:Q)}{\in}{\Gamma}{Premise}
        \Pf{P}{\neq}{\exists, \land}{Premise}
        \Hand\semideclchkvalPf{\Theta}{\Gamma}{x}{P}{W'}{By \SemiDeclChkValVar}
      \end{llproof} 

      \ProofCaseRule{\DeclChkValPair}
      Straightforward. Use \Lemmaref{lem:append-probs}.

      \DerivationProofCase{\DeclChkValWith}
      {
        \judgechkval{\Theta}{\Gamma}{v}{P_0}
        \\
        \judgeentail{\Theta}{\phi}
      }
      { \judgechkval{\Theta}{\Gamma}{v}{P_0 \land \phi} }
      \begin{llproof}
        \judgechkvalPf{\Theta}{\Gamma}{v}{P_0}{Subderivation}
        \semideclchkvalPf{\Theta}{\Gamma}{v}{P_0}{\chi}{By \ih}
        \semideclnegPf{\Theta}{\Gamma}{\chi}{\ditto}
        \judgeentailPf{\Theta}{\phi}{Subderivation}
        \semideclentailwahPf{\Theta}{\phi}{By \SemiDeclWTrueProp}
        \Hand\semideclnegPf{\Theta}{\Gamma}{\phi, \chi}{By \SemiChkProblemsEmpty and \SemiChkProblemsWah}
        \Hand\semideclchkvalPf{\Theta}{\Gamma}{v}{P_0 \land \phi}{\phi, \chi}{By \SemiDeclChkValWith}
      \end{llproof} 

      \DerivationProofCase{\DeclChkValDownshift}
      { \judgechkexp{\Theta}{\Gamma}{e}{N} }
      { \judgechkval{\Theta}{\Gamma}{\thunk{e}}{\downshift{N}} }
      \begin{llproof}
        \judgechkexpPf{\Theta}{\Gamma}{e}{N}{Subderivation}
        \semideclchkexpPf{\Theta}{\Gamma}{e}{N}{By \ih}
        \Hand\semideclchkvalPf{\Theta}{\Gamma}{\thunk{e}}{\downshift{N}}{(e <= N)}{By \SemiDeclChkValDownshift}
        \Hand\semideclnegPf{\Theta}{\Gamma}{(e <= N)}{By \SemiChkProblemsEmpty and \SemiChkProblemsNegChk}
      \end{llproof}

      \item The remaining cases are straightforward.
    \end{itemize}
  \item
    \begin{itemize}
      \DerivationProofCase{\DeclChkExpRec}
      {
        \arrayenvb{
          \simple{\Theta}{N}
          \\
          \judgesub[-]{\Theta}{\alltype{a:\kindnat} M}{N}
        }
        \\
        \judgechkexp{\Theta, a:\kindnat}{\Gamma, x:\downshift{\alltype{a':\kindnat} a' < a \implies [a'/a]M}}{e_0}{M}
      }
      {
        \judgechkexp{\Theta}{\Gamma}{\rec{x : (\alltype{a:\kindnat} M)}{e_0}}{N}
      }
      \begin{llproof}
        \judgesubPf[-]{\Theta}{\alltype{a:\kindnat} M}{N}{Subderivation}
        \semideclsubPf[+]{\Theta}{\downshift{\alltype{a:\kindnat} M}}{\downshift{N}}{W'}{By \Lemmaref{lem:semidecl-sub-complete}}
        \semideclentailwahPf{\Theta}{W'}{\ditto}
        \judgeextractPf[-]{\Theta}{N}{N'}{\Theta_N}{By inversion on \SemiDeclSubPosDownshift}
        \Pf{W'}{=}{\Theta_N \implies^\ast \negsubprob{\alltype{a:\kindnat} M}{N'}}{\ditto}
        \simplePf{\Theta}{N}{Subderivation}
        \eqPf{N'}{N}{By \Lemmaref{lem:extract-determinism}}
        \eqPf{\Theta_N}{\cdot}{\ditto}
        \semideclentailwahPf{\Theta}{\negsubprob{\alltype{a:\kindnat} M}{N}}{Rewrite above}
        \semideclsubPf[-]{\Theta}{\alltype{a:\kindnat} M}{N}{W}{By inversion on \SemiDeclWTrueNegSub}
        \semideclentailwahPf{\Theta}{W}{\ditto}
        \decolumnizePf
        \judgechkexpPf{\Theta, a:\kindnat}{\Gamma, x:\downshift{\alltype{a':\kindnat} a' < a \implies [a'/a]M}}{e_0}{M}{Subderivation}
        \semideclchkexpPf{\Theta, a:\kindnat}{\Gamma, x:\downshift{\alltype{a':\kindnat} a' < a \implies [a'/a]M}}{e_0}{M}{By \ih}
        \Hand\semideclchkexpPf{\Theta}{\Gamma}{\rec{x : (\alltype{a:\kindnat} M)}{e_0}}{N}{By \SemiDeclChkExpRec}
      \end{llproof}
      
      \item The remaining cases are straightforward.
    \end{itemize}
  \item 
    \begin{itemize}
      \ProofCaseRule{\DeclSpineAll}
      Similar to \DeclChkValExists case of part (3), which is straightforward.

      \ProofCaseRule{\DeclSpineImplies}
      Similar to \DeclChkValWith case of part (3).

      \ProofCaseRule{\DeclSpineApp}
      Similar to \DeclChkValPair case of part (3).

      \ProofCaseRule{\DeclSpineNil}
      Straightforward.
    \end{itemize}
  \item Straightforward.
    \qedhere
  \end{enumerate}
\end{proof}

\begin{lemma}[Probs.\ Equiv.\ Resp.\ Entail.]
  \label{lem:probs-equiv-respects-entail}
  If $\semideclneg{\Theta}{\Gamma}{\chi}$ and $\chiequiv{\Theta}{\Gamma}{\chi}{\chi'}$,
  then $\semideclneg{\Theta}{\Gamma}{\chi'}$.
\end{lemma}
\begin{proof}
  By structural induction on the given $\chi$-equivalence derivation.
  \begin{itemize}
    \DerivationProofCase{\ChiEquivEmpty}
    {}
    {
      \chiequiv{\Theta}{\Gamma}{\cdot}{\cdot}
    }
    \begin{llproof}
      \Pf{\chi'}{=}{\cdot}{By inversion}
      \semideclnegPf{\Theta}{\Gamma}{\underbrace{\cdot}_{\chi'}}{By \SemiChkProblemsEmpty}
    \end{llproof}

    \DerivationProofCase{\ChiEquivWah}
    {
      \wahequiv{\Theta}{W}{W'}
      \\
      \chiequiv{\Theta}{\Gamma}{\chi_0}{\chi_0'}
    }
    {
      \chiequiv{\Theta}{\Gamma}{W, \chi_0}{W', \chi_0'}
    }
    \begin{llproof}
      \semideclnegPf{\Theta}{\Gamma}{W, \chi_0}{Given}
      \semideclentailwahPf{\Theta}{W}{By inversion on \SemiChkProblemsWah}
      \semideclnegPf{\Theta}{\Gamma}{\chi_0}{\ditto}
      \wahequivPf{\Theta}{W}{W'}{Subderivation}
      \semideclentailwahPf{\Theta}{W'}{By \Lemmaref{lem:equiv-respects-entail}}
      \chiequivPf{\Theta}{\Gamma}{\chi_0}{\chi_0'}{Subderivation}
      \semideclnegPf{\Theta}{\Gamma}{\chi_0'}{By \ih}
      \semideclnegPf{\Theta}{\Gamma}{W', \chi_0'}{By \SemiChkProblemsWah}
    \end{llproof}

    \DerivationProofCase{\ChiEquivChk}
    {
      \judgeequiv[-]{\Theta}{N}{N'}
      \\
      \chiequiv{\Theta}{\Gamma}{\chi_0}{\chi_0'}
    }
    {
      \chiequiv{\Theta}{\Gamma}{(e <= N), \chi_0}{(e <= N'), \chi_0'}
    }
    \begin{llproof}
      \semideclnegPf{\Theta}{\Gamma}{(e <= N), \chi_0}{Given}
      \semideclchkexpPf{\Theta}{\Gamma}{e}{N}{By inversion on \SemiChkProblemsNegChk}
      \semideclnegPf{\Theta}{\Gamma}{\chi_0}{\ditto}
      \judgeequivPf[-]{\Theta}{N}{N'}{Subderivation}
      \judgechkexpPf{\Theta}{\Gamma}{e}{N}{By \Lemmaref{lem:semidecl-typing-sound}}
      \judgeequivPf[]{\Theta}{\Gamma}{\Gamma}{By repeated \Lemmaref{lem:refl-equiv-tp-fun}}
      \judgechkexpPf{\Theta}{\Gamma}{e}{N'}{By \Lemmaref{lem:prog-typing-respects-equiv}}
      \semideclchkexpPf{\Theta}{\Gamma}{e}{N'}{By \Lemmaref{lem:semidecl-typing-complete}}
      \chiequivPf{\Theta}{\Gamma}{\chi_0}{\chi_0'}{Subderivation}
      \semideclnegPf{\Theta}{\Gamma}{\chi_0'}{By \ih}
      \semideclnegPf{\Theta}{\Gamma}{(e <= N'), \chi_0'}{By \SemiChkProblemsNegChk}
    \end{llproof}
    \qedhere
  \end{itemize}
\end{proof}

\begin{lemma}[Typing Solves Val-det.]
  \label{lem:typing-solves-val-det}
  ~
  \begin{enumerate}
  \item
    Suppose $\Delta = \Delta_1, \ahat:\tau, \Delta_2$.
    If $\judgeunroll{\Xi}{\Theta; \Delta}{\nu:G[\mu F]}{\beta}{G\; \Fold{F}{\alpha}\;\nu}{\ahat}{Q}{\tau}$\\
    and $\judgefunctor{\Theta; \Delta}{G}{\Xi_G}$
    and $(\ahat:\tau) \notin \Xi_G$\\
    and $\algchk{\Theta; \Delta}{\Gamma}{v}{Q}{\chi}{\Delta'}$\\
    and $[\Delta]Q = Q$
    and $\rextend{\Theta}{\Delta'}{\Omega}$
    and $\semideclneg{\Theta}{\Gamma}{[\Omega]\chi}$,\\
    then there exists $t$ such that $\judgeterm{\Theta}{t}{\tau}$
    and $(\hypeq{\ahat}{\tau}{t}) \in \Delta'$.
  \item
    If $\algchk{\Theta; \Delta}{\Gamma}{v}{P}{\chi}{\Delta'}$\\
    and $\judgetp{\Theta; \Delta}{P}{\Xi_P}$
    and $[\Delta]P = P$
    and $\rextend{\Theta}{\Delta'}{\Omega}$
    and $\semideclneg{\Theta}{\Gamma}{[\Omega]\chi}$,\\
    then for all $(\ahat : \tau) \in \Xi_P$,
    there exists $t$ such that $\judgeterm{\Theta}{t}{\tau}$
    and $(\hypeq{\ahat}{\tau}{t}) \in \Delta'$.
  \item
    If $\algspine{\Theta; \Delta}{\Gamma}{s}{N}{\upshift{P}}{\chi}{\Delta'}$\\
    and $\judgetp{\Theta; \Delta}{N}{\Xi_N}$
    and $[\Delta]N = N$
    and $\rextend{\Theta}{\Delta'}{\Omega}$
    and $\semideclneg{\Theta}{\Gamma}{[\Omega]\chi}$,\\
    then for all $(\ahat : \tau) \in \Xi_N$,
    there exists $t$ such that $\judgeterm{\Theta}{t}{\tau}$
    and $(\hypeq{\ahat}{\tau}{t}) \in \Delta'$.
  \end{enumerate}
\end{lemma}
\begin{proof}
  Parts (1) and (2) are mutually recursive.
  In each part, we may elide straightforward checks for
  ``$[\Delta]A = A$'',
  ``$\rextend{\Theta}{\Delta}{\Omega}$'',
  and ``$\semideclneg{\Theta}{\Gamma}{[\Omega]\chi}$''
  whenever we use the \ih.
  \begin{enumerate}
  \item
    By structural induction on $v$.
    We case analyze $G$.
    For all cases of $G$ except $\Const{P} \otimes \hat{P}$,
    there is exactly one rule that can conclude the given unrolling derivation;
    we write each such case as the given unrolling derivation specialized
    to that case.
    \begin{itemize}
      \DerivationProofCase{\AlgUnrollSum}
      {\arrayenvbl{
          \composeinj{1}{\beta}{\beta_1}
          \\
          \composeinj{2}{\beta}{\beta_2}
        }
        \\
        \arrayenvbl{
          \judgeunroll{\Xi}{\Theta; \Delta}{\nu:G_1[\mu F]}{\beta_1}{G_1\;\Fold{F}{\alpha}\;\nu}{\ahat}{Q_1}{\tau}
          \\
          \judgeunroll{\Xi}{\Theta; \Delta}{\nu:G_2[\mu F]}{\beta_2}{G_2\;\Fold{F}{\alpha}\;\nu}{\ahat}{Q_2}{\tau}
        }
      }
      { \judgeunroll{\Xi}{\Theta; \Delta}{\nu:(G_1 \oplus G_2)[\mu F]}{\beta}{(G_1 \oplus G_2)\;\Fold{F}{\alpha}\;\nu}{\ahat}{Q_1 + Q_2}{\tau} }
      We case analyze the concluding rule of the given value typing derivation:
      \begin{itemize}
        \DerivationProofCase{\AlgChkValVar}
        {
          Q_1 + Q_2 \neq \exists, \with
          \\
          (v:P)\in\Gamma
          \\
          \algsub[+]{\Theta; \Delta}{P}{Q_1 + Q_2}{\Wah}{\Delta'}
        }
        {
          \algchk{\Theta; \Delta}{\Gamma}{v}{Q_1 + Q_2}{\underbrace{\Wah, \cdot}_\chi}{\Delta'}
        }
        \begin{llproof}
          \inPf{(v:P)}{\Gamma}{Subderivation}
          \simplePf{\Theta}{P}{By inversion on $\Gamma$ WF}
          \groundPf{P}{\ditto}
          \algsubPf[+]{\Theta; \Delta}{P}{Q_1 + Q_2}{\Wah}{\Delta'}{Subderivation}
          \eqPf{P}{P_1 + P_2}{By inversion on \AlgSubPosSum}
          \eqPf{W}{[\Delta']W_1 \land W_2}{\ditto}
          \algequivPf[+]{\Theta; \Delta}{P_1}{Q_1}{W_1}{\Delta''}{\ditto}
          \algequivPf[+]{\Theta; \Delta''}{P_2}{[\Delta'']Q_2}{W_2}{\Delta'}{\ditto}
          \judgeunrollPf{\Xi}{\Theta; \Delta}{\nu:G_1[\mu F]}{\beta_1}{G_1\;\Fold{F}{\alpha}\;\nu}{\ahat}{Q_1}{\tau}{Subderivation}
          \decolumnizePf
          \extendPf{\Theta}{\Delta''}{\Delta'}{By \Lemmaref{lem:alg-equiv-extends}}
          \rextendPf{\Theta}{\Delta''}{\Delta'}{By \Lemmaref{lem:extension-sound}}
          \rextendPf{\Theta}{\Delta'}{\Omega}{Given}
          \rextendPf{\Theta}{\Delta''}{\Omega}{By \Lemmaref{lem:ext-trans}}
          \semideclnegPf{\Theta}{\Gamma}{[\Omega]([\Delta']W_1 \land W_2, \cdot)}{Given}
          \semideclnegPf{\Theta}{\Gamma}{([\Omega]([\Delta']W_1) \land [\Omega]W_2, \cdot)}{By \defn of \defsubst}
          \semideclentailwahPf{\Theta}{[\Omega]([\Delta']W_1) \land [\Omega]([\Delta']W_2)}{By inversion}
          \semideclentailwahPf{\Theta}{[\Omega]([\Delta']W_1)}{By inversion}
          \wahequivPf{\Theta}{[\Omega]W_1}{[\Omega]([\Delta']W_1)}{By \Lemmaref{lem:wah-sandwich}}
          \wahequivPf{\Theta}{[\Omega]([\Delta']W_1)}{[\Omega]W_1}{By \Lemmaref{lem:wah-equiv-symmetric}}
          \semideclentailwahPf{\Theta}{[\Omega]W_1}{By \Lemmaref{lem:equiv-respects-entail}}
          \inPf{\hypeq{\ahat}{\tau}{t}}{\Delta''}{By \Lemmaref{lem:equiv-solves-unrolled-evar}}
          \Hand\judgetermPf{\Theta}{t}{\tau}{\ditto}
          \Hand\inPf{\hypeq{\ahat}{\tau}{t}}{\Delta'}{By inversion on extension}
        \end{llproof} 
      \end{itemize}
      ~\\
      \begin{itemize}
        \DerivationProofCase{\AlgChkValIn{k}}
        {
          \algchk{\Theta; \Delta}{\Gamma}{v_0}{Q_k}{\chi}{\Delta'}
        }
        {
          \algchk{\Theta; \Delta}{\Gamma}{\underbrace{\inj{k}{v_k}}_v}{(Q_1 + Q_2)}{\chi}{\Delta'}
        }
        \begin{llproof}
          \algchkPf{\Theta; \Delta}{\Gamma}{v_0}{Q_k}{\chi}{\Delta'}{Subderivation}
          \judgeunrollPf{\Xi}{\Theta; \Delta}{\nu:G_k[\mu F]}{\beta_k}{G_k\;\Fold{F}{\alpha}\;\nu}{\ahat}{Q_k}{\tau}{Subderivation}
          \judgefunctorPf{\Theta; \Delta}{G_k}{\Xi_{G_k}}{Presupposed derivation}
        \end{llproof}
        \begin{itemize}
        \item \textbf{Case} $(\ahat:\tau) \notin \Xi_{G_k}$:\\
          \begin{llproof}
            \Hand\Pf{(\hypeq{\ahat}{\tau}{t})}{\in}{\Delta'}{By \ih (part (1))}
            \Hand\judgetermPf{\Theta}{t}{\tau}{\ditto}
          \end{llproof}
        \item \textbf{Case} $(\ahat:\tau) \in \Xi_{G_k}$:\\
          \begin{llproof}
            \judgetpPf{\Theta; \Delta}{Q_k}{\Xi_{Q_k}}{By \Lemmaref{lem:alg-unroll-output-wf}}
            \Pf{\Xi_{Q_k}}{\supseteq}{\Xi_{G_k}}{\ditto}
            \Pf{}{\ni}{(\ahat:\tau)}{Given subcase}
            \Hand\Pf{(\hypeq{\ahat}{\tau}{t})}{\in}{\Delta'}{By \ih (part (2))}
            \Hand\judgetermPf{\Theta}{t}{\tau}{\ditto}
          \end{llproof} 
        \end{itemize}
      \end{itemize}

      \DerivationProofCase{\AlgUnrollId}
      {
        \judgeunroll{\Xi,a':\tau}{\Theta, a':\tau; \Delta}{\nu:\hat{P}[\mu F]}{(\clause{q}{t'})}{\hat{P}\;\Fold{F}{\alpha}\;\nu}{\ahat}{Q_0}{\tau} 
      }
      {
        \judgeunroll*{\Xi}{\Theta; \Delta}{\nu : (\Id\otimes\hat{P})[\mu F]}{(\clause{(a',q)}{t'})}{(\Id\otimes\hat{P})\;\Fold{F}{\alpha}\;\nu}{\ahat}{\extype{a':\tau}{\comprehend{\nu:\mu F}{ \Fold{F}{\alpha}\,{\nu} =_\tau a' } \times Q_0}}{\tau}
      }
      \begin{llproof}
        \algchkPf{\Theta; \Delta}{\Gamma}{v}{\extype{a':\tau}{\comprehend{\nu:\mu F}{ \Fold{F}{\alpha}\,{\nu} =_\tau a' } \times Q_0}}{\chi}{\Delta'}{Given}
        \algchkPf{\Theta; \Delta, \ahat':\tau}{\Gamma}{v}{[\ahat'/a'](\comprehend{\nu:\mu F}{ \Fold{F}{\alpha}\,{\nu} =_\tau a' } \times Q_0)}{\chi}{\Delta', \hypeq{\ahat'}{\tau}{t_0}}{By inversion}
        \trailingjust{(on \AlgChkValExists)}
      \end{llproof}
      By \defn of \defsubst and because $\ground{\alpha}$ and $a'$ is not free in $F$,
      \[
        \algchk{\Theta; \Delta, \ahat':\tau}{\Gamma}{v}{\comprehend{\nu:\mu F}{ \Fold{F}{\alpha}\,{\nu} =_\tau \ahat' } \times [\ahat'/a']Q_0}{\chi}{\Delta', \hypeq{\ahat'}{\tau}{t_0}}
      \]
      \begin{llproof}
        \Pf{v}{=}{\pair{v_1}{v_2}}{By inversion on \AlgChkValPair}
        \eqPf{\chi}{[\Delta', \hypeq{\ahat'}{\tau}{t_0}]\chi_1, \chi_2}{\ditto}
        \algchkPf{\Theta; \Delta, \ahat':\tau}{\Gamma}{v_1}{\comprehend{\nu:\mu F}{ \Fold{F}{\alpha}\,{\nu} =_\tau \ahat' }}{\chi_1}{\Delta''}{\ditto}
        \algchkPf{\Theta; \Delta''}{\Gamma}{v_2}{[\Delta'']([\ahat'/a']Q_0)}{\chi_2}{\Delta', \hypeq{\ahat'}{\tau}{t_0}}{\ditto}
        \decolumnizePf
        \judgefunctorPf{\Theta; \Delta}{F}{\Xi_F}{Presupposed derivation}
        \judgealgebraPf{\Xi}{\Theta; \Delta}{\alpha}{F}{\tau}{Presupposed derivation}
        \judgefunctorPf{\Theta; \Delta, \ahat':\tau}{F}{\Xi_F}{\Lemmaref{lem:alg-ctx-weakening}}
        \judgealgebraPf{\Xi}{\Theta; \Delta, \ahat':\tau}{\alpha}{F}{\tau}{\Lemmaref{lem:alg-ctx-weakening}}
        \judgetpPf{\Theta; \Delta, \ahat':\tau}{\comprehend{\nu:\mu F}{ \Fold{F}{\alpha}\,{\nu} =_\tau \ahat'}}{\Xi_F \cup \ahat':\tau}{By \AlgTpFixEVar}
        \judgetpPf{\Theta; \Delta, \ahat':\tau}{\comprehend{\nu:\mu F}{ \Fold{F}{\alpha}\,{\nu} =_\tau \ahat'}}{\Xi_F, \ahat':\tau}{$\ahat'$ not free in $F$}
        \decolumnizePf
        \semideclnegPf{\Theta}{\Gamma}{[\Omega]([\Delta', \hypeq{\ahat'}{\tau}{t_0}]\chi_1, \chi_2)}{Given}
        \semideclnegPf{\Theta}{\Gamma}{[\Omega]([\Delta', \hypeq{\ahat'}{\tau}{t_0}]\chi_1)}{Straightforward}
        \rextendPf{\Theta}{\Delta'}{\Omega}{Given}
        \rextendPf{\Theta}{\Delta', \hypeq{\ahat'}{\tau}{t_0}}{\Omega, \hypeq{\ahat'}{\tau}{t_0}}{By \Lemmaref{lem:extend-relaxed-reflexively}}
        \semideclnegPf{\Theta}{\Gamma}{[\Omega, \hypeq{\ahat'}{\tau}{t_0}]([\Delta', \hypeq{\ahat'}{\tau}{t_0}]\chi_1)}{By \Lemmaref{lem:apply-idempotent}}
        \decolumnizePf
        \chiequivPf{\Theta}{\Gamma}{[\Omega, \hypeq{\ahat'}{\tau}{t_0}]\chi_1}{[\Omega, \hypeq{\ahat'}{\tau}{t_0}]([\Delta', \hypeq{\ahat'}{\tau}{t_0}]\chi_1)}{By \Lemref{lem:probs-sandwich}}
        \chiequivPf{\Theta}{\Gamma}{[\Omega, \hypeq{\ahat'}{\tau}{t_0}]([\Delta', \hypeq{\ahat'}{\tau}{t_0}]\chi_1)}{[\Omega, \hypeq{\ahat'}{\tau}{t_0}]\chi_1}{By \Lemref{lem:chi-equiv-symmetric}}
        \decolumnizePf
        \semideclnegPf{\Theta}{\Gamma}{[\Omega, \hypeq{\ahat'}{\tau}{t_0}]\chi_1}{By \Lemmaref{lem:probs-equiv-respects-entail}}
        \decolumnizePf
        \rextendPf{\Theta}{\Delta''}{\Delta', \hypeq{\ahat'}{\tau}{t_0}}{By \Lemmaref{lem:typing-extends}}
        \rextendPf{\Theta}{\Delta''}{\Omega, \hypeq{\ahat'}{\tau}{t_0}}{By \Lemmaref{lem:ext-trans}}
      \end{llproof} 
      ~\\
      By the \ih (part (2)),
      $\Delta''$ solves all existential variables in $(\Xi_F, \ahat':\tau)$.
      Therefore, there exist $\Delta_0''$, $\Delta_1''$ and $t_0'$
      such that $\Delta'' = \Delta_0'', \hypeq{\ahat'}{\tau}{t_0'}, \Delta_1''$.
      ~\\
      \begin{llproof} 
        \extendPf{\Theta}{\Delta_0'', \hypeq{\ahat'}{\tau}{t_0'}, \Delta_1''}{\Delta', \hypeq{\ahat'}{\tau}{t_0}}{By \Lemmaref{lem:typing-extends}}
        \Pf{\Delta_1''}{=}{\cdot}{By inversion on extension}
        \eqPf{t_0'}{t_0}{\ditto}
        \extendPf{\Theta}{\Delta_0''}{\Delta'}{\ditto}
        \Pf{\Delta''}{=}{\Delta_0'', \hypeq{\ahat'}{\tau}{t_0}}{List append property}
        \extendPf{\Theta}{\Delta, \ahat':\tau}{\Delta_0'', \hypeq{\ahat'}{\tau}{t_0}}{By \Lemmaref{lem:typing-extends}}
        \extendPf{\Theta}{\Delta}{\Delta_0''}{By inversion}
        \decolumnizePf
        \algchkPf{\Theta; \Delta_0'', \hypeq{\ahat'}{\tau}{t_0}}{\Gamma}{v_2}{[\Delta_0'', \hypeq{\ahat'}{\tau}{t_0}]([\ahat'/a']Q_0)}{\dontcare}{\Delta', \hypeq{\ahat'}{\tau}{t_0}}{Rewrite above}
        \algchkPf{\Theta; \Delta_0'', \hypeq{\ahat'}{\tau}{t_0}}{\Gamma}{v_2}{[\Delta_0'']([t_0/\ahat']([\ahat'/a']Q_0))}{\dontcare}{\Delta', \hypeq{\ahat'}{\tau}{t_0}}{By \defn of \defsubst}
        \algchkPf{\Theta; \Delta_0'', \hypeq{\ahat'}{\tau}{t_0}}{\Gamma}{v_2}{[\Delta_0'']([t_0/a']Q_0)}{\dontcare}{\Delta', \hypeq{\ahat'}{\tau}{t_0}}{By property of subst.}
      \end{llproof} 
      ~\\
      By \Lemmaref{lem:typing-extends} and inversion on extension,
      if $\Delta_0''$ solves $\ahat$, then so does $\Delta'$, and we're done.
      We assume $\Delta_0''$ does not solve $\ahat$.\\

      Subderivation:
      \[
        \judgeunroll{\Xi,a':\tau}{\Theta, a':\tau; \Delta}{\nu:\hat{P}[\mu F]}{(\clause{q}{t'})}{\hat{P}\;\Fold{F}{\alpha}\;\nu}{\ahat}{Q_0}{\tau}
      \]
      By \Lemmaref{lem:syn-subs-unroll} but for the algorithmic unrolling judgment,
      \defn of \defsubst,
      and because $a'$ not free in $\hat{P}$ or $F$,
      \[
        \judgeunroll{\dontcare}{\Theta; \Delta}{\nu:\hat{P}[\mu F]}{(\clause{q}{[t_0/a']t'})}{\hat{P}\;\Fold{F}{\alpha}\;\nu}{\ahat}{[t_0/a']Q_0}{\tau}
      \]
      By \Lemmaref{lem:ext-weak-unroll},
      \[
        \judgeunroll{\dontcare}{\Theta; \Delta_0''}{\nu:\hat{P}[\mu F]}{(\clause{q}{[t_0/a']t'})}{\hat{P}\;\Fold{F}{\alpha}\;\nu}{\ahat}{[t_0/a']Q_0}{\tau}
      \]
      By \Lemmaref{lem:right-hand-subst-unroll}
      and since $\Delta_0''$ does not solve $\ahat$,
      \[
        \judgeunroll*{\dontcare}{\Theta; \Delta_0''}{\nu:[\Delta_0'']\hat{P}[\mu [\Delta_0'']F]}{(\clause{q}{[t_0/a']t'})}{[\Delta_0'']\hat{P}\;\Fold{[\Delta_0'']F}{\alpha}\;\nu}{\ahat}{[\Delta_0'']([t_0/a']Q_0)}{\tau}
      \]
      By \Lemmaref{lem:alg-ctx-weakening},
      \[
        \judgeunroll*{\dontcare}{\Theta; \Delta_0'', \hypeq{\ahat'}{\tau}{t_0}}{\nu:[\Delta_0'']\hat{P}[\mu [\Delta_0'']F]}{(\clause{q}{[t_0/a']t'})}{[\Delta_0'']\hat{P}\;\Fold{[\Delta_0'']F}{\alpha}\;\nu}{\ahat}{[\Delta_0'']([t_0/a']Q_0)}{\tau}
      \]
      \begin{llproof}
        \judgefunctorPf{\Theta; \Delta}{\Id \otimes \hat{P}}{\Xi_G}{Given}
        \judgefunctorPf{\Theta; \Delta}{\hat{P}}{\Xi_G}{By inversion}
        \judgefunctorPf{\Theta; \Delta_0''}{\hat{P}}{\Xi_G}{By \Lemmaref{lem:ext-weak-tp}}
        \judgefunctorPf{\Theta; \Delta_0''}{[\Delta_0'']\hat{P}}{\Xi_G'}{By \Lemmaref{lem:right-hand-subst}}
        \supseteqPf{\Xi_G'}{[\Delta_0'']\Xi_G}{\ditto}
        \Pf{(\ahat:\tau)}{\notin}{\Xi_G' \setminus [\Delta_0'']\Xi_G}{\ditto}
        \Pf{(\ahat:\tau)}{\notin}{\Xi_G}{Given}
        \Pf{(\ahat:\tau)}{\notin}{[\Delta_0'']\Xi_G}{By (obvious extension of) \Defnref{def:Xi-subs}}
        \trailingjust{(solutions in $\Delta_0''$ ground)}
        \Pf{(\ahat:\tau)}{\notin}{\Xi_G'}{By set theory}
        \Pf{(\hypeq{\ahat}{\tau}{t})}{\in}{(\Delta', \hypeq{\ahat'}{\tau}{t_0})}{By \ih (part (1))}
        \Hand\judgetermPf{\Theta}{t}{\tau}{\ditto}
        \Hand\Pf{(\hypeq{\ahat}{\tau}{t})}{\in}{\Delta'}{$\because\ahat \neq \ahat'$}
      \end{llproof} 

      \item \textbf{Case} $G = \Const{P} \otimes \hat{P}$\\
        \begin{llproof}
          \judgeunrollPf{\Xi}{\Theta; \Delta}{\nu:(\Const{P} \otimes \hat{P})[\mu F]}{\beta}{(\Const{P} \otimes \hat{P})\; \Fold{F}{\alpha}\;\nu}{\ahat}{Q}{\tau}{Given}
          \Pf{\beta}{=}{(\clause{q}{t'})}{By inversion}
          \judgeunrollPf{\Xi}{\Theta; \Delta}{\nu:(\Const{P} \otimes \hat{P})[\mu F]}{(\clause{q}{t'})}{(\Const{P} \otimes \hat{P})\; \Fold{F}{\alpha}\;\nu}{\ahat}{Q}{\tau}{By equation}
          \judgeunrollPf{\Xi, \Xi'}{\Theta, \Xi'; \Delta}{\nu:\hat{P}[\mu F]}{(\clause{q'}{t'})}{\hat{P}\; \Fold{F}{\alpha}\;\nu}{\ahat}{Q'}{\tau}{By \Lemref{lem:pack-unroll-inversion}}
          \judgetpPf{\Theta, \Xi'; \Delta}{P'}{\dontcare, \Xi'}{\ditto}
          \Pf{Q}{=}{\extype{\Xi'} P' \times Q'}{\ditto}
          \decolumnizePf
          \algchkPf{\Theta; \Delta}{\Gamma}{v}{\extype{\Xi'} P' \times Q'}{\chi}{\Delta'}{Rewrite given}
          \algchkPf{\Theta; \Delta, \Delta_{\Xi'}}{\Gamma}{v}{[\Delta_{\Xi'}/\Xi'](P' \times Q')}{\chi}{\Delta', \Omega_{\Xi'}}{By \Lemref{lem:aux-inversion}}
          \algchkPf{\Theta; \Delta, \Delta_{\Xi'}}{\Gamma}{v}{[\Delta_{\Xi'}/\Xi']P' \times [\Delta_{\Xi'}/\Xi']Q'}{\chi}{\Delta', \Omega_{\Xi'}}{By \defn of \defsubst}
          \decolumnizePf
          \judgetpPf{\Theta; \Delta, \Delta_{\Xi'}}{[\Delta_{\Xi'}/\Xi']P'}{\dontcare, \Delta_{\Xi'}}{By \Lemmaref{lem:value-determined-evar-rename}}
          \trailingjust{(repeated)}
          \extendPf{\Theta}{\Delta, \Delta_{\Xi'}}{\Delta', \Omega_{\Xi'}}{By \Lemmaref{lem:typing-extends}}
          \extendPf{\Theta}{\Delta_{\Xi'}}{\Omega_{\Xi'}}{By inversion on extension}
        \end{llproof}

        \begin{itemize}
          \DerivationProofCase{\AlgChkValVar}
          {
            (P' \times Q') \neq \exists, \with
            \\
            (x:\tilde{Q})\in\Gamma
            \\
            \algsub[+]{\Theta; \Delta, \Delta_{\Xi'}}{\tilde{Q}}{[\Delta_{\Xi'}/\Xi']P' \times [\Delta_{\Xi'}/\Xi']Q'}{\Wah}{\Delta', \Omega_{\Xi'}}
          }
          {
            \algchk{\Theta; \Delta, \Delta_{\Xi'}}{\Gamma}{\underbrace{x}_v}{[\Delta_{\Xi'}/\Xi']P' \times [\Delta_{\Xi'}/\Xi']Q'}{\underbrace{\Wah, \cdot}_\chi}{\Delta', \Omega_{\Xi'}}
          }
          \begin{llproof}
            \algsubPf[+]{\Theta; \Delta, \Delta_{\Xi'}}{\tilde{Q}}{[\Delta_{\Xi'}/\Xi']P' \times [\Delta_{\Xi'}/\Xi']Q'}{\Wah}{\Delta', \Omega_{\Xi'}}{Subderivation}
            \eqPf{\tilde{Q}}{\tilde{Q_1}\times\tilde{Q_2}}{By inversion on \AlgSubPosProd}
            \eqPf{W}{[\Delta', \Omega_{\Xi'}]W_1 \land W_2}{\ditto}
            \algsubPf[+]{\Theta; \Delta, \Delta_{\Xi'}}{\tilde{Q_1}}{[\Delta_{\Xi'}/\Xi']P'}{W_1}{\Delta''}{\ditto}
            \algsubPf[+]{\Theta; \Delta''}{\tilde{Q_2}}{[\Delta'']([\Delta_{\Xi'}/\Xi']Q')}{W_2}{\Delta', \Omega_{\Xi'}}{\ditto}
            \judgetpPf{\Theta; \Delta, \Delta_{\Xi'}}{[\Delta_{\Xi'}/\Xi']P'}{\dontcare, \Delta_{\Xi'}}{Above}
            \decolumnizePf
            \ForallPf{(\ahat_0:\tau_0)}{\Delta_{\Xi'},~\text{there exists}~(\hypeq{\ahat_0}{\tau_0}{t_0}) \in \Delta''}{By \Lemmaref{lem:sub-solves-val-det}}
            \extendPf{\Theta}{\Delta''}{\Delta', \Omega_{\Xi'}}{By \Lemmaref{lem:alg-sub-extends}}
            \eqPf{\Delta''}{\Delta_0'', \Omega_{\Xi'}}{By inversion on extension}
            \extendPf{\Theta}{\Delta_0''}{\Delta'}{\ditto}
            \extendPf{\Theta}{\Delta, \Delta_{\Xi'}}{\Delta_0'', \Omega_{\Xi'}}{By \Lemmaref{lem:alg-sub-extends}}
            \extendPf{\Theta}{\Delta}{\Delta_0''}{By inversion on extension}
            \decolumnizePf
            \judgeunrollPf{\Xi, \Xi'}{\Theta, \Xi'; \Delta}{\nu:\hat{P}[\mu F]}{(\clause{q'}{t'})}{\hat{P}\; \Fold{F}{\alpha}\;\nu}{\ahat}{Q'}{\tau}{Above}
          \end{llproof} 

          By the obvious algorithmic version of \Lemmaref{lem:syn-subs-unroll},
          \[
            \judgeunroll{\dontcare}{\Theta; \Delta}{\nu:\hat{P}[\mu F]}{(\clause{q'}{[\Omega_{\Xi'}/\Xi']t'})}{\hat{P}\; \Fold{F}{\alpha}\;\nu}{\ahat}{[\Omega_{\Xi'}/\Xi']Q'}{\tau}
          \]

          By \Lemmaref{lem:ext-weak-unroll},
          \[
            \judgeunroll{\dontcare}{\Theta; \Delta_0''}{\nu:\hat{P}[\mu F]}{(\clause{q'}{[\Omega_{\Xi'}/\Xi']t'})}{\hat{P}\; \Fold{F}{\alpha}\;\nu}{\ahat}{[\Omega_{\Xi'}/\Xi']Q'}{\tau}
          \]
          By \Lemmaref{lem:right-hand-subst-unroll} and \defn of \defsubst and $\because \ground{\alpha}$,
          \[
            \judgeunroll*{\dontcare}{\Theta; \Delta_0''}{\nu:[\Delta_0'']\hat{P}[\mu [\Delta_0'']F]}{(\clause{q'}{[\Omega_{\Xi'}/\Xi']t'})}{[\Delta_0'']\hat{P}\; \Fold{[\Delta_0'']F}{\alpha}\;\nu}{[\Delta_0'']\ahat}{[\Delta_0'']([\Omega_{\Xi'}/\Xi']Q')}{\tau}
          \]
          By \Lemmaref{lem:alg-ctx-weakening},
          \[
            \judgeunroll*{\dontcare}{\Theta; \underbrace{\Delta_0'', \Omega_{\Xi'}}_{\Delta''}}{\nu:[\Delta_0'']\hat{P}[\mu [\Delta_0'']F]}{(\clause{q'}{[\Omega_{\Xi'}/\Xi']t'})}{[\Delta_0'']\hat{P}\; \Fold{[\Delta_0'']F}{\alpha}\;\nu}{[\Delta_0'']\ahat}{[\Delta_0'']([\Omega_{\Xi'}/\Xi']Q')}{\tau}
          \]

          If $\ahat$ is solved in $\Delta''$,
          then by inversion on 
          $\extend{\Theta}{\Delta''}{(\Delta', \Omega_{\Xi'})}$
          it is solved in $\Delta'$ (as desired).
          Suppose $\ahat$ is unsolved in $\Delta''$;
          then it is unsolved in $\Delta_0''$.
          Therefore, $[\Delta_0'']\ahat = \ahat$, so
          \[
            \judgeunroll*{\dontcare}{\Theta; \underbrace{\Delta_0'', \Omega_{\Xi'}}_{\Delta''}}{\nu:[\Delta_0'']\hat{P}[\mu [\Delta_0'']F]}{(\clause{q'}{[\Omega_{\Xi'}/\Xi']t'})}{[\Delta_0'']\hat{P}\; \Fold{[\Delta_0'']F}{\alpha}\;\nu}{\ahat}{[\Delta_0'']([\Omega_{\Xi'}/\Xi']Q')}{\tau}
          \]

          \begin{llproof}
            \algsubPf[+]{\Theta; \Delta''}{\tilde{Q_2}}{[\Delta'']([\Delta_{\Xi'}/\Xi']Q')}{W_2}{\Delta', \Omega_{\Xi'}}{Above}
            \algsubPf[+]{\Theta; \Delta''}{\tilde{Q_2}}{[\Delta_0'']([\Omega_{\Xi'}/\Xi']Q')}{W_2}{\Delta', \Omega_{\Xi'}}{By subst.\ props.}
            \trailingjust{(using \Defnref{def:solve-evar-rename})}
            \simplePf{\Theta}{\tilde{Q_1} \times \tilde{Q_2}}{By inversion on $\Gamma$ WF}
            \groundPf{\tilde{Q_1} \times \tilde{Q_2}}{\ditto}
            \simplePf{\Theta}{\tilde{Q_2}}{Straightforward}
            \groundPf{\tilde{Q_2}}{Straightforward}
            \decolumnizePf
            \inPf{\hypeq{\ahat}{\tau}{t}}{\Delta', \Omega_{\Xi'}}{By \Lemmaref{lem:sub-solves-unrolled-evar}}
            \Hand\judgetermPf{\Theta}{t}{\tau}{\ditto}
            \Hand\inPf{\hypeq{\ahat}{\tau}{t}}{\Delta'}{Because $\ahat \notin \dom{\Omega_{\Xi'}}$}
          \end{llproof}

          \DerivationProofCase{\AlgChkValPair}
          {
            \arrayenvbl{
              \algchk{\Theta; \Delta, \Delta_{\Xi'}}{\Gamma}{v_1}{[\Delta_{\Xi'}/\Xi']P'}{\chi_1}{\Delta''}
              \\
              \algchk{\Theta; \Delta''}{\Gamma}{v_2}{[\Delta'']([\Delta_{\Xi'}/\Xi']Q')}{\chi_2}{\Delta', \Omega_{\Xi'}}
            }
          }
          {
            \algchk{\Theta; \Delta, \Delta_{\Xi'}}{\Gamma}{\underbrace{\pair{v_1}{v_2}}_v}{[\Delta_{\Xi'}/\Xi']P' \times [\Delta_{\Xi'}/\Xi']Q'}{\underbrace{[\Delta', \Omega_{\Xi'}]\chi_1, \chi_2}_\chi}{\Delta', \Omega_{\Xi'}}
          }
          \begin{llproof}
            \algchkPf{\Theta; \Delta, \Delta_{\Xi'}}{\Gamma}{v_1}{[\Delta_{\Xi'}/\Xi']P'}{\chi_1}{\Delta''}{Subderivation}
            \judgetpPf{\Theta; \Delta, \Delta_{\Xi'}}{[\Delta_{\Xi'}/\Xi']P'}{\dontcare, \Delta_{\Xi'}}{Above}
            \decolumnizePf
            \eqPf{[\Delta](\extype{\Xi'} P' \times Q')}{\extype{\Xi'} P' \times Q'}{Given}
            \eqPf{[\Delta]P'}{P'}{Straightforward}
            \eqPf{[\Delta, \Delta_{\Xi'}]([\Delta_{\Xi'}/\Xi']P')}{[\Delta_{\Xi'}/\Xi']P'}{Straightforward}
            \decolumnizePf
            \algchkPf{\Theta; \Delta''}{\Gamma}{v_2}{[\Delta'']([\Delta_{\Xi'}/\Xi']Q')}{\chi_2}{\Delta', \Omega_{\Xi'}}{Subderivation}
            \extendPf{\Theta}{\Delta''}{\Delta', \Omega_{\Xi'}}{By \Lemmaref{lem:alg-equiv-extends}}
            \rextendPf{\Theta}{\Delta''}{\Delta', \Omega_{\Xi'}}{By \Lemmaref{lem:extension-sound}}
            \rextendPf{\Theta}{\Delta'}{\Omega}{Given}
            \decolumnizePf
            \rextendPf{\Theta}{\Delta', \Omega_{\Xi'}}{\Omega, \Omega_{\Xi'}}{By \Lemmaref{lem:extend-relaxed-reflexively}}
            \rextendPf{\Theta}{\Delta''}{\Omega, \Omega_{\Xi'}}{By \Lemmaref{lem:ext-trans}}
            \decolumnizePf
            \semideclnegPf{\Theta}{\Gamma}{[\Omega]([\Delta', \Omega_{\Xi'}]\chi_1, \chi_2)}{Given}
            \semideclnegPf{\Theta}{\Gamma}{[\Omega]([\Delta', \Omega_{\Xi'}]\chi_1), [\Omega]\chi_2}{By \defn of \defsubst}
            \semideclnegPf{\Theta}{\Gamma}{[\Omega]([\Delta', \Omega_{\Xi'}]\chi_1)}{Straightforward}
            \semideclnegPf{\Theta}{\Gamma}{[\Omega, \Omega_{\Xi'}]([\Delta', \Omega_{\Xi'}]\chi_1)}{By \Lemmaref{lem:apply-idempotent}}
            \chiequivPf{\Theta}{\Gamma}{[\Omega, \Omega_{\Xi'}]\chi_1}{[\Omega, \Omega_{\Xi'}]([\Omega, \Delta']\chi_1)}{By \Lemmaref{lem:probs-sandwich}}
            \chiequivPf{\Theta}{\Gamma}{[\Omega, \Omega_{\Xi'}]([\Omega, \Delta']\chi_1)}{[\Omega, \Omega_{\Xi'}]\chi_1}{By \Lemmaref{lem:chi-equiv-symmetric}}
            \semideclnegPf{\Theta}{\Gamma}{[\Omega, \Omega_{\Xi'}]\chi_1}{By \Lemmaref{lem:probs-equiv-respects-entail}}
            \decolumnizePf
            \ForallPf{(\ahat:\tau)}{\Delta_{\Xi'},~\text{there exists}~(\hypeq{\ahat}{\tau}{t}) \in \Delta''}{By \ih (part (2))}
            \decolumnizePf
            \extendPf{\Theta}{\Delta''}{\Delta', \Omega_{\Xi'}}{By \Lemmaref{lem:typing-extends}}
            \eqPf{\Delta''}{\Delta_0'', \Omega_{\Xi'}}{By inversion}
            \extendPf{\Theta}{\Delta_0''}{\Delta'}{\ditto}
            \extendPf{\Theta}{\Delta, \Delta_{\Xi'}}{\Delta_0'', \Omega_{\Xi'}}{By \Lemmaref{lem:typing-extends}}
            \extendPf{\Theta}{\Delta}{\Delta_0''}{By inversion}
            \extendPf{\Theta}{\Delta_{\Xi'}}{\Omega_{\Xi'}}{\ditto}
            \algchkPf{\Theta; \Delta''}{\Gamma}{v_2}{[\Delta_0'', \Omega_{\Xi'}]([\Delta_{\Xi'}/\Xi']Q')}{\dontcare}{\Delta', \Omega_{\Xi'}}{Rewrite above}
            \algchkPf{\Theta; \Delta''}{\Gamma}{v_2}{[\Delta_0'']([\Omega_{\Xi'}/\Xi']Q')}{\dontcare}{\Delta', \Omega_{\Xi'}}{By subst.\ properties}
          \end{llproof}

          If $\ahat$ is solved in $\Delta''$,
          then by inversion on 
          $\extend{\Theta}{\Delta''}{(\Delta', \Omega_{\Xi'})}$
          it is solved in $\Delta'$ (as desired).
          Suppose $\ahat$ is unsolved in $\Delta''$;
          then it is unsolved in $\Delta_0''$.
          Therefore, $[\Delta_0'']\ahat = \ahat$,
          so by the same reasoning in the previous subcase, that is \AlgChkValVar,
          we have:
          \[
            \judgeunroll*{\dontcare}{\Theta; \underbrace{\Delta_0'', \Omega_{\Xi'}}_{\Delta''}}{\nu:[\Delta_0'']\hat{P}[\mu [\Delta_0'']F]}{(\clause{q'}{[\Delta_0'']([\Omega_{\Xi'}/\Xi']t')})}{[\Delta_0'']\hat{P}\; \Fold{[\Delta_0'']F}{\alpha}\;\nu}{\ahat}{[\Delta_0'']([\Omega_{\Xi'}/\Xi']Q')}{\tau}
          \]
          \begin{llproof}
            \judgefunctorPf{\Theta; \Delta}{\Const{P} \otimes \hat{P}}{\Xi_G}{Given}
            \judgefunctorPf{\Theta; \Delta}{\hat{P}}{\Xi_{\hat{P}}}{By inversion}
            \judgefunctorPf{\Theta; \Delta_0''}{\hat{P}}{\Xi_{\hat{P}}}{By \Lemmaref{lem:ext-weak-tp}}
            \judgefunctorPf{\Theta; \Delta_0''}{[\Delta_0'']\hat{P}}{\Xi_{\hat{P}}'}{By \Lemmaref{lem:right-hand-subst}}
            \supseteqPf{\Xi_{\hat{P}}'}{[\Delta_0'']\Xi_{\hat{P}}}{\ditto}
            \notinPf{(\ahat:\tau)}{\Xi_{\hat{P}}' \setminus [\Delta_0'']\Xi_{\hat{P}}}{\ditto}
            \notinPf{(\ahat:\tau)}{\Xi_G}{Given}
            \subseteqPf{\Xi_{\hat{P}}}{\Xi_G}{By inversion}
            \notinPf{(\ahat:\tau)}{\Xi_{\hat{P}}}{By set theory}
            \notinPf{(\ahat:\tau)}{[\Delta_0'']\Xi_{\hat{P}}}{By (obvious extension of) \Defnref{def:Xi-subs}}
            \trailingjust{(does not add evars.)}
            \notinPf{(\ahat:\tau)}{\Xi_{\hat{P}}'}{By set theory}
            \inPf{(\hypeq{\ahat}{\tau}{t})}{(\Delta', \Omega_{\Xi'})}{By \ih (part (1))}
            \Hand\judgetermPf{\Theta}{t}{\tau}{\ditto}
            \Hand\inPf{(\hypeq{\ahat}{\tau}{t})}{\Delta'}{$\because\ahat \notin \dom{\Omega_{\Xi'}}$}
          \end{llproof}
        \end{itemize}

      \DerivationProofCase{\AlgUnrollUnit}
      { }
      { \judgeunroll{\Xi}{\Theta; \Delta}{\nu:I[\mu F]}{(\clause{\unitexp}{t'})}{I\;\Fold{F}{\alpha}\;\nu}{\ahat}{1 \land (\ahat = t')}{\tau} }
      \begin{llproof}
        \groundPf{t'}{By \Lemmaref{lem:algebras-are-ground}}
        \algchkPf{\Theta; \Delta}{\Gamma}{v}{1 \land (\ahat = t')}{\dontcare}{\Delta'}{Given}
        \algchkPf{\Theta; \Delta}{\Gamma}{v}{1}{\dontcare}{\Delta''}{By inversion on \AlgChkValWith}
        \alginstPf{\Theta; \Delta''}{[\Delta''](\ahat = t')}{\Delta'}{\ditto}
        \Pf{\Delta''}{=}{\Delta}{By inversion on \AlgChkValUnit}
        \Pf{[\Delta''](\ahat = t')}{=}{[\Delta](\ahat = t')}{By equation}
        \Pf{}{=}{(\ahat = t')}{$\because (\ahat:\tau)\in\Delta$ and $\ground{t'}$}
        \alginstPf{\Theta; \Delta}{(\ahat = t')}{\Delta'}{By equations}
        \Hand\Pf{\Delta'}{=}{\Delta_1, \hypeq{\ahat}{\tau}{t'}, \Delta_2}{By inversion on \RuleInst}
        \Hand\judgetermPf{\Theta}{t'}{\tau}{\ditto}
      \end{llproof}
    \end{itemize}
  \item
    By structural induction on $v$, case analyzing the given value typing judgment.
    Use \Lemmaref{lem:sub-solves-val-det} in the \AlgChkValVar case.
    We show the hardest case:
    \begin{itemize}
      \DerivationProofCase{\AlgChkValFix}
      {
        \judgeunroll{\cdot}{\Theta; \Delta}{\nu:F[\mu F]}{\alpha}{F\; \Fold{F}{\alpha}\;\nu}{t}{Q}{\tau}
        \\
        \algchk{\Theta; \Delta}{\Gamma}{v_0}{Q}{\chi}{\Delta'}
      }
      {
        \algchk{\Theta; \Delta}{\Gamma}{\into{v_0}}{\comprehend{\nu : \mu F}{\Fold{F}{\alpha}\,{\nu} =_\tau t}}{\chi}{\Delta'}
      }
      \begin{llproof}
        \judgeunrollPf{\cdot}{\Theta; \Delta}{\nu:F[\mu F]}{\alpha}{F\; \Fold{F}{\alpha}\;\nu}{t}{Q}{\tau}{Subderivation}
        \judgetpPf{\Theta; \Delta}{\comprehend{\nu : \mu F}{\Fold{F}{\alpha}\,{\nu} =_\tau t}}{\dontcare}{Presupposed derivation}
        \judgefunctorPf{\Theta; \Delta}{F}{\Xi_F}{By inversion}
        \decolumnizePf
        \judgetpPf{\Theta; \Delta}{Q}{\Xi_Q}{By \Lemmaref{lem:alg-unroll-output-wf}}
        \Pf{\Xi_F}{\subseteq}{\Xi_Q}{\ditto}
        \algchkPf{\Theta; \Delta}{\Gamma}{v_0}{Q}{\chi}{\Delta'}{Subderivation}
        \decolumnizePf
        \eqPf{[\Delta]\comprehend{\nu : \mu F}{\Fold{F}{\alpha}\,{\nu} =_\tau t}}{\comprehend{\nu : \mu F}{\Fold{F}{\alpha}\,{\nu} =_\tau t}}{Given}
        \decolumnizePf
        \eqPf{[\Delta]F}{F}{Straightforward}
        \eqPf{[\Delta]t}{t}{Straightforward}
        \eqPf{[\Delta]Q}{Q}{By \Lemmaref{lem:unroll-applied}}
      \end{llproof}

      By the \ih for part (2), for all $(\ahat : \tau) \in \Xi_Q$,
      there exists $t_0$ such that $(\hypeq{\ahat}{\tau}{t_0}) \in \Delta'$.
      Because $\Xi_F \subseteq \Xi_Q$, for all $(\ahat : \tau) \in \Xi_F$,
      there exists $t_0$ such that $(\hypeq{\ahat}{\tau}{t_0}) \in \Delta'$.
      If $t$ is not an existential variable
      or it is an existential variable in $\dom{\Xi_Q}$, then we are done.
      Suppose $t$ is an existential variable $\ahat$ not in $\dom{\Xi_Q}$
      (hence not in $\dom{\Xi_F}$);
      this with \AlgTpFixEVar implies $\Xi_P = (\Xi_F, \ahat:\tau)$.
      To complete the proof,
      it suffices to show there exists $t_0$
      such that $\judgeterm{\Theta}{t}{\tau}$
      and $(\hypeq{\ahat}{\tau}{t_0}) \in \Delta'$.
      But this follows from the \ih for part (1)
      with the relevant information above
      and the given $\rextend{\Theta}{\Delta}{\Omega}$
      and $\semideclneg{\Theta}{\Gamma}{[\Omega]\chi}$.
    \end{itemize}
  \item By structural induction on $s$,
    case analyzing the given spine typing judgment.
    Use part (2) in the \AlgSpineApp case.
    \qedhere
  \end{enumerate}
\end{proof}

\begin{lemma}[Aux.\ Alg.\ Typing Complete]
  \label{lem:aux-alg-typing-complete}
  ~
  \begin{enumerate}
    \item
      If $\semideclneg{\Theta}{\Gamma}{\chi}$,
      then $\algneg{\Theta}{\Gamma}{\chi}$.
    \item
      If $\semideclsynhead{\Theta}{\Gamma}{h}{P}$,
      then $\algsynhead{\Theta}{\Gamma}{h}{P}$.
    \item
      If $\semideclsynexp{\Theta}{\Gamma}{\be}{\upshift{P}}$,
      then there exists $P'$
      such that $\algsynexp{\Theta}{\Gamma}{\be}{\upshift{P'}}$
      and $\judgeequiv[+]{\Theta}{P}{P'}$.
    \item
      If $\semideclchkval{\Theta}{\Gamma}{v}{[\Omega]P}{\chi}$
      and $\judgetp{\Theta; \Delta}{P}{\Xi}$
      and $[\Delta]P = P$
      and $\semideclneg{\Theta}{\Gamma}{\chi}$
      and $\rextend{\Theta}{\Delta}{\Omega}$,
      then there exist $\chi'$ and $\Delta'$
      such that $\algchk{\Theta; \Delta}{\Gamma}{v}{P}{\chi'}{\Delta'}$
      and $\rextend{\Theta}{\Delta'}{\Omega}$
      and $\chiequiv{\Theta}{\Gamma}{\chi}{[\Omega]\chi'}$.
    \item
      If $\semideclchkexp{\Theta}{\Gamma}{e}{N}$,
      then $\algchkneg{\Theta}{\Gamma}{e}{N}$.
    \item
      If $\semideclchkmatch{\Theta}{\Gamma}{P}{\clauses{\pa}{e}{i}{I}}{N}$,
      then $\algchkmatch{\Theta}{\Gamma}{P}{\clauses{\pa}{e}{i}{I}}{N}$.
    \item
      If $\semideclspine{\Theta}{\Gamma}{s}{[\Omega]N}{\upshift{P}}{\chi}$
      and $\judgetp{\Theta; \Delta}{N}{\Xi}$
      and $[\Delta]N = N$
      and $\semideclneg{\Theta}{\Gamma}{\chi}$
      and $\rextend{\Theta}{\Delta}{\Omega}$,
      then there exist $P'$, $\chi'$, and $\Delta'$
      such that $\algspine{\Theta; \Delta}{\Gamma}{s}{N}{\upshift{P'}}{\chi'}{\Delta'}$
      and $\rextend{\Theta}{\Delta'}{\Omega}$
      and $\chiequiv{\Theta}{\Gamma}{\chi}{[\Omega]\chi'}$
      and $\judgeequiv[+]{\Theta}{P}{[\Omega]P'}$.
  \end{enumerate}
\end{lemma}
\begin{proof}
  By lexicographic induction on,
  first, the size (\Figureref{fig:size-prog-chi}) of $\chi$
  or the size of the program term of the typing derivation
  ($h$, $\be$, $v$, $e$, $\clauses{\pa}{e}{i}{I}$, or $s$),
  and, second, the size (\Figureref{fig:size}) of the type
  $P$ in parts (4) and (6),
  $N$ in parts (5) and (7),
  or the number of $W$ constraints in the input $\chi$ of part (1).
  All parts are mutually recursive.
  \begin{enumerate}
  \item
    \begin{itemize}
      \DerivationProofCase{\SemiChkProblemsEmpty}
      {
      }
      {
        \semideclneg{\Theta}{\Gamma}{\cdot}
      }
      \begin{llproof}
        \algnegPf{\Theta}{\Gamma}{\cdot}{By \ChkProblemsEmpty}
      \end{llproof}

      \DerivationProofCase{\SemiChkProblemsNegChk}
      {
        \semideclchkexp{\Theta}{\Gamma}{e}{N}
        \\
        \semideclneg{\Theta}{\Gamma}{\chi_0}
      }
      {
        \semideclneg{\Theta}{\Gamma}{\underbrace{(e <= N), \chi_0}_{\chi}}
      }
      \begin{llproof}
        \semideclchkexpPf{\Theta}{\Gamma}{e}{N}{Subderivation}
        \algchknegPf{\Theta}{\Gamma}{e}{N}{By \ih ($\size{e} < \size{\chi}$)}
        \semideclnegPf{\Theta}{\Gamma}{\chi_0}{Subderivation}
        \algnegPf{\Theta}{\Gamma}{\chi_0}{By \ih ($\size{e} < \size{\chi}$)}
        \algnegPf{\Theta}{\Gamma}{(e <= N), \chi_0}{By \ChkProblemsNegChk}
      \end{llproof}

      \DerivationProofCase{\SemiChkProblemsWah}
      {
        \semideclentailwah{\Theta}{W}
        \\
        \semideclneg{\Theta}{\Gamma}{\chi_0}
      }
      {
        \semideclneg{\Theta}{\Gamma}{W, \chi_0}
      }
      \begin{llproof}
        \semideclentailwahPf{\Theta}{W}{Subderivation}
        \entailwahPf{\Theta}{W}{By \Lemmaref{lem:alg-entail-complete}}
        \semideclnegPf{\Theta}{\Gamma}{\chi_0}{Subderivation}
        \algnegPf{\Theta}{\Gamma}{\chi_0}{By \ih ($\size{\chi_0} = \size{\chi}$, but $\chi_0$ has one less $W$)}
        \algnegPf{\Theta}{\Gamma}{W, \chi_0}{By \ChkProblemsWah}
      \end{llproof}
    \end{itemize}
  \item
    \begin{itemize}
      \DerivationProofCase{\SemiDeclSynHeadVar}
      {
        (x : P) \in \Gamma
      }
      {
        \semideclsynhead{\Theta}{\Gamma}{x}{P}
      }
      \begin{llproof}
        \Pf{(x:P)}{\in}{\Gamma}{Subderivation}
        \algsynheadPf{\Theta}{\Gamma}{x}{P}{By \AlgSynHeadVar}
      \end{llproof}

      \DerivationProofCase{\SemiDeclSynValAnnot}
      {
        \judgetp{\Theta}{P}{\Xi}
        \\
        \semideclchkval{\Theta}{\Gamma}{v}{P}{\chi}
        \\
        \semideclneg{\Theta}{\Gamma}{\chi}
      }
      {
        \semideclsynhead{\Theta}{\Gamma}{\annoexp{v}{P}}{P}
      }
      \begin{llproof}
        \semideclchkvalPf{\Theta}{\Gamma}{v}{P}{\chi}{Subderivation}
        \judgetpPf{\Theta}{P}{\Xi}{Subderivation}
        \judgetpPf{\Theta; \cdot}{P}{\Xi}{\Lemmaref{lem:decl-to-ground-alg-wf}}
        \Pf{[\cdot]P}{=}{P}{By \defn of \defsubst}
        \rextendPf{\Theta}{\cdot}{\cdot}{By \RExtEmpty}
        \semideclchkvalPf{\Theta}{\Gamma}{v}{[\cdot]P}{\chi}{By equation}
        \semideclnegPf{\Theta}{\Gamma}{\chi}{Subderivation}
        \algchkPf{\Theta; \cdot}{\Gamma}{v}{P}{\chi'}{\Delta'}{By \ih}
        \rextendPf{\Theta}{\Delta'}{\cdot}{\ditto}
        \chiequivPf{\Theta}{\Gamma}{\chi}{[\cdot]\chi'}{\ditto}
        \chiequivPf{\Theta}{\Gamma}{\chi}{\chi'}{By \defn of \defsubst}
        \semideclnegPf{\Theta}{\Gamma}{\chi'}{By \Lemmaref{lem:probs-equiv-respects-entail}}
        \Pf{\size{\chi'}}{\leq}{\size{v}}{By \Lemmaref{lem:program-typing-shrinks-problems}}
        \Pf{}{<}{\size{v} + 1}{}
        \Pf{}{=}{\size{\annoexp{v}{P}}}{By \defn of \defsize}
        \algnegPf{\Theta}{\Gamma}{\chi'}{By \ih}
        \Pf{\Delta'}{=}{\cdot}{By inversion on relaxed extension}
        \algchkPf{\Theta; \cdot}{\Gamma}{v}{P}{\chi'}{\cdot}{By equation}
        \Hand\algsynheadPf{\Theta}{\Gamma}{\annoexp{v}{P}}{P}{By \AlgSynValAnnot}
      \end{llproof}
    \end{itemize}
  \item
    \begin{itemize}
      \DerivationProofCase{\SemiDeclSynSpineApp}
      {
        \semideclsynhead{\Theta}{\Gamma}{h}{\downshift{N}}
        \\
        \semideclspine{\Theta}{\Gamma}{s}{N}{\upshift{P}}{\chi}
        \\
        \semideclneg{\Theta}{\Gamma}{\chi}
      }
      {
        \semideclsynexp{\Theta}{\Gamma}{h(s)}{\upshift{P}}
      }
      \begin{llproof}
        \semideclsynheadPf{\Theta}{\Gamma}{h}{\downshift{N}}{Subderivation}
        \algsynheadPf{\Theta}{\Gamma}{h}{\downshift{N}}{By \ih}
        \semideclspinePf{\Theta}{\Gamma}{s}{N}{\upshift{P}}{\chi}{Subderivation}
        \Pf{[\cdot]N}{=}{N}{By \defn of \defsubst}
        \judgespinePf{\Theta}{\Gamma}{s}{[\cdot]N}{\upshift{P}}{By equation}
        \judgetpPf{\Theta}{\downshift{N}}{\dontcare}{Output WF}
        \judgetpPf{\Theta}{N}{\dontcare}{By inversion}
        \judgetpPf{\Theta; \cdot}{N}{\dontcare}{By \Lemmaref{lem:decl-to-ground-alg-wf}}
        \rextendPf{\Theta}{\cdot}{\cdot}{By \RExtEmpty}
        \semideclnegPf{\Theta}{\Gamma}{\chi}{Subderivation}
        \algspinePf{\Theta; \cdot}{\Gamma}{s}{N}{\upshift{P'}}{\chi'}{\Delta'}{By \ih}
        \rextendPf{\Theta}{\Delta'}{\cdot}{\ditto}
        \chiequivPf{\Theta}{\Gamma}{\chi}{[\cdot]\chi'}{\ditto}
        \judgeequivPf[+]{\Theta}{P}{[\cdot]P'}{\ditto}
        \Hand\judgeequivPf[+]{\Theta}{P}{P'}{By \defn of \defsubst}
        \chiequivPf{\Theta}{\Gamma}{\chi}{\chi'}{By \defn of \defsubst}
        \Pf{\Delta'}{=}{\cdot}{By inversion on relaxed extension}
        \algspinePf{\Theta; \cdot}{\Gamma}{s}{N}{\upshift{P'}}{\chi'}{\cdot}{By equation}
        \semideclnegPf{\Theta}{\Gamma}{\chi'}{By \Lemmaref{lem:probs-equiv-respects-entail}}
        \Pf{\size{\chi'}}{\leq}{\size{s}}{By \Lemmaref{lem:program-typing-shrinks-problems}}
        \Pf{}{<}{\size{h} + \size{s} + 1}{}
        \Pf{}{=}{\size{h(s)}}{By \defn of \defsize}
        \algnegPf{\Theta}{\Gamma}{\chi'}{By \ih}
        \Hand\algsynexpPf{\Theta}{\Gamma}{h(s)}{\upshift{P'}}{By \AlgSynSpineApp}
      \end{llproof}

      \ProofCaseRule{\SemiDeclSynExpAnnot}
      Straightforward.
      Use \Lemmaref{lem:refl-equiv-tp-fun}.
    \end{itemize}
  \item
    \begin{itemize}
      \DerivationProofCase{\SemiDeclChkValVar}
      {
        [\Omega]P \neq \exists, \land
        \\
        (x:Q) \in \Gamma
        \\
        \semideclsub[+]{\Theta}{Q}{[\Omega]P}{W}
      }
      {
        \semideclchkval{\Theta}{\Gamma}{x}{[\Omega]P}{W}
      }
      \begin{llproof}
        \semideclsubPf[+]{\Theta}{Q}{[\Omega]P}{W}{Subderivation}
        \rextendPf{\Theta}{\Delta}{\Omega}{Given}
        \judgetpPf{\Theta; \Delta}{P}{\Xi}{Given}
        \Pf{[\Delta]P}{=}{P}{Given}
        \judgectxPf{\Theta}{\Gamma}{Presupposed derivation}
        \Pf{(x:Q)}{\in}{\Gamma}{Subderivation}
        \judgetpPf{\Theta}{Q}{\dontcare}{By inversion on $\Gamma$ WF}
        \Pf{}{}{\ground{Q}}{By \Lemmaref{lem:decl-wf-ground}}
        \semideclnegPf{\Theta}{\Gamma}{W}{Given}
        \semideclentailwahPf{\Theta}{W}{By inversion}
        \algsubPf[+]{\Theta; \Delta}{Q}{P}{W'}{\Delta'}{By \Lemmaref{lem:aux-alg-sub-complete}}
        \Hand\rextendPf{\Theta}{\Delta'}{\Omega}{\ditto}
        \wahequivPf{\Theta}{W}{[\Omega]W'}{\ditto}
        \chiequivPf{\Theta}{\Gamma}{\cdot}{\cdot}{By \ChiEquivEmpty}
        \Hand\chiequivPf{\Theta}{\Gamma}{W}{[\Omega]W'}{By \ChiEquivWah}
        \Pf{[\Omega]P}{\neq}{\exists, \land}{Premise}
        \Pf{P}{\neq}{\exists, \land}{Straightforward}
        \Hand\algchkPf{\Theta; \Delta}{\Gamma}{x}{P}{W'}{\Delta'}{By \AlgChkValVar}
      \end{llproof}

      \DerivationProofCase{\SemiDeclChkValUnit}
      { }
      {
        \semideclchkval{\Theta}{\Gamma}{\unit}{\unitty}{\cdot}
      }
      \begin{llproof}
        \algchkPf{\Theta; \Delta}{\Gamma}{\unit}{\unitty}{\cdot}{\Delta}{By \AlgChkValUnit}
        \rextendPf{\Theta}{\Delta}{\Omega}{Given}
        \chiequivPf{\Theta}{\Gamma}{\cdot}{\cdot}{By \ChiEquivEmpty}
      \end{llproof}

      \DerivationProofCase{\SemiDeclChkValPair}
      {
        \semideclchkval{\Theta}{\Gamma}{v_1}{[\Omega]P_1}{\chi_1}
        \\
        \semideclchkval{\Theta}{\Gamma}{v_2}{[\Omega]P_2}{\chi_2}
      } 
      {
        \semideclchkval{\Theta}{\Gamma}{\pair{v_1}{v_2}}{[\Omega]P_1 \times [\Omega]P_2}{\chi_1, \chi_2}
      }
      \begin{llproof}
        \semideclchkvalPf{\Theta}{\Gamma}{v_1}{[\Omega]P_1}{\chi_1}{Subderivation}
        \semideclnegPf{\Theta}{\Gamma}{\chi_1, \chi_2}{Given}
        \semideclnegPf{\Theta}{\Gamma}{\chi_1}{By inversion}
        \semideclnegPf{\Theta}{\Gamma}{\chi_2}{\ditto}
        \algchkPf{\Theta; \Delta}{\Gamma}{v_1}{P_1}{\chi_1'}{\Delta''}{By \ih}
        \rextendPf{\Theta}{\Delta''}{\Omega}{\ditto}
        \chiequivPf{\Theta}{\Gamma}{\chi_1}{[\Omega]\chi_1'}{\ditto}
        \judgetpPf{\Theta; \Delta}{P_1 \times P_2}{\Xi}{Given}
        \judgetpPf{\Theta; \Delta}{P_2}{\dontcare}{By inversion}
        \extendPf{\Theta}{\Delta}{\Delta''}{By \Lemmaref{lem:typing-extends}}
        \judgetpPf{\Theta; \Delta''}{P_2}{\dontcare}{By \Lemmaref{lem:ext-weak-tp}}
        \judgetpPf{\Theta; \Delta''}{[\Delta'']P_2}{\dontcare}{By \Lemmaref{lem:right-hand-subst}}
        \Pf{[\Delta'']P_2}{=}{[\Delta'']([\Delta'']P_2)}{By \Lemmaref{lem:apply-idempotent}}
        \semideclchkvalPf{\Theta}{\Gamma}{v_2}{[\Omega]P_2}{\chi_2}{Subderivation}
        \semideclchkvalPf{\Theta}{\Gamma}{v_2}{[\Omega]([\Delta'']P_2)}{\chi_2'}{By \Lemmaref{lem:typing-sandwich}}
        \chiequivPf{\Theta}{\Gamma}{\chi_2}{\chi_2'}{\ditto}
        \algchkPf{\Theta; \Delta''}{\Gamma}{v_2}{[\Delta'']P_2}{\chi_2''}{\Delta'}{By \ih}
        \Hand\rextendPf{\Theta}{\Delta'}{\Omega}{\ditto}
        \chiequivPf{\Theta}{\Gamma}{\chi_2'}{[\Omega]\chi_2''}{\ditto}
        \Hand\algchkPf{\Theta; \Delta}{\Gamma}{\pair{v_1}{v_2}}{P_1 \times P_2}{[\Delta']\chi_1', \chi_2''}{\Delta'}{By \AlgChkValPair}
        \decolumnizePf
        \chiequivPf{\Theta}{\Gamma}{[\Omega]\chi_1'}{[\Omega]([\Delta]\chi_1')}{By \Lemmaref{lem:probs-sandwich}}
        \chiequivPf{\Theta}{\Gamma}{\chi_1}{[\Omega]([\Delta]\chi_1')}{By \Lemmaref{lem:chi-equiv-trans}}
        \chiequivPf{\Theta}{\Gamma}{\chi_2}{[\Omega]\chi_2''}{By \Lemmaref{lem:chi-equiv-trans}}
        \chiequivPf{\Theta}{\Gamma}{\chi_1, \chi_2}{[\Omega]([\Delta]\chi_1'), [\Omega]\chi_2''}{Append}
        \Hand\chiequivPf{\Theta}{\Gamma}{\chi_1, \chi_2}{[\Omega]([\Delta]\chi_1', \chi_2'')}{By \defn of \defsubst}
      \end{llproof}

      \ProofCaseRule{\SemiDeclChkValIn{k}}
      Straightforward.

      \DerivationProofCase{\SemiDeclChkValExists}
      {
        \semideclchkval{\Theta}{\Gamma}{v}{[t/a]([\Omega]P_0)}{\chi}
        \\
        \judgeterm{\Theta}{t}{\tau}
      }
      {
        \semideclchkval{\Theta}{\Gamma}{v}{(\extype{a:\tau} [\Omega]P_0)}{\chi}
      }
      \begin{llproof}
        \semideclchkvalPf{\Theta}{\Gamma}{v}{[t/a]([\Omega]P_0)}{\chi}{Subderivation}
        \semideclchkvalPf{\Theta}{\Gamma}{v}{[\Omega, \hypeq{\ahat}{\tau}{t}]([\ahat/a]P_0)}{\chi}{By properties of subst.}
        \judgetpPf{\Theta; \Delta}{\extype{a:\tau} P_0}{\Xi}{Given}
        \judgetpPf{\Theta, a:\tau; \Delta}{P_0}{\dontcare, a:\tau}{By inversion}
        \judgetpPf{\Theta; \Delta, \ahat:\tau}{[\ahat/a]P_0}{\dontcare, \ahat:\tau}{By \Lemmaref{lem:value-determined-evar-rename}}
        \decolumnizePf
        \Pf{[\Delta]P}{=}{P}{Given}
        \Pf{[\Delta, \ahat:\tau]P}{=}{P}{By \defn of \defsubst}
        \semideclnegPf{\Theta}{\Gamma}{\chi}{Given}
        \rextendPf{\Theta}{\Delta}{\Omega}{Given}
        \rextendPf{\Theta}{\Delta, \ahat:\tau}{\Omega, \hypeq{\ahat}{\tau}{t}}{By \RExtSolve}
        \algchkPf{\Theta; \Delta, \ahat:\tau}{\Gamma}{v}{[\ahat/a]P_0}{\chi'}{\Delta''}{By \ih}
        \rextendPf{\Theta}{\Delta''}{\Omega, \hypeq{\ahat}{\tau}{t}}{\ditto}
        \chiequivPf{\Theta}{\Gamma}{\chi}{[\Omega, \hypeq{\ahat}{\tau}{t}]\chi'}{\ditto}
        \semideclnegPf{\Theta}{\Gamma}{[\Omega, \hypeq{\ahat}{\tau}{t}]\chi'}{By \Lemref{lem:probs-equiv-respects-entail}}
        \Pf{\hypeq{\ahat}{\tau}{t'}}{\in}{\Delta''}{By \Lemmaref{lem:typing-solves-val-det}}
        \decolumnizePf
        \Pf{\Delta''}{=}{\Delta_0'', \hypeq{\ahat}{\tau}{t'}}{By inversion on \RExtSolved}
        \judgeentailPf{\Theta}{t'=t}{\ditto}
        \Hand\rextendPf{\Theta}{\Delta_0''}{\Omega}{\ditto}
        \algchkPf{\Theta; \Delta, \ahat:\tau}{\Gamma}{v}{[\ahat/a]P_0}{\chi'}{\Delta_0'', \hypeq{\ahat}{\tau}{t'}}{By equation}
        \Hand\algchkPf{\Theta; \Delta}{\Gamma}{v}{\extype{a:\tau}P_0}{\chi'}{\Delta_0''}{By \AlgChkValExists}
        \Pf{[\Delta_0'', \hypeq{\ahat}{\tau}{t'}]\chi'}{=}{\chi'}{By \Lemmaref{lem:output-applied}}
        \Pf{[\Omega, \hypeq{\ahat}{\tau}{t}]\chi'}{=}{[\Omega]\chi'}{Solution to $\ahat$ already applied}
        \Hand\chiequivPf{\Theta}{\Gamma}{\chi}{[\Omega]\chi'}{By equation}
      \end{llproof}

      \DerivationProofCase{\SemiDeclChkValWith}
      {
        \semideclchkval{\Theta}{\Gamma}{v}{[\Omega]P_0}{\chi_0}
      }
      {
        \semideclchkval{\Theta}{\Gamma}{v}{[\Omega]P_0 \land [\Omega]\phi}{[\Omega]\phi, \chi_0}
      }
      \begin{llproof}
        \semideclchkvalPf{\Theta}{\Gamma}{v}{[\Omega]P_0}{\chi_0}{Subderivation}
        \judgetpPf{\Theta; \Delta}{P_0 \land \phi}{\Xi}{Given}
        \judgetpPf{\Theta; \Delta}{P_0}{\Xi}{By inversion}
        \judgetermPf{\Theta; \Delta}{\phi}{\Booltype}{\ditto}
        \algchkPf{\Theta; \Delta}{\Gamma}{v}{P_0}{\chi_0'}{\Delta'}{By \ih}
        \rextendPf{\Theta}{\Delta'}{\Omega}{\ditto}
        \chiequivPf{\Theta}{\Gamma}{\chi_0}{[\Omega]\chi_0'}{\ditto}
        \extendPf{\Theta}{\Delta}{\Delta'}{By \Lemmaref{lem:typing-extends}}
      \end{llproof}
      \begin{itemize}
      \item \textbf{Case }$[\Delta']\phi = (\ahat = t)$ where $\judgeterm{\Theta}{t}{\tau}$:\\
        \begin{llproof}
          \Pf{\Delta'}{=}{\Delta_1', \ahat:\tau, \Delta_2'}{By inversion}
          \alginstPf{\Theta; \Delta'}{[\Delta']\phi}{\underbrace{\Delta_1', \hypeq{\ahat}{\tau}{t}, \Delta_2'}_{\Delta''}}{By \RuleInst}
          \Hand\algchkPf{\Theta; \Delta}{\Gamma}{v}{P_0 \land \phi}{[\Delta'']\phi, \chi_0'}{\Delta''}{By \AlgChkValWith}
          \semideclnegPf{\Theta}{\Gamma}{[\Omega]\phi, \chi_0}{Given}
          \judgeentailPf{\Theta}{[\Omega](\ahat = t)}{By inversion}
          \judgeentailPf{\Theta}{[\Omega]\ahat = t}{By \defn and $\because\ground{t}$}
          \decolumnizePf
          \judgeentailPf{\Theta}{t = [\Omega]\ahat}{By \Lemmaref{lem:equivassert}}
          \Hand\rextendPf{\Theta}{\Delta''}{\Omega}{By \Lemmaref{lem:relaxed-deep-entry}}
          \wahequivPf{\Theta}{[\Omega]\phi}{[\Omega]([\Delta'']\phi)}{By \Lemmaref{lem:wah-sandwich}}
          \chiequivPf{\Theta}{\Gamma}{[\Omega]\phi, \chi_0}{[\Omega]([\Delta'']\phi), [\Omega]\chi_0}{By \ChiEquivWah}
          \Hand\chiequivPf{\Theta}{\Gamma}{[\Omega]\phi, \chi_0}{[\Omega]([\Delta'']\phi, \chi_0)}{By \defn of \defsubst}
        \end{llproof} 
      \item \textbf{Case }$[\Delta']\phi \neq (\ahat = t)$ where $\judgeterm{\Theta}{t}{\tau}$:\\
        \begin{llproof}
          \alginstPf{\Theta; \Delta'}{[\Delta']\phi}{\Delta'}{By \RuleNoInst}
          \algchkPf{\Theta; \Delta}{\Gamma}{v}{P_0 \land \phi}{[\Delta']\phi, \chi_0'}{\Delta'}{By \AlgChkValWith}
          \wahequivPf{\Theta}{[\Omega]\phi}{[\Omega]([\Delta']\phi)}{By \Lemmaref{lem:wah-sandwich}}
          \chiequivPf{\Theta}{\Gamma}{[\Omega]\phi, \chi_0}{[\Omega]([\Delta']\phi), [\Omega]\chi_0}{By \ChiEquivWah}
          \Hand\chiequivPf{\Theta}{\Gamma}{[\Omega]\phi, \chi_0}{[\Omega]([\Delta']\phi, \chi_0)}{By \defn of \defsubst}
          \Hand\rextendPf{\Theta}{\Delta'}{\Omega}{Above}
        \end{llproof} 
      \end{itemize}

      \item \textbf{Case}
      \[
        \Infer{\SemiDeclChkValFix}
        {
          \judgeunroll{\cdot}{\Theta}{\nu:([\Omega]F)[\mu [\Omega]F]}{\alpha}{([\Omega]F)\;\Fold{[\Omega]F}{\alpha}\;\nu}{[\Omega]t}{Q}{\tau}
          \\ 
          \semideclchkval{\Theta}{\Gamma}{v_0}{Q}{\chi}
        }
        {
          \semideclchkval{\Theta}{\Gamma}{\roll{v_0}}{\underbrace{\comprehend{\nu:\mu ([\Omega]F)}{\Fold{[\Omega]F}{\alpha}\,{\nu} =_\tau [\Omega]t}}_{[\Omega]P}}{\chi}
        }
      \]
      (Note that we implicitly used \Lemmaref{lem:algebras-are-ground}.)\\
      \begin{llproof}
        \rextendPf{\Theta}{\Delta}{\Omega}{Given}
        \judgeunrollPf{\cdot}{\Theta}{\nu:([\Omega]F)[\mu [\Omega]F]}{\alpha}{([\Omega]F)\;\Fold{[\Omega]F}{\alpha}\;\nu}{[\Omega]t}{Q}{\tau}{Subderivation}
        \decolumnizePf
        \judgeunrollPf{\cdot}{\Theta; \Delta}{\nu:F[\mu F]}{\alpha}{F\;\Fold{F}{\alpha}\;\nu}{t}{Q'}{\tau}{By \Lemmaref{lem:uncomplete-alg-unroll}}
        \Pf{[\Omega]Q'}{=}{Q}{\ditto}
        \decolumnizePf
        \semideclchkvalPf{\Theta}{\Gamma}{v_0}{Q}{\chi}{Subderivation}
        \semideclchkvalPf{\Theta}{\Gamma}{v_0}{[\Omega]Q'}{\chi}{By equation}
        \judgetpPf{\Theta; \Delta}{Q'}{\dontcare}{By \Lemmaref{lem:alg-unroll-output-wf}}
        \Pf{[\Delta]P}{=}{P}{Given}
        \Pf{[\Delta]F}{=}{F}{By inversion}
        \Pf{[\Delta]t}{=}{t}{By inversion}
        \Pf{[\Delta]Q'}{=}{Q'}{By \Lemmaref{lem:unroll-applied}}
        \semideclnegPf{\Theta}{\Gamma}{\chi}{Given}
        \rextendPf{\Theta}{\Delta}{\Omega}{Given}
        \algchkPf{\Theta; \Delta}{\Gamma}{v_0}{Q'}{\chi'}{\Delta'}{By \ih}
        \Hand\rextendPf{\Theta}{\Delta'}{\Omega}{\ditto}
        \Hand\chiequivPf{\Theta}{\Gamma}{\chi}{[\Omega]\chi'}{\ditto}
        \decolumnizePf
        \Hand\algchkPf{\Theta; \Delta}{\Gamma}{\roll{v_0}}{\comprehend{\nu:\mu F}{\Fold{F}{\alpha}\,{\nu} =_\tau t}}{\chi'}{\Delta'}{By \AlgChkValFix}
      \end{llproof}

      \DerivationProofCase{\SemiDeclChkValDownshift}
      {
      }
      {
        \semideclchkval{\Theta}{\Gamma}{\thunk{e}}{\downshift{[\Omega]N}}{(e <= [\Omega]N)}
      }
      \begin{llproof}
        \Hand\algchkPf{\Theta; \Delta}{\Gamma}{\thunk{e}}{\downshift{N}}{(e <= N)}{\Delta}{By \AlgChkValDownshift}
        \judgetpPf{\Theta; \Delta}{\downshift{N}}{\dontcare}{Given}
        \judgetpPf{\Theta; \Delta}{N}{\dontcare}{By inversion}
        \Hand\rextendPf{\Theta}{\Delta}{\Omega}{Given}
        \judgetpPf{\Theta}{[\Omega]N}{\dontcare}{By \Lemmaref{lem:alg-to-decl-wf}}
        \judgeequivPf[-]{\Theta}{[\Omega]N}{[\Omega]N}{By \Lemmaref{lem:refl-equiv-tp-fun}}
        \chiequivPf{\Theta}{\Gamma}{\cdot}{\cdot}{By \ChiEquivEmpty}
        \Hand\chiequivPf{\Theta}{\Gamma}{(e <= [\Omega]N)}{(e <= [\Omega]N)}{By \ChiEquivChk}
      \end{llproof}
    \end{itemize}
  \item
    \begin{itemize}
      \DerivationProofCase{\SemiDeclChkExpUpshift}
      {
        \semideclchkval{\Theta}{\Gamma}{v}{P}{\chi}
        \\
        \semideclneg{\Theta}{\Gamma}{\chi}
      }
      {
        \semideclchkexp{\Theta}{\Gamma}{\Return{v}}{\upshift{P}}
      }
      \begin{llproof}
        \semideclchkvalPf{\Theta}{\Gamma}{v}{P}{\chi}{Subderivation}
        \Pf{[\cdot]P}{=}{P}{By \defn of \defsubst}
        \semideclchkvalPf{\Theta}{\Gamma}{v}{[\cdot]P}{\chi}{By equation}
        \rextendPf{\Theta}{\cdot}{\cdot}{By \RExtEmpty}
        \judgetpPf{\Theta}{\upshift{P}}{\dontcare}{Presupposed derivation}
        \judgetpPf{\Theta}{P}{\dontcare}{By inversion}
        \judgetpPf{\Theta; \cdot}{P}{\dontcare}{By \Lemmaref{lem:decl-to-ground-alg-wf}}
        \semideclnegPf{\Theta}{\Gamma}{\chi}{Subderivation}
        \algchkPf{\Theta; \cdot}{\Gamma}{v}{P}{\chi'}{\Delta'}{By \ih}
        \rextendPf{\Theta}{\Delta'}{\cdot}{\ditto}
        \chiequivPf{\Theta}{\Gamma}{\chi}{[\cdot]\chi'}{\ditto}
        \chiequivPf{\Theta}{\Gamma}{\chi}{\chi'}{By \defn of \defsubst}
        \Pf{\Delta'}{=}{\cdot}{By inversion}
        \algchkPf{\Theta; \cdot}{\Gamma}{v}{P}{\chi'}{\cdot}{By equation}
        \semideclnegPf{\Theta}{\Gamma}{\chi'}{By \Lemmaref{lem:probs-equiv-respects-entail}}
        \Pf{\size{\chi'}}{\leq}{\size{v}}{By \Lemmaref{lem:program-typing-shrinks-problems}}
        \Pf{}{<}{\size{v} + 1}{}
        \Pf{}{=}{\size{\Return{v}}}{By \defn of \defsize}
        \algnegPf{\Theta}{\Gamma}{\chi'}{By \ih}
        \algchknegPf{\Theta}{\Gamma}{\Return{v}}{\upshift{P}}{By \AlgChkExpUpshift}
      \end{llproof}

      \item \textbf{Case}\\
      \[
        \Infer{\SemiDeclChkExpLet}
        {
          \simple{\Theta}{N}
          \\
          \semideclsynexp{\Theta}{\Gamma}{\be}{\upshift{P}}
          \\
          \judgeextract[+]{\Theta}{P}{P'}{\Theta_P}
          \\
          \semideclchkexp{\Theta, \Theta_P}{\Gamma, x:P'}{e_0}{N}
        }
        {
          \semideclchkexp{\Theta}{\Gamma}{\Let{x}{\be}{e_0}}{N}
        }
      \]
      \begin{llproof}
        \simplePf{\Theta}{N}{Subderivation}
        \semideclsynexpPf{\Theta}{\Gamma}{\be}{\upshift{P}}{Subderivation}
        \algsynexpPf{\Theta}{\Gamma}{\be}{\upshift{Q}}{By \ih}
        \judgeequivPf[+]{\Theta}{P}{Q}{\ditto}
        \judgeextractPf[+]{\Theta}{P}{P'}{\Theta_P}{Subderivation}
        \judgeextractPf[+]{\Theta}{Q}{Q'}{\Theta_Q}{By \Lemmaref{lem:equiv-extract}}
        \judgeequivPf[+]{\Theta, \Theta_P}{P'}{Q'}{\ditto}
        \judgeequivPf[]{\Theta}{\Theta_P}{\Theta_Q}{\ditto}
        \judgeequivPf[+]{\Theta, \Theta_P}{Q'}{P'}{By \Lemmaref{lem:symmetric-equiv-tp-fun}}
        \semideclchkexpPf{\Theta, \Theta_P}{\Gamma, x:P'}{e_0}{N}{Subderivation}
        \judgechkexpPf{\Theta, \Theta_P}{\Gamma, x:P'}{e_0}{N}{By \Lemmaref{lem:semidecl-typing-sound}}
        \judgeequivPf[+]{\Theta}{\Gamma}{\Gamma}{By \Lemmaref{lem:refl-equiv-tp-fun}}
        \trailingjust{(repeated)}
        \judgeequivPf[+]{\Theta, \Theta_P}{\Gamma}{\Gamma}{By \Lemmaref{lem:ix-level-weakening}}
        \trailingjust{(repeated)}
        \judgeequivPf[+]{\Theta, \Theta_P}{\Gamma, x:Q'}{\Gamma, x:P'}{Add entry}
        \judgetpPf{\Theta}{N}{\dontcare}{Presupposed derivation}
        \judgetpPf{\Theta, \Theta_P}{N}{\dontcare}{By \Lemmaref{lem:ix-level-weakening}}
        \judgeequivPf[-]{\Theta, \Theta_P}{N}{N}{By \Lemmaref{lem:refl-equiv-tp-fun}}
        \decolumnizePf
        \judgechkexpPf{\Theta, \Theta_P}{\Gamma, x:Q'}{e_0}{N}{By \Lemmaref{lem:prog-typing-respects-equiv}}
        \semideclchkexpPf{\Theta, \Theta_P}{\Gamma, x:Q'}{e_0}{N}{By \Lemmaref{lem:semidecl-typing-complete}}
        \judgechkexpPf{\Theta, \Theta_P}{\Gamma, x:Q'}{e_0}{N}{By \Lemmaref{lem:semidecl-typing-sound}}
        \judgechkexpPf{\Theta, \Theta_Q}{\Gamma, x:Q'}{e_0}{N}{By \Lemmaref{lem:ctx-equiv-compat}}
        \semideclchkexpPf{\Theta, \Theta_Q}{\Gamma, x:Q'}{e_0}{N}{By \Lemmaref{lem:semidecl-typing-complete}}
        \algchknegPf{\Theta, \Theta_Q}{\Gamma, x:Q'}{e_0}{N}{By \ih (smaller program term)}
        \Hand\algchknegPf{\Theta}{\Gamma}{\Let{x}{\be}{e_0}}{N}{By \AlgChkExpLet}
      \end{llproof}

      \DerivationProofCase{\SemiDeclChkExpRec}
      {
        \arrayenv{
          \simple{\Theta}{N}
          \\
          \semideclsub[-]{\Theta}{\alltype{a:\kindnat} M}{N}{W}
          \\
          \semideclentailwah{\Theta}{W}
          \\
          \semideclchkexp{\Theta, a:\kindnat}{\Gamma, x:\downshift{\alltype{a':\kindnat} a' < a \implies [a'/a]M}}{e_0}{M}
        }
      }
      {
        \semideclchkexp{\Theta}{\Gamma}{\rec{x : (\alltype{a:\kindnat} M)}{e_0}}{N}
      }
      \begin{llproof}
        \simplePf{\Theta}{N}{Subderivation}
        \proofsep
        \semideclsubPf[-]{\Theta}{\alltype{a:\kindnat} M}{N}{W}{Subderivation}
        \semideclsubPf[-]{\Theta}{[\cdot](\alltype{a:\kindnat} M)}{N}{W}{By \defn of \defsubst}
        \judgetpPf{\Theta}{\alltype{a:\kindnat} M}{\dontcare}{Presupposed derivation}
        \judgetpPf{\Theta; \cdot}{\alltype{a:\kindnat} M}{\dontcare}{By \Lemmaref{lem:decl-to-ground-alg-wf}}
        \rextendPf{\Theta}{\cdot}{\cdot}{By \RExtEmpty}
        \eqPf{[\cdot](\alltype{a:\kindnat} M)}{\alltype{a:\kindnat} M}{By \defn of \defsubst}
        \groundPf{N}{By \Lemmaref{lem:decl-wf-ground}}
        \algsubPf[-]{\Theta; \cdot}{\alltype{a:\kindnat} M}{N}{W'}{\Delta'}{By \Lemmaref{lem:aux-alg-sub-complete}}
        \rextendPf{\Theta}{\Delta'}{\cdot}{\ditto}
        \wahequivPf{\Theta}{W}{[\cdot]W'}{\ditto}
        \eqPf{\Delta'}{\cdot}{By inversion}
        \algsubPf[-]{\Theta; \cdot}{\alltype{a:\kindnat} M}{N}{W'}{\cdot}{Rewrite above}
        \proofsep
        \semideclentailwahPf{\Theta}{W}{Subderivation}
        \wahequivPf{\Theta}{W}{W'}{By \defn of \defsubst}
        \semideclentailwahPf{\Theta}{W'}{By \Lemmaref{lem:equiv-respects-entail}}
        \entailwahPf{\Theta}{W'}{By \Lemmaref{lem:alg-entail-complete}}
        \decolumnizePf
        \semideclchkexpPf{\Theta, a:\kindnat}{\Gamma, x:\downshift{\alltype{a':\kindnat} a' < a \implies [a'/a]M}}{e_0}{M}{Subderivation}
        \algchknegPf{\Theta, a:\kindnat}{\Gamma, x:\downshift{\alltype{a':\kindnat} a' < a \implies [a'/a]M}}{e_0}{M}{By \ih}
        \decolumnizePf
        \Hand\algchknegPf{\Theta}{\Gamma}{\rec{x : (\alltype{a:\kindnat} M)}{e_0}}{N}{By \AlgChkExpRec}
      \end{llproof}

      \item The remaining cases for expression typing are straightforward.
    \end{itemize}
  \item Straightforward.
  \item 
    \begin{itemize}
      \DerivationProofCase{\SemiDeclSpineAll}
      {
        \judgeterm{\Theta}{t}{\tau}
        \\
        \semideclspine{\Theta}{\Gamma}{s}{[t/a]([\Omega]N_0)}{\upshift{P}}{\chi}
      }
      {
        \semideclspine{\Theta}{\Gamma}{s}{(\alltype{a:\tau}{[\Omega]N_0})}{\upshift{P}}{\chi}
      }
      \begin{llproof}
        \semideclspinePf{\Theta}{\Gamma}{s}{[t/a]([\Omega]N_0)}{\upshift{P}}{\chi}{Subderivation}
        \semideclspinePf{\Theta}{\Gamma}{s}{[\Omega, \hypeq{\ahat}{\tau}{t}]([\ahat/a]N_0)}{\upshift{P}}{\chi}{By props.\ of subst.}
        \judgetpPf{\Theta; \Delta}{\alltype{a:\tau}N_0}{\Xi}{Given}
        \judgetpPf{\Theta, a:\tau; \Delta}{N_0}{\dontcare, a:\tau}{By inversion}
        \judgetpPf{\Theta; \Delta, \ahat:\tau}{[\ahat/a]N_0}{\dontcare, \ahat:\tau}{By \Lemmaref{lem:value-determined-evar-rename}}
        \decolumnizePf
        \Pf{[\Delta]N}{=}{N}{Given}
        \Pf{[\Delta, \ahat:\tau]N}{=}{N}{By \defn of \defsubst}
        \semideclnegPf{\Theta}{\Gamma}{\chi}{Given}
        \rextendPf{\Theta}{\Delta}{\Omega}{Given}
        \rextendPf{\Theta}{\Delta, \ahat:\tau}{\Omega, \hypeq{\ahat}{\tau}{t}}{By \RExtSolve}
        \algspinePf{\Theta; \Delta, \ahat:\tau}{\Gamma}{s}{[\ahat/a]N_0}{\upshift{P'}}{\chi'}{\Delta''}{By \ih}
        \rextendPf{\Theta}{\Delta''}{\Omega, \hypeq{\ahat}{\tau}{t}}{\ditto}
        \chiequivPf{\Theta}{\Gamma}{\chi}{[\Omega, \hypeq{\ahat}{\tau}{t}]\chi'}{\ditto}
        \judgeequivPf[+]{\Theta}{P}{[\Omega, \hypeq{\ahat}{\tau}{t}]P'}{\ditto}
        \decolumnizePf
        \semideclnegPf{\Theta}{\Gamma}{[\Omega, \hypeq{\ahat}{\tau}{t}]\chi'}{By \Lemmaref{lem:probs-equiv-respects-entail}}
        \Pf{(\hypeq{\ahat}{\tau}{t'})}{\in}{\Delta''}{By \Lemmaref{lem:typing-solves-val-det}}
        \Pf{\Delta''}{=}{\Delta', \hypeq{\ahat}{\tau}{t'}}{By inversion on \RExtSolved}
        \judgeentailPf{\Theta}{t'=t}{\ditto}
        \Hand\rextendPf{\Theta}{\Delta'}{\Omega}{\ditto}
        \decolumnizePf
        \algspinePf{\Theta; \Delta, \ahat:\tau}{\Gamma}{s}{[\ahat/a]N_0}{\upshift{P'}}{\chi'}{\Delta', \hypeq{\ahat}{\tau}{t'}}{By equation}
        \Hand\algspinePf{\Theta; \Delta}{\Gamma}{s}{\alltype{a:\tau}N_0}{\upshift{P'}}{\chi'}{\Delta'}{By \AlgSpineAll}
        \Pf{[\Delta', \hypeq{\ahat}{\tau}{t'}]\chi'}{=}{\chi'}{By \Lemmaref{lem:output-applied}}
        \Pf{[\Delta', \hypeq{\ahat}{\tau}{t'}]P'}{=}{P'}{\ditto}
        \Pf{[\Omega, \hypeq{\ahat}{\tau}{t}]\chi'}{=}{[\Omega]\chi'}{Solution to $\ahat$ already applied}
        \Pf{[\Omega, \hypeq{\ahat}{\tau}{t}]P'}{=}{[\Omega]P'}{Solution to $\ahat$ already applied}
        \Hand\chiequivPf{\Theta}{\Gamma}{\chi}{[\Omega]\chi'}{By equation}
        \Hand\judgeequivPf[+]{\Theta}{P}{[\Omega]P'}{By equation}
      \end{llproof}

      \ProofCaseRule{\SemiDeclSpineImplies}
      Similar to \SemiDeclChkValWith case of part (3).

      \DerivationProofCase{\SemiDeclSpineApp}
      {
        \semideclchkval{\Theta}{\Gamma}{v}{[\Omega]Q}{\chi_1}
        \\
        \semideclspine{\Theta}{\Gamma}{s_0}{[\Omega]N_0}{\upshift{P}}{\chi_2}
      }
      {
        \semideclspine{\Theta}{\Gamma}{v,s_0}{[\Omega]Q \to [\Omega]N_0}{\upshift{P}}{\chi_1, \chi_2}
      }
      \begin{llproof}
        \semideclchkvalPf{\Theta}{\Gamma}{v}{[\Omega]Q}{\chi_1}{Subderivation}
        \judgetpPf{\Theta; \Delta}{Q \to N_0}{\Xi}{Given}
        \judgetpPf{\Theta; \Delta}{Q}{\dontcare}{By inversion}
        \judgetpPf{\Theta; \Delta}{N_0}{\dontcare}{\ditto}
        \Pf{[\Delta](Q \to N_0)}{=}{Q \to N_0}{Given}
        \Pf{[\Delta]Q}{=}{Q}{By inversion}
        \semideclnegPf{\Theta}{\Gamma}{\chi_1, \chi_2}{Given}
        \semideclnegPf{\Theta}{\Gamma}{\chi_1}{Straightforward}
        \semideclnegPf{\Theta}{\Gamma}{\chi_2}{\ditto}
        \rextendPf{\Theta}{\Delta}{\Omega}{Given}
        \proofsep
        \algchkPf{\Theta; \Delta}{\Gamma}{v}{Q}{\chi_1'}{\Delta''}{By \ih}
        \rextendPf{\Theta}{\Delta''}{\Omega}{\ditto}
        \chiequivPf{\Theta}{\Gamma}{\chi_1}{[\Omega]\chi_1'}{\ditto}
        \decolumnizePf
        \semideclspinePf{\Theta}{\Gamma}{s_0}{[\Omega]N_0}{\upshift{P}}{\chi_2}{Subderivation}
        \extendPf{\Theta}{\Delta}{\Delta''}{By \Lemmaref{lem:typing-extends}}
        \judgetpPf{\Theta; \Delta''}{N_0}{\dontcare}{By \Lemmaref{lem:ext-weak-tp}}
        \judgetpPf{\Theta; \Delta''}{[\Delta'']N_0}{\dontcare}{By \Lemmaref{lem:right-hand-subst}}
        \Pf{[\Delta'']([\Delta'']N_0)}{=}{[\Delta'']N_0}{By \Lemmaref{lem:apply-idempotent}}
        \algspinePf{\Theta; \Delta''}{\Gamma}{s_0}{[\Delta'']N_0}{\upshift{P'}}{\chi_2'}{\Delta'}{By \ih}
        \Hand\rextendPf{\Theta}{\Delta'}{\Omega}{\ditto}
        \chiequivPf{\Theta}{\Gamma}{\chi_2}{[\Omega]\chi_2'}{\ditto}
        \Hand\judgeequivPf[+]{\Theta}{P}{[\Omega]P'}{\ditto}
        \decolumnizePf
        \Hand\algspinePf{\Theta; \Delta}{\Gamma}{v, s_0}{Q \to N_0}{\upshift{P'}}{[\Delta']\chi_1', \chi_2'}{\Delta'}{By \AlgSpineApp}
        \decolumnizePf
        \chiequivPf{\Theta}{\Gamma}{[\Omega]\chi_1'}{[\Omega]([\Delta']\chi_1')}{By \Lemmaref{lem:probs-sandwich}}
        \chiequivPf{\Theta}{\Gamma}{\chi_1}{[\Omega]([\Delta']\chi_1')}{By \Lemmaref{lem:chi-equiv-trans}}
        \chiequivPf{\Theta}{\Gamma}{\chi_1, \chi_2}{[\Omega]([\Delta']\chi_1'), [\Omega]\chi_2'}{Append lists}
        \Hand\chiequivPf{\Theta}{\Gamma}{\chi_1, \chi_2}{[\Omega]([\Delta']\chi_1', \chi_2')}{By \defn of \defsubst}
      \end{llproof} 

      \ProofCaseRule{\SemiDeclSpineNil}
      Straightforward.
      Use \Lemmaref{lem:refl-equiv-tp-fun}
      and \ChiEquivEmpty and \Lemmaref{lem:wah-equiv-refl}.
      \qedhere
    \end{itemize}
  \end{enumerate}
\end{proof}

\begin{theorem}[Alg.\ Typing Complete]
  \label{thm:alg-typing-complete}
  ~
  \begin{enumerate}
    \item
      If $\judgesynhead{\Theta}{\Gamma}{h}{P}$,
      then $\algsynhead{\Theta}{\Gamma}{h}{P}$.
    \item
      If $\judgesynexp{\Theta}{\Gamma}{\be}{\upshift{P}}$,
      then there exists $P'$
      such that $\algsynexp{\Theta}{\Gamma}{\be}{\upshift{P'}}$
      and $\judgeequiv[+]{\Theta}{P}{P'}$.
    \item
      If $\judgechkval{\Theta}{\Gamma}{v}{[\Omega]P}$
      and $\judgetp{\Theta; \Delta}{P}{\Xi}$
      and $[\Delta]P = P$
      and $\rextend{\Theta}{\Delta}{\Omega}$,\\
      then there exist $\chi$ and $\Delta'$
      such that $\algchk{\Theta; \Delta}{\Gamma}{v}{P}{\chi}{\Delta'}$
      and $\rextend{\Theta}{\Delta'}{\Omega}$
      and $\algneg{\Theta}{\Gamma}{[\Omega]\chi}$.
    \item
      If $\judgechkexp{\Theta}{\Gamma}{e}{N}$,
      then $\algchkneg{\Theta}{\Gamma}{e}{N}$.
    \item
      If $\judgechkmatch{\Theta}{\Gamma}{P}{\clauses{\pa}{e}{i}{I}}{N}$,
      then $\algchkmatch{\Theta}{\Gamma}{P}{\clauses{\pa}{e}{i}{I}}{N}$.
    \item
      If $\judgespine{\Theta}{\Gamma}{s}{[\Omega]N}{\upshift{P}}$
      and $\judgetp{\Theta; \Delta}{N}{\Xi}$
      and $[\Delta]N = N$
      and $\rextend{\Theta}{\Delta}{\Omega}$,\\
      then there exist $P'$, $\chi$, and $\Delta'$
      such that $\algspine{\Theta; \Delta}{\Gamma}{s}{N}{\upshift{P'}}{\chi}{\Delta'}$
      and $\rextend{\Theta}{\Delta'}{\Omega}$
      and $\algneg{\Theta}{\Gamma}{[\Omega]\chi}$
      and $\judgeequiv[+]{\Theta}{P}{[\Omega]P'}$.
  \end{enumerate}
\end{theorem}
\begin{proof}
  ~
  \begin{enumerate}
  \item ~\\
    \begin{llproof}
      \judgesynheadPf{\Theta}{\Gamma}{h}{P}{Given}
      \semideclsynheadPf{\Theta}{\Gamma}{h}{P}{By \Lemmaref{lem:semidecl-typing-complete}}
      \algsynheadPf{\Theta}{\Gamma}{h}{P}{By \Lemmaref{lem:aux-alg-typing-complete}}
    \end{llproof} 
  \item ~\\
    \begin{llproof}
      \judgesynexpPf{\Theta}{\Gamma}{\be}{\upshift{P}}{Given}
      \semideclsynexpPf{\Theta}{\Gamma}{\be}{\upshift{P}}{By \Lemmaref{lem:semidecl-typing-complete}}
      \Hand\algsynexpPf{\Theta}{\Gamma}{\be}{\upshift{P'}}{By \Lemmaref{lem:aux-alg-typing-complete}}
      \Hand\judgeequivPf[+]{\Theta}{P}{P'}{\ditto}
    \end{llproof} 
  \item ~\\
    \begin{llproof}
      \judgechkvalPf{\Theta}{\Gamma}{v}{[\Omega]P}{Given}
      \semideclchkvalPf{\Theta}{\Gamma}{v}{[\Omega]P}{\chi}{By \Lemmaref{lem:semidecl-typing-complete}}
      \semideclnegPf{\Theta}{\Gamma}{\chi}{\ditto}
      \judgetpPf{\Theta; \Delta}{P}{\Xi}{Given}
      \Pf{[\Delta]P}{=}{P}{Given}
      \rextendPf{\Theta}{\Delta}{\Omega}{Given}
      \Hand\algchkPf{\Theta; \Delta}{\Gamma}{v}{P}{\chi'}{\Delta'}{By \Lemmaref{lem:aux-alg-typing-complete}}
      \Hand\rextendPf{\Theta}{\Delta'}{\Omega}{\ditto}
      \chiequivPf{\Theta}{\Gamma}{\chi}{[\Omega]\chi'}{\ditto}
      \semideclnegPf{\Theta}{\Gamma}{[\Omega]\chi'}{By \Lemmaref{lem:probs-equiv-respects-entail}}
      \Hand\algnegPf{\Theta}{\Gamma}{[\Omega]\chi'}{By \Lemmaref{lem:aux-alg-typing-complete}}
    \end{llproof} 
  \item Similar to part (1).
  \item Similar to part (1).
  \item ~\\
    \begin{llproof}
      \judgespinePf{\Theta}{\Gamma}{s}{[\Omega]N}{\upshift{P}}{Given}
      \judgetpPf{\Theta; \Delta}{N}{\Xi}{Given}
      \Pf{[\Delta]N}{=}{N}{Given}
      \rextendPf{\Theta}{\Delta}{\Omega}{Given}
      \semideclspinePf{\Theta}{\Gamma}{s}{[\Omega]N}{\upshift{P}}{\chi}{By \Lemmaref{lem:semidecl-typing-complete}}
      \semideclnegPf{\Theta}{\Gamma}{\chi}{\ditto}
      \Hand\algspinePf{\Theta; \Delta}{\Gamma}{s}{N}{\upshift{P'}}{\chi'}{\Delta'}{By \Lemmaref{lem:aux-alg-typing-complete}}
      \Hand\rextendPf{\Theta}{\Delta'}{\Omega}{\ditto}
      \chiequivPf{\Theta}{\Gamma}{\chi}{[\Omega]\chi'}{\ditto}
      \Hand\judgeequivPf[+]{\Theta}{P}{[\Omega]P}{\ditto}
      \semideclnegPf{\Theta}{\Gamma}{[\Omega]\chi'}{By \Lemmaref{lem:probs-equiv-respects-entail}}
      \Hand\algnegPf{\Theta}{\Gamma}{[\Omega]\chi'}{By \Lemmaref{lem:aux-alg-typing-complete}}
    \end{llproof} 
    \qedhere
 \end{enumerate}
\end{proof}

\end{document}